\documentclass[10pt,reqno]{amsart}
\usepackage{amsmath,amsthm,amssymb,mathrsfs,amsfonts,bm}
\usepackage{enumerate}
\usepackage{graphicx}
\usepackage{tikz-cd}
\usepackage{tikz}
\usetikzlibrary{decorations.markings}
\usetikzlibrary{decorations.pathreplacing}

\usepackage{color}
\usepackage{xcolor} % A package to add color.
\usepackage[letterpaper, margin=1in]{geometry}
\usepackage[titletoc,title]{appendix}

\usepackage{slashed}
\usepackage{cancel}
\usepackage{cite}
\usepackage[super]{nth}
\usepackage[linktocpage=true]{hyperref}
\usepackage{marginnote}
\usepackage{verbatim}

\usepackage[notocbasic]{nomencl}
\usepackage{xpatch}
\usepackage{etoolbox}
\makenomenclature

\renewcommand{\nompreamble}{We list the notations used frequently in this paper and point out their first appearance:}
\makeatletter
\xpatchcmd{\thenomenclature}{\section*}{\subsection}{}{}
\makeatother

\makeatletter
\renewcommand{\tocsection}[3]{%
	\indentlabel{\@ifnotempty{#2}{ % for numbered sections
			\ignorespaces\bfseries{#2. #3}}}
	\indentlabel{\@ifempty{#2}{\ignorespaces\bfseries{#3}}{}} % for unnumbered sections
	\vspace{1.5pt}}

\renewcommand{\tocsubsection}[3]{%
	\indentlabel{\@ifnotempty{#2}{
			\ignorespaces#2. #3}}
	\indentlabel{\@ifempty{#2}{\ignorespaces #3}{}}
	\vspace{1.5pt}}
\def\@dotsep{2}
\def\@tocline#1#2#3#4#5#6#7{\relax
	\ifnum #1>\c@tocdepth % then omit
	\else
	\par \addpenalty\@secpenalty\addvspace{#2}%
	\begingroup \hyphenpenalty\@M
	\@ifempty{#4}{%
		\@tempdima\csname r@tocindent\number#1\endcsname\relax
	}{%
		\@tempdima#4\relax
	}%
	\parindent\z@ \leftskip#3\relax \advance\leftskip\@tempdima\relax
	\rightskip\@pnumwidth plus4em \parfillskip-\@pnumwidth
	#5\leavevmode\hskip-\@tempdima
	\ifcase #1 \vskip 1em 
	\or\or \hskip 1em \or \hskip 2em \else \hskip 3em \fi%
	#6\nobreak\relax
	\dotfill\hbox to\@pnumwidth{\@tocpagenum{#7}}\par
	\nobreak
	\endgroup
	\fi}
\makeatother

\numberwithin{equation}{section}
\newtheorem{thm}{Theorem}[section]
\newtheorem{lem}[thm]{Lemma}
\newtheorem{prop}[thm]{Proposition}

\newtheorem{cor}[thm]{Corollary}

\theoremstyle{definition}
\newtheorem{defn}[thm]{Definition}

\theoremstyle{remark}
\newtheorem{rem}[thm]{Remark}
\newtheorem*{rem*}{Remark}
\newcommand{\norm}[1]{\Vert#1\Vert}
\newcommand{\abs}[1]{\vert#1\vert}
\newcommand{\jb}[1]{\langle #1\rangle}
\newcommand{\angles}[2]{\langle #1,#2\rangle}
\newcommand{\wt}[1]{\widetilde{#1}}
\newcommand{\BR}{\mathbb{R}}
\newcommand{\BS}{\mathbb{S}}
\newcommand{\BN}{\mathbb{N}}
\newcommand{\BZ}{\mathbb{Z}}
\newcommand{\BC}{\mathbb{C}}
\newcommand{\BQ}{\mathbf{Q}}
\newcommand{\Bm}{\mathbf{m}}
\newcommand{\Ba}{\mathbf{a}}
\newcommand{\BFS}{\mathbf{S}}
\newcommand{\BFV}{\mathbf{V}}
\newcommand{\bop}{\mathrm{b}}
\newcommand{\scop}{\mathrm{sc}}
\newcommand{\cop}{\mathrm{c}}

\newcommand{\Ric}{\mathrm{Ric}}
\newcommand{\Ran}{\mathrm{ran }}
\newcommand{\al}{\alpha}
\newcommand{\be}{\beta}
\newcommand{\m}{\mu}
\newcommand{\n}{\nu}
\newcommand{\ka}{\kappa}
\newcommand{\la}{\lambda}
\newcommand{\cs}{\Gamma}
\newcommand{\ann}{\mathrm{ann}}
\newcommand{\sgn}{\mathrm{sgn\,}}

\newcommand{\IM}{\mathrm{Im\,}}
\newcommand{\RE}{\mathrm{Re\,}}

\newcommand{\sH}{\dot{H}}

\newcommand{\sHb}{\dot{H}_{\bop}}
\newcommand{\eHb}{\bar{H}_{\bop}}
\newcommand{\mPsi}{\Psi_{\scop,\bop}}
\newcommand{\C}{\mathcal{C}}

\newcommand{\CX}{\bar{X}}
\newcommand{\CM}{\bar{M}}

\newcommand{\tr}{\mathrm{tr}}
\newcommand{\sg}{\slashed{g}}
\newcommand{\sd}{\slashed{d}}
\newcommand{\sdelta}{\slashed{\delta}}
\newcommand{\sstar}{\slashed{*}}
\newcommand{\str}{\slashed{\tr}}

\newcommand{\scs}{\slashed{\Gamma}}
\newcommand{\sL}{\slashed{\Delta}}

\newcommand{\sR}{\slashed{R}}

\newcommand{\rsH}{\slashed{H}}

\newcommand{\hX}{\mathring{X}}
\newcommand{\hg}{\mathring{g}}
\newcommand{\hpi}{\mathring{\pi}}
\newcommand{\hcs}{\mathring{\Gamma}}
\newcommand{\hBox}{\mathring{\Box}}
\newcommand{\hR}{\mathring{R}}
\newcommand{\hRic}{\mathring{\Ric}}
\newcommand{\htr}{\mathring{\tr}}
\newcommand{\hd}{\mathring{d}}
\newcommand{\hdelta}{\mathring{\delta}}
\newcommand{\hstar}{\mathring{*}}
\newcommand{\hvol}{\mathring{\mbox{vol}}}

\newcommand{\scform}{{}\widetilde{^{\scop}T^*}\bar{X}}
\newcommand{\cscform}{{}\overline{^{\scop}T^*}\!\bar{X}}
\newcommand{\fscform}{{}^{\scop}T^*\bar{X}}
\newcommand{\fiscform}{{}^{\scop}S^*\bar{X}}
\newcommand{\siscform}{{}\overline{^{\scop}T^*}_{\pa_+\CX}\bar{X}}
\newcommand{\fsiscform}{{}^{\scop}T^*_{\pa_+\CX}\bar{X}}
\newcommand{\abscform}{{}\overline{^{\scop}T^*}_{\pa_-\CX}\bar{X}}
\newcommand{\cbform}{{}\overline{^{\bop}T^*}\!\bar{X}}
\newcommand{\fbform}{{}^{\bop}T^*\bar{X}}
\newcommand{\fibform}{{}^{\bop}S^*\bar{X}}
\newcommand{\fisibform}{{}^{\bop}S^*_{\pa_+\CX}\bar{X}}
\newcommand{\sibform}{{}\overline{^{\bop}T^*}_{\pa_+\CX}\bar{X}}
\newcommand{\fsibform}{{}^{\bop}T^*_{\pa_+\CX}\bar{X}}
\newcommand{\abbform}{{}\overline{^{\bop}T^*}_{\pa_-\CX}\bar{X}}
\newcommand{\scsym}{S^2{}\widetilde{^{\scop}T^*}\bar{X}}
\newcommand{\Vb}{\mathcal{V}_\bop}
\newcommand{\Vsc}{\mathcal{V}_\scop}
\newcommand{\fscvec}{{}^{\scop}T\bar{X}}
\newcommand{\fbvec}{{}^{\bop}T\bar{X}}

\newcommand{\dg}{\dot{g}}
\newcommand{\dA}{\dot{A}}
\newcommand{\dF}{\dot{F}}
\newcommand{\sfS}{\mathsf{S}}
\newcommand{\sfV}{\mathsf{V}}
\newcommand{\bdelta}{\boldsymbol{\delta}}

\newcommand{\pa}{\partial}

\newcommand{\hnabla}{\mathring{\nabla}}
\newcommand{\snabla}{\slashed{\nabla}}

\newcommand{\weight}{\mu_{b_0}}
\newcommand{\ehRN}{r_{b_0}}
\newcommand{\ehKN}{r_{b}}

\newcommand{\Romanupper}[1]
{\MakeUppercase{\romannumeral #1}}

\begin{document}
	
	\title[Linear stability of Kerr-Newman family]{The linear stability of weakly charged and slowly rotating Kerr-Newman family of charged black holes}
	
	\author{Lili He}
	\address{Department of Mathematics \\ Johns Hopkins University \\ Baltimore, MD 21218, USA}
	\email{lhe31@jhu.edu}
	\thanks{}

	\begin{abstract}
In this paper, we prove the linear stability of weakly charged and slowly rotating Kerr-Newman black holes under coupled gravitational and electromagnetic perturbations. We show that the solutions to the linearized Einstein-Maxwell equations decay at an inverse polynomial rate to a linearized Kerr-Newman solution plus a pure gauge term. This work builds on the framework developed in \cite{HHV21} for the study of the Einstein vacuum equations. We work in the generalized wave map and Lorenz gauge. The proof involves the analysis of the resolvent of the Fourier transformed linearized Einstein-Maxwell operator on asymptotically flat spaces, which relies on recent advances in microlocal analysis and non-elliptic Fredholm theory developed in \cite{Vas13}. The most delicate part of the proof is the description of the resolvent at low frequencies.
	\end{abstract} 
	
	\maketitle

	\tableofcontents
		%\addtocontents{toc}{\protect\setcounter{tocdepth}{2}}

\section{Introduction}
The \emph{Einstein-Maxwell system}, which describes the interaction of gravity and electromagnetism, is given by
\begin{equation}\label{EqIntroEM}
	\Ric(g)_{\mu\nu}=2T_{\mu\nu}(g,F),\quad dF=0,\quad \delta_g F=0,
\end{equation}
where $g$ is a Lorentzian metric $g$ with signature $(-,+,+,+)$, $\Ric(g)$ is the Ricci curvature tensor and $T_{\mu\nu}(g,F):= F_{\mu\al}F_{\nu}^{\ \al}-\frac{1}{4}g_{\mu\nu}F_{\al\be}F^{\al\be}$ is the energy momentum tensor associated with an electromagnetic field represented by a $2$-form $F$. Therefore, the first equation in \eqref{EqIntroEM}, Einstein's field equation, describes the relation between the geometry of the spacetime $g$ and the energy momentum of an electromagnetic field $F$, and the last two equations, Maxwell's equations, describes how electromagnetic waves propagate in the spacetime $g$.

Among the most interesting solutions to Einstein-Maxwell equations are the families of black hole solutions. The Kerr-Newman (KN) family of solutions of \eqref{EqIntroEM} describes stationary, charged, and rotating black holes. It was discovered by \cite{DisofKN}, following the discovery of the Kerr \cite{Kerr63}. The family of KN solutions $(g_b,F_b)$ to the Einstein-Maxwell equations are characterized by the parameters
$b=(\Bm,\Ba,\BQ_e,\BQ_m) \in \BR^6$ where $\Bm>0, \Ba\in\BR^3, \BQ_e,\BQ_m$ are interpreted as the mass, angular momentum, electric charge and magnetic charge respectively. The KN spacetime reduces to the Reissner-Nordstr\"{o}m (RN) spacetime for $\Ba=0$, and RN spacetime reduces to Schwarzschild spacetime when $\BQ_e=\BQ_m=0$.

\subsection{Statement of the main result}
The \emph{stability problem} for black holes solutions concerns the long time behavior of solutions to the Einstein-Maxwell system with initial data close to known black hole solutions. The problem of stability of black hole solutions can be roughly divided into three formulations, each of increasing difficulty: mode stability of linearized equations, linear stability, and full nonlinear stability. The problem of mode stability of the linearized equations involves separating the solutions into modes $\psi(t,r,\theta,\phi)=e^{-i\omega t}e^{im\phi}R(r)S(\theta)$ and aiming at proving the lack of exponentially growing (in $t$) modes for all metric or curvature components. The proof of linear stability of Einstein-Maxwell system means proving boundedness and decay for the solutions of the linearized Einstein-Maxwell equations. Finally, the nonlinear stability problem for Einstein-Maxwell equations aims at showing that a small perturbation of a KN black hole converges to another member of the KN family.

In order to understand nonlinear stability, one must first understand linear stability. In this paper, we consider the linear stability problem of the linearized Einstein-Maxwell equations under coupled gravitational and electromagnetic perturbations for weakly charged and slowly rotating Kerr-Newman black holes, that is, the case $\abs{\Ba}+\abs{\BQ_e}+\abs{\BQ_m}\ll\Bm$.

To describe our result, we first recall the initial value problem formalism of Einstein-Maxwell equations (see \S\S\ref{SubsecBasicNL}--\ref{SubsecBasicLin} for detail). Given $5$-tuple $(\Sigma_0, h, k, \mathbf{E}, \mathbf{H})$ where $\Sigma_0$ is a smooth $3$-manifold, $h$ is a Riemannian metric on $\Sigma_0$, $k$ is a symmetric $2$-tensor on $\Sigma_0$, and $\mathbf{E}, \mathbf{H}$ are $1$-forms on $\Sigma_0$.
One seeks $(M, g, F)$ such that $(g,F)$ solve the Einstein-Maxwell system and $(h,k)$ are the metric and second fundamental form induced by $g$, and $(\mathbf{E},\mathbf{H})$ are the electric and magnetic field induced by $(g,F)$ on $\Sigma_0$. A necessary and sufficient condition for local solvability of Einstein-Maxwell equations (see \cite{CB52, CBG69} and \cite[\S6.10]{CB09}) is that $(h,k,\mathbf{E},\mathbf{H})$ satisfy the constraint equations, which consist of Gauss-Codazzi equation and the pull-back of the Maxwell equations to $\Sigma_0$. Linearizing the initial value problem gives rise to the corresponding initial value problem of linearized Einstein-Maxwell equations.

We now state a first version of our linear stability result. For a more formal statement, see Theorem \ref{ThmLinstaiblityKN}.

\begin{thm}
	\label{ThmmainIntro}
Let $b=(\Bm, \Ba,\BQ_e,\BQ_m)$ be the black hole parameters satisfying $\abs{\Ba}+\abs{\BQ_e}+\abs{\BQ_m}\ll\Bm$. Let $0<\alpha<1$. Let $(\dot{h}, \dot{k}, \dot{\mathbf{E}}, \dot{\mathbf{H}})$ be an initial data set on a $3$-manifold $\Sigma_0$ satisfying the linearized constraint equation, and having decay 
\[	\abs{\dot{h}}\lesssim r^{-1-\alpha},\quad  \abs{\dot{k}}, \abs{\dot{\mathbf{E}}},\abs{\dot{\mathbf{H}}}\lesssim r^{-2-\alpha}
\]
 and similar bounds after applying a few $r\pa_r, \pa_\omega$ derivatives. Let $(\dg, \dF)$ a solution to the linearized Einstein-Maxwell equations and attain the initial data $(\dot{h}, \dot{k}, \dot{\mathbf{E}}, \dot{\mathbf{H}})$. Then $(\dg, \dF)$ decays at an inverse polynomial rate in time $t_{b,*}$ to a linearized Kerr-Newman solution plus a pure gauge solution. That is, there exists linearized black hole parameters $\dot{b}=(\dot{\Bm}, \dot{\Ba}, \dot{\BQ}_e,\dot{\BQ}_m)\in\BR\times\BR^3\times\BR\times\BR$ and a vector field $V$ such that 
	\begin{align*}
		\dg= \dg_b(\dot{b})+\mathcal{L}_Vg_b+\tilde{g},\quad
		\dF=\dF_b(\dot{b})+\mathcal{L}_VF_b+\tilde{F}
	\end{align*}
	where 
	\[(\dg_b(\dot{b}),\dF_b(\dot{b})):=\frac{d}{ds}\Big\lvert_{s=0}(g_{b+s\dot{b}},\ F_{b+s\dot{b}})
	\]
	is the linearized Kerr-Newman solutions and the tail $(\tilde{g}, \tilde{F})$ satisfies the bound 
	\[
	\abs{\tilde{g}}+\abs{\tilde{F}}\lesssim t_{b,*}^{-\alpha-1+\epsilon},\quad \epsilon>0.
	\]
\end{thm}
\begin{rem}
	Using the rotation invariance of the Maxwell equation, one can always arrange for the initial data to have vanishing magnetic charge, i.e., $\BQ_m=0$, and thus the electromagnetic $2$-form can be written as $F=dA$ (see \S\ref{SubsecBasicDer} for detail). In this paper, we will restrict to the case where the magnetic charge is $0$ and study Einstein-Maxwell equations of the following form
	\[
	\Ric(g)_{\m\n}=2T_{\m\n}(g,dA),\quad \delta_gd A=0.
	\]
	We also refer the readers to Remark \ref{rem:KNMagSols} for the relation between the non-magnetically charged RN spacetime $(g_b, F_b=dA_b)$ and the magnetically charged RN spacetime $(g_{b,m}, F_{b,m})$.
\end{rem}
\begin{rem}
	The hypersurface $\{t_{b,*}=c\}$ terminates at null infinity. We also note that for bounded $r$, the tail $(\tilde{g},\tilde{F})$ satisfies the bound $\abs{\tilde{g}}+\abs{\tilde{F}}\lesssim \mathfrak{t}^{-1-\alpha+\epsilon}$ for all $\epsilon>0$ where $\mathfrak{t}=t$ for large $r$.	
\end{rem}
\begin{rem}
	 By imposing linearized generalized wave map gauge condition on $\dg$ and replacing $(\dot g_b(\dot b), \dot{F}_b(\dot{b}))$ by its gauge-fixed version, one can choose $V$ such that $V=V_1+V_2$ with $V_1$ lying in a $6$-dimensional space (only depending on $b$ ) of smooth vector fields (asymptotic to a linear combination of spatial translations and boosts) on $M$, and $V_2$ decaying (in $t_{b,*}$) at the rate $t_{b,*}^{-\alpha+\epsilon},\,\epsilon>0$.	
\end{rem}

For charged asymptotically flat black hole spacetimes (i.e. the cosmology constant $\Lambda=0$), the study of mode stability of Reissner-Nordstr\"{o}m black holes was initiated by Moncrief \cite{Mon74, Mon74_2, Mon75} by considering the metric and electromagnetic $4$-potential perturbations. Chandrasekhar \cite{Ch79} and Chandrasekhar-Xanthopoulos \cite{CX79} studied the perturbations of Reissner-Nordstr\"{o}m black holes in the Newman-Penrose formalism. We will work with the formalism of Kodama-Ishibashi \cite{KI04}. The linear stability of full subextremal Reissner-Nordstr\"{o}m spacetime was proved by Giorgi \cite{G20,G20_2}. As for Kerr-Newman spacetime, the mode stability of wave equations on the Kerr-Newman metric was established in \cite{Civin14}. The Teukolsky and Regge-Wheeler equations governing the linear stability of Kerr-Newman spacetime to coupled electromagnetic and gravitational perturbations were derived in \cite{G22}. The Carter tensor and the physical-space analysis in perturbations of Kerr-Newman spacetime were discussed in \cite{G21}. Dirac waves on Kerr-Newman spaceitmes were considered by Finster-Kamran-Smoller-Yau \cite{FKSY03}. We also mention the numerical works on the mode stability \cite{PBG13, DGS15} and nonlinear stability \cite{ZCHLS14} of Kerr-Newman spacetime. For positive cosmology case $\Lambda>0$, the nonlinear stability of slowly rotating Kerr-Newman-de Sitter black holes was proved by Hintz \cite{H18}.

\subsection{Related works}
Besides the above references for charged black holes, we also mention the closely related results on Einstein vacuum equations. The nonlinear stability of Minkowski spacetime was first proved by Christodoulou-Klainerman \cite{CK93} and later an alternative proof was given by Lindblad-Rodnianski \cite{LR03,LR05, LR10} using wave coordinates. For the case $\Lambda>0$, see Friedrich \cite{Friedrich86} for the nonlinear stability of de Sitter spacetime.

There is a large amount of literature on the scalar waves on Schwarzschild and Kerr spacetimes. The study of scalar waves on Schwarzschild metrics was initiated by Wald \cite{W79} and Kay-Wald \cite{KW87}, where the boundedness of the scalar waves was established. Blue-Soffer \cite{BS03} proved spacetime Morawetz type estimates and local decay for the solutions of semilinear wave equations on Schwarzschild manifold, with refinements and extensions due to Blue-Sterbenz \cite{BS06} and Blue-Soffer \cite{BS09}. Dafermos-Rodnianski \cite{DR09, DR13} proved the boundedness and decay of scalar waves on Schwarzschild background by quantifying the celebrated redshift effect. Marzuola-Metcalfe-Tataru-Tohaneanu \cite{MMTT10} established local energy estimates, as well as Strichartz estimates, for the scalar waves on Schwarzschild spacetimes. We also mention the work by Lindblad-Tohaneanu \cite{LT18} on the scalar wave equations on perturbations of Schwarzschild metrics, as well as the work by Holzegel-Kauffman \cite{HK20} on the wave equations on Schwarzschild spacetimes with small non-decaying first order terms. The boundedness and decay of solutions of the scalar wave equations on Kerr spacetimes was studied in Dafermos-Rodnianski \cite{DR11, DR10un}, Tataru-Tohaneanu \cite{TT11} and Andersson-Blue \cite{AB15} for $\Ba\ll\Bm$, as well as in Dafermos-Rodnianski-Shlapentokh-Rothman \cite{DRSR16} for the full subextremal range $\abs{\Ba}<\Bm$. We also mention the work Lindblad-Tohaneanu \cite{LT20} on the scalar wave equations on metrics close to slowly rotating Kerr spacetimes. 

As for non scalar fields on Schwarzschild and Kerr background, we refer to the work by Blue \cite{B08}, Sterbenz-Tataru \cite{ST15} and 	Andersson-B\"{a}chdahl \cite{ABB16} on Maxwell’s equation on Schwarzschild spacetimes. Maxwell’s equation on Kerr spacetimes was discussed in Andersson-Blue \cite{AB12MK} and Andersson-B\"{a}ckdahl-Blue \cite{ABB17}. Dirac waves on Kerr spacetimes were considered by Finster-Kamran-Smoller-Yau \cite{FKSY06}. 

The study of mode stability of Schwarzschild black holes was initiated by Reege-Wheeler \cite{RW57}, followed by the work of Zerilli \cite{Zerilli70} and Vishveshwara \cite{Vish70} in metric perturbations. Bardeen-Press \cite{BP73} discussed the perturbations of Schwarzschild black holes using Newman-Penrose formalism. We also mention the work by Kodama-Ishibashi \cite{KI03}  on the mode stability of of Schwarzschild black holes. Dafermos-Holzegel-Rodnianski \cite{DHR19} proved the linear stability of the Schwarzschild metric in double null gauge. Later, the linear stability of Schwarzschild metric in (generalized) harmonic gauge was studied by Hung \cite{ Hung18, Hung19}, Hung-Keller-Wang \cite{HKW20} and Johnson \cite{Johnson19}. For progress in the nonlinear stability, Klainerman-Szeftel \cite{KS20} proved the nonlinear stability of the Schwarzschild metric under axially symmetric and polarized perturbations. Recently, Dafermos-Holzegel-Rodnianski-Taylor \cite{DHRT21} proved the nonlinear stability of Schwarzschild spacetimes under a restrictive symmetry class, which excludes rotating Kerr solutions as final state of the evolution. 

Teukolsky \cite{Teu73} derived the equations governing the perturbations of Kerr black holes in the Newman-Penrose formalism. Whiting \cite{W89} established the mode stability of the Teukolsky equation for Kerr spacetimes, followed by refinements and extensions by Shlapentokh-Rothman \cite{SR15} and Andersson-Ma-Paganini-Whiting \cite{AMPW17}. Dafermos-Holzegel-Rodnianski \cite{DHR19_2} proved boundedness and decay for the Teukolsky equation on Kerr spacetimes with $\abs{\Ba}\ll\Bm$. Andersson-B\"{a}ckdahl-Blue-Ma \cite{ABBM19} proved the linear stability of slowly rotating Kerr metric for initial data with strong decay using Newman-Penrose formalism. H\"{a}fner-Hintz-Vasy \cite{HHV21} proved the linear stability of slowly rotating Kerr spacetimes using wave map gauge.  Recently, the nonlinear stability of slowly rotating Kerr spacetimes was proved by a series of work of Klainerman-Szeftel and Giorgi-Klainerman-Szeftel \cite{KS19_1, KS19_2, GKS20, KS21, GKS22}. For large $\Ba$ case, Shlapentokh-Rothman-Costa \cite{DCSR20} established the boundedness and decay for the Teukolsky equation on Kerr spacetimes. Andersson-H\"{a}fner-Whiting \cite{AHW22} gave the mode analysis for the linearized equations on subextremal Kerr metric.

In the case of the Einstein equations with a positive cosmological constant, decay rate (to constants) of the solutions of the scalar wave equations on Schwarzschild-de Sitter background was discussed by Bony-H\"{a}fner \cite{BH08}, Dafermos-Rodnianski \cite{DR07SdS} and Melrose-S\`{a} Barreto-Vasy \cite{MSV14}. Dyatlov \cite{D11_1, D11, D12} proved that the solutions of the scalar wave equations on Kerr-de Sitter spacetimes decay exponentially to constants. We also mention Mavrogiannis's works \cite{Mav21, Mav21_2, Mav22} on quasilinear wave equations on cosmological black hole background. The global nonlinear stability of slowly rotating Kerr-de Sitter spacetimes was proved by Hintz-Vasy \cite{HV20}, followed by the work of Fang \cite{Fang21, Fang22}.

\subsection{Main ideas of the proof}
The proof of Theorem~\ref{ThmmainIntro} is based on the framework developed in \cite{HHV21} where the Einstein vacuum equations were studied. The presence of electromagnetic term in the right hand side of Einstein equations adds new difficulties to the analysis of the problem.

\begin{enumerate}
	\item As in the study of Einstein vacuum equations in \cite{HHV21}, we can exploit the diffeomorphism invariance of Einstein-Maxwell equations and impose a generalized wave map gauge. In addition to this, we also need to choose a gauge (generalized Lorenz gauge) for the electromagnetic 4-potential $A$ in order to express the Einstein-Maxwell equations \eqref{EqIntroEM} as a system of nonlinear hyperbolic equations, which is defined as the gauge-fixed Einstein-Maxwell equations. Correspondingly, the linearization of the gauge-fixed Einstein-Maxwell system gives rise to a system of linear hyperbolic equations.
	\item We need to establish the generalized mode stability for the \emph{ungauged} linearized Einstein-Maxwell system \eqref{EqBasicLinEinsteinMaxwell} around a RN solution: all generalized mode solutions of the \emph{ungauged} linearized Einstein-Maxwell system \eqref{EqBasicLinEinsteinMaxwell} are sums of linearized KN solutions and pure gauge solutions. In this work, we give a self-contained proof in full detail of this generalized mode stability, which completes and extends the proof by Kodama and Ishibashi \cite{KI04}.
	\item The fact that the KN metric is not Ricci-flat creates the major difficulty in the mode analysis of the wave type operator $\delta_{g_b} G_{g_b}\delta^*_{g_b}$ (where $(\delta_{g_b}^*u)_{\al\be}=\frac12(\nabla_\al u_\be+\nabla_\be u_\al)$ is the symmetric gradient operator and $(\delta_{g_b}h)_\be=-\nabla^\al h_{\al\be}$ is the negative divergence operator, and $G_{g_b}=\mbox{Id}-\frac{1}{2}g_b\tr_{g_b}$ is the trace reversal operator) acting on $1$-forms. We note that the operator $\delta_{g_b} G_{g_b}\delta^*_{g_b}$ (or its modifications) occurs as both the gauge propagation operator and the gauge potentials operator for the Einstein equations. To deal with this issue, we restrict to the weakly charged case, where we are able to exploit a perturbation argument to extend the invertibility of relevant operators on Schwarzschild metric to weakly charged Reissner-Nordstr\"{o}m spacetime. 
	
	Concretely, we first employ the generalized wave map gauge 
	\begin{equation}
		\wt{\Upsilon}^E(g;g_b):=	\Upsilon^E(g;g_b)-\theta(g;g_b)=0,\quad \Upsilon^E(g;g_b)=g_{\mu\nu}g^{\kappa\lambda}\left(\Gamma(g)^\mu_{\kappa\lambda}-\Gamma(g_b)^\mu_{\kappa\lambda}\right)
	\end{equation}
	where $\theta(g;g_b)$ is linear in $g$ and $\theta(g_b;g_b)=0$. By introducing this type of gauge condition with a suitable choice of $\theta$, the gauge potential operator is given by $\tilde{\delta}_{g_b}G_{g_b}\delta_{g_b}^*$ where $\tilde{\delta}_{g_b}$ is a zero order modification of $\delta_g$. In contrast to $\delta_{g_b}G_{g_b}\delta_{g_b}^*$ which is the gauge potential operator corresponding to the standard wave map gauge $\Upsilon^E(g;g_b)=0$, the operator $\tilde{\delta}_{g_b}G_{g_b}\delta_{g_b}^*$ is invertible (on suitable function spaces) in the Schwarzschild case $b=(\Bm, 0,0)$,  and thus a perturbation argument enables us to extend this invertibility to the weakly charged and slowly rotating KN metrics with sufficiently small charge. Moreover, the generalized wave map gauge allows us to exclude a pure gauge solution which has no geometric meaning.
	
	Next, we implement the constraint damping, which was first discussed in \cite{GCHM05} and played a central role both in numerical work \cite{Pre05} and the stability proofs \cite{HV18, H18, HV20, HHV21}. 	The reason we introduce the constraint damping, i.e., use $\tilde{\delta}_{g_b}^*$ in the construction of linearized gauge-fixed Einstein-Maxwell equations, is that we need to use the invertibility of the corresponding gauge propagation operator $\delta_{g_b}G_{g_b}\tilde{\delta}^*_{g_b}$ to exclude the generalized modes, which grows linearly in $t_{b,*}$ and whose leading term is given by linearized (in $(\dot{\Bm}, 0,\dot{\BQ})$) KN solutions, to the linearized gauge-fixed Einstein-Maxwell equations $L_b(\dg,\dA)=0$. We also note that we use a perturbation argument to %extend \cite[Proposition 10.3 and 10.12]{HHV21} to the 
	prove the invertibility of the gauge propagation operator $\delta_{g_b}G_{g_b}\tilde{\delta}^*_{g_b}$ on the weakly charged and slowly rotating KN metrics $g_b$. 
	\end{enumerate}

\subsubsection{Gauge fixing}
\label{subsubsec:gaugefixing}
In this paper, we introduce the generalized wave map gauge and generalized Lorenz gauge.
	
	Concretely, one fixes a background metric $g^0$, and requires that the identity map $(M,g)\to(M,g^0)$ be a wave map. This is equivalent to
	\[
	\Upsilon^E(g;g^0)=g_{\mu\nu}g^{\kappa\lambda}\left(\Gamma(g)^\mu_{\kappa\lambda}-\Gamma(g^0)^\mu_{\kappa\lambda}\right)=0.
	\]
We define the \emph{generalized wave map gauge} as follows
\[\wt{\Upsilon}^E(g;g^0):=	\Upsilon^E(g;g^0)-\theta(g;g^0)=0
\]
where $\theta(g;g^0)$ is linear in $g$, contains no derivatives of $g$ and $\theta(g^0, g^0)=0$. Then we consider the gauge-fixed Einstein equations
	\[
			P^E(g,A):=\Ric(g)-2T(g, dA)-\wt{\delta}_g^*\wt{\Upsilon}(g;g^0)=0.	
		\]
	At the linearized level, we take $g_b$, around which we linearize the equations, as the background metric, and thus obtain the linearized gauge-fixed Einstein equations
	\[
			L_b^E(\dg,\dA):=D_{g_b}P^E(\dg,\dA)=-\frac12\Box_{g_b}\dg+\mbox{l.o.t}=0
			\] 
			which is a system of linear wave equations.
	
As for the Maxwell equations, we introduce the generalized Lorenz gauge. We fix a background metric $g^0$, and then a (modified) Lorenz gauge reads
	\[
	\Upsilon^M(g,A;g^0)=\tr_g\delta_{g^0}^* A=0.
	\]
	We fix a background $1$-form $A^0$ and define the \emph{generalized Lorenz gauge} as \[\wt{\Upsilon}^M(g,A;g^0,A^0):=	\Upsilon^M(g,A;g^0)-\Upsilon^M(g,A^0;g^0),
	\]
	and then obtain the gauge-fixed Maxwell equations
		\[
			P^M(g,A):=\delta_gdA-d\wt{\Upsilon}^M(g,A;g^0,A^0)=0.
		\]
	Again, at the linearized level, we take $(g_b, A_b)$, around which we linearize the equations, as the background metric and $1$-form, and thus have
		\[		
			L_b^M(\dg,\dA):=D_{g_b}P^M(\dg,\dA)=-\Box_{g_b}\dA+\mbox{l.o.t}=0
			\]
			which is a system of linear wave equations.

	We see that a solution of the ungauged equations satisfying the gauge conditions gives rise to a solution of the gauge-fixed equations. As for the other direction, suppose that $(g, A)$ solves the gauge-fixed Einstein-Maxwell system
	 \[
		\left(P^E(g, A), P^M(g, A)\right)=0,
		\]
		and satisfies 
		\[(\wt{\Upsilon}^E(g; g_b), \wt{\Upsilon}^M(g,A; g_b,A_b)=0\quad \mbox{at } \Sigma_0
		\]
		and
		\[
		(\mathcal{L}_{\pa_\mathfrak{t}}\wt{\Upsilon}^E(g; g_b), \mathcal{L}_{\pa_{\mathfrak{t}}}\wt{\Upsilon}^M(g,A; g_b,A_b)=0\quad \mbox{at } \Sigma_0
		\] 
		where $\Sigma_0=\{\mathfrak{t}=0\}$ is the spacelike Cauchy hypersurface. Applying $\delta_g$ to \[P^M(g,A)=\delta_gdA-d\wt{\Upsilon}^M(g,A;g_b,A_b)=0
		\]
		 gives 
		\[
		-\delta d\wt{\Upsilon}^M(g,A;g_b,A_b)=\Box_{g}\wt{\Upsilon}^M(g,A;g_b,A_b)=0.
		\]
		and thus according to the vanishing initial data of $\wt{\Upsilon}^M(g,A;g_b,A_b)$, we obtain that $\wt{\Upsilon}^M(g,A;g_b,A_b)=0$.
		
		Applying $\delta_gG_g$ to $P^E(g,A)=	\Ric(g)-2T(g, dA)-\wt{\delta}_g^*\wt{\Upsilon}(g;g_0)=0$ and using the second Bianchi identity and the fact $\delta_g T=0$ yields
		\[
		-\delta_g G_g\wt{\delta}_g^*\wt{\Upsilon}^E(g;g_b)=0
		\]
		which is a wave-type equation on the $1$-form $\wt{\Upsilon}^E(g;g_b)$ and thus $\wt{\Upsilon}^E(g;g_b)=0$. We conclude that $(g, A)$ satisfies the gauge conditions, and thus solves the Einstein-Maxwell system. Correspondingly, this direction holds on the linearized level as well.

\subsubsection{Fredholm framework setup}
According to the above discussion, it suffices to consider the equation $L_b(\dg,\dA):=(2L^E_{b}, L^M_b)(\dg,\dA)=0$. A simple linear theory presented in \cite{HV20} reduces the initial value problem for $L_b (\dg,\dA)=0$ to the following inhomogeneous problem.

Concretely, let $b=(\Bm,\Ba,\BQ)$ be close to $b_0=(\Bm_0,0,\BQ_0)$ with $\abs{\BQ_0}\ll\Bm_0$. Fix a function $t_{b,*}$ which equals $t+r_{(\Bm_0,0,\BQ_0),*}$ near event horizon and $t-r_{(\Bm,0,\BQ),*}$ near null infinity $\mathscr{I}^+$, where $r_{b,*}$ is the tortoise coordinate. Then we consider the following inhomogeneous equation
\[
L_b(\dg, \dA)=(2L_b^E(\dg,\dA), L_b^M(\dg, \dA))=f
\]
where $f$ has compact support in $t_*$ and $f$ decays in $r$ at the rate $r^{-2-\alpha}$. Our strategy is to take Fourier transform of the inhomogeneous equation $L_b(\dg,\dA)=f$ in $t_{b,*}$. We define the Fourier transform of $(\dg, \dA)$ as 
\begin{equation}
	(\hat{\dg}(\sigma), \hat{\dA}(\sigma)):=\int_{\BR}e^{it_{b,*}\sigma}(\dg, \dA)(t_{b,*})\,dt_{b,*}
\end{equation}
and likewise for $f$.

According to a crude energy estimate in Proposition \ref{prop:energyestimate}, we have the following integral representation 
\begin{equation}\label{eq:integralrepsolnIntro}
	(\dg(t_{b,*}), \ \dA(t_{b,*}))=
	\frac{1}{2\pi}\int_{\IM\sigma=M+1}\!e^{-i\sigma t_{b,*}}\widehat{L_{b}}(\sigma)^{-1}\hat{f}(\sigma)\,d\sigma,\quad M\gg 1.
\end{equation}
We expect to shift contour of the integration  to $\IM\sigma=c$ for any $c>0$. To this end, we need to understand the location of the poles of $\widehat{L_{b}}(\sigma)^{-1}$ and expect invertibility of $\widehat{L_{b}}(\sigma)^{-1}$ when $\abs{\RE\sigma}\gg1$. Concretely, we need to establish the uniform Fredholm estimates (down to $\sigma=0$) and high energy estimates. The uniform Fredholm estimates allow us to perform the perturbation argument (which is used a lot in this paper) while the high energy estimates imply the invertibility of $\widehat{L_{b}}(\sigma)^{-1}$ when $\abs{\RE\sigma}\gg1$. We point out that the whole section \S\ref{sec:microandsemisetup} is devoted to using microlocal tools to obtain the aforementioned uniform Fredholm estimates and high energy estimates. In brief, the proof of uniform Fredholm estimates uses the non-elliptic Fredholm framework of \cite{Vas13}, radial point estimates at the event horizon \cite{Vas13}, scattering radial point estimates at infinity (\cite{M92, Vas21a} for non-zero $\sigma$ and \cite{Vas21b} for $\sigma$ near zero), propagation of singularities estimates \cite{DH72}, and (for $\sigma=0$) elliptic b-theory \cite{Mel93}. As for the proof of high energy estimates, in addition to the aforementioned estimates at a semiclassical version, it uses estimates at normally hyperbolic trapping \cite{WZ11,D16,HV16} and high energy estimates at infinity \cite{VZ00,Vas21a}.

%%%%%%%%%%%%%%%%%%%%%%%%%%%%%%%%%%%%%%%%%%%%%%%%%%%%%%%%%%%%%%%%%%%%%%%%%%%%%%%%%%%%
\subsubsection{Mode stability of the gauge-fixed Einstein-Maxwell operator for $C^{-1}\leq \abs{\sigma}\leq C, \IM\sigma\geq 0$ with $	C>0$}
In view of the high energy estimates, we are left with the analysis of $\widehat{L_{b}}(\sigma)$ for $\sigma$ in a bounded region in the closed upper half plane. For the analysis in the region $C^{-1}\leq \abs{\sigma}\leq C, \IM\sigma\geq 0$ with $	C>0$, we prove that $\widehat{L_{b_0}}(\sigma)$, with $b_0=(\Bm_0,0,\BQ_0)$ and $\BQ_0\ll\Bm_0$  being a Reissner-Nordstr\"{o}m parameter, is invertible and then use a perturbation argument (in this compact (in $\sigma$) region) to extend the invertibility to $\widehat{L_{b}}(\sigma)$ for nearby $b$.

For the proof of invertibility of $\widehat{L_{b_0}}(\sigma)$, we make use of the geometric mode stability result of Reissner-Nordstr\"{o}m black holes and the invertibility of the gauge progation operator and gauge potential operator. Concretely, suppose that $\widehat{L_{b_0}}(\underline{\dg},\underline{\dA})=0$. Then $(\dg,\dA)=e^{-it_{b_0,*}\sigma}(\underline{\dg},\underline{\dA})$ is a mode solution of  $L_{b_0}(\dg, \dA)=0$.
According to the discussion in \S\ref{subsubsec:gaugefixing}, we see that the solution satisfies 
\[
\delta_{g_{b_0}}d\Big(D_{(g_{b_0}, A_{b_0})}\widetilde{\Upsilon}^M(\dg,\dA;g_{b_0},A_{b_0})\Big)=0.
\]
According to the mode stability of $\Box_{g_{b_0},0}$, we see that $D_{(g_{b_0}, A_{b_0})}\widetilde{\Upsilon}^M(\dg,\dA;g_{b_0},A_{b_0})=0$. Therefore, $(\dg,\dA)$ solves the Maxwell equations. By the discussion in \S\ref{subsubsec:gaugefixing} again, we obtain that 
\[
\delta_{g_{b_0}}G_{g_{b_0}}\tilde{\delta}^*_{g_{b_0}}\Big(D_{g_{b_0}}\widetilde{\Upsilon}^E(\dg;g_{b_0})\Big)=0.
\]
Using the mode stability of the gauge propagation operator (this is where we need the \emph{smallness} of the charge), we see that $D_{g_{b_0}}\widetilde{\Upsilon}^E(\dg;g_{b_0})=0$. Therefore, $(\dg,\dA)$ is a mode solution of the \emph{ungauged} linearized Einstein-Maxwell system. Using the geometric mode stability of Reissner-Nordstr\"{o}m solutions, we conclude that $(\dg,\dA)$ is a pure gauge mode solution, that is, \[
(\dot{g}, \dot{A})=\left(2\delta_g^*\omega, ~\mathcal{L}_{\omega^\sharp}A+d\phi\right)
\]
where $\omega$ is a $1$-form and $\phi$ is a scalar function, both of which are modes. Plugging them back into the linearized gauge condition, and with our choice of the generalized gauge condition, we have
\[
\tilde{\delta}_{g_{b_0}}G_{g_{b_0}}\delta^*_{g_{b_0}}\omega=0,\quad \delta_{g_{b_0}}d\Big(\mathcal{L}_{\omega^\sharp}A+d\phi\Big)=0.
\]
First by the mode stability of the gauge potential operator (this is again where we need the \emph{smallness} of the charge) and next  the mode stability of $\Box_{g_{b_0},0}$, we see that $(\omega, \phi)=0$. This proves the injectivity of $\widehat{L_{b_0}}(\sigma)$. Since $\widehat{L_{b_0}}(\sigma)$ has index $0$, it follows that $\widehat{L_{b_0}}(\sigma)$ is invertible.

%%%%%%%%%%%%%%%%%%%%%%%%%%%%%%%%%%%%%%%%%%%%%%%%%%%%%%%%%%%%%%%%%%%%%%%%%%%%%%%%%
\subsubsection{ The analysis of $\widehat{L_{b}}(\sigma)^{-1}$ near $\sigma=0$}
According to Theorem \ref{thm:modestabilityofmlgEM}, the space $\mathcal{K}_{b}$ of zero energy modes of $L_{b}$ is $8$-dimensional; it is the sum of a $5$-dimensional space of gauge-fixed version of linearized Kerr-Newman solutions, and a 3-dimensional space of pure gauge solutions $(2\delta_{g_b}^*\omega_b, \mathcal{L}_{\omega_{b}^{\sharp}}A_b+d\phi_b)$ with $\omega_b$ asymptotic to a spatial translation. According to Proposition \ref{prop:gezeromodess1}, the space $\widehat{\mathcal{K}}_{b}$ of generalized zero energy modes, which have $o(1)$ decay as $r\to\infty$ for fixed $t_{b,*}$ but allow for polynomial growth in $t_{b,*}$), of $L_{b}$ is the sum of $\mathcal{K}_{b}$ and a 3-dimensional space of pure gauge solutions $(2\delta_{g_b}^*\hat{\omega}_b, \mathcal{L}_{\hat{\omega}_{b}^{\sharp}}A_b+d\hat{\phi}_b)$ with $\hat{\omega}_b$ asymptotic to a Lorentz boost and the coefficient of $t_{b,*}$ being a stationary pure gauge solution $(2\delta_{g_b}^*\omega_b, \mathcal{L}_{\omega_{b}^{\sharp}}A_b+d\phi_b)$.

Therefore, $\widehat{L_{b}}(\sigma)$ is more singular when acting on the $3$-dimensional space of stationary pure gauge solutions because of the existence of the generalized zero energy solutions of $L_{b}(\dg,\dA)=0$ with linear growth in $t_{b,*}$ and leading order terms given as a stationary pure gauge solution. Motivated by this fact, we write $\widehat{L_{b}}(\sigma)$ (actually after multiplied by an operator which is invertible near $\sigma=0$) as a $3\times 3$ block matrix. 

The basic mechanism underlying the analysis of $\widehat{L_{b}}(\sigma)^{-1}$ near $\sigma=0$, is the formal expansion of the resolvent near $\sigma=0$
\begin{align*}
	\widehat{L_b}(\sigma)^{-1}f &= \widehat{L_b}(0)^{-1}f + (\widehat{L_b}(\sigma)^{-1}-\widehat{L_b}(0)^{-1})f \\
	&= u_0 + \sigma \widehat{L_b}(\sigma)^{-1}f_1,
\end{align*}
where we use the formal resolvent identity $\widehat{L_b}(\sigma)^{-1}-\widehat{L_b}(0)^{-1}=-\widehat{L_b}(\sigma)^{-1}(\widehat{L_b}(\sigma)-\widehat{L_b}(0))\widehat{L_{b}}(0)^{-1}$ and thus obtain
\[
u_0 = \widehat{L_b}(0)^{-1}f,\quad
f_1 = -\sigma^{-1}\bigl(\widehat{L_b}(\sigma)-\widehat{L_b}(0)\bigr)u_0.
\]
The first term $u_0$ is $\sigma$-independent. As for the second term $\sigma \widehat{L_b}(\sigma)^{-1}f_1$, we gain a factor of $\sigma$ (i.e. more regularity in $\sigma$ and thus more decay in $t_{b,*}$ after taking inverse Fourier transform). However, since $\widehat{L_b}(0)^{-1}$ loses two orders of decay in $r$ while $\sigma^{-1}(\widehat{L_b}(\sigma)-\widehat{L_b}(0))=2i\rho(\rho\pa_{\rho}-1)+\rho^2C^\infty$ with $\rho=1/r$ only gains back one order decay in $r$, it follows that $f_1$ has one order of decay less than $f$. %That is, \emph{the $\sigma$-gain comes at the cost of losing one order of decay} in the argument of $L(\sigma)^{-1}$.
We expect to do the above expansion iteratively 
\[
	\widehat{L_b}(\sigma)^{-1}f = u_0 + \sigma u_1+\cdots+\sigma^ku_k
\]
where
\[
	u_k = \widehat{L_b}(0)^{-1}f_k,\quad
	f_{k+1}=-\sigma^{-1}\bigl(\widehat{L_b}(\sigma)-\widehat{L_b}(0)\bigr)u_k
\]
as often as possible. However, this iteration requires that $u_k=\widehat{L_b}(0)^{-1}f_k$ has a certain asymptotic behavior or decay rate as $r\to \infty$ such that $\sigma^{-1}(\widehat{L_b}(\sigma)-\widehat{L_b}(0))u_k$ lies in the domain of $\widehat{L_b}(\sigma)^{-1}$. That is, we have to stop the iteration once $u_k=\widehat{L_b}(0)^{-1}f_k$ does not satisfy this requirement and thus we obtain an error term $\sigma^k L(\sigma)^{-1}f_k$ in the expansion.

Specifically, a detailed analysis in \S\ref{subsec:structureof rel} implies that $\widehat{L_b}(\sigma)^{-1}$ (up to a multiplication of an operator invertible near $\sigma=0$) is equal to
\[
\begin{pmatrix}
	L_{00}(b,\sigma)& \sigma^2\wt{L}_{01}(b,\sigma)&\sigma\wt{L}_{02}(b,\sigma)\\
	\sigma\wt{L}_{10}(b,\sigma)& \sigma^2\wt{L}_{11}(b,\sigma)&\sigma^2\wt{L}_{12}(b,\sigma)\\
	\sigma\wt{L}_{20}(b,\sigma)& \sigma^2\wt{L}_{21}(b,\sigma)&\sigma\wt{L}_{22}(b,\sigma)\\
\end{pmatrix}
\]
where $L_{00}(b,\sigma),\ \tilde{L}_{11}(b,\sigma), \ \tilde{L}_{22}(b,\sigma)$ are invertible near $\sigma=0$. Therefore, one obtain
\[
\widehat{L_{b}}(\sigma)^{-1}=\sigma^{-2}P_2(b)+\sigma^{-1}P_1(b)+R^1(b,\sigma)+R^2(b,\sigma),\quad R^{1}(b,\sigma)\sim\abs{\sigma}^{\alpha-1},\quad R^{2}(b,\sigma)\sim\abs{\sigma}^{\alpha}
\]
where $P_1(b), P_2(b)$ are independent of $\sigma$ and of finite rank with explicit range: linearized KN solutions and pure gauge solutions. 
%The presence of this more singular term $R^1(\sigma)$ leads to less decay of the tail of the solution. However, 
The range of $R^1(b,\sigma)$ consists of pure gauge solutions and this allow us to rewrite the inverse Fourier transform of $R^1(b,\sigma)f$ as a pure gauge solution plus a remainder term decaying faster in time $t_{b,*}$ (in fact at the rate $t_{b,*}^{-1-\alpha}$).
%%%%%%%%%%%%%%%%%%%%%%%%%%%%%%%%%%%%%%%%%%%%%%%%%%%%%%%%%%%%%%%%%%%%%%%%%%%%%%%%%%%%
\subsection{Outline}
The paper is organized as follows.
\begin{itemize}
	\item  In \S \ref{sec:bscattering}, we introduce the b- and scattering geometric structures on manifolds with boundaries or corners (in this paper, we work on the compactification of the spatial slice $\Sigma$ of the spacetime $M$), and corresponding function spaces.
	\item In \S \ref{sec:IVP}, we discuss the initial value problems formalism for Einstein-Maxwell equations. We also introduce our choice of gauge, generalized wave map gauge and generalized Lorenz gauge, and then derive the gauge-fixed Einstein-Maxwell system. We include the discussion at both nonlinear and linearized levels.
	\item In \S \ref{sec:KNblackholes}, we realize the family of slowly rotating KN metric as a smooth family of stationary metrics on a fixed manifold $M$. We also introduce several vector bundles over $M$ and discuss how to construct the gauged Cauchy data for the gauge-fixed Einstein-Maxwell system  from any given initial data $(h, k,\mathbf{E},\mathbf{H})$ satisfying the constraint equations.
	\item In \S \ref{sec:microandsemisetup}, we set up the general Fredholm framework, based on tools from microlocal analysis. Concretely, we prove the uniform Fredholm estimates and high energy estimates for the relevant wave type operators (acting on both scalar functions and tensor bundles). We also establish a crude energy estimate which gives an exponentially growing bound on the energy (over a spacelike hypersurface terminating at null infinity) of solutions to the wave equations on slowly rotating Kerr-Newman metric.
	\item In \S\ref{sec:basiccalculation}, we introduce the spherical harmonic decompositions, which will be used in the proof of the geometric mode stability of Reissner-Nordstr\"{o}m spacetime in \S\ref{sec:gmodestability}.
	\item In \S \ref{sec:gmodestability}, we prove the version of the geometric mode stability of Reissner-Nordstr\"{o}m spacetime used in the subsequent sections.
	\item In \S \ref{sec:modeanalysisofscalar}, we discuss the modes of the scalar wave operator on RN and slowly rotating KN spacetimes.
	\item In \S\ref{sec:modeanalysisof1form}, we analyze the modes on various function spaces (which correspond to different decay rate at spatial infinity) of the gauge propagation operator $\mathcal{P}_{b,\gamma}=\delta_{g_b} G_{g_b}\wt{\delta}^*_{g_b,\gamma}$ and the gauge potential wave operator $\mathcal{W}_{b,\gamma}=\wt{\delta}_{g_b,\gamma} G_{g_b}\delta_{g_b}^*$, both of which are wave-type operators acting on $1$-form. From this section on, we are restricted in the small charge case.
	\item In \S\ref{sec:modeanalsysiofGFEM}, we prove that the linearized gauge-fixed Einstein-Maxwell system $L_{g_{b_0},\gamma}$ (linearized around  weakly charged RN metrics and $4$-electromagnetic potentials $(g_{b_0},A_{b_0})$) has no non-zero modes in the closed upper half plane. Moreover, we describe the space of zero modes and generalized zero modes (with linearly growth in $t_{b,*}$) of $L_{b,\gamma}$ for both RN case and slowly rotating KN case with small charge.
	\item In \S\ref{sec:structureofmlEM}, we prove the mode stability of the linearized gauge-fixed Einstein-Maxwell operator $L_{b,\gamma}(\sigma)$ for $b=(\Bm,\Ba,\BQ)$ near $b_0=(\Bm_0,0,\BQ_0),\abs{\BQ_0}\ll\Bm_0$ and $\IM\sigma\geq 0, \sigma\neq 0$ (i.e. the invertibility of the operator $\widehat{L_{b, \gamma}}(\sigma)$ on suitable function spaces for $\IM\sigma\geq 0, \sigma\neq 0$), and also provide a description of the structure of the resolvent $\widehat{L_{b, \gamma}}(\sigma)^{-1}$ near $\sigma=0$.
	\item In \S \ref{sec:regularityofmlEM}, we study the higher regularity of the resolvent $(\widehat{L_{b, \gamma}}(\sigma))^{-1}$ of the linearized gauge-fixed Einstein-Maxwell operator.
	\item In \S \ref{sec:decayestimates}, we put all the previous sections together to study the asymptotic behavior of the solutions $(\dg, \dA)$ to the inhomogeneous linearized gauge-fixed Einstein-Maxwell equations $L_{b,\gamma}(\dg,\dA)=f$.
	\item In \S\ref{sec:mainthm}, we reduce the initial value problem for Einstein-Maxwell equations to the inhomogeneous problem studied in the previous section.
	\item In Appendix \S\ref{sec:appenA}, we provide a detailed calculation of the linearized Einstein-Maxwell system around RN black holes, which is needed in the proof of the geometric mode stability of Reissner-Nordstr\"{o}m spacetime in \S\ref{sec:gmodestability}. In Appendix \ref{app:trap}, we present the calculation of the subprincipal operator of the wave operator acting on tensor bundles at trapping. In Appendix \ref{app:Rad}, we include the computation of the subprincipal operator of the wave operator acting on tensor bundles at radial points at event horizon, which is needed in the radial point estimates at event horizons.
\end{itemize}

\nomenclature[01]{$\mathcal{A}^\ell$}{weighted $L^\infty$ conormal function space, see \eqref{eq:conormalfunctionspace}}
\nomenclature[02]{$A_{b_0}$}{Reissner-Nordstr\"{o}m electromagnetic $4$-potentials with parameters $b_0$, see \eqref{EqRNpotential}}
\nomenclature[03]{$A_{b}$}{Kerr-Newman electromagnetic $4$-potentials with parameters $b$, see \eqref{EqKNMetricEmbed}}
\nomenclature[04]{$b_0$}{fixed Reissner-Nordstr\"{o}m parameters, see \eqref{EqRNParams}}
\nomenclature[05]{$b$}{Kerr-Newman parameters, see \eqref{EqKNparameter}}
\nomenclature[06]{$\delta_g$}{negative divergence, $(\delta_g u)_{\mu_1\ldots\mu_N}=-\nabla^\lambda u_{\lambda\mu_1\ldots\mu_N}$}
\nomenclature[07]{$\delta_g^*$}{symmetric gradient, $(\delta_g^*u)_{\mu\nu}=\frac{1}{2}(u_{\mu;\nu}+u_{\nu;\mu})$}
\nomenclature[08]{$\tilde{\delta}_{g,\gamma}$}{modification of the negative divergence, see \eqref{eq:modifieddeltaFirst}}
\nomenclature[09]{$\tilde{\delta}_{g,\gamma}^*$}{modification of the symmetric gradient, see \eqref{eq:modifieddeltaFirst}}
\nomenclature[10]{$g_{b_0}$}{Reissner-Nordstr\"{o}m metrics with parameter $b_0$, see \eqref{EqKNMetricEmbed}}
\nomenclature[11]{$g_b$}{Kerr-Newman metrics with parameters $b$, see \eqref{EqKNMetricEmbed}}
\nomenclature[12]{$\gamma_0$}{map assigning to a function on $M$ its Cauchy data at $\Sigma_0$, see \eqref{EqIniMapgaugeed}}
\nomenclature[13]{$\eHb^{s,\ell}$}{weighted $\bop$-Sobolev space of extendible distributions, see \eqref{EqBExtSupp}}
\nomenclature[14]{$\sHb^{s,\ell}$}{weighted $\bop$-Sobolev space of supported distributions, see \eqref{EqBExtSupp}}
\nomenclature[15]{$\bar{H}_{\bop,h}^{s,\ell}$}{semiclassical weighted $\bop$-Sobolev space of extendible distributions, see \eqref{Eqseminorm}}
\nomenclature[16]{$\dot{H}_{\bop,h}^{s,\ell}$}{semiclassical weighted $\bop$-Sobolev space of supported distributions, see \eqref{Eqseminorm}}
\nomenclature[17]{$i_b$}{map constructing Cauchy data from geometric initial data, see Proposition \ref{PropKNIni}}
\nomenclature[18]{$\mathcal{K}_b$}{the space of zero energy modes of $L_{b,\gamma}$, see \eqref{eq:mlgEMzeromode}}
\nomenclature[19]{$\widehat{\mathcal{K}}_b$}{the space of generalized zero energy modes of $L_{b,\gamma}$, see Proposition \ref{prop:gezeromodess1}}
\nomenclature[20]{$\mathcal{K}_b^*$}{the space of dual zero energy modes of $L^*_{b,\gamma}$, see \eqref{eq:mlgEMdualzeromode}}
\nomenclature[21]{$\widehat{\mathcal{K}}^*_b$}{the space of dual generalized zero energy modes of $L^*_{b,\gamma}$, see Proposition \ref{prop:gezeromodess1}}
\nomenclature[22]{$\mathcal{L}_V$}{Lie derivative along the vector field $V$}
\nomenclature[23]{$\widetilde{\mathcal{L}}_TV$}{equal to $\mathcal{L}_VT$, see Definition \ref{DefBasicLinLieDerivative}}
\nomenclature[24]{$L_{b,\gamma}$}{linearized gauge-fixed Einstein-Maxwell equations $(2L_{b,\gamma}^E, L_{b,\gamma}^M)$}
\nomenclature[25]{$L^E_{b,\gamma}$}{linearized gauge-fixed Einstein equations, see \eqref{EqBasicLingAEqgamma}}
\nomenclature[26]{$L^M_{b,\gamma}$}{linearized gauge-fixed Maxwell equations, see \eqref{EqBasicLingAEqgamma}}
\nomenclature[27]{$\mathcal{M}$}{the static region of a fixed Reissner-Nordstr\"{o}m spacetime, see \eqref{EqRNStaticRegion}}
\nomenclature[28]{$M$}{extended manifold where $g_b$ is defined, see \eqref{EqRNExt}}
\nomenclature[29]{$\mathcal{P}_{b,\gamma}$}{gauge propagation operator for the generalized wave map gauge $1$-form $\widetilde{\Upsilon}^E$, see \eqref{EqdetailedPW}}
\nomenclature[30]{$\Sigma_0$}{a spacelike Cauchy hypersurface of $M$, see \eqref{EqKNSigma0Ex}}
\nomenclature[31]{$t$}{static time coordinate, see \eqref{EqRNMetric}, or Boyer-Lindquist coordinate, see \eqref{EqKNMetric}}
\nomenclature[32]{$t_0$}{incoming Eddington-Finkelstein coordinate on a fixed Reissner-Nordstr\"{o}m manifold, see \eqref{EqRNNullCoord}}
\nomenclature[33]{$t_{\chi_0}$}{a time function interpolating $t_0$ near event horizon and $t$ near $r=\infty$, see \eqref{EqRNTimefcnChi}}
\nomenclature[34]{$t_{b,*}$}{a time function which is smooth across event horizon and transverses to null infinity near $r=\infty$, see \eqref{EqKNTimeFn}}
\nomenclature[35]{$t_{*}$}{equal to $t_{b_0,*}$, see \eqref{EqRNTimefcn}}
\nomenclature[36]{$\mathfrak{t}$}{a timelike function with respect to Kerr-Newman metric $g_b$, which is equal to $t$ near $r=\infty$, see \eqref{Eqtimeliket}}
\nomenclature[37]{$\theta$}{gauge source function for the generalized wave map gauge, see \eqref{Eqgeneralizedwavegauge}}
\nomenclature[38]{$\mathcal{U}_{b_0}$}{parameter space for slowly rotating Kerr-Newman black holes, see Lemma \ref{lem:smoothdependenceonb}}
\nomenclature[39]{$\mathcal{U}_{b_0,m}$}{parameter space for slowly rotating Kerr-Newman black holes allowing for magnetic charges, see \eqref{EqKNMagParameter}}
\nomenclature[40]{$\Upsilon^E$}{wave map gauge $1$-form, see \eqref{EqBasicNLgaugepo}}
\nomenclature[41]{$\Upsilon^M$}{modified Lorenz gauge function, see \eqref{EqBasicNLgaugefcn}}
\nomenclature[42]{$\widetilde{\Upsilon}^E$}{generalized wave map gauge $1$-form, see \eqref{Eqgeneralizedwavegauge}}
\nomenclature[43]{$\widetilde{\Upsilon}^M$}{generalized Lorenz gauge function, see \eqref{EqgeneralizedLorengauge}}
\nomenclature[44]{$\mathcal{W}_{b,\gamma}$}{gauge potential operator for Einstein equations, see \eqref{EqdetailedPW}}
\nomenclature[45]{$\mathcal{X}$}{the spatial slice of the static region $\mathcal{M}$, see \eqref{EqRNStaticRegion}}
\nomenclature[46]{$X$}{the spatial slice of the extended manifold $M$, see \eqref{EqRNExt}}
\nomenclature[47]{$\CX$}{compactification of $X$ at $r=\infty$, see \eqref{EqradialcomofX}}
\nomenclature[48]{$\pa_-\CX$}{an artificial boundary of $\CX$, see \eqref{EqradialcomofX}}
\nomenclature[49]{$\pa_+\CX$}{the boundary at spatial infinity $r=\infty$ of $\CX$, see \eqref{EqradialcomofX}}
\nomenclature[50]{$\widehat{\bullet}(\sigma)$}{Fourier transform $e^{it_{b,*}\sigma}\bullet e^{-it_{b,*}\sigma}$ of $\bullet$ with respect to the time function $t_{b,*}$; in this paper, $\bullet=\Box_{g_b,0}, \mathcal{P}_{b,\gamma}, \mathcal{W}_{b,\gamma}, L_{b,\gamma}$}

\printnomenclature

%%%%%%%%%%%%%%%%%%%%%%%%%%%%%%%%%%%%%%%%%%%%%%%%%%%%%%%%%%%%%%%%%%%%%%%%%%%%%%%%%
\subsection*{Acknowledgements}
The author would like to thank her advisor, Hans Lindblad, for his constant support and encouragement. The author is also grateful to Peter Hintz and Andr\'{a}s Vasy for helpful discussions.

\section{b- and scattering geometry}
\label{sec:bscattering}

In this section, we will recall the b- and scattering geometry and analysis on manifolds with boundaries or corners. Here, we follow the discussion in \cite[\S2]{HHV21}. In \S\ref{subsec:bscbndles}, we recall the notions of b- and scattering vector bundles. We refer the readers to \cite[\S 2]{Mel93} and \cite[\S 2]{Mel95} for a more detailed exposition. In \S\ref{subsec:radialcom}, we recall the notion of radial compactification. In \S\ref{subsec:functionspaces}, we define the relevant Sobolev, conormal and polyhomogeneous functions defined on manifolds with boundaries or corners.

Let $\CX$ be a general compact manifold with boundary $\pa \CX$, and let $\rho\in C^\infty(X)$ be a defining function for the boundary $\pa\CX$, that is, $\rho\geq 0, \pa \CX=\rho^{-1}(0)$ and $d\rho\neq 0$ on $\pa \CX$.

\subsection{b- and scattering vector bundles}\label{subsec:bscbndles}
The space of \emph{b-vector fields} $\mathcal{V}_{\bop}(\CX)$ and \emph{scattering vector fields} $\mathcal{V}_{\scop}(\CX)$ are defined as
\begin{equation}
	\label{EqbandscV}
	\mathcal{V}_\bop(\CX) = \{ V\in\mathcal{V}(\CX)\mid V\ \text{is tangent to the boundary}\ \pa\CX \}, \quad
	\mathcal{V}_\scop(\CX) = \rho\mathcal{V}_\bop(\CX).
\end{equation}
Let $\rho\in C^\infty(X)$ be a boundary defining function and let $y_1, \cdots, y_{n-1}$ be additional coordinates near the boundary. Then  $\mathcal{V}_\bop(\CX)$ is spanned, over $C^\infty(\CX)$, by $\rho\pa_{\rho}$ and $\pa_{y_i}$, while $\mathcal{V}_\scop(\CX)$ is spanned, over $C^\infty(\CX)$, by $\rho^2\pa_{\rho}$ and $\rho\pa_{y_i}$.

Correspondingly, we have the natural b-tangent bundle $\fbvec$ and scattering tangent bundle $\fscvec$ over $\CX$ such that 
\[\mathcal{V}_{\bop}(\CX)=C^\infty(\CX;\fbvec),\quad \mathcal{V}_\scop(\CX)=C^\infty(\CX;\fscvec).
\]
Restriction to the interior $X$ gives rise to the smooth bundle maps $\fbvec\to T X$ and $\fscvec\to T X$, which are isomorphisms over the interior $X$ and vanish at the boundary $\pa\CX$. The space $\mbox{Diff}_\bop^k(\CX)$, resp. $\mbox{Diff}_\scop^k(X)$ of b-, resp. scattering differential operators of degree $k$ consists of finite sums of up to $k$-fold products of b-, resp. scattering vector fields.

The dual bundles is denoted by $\fbform$, resp. $\fscform$ and called \emph{b-cotangent bundle}, resp. \emph{scattering cotangent bundle}. Now $\{d\rho/\rho, dy_i\}$, resp. $\{d\rho/\rho^2, dy_i/\rho\}$ gives a local basis of $\fbform$, resp. $\fscform$.
 A \emph{scattering metric} is a section $g\in C^\infty(\CX;S^2\,\fscform)$, which defines a non-degenerate fiber metric on ${}^{\scop}T\bar{X}$, while a \emph{b-metric} is a section $g\in C^\infty(\CX;S^2\,\fbform)$, which defines a non-degenerate fiber metric on ${}^{\bop}T\bar{X}$

\subsection{Radial compactification}\label{subsec:radialcom}
These aforementioned b- and scattering structures arise naturally on compactifications of non-compact manifolds (see \cite[\S 1]{Mel95}). Now we carry out the \emph{radial compactification} of the Euclidean space $\BR^n$.
\begin{defn}\label{defn:radialcom}
The \emph{radial compactification} of the Euclidean space $\BR^n$ is defined as 
\begin{equation}
	\label{Eqradialcom}
	\overline{\BR^n} := \bigl(\BR^n \sqcup ([0,1)_\rho\times\BS^{n-1})\bigr) / \sim
\end{equation}	
where $\sim$ identifies $(\rho, \omega)\in[0,1)_\rho\times\BS^{n-1}$ with the point $\rho^{-1}\omega\in\BR^n$. The quotient carries the smooth structure where being smooth over $	\overline{\BR^n} $ means smoothness over the interior $R^n$ in the usual sense, and smoothness over $[0,1)_\rho\times\BS^{n-1}$ in $(\rho,\omega)$.
\end{defn}

Near the boundary $\rho^{-1}(0)$ with $\rho=r^{-1}$, the space of b-vector fields is spanned over $C^\infty(\overline{\BR^n})$ by $\rho\pa_\rho=-r\pa_r$ and $\mathcal{V}(\BS^{n-1})$, while the space of scattering vector fields is spanned over $C^\infty(\overline{\BR^n})$ by $\rho^2\pa_\rho=-\pa_r$ and $\rho\mathcal{V}(\BS^{n-1})$. Returning to the standard coordinates $x^1,\ldots,x^n$ on $\BR^n$, a direct calculation implies that $\{\pa_{x^1},\ldots,\pa_{x^n}\}$ extend to the radial compactification $\overline{\BR^n}$ as smooth scattering vector fields and they form a basis of ${}^{\scop}T\overline{\BR^n}$, that is, the space $\mathcal{V}_{\scop}(\overline{\BR^n})$ of the scattering vector fields is spanned, over $C^\infty(\overline{\BR^n})$, by $\{\pa_{x^1},\ldots,\pa_{x^n}\}$. We also note that the linear vector fields $\pa_{x^1},\ldots,\pa_{x^n}$, and $x^i\pa_{x^j}$, $1\leq i,j\leq n$ lift to span, over $C^\infty(\overline{\BR^n})$, the space $\mathcal{V}_{\bop}(\overline{\BR^n})$ of b-vector fields.

As for the dual bundles, correspondingly we note that the differentials of the standard coordinates $d x^i$, $1\leq i\leq n$ are a basis of ${}^{\scop}T^*\overline{\BR^n}$. Therefore, the Euclidean metric $(d x^1)^2+\cdots+(d x^n)^2$ is a particular example of the \emph{scattering metric}, i.e., an element in $C^\infty(\overline{\BR^n};S^2\,{}^{\scop}T^*\overline{\BR^n})$.

In this work, we are particularly interested in the behavior of the Hamiltonian vector field associated to a smooth function $p\in C^\infty(\fscform)$ where $\CX$ is a manifold with boundaries. Concretely, given a smooth functions $p\in C^\infty(\fscform)$, the Hamilton vector fields $H_p$ extends from the interior to a scattering vector field on $\fscform$, which is again a manifold with boundary ${}^{\scop}T^*_{\pa\CX}\bar{X}$. In this paper, we consider the Hamilton vector field $H_p$ where $p(x,\xi):=|\xi|_{g(x)^{-1}}^2$ and $g\in C^\infty(\CX;S^2\,\fscform)$ is a scattering metric. 

To give a uniform discussion of the behavior of the Hamiltonian vector field $H_p\in\Vsc(\fscform)$ near the boundary of $\pa\CX$ and near the infinity in the fibers of the scattering cotangent bundle $\fscform$, we consider the compactified scattering cotangent bundle $\cscform$ as our basic microlocal space. We note that $\cscform$ is a compact manifold with corners obtained by the radial compactification of the fibers of $\fscform$ (see Figure \ref{fig:simplemicrolocalspace}). Let $\rho_{\mathrm{total}}=x\rho_\xi$, where $x$, resp. $\rho_\xi$ is the boundary defining function of the boundary of $\CX$, resp. the fiber infinity, be the total boundary defining function of $\cscform$. Then we have $\Vsc(\cscform)=\rho_{\mathrm{total}}\Vb(\cscform)$, and thus the rescaled Hamiltonian vector field $\rho_{\mathrm{total}}^{-1}H_p$, where $p\in C^\infty(\fscform)$, is a b-vector field on $\cscform$.
\begin{figure}[!h]
	\begin{tikzpicture}
		\draw(0,0) rectangle (4,4);
		\draw [thick, dashed](0,2)--(4,2);
		%\draw [very thick] (3,1.5)--(3,2.3);
		%\node [left] at (0, 2) {$\abscform$};
		\node [right] at (4, 2) {${}^{\scop}T^*_{\pa\CX}\bar{X}=\{x=0\}$};
		\node [above] at (2, 4) {$\fiscform=\{\rho_\xi=0\}$};
		\node [below] at (2, 0) {$\fiscform=\{\rho_\xi=0\}$};
	%	\node [above ] at (2, 2) {$\CX\times\{0\}$};
		\node [above] at (2, 2) {$0$ section};
	%	\node [circle, inner sep=1pt, fill=black,label= below left:{$R(0)$}] at (3, 1.5) {};
	%	\node [circle, inner sep=1pt, fill=black,label=left:{$R(\sigma)$} ] at (3, 2.3) {};
		%\node  at (1.5, -0.5) {$\sigma>0$};
	\end{tikzpicture}

	\caption{The microlocal space $\cscform$.} 
	\label{fig:simplemicrolocalspace}
\end{figure}

\
\subsection{Function spaces}\label{subsec:functionspaces}
We next recall the notion of b- and scattering Sobolev spaces. Fixing a volume \emph{scattering density} $\mu$ on $\CX$, which in the local coordinates $x\geq 0, y\in\BR^{n-1}$ is a positive multiple of $|\tfrac{d x}{x^2}\frac{d y}{x^{n-1}}|$. With respect to this scattering density, we define the space $L^2(\CX)=L^2_{\scop}$, and since $\CX$ is compact, any two choices of scattering density give rise to equivalent norms on $L^2(\CX)$. We note that using b-density is the usual convention for the definition of b-Sobolev space, however, \emph{we emphasize that in this paper, we use the scattering density even for b-Sobolev spaces}. For $s\in\BN_0$, we define \emph{b-} or \emph{scattering Sobolev space} as follows
\[
H_\bullet^s(\CX) := \{ u\in L^2(\CX,\nu)\mid V_1\cdots V_j u\in L^2(\CX,\nu),\ V_i\in\mathcal{V}_\bullet(\CX),\ 1\leq i\leq j\leq s \},\quad
\bullet=\bop,\scop,
\]
which can be extended to $s\in\BR$ by duality and interpolation. \emph{Weighted b- and scattering Sobolev spaces} are defined by
\[
H_\bullet^{s,\ell}(\CX)= \rho^\ell H_\bullet^s(\CX):= \{ u\mid\rho^{-\ell}u \in H_\bullet^s(\CX) \}.
\]
The space 
\[
H_\bop^{\infty,\ell}(\CX) = \bigcap_{s\in\BR}\eHb^{s,\ell}(\CX).
\]
is a Fr\'{e}chet space and we refer to their elements as \emph{weighted ($L^2$-)conormal functions}. Dually, we define 
\[
H_\bop^{-\infty,\ell}(\CX) = \bigcup_{s\in\BR}\eHb^{s,\ell}(\CX).
\]
We note that $\eHb^{s,\ell}(\CX)\subset C^{-\infty}(\CX)$ where $C^{-\infty}(\CX)$ is the space of extendible distributions at $\CX$ in the sense of \cite[Appendix B]{H07}, and it is also the dual to the space $\dot{C}^\infty(\CX)$ (which is the space of smooth functions vanishing, with all derivatives, at the boundary $\pa \CX$). We also define the spaces of the form
\begin{equation}
	\label{EqbSobolevweight}
	H_\bop^{s,\ell+}(\CX):= \bigcup_{\epsilon>0} H_\bop^{s,\ell+\epsilon}(\CX),\quad
	H_\bop^{s,\ell-}(\CX) := \bigcap_{\epsilon>0} H_\bop^{s,\ell-\epsilon}(\CX)
\end{equation}
for $s\in\BR\cup\{\pm\infty\}$. 

We also use another the \emph{weighted $L^\infty$-conormal functions} space $\mathcal{A}^\ell(\CX)$ to describe the behavior of functions near the boundary $\pa\CX=\{\rho=0\}$
\begin{equation}\label{eq:conormalfunctionspace}
\mathcal{A}^\ell(\CX) := \{ u\in\rho^\ell L^\infty(\CX) \colon \mbox{Diff}_\bop(\CX)u\subset\rho^\ell L^\infty(\CX) \}.
\end{equation}
The relation between $H^{\infty,\ell}(\CX)$ and $\mathcal{A}^{\ell}$ is as follows: for $\CX=\overline{\BR^3}$, a direct calculation implies $\mathcal{A}^\ell(\overline{\BR^3}) \subset H_\bop^{\infty,\ell-3/2-}(\overline{\BR^3})$, and Sobolev embedding theorem gives $H_\bop^{\infty,\ell}(\overline{\BR^3}) \subset \mathcal{A}^{\ell+3/2}(\overline{\BR^3})$.
We note that since we use the scattering density (instead of the b-density) to define the b-Sobolev space, there is a shift $\tfrac32$ in the weight. More specifically, for $s>\frac32$, we have 
\begin{equation}
	\label{EqBHbSobolevEmb}
	H_\bop^{s,\ell}(\overline{\BR^3};|d x^1\,d x^2\,d x^3|)=H_\bop^{s,\ell+3/2}(\overline{\BR^3};\jb{r}^{-3}|d x^1\,d x^2\,d x^3|) \hookrightarrow \jb{r}^{-\ell-3/2}L^\infty(\overline{\BR^3}).
\end{equation}
The spaces $\mathcal{A}^{\ell+}(\CX)$ and $\mathcal{A}^{\ell-}(\CX)$ can be defined analogously to~\eqref{EqbSobolevweight}. The above spaces of sections of vector bundles are defined using local trivializations.

We now recall the notion of \emph{$\mathcal{E}$-smoothness} and its basic properties (see \cite[\S 5]{Mel93}). 
\begin{defn}[{\cite[Definition 5.23]{Mel93}}]
An \emph{index set} is a discrete subset $\mathcal{E}\subset\BC\times\BN_0$ satisfying
\begin{enumerate}
	\item $(z_j,k_j)\in\mathcal{E}, \ \abs{(z_j, k_j)}\to\infty\Longrightarrow \IM z_j\to-\infty$;
	\item $(z,k)\in\mathcal{E}\Longrightarrow (z,l)\in\mathcal{E},\ l\in\BN_0,\ 0\leq l\leq k$;
	\item $(z,k)\in\mathcal{E}\Longrightarrow (z-j, k)\in\mathcal{E},\ j\in\BN_0$.
\end{enumerate}
\end{defn}
\begin{defn}[{\cite[Equation 5.73]{Mel93}}]
Let $\mathcal{E}$ be an index set. A \emph{polyhomogeneous function} on $\CX$ with index set $\mathcal{E}$ is a smooth (in the interior of $\CX$) function $u$  for which there are $a_{(z,j)}\in C^\infty(\pa \CX)$, $(z,j)\in\mathcal{E}$, such that
\begin{equation}
	\label{Eqdefofpoly}
	u - \sum_{\substack{(z,j)\in\mathcal{E}\\\IM z\geq -N}} \rho^{i z}(\log\rho)^j a_{(z,j)} \in \mathcal{A}^N(\CX) \quad\forall\,N\in\BR.
\end{equation}
The space of polyhomogeneous functions on $\CX$ with index set $\mathcal{E}$ is denoted by $\mathcal{A}^\mathcal{E}_{\mathrm{phg}}(\CX)$.
\end{defn}
Another characterization of polyhomogeneous functions with index set $\mathcal{E}$ is given by 
\[
u\in \mathcal{A}^\mathcal{E}_{\mathrm{phg}}(\CX)\Longleftrightarrow\prod_{\substack{(z,j)\in\mathcal{E}\\\IM z\geq -N}} (\rho D_\rho-z)u \in \mathcal{A}^N(\CX) \quad \forall\,N\in\BR,
\]
where $D_\rho=i^{-1}\pa_\rho$ and $\rho$ is the boundary defining function of $\CX$.

\begin{figure}[!ht]
\begin{tikzpicture}
	\filldraw [fill=gray!50] (0,0) circle (1.5cm);
\filldraw [fill=gray!20](0,0) circle (0.5cm);
	\node [above] at (0, 0.4) {$\pa_-\CX$};
		\node [above] at (0, 1.4) {$\pa_+\CX=\pa\CX'$};
\end{tikzpicture}
	\caption{$\CX$ (dark gray annulus) is a submanifold of $\CX'$ (the union of the dark gray annulus and the light gray disk) with two components of boundary $\pa_+\CX=\pa \CX'$ and $\pa_-\CX\subset(\CX')^\circ$. The function spaces such as $\eHb^{s,\ell}(\CX)$ measure b-regularity of degree $s$ and decay rate $\ell$ at $\pa_+\CX$, while standard regularity at $\pa_-\CX$.}
	\label{Figtwoboundaries}
\end{figure}
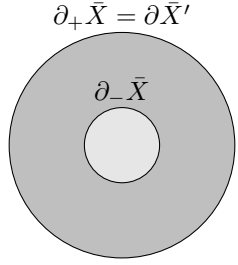

Let $\CX'$ is a compact manifold with boundary, and let $\CX\subset \CX'$ be a submanifold with boundaries. Suppose that the boundaries of $\CX$ has two components
\begin{equation}
	\label{Eqtwoboundaries}
	\pa \CX = \pa_-\CX \sqcup \pa_+\CX,\qquad \pa_-\CX=\pa \CX\setminus\pa \CX',\quad \pa_+X=\pa \CX';
\end{equation}
we call $\pa_+\CX$ the boundary `at infinity', and $\pa_-\CX$ the `artificial' boundary in the interior of $\CX'$. Working on such $\CX$, we impose b- and scattering structure near $\pa_+\CX$ (i.e. the boundary `at infinity'). That is, we define (by a slight abuse of notation)
\[
\Vb(\CX) := \{ V|_{\CX} \mid V\in\Vb(\CX') \}, \quad
\Vsc(X) := \{ V|_{\CX} \mid V\in\Vsc(\CX') \}.
\]
 A typical example of the setting \eqref{Eqtwoboundaries} is $\CX'=\overline{\BR^n}$ and  $\CX=\overline{\{r\geq 1\}}\subset \CX'$. In this example, we see that $\pa_+\CX=\pa X'$ is the boundary 'at infinity' and $\pa_-\CX=\{r=1\}$ is the boundary in the interior of $\CX'=\overline{\BR^n}$. See Figure \ref{Figtwoboundaries}.

Because of the existence of the `artificial' boundary $\pa_-\CX$, we further introduce the following two classes of Sobolev spaces: \emph{extendible Sobolev spaces} $\bar{H}_\bullet^{s,\ell}(\CX)$ and \emph{supported Sobolev spaces} $\dot{H}_\bullet^{s,\ell}(\CX)$
\begin{equation}
	\label{EqBExtSupp}
	\bar{H}_\bullet^{s,\ell}(\CX) := \{ u|_{\CX^\circ} \mid u\in H_\bullet^{s,\ell}(\CX') \},\quad	\dot{H}_\bullet^{s,\ell}(\CX) := \{ u \mid u\in H_\bullet^{s,\ell}(\CX'),\ \mbox{supp}\,u\subset \CX \},\quad\bullet=\bop,\scop.
\end{equation}
Away from $\pa_-\CX$, these are the same as the standard spaces $H_\bullet^{s,\ell}(\CX)$. Therefore, the extendibility/support considerations arise only at $\pa_-\CX$, not at $\pa_+\CX$.

Let $M=\BR_\mathfrak{t}\times \CX$ be a stationary spacetime with `spatial part' $\CX$ and define the projection $\pi_{\CX}\colon M\to \CX$. Let $\mathcal{F}(\CX)$ be  a function space on $\CX$ such as $\mathcal{F}(\CX)=\eHb^{\infty,\ell}(\CX)$ or $\sHb^{\infty,\ell}(\CX)$, and then $\pi_{\CX}^*\mathcal{F}(\CX)$ is a $\mathfrak{t}$-independent function space on $M$. For later use, we further introduce the space of \emph{generalized zero modes} (which depends on $\mathfrak{t}$ polynomially)
\begin{equation}
	\label{EqtimedependSpace}
	\mbox{Poly}(\mathfrak{t})\mathcal{F}(\CX) := \bigcup_{k\in\BN}\mbox{Poly}^k(\mathfrak{t})\mathcal{F}(\CX), \quad
	\mbox{Poly}^k(\mathfrak{t})\mathcal{F}(\CX) := \left\{ \sum_{j=0}^k \mathfrak{t}^j a_j \colon a_j\in\mathcal{F}(\CX) \right\}.
\end{equation}
Here, $\CX$ (or $d\mathfrak{t}$) does not have to be spacelike (or timelike).

For the high energy estimates, we need to work with \emph{semiclassical b-Sobolev spaces} $H_{\bop,h}^{s,\ell}(\CX)$. The semiclassical b-Sobolev norms are defined for $s\in\BN_0,\ u\in H_\bop^{s,\ell}(\CX),\ h\in(0,1]$ by 
\begin{equation}\label{Eqseminorm}
\|u\|_{H_{\bop,h}^s(\CX)}^2 := \sum_{\substack{1\leq i_1,\ldots,i_j\leq N\\0\leq j\leq k}} \norm{(h V_{i_1})\cdots(h V_{i_j})u}_{L^2(\CX)}^2,\quad
\|u\|_{H_{\bop, h}^{k,\ell}(\CX)} := \|\rho^{-\ell}u\|_{H_{\bop,h}^k(\CX)},
\end{equation}
where $V_1,\ldots,V_N\in\Vb(\CX)$ spans $\Vb(\CX)$ over $C^\infty(\CX)$. This definition of the semiclassical norm can be extended to $s\in\BR$ by duality and interpolation. (Alternatively,  $H_{\bop,h}^{s,\ell}(\CX)$ can be defined by using semiclassical b-pseudodifferential operators, see \cite[Appendix~A]{HV18}.) If $\CX$ is a compact manifod as in the setting \eqref{Eqtwoboundaries}, one can also define semiclassical extendible and supported b-Sobolev spaces $\bar H_{\bop,h}^{s,\ell}(X)$ and $\dot H_{\bop,h}^{s,\ell}(X)$ analogously to \eqref{EqBExtSupp}.

\section{Initial value problems formalism of Einstein-Maxwell equations}\label{sec:IVP}
In this section, we shall discuss the initial value problems formalism for Einstein-Maxwell equations. Here, we follow the discussion in \cite[\S3]{H18}. As a general reference, we refer the readers to \cite[\S 6.10]{CB09}. In \S \ref{SubsecBasicDer}, we reduce the Einstein-Maxwell equations for $(g, F)$ in the form \eqref{EqBasicEMCurv1}--\eqref{EqBasicEMCurv2} to the study of the equations for $(g, A)$ of the form \eqref{EqBasicEM}. In \S \ref{SubsecBasicNL}, we formulate the initial value problems for Einstein-Maxwell equations and then write the Einstein-Maxwell equations with respect to a \emph{generalized wave map gauge} and \emph{Lorenz gauge}. In \S \ref{SubsecBasicLin}, we discuss the linearization around a special solution of the Einstein-Maxwell equations, as well as the linearization of the gauge-fixed Einstein-Maxwell equations.

%%%%%%%%%%%%%%%%%%%%%%%%%%%%%%%%%%%%%%%%%%%%%%%%%%
\subsection{Einstein-Maxwell equations}
\label{SubsecBasicDer}
The Einstein-Maxwell equations read
\begin{gather}	\label{EqBasicEMCurv1}
	\Ric(g)_{\m\n}=2 T(g,F)_{\m\n},\\
	\label{EqBasicEMCurv2}  d F=0,\quad \delta_g F=0.
\end{gather}
where $g$ is a Lorentzian metric $g$ with signature $(-,+,+,+)$, $\Ric(g)$ is the Ricci curvature tensor, $T_{\mu\nu}(g,F):= F_{\mu\al}F_{\nu}^{\ \al}-\frac{1}{4}g_{\mu\nu}F_{\al\be}F^{\al\be}$ is the energy momentum tensor associated with an electromagnetic field represented by a $2$-form $F$ and $\delta_gF=-\nabla^\alpha F_{\al\be}$ is the negative divergence. A direct calculation implies that if $F$ solves \eqref{EqBasicEMCurv2}, then we have $\delta_g T(g, F)=0$.

Let $(g,F)$ be a smooth solution of \eqref{EqBasicEMCurv1}--\eqref{EqBasicEMCurv2}. Let $i\colon\Sigma_0\hookrightarrow M$ be a spacelike hypersurface with future timelike unit normal field $n$, and let $h=g|_{\Sigma_0}$ be the induced metric on $\Sigma_0$ with respect ot $g$. The electromagnetic initial data on $\Sigma_0$ are the \emph{magnetic field} $\mathbf{H}$ and \emph{electric field} $\mathbf{E}$. The 1-form $\mathbf{H}$ is the form induced on $\Sigma_0$ by the electromagnetic field $F$, while $\mathbf{E}$ is the $1$-form on $\Sigma_0$  relative to the future timelike unit normal $n$ to $\Sigma_0$ in the spacetime metric $g$. 
Specifically, the electric and magnetic fields $\mathbf{E}, \mathbf{H}$, as measured by observers with 4-velocity $n$, are defined as follows
\begin{equation}
	\label{EqBasicEandHFields}
	\mathbf{E} := -i^* \iota_n F = \star_h i^*\star_g F, \quad \mathbf{H}:= i^* \iota_n\star_g F = \star_h i^*F  
	%\in C^\infty(\Sigma_0;T^*\Sigma_0)
\end{equation}
where $\star$ is the Hodge star operator.

A direct calculation implies that for $T_{g,F}:= T(g,F)$
\begin{equation}
	\label{EqBasicEMTensorN}
	\begin{split}
		T_{g,F}(n,n) &= \frac{1}{2}(|\mathbf{E}|_h^2 + |\mathbf{H}|_h^2), \\
		T_{g,F}(n,\cdot) &= \star_h(\mathbf{H}\wedge\mathbf{E})\quad\mbox{on}\quad T\Sigma_0,
	\end{split}
\end{equation}
where $T_{g,F}(n,n)$ is interpreted as the energy density of the electromagnetic field measured by an observer with $4$-velocity $n$.

We now define the following rotation of $F, \mathbf{E}, \mathbf{H}$:
\begin{align}
	\label{EqBasicEMRotationForm}
	F_\theta &:= \cos(\theta)F + \sin(\theta)\star_g F,\\\label{EqBasicEMRotation}
(\mathbf{E}_\theta,\mathbf{H}_\theta)&:=(\cos(\theta)\mathbf{E}-\sin(\theta)\mathbf{H},\, \sin(\theta)\mathbf{E}+\cos(\theta)\mathbf{H}).
\end{align}
Then 	$\mathbf{E}_\theta$ and $\mathbf{H}_\theta$ are the electric and magnetic field, respectively, induced by $(g,F_\theta)$. Now we establish the following lemma.
\begin{lem}
	\label{LemmaBasicEMRotation}
	If $(g,F)$ solves the Einstein-Maxwell equations \eqref{EqBasicEMCurv1}--\eqref{EqBasicEMCurv2}, then so does $(g,F_\theta)$ for all $\theta\in\BR$, where $F_\theta$ is defined as in \eqref{EqBasicEMRotationForm}.
\end{lem}
\begin{proof}
	According to the facts that $\star_g\star_g=-\mbox{Id}$ and $\delta_g=\star_g d\star_g$ when acting on $2$-forms, the equations \eqref{EqBasicEMCurv2} are equivalent to $dF=0, d\star_g F=0$. Then it follows that 
	\[
	d F_\theta=\cos(\theta) dF+\sin(\theta) d\star_g F=0,\quad \delta_gF_\theta=\star_g(\cos(\theta) d\star_g F-\sin(\theta) dF)=0.
	\] 
Since the energy momentum tensor $T(g, F)$ can be rewritten as 
\[
T(g, F)_{\m\n}=\frac 12 g^{\lambda\sigma}\Big(F_{\m\lambda}F_{\n\sigma}+(\star_gF)_{\m\lambda}(\star_gF)_{\n\sigma}\Big),
\]
 it follows that $
T(g, F_\theta)_{\m\n}=T(g, F)_{\m\n}$. This proves that $(g,F_\theta)$ solves the Einstein-Maxwell equations \eqref{EqBasicEMCurv1}--\eqref{EqBasicEMCurv2}.
\end{proof}

For simplicity, we may assume that 
\begin{equation}
	\label{EqBasicEMHole}
	M \cong \BR_{\mathfrak{t}}\times I_r \times\BS^2,\quad \Sigma_0 = \{\mathfrak{t}=0\}\subset M,
\end{equation}
where $I\subset\BR$ is a non-empty open interval, and $\Sigma_0$ is spacelike. Then given a solution $(g,F)$ of \eqref{EqBasicEMCurv1}--\eqref{EqBasicEMCurv2}, and a $2$-sphere $S=\{\mathfrak{t}=\mathfrak{t}_0,\ r=r_0\}\subset M$ (with $t_0\in\BR$, $r_0\in I$), we can define the \emph{electric} and \emph{magnetic charges}
\begin{equation}
	\label{EqBasicEHCharges}
	Q_e(g,F) := \frac{1}{4\pi}\int_S \star_g F,\qquad Q_m(F) := \frac{1}{4\pi}\int_S F.
\end{equation}
By Stokes' theorem and the equations \eqref{EqBasicEMCurv2} $dF=0, d\star_g F=0$, $Q_e$ and $Q_m$ are independent of the specific choice of $S$. In particular, we can choose $S\subset\Sigma_0$, and use \eqref{EqBasicEandHFields} and $\star_h\star_h=\mbox{Id}$ when acting on $1$-form to find that
\begin{align*}
Q_e(g,F) &= Q_e(h,\mathbf{E}) := \frac{1}{4\pi}\int_S \star_h\mathbf{E} = \frac{1}{4\pi}\int_S \langle\mathbf{E},\nu\rangle d\sigma,\\
Q_m(F)&=Q_m(\mathbf{H}):=\frac{1}{4\pi}\int_S\star_h\mathbf{H}= \frac{1}{4\pi}\int_S \langle\mathbf{H},\nu\rangle d\sigma,
\end{align*}
with $\nu$ the outward pointing unit normal to $S$ (with respect to $h$) and $d\sigma$ the volume element of $S$.

Since $I_r\times \BS^2$ is not a simply connected domain, $dF=0$ does not imply that $F=dA$ for some $1$-form $A$ globally. In fact, the de Rham cohomology group $H_{\mathrm{dR}}^2(I_r\times\BS^2, \BR)\cong\BR$ and $F\mapsto Q_m(F)$ induces an isomorphism. Therefore, the vanishing of $Q_m(F)$ is equivalent to the condition that $F=dA$ for some $1$-form $A$. Now we will show that the vanishing of $Q_m(F)$ can always be arranged by the rotation transformation \eqref{EqBasicEMRotationForm}.

\begin{lem}
	\label{LemmaBasicEMRotationCharge}
Let $(g,F)$ solve the Einstein-Maxwell equations \eqref{EqBasicEMCurv1}--\eqref{EqBasicEMCurv2}. In the setting \eqref{EqBasicEMHole}, there exists $\theta\in\BR$ such that $Q_m(F_\theta)=0$, with $F_\theta$ defined in \eqref{EqBasicEMRotationForm}. Moreover, for such $\theta$, we have
	\[
	Q_e(g,F_\theta)^2 = Q_m(F)^2 + Q_e(g,F)^2.
	\]
\end{lem}
\begin{proof}
	With $F_\theta=\cos(\theta)F+\sin(\theta)\star_g F$, we have $Q_m(F_\theta)=\cos(\theta)Q_m(F)+\sin(\theta)Q_e(g,F)$. Choosing $\theta\in\BR$ such that $(\cos\theta,\sin\theta)\perp(Q_m(F),Q_e(g,F))$ finishes the proof.
\end{proof}

In view of Lemma \ref{LemmaBasicEMRotationCharge}, we may assume $Q_m(F)=0$, which is equivalent to the fact that $F=d A$ for some 1-form $A$. This reduces the study of the Einstein-Maxwell equations on $M$ in the form \eqref{EqBasicEMCurv1}--\eqref{EqBasicEMCurv2} to the study of the following system in $(g,A)$
\begin{equation}
		\label{EqBasicEM}
		\Ric(g)  = 2 T(g,d A),\qquad \delta_g d A = 0,
	\end{equation}
which we continue to call Einstein-Maxwell equations. In this work, we will study the Einstein-Maxwell equations of the above form \eqref{EqBasicEM}.

\subsection{Initial value problems for the nonlinear system}
\label{SubsecBasicNL}

We now discuss the initial value problem of Einstein-Maxwell equations \eqref{EqBasicEM}. Given any initial data
\begin{equation}
	\label{EqBasicNLinitialdata}
	(\Sigma_0,h,k,\mathbf{E},\mathbf{H}),
\end{equation}
where $\Sigma_0$ is a smooth 3-manifold equipped with Riemannian metric $h$, a symmetric 2-tensor $k$, and $1$-forms $\mathbf{E},\mathbf{H}$, one seeks a solution $(M,g,A)$ to \eqref{EqBasicEM} with data $(\Sigma_0, h, k, \mathbf{E}, \mathbf{H})$ in the sense that  $\Sigma_0\hookrightarrow M$ is a spacelike embedding hypersurface with respect to $g$, and $h, k$ are respectively the induced metric and second fundamental form of $\Sigma_0$ with respect to $g$, and $F=dA$ induces the given fields $\mathbf{E},\mathbf{H}$ on $\Sigma_0$ according to \eqref{EqBasicEandHFields}. We define the initial data map 
\begin{equation}
\label{Eqinitialmap}
\tau: (g,F)\mapsto (h,k,\mathbf{E},\mathbf{H})
\end{equation}
if $(g,F)$ induces the data $(h,k,\mathbf{E},\mathbf{H})$ at $\Sigma_0$ in the above sense. We note that the data $(h, k, \mathbf{E}, \mathbf{H})$ cannot be prescribed freely and they must satisfy the following constraint equations, which are obtained by evaluating $G_g(\Ric(g)-2T(g,F))$ on $(n,n)$ where $G_g=\mbox{Id}-\frac12g\tr_g$ and $n$ is future timelike unit normal to $\Sigma_0$, and on $(n,X)$ with $X\in T\Sigma_0$, and pulling back $dF=0, d\star_g F=0$ to $\Sigma_0$, and finally adding the constraint that the magnetic charge $Q_m(F)=Q_m(\mathbf{H})=0$ vanishes.
\begin{gather}
	\begin{gathered}
		\label{EqBasicNLconstraints1}
		R_h - |k|_h^2+(\tr_h k)^2 = 2(|\mathbf{E}|_h^2+|\mathbf{H}|_h^2), \\
		\delta_h k + d\tr_h k = 2\star_h(\mathbf{H}\wedge\mathbf{E});
	\end{gathered} \\
	\label{EqBasicNLconstraints2}
	\delta_h\mathbf{E}= 0,\quad \delta_h\mathbf{H}= 0;\\
	\label{Eqvanishingmcharge}
	Q_m(\mathbf{H})=\frac{1}{4\pi}\int_S\star_h\mathbf{H}=0.
\end{gather}
The constraints \eqref{EqBasicNLconstraints1}--\eqref{EqBasicNLconstraints2} are necessary and sufficient conditions for the local solvability of the Einstein-Maxwell equations (see \cite[\S6.10]{CB09}). 
\begin{rem}
	In the setting $\Sigma_0\cong I_r\times\BS^2$ as in \eqref{EqBasicEMHole}, in view of the discussion around \eqref{EqBasicEMRotationForm} and Lemma~\ref{LemmaBasicEMRotationCharge}, the constraint equations are invariant with $(\mathbf{E},\mathbf{H})$ replaced by $(\mathbf{E}_\theta,\mathbf{H}_\theta)$, and by choosing suuitable $\theta$ we have $Q_m(\mathbf{H}_\theta)=0$. Therefore, we can solve the Einstein-Maxwell equations for $(g,A)$ with initial data $(\mathbf{E}_\theta,\mathbf{H}_\theta)$, and then $F:=(d A)_{-\theta}$ gives rise to a solution of \eqref{EqBasicEMCurv1}--\eqref{EqBasicEMCurv2} with initial data $(\mathbf{E},\mathbf{H})$.	
\end{rem}

\emph{Under the additional assumption} $Q_m(\mathbf{H})=0$, we can solve the Einstein-Maxwell equations of the form \eqref{EqBasicEM} for $(g,A)$, with electromagnetic field then given by $F=dA$. Due to the diffeomorphism invariance of Einstein-Maxwell equations, one has the gauge freedom and thus can fix a gauge. Here, we employ the \emph{generalized wave map gauge}. Concretely, we fix a Lorentzian background metric $g^0$ and define the \emph{gauge 1-form}
\begin{equation}
	\label{EqBasicNLgaugepo}
	\Upsilon^E(g;g^0) := g (g^0)^{-1}\delta_g G_g g^0.
\end{equation}
In local coordinates, the above gauge $1$-form can be written as
\begin{equation}
		\Upsilon^E(g;g^0)_\alpha
		=g_{\alpha\kappa}g^{\mu\nu}(\Gamma(g)_{\mu\nu}^\kappa-\Gamma(g^0)_{\mu\nu}^\kappa).
\end{equation}
We note that $\Upsilon^E(g)\equiv 0$ if and only if the identity map $\mbox{Id}\colon(M,g)\to(M,g^0)$ is a wave map, and thus $\Upsilon^E(g)\equiv 0$ gives rise to the wave map gauge (see \cite{De81}). The \emph{generalized wave map gauge} is defined as
\begin{equation}
	\label{Eqgeneralizedwavegauge}
	\widetilde{\Upsilon}^E(g;g^0)=\Upsilon^E(g;g^0)-\theta(g;g^0)=0
	\end{equation}
where $\theta$ is a $1$-form linear in $g$ and independent of the derivatives of $g$, and satisfies that $\theta(g^0;g^0)=0$. We also point out that if the background metric $g^0$ is Minkowski metric, one obtains the generalized wave coordinates for $\theta(g;g^0)\neq 0$ (see \cite{F85}) and wave coordinates for $\theta(g;g^0)=0$ (see \cite[\S 6]{CB09}), respectively.

The system \eqref{EqBasicEM} has an additional gauge freedom, namely, Einstein-Maxwell equations are invariant under the gauge transformation 
\begin{equation}
	\label{EqBasicNLgaugetrans}
	(g,A) \mapsto (\phi_*g,\phi_*A+d a).
\end{equation}
for any diffeomorphism $\phi:M\to M$ and smooth function $a\in C^\infty(M)$. A typical gauge for $A$ is \emph{Lorenz gauge}, which is given by  $\delta_g A=0$. In fact, it is more convenient to write $\delta_g=-\tr_g\delta_g^*$ and to introduce the \emph{gauge function}
\begin{equation}
	\label{EqBasicNLgaugefcn}
	\Upsilon^M(g,A;g^0) := \tr_g\delta_{g^0}^* A,
\end{equation}
where $g^0$ is a fixed background metric as above. Therefore, $\Upsilon^M(g,A)$ equals $-\delta_g A$ up to terms of order $0$ in $A$. The \emph{generalized Lorenz gauge} is defined as
\begin{equation}
	\label{EqgeneralizedLorengauge}
	\widetilde{\Upsilon}^M(g,A;g^0,A^0) 	=\Upsilon^M(g,A;g^0)-	\Upsilon^M(g,A^0;g^0)=0
\end{equation}
where $A_0$ is a fixed $1$-form.

\begin{rem}\label{rem:gaugerep}
	We claim that any solution $(g,A)$ to the Einstein-Maxwell equations can be transformed into another solution $(\phi^*g, \phi^*A+da)$ which satisfies the generalized wave map gauge and generalized Lorenz gauge conditions. Concretely, we define new coordinates $(y^\alpha)$ by solving
	\[
	\begin{cases}
		\Box_g y^\alpha-\theta(g;g^0)_\beta \nabla^\beta y=0\\
		(y^\alpha, dy^\alpha)|_{\Sigma_0}=(x^\alpha, dx^\alpha)|_{\Sigma_0}.
	\end{cases}
	\]
	Then the pull back $(\phi^*g, \phi^*A)$ of $(g,A)$ with respect to the diffeomorphism $\phi(y)=x$ satisfies the generalized wave map gauge condition and is also a solution to Einstein-Maxwell equations. Then with $(\phi^*g, \phi^*A)$, one continues to solve the equation $\Upsilon^M(\phi^*g,\phi^*A+da;g^0)=\Upsilon^M(\phi^*g,A_0;g^0)$, which is a linear scalar wave equation for $a$, and thus $(\phi^*g,\phi^* A+da)$ satisfies the generalized Lorenz gauge condition and solves the Einstein-Maxwell equations.
\end{rem}

We now consider the gauge-fixed system 
\begin{align}
	\label{EqBasicNLgaugefixE}
	P^E(g,A)&:= \Ric(g) - \tilde{\delta}_g^*\widetilde{\Upsilon}^E(g;g^0) - 2 T(g,dA)=0\\
		\label{EqBasicNLgaugefixM}
 P^M(g,A) &:= \delta_g d A - d\widetilde{\Upsilon}^M(g,A;g^0, A^0)=0
\end{align}
where $\tilde{\delta}_g^*$ is zero order modification of $\delta_g^*$, i.e., $\tilde{\delta}_g^*=\delta_g^*+B$ with $B\in C^\infty(M;\mbox{Hom}(S^2T^*M, T^*M))$, and  $(\delta_g^*u)_{\mu\nu}=\frac{1}{2}(u_{\mu;\nu}+u_{\nu;\mu})$ is the symmetric gradient. Now, \eqref{EqBasicNLgaugefixE}--\eqref{EqBasicNLgaugefixM} is a quasilinear hyperbolic system for $(g,	A)$ (which is principally scalar if one multiplies the first equation by $2$).

In order to relate the initial value problem for the Einstein-Maxwell system for $(g,A)$ to that of the quasilinear hyperbolic system \eqref{EqBasicNLgaugefixE}--\eqref{EqBasicNLgaugefixM}, we need the following lemma.
\begin{lem}\label{lem:gaugeungauge}
		Suppose $(g,A)$ solves the gauge-fixed Einstein-Maxwell equation \[
		P(g,A)=(P^E(g,A),\, P^M(g,A))=0
		\]
		and attains the given initial data $(g_0,g_1,A_0,A_1)$ at the initial hypersurface $\Sigma_0=\{\mathfrak{t}=0\}$ which satisfy the constraint equations \eqref{EqBasicNLconstraints1}--\eqref{Eqvanishingmcharge}. If $(g,A)$ satisfies the generalized wave map gauge and Lorenz gauge conditions initially at $\Sigma_0$, that is , \begin{equation}\label{eq:gaugeinitial}
			\widetilde{\Upsilon}^E(g;g^0)=0,\qquad \widetilde{\Upsilon}^M(g,A;g^0,A^0)=0\quad \mbox{at}\quad\Sigma_0=\{\mathfrak{t}=0\},
	\end{equation}
	then $(g,A)$ solves the Einstein-Maxwell equations with the same initial data.
\end{lem}
\begin{proof}
	We first prove that if $(g,A)$ solves $P(g,A)=0$ and satisfies the constraint equations and the gauge conditions at $\Sigma_0$, then
	\begin{equation}\label{eq:gaugeinitialder}
	\mathcal{L}_{\pa_{\mathfrak{t}}}\widetilde{\Upsilon}^E(g;g^0)=0,\qquad
	\mathcal{L}_{\pa_{\mathfrak{t}}}\widetilde{\Upsilon}^M(g,A;g^0, A^0)=0\quad \mbox{at}\quad \Sigma_0.
\end{equation}
Since the constraint equation \eqref{EqBasicNLconstraints1} is equivalent to evaluating $G_g(\Ric{g}-2T(g,dA))$ on $(n,n)$ and $(n,X)$ on $\Sigma_0$ where $n$ is the future timelike unit normal to $\Sigma_0$ and $X\in T\Sigma_0$, then evaluating $G_gP^M(g,A)$ on $(n,X)$ and $(n,n)$ and using the fact that $\widetilde{\Upsilon}^E(g,A;g^0)=0$ at $\Sigma_0$ yield $\mathcal{L}_{\pa_{\mathfrak{t}}}\widetilde{\Upsilon}^E(g,A;g^0)=0$ at $\Sigma_0$. As for the Maxwell part, the constraint equation implies that the evaluation of $\delta_g dA$ on $n$ is zero at $\Sigma_0$ and thus we have $\mathcal{L}_{\pa_{\mathfrak{t}}}\widetilde{\Upsilon}^M(g,A;g^0,A^0)=0$ at $\Sigma_0$.

Applying the negative divergence operator $\delta_g$ on the equation $P^M(g,A)=0$ yields
\[
\delta_g d\widetilde{\Upsilon}^M(g,A;g^0,A^0)=0.
\]
Then we obtain a linear hyperbolic system for $\widetilde{\Upsilon}^M(g,A;g^0,A^0)$ with vanishing initial data. Therefore, $\widetilde{\Upsilon}^M(g,A;g^0,A^0)=0$ on $M$, which implies that $(g,A)$ solve the Maxwell part $\delta_g dA=0$ of the Einstein-Maxwell equations. According to the second Bianchi identity ($\delta_g G_g\Ric(g)=0$) and the fact that $\delta_gdA=0$ (and thus $\delta_gG_gT(g,dA)=\delta_gT(g,dA)=0$), applying first the trace reversal operator $G_g=\mbox{Id}-\frac12 g\tr_g$ and then the negative divergence operator $\delta_g$ on $P^E(g,A)=0$ yields
\[
\delta_g G_g\tilde{\delta}^*_g\widetilde{\Upsilon}^E(g;g^0)=0.
\]
We again obtain a linear hyperbolic system for $\widetilde{\Upsilon}^E(g;g^0)$ with vanishing initial data, and thus $\widetilde{\Upsilon}^E(g;g^0)=0$ on $M$, which implies that $(g,A)$ solves Einstein part $\Ric(g)-2T(g,dA)=0$ of the Einstein-Maxwell equations. This completes the proof.
\end{proof}

	Owing to Lemma \ref{lem:gaugeungauge}, in order to solve the initial value problem for Einstein-Maxwell equations, it suffices to construct the gauged initial data satisfying the conditions \eqref{eq:gaugeinitial} and solve the gauge-fixed Einstein-Maxwell equations \eqref{EqBasicNLgaugefixE}--\eqref{EqBasicNLgaugefixM} with the constructed gauged initial data. We postpone the construction of the gauged initial data to \S\ref{SsKNGd}.

\subsection{Initial value problems for the linearized system}
\label{SubsecBasicLin}

In this work, we are interested in understanding long time behavior of a general solution to the linearize Einstein-Maxwell equations.

Suppose $(g,F)$ is a solution to the Einstein-Maxwell system \eqref{EqBasicEMCurv1}--\eqref{EqBasicEMCurv2} on a spacetime $M=\BR\times\Sigma_0$. Then around $(g,F)$, the \emph{linearized Einstein-Maxwell equations} are defined as
\begin{equation}
	\label{EqBasicLinEM}
	\begin{gathered}
	D_{g}\Ric(\dg)-2 D_{g,F}T(\dg,\dot{F}):=\frac{d}{ds}\Big\lvert_{s=0}\Big((\Ric-2T)(g+s\dg,F+s\dF)\Big)=0, \\
		d\dot{F}=0,\quad D_{g,F}(\delta_{(\cdot)}(\cdot))(\dg,\dot{F}):=\frac{d}{ds}(\delta_{g+s\dg}(F+s\dF))\Big|_{s=0} = 0.
	\end{gathered}
\end{equation}

Suppose $\Sigma_0$ is spacelike for $g$. We denote by $(\dot{h},\dot{k},\dot{\mathbf{E}},\dot{\mathbf{H}})$ the linearization of the initial data map $\tau$ around $(g,F)$. We note that $(\dot{h},\dot{k},\dot{\mathbf{E}},\dot{\mathbf{H}})$ are the initial data we impose in the initial value problem for the linearized Einstein-Maxwell equations \eqref{EqBasicLinEM}. As in the case of the non-linear Einstein-Maxwell equations where the initial data should satisfy the constraint equations, the initial data $(\dot{h},\dot{k},\dot{\mathbf{E}},\dot{\mathbf{H}})$ for the linearized Einstein-Maxwell equations are required to satisfy the \emph{linearized constraint equations} (see \cite[\S 7.2]{CB09}), which is defined as the linearization of  \eqref{EqBasicNLconstraints1}--\eqref{EqBasicNLconstraints2} around the data $(h,k,\mathbf{E},\mathbf{H})$ induced by $(g,F)$.

Motivated from \eqref{EqBasicEMRotationForm} and \eqref{EqBasicEMRotation}, we introduce the following transformation 
\begin{align}
	\label{EqBasicEMLinRotationForm}
	\dot{F}_\theta &:= \cos(\theta)\dF + \sin(\theta)\frac{d}{ds}\Big\lvert_{s=0}\Big(\star_{g+s\dg} (F+s\dF)\Big),\\\label{EqBasicEMLinRotation}
	(\dot{\mathbf{E}}_\theta,\dot{\mathbf{H}}_\theta)&:=(\cos(\theta)\dot{\mathbf{E}}-\sin(\theta)\dot{\mathbf{H}},\, \sin(\theta)\dot{\mathbf{E}}+\cos(\theta)\dot{\mathbf{H}}).
\end{align}  Then we have
 \[	(h,k,\dot{\mathbf{E}}_\theta,\dot{\mathbf{H}}_\theta)=D_{(g,F_\theta)}\tau(\dg,\dF_\theta).
 \]
 Now we prove the analogue of Lemma \ref{LemmaBasicEMRotation} at the linearized level.
 \begin{lem}
 		\label{LemmaBasicEMLinRotation}
 	If $(\dg,\dF)$ solves the linearized Einstein-Maxwell equations \eqref{EqBasicLinEM} linearized around $(g,F)$, then $(\dg,\dF_\theta+c F_{\theta+\frac{\pi}{2}})$ solves \eqref{EqBasicLinEM} linearized around $(g,F_\theta)$ for any $\theta, c\in\BR$.
 \end{lem}
\begin{proof}
	Since $dF=0,d\Big(\frac{d}{ds}\Big\lvert_{s=0}\big(\star_{g+s\dg}(F+s\dF)\big)\Big)=0$, it follows that 
	\[
	d(\dF_\theta+c F_{\theta+\frac{\pi}{2}})=\cos(\theta)d\dF + \sin(\theta)d\Big(\frac{d}{ds}\Big\lvert_{s=0}\big(\star_{g+s\dg} (F+s\dF)\big)\Big)+cd F_{\theta+\frac{\pi}{2}}=0.
	\] 
	We next calculate
	\begin{align*}
	\frac{d}{ds}\Big\lvert_{s=0}\Big(\star_{g+s\dg}\big(F_\theta+s(\dF_\theta+c F_{\theta+\frac{\pi}{2}})\big)\Big)&=	\frac{d}{ds}\Big\lvert_{s=0}\Big(\star_{g+s\dg}\big(F_\theta+s\dF_\theta\big)\Big)+c\star_g F_{\theta+\frac{\pi}{2}}\\
	&=\cos(\theta)\frac{d}{ds}\Big\lvert_{s=0}\Big(\star_{g+s\dg} (F+s\dF)\Big)-\sin(\theta)\dF+c\star_g F_{\theta+\frac{\pi}{2}}
	\end{align*}
	and this implies 
	\[
	\frac{d}{ds}\Big\lvert_{s=0}\Big(\delta_{g+s\dg}\big(F_\theta+s(\dF_\theta+c F_{\theta+\frac{\pi}{2}})\big)\Big)=0.
	\]
	According to the calculation for $T(g,F)$ in the proof of Lemma \ref{LemmaBasicEMRotation}, with $(F+s\dF)_{\theta}=\cos(\theta)(F+s\dF)+\sin(\theta)\star_{g+s\dg}(F+s\dF)$ we have
	\[
	T(g+s\dg,F+s\dF)=T(g+s\dg, (F+s\dF)_{\theta})=T(g+s\dg, F_\theta+s\dF_{\theta})+\mathcal{O}(s^2)
		\]
	and thus
	\begin{align*}
D_{(g,F_\theta)}T(\dg, \dF_\theta+cF_{\theta+\frac{\pi}{2}})&=	D_{(g,F)}T(\dg,\dF)+\frac{c}{2} g^{\lambda\sigma}((F_\theta)_{\mu\lambda}(F_{\theta+\frac{\pi}{2}})_{\nu\sigma}+(F_{\theta+\frac{\pi}{2}})_{\mu\lambda}(F_{\theta})_{\nu\sigma})\\&\qquad+\frac{c}{2} g^{\lambda\sigma}((\star_gF_\theta)_{\mu\lambda}(\star_gF_{\theta+\frac{\pi}{2}})_{\nu\sigma}+(\star_gF_{\theta+\frac{\pi}{2}})_{\mu\lambda}(\star_gF_{\theta})_{\nu\sigma})\\
&=D_{(g,F)}T(\dg,\dF).
	\end{align*}
This implies that $(\dg,\dF_\theta+c F_{\theta+\frac{\pi}{2}})$ solves the linearization of the equation $\Ric-2T=0$ around $(g, F_\theta)$.
\end{proof}

If $M\cong\BR\times\Sigma_0$, we can define linearized charges as follows
\begin{equation}\label{EqBasicLinEHCharges} 
	\dot{Q}_e(\dg,\dF) := \frac{1}{4\pi}\int_S \frac{d}{ds}\Big\lvert_{s=0}\star_{g+s\dg} (F+s\dF),\qquad \dot{Q}_m(\dF) := \frac{1}{4\pi}\int_S \frac{d}{ds}\Big\lvert_{s=0}(F+s\dF)=\frac{1}{4\pi}\int_S \dF.
\end{equation}
If $(\dg,\dF)$ solves the linearized Einstein-Maxwell equations, by Stokes' theorem and the linearized equations \eqref{EqBasicLinEM} again, $\dot{Q}_e$ and $\dot{Q}_m$ are independent of the specific choice of $S$. In particular, we can choose $S\subset\Sigma_0$, and use the linearization of \eqref{EqBasicEandHFields} around $(g,F)$ to find that 
\begin{align*}
	\dot{Q}_e(\dg,\dF) &= \dot{Q}_e(\dot{h},\dot{\mathbf{E}}) := \frac{1}{4\pi}\int_S \frac{d}{ds}\Big\lvert_{s=0}\star_{h_s}\mathbf{E}_s,\\
	\dot{Q}_m(\dF)&=Q_m(\dot{\mathbf{H}}):=\frac{1}{4\pi}\int_S\frac{d}{ds}\Big\lvert_{s=0}\star_{h_s}\mathbf{H}_s.
\end{align*}
We note that the analogue of Lemma~\ref{LemmaBasicEMRotationCharge} holds for the linearized charges.
\begin{lem}
\label{LemmaBasicEMLinRotationCharges}
Let $(\dg,\dF)$ solve the linearized Einstein-Maxwell equations around $(g,F)$. In the setting \eqref{EqBasicEMHole}, there exist $c,\theta\in\BR$ such that $Q_m(F_\theta)=0$ and $\dot{Q}_m(\dF_\theta+cF_{\theta+\frac{\pi}{2}})=0$, with $F_{\theta+\frac{\pi}{2}}$ and $\dot{F}_\theta$ defined in \eqref{EqBasicEMRotationForm} and \eqref{EqBasicEMLinRotationForm}, respectively.
\end{lem} 
\begin{proof}
	First, according to Lemma \ref{LemmaBasicEMRotationCharge}, there exists $\theta\in\BR$ such that $Q_m(F_\theta)=0$. Since 
	\[
	Q^2_m(F_{\theta+\frac{\pi}{2}})=Q_e(g, F_\theta)^2=Q_m(F)^2+Q_e(g,F)^2\neq 0,
	\] 
	we can choose
	\[
	c=-\frac{\dot{Q}_m(\dF_\theta)}{Q_m(F_{\theta+\frac{\pi}{2}})},
	\] 
	with which we have $\dot{Q}_m(\dF_\theta+cF_{\theta+\frac{\pi}{2}})=\dot{Q}_m(\dF_\theta)+cQ_m(F_{\theta+\frac{\pi}{2}})=0$. This finishes the proof.	
\end{proof}
Therefore, in addition to the linearized constraint equation, we may assume that $\dot{Q}_m(\dF)=0$ which is an analogue of condition \eqref{Eqvanishingmcharge} and can always be arranged by Lemma \ref{LemmaBasicEMLinRotationCharges}. Concretely, without loss of generality, suppose that $F$ satisfies $Q_m(F)=0$. Then for any $(\dot{h},\dot{k}, \dot{\mathbf{E}}, \dot{\mathbf{H}})$ satisfying the linear constraint equations, by choosing a suitable $c\in\BR$ such that $\dot{Q}_m(\dot{\mathbf{H}}_\theta)-cQ_m(\mathbf{E})=0$, the condition \eqref{Eqvanishingmcharge} is arranged.

In the context of the present paper, given initial data satisfying the linearized constraints and having vanishing linearized magnetic charge, we solve the following linearized Einstein-Maxwell system
\begin{equation}
	\label{EqBasicLinEinsteinMaxwell}
	\mathscr{L}(\dg,\dA) =\bigl(\mathscr{L}_1(\dg,\dA),\mathscr{L}_2(\dg,\dA)\bigr)= 0,
\end{equation}
where
\begin{equation}
	\label{EqBasicLinEinsteinMaxwellexp}
	\begin{split}
		&\qquad \mathscr{L}_1(\dg,\dA) = D_g\Ric(\dg) - 2 D_{(g,d A)}T(\dg,\dA), \\
		&\qquad \mathscr{L}_2(\dg,\dA) = D_{(g,A)}(\delta_{(\cdot)}d(\cdot))(\dg,\dA).
	\end{split}
\end{equation}
By choosing $(g,A)$, around which we linearize the Einstein-Maxwell equations, as the background Lorentzian metric and $1$-form in the gauge $1$-form $\Upsilon^E$, gauge source function $\theta$ and gauge function $\Upsilon^M$, one can consider the initial value problem for the corresponding linearized gauge-fixed Einstein-Maxwell equations
\begin{equation}
	\label{EqBasicLingaugefixEq}
	\begin{gathered}
	L^E(\dg,\dA):=	D_g(\Ric)(\dg) - \tilde{\delta}_g^*D_g\Big(\Upsilon^E-\theta\Big)(\dg;g) -2 D_{(g,d A)}T(\dg,d\dA)=0, \\
		L^M(\dg,\dA):=D_{(g,A)}(\delta_{(\cdot)} d(\cdot))(\dg,\dA) - d\Bigl(D_{(g,A)}\Upsilon^M(\dg,\dA;g)-D_{g}\Upsilon^M(\dg,A;g)\Bigr) = 0.
	\end{gathered}
\end{equation}
We recall from  \cite[\S 7.5]{W84} and \cite[\S3]{GL91} that
\begin{equation}
	\label{EqBasicLinRic}
	(D_g\Ric)(\dg) = -\frac{1}{2}\Box_g\dg - \delta_g^*\delta_g G_g\dg + \mathscr{R}_g(\dg),
\end{equation}
where $\Box_g=\nabla^\alpha\nabla_\alpha$ is the tensor wave operator, $\mathscr{R}_g(\dg)_{\mu\nu}=R(g)^{\kappa\ \ \lambda}_{\ \mu\nu}\dg_{\kappa\lambda} + \frac{1}{2}(\Ric(g)_{\mu\ }^{\kappa}\dg_{\nu\kappa}+\Ric(g)_{\nu\ }^{\kappa}\dg_{\mu\kappa})$, and
\begin{equation}
	\label{EqBasicLinGaugeEexp}
	D_g\Upsilon^E(\dg;g) = -\delta_g G_g\dg.
\end{equation}
Since $\tilde{\delta}_g^*$ is a zero order modification of $\delta_g^*$ and $D_{(g,dA)}T(\dg,\dA)$ is of order $0$ in $\dg$ and order $1$ in $A$, it follows that the first equation in \eqref{EqBasicLingaugefixEq} has the principal part $-\frac12\Box_g\dg$. We define the \emph{linearized generalized wave map gauge} as
\begin{equation}
\label{EqBasicLingwavemapgauge}
D_g\widetilde{\Upsilon}^E(\dg;g)=D_g(\Upsilon^E-\theta)(\dg;g)=0.
\end{equation}
Analogously, we define the \emph{linearized generalized Lorenz gauge} as
\begin{equation}
	\label{EqBasicLingLorenzgauge}
D_{(g,A)}\widetilde{\Upsilon}^M(\dg,\dA;g,A)=	D_{(g,A)}\Upsilon^M(\dg,\dA;g)-D_{g}\Upsilon^M(\dg,A;g)=-\delta_g\dA=0,
\end{equation}
and the second equation in \eqref{EqBasicLingaugefixEq} also has principal part $(\delta_g d+d\delta_g)\dA$, and only involves up to first derivatives of $\dg$. Thus, the linearized system \eqref{EqBasicLingaugefixEq} is linear hyperbolic system for $(\dg,\dA)$, whose initial value problem we are able to solve.

As discussed in the non-linear case, in order to relate the initial value problem for the linearized Einstein-Maxwell system for $(\dg,\dA)$ to that of the linear hyperbolic system \eqref{EqBasicLingaugefixEq}, we need the an analogue of Lemma \ref{lem:gaugeungauge} at the linearized level.
\begin{lem}\label{lem:gaugeungaugeLin}
	Suppose $(\dg,\dA)$ solves the linearized gauge-fixed Einstein-Maxwell equation \eqref{EqBasicLingaugefixEq}
	 \[
	L(\dg,\dA)=(L^E(\dg,\dA),\, L^M(\dg,\dA))=0
	\]
	and attains the given initial data $(\dot{g}_0,\dot{g}_1,\dot{A}_0,\dot{A}_1)$ at the initial hypersurface $\Sigma_0=\{\mathfrak{t}=0\}$ which satisfy the linearized constraint equations. If $(\dg,\dA)$ satisfies the linearized generalized wave map gauge and Lorenz gauge conditions initially at $\Sigma_0$, that is,
	\begin{equation}\label{eq:gaugeinitialLin}
	D_g\widetilde{\Upsilon}^E(\dg;g)=0,\qquad \delta_g\dA=0\quad \mbox{at}\quad\Sigma_0,
	\end{equation}
	then $(\dg,\dA)$ solves the Einstein-Maxwell equations with the same initial data.
\end{lem}

\begin{proof}
	The proof is completely analogous to that of Lemma \ref{lem:gaugeungauge}.
\end{proof}
Therefore, in order to solve the initial value problem for linearized Einstein-Maxwell equations, we again need to construct the gauged initial data satisfying the conditions \eqref{eq:gaugeinitialLin}. We postpone the construction of the gauged initial data to \S\ref{SsKNGd}.

Finally, we discuss the pure gauge solution generated from the invariance of Einstein-Maxwell system under the gauge transformation introduced in the non-linear setting $(g,A)\mapsto(\phi^*g, \phi^*A+da)$. Linearizing around $(\phi,a)=(\mbox{Id},0)$ the gauge transformation on a solution $(\dg,\dA)$ of \eqref{EqBasicLinEinsteinMaxwell} yields the following transformation acting on $(\dg,\dA)$
\begin{equation}
	\label{EqBasicLinGaugetrans}
	(\dg,\dA) \mapsto (\dg + \mathcal{L}_V g, \dA+ \mathcal{L}_V A + d a),
\end{equation}
and the linearized Einstein-Maxwell system is invariant under the above transformation \eqref{EqBasicLinGaugetrans}. We note that writing $V=\omega^\sharp$, we have $\mathcal{L}_{\omega^\sharp}g=2\delta_g^*\omega$. For notational convenience, we introduce the following definition:
\begin{defn}[{\cite[Definition 2.4]{H18}}]
	\label{DefBasicLinLieDerivative}
	For a vector field $V$ and a tensor $T$, define $\tilde{\mathcal{L}}_T V := \mathcal{L}_V T$.
\end{defn}
According to Remark \ref{rem:gaugerep} in the non-linear setting, any solution $(g,A)$ can be transformed into another one satisfying the generalized wave map and Lorenz gauge conditions. Here we state its analogue in the linear case.
\begin{rem}
Any solution $(\dg,\dA)$ to the Einstein-Maxwell equations can be transformed into another solution $(\dg + \mathcal{L}_V g, \dA+ \mathcal{L}_V A + d a)$ which satisfies the linearized generalized wave map gauge and Lorenz gauge conditions. Concretely, we first solve define new coordinates $(y^\alpha)$ by solving the equation
\begin{equation}
	\label{EqBasicLinGaugeEinEq}
	D_g\widetilde{\Upsilon}^E(\dg+\mathcal{L}_V g;g)=0
\end{equation}
for the vector field $V$. The equation \eqref{EqBasicLinGaugeEinEq} is a wave equation since its principal part is given by $-\Box_g V^\flat$. Then we solve
\begin{equation}
	\label{EqBasicLinGaugeMaxEq}
\delta_g(\dA+\mathcal{L}_VA+da)=0,
\end{equation}
whose principal part is $-\Box_g a$, for the scalar function $a$. If we choose trivial Cauchy data for $V$ and $a$ on $\Sigma_0$, then $(\dg+\mathcal{L}_V g,\dA+\mathcal{L}_V A+d a)$ solves \eqref{EqBasicLinEinsteinMaxwell}, satisfies the linearized generalized wave map and Lorenz gauge conditions, and attains the same initial data $(\dot{h}, \dot{k},\dot{\mathbf{E}},\dot{\mathbf{H}})$ as $(\dg,\dA)$.
\end{rem}

\section{Kerr-Newman black holes}\label{sec:KNblackholes}

In \S\ref{SsRN}, we introduce the Reissner-Nordstr\"{o}m (RN) family of black holes, which is parametrized by $(\Bm_0, 0,\BQ_0)\in\BR\times\BR^3\times\BR$ with $\Bm_0>0, \abs{\BQ_0}<\Bm_0$. In \S\ref{SsKN}, we introduce the Kerr-Newman (KN) family of black holes with parameters $(\Bm,\Ba,\BQ)\in\BR\times\BR^3\times\BR$ close to $(\Bm_0,0,\BQ_0)$ and define this family of slowly rotating KN metric as a smooth family of stationary metrics on a fixed manifold $M$. In \S\ref{SsKNSt}, we introduce several vector bundles over $M$ and analyze the structure of differential operators, which will be used in the subsequent sections, near infinity. In \S\ref{SsKNGd}, we will discuss how to construct the gauged Cauchy data for the gauge-fixed Einstein-Maxwell system  from the given initial data $(h, k,\mathbf{E},\mathbf{H})$ and also consider its linearized version.

%%%%%%%%%%%%%%%%%%%%%%%%%%%%%%%%%%%%%%%%%%%%%%%%%%
\subsection{The Reissner-Nordstr\"{o}m family}
\label{SsRN}

Given a set of parameters 
\begin{equation}
	\label{EqRNParams}
	b_0=(\Bm_0, 0,\BQ_0)\in\BR\times \BR^3\times\BR,\quad \Bm_0>0,\  \abs{\BQ_0}<\Bm_0,
	\end{equation}
where the parameter $\Bm_0>0$ denote the mass and $\BQ_0$ denotes the charge, the Reissner-Nordstr\"{o}m (RN) solution to the Einstein-Maxwell equations is then given by the spherically symmetric and static RN metric
\begin{equation}
	\label{EqRNMetric}
	g_{b_0} =- \mu_{b_0}(r)\,dt^2 + \mu_{b_0}(r)^{-1}\,dr^2 +r^2\sg,
\end{equation}
where $\sg$ is the round metric on $\BS^2$ and 
\begin{equation}
	\label{EqRNMu}
	\mu_{b_0}(r) = 1-\frac{2\Bm_0}{r}+ \frac{\BQ_{0}^2}{r^2},
\end{equation}
and the RN 4-potential for the electromagnetic field
\begin{equation}
	\label{EqRNPotentialStatic}
	\tilde A_{b_0} = \frac{\BQ_{0}}{r}dt,\quad \tilde F_{b_0}=d\tilde A_{b_0} = -\frac{\BQ_0}{r^2}dr\wedge dt.
\end{equation}
When $\abs{\BQ_0}<\Bm_0$, the expression $\mu_{b_0}$ has two distinct positive roots $\Bm_0-\sqrt{\Bm_0^2-\BQ_0^2}=r_c<r_{b_0}=\Bm_0+\sqrt{\Bm_0^2-\BQ_0^2}$. There is a Cauchy horizon at $r=r_c$ and an \emph{event horizon} $\mathcal{H}$ at $\ehRN$. The expressions \eqref{EqRNMetric} and \eqref{EqRNPotentialStatic} are valid in the \emph{static region}, which sometimes is also called exterior region or domain of outer communication
\begin{equation}
	\label{EqRNStaticRegion}
	\mathcal{M} = \BR_t \times \mathcal{X},\quad \mathcal{X} = (\ehRN, \infty)_r \times \BS^2.
\end{equation}
The singularity of the expression \eqref{EqRNMetric} at $r=\ehRN$ is a coordinate singularity and can be removed by change of coordinates. We define a new coordinate system $(t_0, r,\theta, \varphi)$, which is called \emph{incoming Eddington-Finkelstein coordinates} and make use of the function $t_0$
\begin{equation}
	\label{EqRNNullCoord}
	t_0 = t +r_{b_0,*},\quad \frac{dr_{b_0,*}}{dr}=\frac{1}{\mu_{b_0}},
\end{equation}
where $r_{b_0,*}$ is called the \emph{tortoise coordinate} or \emph{Regge-Wheeler radial coordinate}. In this coordinate system, the RN metric is given by
\begin{equation}
	\label{EqRNMetricNull}
	g_{b_0} = -\mu_{b_0}(r)dt_0^2+2\,dt_0dr+r^2\sg.
\end{equation}
Now the expression \eqref{EqRNMetricNull} for $g_{b_0}$ extends analytically beyond the event horizon $\mathcal{H}=\{r=\ehRN\}$, and thus $g_{b_0}$ is smooth and non-degenerate metric on the extended manifold
\begin{equation}
	\label{EqRNExt}
	M= \BR_{t_0}\times X \supset \mathcal{M},\quad X=[r_-,\infty)_r\times\BS^2 \supset \mathcal{X},
\end{equation}
where $r_-\in(r_c, \ehRN)$. At the same time, we can take the RN $4$-potential to be $\BQ_0 r^{-1}\,dt_0$, which differs from $\tilde A_{b_0}$ by an exact form $d(\int\!\BQ_0 r^{-1}\mu^{-1}\,dr)$.

Since $dt$ is timelike when $r$ is sufficiently large, to capture this useful structure, we introduce another coordinate
\begin{equation}
	\label{EqRNTimefcnChi}
	t_{\chi_0} := t + \int \frac{\chi_0(r)}{\mu_{b_0}(r)}\,d r,
\end{equation}
where $\chi_0(r)\in C^\infty(\BR)$ is identically $1$ for $r\leq 4\ehRN/3$, and vanishes for $r\geq 3\ehRN/2$. Therefore, $t_{\chi_0}$ is a smooth interpolation between $t_0$ near event horizon and $t$ near infinity $r=\infty$ with a suitable choice of the constant of integration. Then we define the RN 4-potential as 
\begin{equation}\label{EqRNpotential}
	A_{b_0}=\frac{\BQ_0}{r}dt_{\chi_0}.
\end{equation}

In order to describe the waves near the future event horizon and future null infinity, we introduce the following function
\begin{equation}
	\label{EqRNTimefcn}
	t_* =t_{b_0,*}:= t +r_{b_0,*}\chi_0(r) - r_{b_0,*}(1-\chi_0(r)),
\end{equation}
where $\chi_0(r)$ is defined as in \eqref{EqRNTimefcnChi}; it smoothly interpolates between $t+r_{b_0,*}$ near the event horizon and $t-r_{b_0,*}$ near future null infinity. Later on, we will study forcing problems for wave equations of the type $\Box_{g_{(\Bm_0,0,\BQ_0)}}u=f$ on the manifold \[
\BR_{t_*}\times X, \quad X=[r_-,\infty)_r\times\BS^2,
\]
where $f$ is supported in $t_*\geq 0$.

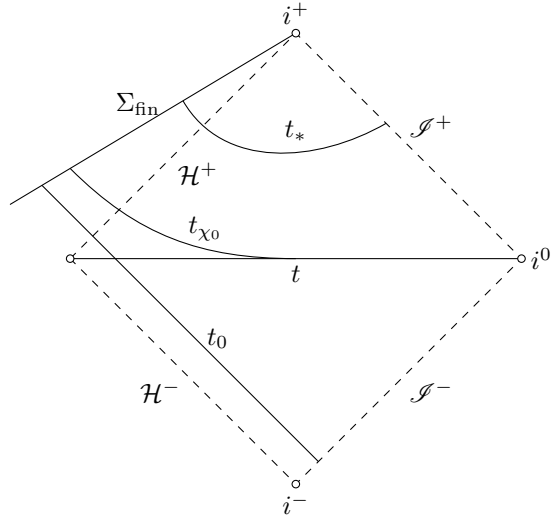
\begin{figure}[!h]
	\begin{tikzpicture}
		\draw [dashed] (-3,0)-- (0,3)--(3,0)--(0,-3)--cycle;
		\draw (-3,0)--(3,0) node [black,midway,xshift=0cm,yshift=-0.2cm
		] 
		{$t$};
		\draw (0,3)--(-3.8, 0.72) node [black,midway,xshift=-0.2cm,yshift=0.2cm
		] 
		{$\Sigma_{\mathrm{fin}}$};
	\draw (-3.375, 0.975)--(0.3, -2.7) node [black,midway,xshift=0.5cm,yshift=-0.2cm
	] 
	{$t_0$};
		\draw (-3, 1.2) to [out=-45, in=180] (0,0) node [black,midway,xshift=-1.2cm,yshift=0.4cm
	] 
	{$t_{\chi_0}$};
		\draw (-1.5, 2.1) to [out=-60, in=-150] (1.2,1.8) node [black,midway,xshift=0cm,yshift=1.7cm
	] 
	{$t_{*}$};
		\draw[fill=white]
		(-3,0) circle (0.5mm) (0,3) circle (0.5mm) 	(3,0) circle (0.5mm) (0,-3) circle (0.5mm);
		\node [above] at (0,3) {$i^+$};
			\node [below] at (0,-3) {$i^-$};
				\node [right] at (3,0) {$i^0$};
				\node [right, above] at(1.8,1.5) {$\mathscr{I}^+$};
					\node [right, below] at(1.8,-1.5) {$\mathscr{I}^-$};
						\node [left, below] at(-1.8,-1.5) {$\mathcal{H}^-$};
							\node [right, below] at(-1.3,1.4) {$\mathcal{H}^+$};
		\end{tikzpicture}
	\caption{ Penrose diagram of the Reissner-Nordstr\"{o}m metric (including future/past event horizon $\mathcal{H}^\pm$,, future/past null infinity $\mathscr{I}^\pm$,  future/past timelike infinity $i^\pm$, and spacelike infinity $i^0$), together with the level sets of the static time function $t$, of the null coordinate $t_0$ from \eqref{EqRNNullCoord}, of the coordinate $t_{\chi_0}$ from \eqref{EqRNTimefcnChi}, and of the function $t_*$ from \eqref{EqRNTimefcn}. Also shown are the boundaries $\Sigma_{\mathrm{fin}}=\BR_{t_*}\times\pa_-\CX$ and $\BR_{t_*}\times\pa_+\CX$ (future null infinity $\mathscr{I}^+$).}
	\label{FigRNTime}
\end{figure}

Recall the definition \ref{defn:radialcom} of radial compactification, now we compactify $X$ in a similar manner
\begin{equation}
	\label{EqradialcomofX}
\CX:=\big(X\sqcup ([0,\ehRN^{-1})_\rho\times\BS^2)\big)/\sim
\end{equation}
where $\sim$ identifies $(\rho, \omega)\in[0,\ehRN)_\rho\times\BS^{n-1}$ with the point $\rho^{-1}\omega\in X$. Therefore, we have $\CX=\overline{\{r\geq r_-\}}\subset\overline{\BR^3}$ with $\pa_-\CX=r^{-1}(r_-)$ and $\pa_+\CX=\pa\overline{\BR^3}=\{\rho=0\}\subset \CX$. Then we see that $\BR_{t_*}\times\pa \CX$ has two components: one is a spacelike hypersurface inside the black hole and given by
\begin{equation}
	\label{EqRNSurfFin}
	\Sigma_{\rm fin} := \BR_{t_*}\times\pa_- \CX,
\end{equation}
 and the other is the future null infinity $\BR_{t_*}\times\pa_+ X$, which is also denoted by $\mathscr{I}^+$. See Figure \ref{FigRNTime} for the level sets of $t,\ t_0,\ t_{\chi_0},\ t_{*}$.

%%%%%%%%%%%%%%%%%%%%%%%%%%%%%%%%%%%%%%%%%%%%%%%%%%
\subsection{The slowly rotating Kerr-Newman family}
\label{SsKN}
Let $b_0=(\Bm_0,0,\BQ_0)$ be a parameter of a RN black hole. Consider Kerr-Newman black hole parameters \begin{equation}
	\label{EqKNparameter}
	b=(\Bm,\Ba,\BQ)\in\BR\times\BR^3\times\BR,
	\end{equation}
where $\Ba\in\BR^3$ is the angular momentum. If $a=|\Ba|\neq 0$, we choose the polar coordinates $(\theta,\varphi)$ on $\BS^2$ such that $\hat\Ba=\Ba/|\Ba|$ is defined by $\theta=0$, and the vector field $\pa_\varphi$ generates counterclockwise rotation around $\Ba/\abs{\Ba}$, with $\BR^3$ carrying the standard orientation. A very convenient coordinate system for the Kerr-Newman metric is Boyer-Lindquist coordinates $(t,r,\theta,\varphi)$ (see \cite{BL67}), in which the Kerr-Newman metric takes the form
\begin{equation}
	\label{EqKNMetric}
	\begin{split}
		g_b^{\mathrm{BL}} &= -\frac{\Delta_b}{\rho_b^2}(d t-a\sin^2\theta\,d\varphi)^2 + \rho_b^2\Bigl(\frac{d r^2}{\Delta_b}+d\theta^2\Bigr) + \frac{\sin^2\theta}{\rho_b^2}\bigl(a\,d t-(r^2+a^2)d\varphi\bigr)^2, \\
			(g_b^{\mathrm{BL}} )^{-1}&= -\frac{1}{\Delta_b\rho_b^2}\bigl((r^2+a^2)\pa_t+a\pa_\varphi\bigr)^2 + \frac{\Delta_b}{\rho_b^2}\pa_r^2 + \frac{1}{\rho_b^2}\pa_\theta^2 + \frac{1}{\rho_b^2\sin^2\theta}(\pa_\varphi+a\sin^2\theta\,\pa_t)^2, \\
		&\quad \Delta_{b} = r^2-2\Bm r+\BQ^2+a^2, \ \ 
		\rho_{b}^2 = r^2+a^2\cos^2\theta,\quad a=\abs{\Ba}.	
	\end{split}
\end{equation}
In the Boyer-Lindquist coordinates, the KN $4$-potential reads
\[
\tilde A_b=\frac{\BQ r}{\rho_b^2}(dt-a\sin^2\theta\,d\varphi)
\]
and the corresponding electromagnetic $2$-form is 
\begin{align*}
\tilde F_b&=-\frac{\BQ}{\rho_b^4}\Big((r^2-a^2\cos^2\theta)dr\wedge(dt-a\sin^2\theta\,d\varphi)\Big)\\
&\quad +\frac{\BQ r}{\rho_b^4}\Big(2 a \cos\theta\sin\theta\,d\theta\wedge(a\,dt-(r^2+a^2)\,d\varphi)\Big).
\end{align*}
We note that $(g_b^{\mathrm{BL}}, \tilde A_b)$ is a solution to the Einstein-Maxwell equations.

In the initial value problem for the Einstein-Maxwell equations, one places asymptotically flat initial data on a Cauchy hypersurface
\[
\label{EqKNSigma0}
\Sigma_0 \subset M,
\]
which is spacelike and equal to $t^{-1}(0)$ where $r$ is large. We introduce the following coordinates
\begin{equation}
	\label{Eqtimeliket}
	\mathfrak{t}=t+\int	\chi_1(r)\Big(\frac{r^2+a^2}{\Delta_b}\sqrt{1-\frac{\Delta_b}{r^2+a^2}}\Big)dr,\quad \varphi_1=\varphi+\int\chi_1(r)\frac{a}{\Delta_b}\,dr
\end{equation}
where $\chi_1(r)$ is a nonnegative smooth function such that $\chi_1(r)=1$ for $r\leq 3\Bm$ and $\chi_1(r)=0$ for $r\geq 4\Bm$. In the new coordinates $(\mathfrak{t}, r, \theta, \varphi_1)$, $g^{\mathrm{BL}}_b$ and $(g^{\mathrm{BL}}_b)^{-1}$ are smooth at event horizon, and furthermore
\[
(g^{\mathrm{BL}}_b)^{-1}(d\mathfrak{t}, d\mathfrak{t})=-1+\frac{\chi^{2}-1}{\rho_b^2}\Big(\frac{(r^2+a^2)^2}{\Delta_b}-r^2+a^2\Big)=\begin{cases}
	-1,&\quad r\leq 3\Bm\\
	\leq -1,&\quad 3\Bm\le r\leq 4\Bm\\
	(g^{\mathrm{BL}}_b)^{-1}(dt,dt),&\quad r\geq 4\Bm
\end{cases}.
\]
By choosing a proper integration constant we have $\mathfrak{t}=t$ for $r\geq 4\Bm$. Then we impose asymptotically flat initial data on the Cauchy hypersurface
\begin{equation}
\label{EqKNSigma0Ex}
\Sigma_0:=\{\mathfrak{t}=0\} \subset M.
\end{equation}

In the present paper, we consider the parameters $(\Bm,\Ba,\BQ)$ close to $(\Bm_0,0,\BQ_0)$, that is, we work on slowly rotating ($\abs{\Ba}\ll\Bm+\abs{\BQ}$) Kerr-Newman black holes. As in the RN case, the singularity of the expression \eqref{EqKNMetric} of the KN metric at
\begin{equation}
	\label{EqKNRadius}
	r=r_b:=\Bm+\sqrt{\Bm^2-(\abs{\BQ}^2+\abs{\Ba}^2)}.
\end{equation}
is a coordinate singularity and can be removed by introducing new coordinates. We let 
\begin{equation}
	\label{EqKNNull}
	t_{b,\chi} = t + \int\frac{r^2+a^2}{\Delta_b}\chi(r)\,d r, \quad
	\varphi_{b,\chi} = \varphi + \int\frac{a}{\Delta_b}\chi(r)\,d r.
\end{equation}
where $\chi(r)\in C^\infty(\BR)$ is identically $1$ near $r\leq 
\Bm+\sqrt{\Bm^2-\BQ^2}$. Then the metric $g_b^{\rm BL}$ and its inverse take the form
\begin{equation}
	\label{EqKNMetric2}
	\begin{split}
		g_b^{\mathrm{BL}} &= -\frac{\Delta_b}{\rho_b^2}(d t_{b,\chi}-a\sin^2\theta\,d\varphi_{b,\chi})^2+2\chi(d t_{b,\chi}-a\sin^2\theta\,d\varphi_{b,\chi})d r \\
		&\hspace{6em} + \frac{\rho_b^2}{\Delta_b}(1-\chi^2)d r^2 + \rho_b^2\,d\theta^2+\frac{\sin^2\theta}{\rho_b^2}\bigl(a\,d t_{b,\chi}-(r^2+a^2)d\varphi_{b,\chi}\bigr)^2,\\
		(g_b^{\mathrm{BL}} )^{-1}&=\rho_b^{-2}\bigg(\Delta_b\pa_r^2+\frac{\chi^2-1}{\Delta_b}\Big((r^2+a^2)\pa_{t_\chi}+a\pa_{\varphi_{b,\chi}}\Big)^2+\pa_\theta^2+\frac{1}{\sin^2\theta}\Big(\pa_{\varphi_{b,\chi}}+a\sin^2\theta\pa_{t_{\chi}}\Big)^2\\
		&\qquad\qquad+2\chi\Big((r^2+a^2)\pa_{t_\chi}+a\pa_{\varphi_{b,\chi}}\Big)\pa_r\bigg),
	\end{split}
\end{equation}
which is smooth and non-degenerate on the extended manifold $M_b=\BR_{t_{b,\chi}}\times [r_-, \infty)_r\times\BS^2_{\theta, \varphi_{b,\chi}}$ with $r_-$ defined in \S \ref{SsRN}. Correspondingly, we have
\[
\tilde A_b=\frac{\BQ r}{\rho_b^2}(dt_{b,\chi}-a\sin^2\theta\,d\varphi_{b,\chi})-\chi\frac{\BQ r}{\Delta_b}dr.
\]
Since the last term is an exact $1$-form which makes no contribution to the electromagnetic $2$-form, it follows that one can define the KN $4$-potential on $M_b$ as
\begin{equation}\label{EqKNPotential}
A^{\mathrm{BL}}_b:=\frac{\BQ r}{\rho_b^2}(dt_{b,\chi}-a\sin^2\theta\,d\varphi_{b,\chi}).
\end{equation}
Near the poles $\theta=0,\pi$ of $\BS^2$, we introduce the smooth coordinates $x=\sin\theta\cos\varphi_{b,\chi}$ and $y=\sin\theta\sin\varphi_{b,\chi}$. Then we have $\sin^2\theta=x^2+y^2$ and 
\[
\pa_{\varphi_{b,\chi}} = x\pa_y-y\pa_x .\quad \pa_\theta=\cos\theta\cos\varphi_{b,\chi}\pa_x+\cos\theta\sin\varphi_{b,\chi}\pa_y.
\]
Therefore, the smoothness of $	(g_b^{\mathrm{BL}} )^{-1}$ near the poles $\theta=0,\pi$ (i.e, $x=y=0$) follows from writing $\pa_\theta^2+\sin^{-2}\theta\pa^2_{\varphi_{b,\chi}}$ as
\begin{equation}\label{eq:smoothnessnearpoles}
\pa_\theta^2+\frac{1}{\sin^2\theta}\pa_{\varphi_{b,\chi}}^2=(1-x^2-y^2)\big(\pa_x^2+\pa_y^2\big)-(x\pa_x+y\pa_y)^2,
\end{equation}
which is smooth near $x=y=0$. Since the volume form is given by
\[
dvol_{g_b^{\mathrm{BL}}}=(r^2+a^2(1-x^2-y^2))(1-x^2-y^2)^{-1/2}\,dt_{b,\chi}\,dr\,dx\,dy
\]
and thus smooth at the poles $\theta=0,\pi$, we conclude that $g^{\mathrm{BL}}_b$ indeed extends smoothly and non-degenerately to the poles $\theta=0,\pi$. 
 
 For $b=(\Bm,\Ba,\BQ)$ close to $(\Bm_0, 0,\BQ_0)$, we take $\chi=\chi_0$ as defined in \eqref{EqRNTimefcnChi} and choose suitable constants of integration such that $t_{b,\chi_0}=t$, $\varphi_{b,\chi_0}=\varphi$ for $r\geq 3\ehRN/2$. Then we define the following diffeomorphism
\begin{equation}
	\begin{split}
\Phi_b \colon M=\BR_{t_{\chi_0}}\times X=\BR_{t_{\chi_0}}\times [r_-, \infty)_r\times\BS^2_{\theta, \varphi}& \to M_b=\BR_{t_{b,\chi_0}}\times [r_-, \infty)_r\times\BS^2_{\theta, \varphi_{b,\chi_0}},\\
(t_{b,\chi_0},r,\theta,\varphi_{b,\chi_0})(\Phi_b(p)) &=(t_{\chi_0},r,\theta,\varphi)(p)
\end{split}
\end{equation}
where $t_{\chi_0}$ is defined in \eqref{EqRNTimefcnChi} and $(\theta,\varphi)$ are polar coordinates on $\BS^2$ satisfying that $\Ba/\abs{\Ba}$ is defined by $\theta=0$. Using this family of diffeomorphisms $\Phi_b$, we define
\begin{equation}
	\label{EqKNMetricEmbed}
	g_b = (\Phi_b)^*(g_b^{\rm BL})\in C^\infty(M;S^2T^*M),\quad 	A_b = (\Phi_b)^*(A^{\mathrm{BL}}_b)\in C^\infty(M;T^*M)
\end{equation}
which is a family of smooth and stationary metric and $1$-form on $M$, respectively. For $b=b_0=(\Bm_0,0,\BQ_0)$, this pullback indeed gives rise to the RN black hole $(g_{b_0}, A_{b_0})$. Following the argument in \cite[Proposition~3.5]{HV18}, one can prove that $(g_b, A_b)$ depends smoothly on $b$ when $b$ is close to $b_0$.

\begin{lem}
	\label{lem:smoothdependenceonb}
	Let $\mathcal{U}_{b_0}$ is a sufficiently small neighborhood of $b_0=(\Bm_0, 0,\BQ_0)$ and let $M=\BR_{t_{\chi_0}}\times X$ with $X=\{p\in\BR^3\mid \abs{p}\geq r_-\}$. Then $(g_b, A_b)\in C^\infty(M;S^2T^*M\oplus T^*M)$ depends smoothly on $b\in\mathcal{U}_{b_0}$.
           \end{lem}
\begin{proof}
	Given $(\Bm, \Ba, \BQ)$ with $a=\abs{\Ba}\neq 0$, let us denote the spherical coordinate system with north pole $\theta=0$ at $\Ba/\abs{\Ba}$ by $(\theta_b, \varphi_b)$, so the pullbacks of the functions $\Delta_b, \rho_b^2$ to $M$ are simply obtained by replacing $\theta$ by $\theta_b$. Then if $(r, \theta_b,\varphi_{b})$ are the polar coordinates of a point $p\in X$, we have
	\[
	r=\abs{p}, \quad a\cos\theta_b=\angles{\Ba}{\frac{p}{\abs{p}}},\quad a^2\sin^2\theta_b=\abs{\Ba}^2-a^2\cos^2\theta_b
	\]
	where $\abs{\cdot}$ and $\langle \cdot,\cdot\rangle$ denote the Euclidean norm and inner product on $X\subset\BR^3$,
	respectively. Since $r\neq 0$, this shows that $r, a\cos\theta_b$ and $a^2\sin^2\theta_b$, and thus the
	pullbacks of $\Delta_b$ and $\rho_b$ are smooth (in $b$) families of smooth functions on $M$. Then the smoothness of $g_b$ and $A_b$ in $b$ follows from that of $a\sin^2\theta_b d\varphi_{b}=(ar^{-2}\pa_{\varphi_{b}})^\flat$ (using the musical isomorphism on Euclidean $\BR^3$), which we will show subsequently.
	
	We now prove the smooth dependence of the inverse metric $g_b^{-1}$: in view of
	the expression \eqref{EqKNMetric2} and the discussion around \eqref{eq:smoothnessnearpoles}, all we need to show is that
	the vector fields $\pa_{t_{\chi_0}}, \pa_r, a\pa_{\varphi_{b}}$ and $a\sin\theta_b\pa_{\theta_b}$ depend smoothly on $\Ba$. For $\pa_{t_{\chi_0}}$ and for the radial vector field $\pa_r=\abs{p}^{-1}p\pa_p$, which do not depend on $b$, the smoothness is clear. Further, we have
	\[
	a\pa_{\varphi_{b}}=\nabla_{\Ba\times p}\quad \mbox{at}\quad p\in X,
	\]
and the expression on the right-hand side is smooth in $\Ba$. Indeed, if $\Ba= a\vec{e_3}:=(0,0,a)$, both sides equal $a(x\pa_y-y\pa_x)$ on $\BR^3_{x,y,z}$, and if $\Ba\in\BR^3$ is any given vector and $R\in SO(\BR^3)$ is a rotation with $R\Ba=a\vec{e}_3$, $a=\abs{\Ba}$, then $(R_*(a\pa_{\varphi_{b}}))|p=(a\pa_{\varphi_{a\vec{e}_3,}})| _{R(p)}$ which we just observed to be equal to $\nabla_{a\vec{e_3}\times R(p)}=R_*\nabla_{\Ba\times p}$.

	In a similar manner, we have
	\[
	a\sin\theta_b\pa_{\theta_b}=\abs{p}^{-1}\nabla_{p\times(p\times\Ba)}\quad \mbox{at} \quad p\in X,
	\]
	and again the expression on the right-hand side is smooth in $\Ba$. This finishes the proof of the Lemma.
\end{proof}

\begin{rem}\label{rem:KNMagSols}
	We consider the enlarged parameter space
	\begin{equation}
		\label{EqKNMagParameter}
		\mathcal{U}_{b_{0,m}}= \{ (\Bm,\Ba,\BQ_e,\BQ_m) \} \subset \BR\times\BR^3\times\BR\times\BR.
	\end{equation}
	Let $\mathcal{U}_{b_{0,m}}$ be a small neighborhood of a RN parameter $
	b_{0,m} = (\Bm_0,0,\BQ_{0,e},\BQ_{0,m})$ with magnetic charge. Now we define the map
	\[
	 \beta\colon \mathcal{U}_{b_{0,m}} \ni (\Bm,\Ba,\BQ_e,\BQ_m) \mapsto (\Bm,\Ba,\sqrt{\BQ_e^2+\BQ_m^2}) \in \mathcal{U}_{b_0}	\]
	and then for $	b_m = (\Bm,\Ba,\BQ_e,\BQ_m) \in \mathcal{U}_{b_{0,m}}$, we define
	\begin{equation}
		\label{EqKNMagSolutions}
	g_{b_m} := g_{\beta(b_m)},\quad F_{b_m} := \BQ_e F_{(\Bm,\Ba,1)} + \BQ_m \star_{g_{b_m}} F_{(\Bm,\Ba,1)}.
	\end{equation}
In view of Lemma~\ref{LemmaBasicEMRotation}, we see that $(g_{b_m},F_{b_m})$ is a solution of the Einstein-Maxwell equations \eqref{EqBasicEMCurv1}--\eqref{EqBasicEMCurv2}. Since the charge always appears in the form of square in the KN metric, it follows that $g_{b_m}$ and $F_{b_m}$ depend smoothly on $b_m\in \mathcal{U}_{b_{0,m}}$.
\end{rem}

\begin{rem}
	The diffeomorphism used to realize the family of KN black holes as a smooth family of black holes on a fixed manifold is not unique. We can also consider coordinates $t_{b,1}$, $\varphi_{b,1}$ (that is, take $\chi\equiv 1$ in the definitions of $t_{b,\chi}$ and $\varphi_{b,\chi}$) and use the following diffeomorphism
	\begin{equation}
		\begin{split}
			\Phi^0_b \colon M=\BR_{t_0}\times [r_-, \infty)_r\times\BS^2_{\theta, \varphi}& \to M_b=\BR_{t_{b,1}}\times [r_-, \infty)_r\times\BS^2_{\theta, \varphi_{b,1}},\\
			(t_{b,1},r,\theta,\varphi_{b,\chi_0})(\Phi^0_b(p)) &=(t_0,r,\theta,\varphi)(p)
		\end{split}
	\end{equation}
to define
	\begin{equation}
		\label{EqKNMetricEmbedalter}
		g^0_b = (\Phi^0_b)^*(g_b^{\rm BL})\in C^\infty(M;S^2T^*M),\quad 	A^0_b = (\Phi^0_b)^*(A^{\mathrm{BL}}_b)\in C^\infty(M;T^*M).
	\end{equation}
Again, $(g^0_b, A^0_b)$ depends smoothly on $b$ when $b$ is close to $b_0$; for $b=b_0=(\Bm_0,0,\BQ_0)$, the above pullback \eqref{EqKNMetricEmbedalter} produces the RN black hole $(g_{b_0}, A_{b_0})$. We point out that this representation is more useful when we do calculations near the event horizon. In particular, for $b=(\Bm, 0,\BQ)$, we have 
	\begin{equation}
		\label{EqKNMetric2RN}
		g_{(\Bm,0,\BQ)}^0 = -\mu_{(\Bm,0,\BQ)}d t_0^2 +2 d t_0\,d r+r^2\sg,\quad A^0_{(\Bm,0,\BQ)}=\frac{\BQ}{r}dt_0,\quad
		\mu_{(\Bm,0,\BQ)}= 1-\frac{2\Bm}{r}+\frac{\BQ}{r^2}.
	\end{equation}
\end{rem}

For $b=(\Bm,\Ba,\BQ)$ close to $b_0=(\Bm_0, 0,\BQ_0)$, we also introduce the following function
\begin{equation}
	\label{EqKNTimeFn}
	t_{b,*} := t+r_{b_0, *}\chi_0(r)-r_{(\Bm, 0,\BQ),*}(1-\chi_0(r)),
\end{equation}
which generalizes $t_*=t_{b_0,*}$ defined in \eqref{EqRNTimefcn}; it equals $t-r_{(\Bm, 0,\BQ),*}$ for $r\geq 3\ehRN/2$.

Now we turn to the discussion of the linearized KN solutions.
\begin{defn}
	\label{DefofKNLin}
	For $\dot{b}\in\BR\times\BR^3\times\BR$, we define the \emph{linearized KN solutions} as 
	\begin{equation}
		\label{EqDefofKNLin}
	\dg_b(\dot{b}) = D_bg_{b}(\dot{b}) = \frac{d}{ds}\Big\lvert_{s=0}g_{b+s \dot{b}},\quad 	\dA_b(\dot{b}) = D_bA_{b}(\dot{b}) = \frac{d}{ds}\Big\lvert_{s=0}A_{b+s \dot{b}}.
	\end{equation}
\end{defn}
Therefore, $(\dg_b(\dot{b}),\dA_b(\dot{b}))$ is a solution of the linearized Einstein-Maxwell equations \eqref{EqBasicLinEinsteinMaxwell} linearized around $(g_b,A_b)$. 

We can also linearize the family $(g_b^0, A^0_b)$ in the parameter $b$ to obtain another version of linearized KN solutions
	\begin{equation}
	\label{EqDefofKNLinalt}
	\dg_b^0(\dot{b}) = D_bg^0_{b}(\dot{b}) = \frac{d}{ds}\Big\lvert_{s=0}g^0_{b+s \dot{b}},\quad 	\dA^0_b(\dot{b}) = D_bA^0_{b}(\dot{b}) = \frac{d}{ds}\Big\lvert_{s=0}A^0_{b+s \dot{b}}.
\end{equation}
Here we list the particular case
\begin{equation}
	\label{EqKNLin2}
	\begin{split}
	\dg_{(\Bm_0,0,\BQ_0)}^0(\dot{\Bm},0,\dot{\BQ})& = \big(\frac{2\dot{\Bm}}{r}-\frac{2\BQ_0\dot{\BQ}}{r^2}\big)\,d t_0^2, \quad 	\dA_{(\Bm_0,0,\BQ_0)}^0(\dot{\Bm},0,\dot{\BQ})=\frac{\dot{\BQ}}{r}dt_0\\
	\dg_{(\Bm_0,0,\BQ_0)}^0(0,\dot\Ba,0) &= ((-\frac{4\Bm_0}{r}+\frac{2\BQ_0}{r^2}) d t_0-2 d r)\sin^2\theta\,d\varphi,\quad 	\dA_{(\Bm_0,0,\BQ_0)}^0(0,\dot\Ba,0) =-\frac{\BQ}{r}\sin^2\theta d\varphi,
	\end{split}
\end{equation}
where $|\dot\Ba|=1$ and $(\theta,\varphi)$ are spherical coordinates adapted to $\dot\Ba$. 
\begin{rem}
	\label{RmkKNLie}
	Since $g_b=((\Phi_b^0)^{-1}\circ\Phi_b)^*g_b^0$, the linearized KN solution $(\dot g_{b}(\dot{b}),\dot A_{b}(\dot{b}))$ can be obtained from $(\dot g^0_{b}(\dot{b}),\dot g^0_{b}(\dot{b}))$ by subtracting off a Lie derivative of $(g^0_b,A^0_b)$ along a vector field generating the flow $(\Phi_b^0)^{-1}\circ\Phi_b$. More specifically, in the coordinates $(t_*,r,\theta,\varphi)$ on $M$, one has
	\begin{subequations}
		\begin{equation}
			\label{EqKNLie}
		(\Phi_b^0)^{-1}\circ\Phi_b \colon (t_*,\,r,\,\theta,\,\varphi) \mapsto \left(t_*+\int\left(\frac{r^2+a^2}{\Delta_b}-\frac{r^2}{\Delta_{b_0}}\right)(1-\chi_0)\,d r,\,r,\,\theta,\,\varphi+\int\frac{a}{\Delta_b}(1-\chi_0)\,d r\right).
		\end{equation}
	Therefore, given $\dot b=(\dot\Bm,\dot\Ba,\dot{\BQ})$, the above two versions of the linearized KN solutions are related by
		\begin{equation}
			\label{EqKNLie2}
			\begin{split}
				\dot g_{b_0}(\dot b) &= \dot g_{b_0}^0(\dot b) - \mathcal{L}_{V(\dot b)} g^0_{b_0},\quad \dot g_{b_0}(\dot b) = \dot g_{b_0}^0(\dot b) - \mathcal{L}_{V(\dot b)} A^0_{b_0},\\
				V(\dot b)&=\frac{d}{d s}\bigg|_{s=0}((\Phi_{b_0+s\dot b}^0)^{-1}\circ\Phi_{b_0+s\dot b})\\
				&= \biggl(\dot\Bm\Bigl(\int_{r_0}^r \frac{2 r^3}{\Delta_{b_0}^2}(1-\chi_0)\,d r\Bigr)-\dot{\BQ}\Bigl(\int_{r_0}^r \frac{ 2\BQ_0r^2}{\Delta_{b_0}^2}(1-\chi_0)\,d r\Bigr)\biggr)\pa_{t_*} + \dot\Ba\Bigl(\int_{r_0}^r\frac{1-\chi_0}{\Delta_{b_0}}d r\Bigr)\pa_\varphi.
			\end{split}
		\end{equation}
	\end{subequations}
	where $r_0$ in the definition of $V(\dot b)$, can be chosen freely.
\end{rem}

%%%%%%%%%%%%%%%%%%%%%%%%%%%%%%%%%%%%%%%%%%%%%%%%%%
\subsection{Stationarity, vector bundles, and geometric operators}
\label{SsKNSt}
Now we introduce some basics of the vector bundles and geometric operators (see \cite[\S 3.3]{HHV21}). In the notation~\eqref{EqRNExt}, let
\[
\pi_X \colon M \to X;
\]
be the projection to the spatial part of the manifold $M$, whose definition is independent of the choice of time function. 
%Suppose that $E_1\to X$ is a vector bundle; then differentiation along $\pa_t=\pa_{t_0}=\pa_{t_*}$ is a well-defined operation on sections of the pullback bundle $\pi_X^*E_1$. The tangent bundle of $M$ is an important example of such a pullback bundle, as
%\[
%T M \cong \pi_X^*(T_{t_0^{-1}(0)}M) \cong \pi_X^*(T_{t_*^{-1}(0)}M),
%\]
%likewise for the cotangent bundle and other tensor bundles.
Let $E_1,\ E_2\to X$ be two vector bundles over $X$, and suppose that $\widehat{L}(0)\in\mbox{Diff}(X;E_1,E_2)$ is a differential operator. Let $\mathfrak{t}=t_*+F$ with $F\in C^\infty(X)$ and $d\mathfrak{t}\neq 0$ everywhere, we then define its \emph{stationary extension} $L$ acting on a section $u\in C^\infty(M;\pi_X^*E_1)$ in the following way
\[
u\mapsto(L u)(\mathfrak{t},-):=\widehat{L} (0)(u(\mathfrak{t},-))\in C^\infty(M;\pi_X^*E_2).
\]
We point out that this stationary extension indeed depends on the choice of the time function $\mathfrak{t}$. However, when acting on a stationary section, the stationary extension defined above is \emph{independent} of the choice of the time function $\mathfrak{t}$ since
\begin{equation}
	\label{EqKNStStat}
	L\pi_X^*=\pi_X^*\widehat{L}(0).
\end{equation}
By making use of the stationary extension, one can regard $\mbox{Diff}_{\bullet}(\CX;E)$ as s subalgebra of $\mbox{Diff}_{\bullet}(M;\pi_X^*E)$ where $\bullet=\bop,\scop$.

Conversely, given a \emph{stationary} differential operator $L\in\mbox{Diff}(M;\pi_X^*E_1,\pi_X^*E_2)$ which commutes with $\pa_t$, one can find a unique operator $\widehat{L} (0)\in\mbox{Diff}(X;E_1,E_2)$ such that the relation~\eqref{EqKNStStat} holds. Moreover, $\widehat{L} (0)$ is independent of the choice of the time function $\mathfrak{t}$. More generally, we consider the formal conjugation of $L$ by $e^{i\sigma\mathfrak{t}}$
\[
\widehat{L} (\sigma) := e^{i\sigma\mathfrak{t}}L e^{-i\sigma\mathfrak{t}} \in \mbox{Diff}(X;E_1,E_2).
\]
We note that using another time function, $\mathfrak{t}+F'$ with $F'\in C^\infty(X)$ in the above formal conjugation means conjugating $\widehat{L}(\sigma)$ by $e^{i\sigma F'}$.

 Motivated by the product decomposition \eqref{EqRNExt}, we use the splitting
 \[
 T^*M \cong T^*\BR_{t_0} \oplus T^*X = \pi_T^*(T^*\BR_{t_0}) \oplus \pi_X^*(T^*X)
 \]
 where $\pi_T\colon M_0\to\BR_{t_0}$ is the projection. Correspondingly, we introduce the following \emph{extended scattering cotangent bundle} of $\CX$
\begin{equation}
	\label{EqKNStExtTsc}
	\scform := \BR d t_0 \oplus {}^{\scop}T^*\CX.
\end{equation}
Here $d t_0$ is simply a notation of the basis of a real line bundle of rank $1$ over $\CX$, which we identify as the differential of the time function $t_0\in C^\infty(M)$ when we look at the pullback bundle $\pi_X^*\scform\to M$. The extended scattering cotangent bundle $\scform$ is spanned, over $C^\infty(\CX)$, by $d t_0$ and the differentials $d x^i$, where $(x^1,x^2,x^3)$ are standard coordinates on $X\subset\BR^3$. We remark that we can also use another time function, for example $t_{b,*}$, in~\eqref{EqKNStExtTsc} because the difference between the differentials of different time functions is given by a smooth scattering 1-form on $\CX$.

Given a stationary metric $g$ on $M$, one can find a unique $\tilde{g}\in C^\infty(X;\scsym)$ such that $\pi_X^*\tilde{g}=g$. Identifying $g$ with $\tilde{g}$ and applying this procedure to the Kerr-Newman family, we obtain
\begin{equation}
	\label{EqKNStMetrics}
	g_b,\ g_b^0 \in C^\infty(\CX;\scsym).
\end{equation}
We also record the facts
\begin{equation}
	\label{EqKNStMetrics2}
	\begin{split}
		&g_b - \underline{g} \in \rho C^\infty, \quad \underline{g}:=-d t_{\chi_0}^2+d r^2+r^2\sg, \\
		&g_{(\Bm,\Ba,\BQ)}-g_{(\Bm,0,\BQ)} \in \rho^2 C^\infty(\CX;\scsym),
	\end{split}
\end{equation}
that is, a KN metric is a $\mathcal{O}(\rho)$, resp. $\mathcal{O}(\rho^2)$ perturbation of the Minkowski metric $\underline{g}$, resp.  the RN metric of the same mass and charge.

We shall discuss some basic geometric operators (which will be used frequently later on) on Kerr-Newman spacetimes, for example,  
\begin{equation}
	\label{EqKNStDel}
	(\delta_g^*\omega)_{\mu\nu}=\frac12(\nabla_\nu\omega_{\mu}+\nabla_\mu\omega_{\nu}), \quad
	(\delta_g h)_\alpha=-g^{\mu\nu}\nabla_\mu h_{\nu\alpha}, \quad
	G_g = \mbox{Id}-\frac12 g\,\tr_g.
\end{equation}
We denote by
\begin{equation}
	\label{EqKNStWave}
	\Box_{g,0},\ \Box_{g,1},\ \Box_{g,2},
\end{equation}
the tensor wave operator $g^{\mu\nu}\nabla_\mu\nabla_\nu$ on scalar functions, 1-forms, and symmetric 2-tensors, respectively. When the bundle is clear from the context, we shall drop the notation of the bundles and simply write $\Box_g$. We use the splitting
\begin{align}
		\label{EqKNStBox2Spl1}
\scform &= \langle d t\rangle \oplus \fscform, \\
\label{EqKNStBox2Spl2}
\scsym&= \langle d t^2\rangle\oplus (2 d t\otimes_s \fscform) \oplus S^2\,\fscform.
\end{align}
and work in the trivialization of $\fscform$, resp. $S^2\,\fscform$ given by $dx^i, 1\leq i\leq3$, resp. $2dx^i\otimes_s dx^j, 1\leq i\leq j\leq 3$.
\begin{prop}
	\label{PropKNStBox}
	Working in the splittings \eqref{EqKNStBox2Spl1} and \eqref{EqKNStBox2Spl2} and the trivializations of $\fscform$ and $S^2\,\fscform$, we write the operators $\Box_{g_b,2}, \Box_{g_b,1}, \Box_{g_b,0}$ as
	\begin{align}
			\label{EqKNStBox0}
		\Box_{g_b,0} &= g_b^{-1}(dt_{\chi_0}, dt_{\chi_0}) \pa_{t_{\chi_0}}^2 + \widehat{\Box_{g_b,0}}(0) + Q_{b,0}^1 \pa_{t_{\chi_0}},\\
	\label{EqKNStBox1}
	\Box_{g_b,1}	&=\Box_{g_b,0}\otimes\mathrm{Id}_{4\times 4}+Q_{b,1}^1\pa_{t_{\chi_0}}+Q_{b,1}^0,\\	
	\label{EqKNStBox2}
	\Box_{g_b,2}	&=\Box_{g_b,0}\otimes\mathrm{Id}_{10\times 10}+Q_{b,2}^1\pa_{t_{\chi_0}}+Q_{b,2}^0.
	\end{align}
	Then we have
	 \begin{gather}
	\widehat{\Box_{g_b,0}}(0)\in\rho^2\mathrm{Diff}_\bop^2(\CX),\quad  Q^1_{b,0}\in\rho^3\mathrm{Diff}_\bop^1(\CX),\\
	 \quad Q_{b,1}^1\in \rho^2\mathrm{Diff}_\bop^0(\CX;\scform),\quad Q_{b,1}^0\in \rho^3\mathrm{Diff}^1_\bop(\CX;\scform),\\
	 \quad Q_{b,2}^1\in \rho^2\mathrm{Diff}_\bop^0(\CX;S^2\,\scform),\quad Q_{b,2}^0\in \rho^3\mathrm{Diff}^1_\bop(\CX;S^2\,\scform).
	\end{gather}
Moreover, 
\begin{equation}
\widehat{\Box_{g_b,j}}(0)-\widehat{\Box_{\underline{g},j}}(0)\in \rho^3\mathrm{Diff}_\bop^2(\CX) ,\quad \underline{g}=-dt_{\chi_0}^2+dx^2,\quad j=0,1,2.
\end{equation}
\end{prop}

Since the metric components are smooth away from $\pa_+\CX$, it suffices to analyze $\Box_{g,\bullet}, \bullet=0,1,2$ near spatial infinity $\pa_+\CX$ where $t_{\chi_0}\equiv t$ is the static time coordinate. Proposition \ref{PropKNStBox} is a consequence of the following lemma (by taking $h=dx^2$).

\begin{lem}
	\label{LemmaKNStBox}
	Suppose $g$ is a stationary Lorentzian metric on $M$ satisfying
	\begin{equation}
		\label{EqKNStBoxStruct}
		\begin{split}
			g(\pa_t,\pa_t) &\in -1+\rho C^\infty(\CX), \\
			g(\pa_t,-) &\in \rho^2 C^\infty(\CX;\fscform), \\
			g|_{\fscvec\times\fscvec} &\in h + \rho C^\infty(\CX;\scsym),
		\end{split}
	\end{equation}
	where $h\in C^\infty(\CX;\scsym)$ is a Riemannian metric. Then the operator $\Box_{g,2},\Box_{g,1}, \Box_{g,0}$ satisfy the statements in Proposition \ref{PropKNStBox} with $Q_{b,j}^0\in \rho^3\mathrm{Diff}^1_\bop(\CX)$ replaced by $Q_{b,j}^0\in \rho^2\mathrm{Diff}^1_\bop(\CX)$ for $j=1,2$.

Moreover, $\widehat{\Box_{g,j}}(0)\bmod\rho^3\mathrm{Diff}_\bop^2(\CX)$ only depends on $h$ for $j=0,1,2$.
\end{lem}
\begin{proof}
	We introduce coordinates
	\[
	x=(x^0,x^1,x^2,x^3),
	\]
	where $x^0=t$, and $x^1,x^2,x^3$ are standard coordinates on $\BR^3$. We use Latin letters $i,j,k,\cdots$ for indices from $1$ to $3$, and Greek letters $\alpha,\beta,\mu,\nu,\cdots$ for indices from $0$ to $3$. We let $\mathcal{O}^k:=\rho^k C^\infty(\CX)$ and interpret $f=\mathcal{O}^k$ as $f\in\mathcal{O}^k$ when $f$ is a smooth function. 
	
	We first compute the Levi-Civita connection of $g$. Using the facts that $g$ is stationary, $\pa_i\in\rho\Vb(\CX)$ and 
	\[
	g^{-1}(dt,dt)\in -1+\rho C^{\infty}(\CX),\quad g^{-1}(dt,-)\in \rho^2 C^\infty(\CX;\fscvec),\quad	g^{-1}|_{\fscform\times\fscform} \in h^{-1} + \rho C^\infty(\CX;S^2\,\fscvec),
	\] 
	we find that 
	\begin{equation}
		\label{EqKStGamma}
		\Gamma_{00}^0=\mathcal{O}^4,\quad \Gamma^{i}_{00}=\mathcal{O}^2,\quad \Gamma_{0i}^0=\mathcal{O}^2,\quad \Gamma_{0i}^j=\mathcal{O}^3,\quad \Gamma_{ij}^0=\mathcal{O}^3,\quad \Gamma_{ij}^k=\Gamma_{ij}^k(h)+\mathcal{O}^2,\quad \Gamma_{ij}^k(h)=\mathcal{O}^1.
	\end{equation}

Since $\Box_{g,0}=g^{\al\be}\pa_{\al}\pa_{\be}-g^{\al\be}\Gamma_{\al\be}^\mu\pa_\mu$ where $g^{\al\be}$ is the inverse to $g$, we see that 
\[
\Box_{g,0}=g^{\al\be}\pa_{\al}\pa_{\be}+f^\mu\pa_\mu,\quad f^0=\mathcal{O}^3,\  f^i=\mathcal{O}^1,
\]
and thus we can write 
\[
\Box_{g,0} = g^{00} \pa_{t}^2 + Q \pa_{t}+\widehat{\Box_{g,0}}(0)\quad \mbox{with}\quad Q\in\rho^3\mathrm{Diff}^1_\bop(\CX),\  \widehat{\Box_{g,0}}(0)\in \rho^2\mathrm{Diff}^2_\bop(\CX).
\]
Moreover, $\widehat{\Box_{g,0}}(0)\bmod\rho^3\mathrm{Diff}_\bop^2(\CX)$ only depends on $h$.

Next, we calculate $\Box_{g,1}$. Since 
\[
(\Box_{g,1}u)_\alpha=\Box_{g,0}u_\alpha-2g^{\mu\nu}\Gamma_{\mu\al}^\kappa\pa_\nu u_{\kappa}+g^{\mu\nu}\Gamma_{\mu\nu}^\kappa\Gamma_{\kappa\al}^\be u_\be+g^{\mu\nu}\Gamma_{\mu\al}^\kappa\Gamma_{\kappa\nu}^\be u_\be-g^{\mu\nu}(\pa_\mu\Gamma_{\nu\al}^\kappa)u_\kappa,
\]
we can write 
\[
\Box_{g,1}=\Box_{g,0}\otimes\mathrm{Id}_{4\times 4}+Q_{g,1}^1\pa_t+Q_{g,1}^0
\]
where $Q_{g,1}^1\in \rho^2\mathrm{Diff}_\bop^0(\CX;\scform)$ and $Q_{g,1}^0\in \rho^2\mathrm{Diff}^1_\bop(\CX;\scform)$. Moreover, $Q_{g,1}^0\in \rho^3\mathrm{Diff}^1_\bop(\CX;\scform)$ if $\Gamma_{ij}^k(h)$ vanishes, and $\widehat{\Box_{g,1}}(0)\bmod\rho^3\mathrm{Diff}_\bop^2(\CX;\scform)$ only depends on $h$.

Finally, we calculate $\Box_{g,2}$. Since 
\[
(\Box_{g,2}h)_{\alpha\be}=\Box_{g,0}h_{\alpha\be}-2g^{\mu\nu}\Gamma_{\mu\al}^\kappa\pa_\nu h_{\kappa\be}-2g^{\mu\nu}\Gamma_{\mu\be}^\kappa\pa_\nu h_{\kappa\al}+S_{g,2}^0
\]
where $S_{g,2}^0\in\mathrm{Diff}_b^0(\CX;\scsym)$ whose elements are linear combinations of terms of the types $g\Gamma\cdot\Gamma$ and $g\pa\Gamma$, it follows that we can write 
\[
\Box_{g,2}=\Box_{g,0}\otimes\mathrm{Id}_{10\times 10}+Q_{g,2}^1\pa_t+Q_{g,2}^0
\]
where $Q_{g,2}^1\in \rho^2\mathrm{Diff}_\bop^0(\CX;\scsym)$ and $Q_{g,2}^0\in \rho^2\mathrm{Diff}^1_\bop(\CX;\scsym)$. Again, $Q_{g,2}^0\in \rho^3\mathrm{Diff}^1_\bop(\CX;\scsym)$ if $\Gamma_{ij}^k(h)$ vanishes, and $\widehat{\Box_{g,2}}(0)\bmod\rho^3\mathrm{Diff}_\bop^2(\CX;\scsym)$ only depends on $h$.
\end{proof}

We next discuss the curvature tensor and other geometric operators associated to $g_b$.
\begin{prop}\label{PropKNStCurvature}
	Let $g$ be a metric of the form \eqref{EqKNStBoxStruct} and $\tilde{g}=-dt^2+h$ where $h\in C^\infty(\CX;\scsym)$ is Riemannian. Then we have
	\begin{align}
			\label{EqKNStCurvature}
		&\mathrm{Riem}(g)-\mathrm{Riem}(\tilde{g})\in\rho^3C^\infty\bigl(\CX;{}\widetilde{^\scop T}\CX \otimes (\scform)^3\bigr),\quad \Ric(g)-\Ric(\tilde{g})\in\rho^3C^\infty\bigl(\CX; \scsym\bigr),\\
		\label{EqKNStND}
		&	\delta_g = -\iota_{dt_{\chi_0}^\sharp}\pa_{t_{\chi_0}} + \widehat{\delta_g}(0), \quad
		\widehat{\delta_g}(0) \in \rho\mathrm{Diff}_\bop^1(\CX), \quad \widehat{\delta_g}(0)-\widehat{\delta_{\tilde{g}}}(0)\in \rho^2\mathrm{Diff}_\bop^1(\CX),\\
		\label{EqKNStSG}
		&	\delta_g^* = dt_{\chi_0} \otimes_s \pa_{t_{\chi_0}} + \widehat{\delta_g^*}(0), \quad
		\widehat{\delta_g^*}(0) \in \rho\mathrm{Diff}_\bop^1(\CX),\quad \widehat{\delta_g^*}(0)-\widehat{\delta_{\tilde{g}}^*}(0)\in \rho^2\mathrm{Diff}_\bop^0(\CX).
	\end{align}
In particular, these hold for $g=g_b$ and $\tilde{g}=\underline{g}=-dt_{\chi_0}^2+dx^2$.
\end{prop}
\begin{proof}
	Since
	\[
		\Gamma_{00}^0(\tilde{g})= \Gamma^{i}_{00}(\tilde{g})= \Gamma_{0i}^0(\tilde{g})=\Gamma_{0i}^j(\tilde{g})= \Gamma_{ij}^0(\tilde{g})=0,\quad \Gamma_{ij}^k(\tilde{g})=\mathcal{O}^1,\quad \Gamma_{\al\be}^\mu(g)=\Gamma_{\al\be}^\mu(\tilde{g})+\mathcal{O}^2,
	\]
	and $g,\tilde{g}$ are stationary, the proposition follows from the formulas
	\begin{gather*}
	\mathrm{Riem}(g)^\nu_{\ \mu\al\be}=\partial_\al\Gamma_{\beta\mu}^\nu(g)-\partial_\be\Gamma_{\al\mu}^\nu(g)+\Gamma_{\al\kappa}^\nu(g)\Gamma^\kappa_{\beta\mu}(g)-\Gamma_{\be\kappa}^\nu(g)\Gamma_{\al\mu}^\kappa(g),\\
	\delta_g u=-g^{\mu\nu}\pa_\mu u_\nu+g^{\m\n}\Gamma_{\mu\nu}^\kappa(g) u_\kappa,\quad (\delta_g h)_\alpha=-g^{\mu\nu}\pa_\mu h_{\nu\al}+g^{\m\n}\Gamma_{\mu\nu}^\kappa(g) h_{\kappa\al}+g^{\m\n}\Gamma_{\mu\al}^\kappa(g) h_{\nu\kappa},\\
	(\delta_g^*u)_{\al\be}=\frac12(\pa_\al u_\be+\pa_\be u_\al)-\Gamma_{\al\be}^\kappa(g) u_\kappa.
	\end{gather*}
\end{proof}

In view of the facts in \eqref{EqKNStMetrics2}, we have the following further leading order control
\begin{lem}
Suppose that $g_1,g_2$ are two metrics of the form \eqref{EqKNStBoxStruct} so that in addition $g_1-g_2\in\rho^2 C^\infty(X;\scsym)$. Then
\begin{gather}
		\label{EqKNStBoxLeading}
	\widehat{\Box_{g_1,j}}(0)-\widehat{\Box_{g_2,j}}(0) \in \rho^4\mathrm{Diff}_\bop^2(\CX),\quad j=0,1,2,\\
		\label{EqKNStCurvatureLeading}
	\mathrm{Riem}(g_2)-\mathrm{Riem}(g_1)\in\rho^4C^\infty\bigl(\CX;{}\widetilde{^\scop T}\CX \otimes (\scform)^3\bigr),\quad \Ric(g_2)-\Ric(g_1)\in\rho^4C^\infty\bigl(\CX; \scsym\bigr),\\
	\label{EqKNStNDLeading}
	\widehat{\delta_{g_1}}(0)-\widehat{\delta_{g_2}}(0) \in \rho^3\mathrm{Diff}^1_\bop(\CX),\\
	\label{EqKNStSGLeading}
	\widehat{\delta_{g_1}^*}(0)-\widehat{\delta_{g_2}^*}(0) \in \rho^3C^\infty(\CX;\mathrm{Hom}(\scform,\scsym)),
\end{gather}
\end{lem}
\begin{proof}
	The proof follows from the facts that $\Gamma_{\al\be}^\mu(g_1)=\mathcal{O}^1$ and $\Gamma_{\al\be}^\mu(g_2)-\Gamma_{\al\be}^\mu(g_1)=\mathcal{O}^3$.
\end{proof}

\subsubsection{Relevant geometric operators in linearized gauge-fixed Einstein-Maxwell equations}
We now study the linearized gauge-fixed Einstein-Maxwell operator arising from the generalized wave map and Lorenz gauge.
\begin{defn}
	Let $B\in C^\infty(M;\mbox{Hom}(T^*M, S^2T^*M)), F\in C^\infty(M;\mbox{Hom}(S^2T^*M, T^*M))$ and $g$ be a Lorentzian metric. Then we define the \textit{modified symmetric gradient} $\wt{\delta}^*_{g, B}$ and \textit{modified negative divergence} $\wt{\delta}_{g,F}$ as follows.
	\begin{equation}\label{eq:defofmodideriFirst}
		\wt{\delta}^*_{g, B}=\delta_g^*+B_g,\quad \wt{\delta}_{g,F}=\delta_g+F_g.
	\end{equation}
\end{defn}
In this paper, we will use $B_g, F_g$ of the following form
\begin{gather}\label{eq:defofBFirst}
	B_g=B(g;\mathfrak{c},\gamma):=2\gamma\mathfrak{c}\otimes_s(\bullet)-\frac{1}{2}\gamma g^{-1}(\mathfrak{c},\bullet)g,\\\label{eq:defofFFirst}
	F_g=F(g;\mathfrak{c},\gamma):=2\gamma\iota^g_{\mathfrak{c}}(\bullet)-\frac{1}{2}\gamma\mathfrak{c}\tr_g(\bullet).
\end{gather}
where $\gamma\in\BR$ is a sufficiently small constant to be determined later on, $\mathfrak{c}\in C^\infty_c(X, \scform)$ is a stationary $1$-form with compact support and $\iota_\mathfrak{c}^g(h)=g^{\mu\nu}\mathfrak{c}_\nu h_{\mu\al}$. With the above choices of $B_g$ and $F_g$ in \eqref{eq:defofBFirst} and \eqref{eq:defofFFirst}, we find that $\wt{\delta}^*_{g,B}$ and $\wt{\delta}_{g,F}$ are formally adjoint to each other. From now on, we denote 
\begin{equation}\label{eq:modifieddeltaFirst}
	\wt{\delta}^*_{g,\gamma}:=\wt{\delta}^*_{g,B}=\delta_g^*+B_g,\quad \wt{\delta}_{g,\gamma}:=\wt{\delta}_{g,F}=\delta_g+F_g.
\end{equation}

By choosing $\tilde{\delta}_{g_b}^*=\wt{\delta}^*_{g_b,\gamma}$ and $\theta(g;g_b)=F(g_b;\mathfrak{c},\gamma)[G_{g_b} g]$, we obtain the \textit{linearized gauge-fixed Einstein-Maxwell system} $L_{b,\gamma}(\dg,\dA)=(2L^E_{b,\gamma}(\dg,\dA),\ L^M_{b,\gamma}(\dg,\dA))=0$ around $(g_b,A_b)$
\begin{equation}
	\label{EqBasicLingAEqgamma}
	\begin{gathered}
		L_{b,\gamma}^E(\dg,\dA):=	D_{g_b}(\Ric)(\dg) - \tilde{\delta}_{g_b,\gamma}^*D_{g_b}\widetilde{\Upsilon}^E(\dg;g_b) -2 D_{(g_b,d A_b)}T(\dg,d\dA),\quad \widetilde{\Upsilon}^E=\Upsilon^E-\theta, \\
		L_{b,\gamma}^M(\dg,\dA):=D_{(g_b,A_b)}(\delta_{(\cdot)} d(\cdot))(\dg,\dA) - d\Big(D_{(g_b,A_b)}\widetilde{\Upsilon}^M(\dg,\dA;g_b,A_b)\Big).
	\end{gathered}
\end{equation}
	According to the calculation \eqref{EqBasicLinRic}--\eqref{EqBasicLingLorenzgauge}, \eqref{eq:DT} and \eqref{eq:L_2}, we find that 
\begin{equation}
	\label{EqdetailedLinEM}
	\begin{split}
		L_{b,\gamma}^E(\dg,\dA)&=-\frac12\Box_{g_b,2}\dg+(\tilde{\delta}_{g_b,\gamma}^*-\delta_{g_b}^*)\delta_{g_b}G_{g_b}\dg+\delta_{g_b}^*(\tilde{\delta}_{g_b,\gamma}-\delta_{g_b})G_{g_b}\dg\\
	&\quad	+(\tilde{\delta}_{g_b,\gamma}^*-\delta_{g_b}^*)(\tilde{\delta}_{g_b,\gamma}-\delta_{g_b})G_{g_b}\dg+\mathscr{R}_{g_b}(\dg)_{\mu\nu}-2 D_{g_b,d A_b}T(\dg,d\dA),\\
		L_{b,\gamma}^M(\dg,\dA)&=(d\delta_{g_b}+\delta_{g_b}d)\dA-(\delta_{g_b}G_{g_b}\dg)^\ka (dA_b)_{\ka\m}+\frac 12(\nabla^{\n}\dg^{\ka}_{~\m}-\nabla^{\ka}\dg^{\n}_{~\m})(dA_b)_{\n\ka}+\dg^{\n\al}\nabla_\al (dA_b)_{\n\m}
	\end{split}
\end{equation}
where $d\delta_{g_b}+\delta_{g_b}d=-\Box_{g_b,1}+\Ric(g_b)$ and 
\begin{equation}\label{EqRT}
	\begin{split}
(\mathscr{R}_{g_b}\dg)_{\m\n}&=\mathrm{Riem}(g_b)^{\al\ \ \beta}_{\ \m\n}\dg_{\al\be}+\frac 12\left(\Ric(g_b)_{\m}^{\ \kappa}\dg_{\n\ka}+\Ric(g_b)_{\n}^{\ \kappa}\dg_{\m\ka}\right),\\
 D_{g_b,d A_b}T(\dg,d\dA)&=\left((g_b)^{\alpha\beta}(d\dA)_{\mu\alpha}(dA_b)_{\nu\beta}+(g_b)^{\alpha\beta}(dA_b)_{\mu\alpha}(d\dA)_{\nu\beta}\right)\!\!-\!\frac 12(g_b)_{\mu\nu}(d\dA)_{\alpha\beta}(dA_b)^{\alpha\beta}\\
 &\quad-\dg^{\alpha\beta}(dA_b)_{\mu\alpha}(dA_b)_{\nu\beta}-\frac 14\left(\dg_{\mu\nu}(dA_b)_{\alpha\beta}(dA_b)^{\alpha\beta}\!\!-\!2(g_b)_{\mu\nu}\dg^{\kappa\alpha}(dA_b)_{\alpha\beta}(dA_b)_{\kappa}^{~\be}\right).
 \end{split}
\end{equation}
We note that the first two terms of the second line of $L_{b,\gamma}^E(\dg,\dA)$, the last two terms of $ D_{g_b,d A_b}T(\dg,d\dA)$ and the last term of $	L_{b,\gamma}^M(\dg,\dA)$ contain no derivatives of $(\dg,\dA)$.

For later use, we also define the gauge propagation operator $\mathcal{P}_{b,\gamma}$ and the gauge potential wave operator  $\mathcal{W}_{b,\gamma}$ for the gauge $1$-form $\widetilde{\Upsilon}^E(\dg;g_b)$:
\begin{equation}
	\label{EqdetailedPW}
	\begin{split}
	\mathcal{P}_{b,\gamma}&:=\delta_{g_b}G_{g_b}\tilde{\delta}_{g_b,\gamma}^*=-\frac12\Box_{g_b,1}-\frac12\Ric(g_b)+\delta_{g_b}G_{g_b}(\tilde{\delta}_{g_b,\gamma}^*-\delta_{g_b}^*),\\	\mathcal{W}_{b,\gamma}&:=\tilde{\delta}_{g_b,\gamma}G_{g_b}\delta_{g_b}^*=-\frac12\Box_{g_b,1}-\frac12\Ric(g_b)+(\tilde{\delta}_{g_b,\gamma}-\delta_{g_b})G_{g_b}\delta_{g_b}^*.
	\end{split}
\end{equation}
Then we have
\begin{lem}
	\label{LemPWL}
	Let $L_{b,\gamma}=(2L_{b,\gamma}^E, L_{b,\gamma}^M), \mathcal{P}_{b,\gamma},\mathcal{W}_{b,\gamma}$ be defined as above. Then we have
	\begin{align}
		\label{EqKNP}
		\mathcal{P}_{b,\gamma}&=-\frac12\Box_{g_b,1}+P^1\pa_{t_{\chi_0}}+P^0,\quad\widehat{\mathcal{P}_{b, \gamma}}(0)=-\frac12\widehat{\Box_{g,1}}(0)+P^0,\\
		\label{EaKNW}
		\mathcal{W}_{b,\gamma}&=-\frac12\Box_{g_b,1}+W^1\pa_{t_{\chi_0}}+W^0,\quad\widehat{\mathcal{W}_{b, \gamma}}(0)=-\frac12\widehat{\Box_{g,1}}(0)+W^0,
	\end{align}
where
\begin{equation}
	\label{EqKNPWLower}
		P^1,W^1\in \rho^\infty\mathrm{Diff}_\bop^0(\CX;\scform),\quad P^0,W^0\in \rho^4\mathrm{Diff}^1_\bop(\CX;\scform).
\end{equation}
We also have
\begin{equation}
	\label{EqKNL}
	L_{b,\gamma}=-\Box_{g_b,0}\otimes\mathrm{Id}_{14\times 14}+L^1\pa_{t_{\chi_0}}+L^0,\quad \widehat{L_{b, \gamma}}(0)=-\widehat{\Box_{g,0}}(0)\otimes\mathrm{Id}_{14\times 14}+L^0,
\end{equation}
where
\begin{equation}
	\label{EqKNLLower}
	L^{1}\in\rho^2\mathrm{Diff}^0_\bop(\CX;\scsym\oplus\scform),\quad L^{0}\in\rho^3\mathrm{Diff}^1_\bop(\CX;\scsym\oplus\scform).
\end{equation}
Moreover, we have
\begin{equation}
	\label{EqLLeading}
	-\widehat{L_{b, \gamma}}(0)-\widehat{\Box_{\underline{g},0}}(0)\otimes\mathrm{Id}_{14\times 14}\in\rho^3\mathrm{Diff}^2_\bop(\CX;\scsym\oplus\scform),\quad \underline{g}=-dt_{\chi_0}^2+dx^2.
\end{equation}
	\end{lem}
\begin{proof}
	Since $\tilde{\delta}_{g_b,\gamma}^*-\delta_{g_b}^*$ and $\tilde{\delta}_{g_b,\gamma}-\delta_{g_b}$ are compactly supported and $\Ric(g_b)=\mathcal{O}^4$,
	then the conclusions \eqref{EqKNP}--\eqref{EqKNPWLower} follow.

According to \eqref{EqRT} and using the facts that $\mathrm{Riem}(g_b)=\mathcal{O}^3, dA_b=\mathcal{O}^2$, we see that $\mathscr{R}_{g_b}\in\rho^3\mathrm{Diff}_\bop^0%(\CX;\scsym)
$ and $D_{(g_b,dA_b)}T=T^1\pa_{t_{\chi_0}}+T^0$ with 
\[ T^1\in\rho^2\mathrm{Diff}^0_\bop(\CX;\scsym\oplus\scform),\quad T^0\in\rho^3\mathrm{Diff}^1_\bop(\CX;\scsym\oplus\scform),
\]
and thus
\[
2L_{b,\gamma}^E(\dg,\dA)=-\Box_{g_b,2}\dg+L^{E,1}\pa_{t_{\chi_0}}+L^{E,0}
\]
with 
\[ L^{E,1}\in\rho^2\mathrm{Diff}^0_\bop(\CX;\scsym\oplus\scform),\quad L^{E,0}\in\rho^3\mathrm{Diff}^1_\bop(\CX;\scsym\oplus\scform).
\]

According to the expression for $L^E_{b,\gamma}(\dg,\dA)$ in \eqref{EqdetailedLinEM} and using Proposition \ref{PropKNStCurvature}, we find that 
\[
L_{b,\gamma}(\dg,\dA)=-\Box_{g_b,1}\dA+L^{M,1}\pa_{t_{\chi_0}}+L^{M,0}
\]
where
\[
L^{M,1}\in\rho^2\mathrm{Diff}^0_\bop(\CX;\scsym\oplus\scform),\quad L^{M,0}\in\rho^3\mathrm{Diff}^1_\bop(\CX;\scsym\oplus\scform).
\]
Combining the above conclusions for $L^E_{b,\gamma}$ and $L^M_{b,\gamma}$ and using the description of $\Box_{g_b,1},\Box_{g_b,2}$ in Proposition \ref{PropKNStBox}, we obtain \eqref{EqKNL}--\eqref{EqLLeading}.
\end{proof}

Now we let
\begin{equation}
	\widehat{L_{b, \gamma}}(\sigma)=e^{i t_{b,*}\sigma}L_{b,\gamma}e^{-i t_{b,*}\sigma},\quad 	\widehat{\mathcal{P}_{b, \gamma}}(\sigma)=e^{i t_{b,*}\sigma}\mathcal{P}_{b,\gamma}e^{-i t_{b,*}\sigma},\quad \widehat{\mathcal{W}_{b, \gamma}}(\sigma)=e^{i t_{b,*}\sigma}\mathcal{W}_{b,\gamma}e^{-i t_{b,*}\sigma}.
\end{equation}
In order to study the property of $	\widehat{L_{b, \gamma}}(\sigma),\ \widehat{\mathcal{P}_{b, \gamma}}(\sigma),\ \widehat{\mathcal{W}_{b, \gamma}}(\sigma)$, we first express $L_{b,\gamma}, \mathcal{P}_{b,\gamma}, \mathcal{W}_{b,\gamma}$ in the coordinates $(t_{b,*},x^i), i=1,2,3$.

\begin{lem}
	\label{LemPWLtStar}
	In the coordinates $(t_{b,*}, x^i), i=1,2,3$, the operator $\Box_{g_b,0}$ take the form
\begin{align}
	\Box_{g_b,0}=Q^2\pa_{t_{b,*}}^2+Q^1\pa_{t_{b,*}}+\widehat{\Box_{g_b,0}}(0) 
\end{align}
where $Q^2\in \rho^2C^\infty(\CX),\  Q^1\in2\rho^2\pa_\rho-2\rho+\rho^3\mathrm{Diff}_{\bop}^1(\CX)$ and $ \widehat{\Box_{g_b,0}}(0)\in\rho^2\mathrm{Diff}^2_\bop(\CX)$.

Moreover, the operators $\mathcal{P}_{b,\gamma}, \mathcal{W}_{b,\gamma}, L_{b,\gamma}$ have the form
\begin{align}
	\mathcal{P}_{b,\gamma}&=-\frac12\Box_{g_b,0}\otimes \mathrm{Id}_{4\times 4}+\tilde{P}^1\pa_{t_{b,*}}+\tilde{P}^0,\\		\mathcal{W}_{b,\gamma}&=-\frac12\Box_{g_b,0}\otimes \mathrm{Id}_{4\times 4}+\tilde{W}^1\pa_{t_{b,*}}+\tilde{W}^0,\\
	L_{b,\gamma}&=-\Box_{g_b,0}\otimes \mathrm{Id}_{14\times 14}+\tilde{L}^1\pa_{t_{b,*}}+\tilde{L}^0,
\end{align}
where
\begin{align*}
	\tilde{P}^1,\tilde{W}^1,\tilde{L}^1\in\rho^2\mathrm{Diff}_\bop^0(\CX),\quad \tilde{P}^0,\tilde{W}^0,\tilde{L}^0\in\rho^3\mathrm{Diff}_\bop^0(\CX).
\end{align*}
	\end{lem}

\begin{proof}
	With parameter $b=(\Bm, \Ba,\BQ)$, near $\pa_+\CX$, we have $t_{\chi_0}=t$ and $t_{b,*}=t-r_{(\Bm, 0,\BQ),*}$. Then changing from $(t_{\chi_0},r)$ coordinates to $(t_{b,*}, r)$ coordinates transforms $\pa_{t_{\chi_0}}, \pa_r$ into 
	\[\pa_{t_{b,*}}, \quad \pa_r-\frac{1}{\mu_{(\Bm,0,\BQ)}}\pa_{t_{b,*}}\quad \mbox{near}\quad \pa_+\CX,\quad \mu_{(\Bm,0,\BQ)}=1-\frac{2\Bm}{r}+\frac{\BQ^2}{r^2}.
	\]
	Since $g_b-g_{(\Bm,0,\BQ)}\in\rho^2 C^\infty(\CX;\scsym)$, using the facts that in the coordinate $(t, x^i)$,  \[\Gamma_{\al\be}^\kappa(g_{(\Bm,0,\BQ)})=\mathcal{O}^2,\quad \Gamma_{\al\be}^\kappa(g_b)-\Gamma_{\al\be}^\kappa(g_{(\Bm,0,\BQ)})\in\mathcal{O}^3,
	\]
	 it follows that 
	\[
	\Box_{g_b,0}-\Box_{g_{(\Bm,0,\BQ)},0}=\rho^2C^\infty(\CX)\pa^2_{t_{\chi_0}}+\rho^3\mathrm{Diff}^1_\bop(\CX)\pa_{t_{\chi_0}}+\rho^4\mathrm{Diff}^2_\bop(\CX)
	\]
	and thus 
		\[
	\Box_{g_b,0}-\Box_{g_{(\Bm,0,\BQ)},0}=\rho^2C^\infty(\CX)\pa^2_{t_{b,*}}+\rho^3\mathrm{Diff}^1_\bop(\CX)\pa_{t_{b,*}}+\rho^4\mathrm{Diff}^2_\bop(\CX).
	\]
	Then it remains to calculate $\Box_{g_{(\Bm,0,\BQ)},0}$ in $(t_{b,*}, r)$ coordinates. Since near $\pa_+\CX$
	\[
	g_{(\Bm,0,\BQ)}=-\mu_{(\Bm,0,\BQ)}dt^2_{b,*}-2dt_{b,*}dr+r^2\sg,
	\]
	we have for $\rho=1/r$
	\begin{align*}
	\Box_{g_{(\Bm,0,\BQ)},0}&=-2\pa_r\pa_{t_{b,*}}+\mu_{(\Bm,0,\BQ)}\pa_r^2+\frac{2r-2\Bm}{r^2}\pa_r-\frac{2}{r}\pa_{t_{b,*}}+\frac{1}{r^2}\sL\\
	&=2\rho^2\pa_\rho\pa_{t_{b,*}}-2\rho\pa_{t_{b,*}}+\rho^2\mathrm{Diff}_\bop^2(\CX)\quad \mbox{near}\quad \pa_+\CX.
	\end{align*}
This completes the proof for $\Box_{g_b,0}$. Then the proof for $\mathcal{P}_{b,\gamma}, \mathcal{W}_{b,\gamma}, L_{b,\gamma}$ follows directly from Proposition \ref{PropKNStBox} and Lemma \ref{LemPWL}.
\end{proof}
 
 Then the following proposition is a direct result of Lemma \ref{LemPWLtStar}.
 \begin{prop}
 	\label{PropFourierPWL}
 	The operator $\Box_{g_b,0}$ take the form
 	\begin{align}
 		\widehat{\Box_{g_b,0}}(\sigma)=-2i\sigma\rho(\rho\pa_\rho-1)+\widehat{\Box_{g_b,0}}(0)+\sigma Q+\sigma^2V
 	\end{align}
 	where $V\in \rho^2C^\infty(\CX),\  Q\in\rho^3\mathrm{Diff}_{\bop}^1(\CX)$ and $ \widehat{\Box_{g_b,0}}(0)\in\rho^2\mathrm{Diff}^2_\bop(\CX)$.
 	
 	Moreover, the operators $,\mathcal{P}_{b,\gamma}, \mathcal{W}_{b,\gamma}, L_{b,\gamma}$ have the form
 	\begin{align}
 		\mathcal{P}_{b,\gamma}&=i\sigma\rho(\rho\pa_\rho-1)\otimes\mathrm{Id}_{4\times4}+\widehat{\mathcal{P}_{b,\gamma}}(0)+\sigma Q_P+\sigma^2V_P,\\		\mathcal{W}_{b,\gamma}&=i\sigma\rho(\rho\pa_\rho-1)\otimes\mathrm{Id}_{4\times4}+\widehat{\mathcal{W}_{b,\gamma}}(0)+\sigma Q_W+\sigma^2V_W,\\
 		L_{b,\gamma}&=2i\sigma\rho(\rho\pa_\rho-1)\otimes\mathrm{Id}_{14\times14}+\widehat{L_{b,\gamma}}(0)+\sigma Q_L+\sigma^2V_L,
 	\end{align}
 	where
 	\begin{align*}
 		V_P, V_W, V_L\in\rho^2\mathrm{Diff}_\bop^0(\CX),\quad Q_P,Q_W,Q_L\in\rho^2\mathrm{Diff}_\scop^1(\CX),\quad \widehat{\mathcal{P}_{b,\gamma}}(0),\widehat{\mathcal{W}_{b,\gamma}}(0),\widehat{L_{b,\gamma}}(0)\in\rho^2\mathrm{Diff}_\bop^2(\CX).
 	\end{align*}
 \end{prop}

%%%%%%%%%%%%%%%%%%%%%%%%%%%%%%%%%%%%%%%%%%%%%%%%%%
\subsection{Constructing gauged initial data}
\label{SsKNGd}
According to Lemmas \ref{lem:gaugeungauge} and \ref{lem:gaugeungaugeLin}, we now show how to construct gauged initial data for the gauge-fixed Einstein-Maxwell system \eqref{EqBasicNLgaugefixE}--\eqref{EqBasicNLgaugefixM} from initial data given on $\Sigma_0$ which are close to the data $(h_{b},k_{b},\mathbf{E}_{b},\mathbf{H}_{b})$ induced by the KN black holes with parameter $b$. 

Recall that we exploit the generalized wave map gauge for the Einstein equations,
\begin{equation}
	\label{EqKNIniGaugeE}
\widetilde{\Upsilon}^E(g;g_b)=\Upsilon^E(g;g_b)-\theta(g;g_b)= g g_{b}^{-1} \delta_g G_g g_{b}-\theta(g;g_b)=0
\end{equation}
and the generalized Lorenz gauge for the Maxwell equations,
\begin{equation}
	\label{EqKNIniGaugeM}
\widetilde{\Upsilon}^M(g,A;g_b,A_b)	=\Upsilon^M(g,A;g_b)-\Upsilon^M(g,A_b;g_b)= \tr_g \delta_{g_{b}}^* A-\tr_g \delta_{g_{b}}^* A_b=0.
\end{equation}
We point out that for a general treatment, we will construct the gauged initial data for any $\theta(g;g_b)$ satisfying that $\theta(g_b;g_b)=0$ and $\theta(g;g_b)$ contains no derivatives of $g$, without being restricted to the specific $\theta(g;g_b)$ we chose in \S\ref{SsKNSt}.

In order to incorporate the vanishing magnetic charge condition, we consider for $s>2,0<\alpha<1$ the subspace of the initial data for ungauged Einstein-Maxwell equations
\begin{equation}
	\label{EqKNIniSpace}
	\begin{split}
		\mathcal{Z}^{s,\alpha} &:= \Bigl\{ (h,k,\mathbf{E},\mathbf{H}) \mid  d\star_h\mathbf{H}=0,\ \int_{\BS^2}\star_h\mathbf{H}=0,\\
	&\hspace{3em}	(h-h_b,k-k_b,\mathbf{E}-\mathbf{E}_b,\mathbf{H}-\mathbf{H}_b)\in \eHb^{s+1,-1/2+\alpha}(\bar{\Sigma}_0;S_{>0}^2 {}^{\scop}T^*\bar{\Sigma}_0)\\
		&\hspace{5em}\times \eHb^{s,1/2+\alpha}(\bar{\Sigma}_0;S^2{}^{\scop}T^*\bar{\Sigma}_0) \times \eHb^{s,1/2+\alpha}(\bar{\Sigma}_0;{}^{\scop}T^*\bar{\Sigma}_0)\times \eHb^{s,1/2+\alpha}(\bar{\Sigma}_0;{}^{\scop}T^*\bar{\Sigma}_0) \Bigr\},
	\end{split}
\end{equation}
where $S^2_{>0}{}^{\scop}T^*\bar{\Sigma}_0$ denotes the fiber bundle of positive definite inner products on ${}^{\scop}T\bar{\Sigma}_0$. We remark that the definition of the space $\mathcal{Z}^{s,\alpha}$ includes \eqref{Eqvanishingmcharge} and one of the constraint equations \eqref{EqBasicNLconstraints2}, which is necessary since we study the Einstein-Maxwell system in terms of the potential $A$ rather than the electromagnetic 2-form $F$.

For the initial value problems of the gauge-fixed Einstein-Maxwell system $P(g,A)=0$, which is a quasilinear hyperbolic system, the initial data induced by $(g,A)$ at the spacelike hypersurface $\Sigma_0=\{\mathfrak{t}=0\}$ is defined as
\begin{equation}
	\label{EqIniMapgaugeed}
	\gamma_0(g,A):=(g|_{\Sigma_0}, \mathcal{L}_{\pa_{\mathfrak{t}}}g|_{\Sigma_0},A|_{\Sigma_0}, \mathcal{L}_{\pa_{\mathfrak{t}}}A|_{\Sigma_0}).
\end{equation} 
In particular, we have $\gamma_0(g_b, A_b)=(g_b|_{\Sigma_0},0,A_b|_{\Sigma_0}, 0)$. Then we define the subspace of the initial data for the gauge-fixed Einstein-Maxwell equations 
\begin{equation}
	\label{EqKNIniSpacegauged}
	\begin{split}
		\mathcal{Y}^{s,\alpha} &:= \Bigl\{(g_0,g_1,A_0,A_1)\mid (g_0,g_1,A_0,A_1)-\gamma_0(g_b,A_b)\in \eHb^{s+1,-1/2+\alpha}(\bar{\Sigma}_0;S^2{}\widetilde{^{\scop} T^*}\bar{\Sigma}_0) \\
		&	\hspace{4em}\times \eHb^{s,1/2+\alpha}(\bar{\Sigma}_0;S^2 {}\widetilde{^{\scop}T^*}{\bar{\Sigma}_0}) 
	 \times \eHb^{s,-1/2+\alpha}(\bar{\Sigma}_0;{}\widetilde{^{\scop}T^*}{\bar{\Sigma}_0}) \times \eHb^{s-1,1/2+\alpha}(\bar{\Sigma}_0;{}\widetilde{^{\scop}T^*}{\Sigma_0})\Bigr\}.
	\end{split}
\end{equation}
We now prove:

\begin{prop}
	\label{PropKNIni}
	There exist a neighborhood
	\begin{align*}
		&(h_{b},k_{b},\mathbf{E}_{b},\mathbf{H}_{b}) \in \mathcal{U} \subset \mathcal{Z}^{2,\alpha}
	\end{align*}
	of KN initial data and a smooth map
	\begin{align*}
		i_b \colon \mathcal{U}\cap \mathcal{Z}^{s,\alpha} \to \mathcal{Y}^{s,\alpha}
	\end{align*}
	for all $s\geq 2, 0<\alpha<1$, such that for $(h,k,\mathbf{E},\mathbf{H})\in\mathcal{U}$, the sections
	\[
	(g_0,g_1,A_0,A_1) = i_b(h,k,\mathbf{E},\mathbf{H})
	\]
	induce the data $(h,k,\mathbf{E},\mathbf{H})$ on $\Sigma_0=\{\mathfrak{t}=0\}$, and they satisfy the gauge conditions \eqref{EqKNIniGaugeE} and \eqref{EqKNIniGaugeM} in the sense that for any section $(g,A)$ of $S^2 T^*M\oplus T^*M$ near $\Sigma_0$ with $\gamma_0(g,A)=(g_0,g_1,A_0,A_1)$, it induces the initial data $(h,k,\mathbf{E},\mathbf{H})$ at $\Sigma_0$ (i.e., $\tau((g,dA))=(h,k,\mathbf{E},\mathbf{H})$) and the conditions \eqref{EqKNIniGaugeE} and \eqref{EqKNIniGaugeM} hold at $\Sigma_0$.
	
	Moreover, for exact KN data with parameter $b$, we have $i_b(h_b,k_b,\mathbf{E}_b,\mathbf{H}_b)=\gamma_0(g_b,A_b)=(g_b|_{\Sigma_0},0,A_b|_{\Sigma_0}, 0)$.
\end{prop}

\begin{proof}
	We will start with the metric components $(g_0,g_1)$ of the map $i_b$. We write
	\[
	g_b=-N_b^2(d\mathfrak{t})^2+g_{b,ij}(dx^i+X_b^id\mathfrak{t})(dx^j+X_b^jd\mathfrak{t})
	\]
	where $N_b\in 1+\rho C^\infty(\Sigma_0)$ and $X_b\in \rho^2C^\infty(\bar{\Sigma}_0; {}^{\scop}T\bar{\Sigma}_0)$. Then we define the component $g_0$ of $i_b(h,k,\mathbf{E},\mathbf{B})$ as
	\[
	g_0:=g|_{\{\mathfrak{t}=0\}}=-N_b^2(d\mathfrak{t})^2+h_{ij}(dx^i+X_b^id\mathfrak{t})(dx^j+X_b^jd\mathfrak{t}).
	\] 
	Moreover, $g_0=g_b|_{\Sigma_0}$ if $h=h_b$ and $g_0$ satisfies the estimate
	\[
	\norm{g_0-g_b|_{\Sigma_0}}_{ \eHb^{s+1,-1/2+\alpha}(\bar{\Sigma}_0;S^2{}\widetilde{^{\scop} T^*}\bar{\Sigma}_0)}\lesssim \norm{h-h_b}_{\eHb^{s+1,-1/2+\alpha}(\bar{\Sigma}_0;S_{>0}^2 {}^{\scop}T^*\bar{\Sigma}_0)}.
	\]
	 We next define $g_1=\mathcal{L}_{\pa_{\mathfrak{t}}}g|_{\Sigma_0}$. Let $\nabla^{g_0}$ be the Levi-Civita connection of $g_0$ and let $n$ resp. $n_b$ be the future timelike unit normal to $\Sigma_0$ with respect to $g_0$ resp. $g_b|_{\Sigma_0}$. Then we have 
	 \[
	 n=\frac{-\nabla^{g_0} \mathfrak{t}}{\sqrt{-g_0(\nabla^{g_0} \mathfrak{t}, \nabla^{g_0} \mathfrak{t})}}=(1+N_b^{-2}h(X_b,X_b))^{-1/2}(\frac{1}{N_b}, -\frac{X_b^i}{N_b}),\quad n-n_b\in \eHb^{s+1, 1/2+\alpha}(\bar{\Sigma}_0;{}\widetilde{^{\scop}T}\bar{\Sigma}_0).
	 \]
	 Since we require
	\[
	k_{ij}=\mathrm{I\!I}_g(\partial_i, \partial_j)|_{\Sigma_0}=-\langle\nabla^g_{\pa_i}n, \pa_j\rangle=-\frac{1}{2}\mathcal{L}_ng_{ij}=-\frac{1}{2}n^\alpha\partial_\alpha g_{ij}-\frac{1}{2}g_{i\alpha}\partial_jn^{\alpha}-\frac{1}{2}g_{j\alpha}\partial_in^{\alpha},
	\]
	where the Greek index $\alpha$ ranges over $0,1,2,3$ with $x^0=\mathfrak{t}$, we must define
	\[
	(g_1)_{ij}=\pa_{\mathfrak{t}}g_{ij}|_{\Sigma_0}=-2(1+N_b^{-2}h(X_b,X_b))^{1/2}N_bk_{ij}+\mathcal{L}_{X_b}h_{ij}
	\]
	Now the initial data for the remaining components of $g_1$, i.e. $(g_1)_{0\alpha}=\partial_{\mathfrak{t}}g_{0\alpha}|_{\Sigma_0}$, are fixed by the generalized wave map gauge condition. More precisely, since $\theta(g;g_b)$ does not contain the derivatives of $g$, it follows that
		\begin{align*}
	0=\widetilde{\Upsilon}^E(g;g_b)_i&=g_{i\kappa}g^{\mu\nu}(\Gamma(g)_{\mu\nu}^\kappa-\Gamma(g_b)_{\mu\nu}^\kappa)-\theta(g;g_b)_i\\
	&=g^{\mu\nu}\partial_{\mu}g_{i\nu}-\frac{1}{2}g^{\mu\nu}\partial_i g_{\mu\nu}-g_{i\kappa}g^{\mu\nu}\Gamma(g^0)_{\mu\nu}^\kappa-\theta(g;g_b)_i\quad \mbox{at}\quad \Sigma_0
	\end{align*}
	fixes $\partial_{\mathfrak{t}}g_{0i}|_{\Sigma_0}$ and
	\begin{align*}
	0=\widetilde{\Upsilon}^E(g;g_b)_0&=g_{0\kappa}g^{\mu\nu}(\Gamma(g)_{\mu\nu}^\kappa-\Gamma(g_b)_{\mu\nu}^\kappa)-\theta(g;g_b)_0\\
	&=g^{\mu\nu}\partial_{\mu}g_{0\nu}-\frac{1}{2}g^{\mu\nu}\partial_0 g_{\mu\nu}-g_{0\kappa}g^{\mu\nu}\Gamma(g_b)_{\mu\nu}^\kappa-\theta(g;g_b)_0\quad \mbox{at}\quad \Sigma_0
	\end{align*}
	fixes $\partial_{\mathfrak{t}}g_{00}|_{\Sigma_0}$. This explicit construction of $g_1$ implies that $g_1=\pa_{\mathfrak{t}}g_b|_{\Sigma_0}=0$ if $(h,k)=(h_b,k_b)$, and 
	\[
		\norm{g_1}_{ \eHb^{s,1/2+\alpha}(\bar{\Sigma}_0;S^2{}\widetilde{^{\scop} T^*}\bar{\Sigma}_0)}\lesssim\norm{h-h_b}_{\eHb^{s+1,-1/2+\alpha}(\bar{\Sigma}_0;S_{>0}^2 {}^{\scop}T^*\bar{\Sigma}_0)}+\norm{k-k_b}_{\eHb^{s,1/2+\alpha}(\bar{\Sigma}_0;S^2{}^{\scop}T^*\bar{\Sigma}_0)}.
	\]
	So we finish the construction of the initial data $(g_0, g_1)$ which induces the data $(h,k)$ on $\Sigma_0$ and satisfies the generalized wave map gauge condition at $\Sigma_0$.
	
	We next turn to the construction of the components $(A_0,A_1)$ of the map $i_b$. Here we closely follow the construction in \cite[\S 3.5]{H18}. Motivated by the relation $F=dA$, we first construct a bounded linear map
	\begin{equation}
		\label{EqKNdSIniInverse}
		\mathcal{B}^\sharp \colon \Bigl\{ u\in \eHb^{s,\ell}(\bar{\Sigma}_0;\Lambda^2 {}^{\scop}T^*\bar{\Sigma}_0) \colon d u=0,\ \int_{\BS^2} u=0 \Bigr\} \to \eHb^{s,\ell-1}(\bar{\Sigma}_0;{}^{\scop}T^*\bar{\Sigma}_0)
	\end{equation}
	such that $d\circ\mathcal{B}^\sharp=\mbox{Id}$ with $\ell>-1/2$. Using the structure $\bar{\Sigma}_0\cong I\times\BS^2$, $I=[0, 1/r_-]_{\rho=1/r}$, of $\bar{\Sigma}_0$, we define $\pi\colon\bar{\Sigma}_0\to\BS^2$ to be the projection onto the second factor, and $i:\colon\BS^2\hookrightarrow\bar{\Sigma}_0$ be the embedding $\omega\mapsto(1/r_-,\omega)$. Then the map $K\colon \eHb^{\sigma,\ell}(\bar{\Sigma}_0;\Lambda {}^{\scop}T^*\bar{\Sigma}_0)\to \eHb^{\sigma,\ell-1}(\bar{\Sigma}_0;\Lambda{}^{\scop}T^*\Sigma_0)$, $\sigma\geq 0,\ell>-1/2$, defined by linear extension from
	\[
	K(f(r,\omega) dr\wedge \pi^*u)(r,\omega) := (\pi^*u)(r,\omega)\int_{r_-}^r f(s,\omega)\,ds,\quad K(f(r,\omega)\pi^*u):=0,
	\]
	for $f\in \eHb^{\sigma,1/2+\alpha}(\bar{\Sigma}_0)$ and $u\in C^\infty(\BS^2;\Lambda T^*\BS^2)$, satisfies $\mbox{Id}-\pi^*j^*=d K+K d$. Therefore, for $u$ satisfying $du=0$, we conclude that $u=d K u + \pi^*(j^* u)$. According to Hodge decomposition theory and the fact that the de Rham cohomology group $H^2_{dR}(\BS^2)\cong \BR$, we see that $u_1:=j^*u\in H^s(\BS^2;\Lambda^2 T^*\BS^2)$ can be written uniquely as $u_1=\sstar u_0+\sd u_2$ where $u_0\in\BR$ and $u_2\in H^{s+1}(\BS^2;T^*\BS^2)$. Here, $\sd,\sdelta,\sstar$ denote the exterior differential, codifferential and Hodge star on $\BS^2$. Since $K(f(r,\omega)\pi^*u)=0$ and $\int_{\BS^2}u=0$, it follows that
	\[0=\int_{\BS^2}\pi^*(j^*u)=\int_{\BS^2}u'=\int_{\BS^2} u_0\,d\mbox{Vol}_{\sg}=4\pi u_0,
	\]
	 and thus $u=d\mathcal{B}^\sharp u$ with $\mathcal{B}^\sharp u:=K u+\pi^* u_2$, which finishes the construction of the map \eqref{EqKNdSIniInverse}.
	
	Then we define
	\[
 A_0=(\iota_{\pa_{\mathfrak{t}}}A_b)\,d\mathfrak{t}+A_0^\sharp,\quad A_0^\sharp:=i^*A_b + \mathcal{A}^\sharp(\star_h\mathbf{H}-\star_{h_b}\mathbf{H}_b)\in i^* A_b+\eHb^{s,-1/2+\alpha}(\bar{\Sigma}_0; {}^{\scop}T^*\bar{\Sigma}_0).
	\]
	 The above definition satisfies $d A_0^\sharp=i^*(\star_h\mathbf{H})$, and $A_0=A_b$ if $(h,\mathbf{H})=(h_b,\mathbf{H}_b)$. Next we assume that
	\[
	A_1= a_1\,d{\mathfrak{t}} + A_1^\sharp
	\]
	with $a_1\in \eHb^{s-1,\ell_1}(\bar{\Sigma}_0)$ and $A_1^\sharp\in \eHb^{s-1, \ell}(\bar{\Sigma}_0;{}^{\scop}T^*\bar{\Sigma}_0)$ to be determined. Recall the future timelike unit normal $n$ to $\Sigma_0$ with respect to $g_0$ as defined above. The requirement that $-i^*\iota_n d A=\mathbf{E}$ at $\Sigma_0$ then reads $\mathbf{E}= n^0(d(\iota_{\pa_{\mathfrak{t}}}A_b)-A_1^\sharp) - \iota_{n^k\pa_k} d A_0^\sharp$. Then we must define
	\[
	 A_1^\sharp:=d(\iota_{\pa_{\mathfrak{t}}}A_b) - \frac{(\mathbf{E} + \iota_{n^k\pa_k} d A_0^\sharp)}{n^0}\in \eHb^{s-1,1/2+\alpha}(\bar{\Sigma}_0; {}^{\scop}T^*\bar{\Sigma}_0),
	\]
	 which satisfies that $A_1^\sharp=0$ if $(h,k,\mathbf{E},\mathbf{H})=(h_b,k_b,\mathbf{E}_b,\mathbf{H}_b)$.
	
	Lastly, the generalized Lorenz gauge condition 
	\begin{align*}
	0=\widetilde{\Upsilon}^M(g,A;g_b,A_b)&=\tr_{g_0}\delta_{g_b}^*(A-A_b)\\
	&=\frac{1}{2}g_0^{\mu\nu}\Big(\pa_\mu(A_\nu-A_{b,\nu})+\pa_\mu(A_\nu-A_{b,\nu})-2\Gamma_{\mu\nu}^\kappa(g_0)(A_\kappa-A_{b,\kappa})\Big)\quad\mbox{at}\quad\Sigma_0,
	\end{align*}
	where $g_0^{\mu\nu}$ is the inverse $g_0$, fixes $a_1=\pa_{\mathfrak{t}}A_0\in \eHb^{s-1,1/2+\alpha}(\bar{\Sigma}_0)$. In particular, we have $a_1=0$ if the data $(h,k,\mathbf{E},\mathbf{H})$ are induced by $(g_b,A_b)$. Moreover, we have 
	\begin{align*}
&	\norm{A_0-A_b|_{\Sigma_0}}_{ \eHb^{s,-1/2+\alpha}(\bar{\Sigma}_0;{}\widetilde{^{\scop} T^*}\bar{\Sigma}_0)}+\norm{A_1}_{ \eHb^{s-1,1/2+\alpha}(\bar{\Sigma}_0;{}\widetilde{^{\scop} T^*}\bar{\Sigma}_0)}\\
&\quad\lesssim\norm{h-h_b}_{\eHb^{s,-1/2+\alpha}(\bar{\Sigma}_0;S_{>0}^2 {}^{\scop}T^*\bar{\Sigma}_0)}
 +\norm{\mathbf{H}-\mathbf{H}_b}_{\eHb^{s,1/2+\alpha}(\bar{\Sigma}_0; {}^{\scop}T^*\bar{\Sigma}_0)}+\norm{\mathbf{E}-\mathbf{E}_b}_{\eHb^{s,1/2+\alpha}(\bar{\Sigma}_0; {}^{\scop}T^*\bar{\Sigma}_0)}	
	\end{align*}
Also, according to the above explicit construction of the map $i_b$, we see that the map $i_b$ is smooth. This completes the proof.
\end{proof}

Now we can make use of the map $i_b$ to construct gauged initial data at the linearized level. Correspondingly, we consider for $s>2,0<\alpha<1$ the subspace of the initial data for ungauged linearized Einstein-Maxwell equations around $(g_b,A_b)$
\begin{equation}
	\label{EqKNIniSpaceLin}
	\begin{split}
		\dot{\mathcal{Z}}^{s,\alpha} &:= \Bigl\{ (\dot{h},\dot{k},\dot{\mathbf{E}},\dot{\mathbf{H}}) \mid  D_{(h_b, \mathbf{H}_b)}(\star_{(\bullet)}(\bullet))(\dot{h},\dot{\mathbf{H}})=0,\ \int_{\BS^2}\frac{d}{ds}|_{s=0}\star_{h_b+s\dot{h}}(\mathbf{H}_b+s\dot{\mathbf{H}})=0,\\
		&\hspace{3em}(\dot{h},\dot{k},\dot{\mathbf{E}},\dot{\mathbf{H}})	\in \eHb^{s+1,-1/2+\alpha}(\bar{\Sigma}_0;S_{>0}^2 {}^{\scop}T^*\bar{\Sigma}_0)\\
		&\hspace{5em}\times \eHb^{s,1/2+\alpha}(\bar{\Sigma}_0;S^2{}^{\scop}T^*\bar{\Sigma}_0) \times \eHb^{s,1/2+\alpha}(\bar{\Sigma}_0;{}^{\scop}T^*\bar{\Sigma}_0)\times \eHb^{s,1/2+\alpha}(\bar{\Sigma}_0;{}^{\scop}T^*\bar{\Sigma}_0) \Bigr\},
	\end{split}
\end{equation}
where $S^2_{>0}{}^{\scop}T^*\bar{\Sigma}_0$ denotes the fiber bundle of positive definite inner products on ${}^{\scop}T\bar{\Sigma}_0$. We again remark that the definition of the space $\dot{\mathcal{Z}}^{s,\alpha}$ incorporates the vanishing of the linearized magnetic charge and the linearization of one of the constraint equations \eqref{EqBasicNLconstraints2}.

We also define the subspace of the initial data for the gauge-fixed linearized Einstein-Maxwell equations 
\begin{equation}
	\label{EqKNIniSpacegaugedLin}
	\begin{split}
		\dot{\mathcal{Y}}^{s,\alpha} &:= \Bigl\{(\dot{g}_0,\dot{g}_1,\dot{A}_0,\dot{A}_1)\mid (\dot{g}_0,\dot{g}_1,\dot{A}_0,\dot{A}_1)\in \eHb^{s+1,-1/2+\alpha}(\bar{\Sigma}_0;S^2{}\widetilde{^{\scop} T^*}\bar{\Sigma}_0) \\
		&	\hspace{4em}\times \eHb^{s,1/2+\alpha}(\bar{\Sigma}_0;S^2 {}\widetilde{^{\scop}T^*}{\bar{\Sigma}_0}) 
		\times \eHb^{s,-1/2+\alpha}(\bar{\Sigma}_0;{}\widetilde{^{\scop}T^*}{\bar{\Sigma}_0}) \times \eHb^{s-1,1/2+\alpha}(\bar{\Sigma}_0;{}\widetilde{^{\scop}T^*}{\Sigma_0})\Bigr\}.
	\end{split}
\end{equation}
Then we have

\begin{cor}
	\label{CorKNIniLin}
For all $s\geq 2, 0<\alpha<1$, the map
\begin{align*}
	D_{(h_b, k_b, \mathbf{E}_b,\mathbf{H}_b)}i_b \colon  \dot{\mathcal{Z}}^{s,\alpha} \to \dot{\mathcal{Y}}^{s,\alpha}
\end{align*}
satisfies that for $(\dot{h},\dot{k},\dot{\mathbf{E}},\dot{\mathbf{H}})\in\dot{\mathcal{Z}}^{s,\alpha}$, the sections
\[
(\dot{g}_0,\dot{g}_1,\dot{A}_0,\dot{A}_1) = 	D_{(h_b, k_b, \mathbf{E}_b,\mathbf{H}_b)}i_b(\dot{h},\dot{k},\dot{\mathbf{E}},\dot{\mathbf{H}})
\]
induce the data $(\dot{h},\dot{k},\dot{\mathbf{E}},\dot{\mathbf{H}})$ on $\Sigma_0=\{\mathfrak{t}=0\}$, and they satisfy the linearized gauge conditions $D_{g_b}\widetilde{\Upsilon}^E(\dg)=0$ and $D_{(g_b, A_b)}\widetilde{\Upsilon}^M(\dg,\dA;g_b,A_b)=-\delta_{g_b}\dot{A}=0$ in the sense that for any section $(\dg,\dA)$ of $S^2 T^*M\oplus T^*M$ near $\Sigma_0$ with $\gamma_0(\dg,\dA)=(\dot{g}_0,\dot{g}_1,\dot{A}_0,\dot{A}_1)$, it induces the initial data $(\dot{h},\dot{k},\dot{\mathbf{E}},\dot{\mathbf{H}})$ at $\Sigma_0$ and the linearized gauge conditions hold at $\Sigma_0$.
\end{cor}
\begin{proof}
Let
\[
h(s) = h_b+s\dot{h},\ k(s)=k_b+s\dot{k},\ \mathbf{E}(s)=\mathbf{E}_b+s\dot{\mathbf{E}},\ \mathbf{B}(s)=\mathbf{H}_b+s\dot{\mathbf{H}}.
\]
For $s$ close to $0$, we find that 
\[
\star_{h+s\dot{h}}\mathbf{H}(s) = \star_h\mathbf{H}+ s D_{(h_b,\mathbf{H}_b)}(\star_{(\cdot)}(\cdot))(\dot{h},\dot{\mathbf{H}})+\mathcal{O}(s^2)
\]
and $d(s):=(h(s),k(s),\mathbf{E}(s),\mathbf{H}(s))\in\mathcal{U}\subset \mathcal{Z}^{2,\alpha}$. Therefore, we have
\[
(\dot{g}_0,\dot{g}_1,\dot{A}_0,\dot{A}_1) = \frac{d}{ds}i_b(d(s))\big|_{s=0} = D_{(h_b,k_b,\mathbf{E}_b,\mathbf{H}_b)}i_b(\dot{h},\dot{k},\dot{\mathbf{E}},\dot{\mathbf{H}})
\]
and $(\dot{g}_0,\dot{g}_1,\dot{A}_0,\dot{A}_1) $ satisfy the linearized gauge conditions in the sense that 
\begin{equation}
	\label{EqKNdSIniLinGauge}
	D_{g_b}\widetilde{\Upsilon}^E(\dot{g})=0, \quad D_{(g_b,A_b)}\widetilde{\Upsilon}^M(\dg,\dot{A};g_b,A_b)=-\delta_{g_b}\dA=0,
\end{equation}
for any $(\dg,\dA)$ with $\gamma_0(\dot{g},\dot{A})=(\dot{g}_0,\dot{g}_1,\dot{A}_0,\dot{A}_1)$.
\end{proof}

%%%%%%%%%%%%%%%%%%%%%%%%%%%%%%%%%%%%%%%%%%%%%%%%%%%%%%%%%%%%%%%%%%%%%%%%%%%
\section{Analysis of the linearized gauge-fixed Einstein-Maxwell operator}
\label{sec:microandsemisetup}
Throughout this section, we continue using the notation from \S\ref{sec:KNblackholes}. 

In \S\ref{subsec:microlocalanalysis}, we will discuss the characteristic set and the global dynamics of the Hamiltonian flow of the principal symbol $p(\sigma)$ of the operator $\widehat{\Box_{g_{b}}}(\sigma):=e^{it_{b,*}\sigma}\Box_{g_b}e^{-it_{b,*}\sigma}$ on full subextremal KN spacetimes (see Figure \ref{fig:microlocalphase} for an illustration). Concretely, first we will show that the characteristic set of $p(\sigma)$ has two parts, one of which lies inside the ergoregion while the other is at the spatial infinity $\pa_+\CX$. Then we prove that there exist radial points, to which the nearby integral curves of the Hamiltonian vector field $H_{p(\sigma)}$ converge in either forward direction or backward direction, at event horizon and spatial infinity $\pa_+\CX$. Finally, we describe the global dynamics of the Hamiltonian flow of the principal symbol $p(\sigma)$.

 In \S\ref{subsec:fredholmestimates}, we combine the global dynamics of the Hamiltonian flow of the principal symbol $p(\sigma)$ established in \S\ref{subsec:microlocalanalysis} with the elliptic estimate, propagation of singularities estimate, the radial point estimate at event horizon, the scattering radial point estimate at spatial infinity $\pa_+\CX$ and hyperbolic estimate to prove the uniform Fredholm estimates for the operators $\widehat{\Box_{g_b}}(\sigma),\widehat{\mathcal{P}_{b,\gamma}}(\sigma), \widehat{\mathcal{W}_{b, \gamma}}(\sigma), \widehat{L_{b,\gamma}}(\sigma)$ for bounded $\sigma$ in the closed upper half plane.
 
 In\S\ref{subsec:desodfernel}, we will give a detailed description of kernel of the following wave operators: $\widehat{\Box_{g_{b}}}(\sigma)$ on scalar functions, $\widehat{\mathcal{P}_{b, \gamma}}(\sigma),\widehat{\mathcal{W}_{b, \gamma}}(\sigma)$ on scattering $1$-forms and the linearized gauge-fixed Einstein-Maxwell operator $\widehat{L_{b,\gamma}}(\sigma)$.
 
 In \S\ref{subsec:semiclassicalanalysis}, we will analyze the characteristic set and the global dynamics of the Hamiltonian flow of the semiclassical principal symbol $p_{h,z}$ of the semiclassical rescaled operator $h^2\widehat{\Box_{g_{b}}}(h^{-1}z)$ on full subextremal KN spacetimes for $z\in\BR\setminus 0$ (see Figure \ref{fig:semiphase} for an illustration). Specifically, we first show that the characteristic set of $p_{h,z}$ can be split into a disjoint union of two components. Then we prove that there exist radial points at event horizon and spatial infinity $\pa_+\CX$. Next, we discuss the trapped set associated to the Hamiltonian flow of the semiclassical symbol $p_{h,z}$ and prove that it is normally hyperbolic. Finally, we describe the global dynamics of the Hamiltonian flow of the semiclassical principal symbol $p_{h,z}$.
 
 In \S\ref{subsec:highenergyestimates}, we combine the global dynamics of the Hamiltonian flow of the semiclassical principal symbol $p_{h,z}$ established in \S\ref{subsec:semiclassicalanalysis} with the elliptic estimate, propagation of singularities estimate, the radial point estimate at event horizon, the scattering radial point estimate at spatial infinity $\pa_+\CX$ and hyperbolic estimate, all of which are in the semiclassical version, together with the estimate at normally hyperbolic trapping to establish the high energy estimates for the operators $\widehat{\Box_{g_b}}(\sigma),\widehat{\mathcal{P}_{b,\gamma}}(\sigma), \widehat{\mathcal{W}_{b, \gamma}}(\sigma), \widehat{L_{b,\gamma}}(\sigma)$ for $\abs{\RE\sigma}\gg1, \IM\sigma\leq C$ in the closed upper half plane. This implies the invertibility of these operators for $\abs{\RE\sigma}\gg1, \IM\sigma\leq C$ in the closed upper half plane.

In \S\ref{subsec:energyestimates}, we shall establish the energy estimates for the solutions to various wave type equations on slowly rotating Kerr-Newman metrics.

Fix KN black hole parameters $b=(\Bm, \Ba, \BQ)$ close to $b_0=(\Bm_0, \Ba_0, \BQ_0)$ and let
\[
g=g_b
\]
Recall that with $a=\abs{\Ba}$
\begin{align*}
	g^{-1}&=\rho_b^{-2}\bigg(\Delta_b\pa_r^2+\frac{\chi^2-1}{\Delta_b}\Big((r^2+a^2)\pa_{t_\chi}+a\pa_\varphi\Big)^2+\pa_\theta^2+\frac{1}{\sin^2\theta}\Big(\pa_\varphi+a\sin^2\theta\pa_{t_{\chi}}\Big)^2\\
	&\qquad\qquad+2\chi\Big((r^2+a^2)\pa_{t_\chi}+a\pa_\varphi\Big)\pa_r\bigg).
	\end{align*}
We define 
\[
\widehat{\Box_{g}}(\sigma):=e^{it_{b,*}\sigma}\Box_{g_b}e^{-it_{b,*}\sigma}.
\]
Let $\cscform$ be the compact manifold with the corners of codimension two obtained by radial compactification of the fibers $\fscform$ (whose detail will be discussed later on). Then its boundary consists of three hypersurfaces
\[
\pa\cscform=\fiscform\cup\abscform\cup\siscform
\]
where $\fiscform=(\fscform\setminus0)/\BR^+$ is called fiber infinity.

\subsection{Microlocal geometry and dynamics of Kerr-Newman spacetimes}\label{subsec:microlocalanalysis}
Now, we are ready to discuss the microlocal geometry of the KN metric $g=g_b$. Away from $\pa_+\CX$, we note that $\fscform$ is isomorphic to $T^*\CX$ and write the covectors as
\[
\xi_r dr+\xi_\theta d\theta+\xi_\varphi d\varphi.
\]
Then the principal symbol of $\widehat{\Box_{g}}(\sigma)$ is given by
\[
p=\sigma_2(\widehat{\Box_{g}}(\sigma))=-\rho_b^{-2}\Big(\Delta_b\xi_r^2+\xi_\theta^2+\big(\frac{a^2}{\Delta_b}(\chi^2-1)+\frac{1}{\sin^2\theta}\big)\xi_\varphi^2+2\chi a \xi_\varphi\xi_r\Big).
\]
Near $\pa_+\CX$, we write the scattering covectors as
\[
\zeta_\rho\frac{d\rho}{\rho^2}+\zeta_{\theta}\frac{d\theta}{\rho}+\zeta_{\varphi}\frac{d\varphi}{\rho},\quad \rho=\frac{1}{r}.
\]
Then the scattering principal symbol (see \cite{M92}) of $\widehat{\Box_{g}}(\sigma)$ at $\pa_+\CX$  is given by
\[
2\sigma\zeta_\rho-\zeta_\rho^2-\zeta_\theta^2-\frac{1}{\sin^2\theta}\zeta_\varphi^2=-(\zeta_\rho-\sigma)^2-\zeta_\theta^2-\frac{1}{\sin^2\theta}\zeta_\varphi^2+\sigma^2
\]
To handle the poles $\{\theta=0\}$ and $\{\theta=\pi\}$, where the spherical coordinates break down, we introduce new coordinates 
\[
x_1=\sin\theta\cos\varphi,\quad x_2=\sin\theta\sin\varphi,
\]
and let $\xi_1$ and $\xi_2$ be dual variables to $x_1$ and $x_2$ respectively. Correspondingly, we have
\[
\xi_\theta=(x_1\xi_1+x_2\xi_2)\cot\theta,\quad \xi_\varphi=x_1\xi_2-x_2\xi_1.
\]
and note that $\xi_\varphi$ is a smooth function in $(x_1, x_2, \xi_1, \xi_2)$. Since 
\begin{align*}
\xi_\theta^2+\frac{1}{\sin^2\theta}\xi_\varphi^2=\frac{1-x_1^2-x_2^2}{x_1^2+x_2^2}(x_1\xi_1+x_2\xi_2)^2+\frac{1}{x_1^2+x_2^2}(x_1\xi_2-x_2\xi_1)^2=\xi_1^2+
\xi_2^2-(x_1\xi_1+x_2\xi_2)^2
\end{align*}
is a smooth function in $(r, x_1,x_2, \xi_r, \xi_1,\xi_2)$ near the poles, so is $p$. Therefore, we can perform all symbol calculations away from these two poles and extend the results to the poles. From now on, we do not emphasize the analysis near these two poles.
\subsubsection{Characteristic set}
We first study the characteristic set. Away from $\pa_+\CX$, since 
\[
p=-\rho_b^{-2}\Big(\Delta_b(\xi_r+\frac{a\chi}{\Delta_b}\xi_{\varphi})^2+\xi_\theta^2+\frac{\Delta_b-a^2\sin^2\theta}{\Delta_b\sin^2\theta}\xi_\varphi^2\Big),
\]
it follows that the characteristic set $\{p=0, \xi\neq 0\}$ of $p$ satisfies
\[
\{p=0, \xi=(\xi_r, \xi_\theta, \xi_\varphi)\neq0\}\subset\{\Delta_b-a^2\sin^2\theta\leq 0, \xi\neq0\}\subset\fscform\setminus0.
\]
Since $\chi=1$ in the region $\Delta_b-a^2\sin^2\theta\leq 0$, the principal symbol can be written as 
\[
p=-\rho_b^{-2}\Big(\Delta_b\xi_r^2+2a\xi_r\xi_\varphi+\tilde{p}\Big)\quad\mbox{where}\quad \tilde{p}=\xi_\theta^2+\frac{1}{\sin^2\theta}\xi_\varphi^2,
\]
and thus $\xi_r\neq 0$ at the characteristic set of $p$. As a consequence, in the fiber radial compactification $\cscform$ of $\fscform$, we use the following coordinates near the fiber infinity $\cscform$,
\begin{equation}\label{eq:coordatfiberinfty}
\rho_\xi=\frac{1}{\abs{\xi_r}}\in[0, \infty), \quad \hat{\xi}=(\hat{\xi}_\theta, \hat{\xi}_\varphi)=\rho_\xi(\xi_\theta, \xi_\varphi).
\end{equation}
We note that $\rho_\xi$ is a boundary defining function of $\cscform$, i.e., $\rho_\xi=0$ at $\pa\cscform=\fiscform$, and $\hat{\xi}$ is a coordinate system on $\fiscform$.

Since $p$ is a homogeneous polynomial of order $2$ in $\xi=(\xi_r, \xi_\theta, \xi_\varphi)$, the rescaled symbol $\rho_\xi^2p$ and the rescaled Hamiltonian vector field $\rho_\xi H_p$ extend smoothly to $\cscform$. Therefore, we study the flow of the rescaled Hamiltonian vector field $\rho_\xi H_p$ on the characteristic set $\{\rho_\xi^2 p=0\}\subset\cscform\setminus0$.
We now split the characteristic set $\{\rho_\xi^2 p=0\}\subset\cscform\setminus 0$ into two components
\[
\Sigma_{\pm}:=\{\rho_\xi^2p=0\}\cap\{\pm\xi_r>0\},\quad  \pa\Sigma_{\pm}=\Sigma_{\pm}\cap\fiscform.
\]

Near $\pa_+\CX$, we have $\chi=0$ and thus
\[
	g^{-1}=\rho_b^{-2}\bigg(\Delta_b\pa_r^2-\frac{1}{\Delta_b}\Big((r^2+a^2)\pa_{t_\chi}+a\pa_\varphi\Big)^2+\pa_\theta^2+\frac{1}{\sin^2\theta}\Big(\pa_\varphi+a\sin^2\theta\pa_{t_{\chi}}\Big)^2\bigg).
\]
Then using $t_\chi=t$ and $t_{b,*}=t-r_{(\Bm, \Ba_0,\BQ),*}$ near $\pa_+\CX$, we calculate
\[
\widehat{\Box_{g}}(\sigma)=-2i\sigma\rho^2\pa_\rho+(\rho^2\pa_\rho)^2+\rho^2\pa_\theta^2+\frac{\rho^2}{\sin^2\theta}\pa_\varphi^2+\rho\mbox{Diff}_{\scop}^2(\CX),\quad \rho=\frac{1}{r}.
\]
Therefore, the scattering principal symbol of $\widehat{\Box_{g}}(\sigma)$ at $\pa_+\CX$ is given by
\[
p_{\scop}(\sigma)=\sigma_{\scop}(\widehat{\Box_{g}}(\sigma))=-(\zeta_\rho-\sigma)^2-\tilde{p}+\sigma^2,\quad \tilde{p}=\zeta_\theta^2+\frac{1}{\sin^2\theta}\zeta_\varphi^2,
\]
and the scattering characteristic set is the surface
\begin{align*}
\Sigma_{\scop}(\sigma)&=\{p_{\scop}(\sigma)=0\}\subset \siscform.
\end{align*}
We note that $\Sigma_{\scop}(\sigma)$ is the scattering zero section $o_{\pa_+\CX}\subset\siscform$ for $\IM\sigma>0$ and $\sigma=0$.

%%%%%%%%%%%%%%%%%%%%%%%%%%%%%%%%%%%%%%%%%%%%%%%%%%%%%%
\subsubsection{The radial points at event horizon}
We now analyze the behavior of the Hamiltonian flow $\exp(s\rho_\xi H_p)$ on the characteristic set $\Sigma=\Sigma_+\cup\Sigma_-\subset\cscform\setminus 0$, that is, we consider the portion whose projection to the base space $(r, \theta, \varphi)$ is away from $\pa_+\CX$. Near $\Sigma=\Sigma_+\cup\Sigma_-$, we see that $\chi=1$ and thus
\[
p=-\rho_b^{-2}\Big(\Delta_b\xi_r^2+2a\xi_r\xi_\varphi+\tilde{p}\Big)\quad\mbox{where}\quad \tilde{p}=\xi_\theta^2+\frac{1}{\sin^2\theta}\xi_\varphi^2.
\]
Then we calculate
\[
H_p=-\rho_b^{-2}\Big((2\Delta_b\xi_r+2a\xi_\varphi)\pa_r+2a\xi_r\pa_\varphi-\frac{\pa_r\Delta_b}{\pa r}\xi_r^2\pa_{\xi_r}+H_{\tilde{p}}\Big)-\rho_b^{-2}pH_{\rho_b^{2}}.
\]
In the coordinates \eqref{eq:coordatfiberinfty} near the fiber infinity $\fiscform$, we compute
\begin{equation}\label{eq:CrescaledHamil}
	\begin{split}
\rho_\xi H_p&=-\rho_b^{-2}\Big((2\Delta_b(\sgn\xi_r)+2a\hat{\xi}_\varphi)\pa_r+2a(\sgn\xi_r)\pa_\varphi-\frac{\pa\Delta_b}{\pa r}(\sgn\xi_r)\xi_r\pa_{\xi_r}+\rho_\xi H_{\tilde{p}}\Big)-\rho_b^{-2}(\rho_\xi p)H_{\rho_b^2}\\
&=-\rho_b^{-2}\Big((2\Delta_b(\sgn\xi_r)+2a\hat{\xi}_\varphi)\pa_r+2a(\sgn\xi_r)\pa_\varphi+2\hat{\xi}_\theta\pa_{\theta}+\frac{2}{\sin^2\theta}\hat{\xi}_\varphi\pa_{\varphi}\Big)\\
&\quad-\rho_b^{-2}\frac{\pa\Delta_b}{\pa r}(\sgn\xi_r)\Big(\rho_\xi\pa_{\rho_\xi}+\hat{\xi}_\theta\pa_{\hat{\xi}_\theta}+\hat{\xi}_\varphi\pa_{\hat{\xi}_\varphi}\Big)-\rho_b^{-2}\frac{2\cos\theta}{\sin^3\theta}\hat{\xi}^2_\varphi\pa_{\hat{\xi}_\theta}-\rho_b^{-2}(\rho^2_\xi p)\rho^{-1}_\xi H_{\rho_b^2}.
\end{split}
\end{equation}
We let
\[
\Lambda_\pm=\{\Delta_b=0, (\xi_{\theta}, \xi_{\varphi})=0, \pm\xi_r>0\}
\]
and
\[
L_\pm:=\pa\Lambda_\pm=\Lambda_\pm\cap\fiscform=\{\Delta_b=0, \rho_\xi=\pm\frac{1}{\xi_r}=0, \hat{\xi}=0\}.
\]

Now we describe the property of the set $L_\pm$. More specifically, $L_\pm$ are \textit{radial sources/sinks} in the following sense (see \cite[Definition E.50]{DZ19}):
\begin{defn}
	Let $\kappa:\cscform\setminus0\to \fiscform=(\fscform\setminus0)/\BR^+$ be the natural projection map, i.e., for each $(r,\theta, \varphi, \xi)\in\fscform\setminus 0$, the ray $(r, \theta, \varphi, s\xi)$ converges to $\kappa((r, \theta, \varphi, \xi))$ as $s\to\infty$. Then for the rescaled Hamiltonian flow 
	\[
	\phi_s:=\exp(s\rho_\xi H_p): \cscform\to\cscform, 
	\]
	we say that a nonempty compact $\phi_s$-invariant set
	\[
	L\subset\{\rho_\xi^2 p=0\}\cap\fiscform
	\]
	is a \textit{radial source} for $p$, if there exists a neighborhood $U\subset\cscform$ of $L$ such that uniformly in $(r, \theta, \varphi,\xi)\in U\cap\fscform$, one have
	\[
	\kappa(\phi_s(r, \theta, \varphi, \xi))\to L\quad \mbox{as}\quad s\to-\infty
	\]
	and 
	\[
	\abs{\phi_s(r ,\theta, \varphi, \xi)}\geq Be^{B\abs{s}}\abs{\xi},\quad s\leq 0
	\]
	for some $B>0$. Here, $\abs{\cdot}$ denotes a norm on the fibers of $\fscform$.
	
	A \textit{radial sink} for $p$ is be definition a radial source for $-p$.
\end{defn}
\begin{lem}\label{lem:radialpointateh}
	$L_\pm$ are invariant under the Hamiltonian flow $\exp(s\rho_\xi H_p)$ on the characteristic set $\Sigma$. Moreover, $L_+$ is a radial sink and $L_-$ is a radial source for the flow $\exp(s\rho_\xi H_p)$, see Figure \ref{fig:charaneareh}.
	\begin{figure}[!h]
		\centering
		\begin{tikzpicture}
			\draw (-2,0)--(2,0);
			\draw (-2,1)--(2,1);
			\draw (-2,2)--(2,2);
			\draw [dashed] (0,2)--(0,0);
			\draw [
			decoration={markings, mark=at position 0.5 with {\arrow{stealth}}},
			postaction={decorate}
			]
			(0,0) .. controls (0.3, 0.3)  .. (-2,0.6);
			\draw [
			decoration={markings, mark=at position 0.5 with {\arrow{stealth reversed}}},
			postaction={decorate}
			]
			(0,2) .. controls (0.3, 1.7)  .. (-2,1.4);
			\node [circle, inner sep=1pt, fill=black, label=above:{$L_+$}] at (0,2) {};
				\node [circle, inner sep=1pt, fill=black, label=below:{$L_-$}] at (0,0) {};
		\end{tikzpicture}
	\caption{The flow of $\exp(s\rho_\xi H_p)$ near $L_\pm$. It is projected to the $(r,\xi_r)$ coordinates and drawn in a fiber-radially compactified view. The horizontal coordinate is $r$; the dashed line is the event horizon $\{r=r_{b}\}$. The vertical coordinate is $\xi_r/\jb{\xi_r}$, so the top and bottom lines correspond to the fiber infinity $\{\rho_\xi=0\}$. The midline $\{\xi_r=0\}$ lies outside the characteristic set $\{\rho^2_\xi p=0\}$.}
	\label{fig:charaneareh}
	\end{figure}
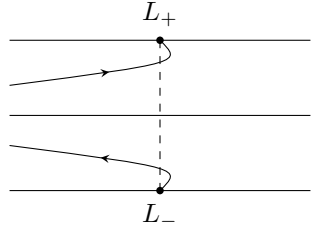
\end{lem}
\begin{proof}
	Since $\{\rho_\xi^2\tilde{p}=0\}\Leftrightarrow\{\hat{\xi}=(\hat{\xi}_\theta, \hat{\xi}_\varphi)=0\}
$, we rewrite
\[
L_\pm=\{\Delta_b=0, \rho_\xi=\pm\frac{1}{\xi_r}=0, \rho_\xi^2\tilde{p}=0\}.
\]
Using the coordinates \eqref{eq:coordatfiberinfty} and expression \eqref{eq:CrescaledHamil}, we find that 
\begin{equation}\label{eq:deriof4quantities}
	\begin{split}
\rho_\xi H_p\rho_\xi&=-\rho_b^{-2}\Big(\frac{\pa\Delta_b}{\pa r}+2r\rho_\xi^2 p\Big)(\sgn\xi_r)\rho_\xi,\\
		\rho_\xi H_p\hat{\xi}_\varphi&=-\rho_b^{-2}\Big(\frac{\pa\Delta_b}{\pa r}+2r\rho_\xi^2 p\Big)(\sgn\xi_r)\hat{\xi}_\varphi,\\
		\rho_\xi H_p\Delta_b&=-2\rho_b^{-2}\frac{\pa\Delta_b}{\pa r}\Big((\sgn \xi_r)\Delta_b+a\hat{\xi}_\varphi\Big),\\
	\rho_\xi H_p(\rho_\xi^2\tilde{p})
			&=-\rho_b^{-2}\Big(2\frac{\pa\Delta_b}{\pa r}+4r\rho_\xi^2p\Big)(\sgn\xi_r)\rho_\xi^2\tilde{p}-2\rho_b^{-2}a^2\sin (2\theta)(\rho_\xi^2p)\hat{\xi}_\theta.
	\end{split}
\end{equation}
This implies that $\rho_\xi H_p$ is tangent to $L_\pm$, so $L_\pm$ is invariant under the flow $\exp(s\rho_\xi H_p)$. Moreover, in a neighborhood of $L_\pm=\{\Delta_b=0, \rho_\xi=\pm\xi_r^{-1}=0, \rho_\xi^2\tilde{p}=0\}\subset\{\rho_\xi^2p=0\}$, we have
\begin{equation}\label{eq:deriof3boundarydefining}
	\begin{split}
	\pm	\rho_\xi H_p\rho_\xi^2&\leq -C_0\rho_\xi^2,\\
		\pm	\rho_\xi H_p\hat{\xi}_\varphi^2&\leq -C_0\hat{\xi}_\varphi^2,\\
		\pm	\rho_\xi H_p\Delta_b^2&\leq -C_0\Delta_b^2+C_1\hat{\xi}_\varphi^2,\\
	\pm	\rho_\xi H_p(\rho_\xi^2\tilde{p})&\leq -C_0\rho_\xi^2\tilde{p}+C_1\big(\Delta_b^2+\hat{\xi}_\varphi^2\big).
	\end{split}
	\end{equation}
where $C_0, C_1>0$. It follows from \eqref{eq:deriof3boundarydefining} that in a sufficiently small neighborhood of $L_{\pm}$
\begin{equation}
		\pm\rho_\xi H_pf\leq -Cf,\quad f=\rho_\xi^2+\hat{\xi}_\varphi^2+C_2\Delta_b^2+C_3\rho^2_\xi\tilde{p}\geq 0.
\end{equation}
for some $C,C_2, C_3>0$. This implies that for $(r, \theta,\varphi, \xi)$ in a sufficiently small neighborhood of $L_\pm$, we have
\[
f(\phi_s((r, \theta,\varphi, \xi)))\leq e^{-C\abs{s}}f((r, \theta,\varphi, \xi))\quad \mbox{for}\quad \pm s>0.
\]
Therefore, we have uniformly in $(r, \theta,\varphi, \xi)$ in a neighborhood of $L_\pm$
\begin{align*}
	\phi_s((r, \theta,\varphi, \xi))\to L_\pm\quad\mbox{as}\quad s\to\pm\infty.
\end{align*}
Moreover, according to the first inequality in \eqref{eq:deriof3boundarydefining}, we have
\[
\rho_\xi(\phi_s((r, \theta,\varphi, \xi)))\leq e^{-C_0\abs{s}}\rho_\xi((r, \theta,\varphi, \xi)) \quad \mbox{for}\quad \pm s\geq 0.
\]
This proves that $L_+$ is a radial sink and $L_-$ is a radial source.
\end{proof}
Finally, we consider the imaginary part of the operator $\widehat{\Box_{g}}(\sigma)$ at $L_\pm$, which is needed to calculate the \textit{threshold quantity} for the order of the Sobolev spaces (here the differential order) in the radial point estimates. Recall that near $L_\pm$, the inverse metric $G$ takes the form 
\[
g^{-1}=\rho_b^{-2}\Big(a^2\sin^2\theta\pa^2_{t_\chi}+2(r^2+a^2)\pa_{t_\chi}\pa_r+2a\pa_{t_\chi}\pa_\varphi+\Delta_b\pa_r^2+2a\pa_r\pa_\varphi+\pa_\theta^2+\frac{1}{\sin^2\theta}\pa_\varphi^2\Big)
\]
and $t_\chi=t_{b,*}$. A direct calculation shows that 
\[
\widehat{\Box_{g}}(\sigma)^*=\widehat{\Box_{g}}(\bar{\sigma}).
\]
where $\widehat{\Box_{g}}(\sigma)^*$ denotes the formal adjoint of $\widehat{\Box_{g}}(\sigma)$ defined by $\angles{\widehat{\Box_{g_{b}}}(\sigma)u, }{v}=\angles{u}{\widehat{\Box_{g}}(\sigma)^*v}$ with respect to $L^2(\CX; \sqrt{\abs{\det g}}drd\theta d\varphi)$ for $u,v\in C_c^\infty(X)$. We first compute
\begin{align*}
p_1:=\sigma_1\big(\IM\widehat{\Box_{g}}(\sigma)\big)=\sigma_1\Big(\frac{\widehat{\Box_{g}}(\sigma)-\widehat{\Box_{g}}(\sigma)^*}{2i}\Big)=\rho_b^{-2}\IM\sigma\Big(2(r^2+a^2)\xi_r+2a\xi_\varphi\Big)
	\end{align*}
Let $\beta_0\in C^\infty(L_\pm)$ be a positive function defined as
\[
\beta_0=\mp\frac{\rho_\xi H_p\rho_\xi}{\rho_\xi}|_{L_\pm}.
\]
Then we define $\tilde{\beta}\in C^\infty(L_\pm)$ as 
\[
\tilde{\beta}=\mp\frac{\rho_\xi p_1}{\beta_0}|_{L_\pm}
\]
Let
\[
\beta_{\sup}=\sup\tilde{\beta},\quad \beta_{\inf}=\inf \tilde{\beta}.
\]
If $\tilde{\beta}$ is a constant along $L_\pm$, we may write
\[
\beta=\beta_{\sup}=\beta_{\inf}.
\]
We note that $\beta_{\sup}$ and $\beta_{\inf}$ are the relevant quantities in the radial point estimates. More specifically, in our setting where we consider the operator $\widehat{\Box_{g}}(\sigma)$ with $\IM\sigma\geq 0$, the threshold regularity condition in the \textit{high regularity radial estimate} is given by
\begin{equation}
	s>\frac{1}{2}+\beta_{\sup},
\end{equation}
while the \textit{low regularity radial estimate} requires 
\begin{equation}
	s<\frac{1}{2}+\beta_{\inf}.
	\end{equation}
Using the first equality in \eqref{eq:deriof4quantities}, we find for the operator $\widehat{\Box_{g}}(\sigma)$ that
\[
\tilde{\beta}=-\IM\sigma\frac{\ehKN^2+a^2}{\ehKN-\Bm},
\]
and thus
\begin{equation}\label{eq:thresholdreg}
\beta=\beta_{\sup}=\beta_{\inf}=-\IM\sigma\frac{\ehKN^2+a^2}{\ehKN-\Bm}\leq 0\quad \mbox{for}\quad \IM\sigma\geq 0.
\end{equation}

%%%%%%%%%%%%%%%%%%%%%%%%%%%%%%%%%%%%%%%%%%%%%%%%%%%%%%
\subsubsection{The radial points at spatial infinity $\pa_+\CX$}
We now turn to the analysis near $\pa_+\CX$. Recall that for the operator $\widehat{\Box_{g}}(\sigma)$ with $\IM\sigma\geq0$, the scattering characteristic set is the surface
\[
\Sigma_{\scop}(\sigma)=\{(\rho=\frac{1}{r}=0, \theta, \varphi,\zeta)):p_{\scop}(\sigma)=0\}\subset \siscform.
\] 
where
\[
p_{\scop}(\sigma)=-(\zeta_\rho-\sigma)^2-\tilde{p}+\sigma^2,\quad \tilde{p}=\zeta_\theta^2+\frac{1}{\sin^2\theta}\zeta_\varphi^2.
\]
Using the identification of $T^*\CX$ and $\fscform$ in $\rho>0$
\[
\zeta_\rho=\rho^2\xi_\rho=-\xi_r, \quad (\zeta_\theta, \zeta_\varphi)=\rho(\xi_\theta, \xi_\varphi),
\]
we find that the scattering Hamiltonian vector field associated to $p$ in $\rho>0$ is given by
\begin{align*}
{}^{\scop}\!H_p&=\rho^2\frac{\pa p}{\pa\zeta_\rho}\Big(\frac{\pa}{\pa\rho}+\frac{2\zeta_\rho}{\rho}\frac{\pa}{\pa\zeta_\rho}+\big(\sum_{\mu=\theta, \varphi}\frac{\zeta_\mu}{\rho}\frac{\pa}{\pa\zeta_\mu}\big)\Big)-\Big(\frac{\pa p}{\pa\rho}+\frac{2\zeta_\rho}{\rho}\frac{\pa p}{\pa\zeta_\rho}+\big(\sum_{\mu=\theta, \varphi}\frac{\zeta_\mu}{\rho}\frac{\pa p}{\pa\zeta_\mu}\big)\Big)\rho^2\frac{\pa}{\pa\zeta_\rho}\\
&\quad+\sum_{\mu=\theta,\varphi}\rho\Big(\frac{\pa p}{\pa\zeta_\mu}\frac{\pa}{\pa\mu}-\frac{\pa p}{\pa \mu}\frac{\pa}{\pa\zeta_\mu}\Big)\\
&=\rho\Big(\frac{\pa p}{\pa\zeta_\rho}\big(\rho\frac{\pa}{\pa\rho}+(\sum_{\mu=\theta, \varphi}\zeta_\mu\frac{\pa}{\pa\zeta_\mu})\big)-\big(\rho\frac{\pa p}{\pa\rho}+(\sum_{\mu=\theta, \varphi}\zeta_\mu\frac{\pa p}{\pa\zeta_\mu})\big)\frac{\pa}{\pa\zeta_\rho}+\sum_{\mu=\theta,\varphi}\big(\frac{\pa p}{\pa\zeta_\mu}\frac{\pa}{\pa\mu}-\frac{\pa p}{\pa \mu}\frac{\pa}{\pa\zeta_\mu}\big)\Big).
\end{align*}
By introducing the following coordinates in the fiber radial compactification $\cscform$ of $\fscform$
\[
\rho_\zeta=(1+\zeta_r^2+\zeta_\theta^2+\frac{1}{\sin^2\theta}\zeta_\varphi^2)^{-\frac12},\quad \hat{\zeta}=(\hat{\zeta}_\rho, \hat{\zeta}_\theta, \hat{\zeta}_\varphi)=\rho_\zeta(\zeta_\rho,\zeta_\theta, \zeta_\varphi),
\]
the rescaled scattering Hamiltonian vector field $\rho_\zeta\rho^{-1}{}^{sc}\!H_p$ can be extended to an element in $\mathcal{V}_b(\cscform)$ taking the following form
\begin{align*}
{}^{\scop}\!H_p^{2,0}&:=\rho_\zeta^{2-1}\rho^{-0-1}{}^{\scop}\!H_p\\
&=\rho_\zeta\Big(\frac{\pa p}{\pa\zeta_\rho}\big(\rho\frac{\pa}{\pa\rho}+(\sum_{\mu=\theta, \varphi}\zeta_\mu\frac{\pa}{\pa\zeta_\mu})\big)-\big(\rho\frac{\pa p}{\pa\rho}+(\sum_{\mu=\theta, \varphi}\zeta_\mu\frac{\pa p}{\pa\zeta_\mu})\big)\frac{\pa}{\pa\zeta_\rho}+\sum_{\mu=\theta,\varphi}\big(\frac{\pa p}{\pa\zeta_\mu}\frac{\pa}{\pa\mu}-\frac{\pa p}{\pa \mu}\frac{\pa}{\pa\zeta_\mu}\big)\Big).
\end{align*}

We now discuss the integral curves of ${}^{\scop}\!H_{p_{\scop}(\sigma)}^{2,0}$ on the scattering characteristic set $\Sigma_{\scop}(\sigma)=\{p_{\scop}(\sigma)=0\}\subset\siscform$ for $\sigma\in\BR\setminus0$ (see \cite{MZ96}). Since $\Sigma_{\scop}(\sigma)$ has no intersection with the fiber infinity $\pa\siscform$, we can drop the factor $\rho_\zeta$ which is non-zero on $\Sigma_{\scop}(\sigma)$. Therefore, with $p_{\scop}(\sigma)=-(\zeta_\rho-\sigma)^2-\tilde{p}+\sigma^2$, we write
\[
{}^{\scop}\!H_{p_{\scop}(\sigma)}^{2,0}=(-2\zeta_\rho+2\sigma)\rho\pa_\rho+2\tilde{p}\pa_{\zeta_\rho}+(-2\zeta_\rho+2\sigma)\sum_{\mu=\theta,\varphi}\zeta_\mu\pa_{\zeta_\mu}-H_{\tilde{p}}\quad\mbox{on}\quad \Sigma_{\scop}(\sigma)
\]
where
\[
H_{\tilde{p}}=2\zeta_\theta\pa_\theta+\frac{2\zeta_\varphi}{\sin^2\theta}\pa_\varphi+\frac{\cos\theta}{\sin^2\theta}\zeta^2_\varphi\pa_{\zeta_\theta}.
\]
\begin{lem}\label{lem:radialpointatinfty}
	For $0\neq\sigma\in\BR$, on the characteristic variety  
	\begin{equation}
		\Sigma_{\scop}(\sigma)=\{p_{\scop}(\sigma)=-(\zeta_\rho-\sigma)^2-\tilde{p}+\sigma^2=0\}\subset\siscform,\quad \tilde{p}=\zeta_\theta^2+\frac{1}{\sin^2\theta}\zeta_\varphi^2,
	\end{equation}
there are two invariant submanifolds under the Hamiltonian flow $\exp(s{}^{\scop}\!H_{p_{\scop}(\sigma)}^{2,0})$, one of which is the zero section $R(0)=\{\rho=0, \zeta_\rho=0,(\zeta_\theta, \zeta_\varphi)=0\}$ and the other is $R(\sigma)=\{\rho=0, \zeta_\rho=2\sigma,(\zeta_\theta, \zeta_\varphi)=0\}$.

For $(\theta^0, \varphi^0, \zeta_\rho^0, \zeta_\theta^0, \zeta^0_\varphi)\in\Sigma_{sc}(\sigma)\setminus (R(0)\cup R(\sigma))$, the integral curves $\exp(s{}^{sc}\!H_{p_{\scop}(\sigma)}^{2,0})(\theta^0, \varphi^0, \zeta^0_\rho, \zeta^0_\theta, \zeta^0_\varphi)$ take the following form
\begin{equation}\label{eq:exofscacurves}
	\begin{split}
		\zeta_\rho(s)&=\sigma+\abs{\sigma}\sin(s+s_0),\quad 	\zeta_\rho(0)=\zeta_\rho^0,\\
		(\zeta_\theta, \zeta_\varphi)&=\abs{\sigma}\cos(s+s_0)(\tilde{\zeta}_\theta, \tilde{\zeta}_\varphi),\quad \abs{\sigma}\cos(s_0)=\tilde{p}(\theta^0, \varphi^0, \zeta^0_\theta, \zeta^0_\varphi)^{\frac12},\\
		(\theta, \varphi, \tilde{\zeta}_\theta, \tilde{\zeta}_\varphi)&=\exp((s+s_0)H_{-\frac{1}{2}\tilde{p}})(\theta^0, \varphi^0, \tilde{\zeta}_\theta^0, \tilde{\zeta}_\varphi^0),\quad (\tilde{\zeta}_\theta^0, \tilde{\zeta}_\varphi^0)=\tilde{p}(\theta^0, \varphi^0, \zeta^0_\theta, \zeta^0_\varphi)^{-\frac12}\cdot(\zeta_\theta^0, \zeta_\varphi^0)
	\end{split}
\end{equation}
where $s_0\in(-\frac{\pi}{2},\frac{\pi}{2})$, $s\in(-\frac{\pi}{2}-s_0, \frac{\pi}{2}-s_0)$ and $\frac{ds}{ds'}=2\tilde{p}(\theta, \varphi, \zeta_\theta, \zeta_\varphi)^{\frac12}$ where $\frac{d}{ds'}:={}^{\scop}\!H_{p_{\scop}(\sigma)}^{2,0}$.

This implies that for $\sigma>0$, the zero section is a radial source and the nonzero one is a radial sink, while for $\sigma<0$, the zero section is a radial  sink and the nonzero one is a radial source,  see Figure \ref{fig:charaatinfty}.

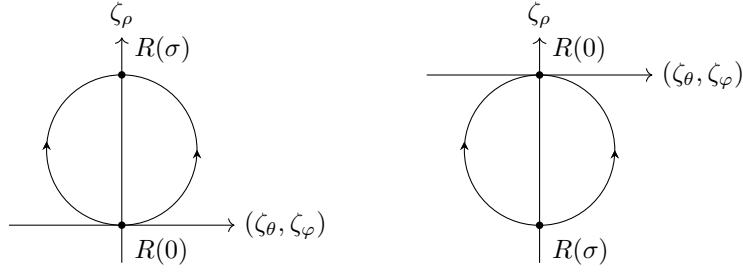
\begin{figure}[h]
	\centering
	\begin{tikzpicture}
		\draw[->](-1.5, 0)--(1.5,0) node[right] {$(\zeta_\theta, \zeta_\varphi)$};
		\draw[->](0,-0.5)--(0,2.5) node[above] {$\zeta_\rho$};
		\draw[
		decoration={markings, mark=at position 0 with {\arrow{stealth}}},
		postaction={decorate}
		]
		[
		decoration={markings, mark=at position 0.5 with {\arrow{stealth reversed}}},
		postaction={decorate}
		]
		(0,1) circle (1cm);
		\node[circle,inner sep=1pt,fill=black, label=above right: $R(\sigma)$] at (0,2) {};
		\node[circle,inner sep=1pt,fill=black, label=below right: $R(0)$] at (0,0) {};
	\end{tikzpicture}
	\qquad\quad
	\begin{tikzpicture}
		\draw[->](-1.5, 0)--(1.5,0) node[right] {$(\zeta_\theta, \zeta_\varphi)$};
		\draw[->](0,-2.5)--(0,0.5) node[above] {$\zeta_\rho$};
		\draw[
		decoration={markings, mark=at position 0 with {\arrow{stealth}}},
		postaction={decorate}
		]
		[
		decoration={markings, mark=at position 0.5 with {\arrow{stealth reversed}}},
		postaction={decorate}
		]
		(0,-1) circle (1cm);
		\node[circle,inner sep=1pt,fill=black, label=below right: $R(\sigma)$] at (0,-2) {};
		\node[circle,inner sep=1pt,fill=black, label=above right: $R(0)$] at (0,0) {};
	\end{tikzpicture}
	\caption{ The integral curves of ${}^{\scop}\!H_{p_{\scop}(\sigma)}^{2,0}$ in $\zeta_\rho$ and $(\zeta_\theta, \zeta_\varphi)$ coordinates. The left hand side is the case $\sigma>0$ and the right hand side is the case $\sigma<0$.}
	\label{fig:charaatinfty}
\end{figure}

\end{lem}

\begin{proof}
	At $\pa\CX$, the rescaled scattering Hamiltonian vector field is 
\[
	{}^{\scop}\!H_{p_{\scop}(\sigma)}^{2,0}=(-2\zeta_\rho+2\sigma)\rho\pa_\rho+2\tilde{p}\pa_{\zeta_\rho}+(-2\zeta_\rho+2\sigma)\sum_{\mu=\theta,\varphi}\zeta_\mu\pa_{\zeta_\mu}-H_{\tilde{p}},
	\]
	 Introducing the polar coordinates with respect to the radial variable in $(\zeta_\theta, \zeta_\varphi)$
\[
	(\tilde{\zeta}_\theta, \tilde{\zeta}_\varphi)=\tilde{p}(\theta, \varphi, \zeta_\theta, \zeta_\varphi)^{-\frac12}\cdot(\zeta_\theta, \zeta_\varphi),\quad \abs{(\zeta_\theta, \zeta_\varphi)}=\tilde{p}(\theta, \varphi, \zeta_\theta, \zeta_\varphi)^{\frac 12}.
\]
Let $\frac{d}{ds'}:={}^{\scop}\!H_{p_{\scop}(\sigma)}^{2,0}$. Then we have
\begin{gather*}
	\frac{d}{ds'}\zeta_\rho=2\tilde{p}(\theta, \varphi, \zeta_\theta, \zeta_\varphi), \quad \frac{d}{ds'}\tilde{p}(\theta, \varphi, \zeta_\theta, \zeta_\varphi)^{1/2}=-2(\zeta_\rho-\sigma)\tilde{p}(\theta, \varphi, \zeta_\theta, \zeta_\varphi)^{\frac12},\\
\frac{d}{ds'}(\tilde{\zeta}_\theta, \tilde{\zeta}_\varphi)=\tilde{p}^{-\frac 12}(\frac{\pa\tilde{p}}{\pa\theta},\frac{\pa\tilde{p}}{\pa\varphi})(\theta, \varphi, \zeta_\theta, \zeta_\varphi),\quad \frac{d}{ds'}(\theta, \varphi)=-(\frac{\pa\tilde{p}}{\pa{\zeta_\theta}},\frac{\pa\tilde{p}}{\pa{\zeta_\varphi}})(\theta, \varphi, \zeta_\theta, \zeta_\varphi).
\end{gather*}
This implies that $\frac{d}{ds'}$ is tangent to $R(0)$ and $R(\sigma)$, so they are invariant under the Hamiltonian flow $\exp(s{}^{\scop}\!H_{p_{\scop}(\sigma)}^{2,0})$.

When $\tilde{p}(\theta, \varphi, \zeta_\theta, \zeta_\varphi)\neq 0$, we introduce a new parameter $s$ satisfying $\frac{ds}{ds'}=2\tilde{p}(\theta, \varphi, \zeta_\theta, \zeta_\varphi)^{\frac12}$ and obtain
\begin{equation}
	\begin{gathered}
		 \frac{d}{ds}(\zeta_\rho-\sigma)=\tilde{p}(\theta, \varphi, \zeta_\theta, \zeta_\varphi)^{\frac{1}{2}},\quad 	\frac{d}{ds}\tilde{p}(\theta, \varphi, \zeta_\theta, \zeta_\varphi)^{\frac{1}{2}}=-(\zeta_\rho-\sigma),\\
	\frac{d}{ds}(\tilde{\zeta}_\theta, \tilde{\zeta}_\varphi)=\frac{1}{2}(\frac{\pa\tilde{p}}{\pa\theta},\frac{\pa\tilde{p}}{\pa\varphi})(\theta,\varphi,\tilde{\zeta}_\theta, \tilde{\zeta}_\varphi ),\quad \frac{d}{ds}(\theta, \varphi)=-\frac{1}{2}(\frac{\pa\tilde{p}}{\pa{\tilde{\zeta}_\theta}},\frac{\pa\tilde{p}}{\pa{\tilde{\zeta}_\varphi}})(\theta,\varphi,\tilde{\zeta}_\theta, \tilde{\zeta}_\varphi ).	
\end{gathered}
\end{equation}
Integrating the above system with respect to the new parameter $s$ gives \eqref{eq:exofscacurves}.
\end{proof}
For the propagation of singularities estimates at these radial points $R(\sigma)$ and $R(0)$, there is a \textit{threshold quantity} for the order of the Sobolev spaces (here the scattering decay order). Let $\beta_0$ be a positive function defined at the radial points as
\[
{}^{\scop}\!H_{p_{\scop}(\sigma)}^{2,0}\rho=\pm\beta_0\rho.
\] 
Here and in what follows, we use $+$ for radial sources and $-$ for radial sinks. We next define $\tilde{\beta}$ at the radial points as 
\[
\sigma_{\scop}(\rho^{-1}\IM\widehat{\Box_{g}}(\sigma))=\sigma_{\scop}(\frac{\widehat{\Box_{g}}(\sigma)-\widehat{\Box_{g}}(\sigma)^*}{2i\rho})=\pm\beta_0\tilde{\beta}
\]
Let
\[
\beta_{\sup}=\sup\tilde{\beta},\quad \beta_{\inf}=\inf \tilde{\beta}
\]
where the supremum and infimum are taken over the radial points set. If $\tilde{\beta}$ is a constant along the radial points set, we may write
\[
\beta=\beta_{\sup}=\beta_{\inf}.
\]
We note that $\beta_{\sup}$ and $\beta_{\inf}$ are the relevant quantities in the radial point estimates. More specifically, in our setting where we consider the operator $\widehat{\Box_{g}}(\sigma)$ with $\sigma\in\BR\setminus0$, the threshold scattering decay condition in the \textit{high scattering decay radial estimate} is given by
\begin{equation}
	r>-\frac{1}{2}-\beta_{\inf},
\end{equation}
while the \textit{low scattering decay radial estimate} requires 
\begin{equation}
	r<-\frac{1}{2}-\beta_{\sup}.
\end{equation}
In our setting, since $\widehat{\Box_{g}}(\sigma)=\widehat{\Box_{g}}(\sigma)^*$ for $\sigma\in\BR\setminus0$, it follows that \begin{equation}\label{eq:thresholdscdecay}
	\beta=\beta_{\sup}=\beta_{\inf}=0\quad \mbox{on}\quad R(\sigma)\quad \mbox {and}\quad R(0).
\end{equation}

%%%%%%%%%%%%%%%%%%%%%%%%%%%%%%%%%%%%%%%%%%%%%%%%%%%%%%
\subsubsection{Global dynamics of the Hamiltonian flow}
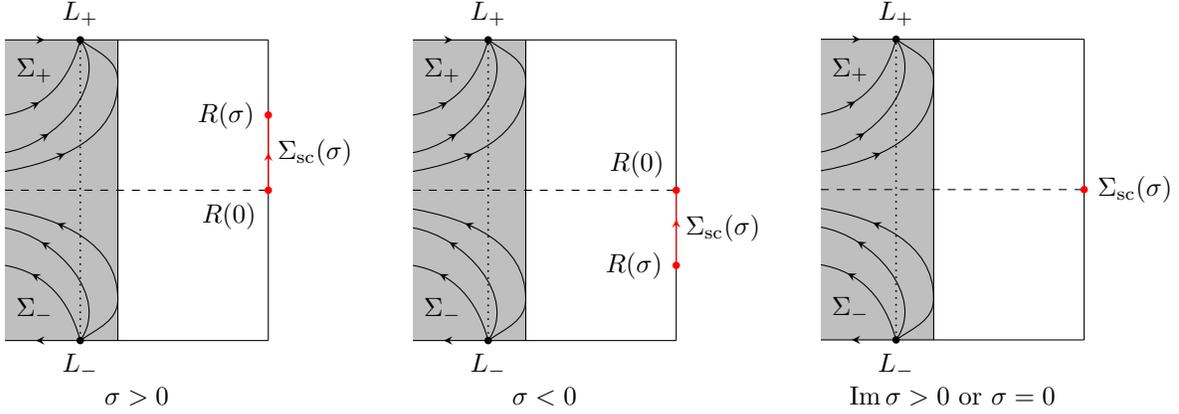
\begin{figure}[!h]
	\begin{tikzpicture}
			\fill [color=gray!50] (0,0)--(0,4)--(-1.5,4)--(-1.5,0)--cycle;
		\draw[
	decoration={markings, mark=at position 0.15 with {\arrow{stealth reversed}}},
	postaction={decorate}]
 (-1.5, 0)--(2,0);
	\draw [dashed](-1.5, 2)--(2,2);
	\draw [
	decoration={markings, mark=at position 0.15 with {\arrow{stealth}}},
	postaction={decorate}]
	(-1.5,4)--(2,4);
	\draw (2,0)--(2,4);
		\draw[red,
	decoration={markings, mark=at position 0.5 with {\arrow{stealth}}},
	postaction={decorate}
	]
	(2,2)--(2,3) node[black, midway, right] {$\Sigma_{\scop}(\sigma)$};
		\node[circle,inner sep=1pt,fill=red,label=left:{$R(\sigma)$}] at (2,3) {};
	\node[circle,inner sep=1pt,fill=red,label=below left:{$R(0)$}] at (2,2) {};
	\draw (0,0)--(0,4);
	\draw [dotted, semithick](-0.5,0)--(-0.5,4);
		\node[circle,inner sep=1pt,fill=black,label=above:{$L_+$}] at (-0.5,4) {};
	\node[circle,inner sep=1pt,fill=black,label=below:{$L_-$}] at (-0.5,0) {};
		\draw[
	decoration={markings, mark=at position 0.75 with {\arrow{stealth reversed}}},
	postaction={decorate}
	] (-0.5, 4) to [out=-35, in=90] (0, 3.45) to [out=-90, in=10] (-1.5, 2.25);
		\draw[
	decoration={markings, mark=at position 0.75 with {\arrow{stealth reversed}}},
	postaction={decorate}
	] (-0.5, 4) to [out=-55, in=10] (-1.5, 2.5);
	\draw[
decoration={markings, mark=at position 0.75 with {\arrow{stealth reversed}}},
postaction={decorate}
] (-0.5, 4) to [out=-90-15, in=10]  (-1.5, 3);
	\draw[
decoration={markings, mark=at position 0.75 with {\arrow{stealth}}},
postaction={decorate}
] (-0.5, 0) to [out=35, in=-90] (0, 0.55) to [out=90, in=-10] (-1.5, 1.75);
\draw[
decoration={markings, mark=at position 0.75 with {\arrow{stealth}}},
postaction={decorate}
] (-0.5, 0) to [out=55, in=-10] (-1.5, 1.5);
\draw[
decoration={markings, mark=at position 0.75 with {\arrow{stealth}}},
postaction={decorate}
] (-0.5, 0) to [out=90+15, in=-10]  (-1.5, 1);
\node[label= below right:{$\Sigma_+$}] at (-1.6,4) {};
\node[label= above right:{$\Sigma_-$}] at (-1.6, 0) {};
\node at (0.25, -0.75) {$\sigma>0$};
	\end{tikzpicture}
\quad 
	\begin{tikzpicture}
	\fill [color=gray!50] (0,0)--(0,4)--(-1.5,4)--(-1.5,0)--cycle;
	\draw[
	decoration={markings, mark=at position 0.15 with {\arrow{stealth reversed}}},
	postaction={decorate}]
	(-1.5, 0)--(2,0);
	\draw [dashed](-1.5, 2)--(2,2);
	\draw [
	decoration={markings, mark=at position 0.15 with {\arrow{stealth}}},
	postaction={decorate}]
	(-1.5,4)--(2,4);
	\draw (2,0)--(2,4);
	\draw[red,
	decoration={markings, mark=at position 0.5 with {\arrow{stealth reversed}}},
	postaction={decorate}
	]
	(2,2)--(2,1) node[black, midway, right] {$\Sigma_{\scop}(\sigma)$};
	\node[circle,inner sep=1pt,fill=red,label=left:{$R(\sigma)$}] at (2,1) {};
	\node[circle,inner sep=1pt,fill=red,label=above left:{$R(0)$}] at (2,2) {};
	\draw (0,0)--(0,4);
	\draw [dotted, semithick](-0.5,0)--(-0.5,4);
	\node[circle,inner sep=1pt,fill=black,label=above:{$L_+$}] at (-0.5,4) {};
	\node[circle,inner sep=1pt,fill=black,label=below:{$L_-$}] at (-0.5,0) {};
	\draw[
	decoration={markings, mark=at position 0.75 with {\arrow{stealth reversed}}},
	postaction={decorate}
	] (-0.5, 4) to [out=-35, in=90] (0, 3.45) to [out=-90, in=10] (-1.5, 2.25);
	\draw[
	decoration={markings, mark=at position 0.75 with {\arrow{stealth reversed}}},
	postaction={decorate}
	] (-0.5, 4) to [out=-55, in=10] (-1.5, 2.5);
	\draw[
	decoration={markings, mark=at position 0.75 with {\arrow{stealth reversed}}},
	postaction={decorate}
	] (-0.5, 4) to [out=-90-15, in=10]  (-1.5, 3);
	\draw[
	decoration={markings, mark=at position 0.75 with {\arrow{stealth}}},
	postaction={decorate}
	] (-0.5, 0) to [out=35, in=-90] (0, 0.55) to [out=90, in=-10] (-1.5, 1.75);
	\draw[
	decoration={markings, mark=at position 0.75 with {\arrow{stealth}}},
	postaction={decorate}
	] (-0.5, 0) to [out=55, in=-10] (-1.5, 1.5);
	\draw[
	decoration={markings, mark=at position 0.75 with {\arrow{stealth}}},
	postaction={decorate}
	] (-0.5, 0) to [out=90+15, in=-10]  (-1.5, 1);
	\node[label= below right:{$\Sigma_+$}] at (-1.6,4) {};
	\node[label= above right:{$\Sigma_-$}] at (-1.6, 0) {};
	\node at (0.25, -0.75) {$\sigma<0$};
\end{tikzpicture}
\quad
	\begin{tikzpicture}
	\fill [color=gray!50] (0,0)--(0,4)--(-1.5,4)--(-1.5,0)--cycle;
	
	\draw[
	decoration={markings, mark=at position 0.15 with {\arrow{stealth reversed}}},
	postaction={decorate}]
	(-1.5, 0)--(2,0);

	\draw [dashed](-1.5, 2)--(2,2);

	\draw [
	decoration={markings, mark=at position 0.15 with {\arrow{stealth}}},
	postaction={decorate}]
	(-1.5,4)--(2,4);

	\draw (2,0)--(2,4);
	
	\node[circle,inner sep=1pt,fill=red,label=right:{$\Sigma_{\scop}(\sigma)$}] at (2,2) {};

	\draw (0,0)--(0,4);
	
	\draw [dotted, semithick](-0.5,0)--(-0.5,4);
	
	\node[circle,inner sep=1pt,fill=black,label=above:{$L_+$}] at (-0.5,4) {};
	
	\node[circle,inner sep=1pt,fill=black,label=below:{$L_-$}] at (-0.5,0) {};

	\draw[
	decoration={markings, mark=at position 0.75 with {\arrow{stealth reversed}}},
	postaction={decorate}
	] (-0.5, 4) to [out=-35, in=90] (0, 3.45) to [out=-90, in=10] (-1.5, 2.25);
	\draw[
	decoration={markings, mark=at position 0.75 with {\arrow{stealth reversed}}},
	postaction={decorate}
	] (-0.5, 4) to [out=-55, in=10] (-1.5, 2.5);
	\draw[
	decoration={markings, mark=at position 0.75 with {\arrow{stealth reversed}}},
	postaction={decorate}
	] (-0.5, 4) to [out=-90-15, in=10]  (-1.5, 3);

	\draw[
	decoration={markings, mark=at position 0.75 with {\arrow{stealth}}},
	postaction={decorate}
	] (-0.5, 0) to [out=35, in=-90] (0, 0.55) to [out=90, in=-10] (-1.5, 1.75);
	\draw[
	decoration={markings, mark=at position 0.75 with {\arrow{stealth}}},
	postaction={decorate}
	] (-0.5, 0) to [out=55, in=-10] (-1.5, 1.5);
	\draw[
	decoration={markings, mark=at position 0.75 with {\arrow{stealth}}},
	postaction={decorate}
	] (-0.5, 0) to [out=90+15, in=-10]  (-1.5, 1);

	\node[label= below right:{$\Sigma_+$}] at (-1.6,4) {};
	\node[label= above right:{$\Sigma_-$}] at (-1.6, 0) {};

	\node at (0.25, -0.75) {$\IM\sigma>0$ or $\sigma=0$};
\end{tikzpicture}
\caption{The flow of $\rho_\xi H_p$ (resp. ${}^{\scop}\!H_{p_{\scop}(\sigma)}^{1,0}$) away from infinity $\pa_+\CX=\{\rho=\frac1 r=0\}$ (resp., at $\pa_+\CX$), which is projected to the coordinates $(r, \xi_r)$ (resp. $(\rho, \zeta_\rho)$) and drawn in a fiber-radially compactified view. The shaded region is the characteristic set $\Sigma=\Sigma_+\cup\Sigma_-=\{\rho_\xi^2p=0\}$ away from $\pa_+\CX$, while the red region is the scattering characteristic set $\Sigma_{\scop}(\sigma)=\{p_{\scop}(\sigma)=0\}$ at $\pa_+\CX$. The horizontal coordinate is $\rho$ and the rightmost vertical line corresponds to $\pa_+\CX$. The dotted line is the even horizon $r=r_b$ and the vertical line in the middle is $\{r=r_{\mathrm{ergo}}\}$ where $r_{\mathrm{ergo}}$ is the equatorial radius (i.e. greatest radius) of the ergosphere. The vertical coordinate is $\xi_r/\jb{\xi_r}$ (resp. $\zeta_\rho/\jb{\zeta_\rho}$), so the top and bottom lines stand for the fiber infinity. The midline $\{\xi_r=0\}$ (resp. $\zeta_\rho=0$) does not intersect $\Sigma$, but intersects $\Sigma_{\scop}(\sigma)$.
}
\label{fig:microlocalphase}
\end{figure}

Now we study the global behavior of the flow of $\rho_\xi H_p$ (resp., ${}^{\scop}\!H_{p_{\scop}(\sigma)}^{2,0}$) away from infinity $\pa_+\CX=\{\rho=\frac1 r=0\}$ (resp., at $\pa_+\CX$). 
\begin{prop}\label{prop:microlocalglobaldy}
The characteristic set of $\widehat{\Box_{g}}(\sigma)$ is a disjoint union $\Sigma\cup\Sigma_{sc}(\sigma)$ (see Figure \ref{fig:microlocalphase}). 
\begin{enumerate}
	\item Let $(r, \theta, \varphi, \xi)\in\Sigma_{\pm}$ and put $\phi(s)=\exp(s\rho_\xi H_{p})(r, \theta, \varphi, \xi)$. Then $\phi(s)\to L_{\pm}$ as $s\to\pm\infty$. Moreover, if $(r, \theta, \varphi, \xi)\notin \Lambda_\pm=\{\Delta_b=0, (\xi_\theta, \xi_\varphi)=0, \pm\xi_r>0\}$, then $\phi(s)$ crosses $\pa_-\CX$ into the inward direction of deceasing $r$ at some $s_0$ with $\pm s_0\leq 0$.
\item For $\sigma\in\BR\setminus 0$, let $(\rho=0, \theta, \varphi, \zeta)\in\Sigma_{\scop}(\sigma)$ and put $\phi(s)=\exp(s{}^{\scop}\!H_{p_{\scop}(\sigma)}^{2,0})( \theta, \varphi, \zeta)$. Then either $\phi(s)\to R(\frac{\sigma}{2}+\frac{\abs{\sigma}}{2})=\{\rho=0, \zeta_\rho=\sigma+\abs{\sigma}, \tilde{p}=0\}$ as $s\to\infty$, or $\phi(s)\to R(\frac{\sigma}{2}-\frac{\abs{\sigma}}{2})=\{\rho=0, \zeta_\rho=\sigma-\abs{\sigma}, \tilde{p}=0\}$ as $s\to-\infty$. Moreover, if $(\rho=0, \theta, \varphi, \zeta)\notin R(0)\cup R(\sigma)$, then $\phi(s)\to R(\frac{\sigma}{2}\pm\frac{\abs{\sigma}}{2})$ as $s\to\pm\infty$.
\end{enumerate}
\end{prop}

\begin{proof}
	We only discuss the case $\Sigma_+$ as the other case $\Sigma_-$ can be handled in a similar manner. Let $(r, \theta, \varphi, \xi)\in\Sigma_+$ and put $\phi(s)=\exp(s\rho_\xi H_{p})(r, \theta, \varphi, \xi)$. 
	Since $\Sigma_+$ is invariant under the flow $\exp(s\rho_\xi H_{p})$, it follows that $\phi(s)\in\Sigma_\pm$ for all $s$ for which it is well-defined. Put
\[
 f(s)=\big(\rho_\xi^2+(\Delta_b+2a\sgn(\xi_r)\hat{\xi}_\varphi)^2+\rho^2_\xi\tilde{p}\big)(\phi(s)),\quad \dot{f}(s)=\rho_\xi H_pf(\phi(s)).
\]
Using the calculation in \eqref{eq:deriof4quantities}, we find that
\[
\dot{f}(s)\leq-2\big(\rho_b^{-2}\frac{\pa\Delta_b}{\pa r}f\big)(\phi(s))\leq -C_0f(s) \quad \mbox{on}\quad \Sigma_+\subset\{r\geq r_-\mid \Delta_b-a^2\sin^2\theta\leq 0\}
\]
for some $C_0>0$. This gives
\[
f(s)\leq e^{-C_0s}f(0)=e^{-C_0s}f((r, \theta, \varphi, \xi))\quad \mbox{for}\quad s\geq 0
\]
and thus
\[
\phi(s)\to L_+\quad \mbox{as}\quad s\to\infty.
\]
Using the last expression in \eqref{eq:deriof4quantities}, we have
\[
\rho_\xi H_p(\rho_\xi^2\tilde{p})(\phi(s))=-2\big(\rho_b^{-2}\frac{\pa\Delta_b}{\pa r}\rho_\xi^2\tilde{p}\big)(\phi(s))\leq -C_0(\rho_\xi^2\tilde{p})(\phi(s)).
\]
for some $C_0>0$. Therefore
\[
(\rho_\xi^2\tilde{p})(\phi(s))\geq e^{-C_0s}(\rho_\xi^2\tilde{p})(r, \theta, \varphi,\xi)
\]
for $s\leq0$ as long as the flow remains in the region $\Sigma_+\subset\{r\geq r_-\mid \Delta_b-a^2\sin^2\theta\leq 0\}$. Since on $\Sigma_+$
\[
2a^2+\frac{1}{2}\hat{\xi}_\varphi^2\geq -2a\hat{\xi}_\varphi =\Delta_b+\rho_\xi^2\tilde{p}\geq \Delta_b+\frac{1}{2}\rho_\xi^2\tilde{p}+\frac{1}{2}\hat{\xi}_\varphi^2
\]
where we use $\Delta_b+2a\hat{\xi}_\varphi+\rho_\xi^2\tilde{p}=0$ on $\Sigma_+$ in the second equality and $\rho_\xi^2\tilde{p}\geq \hat{\xi}_\varphi^2$ in the last inequality, it follows that
\[
2a^2-\Delta_b\geq \frac{1}{2}\rho_\xi^2\tilde{p}
\quad \mbox{on}\quad \Sigma_+.\]
Since\
\[
\{\rho_\xi^2\tilde{p}=0\}\cap\Sigma_+=\Lambda_+,
\]
then if $(r, \theta, \varphi, \xi)\in\Sigma_+\setminus \Lambda_+$, we have 
\[
(2a^2-\Delta_b)(\phi(s))\geq \frac{1}{2}\rho_\xi^2\tilde{p}(\phi(s))\geq Ce^{-C_0s},\quad C, C_0>0
\]
for $s\leq0$ as long as the flow remains in the region $\Sigma_+\subset\{r\geq r_-\mid \Delta_b-a^2\sin^2\theta\leq 0\}$. As a consequence, $\phi(s)$ cross $\pa_-\CX$ into the inward direction of deceasing $r$ at some $s_0$ with $s_0\leq 0$. This proves the part (i).

The proof of part (2) follows from Lemma \ref{lem:radialpointatinfty}.
\end{proof}

\subsubsection{Second microlocalized scattering algebra}\label{subsubsec:secondmic}
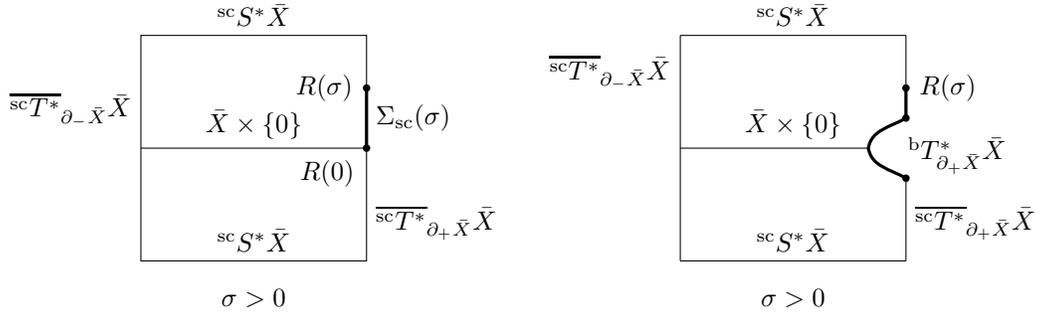
\begin{figure}[!h]
	\begin{tikzpicture}
		\draw(0,0) rectangle (3,3);
		\draw (0,1.5)--(3,1.5);
		\draw [very thick] (3,1.5)--(3,2.3);
		\node [left] at (0, 2) {$\abscform$};
			\node [right] at (3, 0.5) {$\siscform$};
			\node [above] at (1.5, 3) {$\fiscform$};
				\node [above] at (1.5, 0) {$\fiscform$};
			\node [above ] at (1.5, 1.5) {$\CX\times\{0\}$};
				\node [right] at (3, 1.9) {$\Sigma_{\scop}(\sigma)$};
					\node [circle, inner sep=1pt, fill=black,label= below left:{$R(0)$}] at (3, 1.5) {};
					\node [circle, inner sep=1pt, fill=black,label=left:{$R(\sigma)$} ] at (3, 2.3) {};
						\node  at (1.5, -0.5) {$\sigma>0$};
	\end{tikzpicture}
\quad
	\begin{tikzpicture}
	\draw (0,0)--(3,0);
	\draw(0,0)--(0,3);
	\draw (0,3)--(3,3);
	\draw (0,1.5)--(2.5,1.5);
	\draw (3,2.3)--(3,3);
	\draw (3,0)--(3,1.1);
	\draw [very thick](3,1.1) to [out=150, in=-80] (2.5, 1.5) to [out=80,in=-150] (3,1.9)--(3,2.3);
	\node [left] at (0, 2.5) {$\abscform$};
		\node [right] at (3, 0.5) {$\siscform$};
	\node [above] at (1.5, 3) {$\fiscform$};
	\node [above] at (1.5, 0) {$\fiscform$};
	\node [above ] at (1.5, 1.5) {$\CX\times\{0\}$};
\node[circle, inner sep=1pt, fill=black] at (3,1.1) {};
\node[circle, inner sep=1pt, fill=black] at (3,1.9) {};
\node[circle, inner sep=1pt, fill=black, label=right:{$R(\sigma)$}] at (3,2.3) {};
\node [right] at (2.9,1.4) {$\fsibform$};
	\node  at (1.5, -0.5) {$\sigma>0$};
\end{tikzpicture}
	\caption{Second microlocalized $\CX$. The left hand side is the fiber-compactification $\cscform$ of the scattering cotangent bundle $\fscform$, and the right hand side is its blow-up $[\cscform;o_{\pa_+\CX}]$ at the scattering zero section at the boundary $\pa_+\CX$. Also shown is the scattering characteristic set $\Sigma_{\scop}(\sigma)$ at $\pa_+\CX$ of $\widehat{\Box_{g}}(\sigma)$ for $\sigma>0$. The interior of the front face of the blow-up, shown by curved arcs, can be identified with $\fsibform$.} 
	\label{fig:scatteringblowup}
\end{figure}

To do the analysis at the scattering zero section at $\pa_+\CX$, we introduce the second microlocalized $\CX$ by blowing up the zero section $o_{\pa_+\CX}$ at $\pa_+\CX$ (i.e. replacing $o_{\pa_+\CX}$ by its inward pointing spherical normal bundle and thus obtaining a smooth manifold with corners), see Figure \ref{fig:scatteringblowup}. 

\begin{figure}[!h]
\begin{tikzpicture}
	\draw(0,0) rectangle (3,3);
	\draw (0,1.5)--(3,1.5);
	\node [left] at (0, 2) {$\abbform$};
	\node [right] at (3, 0.5) {$\sibform$};
	\node [above] at (1.5, 3) {$\fibform$};
	\node [above] at (1.5, 0) {$\fibform$};
	\node [above ] at (1.5, 1.5) {$\CX\times\{0\}$};
\end{tikzpicture}
\quad
\begin{tikzpicture}
	\draw(0,0)--(0,3);
	\draw (0,1.5)--(3,1.5);
	\draw (0,0)--(2.2,0);
	\draw (0,3)--(2.2,3);
	\draw (2.2,3) to [out=-90, in=180] (3,2.2) -- (3,0.8) to [out=-180,in=90] (2.2,0);
	\node [left] at (0, 2.5) {$\abbform$};
	\node [above] at (1.2, 3) {$\fibform$};
	\node [above] at (1.2, 0) {$\fibform$};
	\node [above ] at (1.2, 1.5) {$\CX\times\{0\}$};
	\node [right] at (3,1.4) {$\sibform$};
	\node [right] at (2.6,2.7) {ff=$[\siscform;o_{\pa_+\CX}]$};
		\node [right] at (2.6,0.3) {ff=$[\siscform;o_{\pa_+\CX}]$};
\end{tikzpicture}	
\caption{The left hand side is the fiber-compactification $\cbform$ of the b-cotangent bundle $\fbform$, and the right hand side is its blow-up $[\cbform;\fisibform ]$ at the fiber infinity at the boundary $\pa_+\CX$. The interior of the front face of the blow-up, shown by curved arcs, can be identified with $[\fsiscform;o_{\pa_+\CX}]$.}
\label{fig:bblowup}
\end{figure}
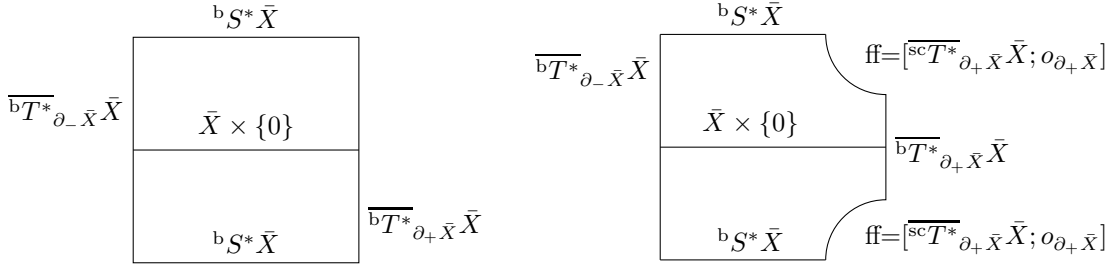

On the other hand, from an analytically better behaved, but geometrically equivalent perspective, one takes $\cbform$ and blows up its corner $\fisibform$, i.e. the fiber infinity at $\pa_+\CX$, see Figure \ref{fig:bblowup}. These two spaces $[\cscform;o_{\pa_+\CX}]$ and $[\cbform;\fisibform ]$ are naturally the same, see \cite[Lemma 5.1]{Vas21}. 

In order to actually do analysis, one needs to introduce the pseudodifferential operator space $\mPsi^{s,r,l}(\CX)$. The three orders of $\mPsi^{s,r,l}(\CX)$ are the sc-diffferential order $s$, the sc-decay order $r$ and b-decay order $l$. The operators $\mPsi^{s,r,l}(\CX)$ arise from the quantization of the symbols in the following class
\[
S^{s,r,l}(\CX):=\rho_{\bop}^{-l}\rho_{\scop}^{-r}\rho_{\infty}^{-s}S^{0,0,0}([\cbform;\fisibform ])=\rho_{\bop}^{-l}\rho_{\scop}^{-r}\rho_{\infty}^{-s}S^{0,0}(\cbform)
\]
where $\rho_\infty$, resp. $\rho_{\scop}$, resp. $\rho_{\bop}$ correspond to the boundary defining functions of the three boundary hypersurfaces of $[\cbform;\fisibform ]$ (or equivalently $[\cscform;o_{\pa_+\CX}]$): the lift of b-fiber infinity (or equivalently the lift of sc-fiber infinity), the new front face of $[\cbform;\fisibform ]$ (or the lift of $\fsiscform$ in $[\cscform;o_{\pa_+\CX}]$), and the lift of $\pa_+\CX$ in $[\cbform;\fisibform ]$ (or the new front face of $[\cscform;o_{\pa_+\CX}]$), respectively. Here, the symbol spaces $S^{0,0}(\cbform)$ (resp. $S^{0,0,0}([\cbform;\fisibform ])$) consist of $L^\infty(\cbform) $ (resp. $L^\infty([\cbform;\fisibform ])$) functions which are smooth away from the boundary hypersurfaces $\fibform\cup\sibform$ (resp. $\fibform\cup\sibform\cup\mbox{ff}$), and they remain so under iterated applications of $C^\infty$ b-vector fields on $\cbform$ (resp. $[\cbform;\fibform]$), namely, vector fields tangent to the aforementioned boundary hypersurfaces. In particular, we have
\[
\mPsi^{s, s+l,l}(\CX)=\Psi_{\bop}^{s,l}(\CX).
\]
We refer the reader to \cite[\S5]{Vas21} for a more detailed description of the operators $\mPsi^{s,r,l}(\CX)$ and the symbols $S^{s,r,l}(\CX)$.

Correspondingly, one can define the second microlocal Sobolev space $H^{s,r,l}_{\scop,\bop}$ as follows. Fix an elliptic operator $A\in\mPsi^{s,r,l}(\CX)$, let
\[
H^{s,r,l}_{\scop,\bop}(\CX)=\{u\in C^{-\infty}(\CX)\mid Au\in L^2\}.
\]
Therefore, we have
\[
H_{\scop,\bop}^{s,s+l,l}(\CX)=H_\bop^{s,l}(\CX).
\]

%%%%%%%%%%%%%%%%%%%%%%%%%%%%%%%%%%%%%%%%%%%%%%%%%%%%%%%%%%%%%%%%%%%%%%%%%%%%%%%%%
\subsection{Uniform Fredholm estimates}\label{subsec:fredholmestimates}
In this section, we prove the uniform Fredholm estimates first for operators $\widehat{\Box_{g_{b}}}(\sigma)$ acting on scalar functions, then for $\widehat{\mathcal{P}_{b, \gamma}}(\sigma), \widehat{\mathcal{W}_{b, \gamma}}(\sigma)$ acting on scattering $1$-forms and the linearized gauge-fixed Einstein-Maxwell operator $\widehat{L_{b,\gamma}}(\sigma)$, as well as their formal adjoints for $\sigma\in\BC, \abs{\sigma}\leq C$.
%with respect to volume density $L^2(\CX;\sqrt{\det\abs{g_b}}drd\theta d\varphi)$. 
To this end, we combine the global dynamics of of the Hamiltonian flow of the principal symbol $p_{\sigma}$ established in \ref{prop:microlocalglobaldy} with the elliptic estimate, propagation of singularities estimate, the radial point estimate at event horizon, the scattering radial point estimate at spatial infinity $\pa_+\CX$ and hyperbolic estimate.
\subsubsection{Uniform Fredholm estimates for scalar wave operators}
We first prove uniform Fredholm estimates for the operator $\widehat{\Box_{g_{b}}}(\sigma)$ acting on scalar functions, as well as its formal adjoint with respect to volume density $L^2(\CX;\sqrt{\det\abs{g_b}}drd\theta d\varphi)$ for $\sigma\in\BC, \abs{\sigma}\leq C$.
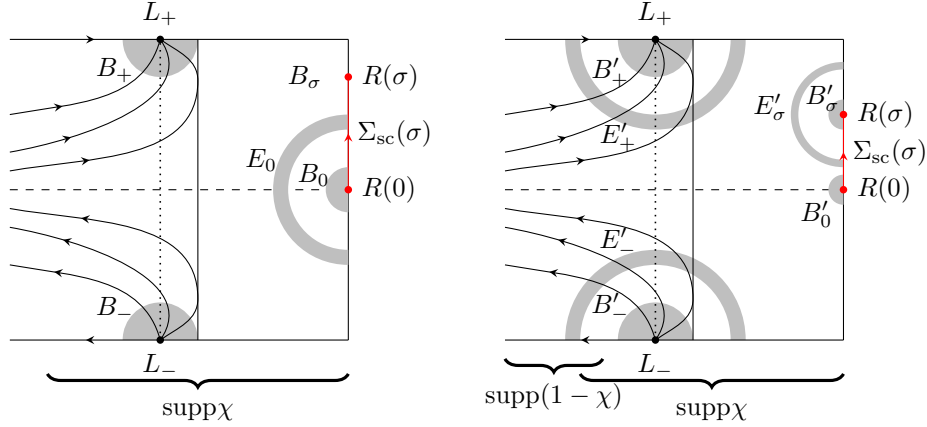
\begin{figure}[!h]
	\begin{tikzpicture}
		\fill [color=gray!50] (-1,4) to [out=-90, in=-180](-0.5,3.5) to[out=0, in=-90] (-0, 4);
			\node[label= below right:{$B_+$}] at (-1.6,4) {};
				\fill [color=gray!50] (-1,0) to [out=90, in=180](-0.5,0.5) to[out=0, in=90] (-0, 0);
					\node[label= above right:{$B_-$}] at (-1.6,0) {};
		\draw[
		decoration={markings, mark=at position 0.25 with {\arrow{stealth reversed}}},
		postaction={decorate}]
		(-2.5, 0)--(2,0);
		\draw [dashed](-2.5, 2)--(2,2);
		\draw [
		decoration={markings, mark=at position 0.25 with {\arrow{stealth}}},
		postaction={decorate}]
		(-2.5,4)--(2,4);
		\draw (2,0)--(2,4);
		
			\fill [color=gray!50] (2,2.3) to [out=180, in=90](1.7,2) to[out=-90, in=180] (2, 1.7);
				\fill [color=gray!50] (2,3) to [out=180, in=90](1,2) to[out=-90, in=180] (2, 1)--(2,1.2) to [out=180, in=-90] (1.2,2) to[ out=90, in=180] (2, 2.8)--(2,3);
				\node[label=above left:{$B_0$}] at (2,1.8) {};
				\node[label=above left:{$E_0$}] at (1.3,2) {};
					\draw[red,
				decoration={markings, mark=at position 0.5 with {\arrow{stealth}}},
				postaction={decorate}
				]
				(2,2)--(2,3.5) node[black, midway, right] {$\Sigma_{\scop}(\sigma)$};
				\node[circle,inner sep=1pt,fill=red,label=right:{$R(\sigma)$}] at (2,3.5) {};
				\node[label=left:{$B_\sigma$}] at (1.9,3.5) {};
				\node[circle,inner sep=1pt,fill=red,label=right:{$R(0)$}] at (2,2) {};
		\draw (0,0)--(0,4);
		\draw [dotted, semithick](-0.5,0)--(-0.5,4);
		\node[circle,inner sep=1pt,fill=black,label=above:{$L_+$}] at (-0.5,4) {};
		\node[circle,inner sep=1pt,fill=black,label=below:{$L_-$}] at (-0.5,0) {};
		\draw[
		decoration={markings, mark=at position 0.75 with {\arrow{stealth reversed}}},
		postaction={decorate}
		] (-0.5, 4) to [out=-35, in=90] (0, 3.45) to [out=-90, in=10] (-2.5, 2.25);
		\draw[
		decoration={markings, mark=at position 0.75 with {\arrow{stealth reversed}}},
		postaction={decorate}
		] (-0.5, 4) to [out=-55, in=10] (-2.5, 2.5);
		\draw[
		decoration={markings, mark=at position 0.75 with {\arrow{stealth reversed}}},
		postaction={decorate}
		] (-0.5, 4) to [out=-90-15, in=10]  (-2.5, 3);
		\draw[
		decoration={markings, mark=at position 0.75 with {\arrow{stealth}}},
		postaction={decorate}
		] (-0.5, 0) to [out=35, in=-90] (0, 0.55) to [out=90, in=-10] (-2.5, 1.75);
		\draw[
		decoration={markings, mark=at position 0.75 with {\arrow{stealth}}},
		postaction={decorate}
		] (-0.5, 0) to [out=55, in=-10] (-2.5, 1.5);
		\draw[
		decoration={markings, mark=at position 0.75 with {\arrow{stealth}}},
		postaction={decorate}
		] (-0.5, 0) to [out=90+15, in=-10]  (-2.5, 1);
	
		 \draw [very thick,decorate,decoration={brace,amplitude=5pt,mirror},xshift=0pt,yshift=0pt]
		(-2,-0.45) -- (2,-0.45) node [black,midway,xshift=0cm,yshift=-0.5cm] 
		{$\mathrm{supp}\chi$};
	\end{tikzpicture}
	\quad 
	\begin{tikzpicture}
		\fill [color=gray!50] (-1,4) to [out=-90, in=-180](-0.5,3.5) to[out=0, in=-90] (-0, 4);
			\fill [color=gray!50] (-1.7,4) to [out=-90, in=-180](-0.5,2.8) to[out=0, in=-90] (0.7,4)--(0.5, 4) to[out=-90, in=0](-0.5, 3) to [out=180, in=-90] (-1.5, 4);
	\node[label= below right:{$B'_+$}] at (-1.6,4) {};
		\node[label= left:{$E'_+$}] at (-0.5,2.7) {};
	\fill [color=gray!50] (-1,0) to [out=90, in=180](-0.5,0.5) to[out=0, in=90] (-0, 0);
		\fill [color=gray!50] (-1.7,0) to [out=90, in=180](-0.5,1.2) to[out=0, in=90] (0.7,0)--(0.5, 0) to[out=90, in=0](-0.5, 1) to [out=-180, in=90] (-1.5, 0);
	\node[label= above right:{$B'_-$}] at (-1.6,0) {};
		\node[label= left:{$E'_-$}] at (-0.5,1.3) {};
		\draw[
		decoration={markings, mark=at position 0.25 with {\arrow{stealth reversed}}},
		postaction={decorate}]
		(-2.5, 0)--(2,0);
		\draw [dashed](-2.5, 2)--(2,2);
		\draw [
		decoration={markings, mark=at position 0.25 with {\arrow{stealth}}},
		postaction={decorate}]
		(-2.5,4)--(2,4);
		\draw (2,0)--(2,4);

		\draw (0,0)--(0,4);
		\draw [dotted, semithick](-0.5,0)--(-0.5,4);
		\node[circle,inner sep=1pt,fill=black,label=above:{$L_+$}] at (-0.5,4) {};		
		\node[circle,inner sep=1pt,fill=black,label=below:{$L_-$}] at (-0.5,0) {};
		\draw[
		decoration={markings, mark=at position 0.75 with {\arrow{stealth reversed}}},
		postaction={decorate}
		] (-0.5, 4) to [out=-35, in=90] (0, 3.45) to [out=-90, in=10] (-2.5, 2.25);
		\draw[
		decoration={markings, mark=at position 0.75 with {\arrow{stealth reversed}}},
		postaction={decorate}
		] (-0.5, 4) to [out=-55, in=10] (-2.5, 2.5);
		\draw[
		decoration={markings, mark=at position 0.75 with {\arrow{stealth reversed}}},
		postaction={decorate}
		] (-0.5, 4) to [out=-90-15, in=10]  (-2.5, 3);
		\draw[
		decoration={markings, mark=at position 0.75 with {\arrow{stealth}}},
		postaction={decorate}
		] (-0.5, 0) to [out=35, in=-90] (0, 0.55) to [out=90, in=-10] (-2.5, 1.75);
		\draw[
		decoration={markings, mark=at position 0.75 with {\arrow{stealth}}},
		postaction={decorate}
		] (-0.5, 0) to [out=55, in=-10] (-2.5, 1.5);
		\draw[
		decoration={markings, mark=at position 0.75 with {\arrow{stealth}}},
		postaction={decorate}
		] (-0.5, 0) to [out=90+15, in=-10]  (-2.5, 1);
	\fill [color=gray!50] (2,2.2) to [out=180, in=90](1.8,2) to[out=-90, in=180] (2, 1.8);
		\fill [color=gray!50] (2,3.2) to [out=180, in=90](1.8,3) to[out=-90, in=180] (2, 2.8);
\fill [color=gray!50] (2,3.7) to [out=180, in=90](1.3,3) to[out=-90, in=180] (2, 2.3)--(2,2.4) to [out=180, in=-90] (1.4,3) to[ out=90, in=180] (2, 3.6)--(2,3.7);
	\draw[red,
decoration={markings, mark=at position 0.5 with {\arrow{stealth}}},
postaction={decorate}
]
(2,2)--(2,3) node[black, midway, right]
 {$\Sigma_{\scop}(\sigma)$};
 	\node[circle,inner sep=1pt,fill=red,label= right:{$R(\sigma)$}] at (2,3) {};
 \node[circle,inner sep=1pt,fill=red,label=right:{$R(0)$}] at (2,2) {};
\node[label=below left:{$B'_0$}] at (2.1,2.1) {};
	\node[label=above left:{$B'_\sigma$}] at (2.2,2.8) {};
	\node[label= left:{$E'_\sigma$}] at (1.5, 3.1) {};
	\draw [very thick,decorate,decoration={brace,amplitude=5pt,mirror},xshift=0pt,yshift=0pt]
	(-1.5,-0.45) -- (2,-0.45) node [black,midway,xshift=0cm,yshift=-0.5cm] 
	{$\mathrm{supp}\chi$};
	\draw [very thick,decorate,decoration={brace,amplitude=5pt,mirror},xshift=0pt,yshift=0pt]
	(-2.5,-0.25) -- (-1.2,-0.25) node [black,midway,xshift=0cm,yshift=-0.5cm] 
	{$\mathrm{supp}(1-\chi)$};
	\end{tikzpicture}
	
		\caption{A phase space picture of the proof of the estimate \eqref{eq:Fredholmsecondmic}, on the left, and \eqref{eq:Fredholmsecondmicdual}, on the right. The coordinates and notations $L_\pm, R(0), R(\sigma)$ are the same as in Figure \ref{fig:microlocalphase}. For \eqref{eq:Fredholmsecondmic}, we use hyperbolic estimates to control $u$ via $\chi u$; $\chi$ is controlled (modulo elliptic estimates) by $B_\pm, B_0,B_\sigma$; $B_0$ is controlled by $E_0$ using low b-decay radial point estimates and $E_0$ is again controlled by $B_\sigma$; finally $B_\pm, B_\sigma$ are controlled using high regularity radial points estimates at $L_\pm$ and high sc-decay radial point estimates at $R(\sigma)$. For \eqref{eq:Fredholmsecondmicdual}, we use hyperbolic estimates to bound $(1-\chi)u$; $\chi$ is controlled (modulo elliptic estimates) by $B'_\pm,  B'_0, B'_\sigma$; $B'_\pm$ is controlled by $E'_\pm$ using low regularity radial point estimates and $E'_\pm$ is controlled by $1-\chi$; $B'_\sigma$ is controlled by $E'_\sigma$ using low sc-decay radial points estimates and $E'_\sigma$ is controlled by $B_0'$; finally $B'_0$ is controlled using high b-decay radial points estimates at $R(0)$.
	}
	\label{fig:micphaseest}
\end{figure}
\begin{thm}\label{thm:Fredholmestimate}
	Let $b_0=(\Bm_0, \Ba_0, \BQ_0)$ with $\abs{\BQ_0}+\abs{\Ba_0}<\Bm_0$. Then there exists $\epsilon>0$ such that for $b=(\Bm, \Ba,\BQ)$ with $\abs{b-b_0}<\epsilon$ and for $s>s_0>\frac{1}{2}, \ell-1\leq\ell_0<\ell<-\frac{1}{2}$ with $s+\ell>s_0+\ell_0>-\frac12$ and $\ell\neq -\frac32$, the following holds. 
	\begin{enumerate}
		 \item For any fixed $C_1>0$, there exist $C>0$ independent of $b$, such that for $\sigma\in\BC, \IM\sigma\geq 0, C_1^{-1}\leq\abs{\sigma}\leq C_1$, we have
	\begin{align}\label{eq:Fredholmenergy1}
		\norm{u}_{\bar{H}_{\bop}^{s,\ell}(\CX)}&\leq C\Big(\norm{\widehat{\Box_{g_{b}}}(\sigma)u}_{\bar{H}_{\bop}^{s,\ell+1}(\CX)}+\norm{u}_{\bar{H}_{\bop}^{s_0,\ell_0}(\CX)}\Big),\\\label{eq:Fredholmenergydual1}
		\norm{u}_{\dot{H}_{\bop}^{-s,-\ell-1}(\CX)}&\leq C\Big(\norm{\widehat{\Box_{g_{b}}}(\sigma)^*u}_{\dot{H}_{\bop}^{-s,-\ell}(\CX)}+\norm{u}_{\dot{H}_{\bop}^{-N,-\ell-3}(\CX)}\Big); \\\label{eq:Fredholmenergy2}
		\norm{u}_{\bar{H}_{\bop}^{s,\ell}(\CX)}&\leq C\Big(\norm{\widehat{\Box_{g_{b}}}(\sigma)u}_{\bar{H}_{\bop}^{s-1,\ell+2}(\CX)}+\norm{u}_{\bar{H}_{\bop}^{s_0,\ell_0}(\CX)}\Big),\\\label{eq:Fredholmenergydual2} \norm{u}_{\dot{H}_{\bop}^{-s+1,-\ell-2}(\CX)}&\leq C\Big(\norm{\widehat{\Box_{g_{b}}}(\sigma)^*u}_{\dot{H}_{\bop}^{-s,-\ell}(\CX)}+\norm{u}_{\dot{H}_{\bop}^{-N,-\ell-3}(\CX)}\Big)
	\end{align}
	where $C$ only depends on $b_0, s,s_0,\ell, \delta,C_1$. Therefore, the operators
	\begin{align}\label{eq:Fredholm1}
	&\widehat{\Box_{g_{b}}}(\sigma):\{u\in\bar{H}_{\bop}^{s,\ell}(\CX)\mid	\widehat{\Box_{g_{b}}}(\sigma)u\in\bar{H}_{\bop}^{s,\ell+1}(\CX)\}\to\bar{H}_{\bop}^{s,\ell+1}(\CX)\\\label{eq:Fredholm2}
	&\widehat{\Box_{g_{b}}}(\sigma):\{u\in\bar{H}_{\bop}^{s,\ell}(\CX)\mid	\widehat{\Box_{g_{b}}}(\sigma)u\in\bar{H}_{\bop}^{s-1,\ell+2}(\CX)\}\to\bar{H}_{\bop}^{s-1,\ell+2}(\CX)
	\end{align}
	are Fredholm for $\IM\sigma\geq 0, \sigma\neq 0$.
	\item Let $\ell\in(-\frac32, -\frac12)$. For any fixed $C_1>0$, there exist $C>0$ independent of $b$, such that for $\sigma\in\BC, \IM\sigma\geq 0, \abs{\sigma}\leq C_1$, we have
	\begin{equation}\label{eq:uniformFredholmenergy}
		\begin{split}
		\norm{u}_{\bar{H}_{\bop}^{s,\ell}(\CX)}&\leq C\Big(\norm{\widehat{\Box_{g_{b}}}(\sigma)u}_{\bar{H}_{\bop}^{s-1,\ell+2}(\CX)}+\norm{u}_{\bar{H}_{\bop}^{s_0,\ell_0}(\CX)}\Big),\\
		\norm{u}_{\dot{H}_{\bop}^{-s+1,-\ell-2}(\CX)}&\leq C\Big(\norm{\widehat{\Box_{g_{b}}}(\sigma)^*u}_{\dot{H}_{\bop}^{-s+1,-\ell-2}(\CX)}+\norm{u}_{\dot{H}_{\bop}^{-N,-\ell-3}(\CX)}\Big)
		\end{split}
	\end{equation}
	where $C$ only depends on $b_0, s,s_0,\ell, \delta,C_1$. Therefore, the operator
	\begin{equation}\label{eq:Fredholm0}
	\widehat{\Box_{g_{b}}}(0):\{u\in\bar{H}_{\bop}^{s,\ell}(\CX)\mid	\widehat{\Box_{g_{b}}}(0)u\in\bar{H}_{\bop}^{s-1,\ell+2}(\CX)\}\to\bar{H}_{\bop}^{s-1,\ell+2}(\CX)
	\end{equation}
	is Fredholm.

	\end{enumerate}
\end{thm}
\begin{proof}
We first prove the statement (1). We claim that it suffices to prove the following estimates for $s>s_0>\frac 12, r>r_0>-\frac12, \ell<-\frac 12,\ell\neq -\frac32$
	\begin{equation}\label{eq:Fredholmsecondmic}
		\norm{u}_{\bar{H}_{\scop,\bop}^{s,r,\ell}(\CX)}\leq C\Big(	\norm{\widehat{\Box_{g_{b}}}(\sigma)u}_{\bar{H}_{\scop,\bop}^{s-1,r+1,\ell+1}(\CX)}+	\norm{u}_{\bar{H}_{\scop,\bop}^{s_0,r_0,\ell-1}(\CX)}\Big)
	\end{equation}
	and
	\begin{equation}\label{eq:Fredholmsecondmicdual}
		\norm{u}_{\dot{H}_{\scop,\bop}^{-s+1,-r-1,-\ell-1}(\CX)}\leq C\Big(	\norm{\widehat{\Box_{g_{b}}}(\sigma)^*u}_{\dot{H}_{\scop,\bop}^{-s,-r,-\ell}(\CX)}+	\norm{u}_{\dot{H}_{\scop,\bop}^{-N,-N,-\ell-3}(\CX)}\Big)
	\end{equation}
Letting $r=s+\ell>-\frac 12$ and $r_0=s_0+\ell_0>-\frac12$, we have 
\[
	\norm{u}_{\bar{H}_{\scop,\bop}^{s,s+\ell,\ell}(\CX)}\leq C\Big(	\norm{\widehat{\Box_{g_{b}}}(\sigma)u}_{\bar{H}_{\scop,\bop}^{s-1,s+\ell+1,\ell+1}(\CX)}+	\norm{u}_{\bar{H}_{\scop,\bop}^{s_0,s_0+\ell_0,\ell_0}(\CX)}\Big)
\]
and
\[
	\norm{u}_{\dot{H}_{\scop,\bop}^{-s+1,-s-\ell-1,-\ell-1}(\CX)}\leq C\Big(	\norm{\widehat{\Box_{g_{b}}}(\sigma)^*u}_{\dot{H}_{\scop,\bop}^{-s,-s-\ell,-\ell}(\CX)}+	\norm{u}_{\dot{H}_{\scop,\bop}^{-N,-N,-\ell-3}(\CX)}\Big).
\]
	Then using the facts $\bar{H}_{\scop,\bop}^{s,s+\ell,\ell}=\bar{H}_{\bop}^{s,l}, \dot{H}_{\scop,\bop}^{s,s+\ell,\ell}=\dot{H}_{\bop}^{s,l}$,
	\[ \bar{H}_{\bop}^{s,\ell+1}=\bar{H}^{s,s+\ell+1, \ell+1}_{\scop,\bop}\subset \bar{H}^{s-1,s+\ell+1,\ell+1}_{\scop,\bop},\quad\bar{H}_{\bop}^{s-1,\ell+2}=\bar{H}^{s-1,s+\ell+1, \ell+2}_{\scop,\bop}\subset \bar{H}^{s-1,s+\ell+1,\ell+1}_{\scop,\bop}
	\]
	and
	\begin{gather*}\dot{H}_{\bop}^{-s,-\ell-1}=\dot{H}^{-s,-s-\ell-1, -\ell-1}_{\scop,\bop}\supset \dot{H}^{-s+1,-s-\ell-1,-\ell-1}_{\scop,\bop},\\ \dot{H}_{\bop}^{-s+1,-\ell-2}=\dot{H}^{-s+1,-s-\ell-1,-\ell-2}_{\scop,\bop}\supset \dot{H}^{-s+1,-s-\ell-1,-\ell-1}_{\scop,\bop},
	\end{gather*}
	we obtain \eqref{eq:Fredholmenergy1}, \eqref{eq:Fredholmenergydual1}, \eqref{eq:Fredholmenergy2} and  \eqref{eq:Fredholmenergydual2}. Now we will show \eqref{eq:Fredholmsecondmic} and \eqref{eq:Fredholmsecondmicdual}. Here we only discuss the case $\RE\sigma>0$ (see Figure \ref{fig:micphaseest} for a phase space illustration of the proof of estimates \eqref{eq:Fredholmsecondmic} and \eqref{eq:Fredholmsecondmicdual}), as the case $\RE\sigma\leq0$ can be handled in a similar manner. 
	\begin{itemize}
		\item \underline{Proof of the estimate \eqref{eq:Fredholmsecondmic}.} For $s>s_0>\frac 12\geq\frac{1}{2}-\IM(\sigma)\frac{\ehKN^2+a^2}{\ehKN^2-\Bm}$ (as calculated \eqref{eq:thresholdreg}), using the high regularity radial point estimate (see \cite[Proposition 2.3]{Vas13}, \cite[Proposition 5.27]{Vas18}), there exist $B_\pm, G_\pm, S_\pm \in\Psi^0(\CX)$ microlocally supported near $L_\pm$ with $L_\pm\subset \mathrm{ell}(B_\pm), L_\pm\subset \mathrm{ell}(S_\pm)$ and satisfying that all the forward (backward) null characteristics from $\mathrm{WF}(B_\pm)$ tend to $L_\pm$, while remaining in the elliptic set of $G_\pm$, such that
		\begin{equation}\label{eq:michighregradiales}
			\norm{B_\pm u}_{H^{s,r,\ell}_{\scop,\bop}(\CX)}\leq C\norm{G_\pm\widehat{\Box_{g_{b}}}(\sigma)u}_{H^{s-1, r+1,\ell+1}_{\scop,\bop}(\CX)}+C\norm{S_\pm u}_{H^{s_0,r_0,\ell_0}_{\scop,\bop}(\CX)}
		\end{equation}
		where $r_0<r, \ell_0<\ell$ and the sc-decay order $r$ and b-decay order $\ell$ are actually irrelevant here.
		
		For $r>r_0>-\frac 12$ (as calculated in \eqref{eq:thresholdscdecay}), using the high sc-decay radial point estimates at the non-zero scattering section $R(\sigma)$ (see \cite[\S 4]{Vas21a}. The proof follows from a positive commutator estimate
		\[
		i(\widehat{\Box_{g_{b}}}(\sigma)^*A-A\widehat{\Box_{g_{b}}}(\sigma))=\IM \widehat{\Box_{g_{b}}}(\sigma)A+A\IM\widehat{\Box_{g_{b}}}(\sigma) +i[\RE \widehat{\Box_{g_{b}}}(\sigma), A]\quad
		\]
		where 
		\[
		\RE \widehat{\Box_{g_{b}}}(\sigma)=\frac{\widehat{\Box_{g_{b}}}(\sigma)+\widehat{\Box_{g_{b}}}(\sigma)^*}{2},\quad\IM \widehat{\Box_{g_{b}}}(\sigma)=\frac{\widehat{\Box_{g_{b}}}(\sigma)-\widehat{\Box_{g_{b}}}(\sigma)^*}{2i}.
		\]
		Let $A\in\Psi_{\scop,\bop}^{2s-1,2r+1,2l+1}(\CX)$ with principal symbol \[
		a=\chi_0(\zeta_\theta^2+\frac{1}{\sin^2\theta}\zeta_\varphi^2)\chi_1((\zeta_\rho-2\RE z)^2)\rho^{-2r+1}(\zeta_\theta^2+\frac{1}{\sin^2\theta}\zeta_\varphi^2+\zeta_\rho^2)^{s-\frac{1}{2}}
		\]
		where $\chi_0, \chi_1$ are identically 0 near $0$ and have compact support sufficiently close to $0$, and $\chi_1$ has relatively large support such that $\mbox{supp}\chi_0\cap \mbox{supp}\chi_1'$ is disjoint from the characteristic set of $\RE P_h(z)$. Then using Lemma 4.1 and the calculation before Lemma 4.8 in \cite{Vas21a}, it follows that $(\IM \widehat{\Box_{g_{b}}}(\sigma)A+A\IM \widehat{\Box_{g_{b}}}(\sigma)+i[\RE \widehat{\Box_{g_{b}}}(\sigma), A])\in \Psi_{\scop,\bop}^{2s,2r,2l}(\CX)$ whose principal symbol is positive definite elliptic near $R(\sigma)$ if $r>-\frac{1}{2}$. This, together with a regularization argument,  proves high sc-decay radial point estimates at $R(\sigma)$), there exist $B_\sigma, G_\sigma, S_\sigma \in\Psi^{0,0}_{\scop}(\CX)$ microlocally supported near $R(\sigma)$ with $R(\sigma)\subset \mathrm{ell}_{\scop}(B_\sigma), R(\sigma)\subset\mathrm{ell}_{\scop}(S_\sigma)$ and satisfying that all the forward null characteristics from $\mathrm{WF}_{\scop}(B_\sigma)$ tend to $R(\sigma)$, while remaining in the scattering elliptic set of $G_\sigma$, such that
		\begin{equation}\label{eq:michighscdecayradiales}
			\norm{B_\sigma u}_{H^{s,r,\ell}_{\scop,\bop}(\CX)}\leq C \norm{G_\sigma\widehat{\Box_{g_{b}}}(\sigma)u}_{H^{s-1, r+1,\ell+1}_{\scop,\bop}(\CX)}+C\norm{S_\sigma u}_{H^{s_0,r_0,\ell_0}_{\scop,\bop}(\CX)}
		\end{equation}
		where $s_0<s, \ell_0<\ell$ and the differential order $s$ and b-decay order $\ell$ are actually irrelevant here.
		
		For $\ell<-\frac 12$ (as calculated in \eqref{eq:thresholdscdecay}), using the low b-decay radial point estimates at the zero scattering section $R(0)$ (see \cite[\S 4]{Vas21a}), there exist $B_0, G_0\in\Psi^{0,0,0}_{\scop,\bop}(\CX)$ microlocally supported near $R(0)$ (in fact near the blown up $R(0)$ in the second microlocalized space introduced in \S\ref{subsubsec:secondmic}) with $R(0)\subset \mathrm{ell}_{\scop,\bop}(B_0)$ and $E_0\in\Psi_{\scop,\bop}^{0,0,0}(\CX), \mathrm{WF}_{\scop,\bop}(E_0)\cap R(0)=\emptyset$, and satisfying that all the forward null characteristics from $\mathrm{WF}_{\scop,\bop}(B_0)\setminus R(0)$ enter $\mathrm{ell}_{\scop,\bop}(E_0)$, while remaining in the second microlocalized scattering elliptic set of $G_0$, such that for any sufficiently large $N>0$
		\begin{equation}\label{eq:miclowbdecayradiales}
			\norm{B_0 u}_{H^{s,r,\ell}_{\scop,\bop}(\CX)}\leq C h^{-1}\norm{G_0\widehat{\Box_{g_{b}}}(\sigma)u}_{H^{s-1, r+1,\ell+1}_{\scop,\bop}(\CX)}+C\norm{E_0 u}_{H^{s,r,\ell}_{\scop,\bop}(\CX)}+C\norm{u}_{\bar{H}^{-N,-N,\ell}_{\scop,\bop}(\CX)}
		\end{equation}
		where the differential order $s$ is actually irrelevant here.

		Let $r_-<r_0<\ehKN$ and $\chi\in C_c^\infty(\CX)$ with $\chi=1$ near $r\geq\ehKN$ and $\mathrm{supp}\chi\subset\{r\geq r_0\}$. By part (1) in Proposition \ref{prop:microlocalglobaldy}, propagation of singularities estimates and elliptic estimates (Concretely, if $(r,\theta, \varphi, \xi) (\mbox{ or } (\rho,\theta, \varphi, \zeta))\in \mathrm{WF}(\chi)\cap(\Sigma\cup\Sigma_{\scop})^c$, we use elliptic estimates. If $(r,\theta, \varphi, \xi)(\mbox{ or }(\rho,\theta, \varphi, \xi))\in \mathrm{WF}(\chi)\cap(\Sigma\cup\Sigma_{\scop})$, then there exists $s\in\BR$ with $\exp(s{}^{\scop}\!H_{p_{h,z}}^{2,0})(\rho,\theta, \varphi,\xi)\in \mathrm{ell}_h(B_\pm)\cup\mathrm{ell}_h(B_1)\cup\mathrm{ell}_h(B_0)\cup\mathrm{ell}_h(B_z)$, and thus we use propagation of singularities estimates. We point out that in the set $\{\RE p_{\scop}(\sigma)=0\}\cap\pa_+\CX$, the scattering symbol has a nonnegative imaginary part, so one can propagate estimates towards $R(0)$. Finally, by using a pseudodifferential partition of unity, $\chi$ can be written as s sum of operators falling into the above two case), we have
		\begin{equation}
			\begin{split}
				\norm{\chi u}_{H^{s,r,\ell}_{\scop,\bop}(\CX)}&\leq C \norm{\widehat{\Box_{g_{b}}}(\sigma)u}_{\bar{H}^{s-1, r+1,\ell+1}_{\scop,\bop}(\CX)}+C\norm{B_+ u}_{H^{s,r,\ell}_{\scop,\bop}(\CX)}+C\norm{B_- u}_{H^{s,r,\ell}_{\scop,\bop}(\CX)}\\
				&+C\norm{B_0 u}_{H^{s,r,\ell}_{\scop,\bop}(\CX)}+C\norm{B_\sigma u}_{H^{s,r,\ell}_{\scop,\bop}(\CX)}+C\norm{u}_{\bar{H}^{-N,-N,\ell}_{\scop,\bop}(\CX)}.
			\end{split}
		\end{equation}
		By the same reasoning as above in the control of $\chi u$, it follows that
		\begin{equation}
			\begin{split}
				\norm{E_0 u}_{H^{s,r,\ell}_{\scop,\bop}(\CX)}\leq C \norm{\widehat{\Box_{g_{b}}}(\sigma)u}_{\bar{H}^{s-1, r+1,\ell+1}_{\scop,\bop}(\CX)}+C\norm{B_\sigma u}_{H^{s,r,\ell}_{\scop,\bop}(\CX)}+C\norm{u}_{\bar{H}^{-N,-N,\ell}_{\scop,\bop}(\CX)}.
			\end{split}
		\end{equation}
		Putting all the above estimates together yields for $s>s_0>\frac 12, r>r_0>-\frac 12, \ell<-\frac 12$ and for any sufficiently large $N>0$
		\begin{equation}\label{eq:micglobales1}
			\norm{\chi u}_{H^{s,r,\ell}_{\scop,\bop}(\CX)}\leq C \norm{\widehat{\Box_{g_{b}}}(\sigma)u}_{\bar{H}^{s-1, r+1,\ell+1}_{\scop,\bop}(\CX)}+C\norm{u}_{\bar{H}^{s_0,r_0,-N}_{\scop,\bop}(\CX)}+C\norm{u}_{\bar{H}^{-N,-N,\ell}_{\scop,\bop}(\CX)}.
		\end{equation}
		
		Since
		\[
		p=-\rho_b^{-2}\Big(\Delta_b\big(\xi_r+\frac{a\xi_\varphi}{\Delta_b}\big)^2
		-\frac{a^2\xi_\varphi^2}{\Delta_b}+\tilde{p}\Big)
		\]
		where
		\[ \tilde{p}=\xi_\theta^2+\frac{1}{\sin^2\theta}\xi_\varphi^2
		\]
		is hyperbolic with respect to $r$ in the region $\{r_-\leq r<\ehKN\}$ (see \cite[definition E.55]{DZ19}), using the hyperbolic estimates (see \cite[Theorem E.56]{DZ19}), we have for all $s,r,\ell\in \BR$
		\begin{equation}\label{eq:michyperesext}
			\norm{(1-\chi) u}_{\bar{H}^{s,r,\ell}_{\scop,\bop}(\CX)}\leq C \norm{\widehat{\Box_{g_{b}}}(\sigma)u}_{\bar{H}^{s-1, r+1,\ell+1}_{\scop,\bop}(\CX)}+\norm{\chi u}_{\bar{H}^{s,r,\ell}_{\scop,\bop}(\CX)}
		\end{equation}
		where the sc-decay order $r$ and b-decay order $\ell$ are actually relevant here. Combining \eqref{eq:micglobales1} with \eqref{eq:michyperesext} yields for $s>s_0>\frac 12, r>r_0>-\frac 12, \ell<-\frac12$ and for any sufficiently large $N>0$ 
		\begin{equation}
			\norm{u}_{\bar{H}^{s,r,\ell}_{\scop,\bop}(\CX)}\leq C \norm{\widehat{\Box_{g_{b}}}(\sigma)u}_{\bar{H}^{s-1, r+1,\ell+1}_{\scop,\bop}(\CX)}+C\norm{u}_{\bar{H}^{s_0,r_0,-N}_{\scop,\bop}(\CX)}+C\norm{u}_{\bar{H}^{-N,-N,\ell}_{\scop,\bop}(\CX)}.
		\end{equation}
	Since near $\pa_+\CX$
	\[
	\widehat{\Box_{g_{b}}}(\sigma)-N(	\widehat{\Box_{g_{b}}}(\sigma))\in \rho^2\mbox{Diff}_\bop^2(\CX), \quad N(	\widehat{\Box_{g_{b}}}(\sigma))=-2i\sigma\rho(\rho\pa_\rho-1),
	\]
	it follows that from \cite[Lemma 4.13 and Proposition 4.16]{Vas21a} that for $\ell<-\frac12$
	\begin{align*}
	\norm{u}_{\bar{H}^{-N,-N,\ell}_{\scop,\bop}(\CX)}&\leq 	\norm{u}_{\bar{H}^{-N-\ell,-N,\ell}_{\scop,\bop}(\CX)}=\norm{u}_{\bar{H}^{-N-\ell,\ell}_{\bop}(\CX)}\leq C\norm{N(\widehat{\Box_{g_{b}}}(\sigma))u}_{\bar{H}^{-N-\ell,\ell+1}_{\bop}(\CX)}\\
	&\leq C\norm{\widehat{\Box_{g_{b}}}(\sigma)u}_{\bar{H}^{-N-\ell,\ell+1}_{\bop}(\CX)}+C\norm{u}_{\bar{H}^{-N-\ell+2,\ell-1}_{\bop}(\CX)}\\
	&=C\norm{\widehat{\Box_{g_{b}}}(\sigma)u}_{\bar{H}^{-N-\ell,-N+1,\ell+1}_{\scop,\bop}(\CX)}+C\norm{u}_{\bar{H}^{-N-\ell+2,-N+1,\ell-1}_{\scop,\bop}(\CX)},
	\end{align*}
and thus  for $s>s_0>\frac 12, r>r_0>-\frac 12, \ell<-\frac12$
\begin{equation}
	\norm{u}_{\bar{H}^{s,r,\ell}_{\scop,\bop}(\CX)}\leq C \norm{\widehat{\Box_{g_{b}}}(\sigma)u}_{\bar{H}^{s-1, r+1,\ell+1}_{\scop,\bop}(\CX)}+C\norm{u}_{\bar{H}^{s_0,r_0,\ell-1}_{\scop,\bop}(\CX)}.
\end{equation}
This finishes the proof of \eqref{eq:Fredholmsecondmic}.
		
		\item \underline{Proof of the estimate \eqref{eq:Fredholmsecondmicdual}.} For $s>\frac 12\geq\frac{1}{2}-\IM(\sigma)\frac{\ehKN^2+a^2}{\ehKN^2-\Bm}$, we have $1-s\leq\frac{1}{2}-\IM(\bar{\sigma})\frac{\ehKN^2+a^2}{\ehKN^2-\Bm}$. Then using the low regularity radial point estimates (see \cite[Proposition 2.4]{Vas13}, \cite[Proposition 5.27]{Vas18}), there exist $B'_\pm, G'_\pm\in\Psi^0(\CX)$ microlocally supported near $L_\pm$ with $L_\pm\subset \mathrm{ell}(B'_\pm)$ and $E'_\pm\in\Psi^0(\CX), \mathrm{WF}(E'_\pm)\cap L_\pm=\emptyset$, and satisfying that all the backward (forward) null characteristics from $\mathrm{WF}(B'_\pm)\setminus L_\pm$ reach $\mathrm{ell}(E'_\pm)$, while remaining in the elliptic set of $G'_\pm$, such that for any $N\in\BR$
		\begin{equation}\label{eq:micregradialesdual}
			\begin{split}
				&\norm{B'_\pm u}_{H^{1-s,-r-1,-\ell-1}_{\scop,\bop}(\CX)}\\
				&\quad\leq C \norm{G'\widehat{\Box_{g_{b}}}(\sigma)^*u}_{H^{-s, -r,-\ell}_{\scop,\bop}(\CX)}+C\norm{E'_\pm u}_{H^{1-s,-r-1,-\ell-1}_{\scop,\bop}(\CX)}+C\norm{u}_{\dot{H}^{-N,-N,-N}_{\scop,\bop}(\CX)}
			\end{split}
		\end{equation}
		where the sc-decay order $r$ and b-decay order $\ell$ are actually irrelevant here.
		
		For $r>-\frac 12$, we have $-r-1<-\frac 12$. Then using the low sc-decay radial point estimates at the non-zero scattering section $R(\sigma)$ (see \cite[\S 4]{Vas21a} and the above discussion about the proof of high sc-decay radial point estimates at $R(\sigma)$. We note that the only difference is that now the term involving $\chi_0'$ does not have the correct sign and needs to be treated as an error term $E_\sigma'u$), there exist $B'_\sigma, G'_\sigma \in\Psi^{0,0}_{\scop}(\CX)$ microlocally supported near $R(\sigma)$ with $R(\sigma)\subset \mathrm{ell}_{\scop}(B_\sigma)$ and $E'_\sigma\in\Psi^{0,0}_{\scop}(\CX), \mathrm{WF}_{\scop}(E'_\sigma)\cap R(\sigma)=\emptyset$, and satisfying that all the backward null characteristics from $\mathrm{WF}_{\scop}(B'_\sigma)\setminus R(\sigma)$ reach $\mathrm{ell}_{\scop,h}(E'_\sigma)$, while remaining in the scattering elliptic set of $G'_\sigma$, such that
		\begin{equation}\label{eq:lmicowscdecayradialesdual}
			\begin{split}
				&\norm{B'_\sigma u}_{H^{1-s,-r-1,-\ell-1}_{\scop,\bop}(\CX)}\\
				&\quad\leq C\norm{G'_\sigma\widehat{\Box_{g_{b}}}(\sigma)^*u}_{H^{-s, -r,-\ell}_{\scop,\bop}(\CX)}+C\norm{E'_\sigma u}_{H^{1-s,-r-1,-\ell-1}_{\scop,\bop}(\CX)}+C\norm{u}_{\dot{H}^{-N,-N,-N}_{\scop,\bop}(\CX)}
			\end{split}
		\end{equation}
		where the differential order $s$ and b-decay order $\ell$ are actually irrelevant here.
		
		For $\ell<-\frac 12$, we have $-\ell-1>-\frac 12$. Then using the high b-decay radial point estimates at the zero scattering section $R(0)$ (see \cite[\S 4]{Vas21a}), there exist $B'_0, G'_0\in\Psi^{0,0,0}_{\scop,\bop}(\CX)$ microlocally supported near $R(0)$ (in fact near the blown up $R(0)$ in the second microlocalized space introduced in \S\ref{subsubsec:secondmic}) with $R(0)\subset \mathrm{ell}_{\scop,\bop}(B'_0)$, and satisfying all the backward null characteristics from $\mathrm{WF}_{\scop,\bop}(B'_0)$ tend to $R(0)$, while remaining in the second microlocalized scattering elliptic set of $G'_0$, such that for any $N\in\BR$
		\begin{equation}\label{eq:micbdecayradialesdual}
			\norm{B'_0 u}_{H^{1-s,-r-1,-\ell-1}_{\scop,\bop}(\CX)}\leq C \norm{G'_0\widehat{\Box_{g_{b}}}(\sigma)^*u}_{H^{-s, -r,-\ell}_{\scop,\bop}(\CX)}+C\norm{u}_{\dot{H}^{-N,-N,-\ell-1}_{\scop,\bop}(\CX)}
		\end{equation}
		where the differential order $s$ is actually irrelevant here.

		By the same reasoning as in the proof of the estimate \eqref{eq:Fredholmsecondmic}, we have
		\begin{equation}
			\begin{split}
				&\norm{\chi u}_{H^{1-s,-r-1,-\ell-1}_{\scop,\bop}(\CX)}\\
				&\quad\leq C \norm{\widehat{\Box_{g_{b}}}(\sigma)^*u}_{\dot{H}^{-s, -r,-\ell}_{\scop,\bop}(\CX)}+C\norm{B_+ u}_{H^{1-s,-r-1,-\ell-1}_{\scop,\bop}(\CX)}\\
				&\quad\quad+C\norm{B'_- u}_{H^{1-s,-r-1,-\ell-1}_{\scop,\bop}(\CX)}
				+C\norm{B'_0 u}_{H^{1-s,-r-1,-\ell-1}_{\scop,\bop}(\CX)}\\
				&\quad\quad+C\norm{B'_\sigma u}_{H^{1-s,-r-1,-\ell-1}_{\scop,\bop}(\CX)}+C\norm{u}_{\dot{H}^{-N,-N,-\ell-1}_{\scop,\bop}(\CX)},
			\end{split}
		\end{equation}
		and
		\begin{equation}
			\begin{split}
				&\norm{E'_\pm u}_{H^{1-s,-r-1,-\ell-1}_{\scop,\bop,h}(\CX)}+\norm{E'_\sigma u}_{H^{1-s,-r-1,-\ell-1}_{\scop,\bop,h}(\CX)}\\
				&\quad\leq C \norm{\widehat{\Box_{g_{b}}}(\sigma)^*u}_{\dot{H}^{-s, -r,-\ell}_{\scop,\bop,h}(\CX)}+C\norm{(1-\chi) u}_{\dot{H}^{1-s,-r-1,-\ell-1}_{\scop,\bop}(\CX)}\\&\quad\quad+C\norm{B'_0 u}_{H^{1-s,-r-1,-\ell-1}_{\scop,\bop}(\CX)}+C\norm{u}_{\dot{H}^{-N,-N,-\ell-1}_{\scop,\bop}(\CX)}.
			\end{split}
		\end{equation}
		Putting all the above estimates together yields for $s>\frac 12, r>-\frac 12, \ell<-\frac 12$
		\begin{equation}\label{eq:micglobales2}
			\begin{split}
				\norm{\chi u}_{H^{1-s,-r-1,-\ell-1}_{\scop,\bop}(\CX)}&\leq C \norm{\widehat{\Box_{g_{b}}}(\sigma)^*u}_{\dot{H}^{-s, -r,-\ell}_{\scop,\bop}(\CX)}+C\norm{(1-\chi) u}_{\dot{H}^{1-s,-r-1,-\ell-1}_{\scop,\bop}(\CX)}\\
				&\quad+C\norm{u}_{\dot{H}^{-N,-N,-\ell-1}_{\scop,\bop}(\CX)},
			\end{split}
		\end{equation}
		Again using the semiclassical hyperbolic estimates (see \cite[Theorem E.56]{DZ19}), we have for all $s,r,\ell\in \BR$
		\begin{equation}\label{eq:michyperessupp}
			\norm{(1-\chi) u}_{\dot{H}^{1-s,-r-1,-\ell-1}_{\scop,\bop,h}(\CX)}\leq C\norm{\widehat{\Box_{g_{b}}}(\sigma)^*u}_{\dot{H}^{-s, -r,-\ell}_{\scop,\bop}(\CX)}
		\end{equation}
		where the sc-decay order $r$ and b-decay order $\ell$ are actually relevant here. Combining \eqref{eq:micglobales2} with \eqref{eq:michyperessupp} yields for $s>\frac 12, r>-\frac 12, \ell<-\frac12$
		\begin{equation}
			\norm{u}_{\dot{H}^{1-s,-r-1,-\ell-1}_{\scop,\bop}(\CX)}\leq C \norm{\widehat{\Box_{g_{b}}}(\sigma)^*u}_{\dot{H}^{-s, -r,-\ell}_{\scop,\bop}(\CX)}+C\norm{u}_{\dot{H}^{-N,-N,-\ell-1}_{\scop,\bop}(\CX)}.
		\end{equation}
By the same argument as in the proof of the estimate \eqref{eq:Fredholmsecondmic}, we have for $s>\frac 12, r>-\frac 12, \ell<-\frac12$
\begin{equation}
	\norm{u}_{\dot{H}^{1-s,-r-1,-\ell-1}_{\scop,\bop}(\CX)}\leq C \norm{\widehat{\Box_{g_{b}}}(\sigma)^*u}_{\dot{H}^{-s, -r,-\ell}_{\scop,\bop}(\CX)}+C\norm{u}_{\dot{H}^{-N,-N,-\ell-2}_{\scop,\bop}(\CX)}.
\end{equation}	
	This proves the estimate \eqref{eq:Fredholmsecondmicdual}.
	\item \underline{Fredholm property of $\widehat{\Box_{g_{b}}}(\sigma)$ for $\IM\sigma\geq 0, \sigma\neq 0$.} We only discuss the proof of \eqref{eq:Fredholm1} in detail as \eqref{eq:Fredholm2} can be handled in a completely analogous manner. We define the space \[
	\mathcal{X}(\sigma):=\{u\in\bar{H}_\bop^{s,\ell}(\CX)\mid \widehat{\Box_{g_{b}}}(\sigma)u\in\bar{H}_\bop^{s,\ell+1}(\CX)\}.
	\]
	We note that $\mathcal{X}(\sigma)$ is a Hilbert space endowed with the norm $\norm{u}_{\mathcal{X}(\sigma)}:=\norm{u}_{\bar{H}_\bop^{s,\ell}}+\norm{\widehat{\Box_{g_{b}}}(\sigma)u}_{\bar{H}_\bop^{s,\ell+1}}$. Let $\{u_j\}$ is a bounded sequence in $\ker\widehat{\Box_{g_{b}}}(\sigma)\subset\mathcal{X}(\sigma)$. Since $\bar{H}_\bop^{s,\ell}\to\bar{H}_\bop^{s_0, \ell_0}$ is a compact embedding, by passing to a subsequence we may assume that $u_j$ converges in $\bar{H}_\bop^{s_0,\ell_0}$. By \eqref{eq:Fredholmenergy1}, we see that $u_j$ is a Cauchy sequence in $\bar{H}_\bop^{s,\ell}$. This implies that $\ker\widehat{\Box_{g_{b}}}(\sigma)$ is finite dimensional. We then show that $\Ran_{\mathcal{X}(\sigma)}\widehat{\Box_{g_{b}}}(\sigma)$ is closed in $\bar{H}^{s,\ell+1}_\bop$. Let $\{f_j\}\subset\Ran_{\mathcal{X}(\sigma)}\widehat{\Box_{g_{b}}}(\sigma)$ be a convergent sequence with $f_j\to f_\infty\in\bar{H}^{s,\ell+1}_\bop$. We may write $f_j=\widehat{\Box_{g_{b}}}(\sigma)u_j$ where $u_j\in\mathcal{X}(\sigma)$ is in the orthogonal complement of $\ker\widehat{\Box_{g_{b}}}(\sigma)$. We claim that $\{u_j\}$ is bounded in $\mathcal{X}(\sigma)$. (Otherwise, suppose that $\{u_j\}$ is unbounded in $\mathcal{X}(\sigma)$. Passing to a subsequence, we may assume that $\norm{u_j}_{\mathcal{X}(\sigma)}\to\infty$. Let $\tilde{u}_j=u_j/\norm{u_j}_{\mathcal{X}(\sigma)}$ and $\tilde{f}_j=\widehat{\Box_{g_{b}}}(\sigma)\tilde{u}_j=f_j/\norm{u_j}_{\mathcal{X}(\sigma)}$. Then $\norm{\tilde{u}_j}_{\mathcal{X}(\sigma)} =1$ and $\norm{\tilde{f}_j}_{\bar{H}_\bop^{s,\ell+1}}\to 0$. By the argument as above, passing to a subsequence, we have that $\tilde{u}_j\to\tilde{u}_\infty\in\mathcal{X}(\sigma)$ and thus $\widehat{\Box_{g_{b}}}(\sigma)\tilde{u}_\infty=0$. On the other hand, since $\tilde{u}_\infty$ is in the orthogonal complement of $\ker\widehat{\Box_{g_{b}}}(\sigma)$, we have $\widehat{\Box_{g_{b}}}(\sigma)\tilde{u}_\infty\neq0$; this gives a contradiction.) Then argue as above, by passing to a subsequence we may assume that $u_j\to u_\infty\in\mathcal{X}(\sigma)$. It follows that $\widehat{\Box_{g_{b}}}(\sigma)u_j\to\widehat{\Box_{g_{b}}}(\sigma)u_\infty\in \bar{H}^{s,\ell+1}_\bop$ and thus $\widehat{\Box_{g_{b}}}(\sigma)u_\infty=f_\infty$. This proves that $f_\infty\in\Ran_{\mathcal{X}(\sigma)}\widehat{\Box_{g_{b}}}(\sigma)$ and thus $\Ran_{\mathcal{X}(\sigma)}\widehat{\Box_{g_{b}}}(\sigma)$ is closed. Finally, it remains to show that $\Ran_{\mathcal{X}(\sigma)}\widehat{\Box_{g_{b}}}(\sigma)$ has finite codimension. Proceeding as in the proof of the finite dimension property of $\ker\widehat{\Box_{g_{b}}}(\sigma)$ and using \eqref{eq:Fredholmenergydual1}, we obtain that $\ker\widehat{\Box_{g_{b}}}(\sigma)^*\subset \dot{H}_b^{-s,-\ell-1}$ is also finite dimensional. Let $f\in\bar{H}_\bop^{s,\ell+1}$ be such that $\angles{f}{v}=0$ for all $v\in\ker\widehat{\Box_{g_{b}}}(\sigma)^*$. We claim that there exists $u\in\bar{H}^{s,\ell}_\bop$ such that $\widehat{\Box_{g_{b}}}(\sigma)u=f$. (This implies that $\Ran_{\mathcal{X}(\sigma)}\widehat{\Box_{g_{b}}}(\sigma)$ has finite codimension.) We define 
	\[
	\mathcal{Y}(\sigma):=\{v\in\dot{H}_\bop^{-s,-\ell-1}(\CX)\mid \widehat{\Box_{g_{b}}}(\sigma)^*v\in\dot{H}_\bop^{-s,-\ell}(\CX)\}.
	\]
	Let $V$ be a complementary space of $\ker\widehat{\Box_{g_{b}}}(\sigma)^*$ in $\dot{H}^{-s,-\ell-1}_\bop$. Then there exists a constant $C_2$ such that 
	\[
	\norm{v}_{\dot{H}^{-s,-\ell-1}_\bop}\leq C_2\norm{\widehat{\Box_{g_{b}}}(\sigma)^* v}_{\dot{H}^{-s,-\ell}_\bop},\quad v\in\mathcal{Y}(\sigma)\cap V.
	\]
	If this were not true, we could find a sequence ${v_j}\subset\mathcal{Y}(\sigma)\cap V$ such that 
	\[
\norm{v_j}_{\dot{H}^{-s,-\ell-1}_\bop}=1,\quad 	\norm{v_j}_{\dot{H}^{-s,-\ell-1}_\bop}\geq j\norm{\widehat{\Box_{g_{b}}}(\sigma)v_j}_{\dot{H}^{-s,-\ell}_\bop}.
	\]
	By \eqref{eq:Fredholmenergydual1} and the compact embedding $\dot{H}^{-s,-\ell-1}_\bop\to\dot{H}^{-N,-\ell-3}_\bop$, passing to a subsequence we have $v_j\to v_\infty$ with $\norm{v_\infty}_{\dot{H}^{-s,-\ell-1}_\bop}=1$ and $v_\infty\in V\cap\ker\widehat{\Box_{g_{b}}}(\sigma)^*$; this gives a contradiction. If $f\in\bar{H}_\bop^{s,\ell+1}$ satisfies $\angles{f}{v}=0$ for all $v\in\ker\widehat{\Box_{g_{b}}}(\sigma)^*$, we have
\[
\abs{\angles{f}{v}}\leq C_2\norm{f}_{\bar{H}_\bop^{s,\ell+1}}\norm{\widehat{\Box_{g_{b}}}(\sigma)^*v}_{\dot{H}_\bop^{-s,-\ell}},\quad v\in\mathcal{Y}(\sigma)\cap V.
\]
Since any $v\in\mathcal{Y}(\sigma)$ can be written as $v=v_1+v_2$ with $v_1\in\ker\widehat{\Box_{g_{b}}}(\sigma)^*, v_2\in \mathcal{Y}(\sigma)\cap V$, it follows that 
\[
\abs{\angles{f}{v}}=\abs{\angles{f}{v_2}}\leq C_2\norm{f}_{\bar{H}_\bop^{s,\ell+1}}\norm{\widehat{\Box_{g_{b}}}(\sigma)^*v_2}_{\dot{H}_\bop^{-s,-\ell}}=C_2\norm{f}_{\bar{H}_\bop^{s,\ell+1}}\norm{\widehat{\Box_{g_{b}}}(\sigma)^*v}_{\dot{H}_\bop^{-s,-\ell}},\quad v\in\mathcal{Y}(\sigma).
\]
By Hahn-Banach Theorem, the anti-linear form $\widehat{\Box_{g_{b}}}(\sigma)^*v\to \angles{f}{v}$ with $v\in\mathcal{Y}(\sigma)$ can be extended to a continuous anti-linear form on $\dot{H}^{-s,-\ell}_\bop$. Since $(\dot{H}^{-s,-\ell}_\bop)^*=\bar{H}^{s,\ell}_\bop$, there exists $u\in\bar{H}^{s,\ell}_\bop$ such that $\angles{u}{\widehat{\Box_{g_{b}}}(\sigma)^*v}=\angles{f}{v}$ for $v\in\mathcal{Y}(\sigma)$. In particular, for $v\in C_c^\infty(X^\circ)$
\[
\angles{\widehat{\Box_{g_{b}}}(\sigma)u}{v}=\angles{u}{\widehat{\Box_{g_{b}}}(\sigma)^*v}=\angles{f}{v}.
\]
This proves that $\widehat{\Box_{g_{b}}}(\sigma)u=f$.

We now prove statement (2). Combining the estimates near spatial infinity $\pa_+\CX$ (see \cite[Proposition 5.3]{Vas21b}) with the radial point estimates at event horizon, propagation of singularity estimates, elliptic estimates and hyperbolic estimates as in the proof of statement (1), we obtain \eqref{eq:uniformFredholmenergy}. The Fredholm property follows as in the nonzero $\sigma$ case.
	\end{itemize}
\end{proof}

%%%%%%%%%%%%%%%%%%%%%%%%%%%%%%%%%%%%%%%%%%%%%%%%%%%%%%
\subsubsection{Uniform Fredholm estimates for tensor wave operators}
Now we prove an analogy to Theorem \ref{thm:Fredholmestimate} for the operators  $\widehat{\mathcal{P}_{b,\gamma}}(\sigma),\  \widehat{\mathcal{W}_{b,\gamma}}(\sigma)$ and $\widehat{L_{b, \gamma}}(\sigma)$ acting on bundles. We note that the formal adjoint of $\mathcal{P}_{b,\gamma}, \mathcal{W}_{b,\gamma}, L_{b,\gamma}$ with respect to the natural inner product induced by $g_b$ and the volume form $dvol_{g_b}$ is 
\[
\mathcal{W}_{b,\gamma},\quad \mathcal{P}_{b,\gamma},\quad \begin{pmatrix}
	G_{g_b}&0\\
	0&4
\end{pmatrix}L_{b,\gamma}\begin{pmatrix}
G_{g_b}&0\\
0&\frac14
\end{pmatrix},
\]
respectively (see \eqref{eq:modifiedlinearizedEinsteinDual} and \eqref{eq:modifiedlinearizedMaxwellDual} for the derivation of the adjoint of $L_{b,\gamma}$).
\begin{thm}\label{thm:FredholmestimateBundle}
	Let $b_0=(\Bm_0, \Ba_0, \BQ_0)$ with $\abs{\BQ_0}+\abs{\Ba_0}<\Bm_0$ and $\abs{\Ba_0}\ll\abs{\Bm_0}+\abs{\BQ_0}$. Then there exists $\epsilon>0$ such that for $b=(\Bm, \Ba,\BQ)$ with $\abs{b-b_0}<\epsilon$ and for $s>s_0>2$ (resp. $s>s_0>3$), $\ell-1\leq\ell_0<\ell<-\frac{1}{2}$ with $s+\ell>s_0+\ell_0>-\frac12$ and $\ell\neq -\frac32$, the conclusions in Theorem \ref{thm:Fredholmestimate} hold for $\widehat{\mathcal{P}_{b,\gamma}}(\sigma)$ and $\widehat{\mathcal{W}_{b,\gamma}}(\sigma)$ (resp. $\widehat{L_{b,\gamma}}(\sigma)$).	\end{thm}
\begin{proof}
	 The proof is analogous to that of Theorem \ref{thm:Fredholmestimate}, except for the computation of threshold regularity in the radial point estimate at event horizon and the threshold decay rate in the radial point estimate at spatial infinity $\pa_+\CX$.
	 
	 As for the calculation of threshold regularity in the radial point estimate at event horizon, the Reissner-Nordstr\"{o}m metric case $g_b=g_{(\Bm_0, 0,\BQ_0)}$ was done in appendix \ref{app:Rad}. According to \eqref{EqRadReg} and \eqref{eq:thresholdreg}, for $\IM\sigma\geq0$, the threshold regularity is $\frac32-\frac{2\gamma}{\kappa}$ for $\widehat{\mathcal{P}_{b_0,\gamma}}(\sigma)$ and  $\frac32+\frac{\gamma}{2\kappa}$ for $\widehat{\mathcal{W}_{b_0,\gamma}}(\sigma)$ acting on 1-forms, and is $\frac52-\frac{2\gamma}{\kappa}$ for $\widehat{L_{b, \gamma}}(\sigma)$ acting on $\scsym\oplus\scform$. This implies that the threshold regularity for nearby Kerr-Newman metrics is close to $\frac32$ for $\widehat{\mathcal{P}_{b, \gamma}}(\sigma)$ and $\widehat{\mathcal{W}_{b, \gamma}}(\sigma)$, and close to $\frac52$ for $\widehat{L_{b, \gamma}}(\sigma)$ as $\gamma$ is a sufficiently small constant.
	
	For $0\neq\sigma\in\BR$, the radial point estimate at $R(0)$ and $R(\sigma)$ for $\widehat{\mathcal{P}_{b, \gamma}}(\sigma),\ \widehat{\mathcal{W}_{b, \gamma}}(\sigma),\ \widehat{L_{b,\gamma}}(\sigma)$ requires the computation of a threshold decay rate relative to $L^2(\CX)$. Concretely, the threshold $-\frac12$ from \cite[Theorems~1.1 and 1.3]{Vas21a} is modified by the subprincipal symbol $\sigma_{\scop}(\frac{1}{2 i\rho}(\widehat{\bullet}(\sigma)-\widehat{\bullet}(\sigma)^*))|_{R(0),R(\sigma)}$ where $\bullet=\mathcal{P}_{b,\gamma},\ \mathcal{W}_{b,\gamma},\ L_{b,\gamma}$. Working in the trivialization of $\scsym\oplus\scform$ given in terms of the differentials of standard coordinates $t,x^1,x^2,x^3$, according to Proposition \ref{PropKNStBox} and Lemma \ref{LemPWL}, we see that 
	\begin{gather*}
\sigma_{\scop}(\frac{1}{2 i\rho}(\widehat{\bullet}(\sigma)-\widehat{\bullet}(\sigma)^*))|_{R(0),R(\sigma)}=\sigma_{\scop}(\frac{1}{2 i\rho}(\widehat{\Box_{g_b,0}}(\sigma)-\widehat{\Box_{g_b,0}}(\sigma)^*)\otimes \mathrm{Id}_{4\times 4})|_{R(0),R(\sigma)}=0,\quad \bullet=\mathcal{P},\mathcal{W},\\ 	\sigma_{\scop}(\frac{1}{2 i\rho}(\widehat{L_{b,\gamma}}(\sigma)-\widehat{L_{b,\gamma}}(\sigma)^*))|_{R(0),R(\sigma)}=\sigma_{\scop}(\frac{1}{2 i\rho}(\widehat{\Box_{g_b,0}}(\sigma)-\widehat{\Box_{g_b,0}}(\sigma)^*)\otimes \mathrm{Id}_{14\times 14})|_{R(0),R(\sigma)}=0.
	\end{gather*}
Therefore, the threshold decay rate is still $-\frac12$.
\end{proof}

%%%%%%%%%%%%%%%%%%%%%%%%%%%%%%%%%%%%%%%%%%%%%%%%%%%%%%%%%%%%%%%%%%%%%%%%%%%%%%%%%
\subsection{Description of the kernel of the wave type operators}
\label{subsec:desodfernel}
In this section, we give a detailed description of kernel of the following wave operators: $\widehat{\Box_{g_{b}}}(\sigma)$ on scalar functions, $\widehat{\mathcal{P}_{b, \gamma}}(\sigma),\widehat{\mathcal{W}_{b, \gamma}}(\sigma)$ on scattering $1$-forms and the linearized gauge-fixed Einstein-Maxwell operator $\widehat{L_{b,\gamma}}(\sigma)$.

\begin{prop}\label{prop:desofkernel}
Let $b=(\Bm, \Ba, \BQ)$ with $\abs{\BQ}+\abs{\Ba}<\abs{\Bm}$ and $a=\abs{\Ba}\ll\abs{\Bm}+\abs{\BQ}$. Suppose $u\in\bar{H}^{s,\ell}_\bop(\CX;\BC)$ with $s>\frac12$.
\begin{enumerate}
	\item If $\widehat{\Box_{g_{b}}}(0)u=0$ and $u\in\bar{H}^{s,\ell}_\bop(\CX;\BC)$ with $\ell\in\BR$, then $u\in\bar{H}^{\infty,\ell}_\bop(\CX;\BC)$ and has a polyhomogeneous expansion with index set contained in $\{(z,k)\mid z\in i\BZ, k\in\BN\}$. In particular, if $-\frac32<\ell<-\frac 12$, then $u\in\mathcal{A}^{1}(\CX;\BC)$. More precisely, there exists $u_0\in C^\infty(\pa_+\CX; \BC)$ such that $u-\frac{u_0}{r}\in\mathcal{A}^{2-}(\CX;\BC)$.
	
	\item If $\widehat{\Box_{g_{b}}}(0)^*u=0$ and $u\in\dot{H}^{-\infty,\ell}_\bop(\CX;\BC)$ with $\ell\in\BR$, then $u\in\dot{H}^{\frac12-,\ell}_\bop(\CX;\BC)$. Near $\pa_+\CX$, $u$ has the same polyhomegeneous expansion as in part (1).
	
	\item If $\sigma\neq 0$ and $u\in\ker\widehat{\Box_{g_{b}}}(\sigma)\cap\bar{H}^{s,\ell}_\bop(\CX;\BC)$ for some $\ell\in\BR$ with $s+\ell>-\frac12$, then $u\in\rho C^\infty(\pa_+\CX;\BC)+\mathcal{A}^{2-}(\CX;\BC)$.
\end{enumerate}
The above statements also hold for $\widehat{\mathcal{P}_{b, \gamma}}(\sigma),\widehat{\mathcal{W}_{b, \gamma}}$ acting on $\omega\in\bar{H}^{s,\ell}_\bop(\CX;\scform)$ with $s>2$ (for statement (2), $\omega\in\sHb^{-\frac12-C(\gamma,a),\ell}$) and $\widehat{L_{b,\gamma}}(\sigma)$ acting on $(\dg, \dA)\in\bar{H}^{s,\ell}_\bop(\CX;\scform\oplus\scsym)$ with $s>3$ (for statement (2), $\omega\in\sHb^{-\frac32-C(\gamma,a),\ell}$) where $C(\gamma,a)>0$ is a sufficiently small constant depending on $\gamma,\ a$.
\end{prop}
\begin{proof}
	We only discuss the proof for the operator $\widehat{\Box_{g_{b}}}(\sigma)$ in detail here as the other operators $\widehat{\mathcal{P}_{b, \gamma}}(\sigma), \widehat{\mathcal{W}_{b,\gamma}}$, $\widehat{L_{b,\gamma}}(\sigma)$ can be handled similarly.
	\begin{itemize}
		\item \underline{Proof of statement (1).} According to the high regularity radial point estimate (since $s>\frac12$ and $\frac12$ is the threshold regularity at the radial points at event horizon), \ref{prop:microlocalglobaldy}, propagation of singularity estimate and elliptic estimate for b-pseudodifferential operators, we conclude that $u\in H^{\infty, \ell}_\bop(\CX;\BC)$.
		
		To prove the polyhomogeneous expansion of $u$, we exploit a typical \textit{normal operator argument} (see \cite[\S 5]{Mel93}). Concretely, since 
		\[
		\widehat{\Box_{g_{b}}}(0)=(\rho^2\pa_\rho)^2-2\rho^3\pa_\rho+\rho^2(\pa_\theta^2+\frac{1}{\sin^2\theta}\pa_\varphi^2)+\rho^3\mbox{Diff}_\bop^2=\rho^2\Big(\rho\pa_\rho(\rho\pa_\rho-1)+\sL\Big)+\rho^3\mbox{Diff}_\bop^2\in \rho^2\mbox{Diff}_\bop^2,
		\]
	the normal operator $N(\widehat{\Box_{g_{b}}}(0))$ of $\widehat{\Box_{g_{b}}}(0)$ is defined by factoring out an overall weight (here is $\rho^2$) and freezing the coefficients at the boundary $\rho=0$. Therefore, we have
	\[
	N(\widehat{\Box_{g_{b}}}(0))=\rho^2\Big(\rho\pa_\rho(\rho\pa_\rho-1)+\sL\Big).
	\]
	Let $\chi(\rho)\in\C^\infty_c(\BR)$ be identically $1$ near $0$ and vanishing when $\rho=1/r\geq 1/3\Bm$. Then we have
	\begin{equation}\label{eq:eqnforchiu}
	\rho^{-2}N(\widehat{\Box_{g_{b}}}(0))(\chi u)=\rho^{-2}\Big(N(\widehat{\Box_{g_{b}}}(0))-\widehat{\Box_{g_{b}}}(0)\Big)(\chi u)+\rho^{-2}[\widehat{\Box_{g_{b}}}(0), \chi]u:=f\in\bar{H}^{\infty, \ell+1}_\bop(\CX;\BC).
	\end{equation}
	For simplicity of notation, we denote $\rho^{-2}N(\widehat{\Box_{g_{b}}}(0))$ by $N$. Using the Mellin transform in $\rho$, defined by (we drop the dependence on the angular variables on $\pa_+\CX$ from the notation)
	\[
	\widehat{\chi u}(\xi):=\int_0^\infty \rho^{-i\xi}(\chi u)(\rho)\,\frac{d\rho}{\rho},
	\]
	the equation $N(\chi u)=f$ becomes
	\[
	\widehat{N}(\xi)\big(\widehat{\chi u}\big)(\xi)=\Big(i\xi(i\xi-1)+\sL\Big)(\widehat{\chi u})(\xi)=\hat{f}(\xi).
	\]
	At this point, we only know that $\widehat{\chi u}(\xi)$ is holomorphic and Schwartz in $\RE\xi$ (with $\IM\xi$ fixed) in the region $\IM\xi>-\ell-\frac 32$ and so is $\hat{f}(\xi)$ in $\IM\xi>-\ell-\frac52$. Therefore, we have the following inverse Mellin transform
	\[
	\chi u(\rho)=\frac{1}{2\pi}\int_{\IM\xi=-\ell-\frac32+\epsilon}\rho^{i\xi}\widehat{\chi u}(\xi)\,d\xi,\quad \epsilon>0.
		\]
	
	We denote by $Y_l^m$ with $l\in\BN, m\in\BZ, \abs{m}\leq l$ the spherical harmonic function of degree $l$ and order $m$ which satisfies $\slashed{\Delta}Y_l^m=-l(l+1)Y_l^m$. We denote by $\BFS_l=\{Y_l^m: \abs{m}\leq l\}$ the space of degree $l$ spherical harmonic functions. Then one can conclude that $\{\BFS_l: l\in\BN\}$ form an orthogonal basis of $L^2(\BS^2)$. Therefore, one can expand $(\widehat{\chi u})(\xi)$ and $\hat{f}(\xi)$ in terms of spherical harmonics $Y_l^m$, i.e., we write $(\widehat{\chi u})(\xi)=\sum u_l^m(\xi)Y^m_l, \hat{f}=\sum f_l^m(\xi)Y_l^m$. Restricted to $\BFS_l$, the inverse \[\widehat{N}(\xi)^{-1}=\Big(i\xi(i\xi-1)+\sL\Big)^{-1}=\Big(i\xi(i\xi-1)-l(l+1))^{-1}
	\]
	is meromorphic in $\xi$ with simple poles given by $\xi=-i(l+1), il$. Then in the inverse Mellin transform
		\[
	\chi u(\rho)=\frac{1}{2\pi}\int_{\IM\xi=-\ell-\frac32+\epsilon}\rho^{i\xi}(\widehat{N}(\xi))^{-1}\hat{f}(\xi)\,d\xi,\quad \epsilon>0, \quad -\ell-\frac 32+\epsilon\notin\BZ,
	\]
	we can shift the contour of integration through the pole at $\xi=ik$ to $\IM\xi=-\ell-\frac52+\epsilon$ where $k\in\BZ$ and $k\in(-\ell-\frac32+\epsilon, -\ell-\frac52+\epsilon)$, and the Residue Theorem gives 
	\begin{equation}\label{eq:partialexpansion}
	\chi u(\rho)=\rho^{-k}u_k+\tilde{u}_k,\quad \tilde{u}_k=\frac{1}{2\pi}\int_{\IM\xi=-\ell-\frac 52+\epsilon}\rho^{i\xi}(\widehat{N}(\xi))^{-1}\hat{f}(\xi)\,d\xi\in \mathcal{A}^{\ell+\frac 52-\epsilon}(\CX;\BC)
	\end{equation}
	where 
	\[
	u_k=\begin{cases}
		\sum_{-k\leq m\leq k}\frac{-1}{2k+1}f_k^m(ik)Y_k^m,\quad &k\geq 0\\
	\sum_{k+1\leq m\leq -k-1}\frac{1}{2k+1}f_{-k-1}^m(-ik-i)Y_{-k-1}^m,\quad &k\leq-1
	\end{cases}.
	\]
To deduce the full polyhomogeneous expansion for $\chi u$, we plug the above partial expansion \eqref{eq:partialexpansion} into the equation \eqref{eq:eqnforchiu} to obtain
\[
N\tilde{u}_k=N(\chi u)=f_1+f_2,\quad f_1\in \rho^{-k+1}C^\infty(\pa_+\CX;\BC)\quad \mbox{and}\quad f_2\in\mathcal{A}^{\ell+\frac72-\epsilon}(\CX,\BC).
\]  
Now we solve the following two equations $N\tilde{u}_k^j=f_j, j=1,2$. First, using the Mellin transform as before, we see that 
\[
\tilde{u}_k^2=\rho^{-k+1}u_{k+1}+\tilde{u}_k\quad{where}\quad u_{k+1}\in\begin{cases}
	&\BFS_{k-1},\quad k-1\geq 0\\
	&\BFS_{-k},\quad k-1\leq-1
\end{cases}\quad \mbox{and}\quad \tilde{u}_{k+1}\in\mathcal{A}^{\ell+\frac72-\epsilon}(\CX;\BC).
\]
As for $u_k^1$, we can solve $N\tilde{u}_k^1=f_1$ more directly. Since any function in $C^\infty(\pa_+\CX)$ can be expressed as a linear combination of $\BFS_l$, it suffices to solve $N\tilde{u}_k^1=f_1$ for $f_1\in\rho^{-k}\BFS_l$. More explicitly, 
\[ 
u_k^1=
\begin{cases}
(k(k+1)-l(l+1))^{-1}f_1,\quad &k(k+1)\neq l(l+1)\\
-(2k+1)^{-1}f_1\ln\rho,\quad&k(k+1)=l(l+1)
\end{cases} .  
\]
Therefore, we have
\[
\chi u=\rho^{-k}u_k+u_k^1+\rho^{-k+1}u_{k+1}+\tilde{u}_{k+1}.
\]
Plugging the partial expansion into the equation \eqref{eq:eqnforchiu} and proceeding as above iteratively give the full polyhomogeneous expansion. 

\item\underline{Proof of statement (2).} Near $\pa_+\CX$, according to the elliptic estimate for b-pseudodifferential operator, we conclude that $u\in\bar{H}^{\infty,\ell}_\bop$ there. Away form spatial infinity $\pa_+\CX$, since $u$ is $0$ in the region $r<\ehKN$, by Proposition \ref{prop:microlocalglobaldy}, propagation of singularity estimate and elliptic estimate we see that $u$ is smooth away from the radial points at event horizon. Finally, using the low regularity radial point estimate yields $u\in\dot{H}^{\frac12-,\ell}_\bop(\CX;\BC)$. The polyhomogeneous expansion of $u$ near $\pa_+\CX$ can be obtained as in the proof of statement (1) by exploiting the normal operator argument.

\item\underline{Proof of statement (3).} Since $s+\ell>-\frac12$, according to the high sc-decay estimate at the radial point $R(\sigma)$, propagation of singularity estimate and elliptic estimate, we see that away from the radial point $R(0)$, $u$ is in $\bar{H}^{\infty, \ell}_\bop(\CX;\BC)$. Near $R(0)$ the low b-decay radial point estimate gives $u\in\bar{H}^{\infty,\min\{-\frac12-, \ell\}}_\bop(\CX;\BC)$ at $R(0)$. Therefore, we have $u\in\bar{H}^{\infty,\min\{-\frac12-, \ell\}}_\bop(\CX;\BC)$. Noe since $\sigma\neq 0$, we have $
\widehat{\Box_{g_{b}}}(\sigma)=-2i\sigma\rho(\rho\pa_\rho-1)+\rho^2\mbox{Diff}_\bop^2$, and thus $N(\widehat{\Box_{g_{b}}}(\sigma))=-2i\sigma\rho(\rho\pa_\rho-1)$. Then using the typical normal operator argument as in the proof of statement (1), we obtain that $u=\rho C^\infty(\pa_+\CX;\CX)+\mathcal{A}^{2-}(\CX;\BC)$.

\item\underline{Proof of $\widehat{\mathcal{P}_{b, \gamma}}(\sigma), \widehat{\mathcal{W}_{b,\gamma}}$ and $\widehat{L_{b,\gamma}}(\sigma)$.} Near the spatial infinity, we work in the standard coordinate trivialization of $\scform$ for $\widehat{\mathcal{P}_{b, \gamma}}(\sigma), \widehat{\mathcal{W}_{b,\gamma}}$ (which is given in terms of the differentials $dt, dx^1, dx^2, dx^3$ of the standard coordinates) and $\scsym\oplus\scform$ for $\widehat{L_{b,\gamma}}(\sigma)$ (which is given in terms of the differentials $dt^2, dtdx^i, dx^idx^j, dt,dx^i1\leq j\leq i\leq 3$ of the standard coordinates). 

Then according to Propositions \ref{PropKNStBox} and \ref{PropFourierPWL}, the normal operator of $2\widehat{\mathcal{P}_{b, \gamma}}(\sigma), 2\widehat{\mathcal{W}_{b,\gamma}}$ (resp. $\widehat{L_{b,\gamma}}(\sigma)$) is, $-\rho^2\Big(\rho\pa_\rho(\rho\pa_\rho-1)+\sL\Big)$ for $\sigma=0$ and $2i\sigma\rho(\rho\pa_\rho-1)$ for $\sigma\neq 0$, tensored with $4\times 4$ identity matrix (reps. $14\times 14$ identity matrix). Therefore, the proof for these operators follows in a completely analogous manner as before.
	\end{itemize}
\end{proof}

%%%%%%%%%%%%%%%%%%%%%%%%%%%%%%%%%%%%%%%%%%%%%%%%%%%%%%%%%%%%%%%%%%%%%%%%%%%%%%%%%
\subsection{Semiclassical behavior of Kerr-Newman spacetimes}\label{subsec:semiclassicalanalysis}
Recall that 
\[
\widehat{\Box_{g}}(\sigma):=e^{it_{b,*}\sigma}\Box_{g_b}e^{-it_{b,*}\sigma}.
\]
where the inverse metric $g^{-1}$ is given by
\begin{equation}\label{eq:inversemetricsemi}
	\begin{split}
	g^{-1}&=\rho_b^{-2}\bigg(\Delta_b\pa_r^2+\frac{\chi^2-1}{\Delta_b}\Big((r^2+a^2)\pa_{t_\chi}+a\pa_\varphi\Big)^2+\pa_\theta^2+\frac{1}{\sin^2\theta}\Big(\pa_\varphi+a\sin^2\theta\pa_{t_{\chi}}\Big)^2\\
	&\qquad\qquad+2\chi\Big((r^2+a^2)\pa_{t_\chi}+a\pa_\varphi\Big)\pa_r\bigg).
	\end{split}
\end{equation}
We introduce the semiclassically rescaled version of the operator $\widehat{\Box_{g}}(\sigma)$:
\begin{equation}
h^2	\widehat{\Box_{g}}(h^{-1}z),\quad h=\frac{1}{\abs{\sigma}}\quad \mbox{and}\quad z=\frac{\RE\sigma}{\abs{\sigma}}+i\frac{\IM\sigma}{\abs{\sigma}}.
	\end{equation}
When $\IM\sigma$ is bounded, we write $z=z_{\mathrm{R}}+ihz_{\mathrm{I}}$ and parametrize $h^2\widehat{\Box_{g}}(h^{-1}z)$ by $z_{\mathrm{R}}$ and $z_{\mathrm{I}}=h^{-1}\IM z$ rather than $\RE z$ and $\IM z$. This still gives a family of operators in $\mbox{Diff}^2_{\scop,h}$ which depends smoothly on $z_{\mathrm{R}}, z_{\mathrm{I}}$, and its semiclassical principal symbol does not depend on $z_{\mathrm{I}}$. Therefore, when analyzing the semiclassical principal symbol of $h^2\widehat{\Box_{g}}(h^{-1}z)$, it suffices to consider the case $z\in\BR$.

For $z\in\BR$, the semiclassical principal symbol of $h^2\widehat{\Box_{g}}(h^{-1}z)$ is given by
\begin{equation}\label{eq:semisymbol}
	p_{h,z}(\xi):=\sigma_h(h^2\widehat{\Box_{g}}(h^{-1}z))=-g^{-1}(-zdt_{b,*}+\xi,-zdt_{b,*}+\xi),\quad \xi\in\pa_{t_{b,*}}^\perp=\fscform.
\end{equation}
As before, away from $\pa_+\CX$, we write the scattering covectors as
\[
\xi:=\xi_r dr+\xi_\theta d\theta+\xi_\varphi d\varphi.
\]
While near $\pa_+\CX$, we write them as
\[
\xi:=\xi_\rho\frac{d\rho}{\rho^2}+\xi_{\theta}\frac{d\theta}{\rho}+\xi_{\varphi}\frac{d\varphi}{\rho},\quad \rho=\frac{1}{r}.
\]
\subsubsection{Characteristic set}
We first study the characteristic set 
\[
\Sigma_{h}:=\{\jb{\xi}^{-2}p_{h,z}(\xi)=0\}\subset\cscform,\quad \jb{\xi}=\sqrt{1+\xi_r^2+\xi_\theta^2+\frac{1}{\sin^2\theta}\xi_\varphi^2}.
\]
We will show that for $z\in\BR\setminus 0$, it splits into two components. To achieve this, we make use of an auxiliary covector $d\mathfrak{t}$ which is timelike everywhere on $\CX$. 
%The construction of $d\bar{t}$ is as follows. Let $\tilde{t}=t_{\chi}+b(r)$ and calculate
%\[
%g^{-1}(d\mathfrak{t}, d\mathfrak{t})=\rho_b^{-2}\Big(\frac{\chi^2-1}{\Delta_b}(r^2+a^2)^2+a^2\sin^2\theta+(b'(r))^2\Delta_b+2\chi b'(r)(r^2+a^2)\Big).
%]
%Then we choose $b(r)$ such that $G(d\tilde{t}, d\tilde{t})<0$; more specifically, $G(d\tilde{t}, d\tilde{t})=-1$ for $r\leq 2\Bm$ while $G(d\tilde{t}, d\tilde{t})=G(dt_\chi, dt_\chi)<0$ for $r\geq 4\Bm$.
\begin{lem}[Splitting of the characteristic set]\label{lem:charasemi}
	There exist closed $z$-dependent sets $\Sigma_h^{\pm}\subset\cscform$ such that
	\begin{equation}
		\{\jb{\xi}^{-2}p_{h,z}(\xi)=0\}=\Sigma_{h}^+\sqcup\Sigma_{h}^-,\quad z\in\BR\setminus 0.
	\end{equation} 
Moreover, 
\begin{equation}\label{eq:splitofcha}
	\pm\jb{\xi}^{-1}g^{-1}(-zdt_{b,*}+\xi, d\mathfrak{t})>0\quad \mbox{on}\quad \Sigma_{h}^{\pm}\quad\mbox{for}\quad z\in\BR\setminus 0.
\end{equation}
Finally, 
\begin{align}\label{eq:charaisemptyatinfty}	(\Sigma_{h}^{\pm}\cap\fiscform)\cap\{\Delta_b-a^2\sin^2\theta>0\}&=\emptyset\quad \mbox{for}\quad  z\in\BR;\\\label{eq:onecomisempty}
	\Sigma_{h}^{\pm}\cap\{\Delta_b-a^2\sin^2\theta>0\}&=\emptyset\quad \mbox{for}\quad \mp z>0.
\end{align}
	\end{lem}
\begin{proof}
	We put
	\[
	\Sigma_{h}^{\pm}:=\{\jb{\xi}^{-2}p_{h,z}(\xi)=0\}\cap \{\pm\jb{\xi}^{-1}g^{-1}(-zdt_{b,*}+\xi, d\mathfrak{t})\geq0\}.
	\]
	It is clear that $\Sigma_{h}^{\pm}$ are closed and their union is the characteristic set. In order to show that $\Sigma_h^+\cap\Sigma_h^-=\emptyset$ and \eqref{eq:splitofcha} for $z\in\BR\setminus0$, we need 
	\[
	\{\jb{\xi}^{-2}p_{h,z}(\xi)=0\}\cap \{\jb{\xi}^{-1}g^{-1}(-zdt_{b,*}+\xi, d\mathfrak{t})=0\}=\emptyset\quad \mbox{for}\quad z\in\BR\setminus 0.
	\]
On the fiber infinity $\fiscform$, $\jb{\xi}^{-1}g^{-1}(\xi, d\mathfrak{t})=0$ implies that $\jb{\xi}^{-1}\xi$ is spacelike or $0$ as $d\mathfrak{t}$ is timelike. If the intersection were not empty at $\fiscform$, $\jb{\xi}^{-1}\xi$ must be $0$, which is impossible. Also, the fact that $\jb{\xi}^{-2}p_{h,z}(\xi)=\jb{\xi}^{-2}p(\xi)=0$ gives \eqref{eq:charaisemptyatinfty}.

In the interior $\fscform$, if the intersection were not empty, since $d\mathfrak{t}$ is timelike, $g^{-1}(-zdt_{b,*}+\xi, d\mathfrak{t})=0$ implies that $-zdt_{b,*}+\xi$ must be spacelike or $0$. Then the equality $g^{-1}(-zdt_{b,*}+\xi, -zdt_{b,*}+\xi)=0$ forces $-zdt_{b,*}+\xi$ to be $0$, which cannot happen for $\xi\in\pa_{t_{b,*}}^\perp$ and $z\in\BR\setminus 0$.

On the region $\{\Delta_b-a^2\sin^2\theta>0\}$, we see that $g^{-1}(\xi, \xi)>0$ for $0\neq \xi\in\fscform$. For any $p\in\{\Delta_b-a^2\sin^2\theta>0\}$, let $\{d\mathfrak{t}, X_1, X_2, X_3\}$ be an orthogonal basis of $T^*_p\CM$ with $G|_p(X_i, X_i)=1$. We write \[-zdt_{b,*}+\xi=(a_0-z)d\mathfrak{t}+\sum_{i=1}^3a_i X_i,
\]
and then $g^{-1}(-zdt_{b,*}+\xi,-zdt_{b,*}+\xi)=0$ gives $(a_0-z)^2G(d\mathfrak{t}, d\mathfrak{t})+\sum_{i=1}^3a_i^2=0$. Since $-zdt_{b,*}+\xi+zd\mathfrak{t}\in T^*_p\CX$, it follows that $g^{-1}(-zdt_{b,*}+\xi+zd\mathfrak{t},-zdt_{b,*}+\xi+zd\mathfrak{t})=a_0^2G(d\mathfrak{t}, d\mathfrak{t})+\sum_{i=1}^3a_i^2>0$. Therefore, we have $z^2-2a_0z>0$. Then 
\[
\pm g^{-1}(-zdt_{b,*}+\xi, d\mathfrak{t})=\pm(a_0-z)g^{-1}(d\mathfrak{t}, d\mathfrak{t})=\pm(a_0-\frac{z}{2}-\frac{z}{2})g^{-1}(d\mathfrak{t}, d\mathfrak{t})>0\quad \mbox{for}\quad \pm z>0.
\]
This finishes the proof of \eqref{eq:onecomisempty}.
\end{proof}

\subsubsection{The radial points at event horizon}Now in the fiber radial compactification $\cscform$ of $\fscform$, we use the following coordinates near the fiber infinity $\fiscform$,
\begin{equation}\label{eq:coordatfiberinftysemi}
	\rho_\xi=(\xi_r^2+\xi_\theta^2+\frac{1}{\sin^2\theta}\xi_\varphi^2)^{-\frac 12}\in[0, \infty), \quad \hat{\xi}=(\hat{\xi}_r,\hat{\xi}_\theta, \hat{\xi}_\varphi)=\rho_\xi(\xi_r, \xi_\theta, \xi_\varphi).
\end{equation}
We note that $\rho_\xi$ is a boundary defining function of $\cscform$, i.e., $\rho_\xi=0$ at $\fiscform$, and \[\hat{\xi}\in\fiscform=\{\hat{\xi}_r^2+\hat{\xi}_\theta^2+\frac{1}{\sin^2\theta}\hat{\xi}_\varphi^2=1\}. 
\]
We now analyze the behavior of the Hamiltonian flow $\exp(s\rho_\xi H_{p_{h,z}})$ on the characteristic set $\Sigma_h=\Sigma_h^+\cup\Sigma_h^-$, concentrating on the behavior away from $\pa_+\CX$. With $t_{b,*}=t_{\chi}+c(r)$, we write
\begin{equation}\label{eq:symbolsemi}
	p_{h,z}=-\rho_b^{-2}\Big(\Delta_b\big(\xi_r-zc'(r)+\frac{\chi\big(a\xi_\varphi-(r^2+a^2)z\big)}{\Delta_b}\big)^2
-\frac{1}{\Delta_b}\big(a\xi_\varphi-(r^2+a^2)z\big)^2+\tilde{p}_{h,z}\Big)
\end{equation}
where
\begin{equation} \tilde{p}_{h,z}=\xi_\theta^2+\frac{1}{\sin^2\theta}(\xi_\varphi-a\sin^2\theta z)^2.
\end{equation} 
Then we calculate
\begin{align*}
H_{p_{h,z}}&=-\rho_b^{-2}\bigg(\Big(2\Delta_b\big(\xi_r-zc'(r)\big)+2\chi\big(a\xi_\varphi-(r^2+a^2)z\big)\Big)\pa_r+\Big(2a\frac{\chi^2-1}{\Delta_b}\big(a\xi_\varphi-(r^2+a^2)z\big)+2\chi a\big(\xi_r-c'(r)z\big)\Big)\pa_\varphi\\
&\qquad\qquad-\frac{\pa(-\rho_b^2 p_{h,z})}{\pa r}\pa_{\xi_r}+H_{\tilde{p}_{h,z}}\bigg)-\rho_b^{-2}p_{h,z}H_{\rho_b^{2}}
\end{align*}
where
\begin{align*}
	\frac{\pa(-\rho_b^2 p_{h,z})}{\pa r}&=2\Delta_b\Big(\xi_r-zc'(r)+\frac{\chi\big(a\xi_\varphi-(r^2+a^2)z\big)}{\Delta_b}\Big)\pa_r\Big(\xi_r-zc'(r)+\frac{\chi\big(a\xi_\varphi-(r^2+a^2)z\big)}{\Delta_b}\Big)\\
	&\qquad +\frac{\pa\Delta_b}{\pa r}\big(\xi_r-zc'(r)+\frac{\chi\big(a\xi_\varphi-(r^2+a^2)z\big)}{\Delta_b}\big)^2-\pa_r\Big(\frac{1}{\Delta_b}\big(a\xi_\varphi-(r^2+a^2)z\big)^2\Big)
\end{align*}
In the coordinates \eqref{eq:coordatfiberinftysemi} near the fiber infinity $\fiscform$, we compute
\begin{equation}\label{eq:CrescaledHamilsemi}
	\begin{split}
		\rho_\xi H_{p_{h,z}}&=-\rho_b^{-2}\bigg(\Big(2\Delta_b\big(\hat{\xi}_r-\rho_\xi zc'(r)\big)-2\rho_\xi\chi(r^2+a^2)z+2\chi a\hat{\xi}_\varphi\Big)\pa_r\\
		&\qquad\qquad+\Big(2a\frac{\chi^2-1}{\Delta_b}\big(a\hat{\xi}_\varphi-\rho_\xi(r^2+a^2)z\big)+2\chi a\big(\hat{\xi}_r-\rho_\xi c'(r)z\big)\Big)\pa_\varphi\\
		&\qquad\qquad\quad-\rho_\xi\frac{\pa(-\rho_b^2 p_{h,z})}{\pa r}\pa_{\xi_r}+\rho_\xi H_{\tilde{p}_{h,z}}\bigg)-\rho_b^{-2}(\rho_\xi^2p_{h,z})\rho_\xi^{-1}H_{\rho_b^{2}}.
	\end{split}
\end{equation}
We let
\[
L_\pm:=\{\rho_\xi=0,  \hat{\xi}_r=\pm1,(\hat{\xi}_\theta, \hat{\xi}_\varphi)=0,\Delta_b=0\}\subset\Big(\Sigma_{h}^{\pm}\cap\fiscform\Big).
\]

Now we describe the property of the set $L_\pm$. 
\begin{lem}\label{lem:radialpointatehsemi}
	$L_\pm$ are invariant under the Hamiltonian flow $\exp(s\rho_\xi H_{p_{h,z}})$ on the characteristic set $\Sigma_{h}$. Moreover, $L_+$ is a radial sink and $L_-$ is a radial source for the flow $\exp(s\rho_\xi H_{p_{h,z}})$.
\end{lem}
\begin{proof}
We rewrite
	\[
	L_\pm=\{\Delta_b=0, \rho_\xi=0,\hat{\xi}_r=\pm 1, \rho_\xi^2\tilde{p}_{h,z}=0\}.
	\]
	Using the coordinates \eqref{eq:coordatfiberinftysemi}, the expression \eqref{eq:CrescaledHamilsemi} and the facts that $c(r)=0, \chi=1$ near $L_\pm$, we find that near $L_\pm$
	\begin{equation}\label{eq:deriof4quantitiessemi}
		\begin{split}
			\rho_\xi H_{p_{h,z}}\rho_\xi&=-\rho_b^{-2}\Big(\frac{\pa\Delta_b}{\pa r}\hat{\xi}_r^3-4r\rho_\xi\hat{\xi}_r^2z+H_{\tilde{p}_{h,z}}\rho_\xi\Big)\rho_\xi-\rho_b^{-2}(\rho_\xi^2p_{h,z})\rho_\xi^{-1}H_{\rho_b^{2}}\rho_\xi,\\
				\rho_\xi H_{p_{h,z}}\hat{\xi}_\varphi&=(H_{p_{h,z}}\rho_\xi)\hat{\xi}_\varphi,\\
			\rho_\xi H_{p_{h,z}}\Delta_b&=-2\rho_b^{-2}\frac{\pa\Delta_b}{\pa r}\Big(\Delta_b\hat{\xi}_r-\rho_\xi(r^2+a^2)z+a\hat{\xi}_\varphi\Big),\\
			\rho_\xi H_{p_{h,z}}(\rho_\xi^2\tilde{p}_{h,z})
			&=2(H_{p_{h,z}}\rho_\xi)\rho_\xi^2\tilde{p}_{h,z}-\rho_b^{-2}(\rho_\xi^2p_{h,z})\rho_\xi H_{\rho_b^{2}}\tilde{p}_{h,z}.
		\end{split}
	\end{equation}
	This implies that $\rho_\xi H_p$ is tangent to $L_\pm$, so $L_\pm$ is invariant under the flow $\exp(s\rho_\xi H_{p_{h,z}})$. Moreover, in a neighborhood of $L_\pm=\{\Delta_b=0, \rho_\xi=0,\hat{\xi}_r=\pm 1, \rho_\xi^2\tilde{p}_{h,z}=0\}\subset\{\rho_\xi^2p_{h,z}=0\}$, we have
	\begin{equation}\label{eq:deriof3boundarydefiningsemi}
		\begin{split}
			\pm	\rho_\xi H_p\rho_\xi^2&\leq -C_0\rho_\xi^2,\\
				\pm	\rho_\xi H_p\hat{\xi}_\varphi^2&\leq -C_0\hat{\xi}_\varphi^2,\\
			\pm	\rho_\xi H_p\Delta_b^2&\leq -C_0\Delta_b^2+C_1(\rho_\xi^2+\hat{\xi}_\varphi^2)\\
			\pm	\rho_\xi H_p(\rho_\xi^2\tilde{p})&\leq -C_0\rho_\xi^2\tilde{p}+C_1(\rho_\xi^2+\Delta_b^2+\hat{\xi}_\varphi^2)
		\end{split}
	\end{equation}
	where $C_0, C_1>0$. It follows from \eqref{eq:deriof3boundarydefining} that in a sufficiently small neighborhood of $L_{\pm}$
	\begin{equation}
		\pm\rho_\xi H_pf\leq -Cf,\quad f=\rho_\xi^2+\hat{\xi}_\varphi^2+C_2\Delta_b^2+	C_3\rho^2_\xi\tilde{p}\geq 0.
	\end{equation}
	for some $C, C_2, C_3>0$. Proceeding as in the proof of Lemma \ref{lem:radialpointateh} finishes the proof.
\end{proof}
Finally, since $H_{p_{h,z}}\rho_\xi=H_p\rho_\xi$ and 
\begin{equation}\label{eq:semithresholdreg}
	\begin{split}
\rho_\xi\sigma_h\Big(h^{-1}\IM h^{2}\widehat{\Box_{g_{b}}}(h^{-1}z)\Big)&=\IM(h^{-1}z)\rho_b^{-2}\Big(2(r^2+a^2)\hat{\xi}_r+2a\hat{\xi}_\varphi)\Big)\\
&=\rho_\xi\sigma_1\Big(\IM \widehat{\Box_{g_{b}}}(h^{-1}z)\Big)\quad \mbox{at}\quad L_\pm,
\end{split}
\end{equation}
the calculation of the threshold regularity at $L_\pm$ in the microlocal case applies here as well.

\subsubsection{The radial points at spatial infinity $\pa_+\CX$}
We now turn to the analysis near $\pa_+\CX$. Recall that near $\pa_+\CX$, the semiclassical principal symbol of $h^{-2}\widehat{\Box_{g}}(h^{-1}z)$ for $z\in\BR$ is given by 
\begin{equation}\label{eq:semisymatinfty}
	\begin{split}
		p_{h,z}(\zeta)&=-g^{-1}(-zdt_{b,*}+\xi,-zdt_{b,*}+\xi)\\
		&=-(\xi_\rho-z)^2+z^2-\tilde{p}+\rho\sum_{\substack{0\leq j,k,m\leq2\\0\leq j+k+m\leq 2}}a_{jkm}(\rho, \theta,\varphi)\xi_\rho^j\xi_\theta^k\xi_\varphi^m,\quad \tilde{p}=\Big(\xi_\theta^2+\frac{1}{\sin^2\theta}\xi_\varphi^2\Big)
	\end{split}
\end{equation}
where $\xi=\xi_\rho\frac{d\rho}{\rho^2}+\xi_\theta\frac{d\theta}{\rho}+\xi_\varphi\frac{d\varphi}{\rho}$ and $a_{jkm}\in\mathcal{A}^0(\CX)$. Then the scattering Hamiltonian vector field associated to $p_{h,z}$ in $\rho>0$ is given by
\begin{align*}
{}^{\scop}\!H_{p_{h,z}}
	&=\rho\Big(\frac{\pa p_{h,z}}{\pa\xi_\rho}\big(\rho\frac{\pa}{\pa\rho}+(\sum_{\mu=\theta, \varphi}\xi_\mu\frac{\pa}{\pa\xi_\mu})\big)-\big(\rho\frac{\pa p_{h,z}}{\pa\rho}+(\sum_{\mu=\theta, \varphi}\xi_\mu\frac{\pa p_{h,z}}{\pa\xi_\mu})\big)\frac{\pa}{\pa\xi_\rho}+\sum_{\mu=\theta,\varphi}\big(\frac{\pa p_{h,z}}{\pa\xi_\mu}\frac{\pa}{\pa\mu}-\frac{\pa p_{h,z}}{\pa \mu}\frac{\pa}{\pa\xi_\mu}\big)\Big).
\end{align*}
We introduce the rescaled scattering Hamiltonian vector field (here we drop the factor $\jb{\rho_\xi}$ again)
\begin{equation}\label{eq:rescaledscatteringHamil}
	\begin{split}
	{}^{\scop}\!H_{p_{h,z}}^{2,0}&:=\rho^{-1}{}^{\scop}\!H_{p_{h,z}}\\
	&=(-2\xi_\rho+2z)\rho\pa_\rho+2\tilde{p}\pa_{\xi_\rho}+(-2\xi_\rho+2z)\sum_{\mu=\theta,\varphi}\xi_\mu\pa_{\xi_\mu}-H_{\tilde{p}}+\rho V, \quad V\in\mathcal{V}_\bop(\fscform).
	\end{split}
\end{equation}

We now discuss the integral curves of ${}^{\scop}\!H_{p_{h,z}}^{2,0}$ on the semiclassical characteristic set $\Sigma_h$ for $z\in\BR\setminus0$ near $\pa_+\CX$. 
\begin{lem}\label{lem:radialpointatinftysemi}
	For $z\in\BR\setminus 0$, we put $\phi_s(\rho,\theta, \varphi,\xi)=\exp(s{}^{\scop}\!H_{p_{h,z}}^{2,0})(\rho,\theta, \varphi,\xi)$. Then there exists a neighborhood $V_\pm\subset\fscform$ of 
	\[
	R(\frac{z}{2}\pm\frac{\abs{z}}{2})=\{\rho=0, \xi=(z\pm\abs{z})\frac{d\rho}{\rho^2}\}	
	\]
	such that uniformly in $(\rho, \theta, \varphi,\xi)\in V_\pm\cap\Sigma_h$, one has
	\[
\phi_s(\rho, \theta, \varphi, \xi)\to 	R(\frac{z}{2}\pm\frac{\abs{z}}{2})\quad \mbox{as}\quad s\to\pm\infty.
	\]
	This implies that $R(\frac{z}{2}+\frac{\abs{z}}{2})$ is a radial  sink while  $R(\frac{z}{2}-\frac{\abs{z}}{2})$ is a radial source.
\end{lem}

\begin{proof}
	Recall that
\[
R(\frac{z}{2}\pm\frac{\abs{z}}{2})=\{ \rho=0, \tilde{p}=0, \xi_\rho=z\pm\abs{z}\}.
\]
Using the expression \eqref{eq:rescaledscatteringHamil}, we find that in a neighborhood $U^{2\epsilon}_\pm=\{(\xi_\rho-(z\pm\abs{z}))^2+\rho^2+\tilde{p}<2\epsilon\}$ of $R(\frac{z}{2}\pm\frac{\abs{z}}{2})$ where $\epsilon>0$ is a sufficiently small constant,
\begin{equation}\label{eq:deriof2quantities}
	\begin{split}
	\pm	{}^{\scop}\!H_{p_{h,z}}^{2,0}\rho^2&\leq -C_0\rho^2,\\
		\pm	{}^{\scop}\!H_{p_{h,z}}^{2,0}\tilde{p}&\leq -C_0\tilde{p}
	\end{split}
\end{equation}
 for some $C_0>0$. According to the expression \eqref{eq:semisymatinfty} of $p_{h,z}$, there exists $\epsilon'>0$ such that \[\{\rho^2+\tilde{p}<\epsilon'\} \cap\Sigma_h\subset U^\epsilon_+\cup U^\epsilon_- 
 \]
  where $U_\pm^\epsilon$ are neighborhoods of $R(\frac{z}{2}\pm\frac{\abs{z}}{2})$ satisfying $U^\epsilon_+\cap U^\epsilon_-=\emptyset$. Then for $(\rho, \theta,\varphi, \xi)\in U^\epsilon_\pm\cap\{\rho^2+\tilde{p}<\epsilon'\} \cap\Sigma_h$, since $p_{h,z}$ is a constant along the integral curves of ${}^{\scop}\!H_{p_{h,z}}^{2,0}$, we have 
 \begin{equation}\label{eq:intercon}
 	\phi_s(\rho, \theta,\varphi, \xi)\subset U^\epsilon_{\pm}\cap\Sigma_h\quad \mbox{for}\quad \pm s\geq 0.
 \end{equation}
Indeed, if \eqref{eq:intercon} failed, we could find $\pm s_0>0$ such that 
\[
\phi_{s_0}(\rho, \theta,\varphi, \xi)\in U^{2\epsilon}_{\pm}\quad \mbox{and}\quad 	\pm	{}^{\scop}\!H_{p_{h,z}}^{2,0}(\rho^2+\tilde{p})(\phi_{s_0}(\rho, \theta,\varphi, \xi))\geq 0
\]
which contradicts \eqref{eq:deriof2quantities}. Proceeding as in the proof of Lemma \ref{lem:radialpointateh}, it follows that 
uniformly in $(\rho, \theta, \varphi,\xi)\in U^\epsilon_\pm\cap \Sigma_h$, one has
\[
\phi_s(\rho, \theta, \varphi, \xi)\to 	\{\rho=0, \tilde{p}=0\}\cap (U^\epsilon_\pm\cap \Sigma_h)=R(\frac{z}{2}\pm\frac{\abs{z}}{2}) \quad\mbox{as}\quad s\to\pm\infty.
\]
\end{proof}
Finally, since $p_{h,z}$ at $\pa_+\CX$ equals to $p_{\scop}(z)$, the calculation of the threshold scattering decay order at the radial points in the microlocal case applies here as well.

\subsubsection{Structure of the trapped sets}
We now study the structure of the trapped sets for KN spacetime with $\abs{\Ba}^2+\BQ	^2<\Bm^2$. First, we define and investigate the incoming and outgoing tails and the trapped set associated to the Hamiltonian flow $\exp(s{}^\scop\!H^{2,0}_{p_{h,z}})$ where $p_{h,z}$ is the semiclassical symbol given by 
\[
	p_{h,z}=-\rho_b^{-2}\Big(\Delta_b\big(\xi_r-zc'(r)+\frac{\chi\big(a\xi_\varphi-(r^2+a^2)z\big)}{\Delta_b}\big)^2
	-\frac{1}{\Delta_b}\big(a\xi_\varphi-(r^2+a^2)z\big)^2+\tilde{p}_{h,z}\Big)
\]
where
\[ \tilde{p}_{h,z}=\xi_\theta^2+\frac{1}{\sin^2\theta}(\xi_\varphi-a\sin^2\theta z)^2.
\]
To define them, we need an escape function.
\begin{prop}\label{prop:escapefcn}
	Let $f(r)\in C^\infty([r_-,\infty))$ be defined as \begin{equation}
		f(r)=\frac{\Delta_b(r)}{r^4}.
		\end{equation}
	Then there exists $\delta_0=\delta_0(\Bm,\Ba,\BQ)>0$ such that for $(r, \theta, \varphi,\xi)\in\cscform\setminus (L^\pm\cup\pa_+\CX)$ and $z\in\BR\setminus 0$
	\begin{equation}\label{eq:escapefcn}
		f(r)<\delta_0,\quad \jb{\xi}^{-2}p_{h,z}(r, \theta, \varphi,\xi)=0,\quad  \jb{\xi}^{-1} H_{p_{h,z}}f(r, \theta, \varphi,\xi)=0\Longrightarrow ( \jb{\xi}^{-1} H_{p_{h,z}})^2f(r, \theta, \varphi,\xi)<0.
	\end{equation}
\end{prop} 
\begin{proof}
	We first show \eqref{eq:escapefcn} on the fiber infinity $\fiscform$ where $\jb{\xi}^{-1}=0$. According to \eqref{eq:charaisemptyatinfty} in Lemma \ref{lem:charasemi}, we know that
	\[\{\jb{\xi}^{-2}p_{h,z}=0\}\cap\fiscform\subset\{r\geq r_-\mid\Delta_b\leq a^2\sin^2\theta\}\subset\{r_-\leq r\leq \Bm+\sqrt{\Bm^2-\BQ^2}\}.
	\]
	A direct calculation gives 
	\[
		f'(r)=\frac{r\pa_r\Delta_b(r)-4\Delta_b(r)}{r^5}
	=-\frac{2r^2-6\Bm r+4a^2+4\BQ^2}{r^5}>0\quad \quad\mbox{when}\quad r\in[r_-,\Bm+\sqrt{\Bm^2-\BQ^2}].	
	\]
	Therefore, $\jb{\xi}^{-1}H_{p_{h,z}}f=0$ implies $\jb{\xi}^{-1}H_{p_{h,z}}r=0$, that is, 
	\[
	\jb{\xi}^{-1}H_{p_{h,z}}r=-\rho_b^{-2}\jb{\xi}^{-1}\Big(2\Delta_b\big(\xi_r-zc'(r)\big)-2\chi(r^2+a^2)z+2\chi a\xi_\varphi\Big)=0\quad \mbox{on}\quad \fiscform\cap\Sigma_h.
	\]
	This is equivalent to 
	\[
	\Delta_b\hat{\xi}_r=-\chi a\hat{\xi}_\varphi
	\]
	where $\hat{\xi}$ is defined as in \eqref{eq:coordatfiberinftysemi}. Plugging this back to $\jb{\xi}^{-2}p_{h,z}=0$ yields
	\[
	\jb{\xi}^{-2}p_{h,z}=-\rho_b^{-2}\Big(-\Delta_b\hat{\xi}_r^2+\frac{\chi^2-1}{\Delta_b}a^2\hat{\xi}_\varphi^2+\hat{\xi}_\theta^2+\frac{1}{\sin^2\theta}\hat{\xi}_\varphi^2\Big)=0.
	\]
	Since $\chi=1$ on $[r_-, \ehKN]$, it follows that \[\{\jb{\xi}^{-2}p_{h,z}=0,\jb{\xi}^{-1}H_{p_{h,z}}f=0\}\cap\fiscform=L_\pm.
	\] 
	On $\{\jb{\xi}^{-2}p_{h,z}=0,\jb{\xi}^{-1}H_{p_{h,z}}f=0, r>\ehKN\}\cap\fiscform$, we calculate
	\[
	(\jb{\xi}^{-1}H_{p_{h,z}})^2f=-2\rho_b^{-2}f'(r)\Delta_b\jb{\xi}^{-2}H_{p_{h,z}}\xi_r=-\frac{2f'(r)}{\rho_b^4\Delta_b}\frac{\pa\Delta_b}{\pa r}a^2\hat{\xi}_\varphi^2<0
	\]
	where we use the fact that $\jb{\xi}^{-1}H_{p_{h,z}}\jb{\xi}^{-1}=0$ on $\fiscform$.

	We next prove that in the region $r_-\leq r\leq\ehKN$ where $f(r)\leq 0$, $\jb{\xi}^{-2}p_{h,z}$ and $\jb{\xi}^{-1}H_{p_{h,z}}f$ cannot vanish at the same time in the interior $\fscform$ (i.e. away from the fiber infinity). Suppose that $H_{p_{h,z}}f=0$ and $p_{h,z}=0$ in $r_-\leq r\leq \ehKN$. Then we have
	\[
	H_{p_{h,z}}f=-\rho_b^{-2}f'(r)\Big(2\Delta_b\big(\xi_r-zc'(r)\big)-2\chi(r^2+a^2)z+2\chi a\xi_\varphi\Big)=0.
	\]
	Since $f'(r)>0$ when $r\in[r_-,\ehKN]$, $H_{p_{h,z}}f=0$ implies
\[
\Delta_b\big(\xi_r-zc'(r)\big)=\chi(r^2+a^2)z-\chi a\xi_\varphi.
\]
Plugging this into $p_{h,z}=0$ and using the fact that $\chi=1$ in $[r_-, \ehKN]$ yield
\[
-\rho_b^{-2}\Big(-\Delta_b\big(\xi_r-zc'(r)\big)^2+\tilde{p}_{h,z}\Big)=0\quad\mbox{where}\quad \tilde{p}_{h,z}=\xi_\theta^2+\frac{1}{\sin^2\theta}(\xi_\varphi-a\sin^2\theta z)^2,
\]
and thus
\[
\Delta_b(\xi_r-c'(r)z)=0,\quad\xi_\theta=0, \quad \xi_\varphi=a\sin^2\theta z\quad \mbox{in} \quad r\in[r_-, \ehKN].
\]
Therefore, for $z\in\BR\setminus0$
\[
H_{p_{h,z}}f=\rho_b^{-2}f'(r)(2r^2z+2a^2\cos^2\theta z)\neq0\quad \mbox{in} \quad r\in[r_-, \ehKN],
\]
which is a contradiction.

We now turn to the region $r>\ehKN$ where $f>0$. We return to the analysis of 
\[
f'(r)=-\frac{2r^2-6\Bm r+4a^2+4\BQ^2}{r^5}.
\]
It is clear that $f'(r)<0$ when $r>3\Bm$ and $f'(r)>0$ when $\ehKN=\Bm+\sqrt{\Bm^2-a^2-\BQ^2}\leq r<\frac{3\Bm+\sqrt{9\Bm^2-8a^2-8\BQ^2}}{2}$, so in the region $f(r)<\delta$ with $\delta=f(3\Bm)$, we have $f'(r)\neq 0$. Therefore, in the region $f(r)<\delta=f(3\Bm)$, $H_{p_{h,z}}f=0$ is equivalent to $H_{p_{h,z}}r=0$, which is given by 
\[
\Delta_b\big(\xi_r-c'(r)z\big)-\chi\big((r^2+a^2)z-a\xi_\varphi\big)=0.
\]   
Under the conditions that $p_{h,z}=0$ and $H_{p_{h,z}}f=0$, we calculate 
\begin{align*}
	H^2_{p_{h,z}}f=-2\rho_b^{-2}f'(r)\Delta_bH_{p_{h,z}}\xi_r=-\frac{2f'(r)}{\rho_b^4\Delta_b}(a\xi_\varphi-(r^2+a^2)z)\Big(\frac{\pa\Delta_b}{\pa r}(a\xi_\varphi-(r^2+a^2)z)+4zr\Delta_b\Big).
	\end{align*}  
Next, we let 
\[
A=a\xi_\varphi-(r^2+a^2)z,\quad B=\xi_\varphi-a\sin^2\theta z.
\]  
Then we have $z=-\rho_b^{-2}(A-a B)$ and
 \[
	H^2_{p_{h,z}}f=-\frac{2f'(r)}{\rho_b^6\Delta_b}A\Big(\frac{\pa\Delta_b}{\pa r}A\rho_b^2-4r\Delta_b(A-a B)\Big)=-\frac{2f'(r)}{\rho_b^6\Delta_b}\Big(\big(\frac{\pa\Delta_b}{\pa r}\rho_b^2-4r\Delta_b\big)A^2+4a r\Delta_bAB\Big).
\]  
Using $p_{h,z}=0$ and $H_{p_{h,z}}f=0$, we obtain
\[
p_{h,z}=-\rho_b^{-2}\Big(-\frac{1}{\Delta_b}A^2+\xi_\theta^2+\frac{B^2}{\sin^2\theta}\Big)=0
\]
and thus
\begin{equation}\label{eq:ABnonzero}
A^2\geq \Delta_bB^2
\end{equation}
where we use the fact that $B=\xi_\varphi=0$ when $\sin\theta=0$. We also note that $p_{h,z}=0$ and $z\in\BR\setminus 0$ imply $A\neq0$ since otherwise $B=0$ and $z=0$.
We now calculate
\begin{align*}
	\big(\frac{\pa\Delta_b}{\pa r}\rho_b^2-4r\Delta_b\big)A^2+4a r\Delta_bAB& <-2r(r^2-3\Bm r+(a^2+2\BQ^2))A^2+4\abs{a}r\Delta_b\abs{AB}\\
	&<-2r(r^2-3\Bm r+(a^2+2\BQ^2))A^2+4\Bm r\Delta_b^{1/2}A^2\\
	&<-2r(r^2-5\Bm r+(a^2+2\BQ^2))A^2
\end{align*}
where we use $\abs{a}<\Bm$ and $A^2\geq \Delta_bB^2$ in the second step and $\Delta_b< r^2$ for $r\geq \ehKN$ in the third step. As a consequence, $H^2_{p_{h,z}}f<0$ when $r>5\Bm$.

We now assume that $\ehKN<r\leq 5\Bm$ and $f(r)<\delta'$. Then we have
\begin{align*}
	\Delta_b(r)&=r^4f(r)\leq (5\Bm)^4f(r)<(5\Bm)^4
\delta'\\
\pa_r\Delta_b&=2r-2\Bm> 2\sqrt{\Bm^2-a^2-\BQ^2}.
\end{align*}
By choosing $\delta'>0$ sufficiently small such that
\[
2\Bm\sqrt{\Bm^2-a^2-\BQ^2}-4(5\Bm)^4\delta'-4\Bm(5\Bm)^2\sqrt{\delta'}\geq 0,
\]
we conclude that on $\{\ehKN<r\leq 5\Bm\}\cap\{f(r)<\delta'\}$
\begin{align*}
r^5f'(r)&=r\pa_r\Delta_b-4\Delta_b>2\Bm\sqrt{\Bm^2-a^2-\BQ^2}-4(5\Bm)^4\delta'\geq 0,\\
\big(\frac{\pa\Delta_b}{\pa r}\rho_b^2-4r\Delta_b\big)A^2+4a r\Delta_bAB&\geq r\big(\frac{\pa\Delta_b}{\pa r}r-4\Delta_b-4\Bm\Delta_b^{1/2}\big)A^2\\
&>\Bm\big(2\Bm\sqrt{\Bm^2-a^2-\BQ^2}-4(5\Bm)^4\delta'-4\Bm(5\Bm)^2\sqrt{\delta'}\big)A^2\geq 0,
\end{align*}
and thus $H^2_{p_{h,z}}f<0$. Then letting $\delta_0=\min\{f(5\Bm), \delta'\}$ finishes the proof.
\end{proof}

We now define the outgoing/incoming tails and the trapped set (see \cite[Definition 2.1]{D15},\cite[Definition 6.1]{DZ19}). Recall that \[{}^\scop\!H^{2,0}_{p_{h,z}}=\rho^{-1}\jb{\xi}^{-1}{}^\scop\!H_{p_{h,z}}, \quad\rho=r^{-1}.
\]
\begin{defn}
	Let $f$ be the function defined as in Proposition \ref{prop:escapefcn}. Let $(r,\theta,\varphi,\xi)\in\cscform$ and $\phi(s)=\exp(s{}^\scop\!H^{2,0}_{p_{h,z}})(r,\theta,\varphi,\xi)$ be a maximally extended integral curve with domain of definition $I\subset\BR$ on $\Sigma_h=\{\jb{\xi}^{-2}p_{h,z}=0\}\subset\cscform$. Let $s_-=\inf I, s_+=\sup I$. We say that $\phi(s)$ is trapped as $s\to\pm\infty$, if there exists $\delta>0$ and $T>0$ such that $f(\phi(s))>\delta$ for all $\pm s\geq T$ (as a consequence, $s_\pm=\pm\infty$). We define the \textit{incoming tail} $\Gamma_-$ and the \textit{outgoing tail} $\Gamma_+$ as 
	\[
	\Gamma_\mp=\{(r,\theta,\varphi,\xi)\in\cscform\mid \phi(s)=\exp(s{}^\scop\!H^{2,0}_{p_{h,z}})(r,\theta,\varphi,\xi) \mbox{ is trapped as } s\to\pm\infty\}.
	\]
Then we define the \textit{trapped set} as $K:=\Gamma_-\cap\Gamma_+$.
\end{defn}
It follows from the definition that $\Gamma_\pm, K$ are invariant under the flow $\exp(s{}^\scop\!H^{2,0}_{p_{h,z}})$. We next study the properties of $\Gamma_{\pm}$ and $K$. We need the following lemmas.

\begin{lem}\label{lem:signofderiofr}
	For $z\in\BR\setminus 0$, we have 
	\begin{equation}
		\pm\jb{\xi}^{-1}H_{p_{h,z}}r>0\quad \mbox{on}\quad (\Sigma^\pm_h\cap\{r_-\leq r\leq \ehKN\})\setminus L_\pm.
	\end{equation}	
\end{lem}

\begin{proof}
	We first consider the case $r_-\leq r<\ehKN$ and finite $\xi$. Since $\chi=1, c'(r)=0$ on $\{r_-\leq r<\ehKN\}$, we write $p_{h,z}=0$ as
	\[
	p_{h,z}=-\rho_b^{-2}\Big(\Delta_b\xi_r^2-2(r^2+a^2)\xi_rz+2a\xi_r\xi_\varphi+\tilde{p}_{h,z}\Big)=0\quad\mbox{where}\quad \tilde{p}_{h,z}=\xi_\theta^2+\frac{1}{\sin^2\theta}(\xi_\varphi-a\sin^2\theta z)^2.
	\]
	This is a quadratic equation in $\xi_r$ with positive discriminant on $\{r_-\leq r<\ehKN\}$ since $\Delta_b<0$ there. As a consequence, it has two distinct roots $\xi_r^-<0<\xi_r^+$ and $\pm\pa_{\xi_r}p_{h,z}>0$ at $\xi_r^\pm$.
	By Lemma \ref{lem:charasemi}, $\Sigma_h^\pm$ is characterized by the conditions $\pm g^{-1}(-zdt_{\chi}+\xi, d\mathfrak{t})>0$ (because $t_{b,*}=t_{\chi}$ in the region $r_-\leq r\leq \ehKN$), that is, with $\mathfrak{t}=t_\chi+b(r)$ (according to the definition of $\mathfrak{t}$ in \eqref{Eqtimeliket}, we see that $b'(r)\leq 0$ in the region $r_-\leq r\leq\ehKN$), we have 
	\begin{equation}\label{eq:valueofxir}
		\pm\xi_r>\pm\frac{b'(r)\big((r^2+a^2)z-a\xi_\varphi\big)+a\big(a\sin^2\theta z-\xi_\varphi\big)}{r^2+ a^2+b'(r)\Delta_b(r)}.
	\end{equation}
	If $g^{-1}(-zdt_{\chi}+\xi, d\mathfrak{t})=0$, then $-zdt_{\chi}+\xi$ is a nonzero spacelike vector and thus $p_{h,z}(\xi)=-g^{-1}(-zdt_{\chi}+\xi, -zdt_{\chi}+\xi)<0$. That is, substituting the right-hand side of \eqref{eq:valueofxir} into $p_{h,z}$ yields a negative number. Therefore, 
	\[
	(r, \theta, \varphi, \xi_r^\pm, \xi_\theta, \xi_\varphi)\in\Sigma_h^\pm\quad \mbox{and}\quad \pm\jb{\xi}^{-1}H_{p_{h,z}}r=\pm\jb{\xi}^{-1}\pa_{\xi_r}p_{h,z}>0\quad \mbox{at}\quad \xi_r^\pm.
	\]

	In case of fiber infinity $\fiscform\cap\{r_-\leq r<\ehKN\}$, using the coordinates \eqref{eq:coordatfiberinftysemi}, we write 
	\[
	\rho_\xi^2p_{h,z}=-\rho_b^{-2}\Big(\Delta_b\hat{\xi}_r^2+2a\hat{\xi}_r\hat{\xi}_\varphi+\hat{\xi}_\theta^2+\frac{1}{\sin^2\theta}\hat{\xi}_\varphi^2\Big)
	\]
	and
	\[
	\rho_\xi g^{-1}(-zdt_\chi+\xi, d\mathfrak{t})=\rho_b^{-2}\Big(\big(r^2+a^2+b'(r)\Delta_b(r)\big)\hat{\xi}_r+\big(a(1+b'(r))\big)\hat{\xi}_\varphi\Big).
	\]
	Then the proof proceeds as in the finite $\xi$ case.
	
	It remains to consider the case $r=\ehKN$ where $\Delta_b=0$. For finite $\xi$, we write 
	\[
	p_{h,z}=-\rho_b^{-2}\Big(-2(r^2+a^2)\xi_rz+2a\xi_r\xi_\varphi+\tilde{p}_{h,z}\Big)=0\quad\mbox{where}\quad \tilde{p}_{h,z}=\xi_\theta^2+\frac{1}{\sin^2\theta}(\xi_\varphi-a\sin^2\theta z)^2.
	\]
	Since $a\xi_\varphi-(r^2+a^2)z\neq 0$ on $\Sigma_h$ (otherwise $a^2\sin^2\theta=r^2+a^2$ at $\ehKN$, which is impossible), the right-hand side of $p_{h,z}$ is a linear equation in $\xi_r$ with nonzero leading coefficient and thus has only one root. Again, by Lemma \ref{lem:charasemi}, $\Sigma_h^\pm$ is characterized by the conditions $\pm g^{-1}(-zdt_{\chi}+\xi, d\mathfrak{t})>0$, i.e., 
	\begin{equation}\label{eq:valueofxirlinear}
		\pm\xi_r>\pm\frac{-\big(a(1+b'(r))\big)\xi_\varphi+z\big(a^2\sin^2\theta+(r^2+a^2)b'(r)\big)}{r^2+a^2}.
	\end{equation}
	Similarly, substituting the right-hand side of \eqref{eq:valueofxirlinear} into $p_{h,z}$ yields a negative number. Therefore, we have either
	\[
	(r, \theta, \varphi, \xi_r, \xi_\theta, \xi_\varphi)\in\Sigma_h^+\quad \mbox{and}\quad \jb{\xi}^{-1}H_{p_{h,z}}r=\jb{\xi}^{-1}\pa_{\xi_r}p_{h,z}>0\quad \mbox{at}\quad \xi_r,
	\]
	or
	\[
	(r, \theta, \varphi, \xi_r, \xi_\theta, \xi_\varphi)\in\Sigma_h^-\quad \mbox{and}\quad \jb{\xi}^{-1}H_{p_{h,z}}r=\jb{\xi}^{-1}\pa_{\xi_r}p_{h,z}<0\quad \mbox{at}\quad \xi_r.
	\]

	In case of fiber infinity $\fiscform\cap\{r=\ehKN\}$, using the coordinates \eqref{eq:coordatfiberinftysemi}, we write 
	\[
	\rho_\xi^2p_{h,z}=-\rho_b^{-2}\Big(2a\hat{\xi}_r\hat{\xi}_\varphi+\hat{\xi}_\theta^2+\frac{1}{\sin^2\theta}\hat{\xi}_\varphi^2\Big).
	\]
	Since $\hat{\xi}_\varphi\neq 0$ on $(\{\jb{\xi}^{-2}p_{h,z}=0, r=\ehKN\}\cap\fiscform)\setminus L_\pm$, the right-hand side of $\rho_\xi p_{h,z}$ s a linear equation in $\hat{\xi}_r$ with nonzero leading coefficient and thus has only one root. 
	Then the proof proceeds as in the finite $\xi$ case.
\end{proof}

\begin{lem}\label{lem:escapetechlem}
	Let $z\in\BR\setminus 0$. Let $\delta_0>0$ and $f(r)$ be defined as in Proposition \ref{prop:escapefcn}. Let $(r,\theta,\varphi,\xi)\in\cscform\setminus(L_\pm\cap \pa_+\CX)$ and $\phi(s)=\exp(s{}^\scop\!H^{2,0}_{p_{h,z}})(r,\theta,\varphi,\xi)$ be a maximally extended flow on $\Sigma_h=\{\jb{\xi}^{-2}p_{h,z}=0\}$. 
		\begin{enumerate}
			\item If $f(\phi)(s_0)<\delta_0$ and $\pm\pa_sf(\phi(s))\leq0$ for some $\pm s_0\geq0$, then $f(\phi(s))<\delta_0$ and $\pm\pa_sf(\phi(s))<0$ for  $\pm s> s_0$ and $\phi(s)$ is not trapped as $\pm s\to\infty$.
			\item If $\phi(s)$ is not trapped as $\pm s\to\infty$, then either $f(\phi(s))<\delta_0$ and $\pm\pa_sf(\phi(s))\leq0$ for all sufficiently large $\pm s$, or $\phi(s)\to L_\pm$ (i.e. $f(s)\to0$) as $\pm s\to\infty$.
		\end{enumerate}
	Moreover, 
	\begin{enumerate}
		\item [(3)] If $\phi(s)$ is not trapped as $s\to\infty$, then for $\phi(s)\subset \Sigma^-_h$, $\phi(s)$ crosses $\pa_-\CX$ into the inward direction of decreasing $r$ at some finite time $s_0<\infty$ or $\rho(\phi(s))\to0$ as $s\to\infty$, while for $\phi(s)\subset\Sigma^+_h$, $\phi(s)\to L_+$ or $\rho(\phi(s))\to0$ as $s\to\infty$.
		\item[(4)] If $\phi(s)$ is not trapped as $s\to-\infty$, then for $\phi(s)\subset \Sigma^-_h$, $\phi(s)\to L_-$ or $\rho(\phi(s))\to0$ as $s\to-\infty$, while for $\phi(s)\subset\Sigma_h^+$, $\phi(s)$ crosses $\pa_-\CX$ into the inward direction of decreasing $r$ at some finite time $s_0>-\infty$ or $\rho(\phi(s))\to0$ as $s\to-\infty$.
	\end{enumerate}
\end{lem}

\begin{proof}
	We only prove the case $\pa_sf(\phi(s))\leq0$ as the other case $-\pa_sf(\phi(s))\leq0$ can be handled similarly. We put \[
	f(s)=f(\phi(s)),\quad \dot{f}(s)=\pa_sf(\phi(s))={}^\scop\!H^{2,0}_{p_{h,z}}f(\phi(s)),\quad \ddot{f}(s)=\pa^2_sf(\phi(s))=({}^\scop\!H^{2,0}_{p_{h,z}})^2f(\phi(s)).
	\]
	 Let $s>s_0$. If $f(s)$ attains its minimum of on the interval $[s_0, s]$ at some point $s_1\in(s_0, s)$, then \[f(s_1)<\delta_0,\quad \dot{f}(s_1)=0, \quad \ddot{f}(s_1)\geq 0,
	 \] 
	 which contradicts \eqref{eq:escapefcn}. If $f(s)$ attains its minimum value at $s_0$, then $\dot{f}(s_0)\geq 0$, and thus $\dot{f}(s_0)=0$. Again, it follows from \eqref{eq:escapefcn} that $s_0$ is a strict local maximum point, which is a contradiction. Therefore, $f(s)$ must attain its minimum value at $s$ and $\dot{f}(s)<0$ (otherwise $\dot{f}(s)=0$ and then by \eqref{eq:escapefcn}, $s$ is a strict local maximum point, which is a contradiction), showing that \[f(s)<\delta_0,\quad \dot{f}(s)<0\quad \mbox{for all}\quad s>s_0.
	 \]
	  Suppose $\phi(s)$ is trapped as $s\to\infty$. Then there exists $\delta>0$ and $T>0$ such that $f(s)>\delta$ for all $s\geq T$. Since $\{f(r)\geq \delta\}\cap\cscform$ is a compact set, we can take a sequence $s_j\to\infty$ such that $\phi(s_j)$ converges to some $(r_\infty, \theta_\infty, \varphi_\infty, \xi_\infty)\in \{\delta\leq f(r)\leq f(s_0)<\delta_0\}$. Since $f(s)$ is non increasing  and bounded for all $s\geq s_0$, $\lim_{s\to\infty}f(s)$ exists. Since $\ddot{f}(s), \dddot{f}(s)$ are bounded for all $s\geq \max\{s_0, T\}$, by Barb\u{a}lat's Lemma \cite{B59}, we have ${}^\scop\!H^{1,0}_{p_{h,z}}f((r_\infty, \theta_\infty, \varphi_\infty, \xi_\infty))=({}^\scop\!H^{1,0}_{p_{h,z}})^2f((r_\infty, \theta_\infty, \varphi_\infty, \xi_\infty))=0$. This contradicts \eqref{eq:escapefcn}.
	  
	  We now prove the statement (2), (3) and (4). We only prove the case $\phi(s)$ is not trapped as $s\to\infty$ in detail as the other case can be handled similarly. If $\phi(s)\subset\{f(r)\leq 0\}$ for all $s\geq0$, then $\phi(s)\subset\{r_-\leq r\leq \ehKN\}$ since $\phi(0)\notin\pa_+\CX$. For the case $\phi(s)\subset\Sigma_h^-$, since $\phi(0)\notin L_-$, it follows from Lemma \ref{lem:signofderiofr} that $\phi(s)$ crosses $\pa_-\CX$ into the inward direction of decreasing $r$ at finite time and thus there must exist $s_0>0$ such that $f(s_0)<\delta_0, \dot{f}(s_0)<0$. Proceeding as in the proof of (1), we see that $f(s)<\delta_0, \dot{f}(s)\leq0$ for all $s\geq s_0$. While for the case $\phi(s)\subset\Sigma_h^+$, we claim that $\phi(s)\to L_+$ as $s\to\infty$. Otherwise, there exists a sufficiently small neighborhood $U_+$ of $L_+$ such that $\phi(s)\subset \Sigma_h^+\setminus U_+$ for all $s>0$. By Lemma \ref{lem:signofderiofr}, there exists $\delta>0$ such that 
	  \begin{equation}\label{eq:signofderiofrext}
	  	\jb{\xi}^{-1}H_{p_{h,z}}r\geq \delta \quad \mbox{on}\quad (\Sigma_h^+\cap\{r_-\leq r\leq \ehKN+\delta\})\setminus U_+
	  \end{equation}
  (Otherwise, if \eqref{eq:signofderiofrext} does not hold for any positive $\delta$, then we can find a sequence $\{(r_j, \theta_j, \varphi_j, \xi_j)\}$ such that $r_-\leq r_j\leq \ehKN+\frac{1}{j}$ and $\jb{\xi}^{-1}H_{p_{h,z}}r((r_j, \theta_j, \varphi_j, \xi_j))< \frac{1}{j}$. Let $j\to\infty$, and then we obtain a limit point \[(r_\infty, \theta_\infty, \varphi_\infty, \xi_\infty)\in (\Sigma^+_h\cap\{r_-\leq r\leq \ehKN\})\setminus L_+
  \]
   such that $\jb{\xi}^{-1}H_{p_{h,z}}r((r_\infty, \theta_\infty, \varphi_\infty, \xi_\infty))\leq 0$, which contradicts Lemma \ref{lem:signofderiofr}). It follows from \eqref{eq:signofderiofrext} that $\phi(s)\subset \{r\geq \ehKN+\delta\}$ for all large enough $s$, which contradicts the assumption that $\phi(s)\subset\{r_-\leq r\leq \ehKN\}$. 
   
   On the other hand, suppose there exists $s_0\geq 0$ such that $f(s_0)>0$. Since $\phi(s)$ is not trapped as $s\to\infty$, we can find $s_2>s_1>0$ such that $f(s_2)<f(s_1)<\delta_0$. Then arguing as in the proof of (1), we see that $f(s)$ attains its minimum value on in the interval $[s_1, s_2]$ at the point $s_2$, and thus $\dot{f}(s_2)\leq 0$. Proceeding again as in the proof of (1), we see that $f(s)<\delta_0, \dot{f}(s)\leq0$ for all $s\geq s_2$. 
   
   Also, since $\phi(s)$ is not trapped as $s\to\infty$, it follows that either $\rho(\phi(s))\to0$ as $s\to\infty$, or $\phi(s)$ enters the region $\{r\leq r_b+\delta\}$ for any $\delta>0$ in finite time.  Using \eqref{eq:signofderiofrext}, we find that $\phi(s)\to L_+$ as $s\to\infty$ if $\phi(s)\subset \Sigma^+_h$. If $\phi(s)\subset\Sigma_h^-$, suppose that $\phi(s)$ never crosses $\pa_-\CX$ for $s\geq 0$. Then $\phi(s)$ is well defined for all $s\geq 0$. Since $(\rho,\theta,\varphi,\xi)\notin L_-$, then there exists a neighborhood $V_-$ of $L_-$ such that $ (\rho,\theta,\varphi,\xi)\notin V_-$. Since $L_-$ is a radial source, there exists a neighborhood $U_-$ of $L_-$ and $T>0$ such that 
   \[
   \phi(s)(U_-)\subset V_-\quad \mbox{for all}\quad s\leq-T.
   \]
   Since $(\rho,\theta,\varphi,\xi)\notin V_-$, it follows that $\phi(s)\notin U_-$ for all $s\geq T$. Proceeding as in the proof of \eqref{eq:signofderiofrext} yields
   \[
   	-\jb{\xi}^{-1}H_{p_{h,z}}r\geq \delta \quad \mbox{on}\quad (\Sigma_h^-\cap\{r_-\leq r\leq \ehKN+\delta\})\setminus U_-
   \]
   for some $\delta>0$. This forces $r(\phi(s))>r_b+\delta$ for all $s>T$ and thus $\rho(\phi(s))\to0$ as $s\to\infty$.
 \end{proof}

According to Lemma \ref{lem:escapetechlem}, it is clear that $K\subset \{f(r)\geq \delta_0\}\subset\{r>\ehKN\}$. Therefore, we now restrict our analysis in the region $r>\ehKN$. Recall the calculation in the proof of Proposition \ref{prop:escapefcn}
\[
\Delta_b>0,\ H_{p_{h,z}}r=0\ \Longrightarrow\  H^2_{p_{h,z}}r=2\rho_b^{-4}\Delta_b\pa_r\Big(\frac{(a\xi_\varphi-(r^2+a^2)z)^2}{\Delta_b(r)}\Big)
\]
We define
\begin{equation}
	F(r):=\frac{(a\xi_\varphi-(r^2+a^2)z)^2}{\Delta_b(r)}.
\end{equation}
The key property of $F(r)$ is given by
\begin{prop}\label{prop:newescapefcn}
	Let $z\in\BR\setminus 0$. For each $r\in(\ehKN, \infty)$, we have\begin{equation}\label{eq:newescapefcn}
		p_{h,z}=0,\ \pa_r F(r)=0\ \Longrightarrow\  \pa_r^2F(r)>0.
	\end{equation}
	Moreover, for each fixed $\xi_\varphi$, there exists a unique $r_{\xi_\varphi}\in(\ehKN, \infty)$ such that $\pa_rF(r_{\xi_\varphi})=0$.
\end{prop}
\begin{proof}
	Since $	p_{h,z}=0$, by the discussion around \eqref{eq:ABnonzero}, we see that $a\xi_\varphi-(r^2+a^2)z\neq 0$. Then
	\[
	\pa_rF(r)=-\frac{(a\xi_\varphi-(r^2+a^2)z)}{\Delta_b^2}\Big(\frac{\pa\Delta_b}{\pa r}(a\xi_\varphi-(r^2+a^2)z)+4zr\Delta_b\Big)=0
	\]
	implies
	\begin{align*}
		\pa_r^2F(r)&=-\frac{(a\xi_\varphi-(r^2+a^2)z)}{\Delta_b^2}\Big(\frac{\pa^2\Delta_b}{\pa r^2}(a\xi_\varphi-(r^2+a^2)z)+2zr\frac{\pa\Delta_b}{\pa r}+4z\Delta_b\Big)\\
		&=-\frac{8z^2r}{\Delta_b(\pa_r \Delta_b)^2}\Big(2\frac{\pa^2\Delta_b}{\pa r^2}r\Delta_b-r(\pa_r\Delta_b)^2-2\Delta_b\pa_r\Delta_b\Big)\\
		&=\frac{32z^2r}{\Delta_b(\pa_r \Delta_b)^2}\Big(r^3-3\Bm r^2+3\Bm^2r-\Bm(a^2+\BQ^2)\Big)>0\quad \mbox{on}\quad \ehKN<r<\infty.
	\end{align*}
	Since $F(r)\to \infty$ as $r\to \ehKN$ and $r\to\infty$, $F(r)$ must attains its global minimum value at $r_{\xi_\varphi}\in(\ehKN, \infty)$ and $\pa_rF(r_{\xi_\varphi})=0$. Moreover, the existence of the critical point of $F$ is unique (otherwise, suppose that there are two points $r_1, r_2\in(\ehKN, \infty)$ with $r_1<r_2$ such that $\pa_r F(r_1)=\pa_rF(r_2)=0$. Then let $D=\{r_1<r<r_2\mid \pa_rF(r)>0\}$. Since $\pa_rF(r_1)=0, \pa_r^2F(r_1)>0$, it follows that $D$ is nonempty and we let $d:=\sup D$. Since $\pa_rF(r_2)=0, \pa^2_rF(r_2)>0$, we see that $d\in(r_1,r_2)$. By continuity, we have $\pa_rF(d)=0$. By the definition of $d$, we have $\pa_r^2F(d)\leq 0$, which is a contradiction).
\end{proof}

Let $(r^0,\theta^0, \varphi^0,\xi^0)\in\Sigma_h$ and $(r(s),\theta(s),\varphi(s),\xi(s))=\exp(sH_{p_{h,z}})((r^0,\theta^0, \varphi^0,\xi^0))$. Then we define
\begin{equation}
	\Phi^0(r):=\Phi(r;\theta^0, \varphi^0,\xi_\theta^0,\xi^0_\varphi)=F(r;\xi_\varphi^0)-\tilde{p}_{h,z}((\theta^0, \varphi^0,\xi_\theta^0,\xi_\varphi^0)).
\end{equation}
Since $\tilde{p}_{h,z}$ and $\xi_\varphi$ are constants along the Hamiltonian flows $\exp(sH_{p_{h,z}})$ on $\Sigma_h$, it follows that $(r(s), \xi_r(s))$ is a Hamiltonian flow trajectory of 
\[
H^0(r, \xi_r):=\Delta_b\big(\xi_r-zc'(r)+\frac{\chi(a\xi^0_\varphi-(r^2+a^2)z)}{\Delta_b}\big)^2-\Phi^0(r)
\]
In particular, $(r(s), \xi_r(s))$ solves the equation $H^0(r,\xi_r)=0$.

\begin{prop}
	In the region $r>\ehKN$, the outgoing tail $\Gamma^+$ and incoming tail $\Gamma^-$ are given by 
	\begin{equation}\label{eq:defoftails}
		\Gamma^{\pm}=\Big\{(r,\theta,\varphi,\xi)\!\mid \!p_{h,z}=0,\  \xi_r-zc'(r)+\frac{\chi(a\xi_\varphi-(r^2+a^2)z)}{\Delta_b}=\mp\sgn(r-r_{\xi_\varphi})\sqrt{\frac{\Phi(r;\theta,\varphi,\xi_\theta,\xi_\varphi)}{\Delta_b}}\Big\}
	\end{equation}
where 
\[
\Phi(r;\theta,\varphi,\xi_\theta,\xi_\varphi)=F(r;\xi_\varphi)-\tilde{p}_{h,z}(\theta,\varphi, \xi_\theta,\xi_\varphi)
\]
and $r_{\xi_\varphi}$ is the only solution to the equations $\Phi(r;\theta,\varphi,\xi_\theta,\xi_\varphi)=0$ and $\pa_r\Phi(r;\theta,\varphi,\xi_\theta,\xi_\varphi)=0$. Moreover, $\Gamma^\pm$ are smooth codimension $1$ submanifolds of $\Sigma_h$ intersecting transversely, and their intersection is the trapped set
\begin{equation}\label{eq:exoftrapping}
	K=\Big\{(r,\theta,\varphi,\xi)\!\mid \!p_{h,z}=0,\  r=r_{\xi_\varphi},\ \xi_r-zc'(r)+\frac{\chi(a\xi_\varphi-(r^2+a^2)z)}{\Delta_b}=0\Big\},
\end{equation}
which is a smooth codimension $2$ submanifold of $\Sigma_h$
\end{prop}
\begin{proof}
	Let $(r^0,\theta^0, \varphi^0,\xi^0)\in\Sigma_h$ and $(r(s),\theta(s),\varphi(s),\xi(s))=\exp(sH_{p_{h,z}})((r^0,\theta^0, \varphi^0,\xi^0))$. By Proposition \ref{prop:newescapefcn}, $\pa_r\Phi^0(r_{\xi^0_\varphi})=0$. If $\Phi^0(r_{\xi^0_\varphi})\neq 0$, we have $\abs{\Phi^0(r(s))}+\abs{\pa_r\Phi^0(r(s))}>0$. Suppose that $(r(s),\theta(s),\varphi(s),\xi(s))$ is  trapped as $s\to\infty$. Since $(r(s),\xi_r(s))$ solves the equation $H^0(r,\xi_r)=0$, arguing as in the proof of the statement (1) of Lemma \ref{lem:escapetechlem}, we can find a sequence $s_j\to\infty$ such that $(r(s_j),\xi_r(s_j))$ converges to some $(r_\infty, (\xi_r)_\infty)$ which satisfies $H^0(r_\infty, (\xi_r)_\infty)=0, H_{H^0}r((r_\infty, (\xi_r)_\infty))=H_{H^0}\xi_r((r_\infty, (\xi_r)_\infty))=0$. This contradicts the fact that the Hamiltonian vector field associated to $H^0$ is nonzero on the set $H^0=0$. This proves that $(r(s),\theta(s),\varphi(s),\xi(s))$ is  not trapped as $s\to\infty$. Similarly, $(r(s),\theta(s),\varphi(s),\xi(s))$ escapes as $s\to-\infty$ as well. As a result, $(r^0,\theta^0, \varphi^0,\xi^0)\notin \Gamma^{\pm}$.
	
	On the other hand, we consider the case $\Phi^0(r_{\xi^0_\varphi})=0$. We note that the set of solutions to the equation $H^0(r,\xi_r)=0$ is given by the union $\Gamma^{+,0}\cup \Gamma^{-,0}$ where
	\[		\Gamma^{\pm,0}=\Big\{\xi_r-zc'(r)+\frac{\chi(a\xi^0_\varphi-(r^2+a^2)z)}{\Delta_b}=\mp\sgn(r-r_{\xi^0_\varphi})\sqrt{\frac{\Phi(r;\theta^0,\varphi^0,\xi^0_\theta,\xi^0_\varphi)}{\Delta_b}}\Big\}.
	\]
	First, $(r^0,\theta^0,\varphi^0,\xi^0)\in\Sigma_h$ is equivalent to $H^0(r^0, \xi_r^0)=0$, i.e., $(r^0,\theta^0,\varphi^0,\xi^0)\in\Sigma_h$ implies $(r^0,\xi_r^0)\in \Gamma^{+,0}\cup \Gamma^{-,0}$. Next, if $(r^0, \xi_r^0)\in \Gamma^{\mp,0}$, we see that $(r(s),\theta(s),\varphi(s),\xi(s))$ is trapped as $s\to\pm\infty$. Therefore, we conclude that $(r(s),\theta(s),\varphi(s),\xi(s))$ is trapped as $s\to\pm\infty$ if and only if $(r^0, \xi_r^0)\in \Gamma^{\mp,0}$. Then we obtain
\begin{align*}
	K=\Gamma^+\cap\Gamma^-&=\Big\{(r,\theta,\varphi,\xi)\!\mid \!p_{h,z}=0,\  r=r_{\xi_\varphi},\ \xi_r-zc'(r)+\frac{\chi(a\xi_\varphi-(r^2+a^2)z)}{\Delta_b}=0\Big\}\\&=\Big\{(r,\theta,\varphi,\xi)\!\mid \!p_{h,z}=0,\  \pa_r F=0,\ H_{p_{h,z}}r=0\Big\}.
\end{align*}
We note that $d(H_{p_{h,z}}r)\in\{dr,d\xi_r, d\xi_\varphi\}$ and the coefficient of $d\xi_r$ is $1$. In view of \eqref{eq:newescapefcn}, $d(\pa_rF)\in\{dr, d\xi_\varphi\}$ and the coefficient of $dr$ is positive on $K$. Since $H_{p_{h,z}}r=0$ and $\pa_rF=0$ imply $\pa_{\xi_r}p_{h,z}=0$ and $\pa_r p_{h,z}=0$ respectively on $K$, we have $dp_{h,z}\in\{d\theta, d\xi_\theta, d\xi_\varphi\}$. Therefore, $dp_{h,z}, d(\pa_rF)$ and $d(H_{p_{h,z}}r)$ are linearly independent on $K$ and $K$ is a smooth codimension $2$ submanifold of $\Sigma_h$. 

For any $(r^0,\theta^0, \varphi^0,\xi^0)\in K$, since  $\Phi^0(r)=\frac{1}{2}\abs{\pa_r^2\Phi^0(r_{\xi_\varphi^0})}(r-r_{\xi_\varphi^0})^2+\mathcal{O}((r-r_{\xi_\varphi^0})^3)$, the defining functions of $\Gamma^{\pm,0}$ are smooth functions in $r, \xi_r$ and thus $\Gamma^{\pm,0}$ is a smooth codimension $1$ submanifold of $T^*(\ehKN,\infty)$. If $(r,\theta,\varphi,\xi)\in\Gamma^\pm$, then its projection onto $(\theta, \varphi, \xi_\theta,\xi_\phi)$ must lie in the set of projection of $K$ onto $(\theta, \varphi, \xi_\theta,\xi_\phi)$. Therefore, $\Gamma^\pm$ are smooth codimension $1$ submanifolds of $\Sigma_h$ intersecting transversely at $K$.
\end{proof}

Since $TK$ is in the kernel of $d(H_{p_{h,z}}r)$ and  $d(\pa_r F)$, we see that the symplectic complement of $TK$ in $\Sigma_h$ is $(TK)^\perp=\{H_{H_{p_{h,z}}r}, H_{\pa_rF}\}$. By a direct calculation, we see that $H_{\pa_r F}(H_{p,z}r)=-2\Delta_b\pa_r^2F\neq 0$ on $K$, which implies $TK\cap (TK)^\perp=0$ and thus $K$ is symplectic. %Then we have the following decomposition
%\begin{equation}\label{eq:decom1}
%	T_{K}\fscform=TK\oplus H_{\varphi_+}|_K\oplus H_{\varphi_-}|_K
%\end{equation}
%where $\varphi_{\pm}$ are the defining functions of $\Gamma^{\pm}\subset\Sigma_h$. We also have 
%\begin{equation}\label{eq:decom2}
%	T_K\Gamma^{\pm}=TK\oplus H_{\varphi_{\pm}}|_K.
%\end{equation}
Now we verify that the trapped set $K$ is normally hyperbolic whose definition will be given below (see \cite[\S6.3]{DZ19}).

\begin{defn}
Let $K$ be given as above, i.e., $K$ is symplectic. Let $\phi_s=\exp(sH_{p_{h,z}}):\fscform\to\fscform$ be the Hamiltonian flow and $\varphi_\pm$ be the defining functions of $\Gamma^\pm$. We say $K$ is normally hyperbolic if there exist $C, \nu>0$ such that for all $(r,\theta,\varphi,\xi)\in K$
\begin{equation}\label{eq:defofnormalhypertra}
\frac{d(\varphi_\pm\circ\phi_{\pm s})(r,\theta,\varphi,\xi)}{d\varphi_\pm(r,\theta,\varphi,\xi)}\leq Ce^{-\nu s},\quad  s\geq0
			\end{equation}
		Define the \textit{minimal expansion rate} $\nu_{\min}>0$ as the supremum of all values of $\nu$ for which there exists a constant $C$ such that \eqref{eq:defofnormalhypertra} holds.
\end{defn}
\begin{prop}\label{prop:normallyhyperbolictrap}
	Let $K$ be defined as in \eqref{eq:exoftrapping}. Then $K$ is normally hyperbolic.
\end{prop}
\begin{proof}
It suffices to prove that the minimal expansion $\nu_{\min}$ rate defined above is positive. By \eqref{eq:defoftails}, we have
\[
\Gamma^{\pm}=\{\varphi_{\pm}=\xi_r-zc'(r)+\frac{\chi(a\xi_\varphi-(r^2+a^2)z)}{\Delta_b}\pm\sgn(r-r_{\xi_\varphi})\sqrt{\frac{\Phi(r;\theta,\varphi,\xi_\theta,\xi_\varphi)}{\Delta_b}}=0\}
\] 
where $(\theta,\varphi,\xi_\theta,\xi_\varphi)\in K$. That is $\varphi_\pm$ are defining function of $\Gamma^\pm$. Since $p_{h,z}=\rho_b^{-2}\Delta_b\varphi_+\varphi_-$, it follows that 
\[
H_{p_{h,z}}\varphi_\pm=\mp \rho_b^{-2}\Delta_b(H_{\varphi_+}\varphi_-)\varphi_\pm+(H_{\rho_b^{-2}\Delta_b}
\varphi_\pm)\varphi_\mp\varphi_\pm:=\mp\nu_\pm\varphi_{\pm}
\]
and $\nu_+|_K=\nu_-|_K=\rho_b^{-2}\Delta_b(H_{\varphi_+}\varphi_-)|_K=\nu$. Since
\[
\pa_{\xi_r}(H_{p_{h,z}}\varphi_\pm)=\pa_{\xi_r}(\mp\nu_\pm\varphi_{\pm})=\mp\big((\pa_{\xi_r}\nu_\pm)\varphi_\pm+\nu_\pm\big),
\]
using Proposition \ref{prop:newescapefcn} we find that 
\begin{align*}
	\nu&=-\pa_{\xi_r}(H_{p_{h,z}}\varphi_+)|_K=\rho_b^{-2}\sqrt{2\Delta_b\pa_r^2F(r_{\xi_\varphi})}>0.
\end{align*}
For $(r,\theta,\varphi,\xi)\in K$, we calculate
\[
\pa_sd(\varphi_\pm\circ\phi_{\pm s})=\mp d((\nu_\pm\varphi_\pm)\circ\phi_{\pm s})
=\mp(\nu_\pm\circ\phi_{\pm s}) d(\varphi_\pm\circ\phi_{\pm s}).
\]
Therefore, for all $T>0$ and $(r,\theta,\varphi,\xi)\in K$, we have 
\[
\frac{d(\varphi_\pm\circ\phi_{\pm T})(r,\theta,\varphi,\xi)}{d\varphi_\pm(r,\theta,\varphi,\xi)}=e^{-\jb{\nu}^\pm_T T},\quad \jb{\nu}^\pm_T=\frac{1}{T}\int_0^T\nu\circ\phi_{\pm s}(r,\theta,\varphi,\xi)\,ds.
\]
As a result, the minimal expansion rate is given by 
\begin{equation}\label{eq:calofmer}
\nu_{\min}=\min\{\ \liminf_{T\to\infty}\inf_{(r,\theta,\varphi,\xi)\in K}\jb{\nu}^+_T,\quad \liminf_{T\to\infty}\inf_{(r,\theta, \varphi,\xi)\in K} \jb{\nu}^-_T\ \}>0.
\end{equation}
\end{proof}
Finally, we further determine the position of the tapped set $K$.
\begin{lem}
	For $z\in\BR\setminus 0$, let $\Sigma_h^{\pm}$ be defined as in Lemma \ref{lem:charasemi}. Then $K\subset \Sigma_h^\pm$ for $\pm z>0$. Moreover, $\Gamma^+\cup\Gamma^-\subset\Sigma^\pm_h$ for $\pm z>0$.
\end{lem}

\begin{proof}
	First, according to \eqref{eq:onecomisempty} in Lemma \ref{lem:charasemi}, we have\[
	\Sigma_h^{\mp}\subset\{r\geq r_-\mid\Delta_b\leq a^2\sin^2\theta\}\subset \{r_-\leq r\leq \Bm+\sqrt{\Bm^2-\BQ^2}\}\quad \mbox{for}\quad \pm z>0
	\]
	and thus
	\[
	K\cap\{r> \Bm+\sqrt{\Bm^2-\BQ^2}\}\subset \Sigma^\pm_h\quad \mbox{for}\quad \pm z>0.
	\]
	Since $c'(r)=0, \chi=1$ on $r_-\leq r\leq \Bm+\sqrt{\Bm^2-\BQ^2}$, then for 
	\[
	(r,\theta,\varphi,\xi)\in K\cap\{r_-\leq r\leq \Bm+\sqrt{\Bm^2-\BQ^2}\}=K\cap\{r_b< r\leq \Bm+\sqrt{\Bm^2-\BQ^2}\},
	\]
	we have 
\[	
	a\xi_\varphi-(r^2+a^2)z=-\frac{4zr\Delta_b}{\pa_r\Delta_b}\quad\mbox{and}\quad \xi_r=\frac{4zr}{\pa_r\Delta_b},
	\]
	and thus
	\begin{align*}
		\pm g^{-1}(-zdt_{b,*}+\xi, d\mathfrak{t})&=\pm\rho_b^{-2}\Big(\xi_r(r^2+a^2+b'(r)\Delta_b)+b'(r)(a\xi_\varphi-(r^2+a^2)z)+(a\xi_\varphi-a^2\sin^2\theta z)\Big)\\
		&=\pm\frac{4zr}{\rho_b^2\pa_r\Delta_b}\Big(r^2+a^2-\Delta_b\Big)\pm z=\pm\frac{4zr}{\rho_b^2\pa_r\Delta_b}\Big(2\Bm r-\BQ^2\Big)\pm z>0\quad \mbox{for}\quad \pm z>0.
		\end{align*}
	This implies that 	
	\[
	K\cap\{r_-\leq r\leq \Bm+\sqrt{\Bm^2-\BQ^2}\}\subset \Sigma^\pm_h\quad \mbox{for}\quad \pm z>0,
	\]
	and thus proves $K\subset\Sigma^\pm_h$ for $\pm z>0$.
	
	In order to prove $\Gamma^+\cup\Gamma^-\subset \Sigma^\pm_h$ for $\pm z>0$, it suffices to show that if $(r,\theta,\varphi,\xi)\in\Gamma^{\pm}$, then $\phi(s)=\exp(s{}^{\scop}\!H_{p_{h,z}}^{2,0})(r,\theta, \varphi,\xi)\to K$ as $s\to\mp\infty$. We only prove the case of $(r,\theta,\varphi,\xi)\in\Gamma^-$ as the other case can be handled in a similar manner. Suppose that $\phi(s)\nrightarrow K$ as $s\to\infty$. Then there exists a sequence $s_j\to\infty$ and a neighborhood $U$ of $K$ such that $\phi(s_j)\notin U$ for all $j$. Since $\phi(s)$ is trapped as $s\to\infty$ (i.e., there exist $\delta, T>0$ such that $f(\phi(s))>\delta$ for all $s>T$), passing to a subsequence we have 
	\[
	\phi(s_j)\to(r_\infty, \theta_\infty, \varphi_\infty, \xi_\infty)\quad \mbox{for some}\quad (r_\infty, \theta_\infty, \varphi_\infty, \xi_\infty)\notin K=\Gamma^+\cap\Gamma^-.
	\]
By Lemma \ref{lem:escapetechlem}, we find that $\Gamma^-$ is closed and thus $(r_\infty, \theta_\infty, \varphi_\infty, \xi_\infty)\in \Gamma^-$. It follows that $(r_\infty, \theta_\infty, \varphi_\infty, \xi_\infty)\notin \Gamma^+$, which means that $\phi(s)$ is not trapped as $s\to-\infty$. Then there exists $T_1>0$ such that $f(\phi_\infty(-T_1))<\delta$ where $\phi_\infty(s)=\exp(s{}^{\scop}\!H_{p_{h,z}}^{2,0})(r_\infty,\theta_\infty, \varphi_\infty,\xi_\infty)$. Since $	\phi(s_j)\to(r_\infty, \theta_\infty, \varphi_\infty, \xi_\infty)$, it follows that \[\exp(-T_1{}^{\scop}\!H_{p_{h,z}}^{2,0})(\phi(s_j))\to\phi_\infty(-T_1).
\]
Therefore, $f(\phi(s_j-T_1))<\delta$ for all sufficiently large $j$. But this contradicts the fact that $\phi(s)$ is trapped as $s\to\infty$.
\end{proof}

\subsubsection{Global dynamics of the Hamiltonian flow}
Now we shall analyze the global behavior of the flow of $	{}^{\scop}\!H_{p_{h,z}}^{2,0}:=\rho^{-1}\jb{\xi}^{-1}{}^{\scop}\!H_{p_{h,z}}$ on the semiclassical characteristic set $\Sigma_h=\{\jb{\xi}^{-2}p_{h,z}=0\}$. We need the following technical lemma.
\begin{lem}\label{lem:technicallemmanearinfty}
	For $z\in\BR\setminus 0$, let $(\rho,\theta, \varphi,\xi)\in\Sigma_h$ and we put $\phi(s)=\exp(s{}^{\scop}\!H_{p_{h,z}}^{2,0})(\rho,\theta, \varphi,\xi)$. If $\rho(\phi( s))\to 0$ as $s\to\pm\infty$, then one has 
\[
\phi(s)\to R(\frac{z}{2}\pm\frac{\abs{z}}{2})=\{\rho=0, \xi=(z\pm\abs{z})\frac{d\rho}{\rho^2}\}\quad \mbox{as}\quad s\to \pm\infty.
\]
\begin{proof}
Recall that 
\[
\jb{\xi}{}^{\scop}\!H_{p_{h,z}}^{2,0}=(-2\xi_\rho+2z)\rho\pa_\rho+2\tilde{p}\pa_{\xi_\rho}+(-2\xi_\rho+2z)\sum_{\mu=\theta,\varphi}\xi_\mu\pa_{\xi_\mu}-H_{\tilde{p}}+\rho V, \quad V\in\mathcal{V}_\bop(\fscform).
\]	
Since $p_{h,z}=0$ along the integral curves of ${}^{\scop}\!H_{p_{h,z}}^{2,0}$, we calculate 
	\begin{align*}
		\jb{\xi}{}^{\scop}\!H_{p_{h,z}}^{2,0}\Big(\frac{\xi_\rho-z}{\rho}\Big)=\jb{\xi}\frac{2(\xi_\rho-z)^2+2\tilde{p}+\rho a_1}{\rho}=\jb{\xi}\frac{z^2+\rho a_2}{\rho},\quad a_1, a_2\in\mathcal{A}^0(\CX).
		\end{align*}
	Since $\rho(\phi( s))\to 0$ as $s\to\pm\infty$, then for any $\epsilon>0$, there exists $T>0$ such that $\rho(\phi( s))<\epsilon$ for all $\pm s>T$. Picking $\epsilon$ sufficiently small, we see that ${}^{\scop}\!H_{p_{h,z}}^{2,0}\Big(\frac{\xi_\rho-z}{\rho}\Big)\geq c_1>0$ for all $\pm s>T$. This implies that $(\xi_\rho-z)/\rho\to\pm\infty$ as $s\to\pm\infty$; in particular $\pm(\xi_\rho-z)\geq 0$ for all $\pm s>T_1$ where $T_1>0$ is a sufficiently large constant.

According to Lemma \ref{lem:radialpointatinftysemi}, there exist neighborhoods $V^{\epsilon'}_\pm=\{(\xi_\rho-(z\pm(\sgn z)z)))^2+\tilde{p}+\rho^2\leq \epsilon'\}$ of $R(\frac{z}{2}\pm\frac{\abs{z}}{2})$ such that once $\phi(s)$ enters $V^{\epsilon'}_\pm$ at some time $\pm s_0>0$, it will converge to $R(\frac{z}{2}\pm\frac{\abs{z}}{2})$ as $\pm s\to\infty$. Suppose that $\phi(s)\notin V^{\epsilon'}_\pm$ for all $\pm s\geq 0$. Since $p_{h,z}=0$ along the integral curves of ${}^{\scop}\!H_{p_{h,z}}^{2,0}$, it follows that 
\begin{equation}\label{eq:awayfromradialpt}
\abs{\xi_\rho-z}\leq \delta_1\quad \tilde{p}>\delta_2\quad \mbox{for all} \quad \pm s>T_2
\end{equation}
for some $\delta_1, \delta_2>0,T_2>T_1$. Then for all $\pm s>T_2$, we have
\[
{}^{\scop}\!H_{p_{h,z}}^{2,0}(\xi_\rho-z)=\jb{\xi}^{-1}(2\tilde{p}+\rho a_2)>c_2>0,\quad a_2\in\mathcal{A}^0(\CX).
	\]
	which contradicts \eqref{eq:awayfromradialpt}. Therefore, $\phi(s)$ must enter $V^{\epsilon'}_\pm$ at some time $\pm s_0>0$ and this finishes the proof of the lemma.
\end{proof}
\end{lem}
Now we are at the position to describe the the global behavior of the flow of $	{}^{\scop}\!H_{p_{h,z}}^{2,0}:=\rho^{-1}\jb{\xi}^{-1}{}^{\scop}\!H_{p_{h,z}}$ on the semiclassical characteristic set $\Sigma_h=\{\jb{\xi}^{-2}p_{h,z}=0\}$.

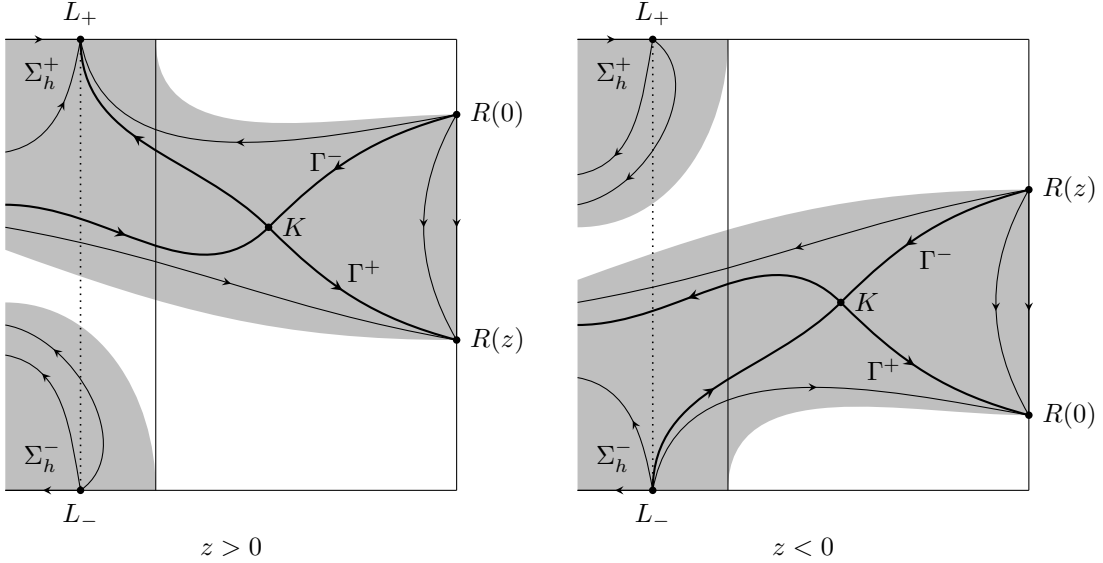
\begin{figure}[!h]
	\begin{tikzpicture}
		\fill [color=gray!50] (-3, 0.2) to [out=-20, in=180] (3,-1)--(3,2) to [out=180, in=-90] (-1,3)--(-3,3)--cycle;
		\fill [color=gray!50] (-1, -3) -- (-3,-3) --(-3,-0.5) to [out=0, in=90] (-1,-3);
		
		\draw (-3,3)--(3,3);
		\draw (-3,-3)--(3,-3);

		\draw (-1,3)--(-1,-3);
		\draw (3,3)--(3,-3);
		\draw [dotted, semithick](-2,3)--(-2,-3);
		
		\node[circle,inner sep=1pt,fill=black,label=above:{$L_+$}] at (-2,3) {};
		\node[circle,inner sep=1pt,fill=black,label=below:{$L_-$}] at (-2,-3) {};
		\node[circle,inner sep=1pt,fill=black,label=right:{$R(z)$}] at (3,-1) {};
		\node[circle,inner sep=1pt,fill=black,label=right:{$R(0)$}] at (3,2) {};
		\node[circle,inner sep=1pt,fill=black,label=right:{$K$}] at (0.5,0.5) {};
		
		\draw[
		decoration={markings, mark=at position 0.5 with {\arrow{stealth}}},
		postaction={decorate}]
		(-2, -3)--(-3,-3);
		
		\draw [
		decoration={markings, mark=at position 0.5 with {\arrow{stealth}}},
		postaction={decorate}]
		(-3,3)--(-2,3);
		
		\draw[ decoration={markings, mark=at position 0.5 with {\arrow{stealth}}},
		postaction={decorate}]
		(3,2)--(3,-1);

		\draw[
		decoration={markings, mark=at position 0.75 with {\arrow{stealth}}},
		postaction={decorate}
		] (-2, -3) to [out=35, in=-10] (-3,-0.8);
		\draw[
		decoration={markings, mark=at position 0.75 with {\arrow{stealth}}},
		postaction={decorate}
		] (-2, -3) to [out=90+10, in=-10]  (-3, -1.2);	
		
		\draw[ decoration={markings, mark=at position 0.5 with {\arrow{stealth}}},
		postaction={decorate}
		]
		(3,2) to [out=-90-30, in=90+30](3,-1); 	
	
		\draw[thick, decoration={markings, mark=at position 0.25 with {\arrow{stealth}}, mark=at position 0.75 with {\arrow{stealth reversed}}},
		postaction={decorate}
		]
		(-3,0.8) to [out=0, in=-90-45] (0.5,0.5) to [out=45, in=-180+15] (3,2);

		\draw[thick, decoration={markings, mark=at position 0.25 with {\arrow{stealth reversed}}, mark=at position 0.75 with {\arrow{stealth}}},
		postaction={decorate}
		]
		(-2,3) to [out=-90, in=180-45] (0.5,0.5) to [out=-45, in=180-15] (3,-1);  
	
		\draw[ decoration={markings, mark=at position 0.5 with {\arrow{stealth reversed}}},
		postaction={decorate}
		]
		(-2,3) to [out=-90+10, in=-180+10]  (3,2);  
		\draw[ decoration={markings, mark=at position 0.5 with {\arrow{stealth reversed}}},
		postaction={decorate}
		]
		(-2,3) to [out=-90-10, in=+10]  (-3,1.5);
		
		\draw[ decoration={markings, mark=at position 0.5 with {\arrow{stealth}}},
		postaction={decorate}
		]
		(-3,0.5) to [out=-10, in=180-10]  (3,-1);  
		
			\node[label= above right:{$\Gamma^-$}] at (0.8, 1) {};	
		\node[label= below right:{$\Gamma^+$}] at (1.3,0.3) {};	
		\node[label= above right:{$\Sigma_h^-$}] at (-3, -3) {};	
		\node[label= below right:{$\Sigma_h^+$}] at (-3, 3) {};	
		\node at (0, -3.75) {$z>0$};
	\end{tikzpicture}
	\quad 
	\begin{tikzpicture}
		\fill [color=gray!50] (-3, -0.2) to [out=20, in=-180] (3,1)--(3,-2) to [out=-180, in=90] (-1,-3)--(-3,-3)--cycle;
		\fill [color=gray!50] (-1, 3) -- (-3,3) --(-3,0.5) to [out=0, in=-90] (-1,3);
		
		\draw (-3,3)--(3,3);
		\draw (-3,-3)--(3,-3);

		\draw (-1,3)--(-1,-3);
		\draw (3,3)--(3,-3);
		\draw [dotted, semithick](-2,3)--(-2,-3);
		
		\node[circle,inner sep=1pt,fill=black,label=above:{$L_+$}] at (-2,3) {};
		\node[circle,inner sep=1pt,fill=black,label=below:{$L_-$}] at (-2,-3) {};
		\node[circle,inner sep=1pt,fill=black,label=right:{$R(z)$}] at (3,1) {};
		\node[circle,inner sep=1pt,fill=black,label=right:{$R(0)$}] at (3,-2) {};
		\node[circle,inner sep=1pt,fill=black,label=right:{$K$}] at (0.5,-0.5) {};
		
		\draw[
		decoration={markings, mark=at position 0.5 with {\arrow{stealth}}},
		postaction={decorate}]
		(-2, -3)--(-3,-3);
		
		\draw [
		decoration={markings, mark=at position 0.5 with {\arrow{stealth}}},
		postaction={decorate}]
		(-3,3)--(-2,3);
		
		\draw[ decoration={markings, mark=at position 0.5 with {\arrow{stealth reversed}}},
		postaction={decorate}]
		(3,-2)--(3,1);

		\draw[
		decoration={markings, mark=at position 0.75 with {\arrow{stealth}}},
		postaction={decorate}
		] (-2, 3) to [out=-35, in=10] (-3,0.8);
		\draw[
		decoration={markings, mark=at position 0.75 with {\arrow{stealth}}},
		postaction={decorate}
		] (-2, 3) to [out=-90-10, in=10]  (-3, 1.2);	
		
		\draw[ decoration={markings, mark=at position 0.5 with {\arrow{stealth reversed}}},
		postaction={decorate}
		]
		(3,-2) to [out=90+30, in=-90-30](3,1);

		\draw[thick, decoration={markings, mark=at position 0.25 with {\arrow{stealth reversed}}, mark=at position 0.75 with {\arrow{stealth}}},
		postaction={decorate}
		]
		(-3,-0.8) to [out=0, in=90+45] (0.5,-0.5) to [out=-45, in=180-15] (3,-2);
		\draw[thick, decoration={markings, mark=at position 0.25 with {\arrow{stealth}}, mark=at position 0.75 with {\arrow{stealth reversed}}},
		postaction={decorate}
		]
		(-2,-3) to [out=90, in=-180+45] (0.5,-0.5) to [out=45, in=-180+15] (3,1);  
		\draw[ decoration={markings, mark=at position 0.5 with {\arrow{stealth}}},
		postaction={decorate}
		]
		(-2,-3) to [out=90-10, in=180-10]  (3,-2);  
		
		\draw[ decoration={markings, mark=at position 0.5 with {\arrow{stealth}}},
		postaction={decorate}
		]
		(-2,-3) to [out=90+10, in=-10]  (-3,-1.5);
		
		\draw[ decoration={markings, mark=at position 0.5 with {\arrow{stealth reversed}}},
		postaction={decorate}
		]
		(-3,-0.5) to [out=10, in=-180+10]  (3,1);  
		
			\node[label= below right:{$\Gamma^+$}] at (0.6, -1) {};	
		\node[label= above right:{$\Gamma^-$}] at (1.3,-0.3) {};	
		\node[label= above right:{$\Sigma_h^-$}] at (-3, -3) {};	
		\node[label= below right:{$\Sigma_h^+$}] at (-3, 3) {};	
		\node at (0, -3.75) {$z<0$};
	\end{tikzpicture}
	\caption{The flow of $\exp(s{}^{\scop}\!H_{p_{h,z}}^{2,0})$, which is projected to the coordinates $(r, \xi_r)$ and drawn in a fiber-radially compactified view. The shaded region is the semiclassical characteristic set $\Sigma_h=\Sigma^+_h\sqcup\Sigma^-_h=\{\jb{\xi}^{-2}p_{h,z}=0\}$. The horizontal coordinate is $r$ and the rightmost vertical line corresponds to $\pa_+\CX$. The dotted line is the even horizon $r=r_b$ and the vertical line in the middle is $\{r=r_{\mathrm{ergo}}\}$ where $r_{\mathrm{ergo}}$ is the equatorial radius (i.e. greatest radius) of the ergosphere. The vertical coordinate is $\xi_r/\jb{\xi_r}$, so the top and bottom lines stand for the fiber infinity.
	}
	\label{fig:semiphase}
\end{figure}
\begin{prop}\label{prop:semiglobaldy}
Let $\pm z>0$. Let $\phi(s)=\exp(s{}^{\scop}\!H_{p_{h,z}}^{2,0})(\rho,\theta, \varphi,\xi)$ be the maximally extended integral curve of ${}^{\scop}\!H_{p_{h,z}}^{2,0}$ on $\Sigma_h$ with the domain of definition $s\in I\subset\BR$. Let $s_-=\inf I, s_+=\sup I$. (see Figure \ref{fig:semiphase}). 
\begin{enumerate}
	\item Let $z>0$.
	\begin{itemize}
		\item [(\romannumeral 1)]
	If $\phi(s)\subset\Sigma^-_h$, then either $\phi(s)\subset L_-$, or $\phi(s)\to L_-$ as $s\to s_-=-\infty$ and $\phi(s)$ crosses $\pa_-\CX$ into the inward direction of deceasing $r$ at finite time $s_+<\infty$.
	\item[(\romannumeral 2)] If $\phi(s)\subset\Sigma^+_h$, then either $\phi(s)\subset L_+\cup R(z)\cup R(0)\cup K$, or $\phi(s)\to L_+\cup R(z)\cap K$ as $s\to s_+=\infty$ and $\phi(s)\to R(0)\cup K$ as $s\to s_-=-\infty$ or $\phi(s)$ crosses $\pa_-\CX$ into the inward direction of deceasing $r$ at finite time $s_->-\infty$. Moreover, if $(\rho,\theta,\varphi,\xi)\notin K$, then $\phi(s)$ cannot converge to $K$ in both the forward and backward direction.
\end{itemize}
\item Let $z<0$.
\begin{itemize}
	\item [(\romannumeral 1)]
If $\phi(s)\subset\Sigma^+_h$, then either $\phi(s)\subset L_+$, or $\phi(s)\to L_+$ as $s\to s_+=\infty$ and $\phi(s)$ crosses $\pa_-\CX$ into the inward direction of deceasing $r$ at finite time $s_->-\infty$.
\item[(\romannumeral 2)] If $\phi(s)\subset\Sigma^-_h$, then either $\phi(s)\subset L_-\cup R(z)\cup R(0)\cup K$, or $\phi(s)\to L_-\cup R(z)\cup K$ as $s\to s_-=-\infty$ and $\phi(s)\to R(0)\cup K$ as $s\to s_+=\infty$ or $\phi(s)$ crosses $\pa_-\CX$ into the inward direction of deceasing $r$ at finite time $s_+<\infty$. Moreover, if $(\rho,\theta,\varphi,\xi)\notin K$, then $\phi(s)$ cannot converge to $K$ in both the forward and backward direction.
\end{itemize}
\end{enumerate}

\begin{proof}
	We only consider the case $z>0$ as the other case $z<0$ can be proved similarly. First, we note that $L_+, L_-, R(0), R(z), K$ are invariant under the flow $\phi(s)$ and if $(\rho,\theta,\varphi,\xi)\in \Gamma^\pm$, then $\phi(s)\to K$ as $\mp s\to\infty$. If $(\rho,\theta,\varphi,\xi)\notin L_+\cup L_-\cup R(0)\cup R(z)\cup K$, then either $(\rho,\theta,\varphi,\xi)\notin\Gamma^-$ or $(\rho,\theta,\varphi,\xi)\notin\Gamma^+$.
	
	If $(\rho,\theta,\varphi,\xi)\notin\Gamma^-$, by Lemma \ref{lem:signofderiofr}, for $\phi(s)\subset \Sigma_h^+$, either $\phi(s)\to L_+$ or $\rho(\phi(s))\to 0$ as $s\to\infty$, while for $\phi(s)\subset \Sigma_h^-$, $\phi(s)$ crosses $\pa_-\CX$ into the inward direction of deceasing $r$ at some finite time $s_0<\infty$. If $\rho(\phi(s))\to 0$ as $s\to\infty$, according to Lemma \ref{lem:technicallemmanearinfty}, we have $\phi(s)\to R(z)$ as $s\to\infty$.
	
		If $(\rho,\theta,\varphi,\xi)\notin\Gamma^+$, by Lemma \ref{lem:signofderiofr}, for $\phi(s)\subset \Sigma_h^+$, either $\rho(\phi(s))\to 0$ as $s\to\infty$ or $\phi(s)$ crosses $\pa_-\CX$ into the inward direction of deceasing $r$ at some finite time $s_0>-\infty$, while for $\phi(s)\subset \Sigma_h^-$, $\phi(s)\to L_-$ as $s\to-\infty$. If $\rho(\phi(s))\to 0$ as $s\to-\infty$, according to Lemma \ref{lem:technicallemmanearinfty}, we have $\phi(s)\to R(0)$ as $s\to-\infty$.
\end{proof}
\end{prop}

\subsection{High energy estimates}\label{subsec:highenergyestimates}
In this section, we prove the high energy estimates first for operators $h^2\widehat{\Box_{g_{b}}}(h^{-1}z)$ acting on scalar functions, then for $h^2\widehat{\mathcal{P}_{b, \gamma}}(h^{-1}z), h^2\widehat{\mathcal{W}_{b, \gamma}}(h^{-1}z)$ acting on scattering $1$-forms and the linearized gauge-fixed Einstein-Maxwell operator $h^2\widehat{L_{b,\gamma}}(h^{-1}z)$, as well as their formal adjoints.
%with respect to volume density $L^2(\CX;\sqrt{\det\abs{g_b}}drd\theta d\varphi)$. 
To this end, we combine the global dynamics of of the Hamiltonian flow of the semiclassical principal symbol $p_{h,z}$ established in \ref{prop:semiglobaldy} with the elliptic estimate, propagation of singularities estimate, the radial point estimate at event horizon, the scattering radial point estimate at spatial infinity $\pa_+\CX$ and hyperbolic estimate, all of which are in the semiclassical version, together with the estimate at normally hyperbolic trapping.

\subsubsection{	High energy estimates for scalar wave operators}
We first prove high energy estimates for the operator $h^2\widehat{\Box_{g_{b}}}(h^{-1}z)$ acting on scalar functions, as well as its formal adjoint with respect to volume density $L^2(\CX;\sqrt{\det\abs{g_b}}drd\theta d\varphi)$.

\begin{figure}[!h]
	\begin{tikzpicture}
		\fill [color=gray!50] (-2.5, 3) to [out=-90, in=180] (-2,2.5) to [out=0, in=-90] (-1.5,3);
			\node[label=below:{$B_+$}] at (-2.5,3) {};
		\fill [color=gray!50] (-2.5, -3) to [out=90, in=180] (-2,-2.5) to [out=0, in=90] (-1.5,-3);
			\node[label=above:{$B_-$}] at (-2.6,-3) {};
		\fill[color=gray!50] (3,2.3)[out=180, in=90] to (2.7, 2) [out=-90, in=180] to (3, 1.7);
		\node[label=left:{$B_0$}] at (3.2,2.4) {};
			\fill[color=gray!50] (3,2.7)--(3,2.8)[out=180, in=90] to (2.2, 2) [out=-90, in=180] to (3, 1.2)--(3,1.3) [out=180, in=-90] to (2.3, 2) [out=90,in=180] to (3,2.7);
		\node[label=left:{$E_0$}] at (2.4,2) {};
		\fill[color=gray!50] (3,-0.5)[out=180, in=90] to (2.5, -1) [out=-90, in=180] to (3, -1.5);
		\node[label=left:{$B_z$}] at (2.9,-1.5) {};
		\filldraw[color=gray!50] (0.5,0.5) circle (8pt);
		\node[label=above:{$B_1$}] at (0.5,0.6) {};
			\filldraw[color=gray!50] (-0.2,1) circle (8pt);
		\node[label=below left:{$B_2$}] at (-0.1,1) {};
			\filldraw[color=gray!50] (1.2,-0.1) circle (8pt);
		\node[label=above right:{$B_2$}] at (1.2,-0.1) {};
		\draw (-3,3)--(3,3);
		\draw (-3,-3)--(3,-3);

		\draw (-1,3)--(-1,-3);
		\draw (3,3)--(3,-3);
		\draw [dotted, semithick](-2,3)--(-2,-3);
		
		\node[circle,inner sep=1pt,fill=black,label=above:{$L_+$}] at (-2,3) {};
		\node[circle,inner sep=1pt,fill=black,label=below:{$L_-$}] at (-2,-3) {};
		\node[circle,inner sep=1pt,fill=black,label=right:{$R(z)$}] at (3,-1) {};
		\node[circle,inner sep=1pt,fill=black,label=right:{$R(0)$}] at (3,2) {};
		\node[circle,inner sep=1pt,fill=black,label=right:{$K$}] at (0.5,0.5) {};
		
		\draw[
		decoration={markings, mark=at position 0.5 with {\arrow{stealth}}},
		postaction={decorate}]
		(-2, -3)--(-3,-3);
		
		\draw [
		decoration={markings, mark=at position 0.5 with {\arrow{stealth}}},
		postaction={decorate}]
		(-3,3)--(-2,3);
		
		\draw[ decoration={markings, mark=at position 0.5 with {\arrow{stealth}}},
		postaction={decorate}]
		(3,2)--(3,-1);

		\draw[
		decoration={markings, mark=at position 0.75 with {\arrow{stealth}}},
		postaction={decorate}
		] (-2, -3) to [out=35, in=-10] (-3,-0.8);
		\draw[
		decoration={markings, mark=at position 0.75 with {\arrow{stealth}}},
		postaction={decorate}
		] (-2, -3) to [out=90+10, in=-10]  (-3, -1.2);	
		
		\draw[ decoration={markings, mark=at position 0.5 with {\arrow{stealth}}},
		postaction={decorate}
		]
		(3,2) to [out=-90-30, in=90+30](3,-1); 	
		\draw[thick, decoration={markings, mark=at position 0.25 with {\arrow{stealth}}, mark=at position 0.75 with {\arrow{stealth reversed}}},
		postaction={decorate}
		]
		(-3,0.8) to [out=0, in=-90-45] (0.5,0.5) to [out=45, in=-180+15] (3,2);
		\draw[thick, decoration={markings, mark=at position 0.25 with {\arrow{stealth reversed}}, mark=at position 0.75 with {\arrow{stealth}}},
		postaction={decorate}
		]
		(-2,3) to [out=-90, in=180-45] (0.5,0.5) to [out=-45, in=180-15] (3,-1);  
		\draw[ decoration={markings, mark=at position 0.5 with {\arrow{stealth reversed}}},
		postaction={decorate}
		]
		(-2,3) to [out=-90+10, in=-180+10]  (3,2);  
		\draw[ decoration={markings, mark=at position 0.5 with {\arrow{stealth reversed}}},
		postaction={decorate}
		]
		(-2,3) to [out=-90-10, in=+10]  (-3,1.5);
		
		\draw[ decoration={markings, mark=at position 0.5 with {\arrow{stealth}}},
		postaction={decorate}
		]
		(-3,0.5) to [out=-10, in=180-10]  (3,-1);  
		
			\node[label= above left:{$\Gamma^-$}] at (-0.9, 0.3) {};	
		\node[label= above left:{$\Gamma^+$}] at (-1,1.1) {};	
		%\node[label= above right:{$\Sigma_h^-$}] at (-3, -3) {};	
		%\node[label= below right:{$\Sigma_h^+$}] at (-3, 3) {};	
		%\node at (0, -3.75) {$z>0$};
		 \draw [very thick,decorate,decoration={brace,amplitude=5pt,mirror},xshift=0pt,yshift=0pt]
		(-2.7,-3.4) -- (3,-3.4) node [black,midway,xshift=0cm,yshift=-0.5cm] 
		{$\mathrm{supp}\chi$};
	\end{tikzpicture}
	\quad 
	\begin{tikzpicture}
	\fill [color=gray!50] (-2.5, 3) to [out=-90, in=180] (-2,2.5) to [out=0, in=-90] (-1.5,3);
	\node[label=below:{$B'_+$}] at (-2.5,3) {};
		\fill[color=gray!50] (-3, 3)--(-3.1,3) to[out=-90,in=180] (-2,1.9)to [out=0,in=-90](-0.9,3)--(-1,3) to[out=-90,in=0] (-2,2) to [out=180,in=-90] (-3,3);
	\node[label=below:{$E'_+$}] at (-3,2.5){};
		\fill [color=gray!50] (-2.5, -3) to [out=90, in=180] (-2,-2.5) to [out=0, in=90] (-1.5,-3);
	\node[label=above:{$B'_-$}] at (-2.5,-3) {};
		\fill[color=gray!50] (-3, -3)--(-3.1,-3) to[out=90,in=180] (-2,-1.9)to [out=0,in=90](-0.9,-3)--(-1,-3) to[out=90,in=0] (-2,-2) to [out=180,in=90] (-3,-3);
		\node[label=above:{$E'_-$}] at (-3,-2.5){};
	\fill[color=gray!50] (3,2.3)[out=180, in=90] to (2.7, 2) [out=-90, in=180] to (3, 1.7);
	\node[label=left:{$B'_0$}] at (3.2,2.4) {};
	\fill[color=gray!50] (3,-0.5)[out=180, in=90] to (2.5, -1) [out=-90, in=180] to (3, -1.5);	
	\node[label=left:{$B'_z$}] at (3.1,-1.5) {};
		\fill[color=gray!50] (3,-0.1)--(3,0)[out=180, in=90] to (2, -1) [out=-90, in=180] to (3, -2)--(3,-1.9) [out=180, in=-90] to (2.1, -1) [out=90,in=180] to (3,-0.1);
	\node[label=left:{$E'_z$}] at (2.8,-2.1) {};
	\filldraw[color=gray!50] (0.5,0.5) circle (8pt);
	\node[label=above:{$B'_1$}] at (0.5,0.6) {};
	\filldraw[color=gray!50] (1.2,1.1) circle (8pt);
	\node[label=below right:{$B'_2$}] at (1.2,1.2) {};
	\filldraw[color=gray!50] (-0.2,0.1) circle (8pt);
	\node[label=above left:{$B'_2$}] at (-0.2,0) {};
	\draw (-4,3)--(3,3);
	\draw (-4,-3)--(3,-3);

	\draw (-1,3)--(-1,-3);
	\draw (3,3)--(3,-3);
	\draw [dotted, semithick](-2,3)--(-2,-3);
	
	\node[circle,inner sep=1pt,fill=black,label=above:{$L_+$}] at (-2,3) {};
	\node[circle,inner sep=1pt,fill=black,label=below:{$L_-$}] at (-2,-3) {};
	\node[circle,inner sep=1pt,fill=black,label=right:{$R(z)$}] at (3,-1) {};
	\node[circle,inner sep=1pt,fill=black,label=right:{$R(0)$}] at (3,2) {};
	\node[circle,inner sep=1pt,fill=black,label=right:{$K$}] at (0.5,0.5) {};
	
	\draw[
	decoration={markings, mark=at position 0.5 with {\arrow{stealth}}},
	postaction={decorate}]
	(-2, -3)--(-4,-3);
	
	\draw [
	decoration={markings, mark=at position 0.5 with {\arrow{stealth}}},
	postaction={decorate}]
	(-4,3)--(-2,3);
	
	\draw[ decoration={markings, mark=at position 0.5 with {\arrow{stealth}}},
	postaction={decorate}]
	(3,2)--(3,-1);

	\draw[
	decoration={markings, mark=at position 0.75 with {\arrow{stealth}}},
	postaction={decorate}
	] (-2, -3) to [out=35, in=-10] (-4,-0.8);
	\draw[
	decoration={markings, mark=at position 0.75 with {\arrow{stealth}}},
	postaction={decorate}
	] (-2, -3) to [out=90+10, in=-10]  (-4, -1.2);	
	
	\draw[ decoration={markings, mark=at position 0.5 with {\arrow{stealth}}},
	postaction={decorate}
	]
	(3,2) to [out=-90-30, in=90+30](3,-1); 	
	\draw[thick, decoration={markings, mark=at position 0.25 with {\arrow{stealth}}, mark=at position 0.75 with {\arrow{stealth reversed}}},
	postaction={decorate}
	]
	(-4,0.8) to [out=0, in=-90-45] (0.5,0.5) to [out=45, in=-180+15] (3,2);
	\draw[thick, decoration={markings, mark=at position 0.25 with {\arrow{stealth reversed}}, mark=at position 0.75 with {\arrow{stealth}}},
	postaction={decorate}
	]
	(-2,3) to [out=-90, in=180-45] (0.5,0.5) to [out=-45, in=180-15] (3,-1);  
	\draw[ decoration={markings, mark=at position 0.5 with {\arrow{stealth reversed}}},
	postaction={decorate}
	]
	(-2,3) to [out=-90+10, in=-180+10]  (3,2);  
	\draw[ decoration={markings, mark=at position 0.5 with {\arrow{stealth reversed}}},
	postaction={decorate}
	]
	(-2,3) to [out=-90-10, in=+10]  (-4,1.5);
	
	\draw[ decoration={markings, mark=at position 0.5 with {\arrow{stealth}}},
	postaction={decorate}
	]
	(-4,0.2) to [out=-10, in=180-10]  (3,-1);  
		\node[label= above left:{$\Gamma^-$}] at (-0.9, 0) {};	
	\node[label= above left:{$\Gamma^+$}] at (-0.8,1) {};	
	
	%\node[label= above right:{$\Sigma_h^-$}] at (-3, -3) {};	
	%\node[label= below right:{$\Sigma_h^+$}] at (-3, 3) {};	
	%\node at (0, -3.75) {$z>0$};
 \draw [very thick,decorate,decoration={brace,amplitude=5pt,mirror},xshift=0pt,yshift=0pt]
(-3.2,-3.4) -- (3,-3.4) node [black,midway,xshift=0cm,yshift=-0.5cm] 
{$\mathrm{supp}\chi$};
\draw [very thick,decorate,decoration={brace,amplitude=5pt,mirror},xshift=0pt,yshift=0pt]
(-4,-3.1)--(-3,-3.1) node [black, near start, xshift=0cm,yshift=-0.6cm] 
{$\mathrm{supp}(1-\chi)$};
\end{tikzpicture}
	\caption{A phase space picture of the proof of the estimate \eqref{eq:highenergysecondmic}, on the left, and \eqref{eq:highenergysecondmicdual}, on the right. The coordinates and notations $L_\pm, R(0), R(z),\Gamma^\pm,K$ are the same as in Figure \ref{fig:semiphase}. For \eqref{eq:highenergysecondmic}, we use semiclassical hyperbolic estimates to control $u$ via $\chi u$; $\chi$ is controlled (modulo elliptic estimates) by $B_\pm, B_1, B_0, B_z$; $B_0$ is controlled by $E_0$ using low b-decay radial point estimates and $E_0$ is again controlled by $B_+, B_1, B_z$; $B_1$ is controlled by $B_2$ by using normally hyperbolic trapping estimate and $B_2$ is controlled by $B_+,B_z$; finally $B_\pm, B_z$ are controlled using high regularity radial points estimates at $L_\pm$ and high sc-decay radial point estimates at $R(z)$. For \eqref{eq:highenergysecondmicdual}, we use semiclassical hyperbolic estimates to bound $(1-\chi)u$; $\chi$ is controlled (modulo elliptic estimates) by $B'_\pm, B'_1, B'_0, B'_z$; $B'_\pm$ is controlled by $E'_\pm$ using low regularity radial point estimates and $E'_\pm$ is controlled by $B'_1, B_0', 1-\chi$; $B'_z$ is controlled by $E'_z$ using low sc-decay radial points estimates and $E'_z$ is controlled by $B_0', B_1', 1-\chi$; $B'_1$ is controlled by $B'_2$ by using normally hyperbolic trapping estimate and $B'_2$ is controlled by $B'_+, 1-\chi$;  finally $B'_0$ is controlled using high b-decay radial points estimates at $R(0)$.
	}
	\label{fig:semiphaseest}
\end{figure}
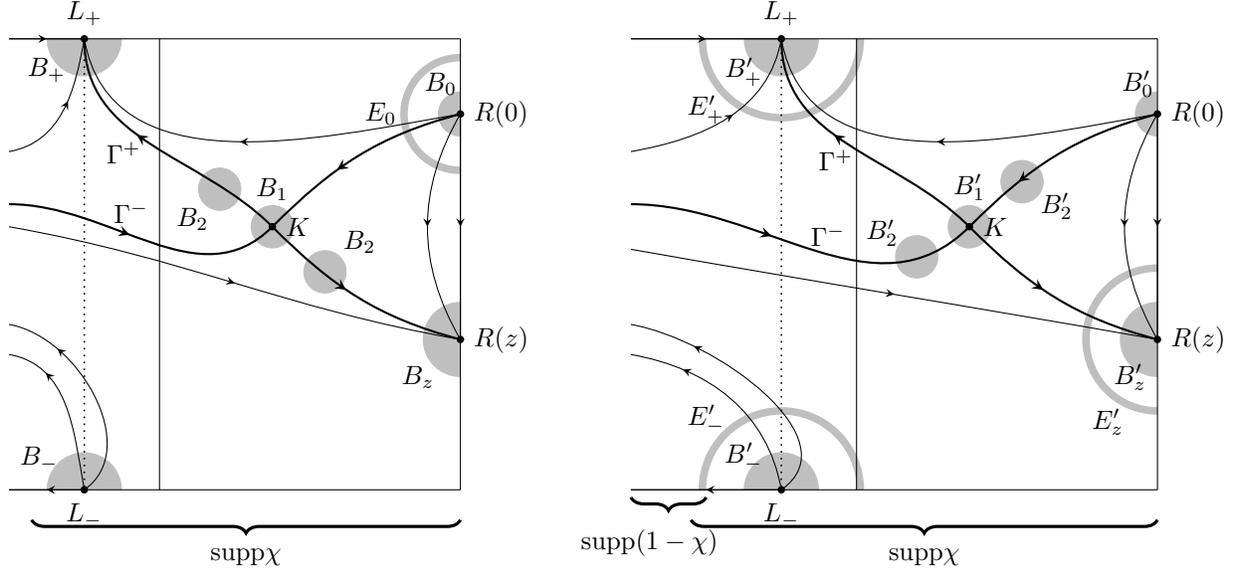
\begin{thm}\label{thm:highenergy}
Let $b_0=(\Bm_0, \Ba_0, \BQ_0)$ with $\abs{\BQ_0}+\abs{\Ba_0}<\Bm_0$. Then there exists $\epsilon>0$ such that for $b=(\Bm, \Ba,\BQ)$ with $\abs{b-b_0}<\epsilon$ and for $s>\frac{1}{2}, \ell<-\frac{1}{2}$ with $s+\ell>-\frac12$, the following holds. For any fixed $C_1>0$, there exist $C_2>1, C>0$ which are independent of $b$, such that for $\sigma\in\BC, \IM\sigma\in[0, C_1], \abs{\RE\sigma}>C_2$, we have
\begin{align}\label{eq:highenergy1}
	\norm{u}_{\bar{H}_{\bop,h}^{s,\ell}(\CX)}&\leq C\norm{\widehat{\Box_{g_{b}}}(\sigma)u}_{\bar{H}_{\bop,h}^{s,\ell+1}(\CX)},\quad\norm{u}_{\dot{H}_{\bop,h}^{-s,-\ell-1}(\CX)}\leq C\norm{\widehat{\Box_{g_{b}}}(\sigma)^*u}_{\dot{H}_{\bop,h}^{-s,-\ell}(\CX)}; \\\label{eq:highenergy2}
	\norm{u}_{\bar{H}_{\bop,h}^{s,\ell}(\CX)}&\leq C\norm{\widehat{\Box_{g_{b}}}(\sigma)u}_{\bar{H}_{\bop,h}^{s-1,\ell+2}(\CX)},\quad \norm{u}_{\dot{H}_{\bop,h}^{-s+1,-\ell-2}(\CX)}\leq C\norm{\widehat{\Box_{g_{b}}}(\sigma)^*u}_{\dot{H}_{\bop,h}^{-s,-\ell}(\CX)},
	\end{align}
where $h=\abs{\sigma}^{-1}$ and $C$ only depends on $b_0, s,\ell, C_1$.

Therefore, \[
\widehat{\Box_{g_{b}}}:\{u\in\bar{H}_{\bop,h}^{s,\ell}(\CX)\mid\bar{H}_{\bop,h}^{s,\ell+1}(\CX)\}\to\bar{H}_{\bop,h}^{s,\ell+1}(\CX)
\quad\mbox{and}\quad
\widehat{\Box_{g_{b}}}:\{u\in\bar{H}_{\bop,h}^{s,\ell}(\CX)\mid\bar{H}_{\bop,h}^{s-1,\ell+2}(\CX)\}\to\bar{H}_{\bop,h}^{s-1,\ell+2}(\CX)
\]
are invertible for $\sigma\in\BC, \IM\sigma\in[0, C_1], \abs{\RE\sigma}>C_2$.
\end{thm}
\begin{proof}
For the simplicity of notations, we let $P_h(z):=h^2\widehat{\Box_{g_{b}}}(h^{-1}z)$.	We first claim that it suffices to prove the following estimates for $s>\frac 12, r>-\frac12, \ell<-\frac 12$
	\begin{equation}\label{eq:highenergysecondmic}
		\norm{u}_{\bar{H}_{\scop,\bop,h}^{s,r,\ell}(\CX)}\leq Ch^{-2}	\norm{P_h(z)u}_{\bar{H}_{\scop,\bop,h}^{s-1,r+1,\ell+1}(\CX)}
		\end{equation}
	and
	\begin{equation}\label{eq:highenergysecondmicdual}
	\norm{u}_{\dot{H}_{\scop,\bop,h}^{-s+1,-r-1,-\ell-1}(\CX)}\leq Ch^{-2}	\norm{P_h(z)^*u}_{\dot{H}_{\scop,\bop,h}^{-s,-r,-\ell}(\CX)}
	\end{equation}
	Since $\bar{H}_{\scop,\bop,h}^{s,s+\ell,\ell}=\bar{H}_{\bop,h}^{s,l}$ and $\dot{H}_{\scop,\bop,h}^{s,s+\ell,\ell}=\dot{H}_{\bop,h}^{s,l}$, letting $r=s+\ell>-\frac 12$ we have 
	\[
	\norm{u}_{\bar{H}_{\scop,\bop,h}^{s,s+\ell,\ell}(\CX)}\leq Ch^{-2}	\norm{P_h(z)u}_{\bar{H}_{\scop,\bop,h}^{s-1,s+\ell+1,\ell+1}(\CX)}
	\]
	and 
	\[
		\norm{u}_{\dot{H}_{\scop,\bop,h}^{-s+1,-s-r-1,-\ell-1}(\CX)}\leq Ch^{-2}	\norm{P_h(z)^*u}_{\dot{H}_{\scop,\bop,h}^{-s,-s-r,-\ell}(\CX)}.
	\]
Then using the facts that 
\begin{gather*} \bar{H}_{\bop,h}^{s,\ell+1}=\bar{H}^{s,s+\ell+1, \ell+1}_{\scop,\bop,h}\subset \bar{H}^{s-1,s+\ell+1,\ell+1}_{\scop,\bop,h},\quad\bar{H}_{\bop,h}^{s-1,\ell+2}=\bar{H}^{s-1,s+\ell+1, \ell+2}_{\scop,\bop,h}\subset \bar{H}^{s-1,s+\ell+1,\ell+1}_{\scop,\bop,h},\\
	\dot{H}_{\bop,h}^{-s,-\ell-1}=\dot{H}^{-s,-s-\ell-1, -\ell-1}_{\scop,\bop,h}\supset \dot{H}^{-s+1,-s-\ell-1,-\ell-1}_{\scop,\bop,h},\\ \dot{H}_{\bop,h}^{-s+1,-\ell-2}=\dot{H}^{-s+1,-s-\ell-1,-\ell-2}_{\scop,\bop,h}\supset \dot{H}^{-s+1,-s-\ell-1,-\ell-1}_{\scop,\bop,h},
\end{gather*}
 we obtain \eqref{eq:highenergy1} and \eqref{eq:highenergy2}.  Now we will show \eqref{eq:highenergysecondmic} and \eqref{eq:highenergysecondmicdual}. Here we only discuss the case $\RE z>0$ (see Figure \ref{fig:semiphaseest} for a phase space illustration of the proof of estimates \eqref{eq:highenergysecondmic} and \eqref{eq:highenergysecondmicdual}), as the case $\RE z<0$ can be handled in a similar manner. 
 \begin{itemize}
 	\item \underline{Proof of the estimate \eqref{eq:highenergysecondmic}.} For $s>s_0>\frac 12\geq\frac{1}{2}-\IM(h^{-1}z)\frac{\ehKN^2+a^2}{\ehKN^2-\Bm}$ (as calculated in \eqref{eq:semithresholdreg} and \eqref{eq:thresholdreg}), using the semiclassical high regularity radial point estimate (see \cite[Proposition 2.10]{Vas13}, \cite[Proposition 5.27]{Vas18}), there exist $B_\pm, G_\pm, S_\pm \in\Psi^0_h(\CX)$ microlocally supported near $L_\pm$ with $L_\pm\subset \mathrm{ell}_h(B_\pm), L_\pm\subset \mathrm{ell}_h(S_\pm)$ and satisfying that all the forward (backward) null characteristics from $\mathrm{WF}_h(B_\pm)$ tend to $L_\pm$, while remaining in the elliptic set of $G_\pm$, such that
 	\begin{equation}\label{eq:highregradiales}
 		\norm{B_\pm u}_{H^{s,r,\ell}_{\scop,\bop,h}(\CX)}\leq C h^{-1}\norm{G_\pm P_h(z)u}_{H^{s-1, r+1,\ell+1}_{\scop,\bop,h}(\CX)}+Ch^{s-s_0}\norm{S_\pm u}_{H^{s_0,r_0,\ell_0}_{\scop,\bop,h}(\CX)}
 	\end{equation}
 where $r_0<r, \ell_0<\ell$ and the sc-decay order $r$ and b-decay order $\ell$ are actually irrelevant here.
 
 For $r>r_0>-\frac 12$ (as calculated in \eqref{eq:thresholdscdecay}), using the semiclassical high sc-decay radial point estimates at the non-zero scattering section $R(z)$ (see \cite{Vas21a}. The proof follows from a positive commutator estimate
 \[
 i(P_h(z)^*A-AP_h(z))=\IM P_h(z)A+A\IM P_h(z)+i[\RE P_h(z), A]\quad
 \]
where 
\[
\RE P_h(z)=\frac{P_h(z)+P_h(z)^*}{2},\quad\IM P_h(z)=\frac{P_h(z)-P_h(z)^*}{2i}.
\]
Let $A\in\Psi_{\scop,\bop,h}^{2s-1,2r+1,2l+1}(\CX)$ with semiclassical principal symbol \[
a=\chi_0(\xi_\theta^2+\frac{1}{\sin^2\theta}\xi_\varphi^2)\chi_1((\xi_\rho-2\RE z)^2)\rho^{-2r+1}(\xi_\theta^2+\frac{1}{\sin^2\theta}\xi_\varphi^2+\xi_\rho^2)^{s-\frac{1}{2}}
\]
where $\chi_0, \chi_1$ are identically 0 near $0$ and have compact support sufficiently close to $0$, and $\chi_1$ has relatively large support such that $\mbox{supp}\chi_0\cap \mbox{supp}\chi_1'$ is disjoint from the characteristic set of $\RE P_h(z)$. Then using Lemma 4.1 and the calculation before Lemma 4.8 in \cite{Vas21a}, it follows that $h^{-1}(\IM P_h(z)A+A\IM P_h(z)+i[\RE P_h(z), A])\in \Psi_{\scop,\bop,h}^{2s,2r,2l}(\CX)$ whose semiclassical principal symbol is positive definite elliptic near $R(z)$ if $r>-\frac{1}{2}$. This, together with a regularization argument,  proves semiclassical high sc-decay radial point estimates at $R(z)$), there exist $B_z, G_z, S_z \in\Psi^{0,0}_{\scop,h}(\CX)$ microlocally supported near $R(z)$ with $R(z)\subset \mathrm{ell}_{\scop,h}(B_z), R(z)\subset\mathrm{ell}_{\scop,h}(S_z)$ and satisfying that all the forward null characteristics from $\mathrm{WF}_{\scop,h}(B_z)$ tend to $R(z)$, while remaining in the scattering elliptic set of $G_z$, such that
 \begin{equation}\label{eq:highscdecayradiales}
 	\norm{B_z u}_{H^{s,r,\ell}_{\scop,\bop,h}(\CX)}\leq C h^{-1}\norm{G_zP_h(z)u}_{H^{s-1, r+1,\ell+1}_{\scop,\bop,h}(\CX)}+Ch^{s-s_0}\norm{S_z u}_{H^{s_0,r_0,\ell_0}_{\scop,\bop,h}(\CX)}
 \end{equation}
 where $s_0<s, \ell_0<\ell$ and the differential order $s$ and b-decay order $\ell$ are actually irrelevant here.
 
  For $\ell<-\frac 12$ (as calculated in \eqref{eq:thresholdscdecay}), using the semiclassical low b-decay radial point estimates at the zero scattering section $R(0)$ (see \cite[\S 4--5]{Vas21a}), there exist $B_0, G_0\in\Psi^{0,0,0}_{\scop,\bop,h}(\CX)$ microlocally supported near $R(0)$ (in fact near the blown up $R(0)$ in the second microlocalized space introduced in \S\ref{subsubsec:secondmic}) with $R(0)\subset \mathrm{ell}_{\scop,\bop,h}(B_0)$ and $E_0\in\Psi_{\scop,\bop,h}^{0,0,0}(\CX), \mathrm{WF}_{\scop,\bop,h}(E_0)\cap R(0)=\emptyset$, and satisfying that all the forward null characteristics from $\mathrm{WF}_{\scop,\bop,h}(B_0)\setminus R(0)$ enter $\mathrm{ell}_{\scop,\bop,h}(E_0)$, while remaining in the second microlocalized scattering elliptic set of $G_0$, such that for any sufficiently large $N>0$
 \begin{equation}\label{eq:lowbdecayradiales}
 	\norm{B_0 u}_{H^{s,r,\ell}_{\scop,\bop,h}(\CX)}\leq C h^{-1}\norm{G_0P_h(z)u}_{H^{s-1, r+1,\ell+1}_{\scop,\bop,h}(\CX)}+C\norm{E_0 u}_{H^{s,r,\ell}_{\scop,\bop,h}(\CX)}+Ch^{N}\norm{u}_{\bar{H}^{-N,-N,\ell}_{\scop,\bop,h}(\CX)}
 \end{equation}
 where the differential order $s$ is actually irrelevant here.
 
Since at the tapped set $K$
\begin{align*}
\sigma_h(h^{-1}\IM P_{h}(z))&=\sigma_h(\frac{P_h(z)-P_h(z)^*}{2ih})=\sigma_h(\frac{P_h(z)-P_h(\bar{z})}{2ih})\\
&=-\rho_b^{-2}\IM\sigma\bigg(-\Big(\xi_r- c'(r)\RE z+\frac{\chi(a \xi_\varphi-(r^2+\Ba^2)\RE z)}{\Delta_b}\Big)\Big(c'(r)+\frac{\chi(r^2+a^2)}{\Delta_b}\Big)\\
&\quad\quad+\frac{2}{\Delta_b}\Big(a\xi_\varphi-(r^2+a^2)\RE z\Big)\Big(r^2+a^2\Big)-2a\Big(\xi_\varphi-a\sin^2\theta \RE z\Big)\bigg)\\
&=-\rho_b^{-2}\IM\sigma\bigg(\frac{2}{\Delta_b}\Big(a\xi_\varphi-(r^2+a^2)\RE z\Big)\Big(r^2+a^2\Big)-2a\Big(\xi_\varphi-a\sin^2\theta \RE z\Big)\bigg)>0
\end{align*}
where we use $\xi_r- c'(r)\RE z+\frac{\chi(a \xi_\varphi-(r^2+a^2)\RE z)}{\Delta_b}=0$ at $K$ in the last equality and \[a\xi_\varphi-(r^2+a^2)\RE z=-\frac{-4 r\Delta_b\RE z}{\pa_r\Delta_b}<0, \quad (a\xi_\varphi-(r^2+a^2)\RE z)^2\geq \Delta_b(\xi_\varphi-a\sin^2\theta\RE z)^2\quad \mbox{at }K,
\]
which follow from Proposition \ref{prop:newescapefcn} and \eqref{eq:ABnonzero}, in the last step, it follows that 
\[
\sigma_h(h^{-1}\IM -P_{h}(z))=\sigma_h(\frac{(-P_h(z))-(-P_h(z))^*}{2ih})<0<\frac{\nu_{\min}}{2}\quad \mbox{at}\quad K.
\] 
Using the hyperbolic trapping estimates (see \cite{D16} and \cite[Theorem 4.7]{HV16}), there exist $B_1, G_1\in\Psi^{0}_{h}(\CX)$ microlocally supported near $K$ with $K\subset \mathrm{ell}_{h}(B_1)$ and $B_2\in\Psi_{h}^{0}(\CX), \mathrm{WF}_{h}(B_2)\cap \Gamma^-=\emptyset$, and satisfying that $
\mathrm{WF}_h(B_1)\cup\mathrm{WF}_h(B_2)$ is contained in the elliptic set of $G_1$, such that for any $N, s',r',\ell'\in\BR$
\begin{equation}\label{eq:normallytrapes1}
	\norm{B_1 u}_{H^{s,r,\ell}_{\scop,\bop,h}(\CX)}\leq C h^{-2}\norm{G_1P_h(z)u}_{H^{s-1, r+1,\ell+1}_{\scop,\bop,h}(\CX)}+C\norm{B_2 u}_{H^{s,r,\ell}_{,h\scop,\bop}(\CX)}+Ch^{N}\norm{u}_{\bar{H}^{s',r',\ell'}_{\scop,\bop,h}(\CX)}
\end{equation}
where the differential order $s'$, the sc-decay order $r'$ and b-decay order $\ell'$ are actually irrelevant here.

Let $r_-<r_0<\ehKN$ and $\chi\in C_c^\infty(\CX)$ with $\chi=1$ near $r\geq\ehKN$ and $\mathrm{supp}\chi\subset\{r\geq r_0\}$. By part (1) in Proposition \ref{prop:semiglobaldy}, semiclassical propagation of singularities estimates and semiclassical elliptic estimates (Concretely, when $(\rho,\theta, \varphi, \xi)\in \mathrm{WF}_h(\chi)\cap\{\jb{\xi}^{-2}p_{h,z}(\xi)\neq0\}$, we use semiclassical elliptic estimates. If $(\rho,\theta, \varphi, \xi)\in \mathrm{WF}_h(\chi)\cap\{\jb{\xi}^{-2}p_{h,z}(\xi)=0\}$, then there exists $s\in\BR$ with $\exp(s{}^{\scop}\!H_{p_{h,z}}^{2,0})(\rho,\theta, \varphi,\xi)\in \mathrm{ell}_h(B_\pm)\cup\mathrm{ell}_h(B_1)\cup\mathrm{ell}_h(B_0)\cup\mathrm{ell}_h(B_z)$, and thus we use semiclassical propagation of singularities estimates. We point out that in the set $\{\RE p_{h}(z)=0\}\cap\pa_+\CX$, the semiclassical symbol has a nonnegative imaginary part, so one can propagate estimates towards $R(0)$. Finally, by using a pseudodifferential partition of unity, $\chi$ can be written as s sum of operators falling into the above two case), we have
\begin{equation}
	\begin{split}
	\norm{\chi u}_{H^{s,r,\ell}_{\scop,\bop,h}(\CX)}&\leq C h^{-1}\norm{P_h(z)u}_{\bar{H}^{s-1, r+1,\ell+1}_{\scop,\bop,h}(\CX)}+C\norm{B_+ u}_{H^{s,r,\ell}_{\scop,\bop,h}(\CX)}+C\norm{B_- u}_{H^{s,r,\ell}_{\scop,\bop,h}(\CX)}\\
	&+C\norm{B_1 u}_{H^{s,r,\ell}_{\scop,\bop,h}(\CX)}+C\norm{B_0 u}_{H^{s,r,\ell}_{\scop,\bop,h}(\CX)}+C\norm{B_z u}_{H^{s,r,\ell}_{\scop,\bop,h}(\CX)}+Ch^{N}\norm{u}_{\bar{H}^{-N,-N,\ell}_{\scop,\bop,h}(\CX)}.
	\end{split}
\end{equation}
By the same reasoning as above in the control of $\chi u$, it follows that
\begin{equation}
	\begin{split}
		\norm{E_0 u}_{H^{s,r,\ell}_{\scop,\bop,h}(\CX)}&\leq C h^{-1}\norm{P_h(z)u}_{\bar{H}^{s-1, r+1,\ell+1}_{\scop,\bop,h}(\CX)}+C\norm{B_+ u}_{H^{s,r,\ell}_{\scop,\bop,h}(\CX)}+C\norm{B_1 u}_{H^{s,r,\ell}_{\scop,\bop,h}(\CX)}\\
		&\quad+C\norm{B_z u}_{H^{s,r,\ell}_{\scop,\bop,h}(\CX)}+Ch^{N}\norm{u}_{\bar{H}^{-N,-N,\ell}_{\scop,\bop,h}(\CX)}
	\end{split}
\end{equation}
and 
\begin{equation}
	\begin{split}
		\norm{B_2 u}_{H^{s,r,\ell}_{\scop,\bop,h}(\CX)}&\leq C h^{-1}\norm{P_h(z)u}_{\bar{H}^{s-1, r+1,\ell+1}_{\scop,\bop,h}(\CX)}+C\norm{B_+ u}_{H^{s,r,\ell}_{\scop,\bop,h}(\CX)}\\
		&\quad+C\norm{B_z u}_{H^{s,r,\ell}_{\scop,\bop,h}(\CX)}+Ch^{N}\norm{u}_{\bar{H}^{-N,-N,\ell}_{\scop,\bop,h}(\CX)}.
		\end{split}
\end{equation}
Putting all the above estimates together yields for $s>s_0>\frac 12, r>r_0>-\frac 12, \ell<-\frac 12$
\begin{equation}\label{eq:semiglobales1}
		\norm{\chi u}_{H^{s,r,\ell}_{\scop,\bop,h}(\CX)}\leq C h^{-2}\norm{P_h(z)u}_{\bar{H}^{s-1, r+1,\ell+1}_{\scop,\bop,h}(\CX)}+Ch^{s-s_0}\norm{u}_{\bar{H}^{s_0,r_0,\ell}_{\scop,\bop,h}(\CX)}.
\end{equation}

Since
\[
	p_{h,z}=-\rho_b^{-2}\Big(\Delta_b\big(\xi_r+\frac{\big(a\xi_\varphi-(r^2+a^2)z\big)}{\Delta_b}\big)^2
	-\frac{1}{\Delta_b}\big(a\xi_\varphi-(r^2+a^2)z\big)^2+\tilde{p}_{h,z}\Big)
\]
where
\[ \tilde{p}_{h,z}=\xi_\theta^2+\frac{1}{\sin^2\theta}(\xi_\varphi-a\sin^2\theta z)^2
\]
  is semiclassical hyperbolic with respect to $r$ in the region $\{r_-\leq r<\ehKN\}$ (see \cite[definition E.55]{DZ19}), using the semiclassical hyperbolic estimates (see \cite[Theorem E.57]{DZ19}), we have for all $s,r,\ell\in \BR$
  \begin{equation}\label{eq:semihyperesext}
  		\norm{(1-\chi) u}_{\bar{H}^{s,r,\ell}_{\scop,\bop,h}(\CX)}\leq C h^{-1}\norm{P_h(z)u}_{\bar{H}^{s-1, r+1,\ell+1}_{\scop,\bop,h}(\CX)}+\norm{\chi u}_{\bar{H}^{s,r,\ell}_{\scop,\bop,h}(\CX)}
  \end{equation}
where the sc-decay order $r$ and b-decay order $\ell$ are actually relevant here. Combining \eqref{eq:semiglobales1} with \eqref{eq:semihyperesext} yields for $s>s_0>\frac 12, r>r_0>-\frac 12, \ell<-\frac12$
\begin{equation}
		\norm{u}_{\bar{H}^{s,r,\ell}_{\scop,\bop,h}(\CX)}\leq C h^{-2}\norm{P_h(z)u}_{\bar{H}^{s-1, r+1,\ell+1}_{\scop,\bop,h}(\CX)}+Ch^{s-s_0}\norm{u}_{\bar{H}^{s_0,r_0,\ell}_{\scop,\bop,h}(\CX)}.
\end{equation}
Taking $h$ small enough, the term $Ch^{s-s_0}\norm{u}_{\bar{H}^{s_0,r_0,\ell}_{h,\scop,\bop}(\CX)}$ can be absorbed into the left-hand side and this proves the estimate \eqref{eq:highenergysecondmic}.

\item \underline{Proof of the estimate \eqref{eq:highenergysecondmicdual}.} For $s>\frac 12\geq\frac{1}{2}-\IM(h^{-1}z)\frac{\ehKN^2+a^2}{\ehKN^2-\Bm}$, we have $1-s\leq\frac{1}{2}-\IM(h^{-1}\overline{z})\frac{\ehKN^2+a^2}{\ehKN^2-\Bm}$. Then using the semiclassical low regularity radial point estimates (see \cite[Proposition 2.11]{Vas13}, \cite[Proposition 5.27]{Vas18}), there exist $B'_\pm, G'_\pm\in\Psi^0_h(\CX)$ microlocally supported near $L_\pm$ with $L_\pm\subset \mathrm{ell}_h(B'_\pm)$ and $E'_\pm\in\Psi^0_h(\CX), \mathrm{WF}_{h}(E'_\pm)\cap L_\pm=\emptyset$, and satisfying that all the backward (forward) null characteristics from $\mathrm{WF}_h(B'_\pm)\setminus L_\pm$ reach $\mathrm{ell}_h(E'_\pm)$, while remaining in the elliptic set of $G'_\pm$, such that for any $N\in\BR$
\begin{equation}\label{eq:highregradialesdual}
	\begin{split}
	&\norm{B'_\pm u}_{H^{1-s,-r-1,-\ell-1}_{\scop,\bop,h}(\CX)}\\
	&\quad\leq C h^{-1}\norm{G'P_h(z)^*u}_{H^{-s, -r,-\ell}_{\scop,\bop,h}(\CX)}+C\norm{E'_\pm u}_{H^{1-s,-r-1,-\ell-1}_{\scop,\bop,h}(\CX)}+Ch^N\norm{u}_{\dot{H}^{-N,-N,-N}_{\scop,\bop,h}(\CX)}
	\end{split}
\end{equation}
where the sc-decay order $r$ and b-decay order $\ell$ are actually irrelevant here.

For $r>-\frac 12$, we have $-r-1<-\frac 12$. Then using the semiclassical low sc-decay radial point estimates at the non-zero scattering section $R(z)$ (see \cite{Vas21a} and the above discussion about the proof of high sc-decay radial point estimates at $R(z)$. We note that the only difference is that now the term involving $\chi_0'$ does not have the correct sign and needs to be treated as an error term $E_z'u$), there exist $B'_z, G'_z \in\Psi^{0,0}_{\scop,h}(\CX)$ microlocally supported near $R(z)$ with $R(z)\subset \mathrm{ell}_{\scop,h}(B_z)$ and $E'_z \in\Psi^{0,0}_{\scop,h}(\CX), \mathrm{WF}_{\scop,h}(E'_z)\cap R(z)=\emptyset$, and satisfying that all the backward null characteristics from $\mathrm{WF}_{\scop,h}(B'_z)\setminus R(z)$ reach $\mathrm{ell}_{\scop,h}(E'_z)$, while remaining in the scattering elliptic set of $G'_z$, such that
\begin{equation}\label{eq:lowscdecayradialesdual}
	\begin{split}
	&\norm{B'_z u}_{H^{1-s,-r-1,-\ell-1}_{\scop,\bop,h}(\CX)}\\
	&\quad\leq C h^{-1}\norm{G'_zP_h(z)^*u}_{H^{-s, -r,-\ell}_{\scop,\bop,h}(\CX)}+C\norm{E'_z u}_{H^{1-s,-r-1,-\ell-1}_{\scop,\bop,h}(\CX)}+Ch^N\norm{u}_{\dot{H}^{-N,-N,-N}_{\scop,\bop,h}(\CX)}
	\end{split}
\end{equation}
where the differential order $s$ and b-decay order $\ell$ are actually irrelevant here.

For $\ell<-\frac 12$, we have $-\ell-1>-\frac 12$. Then using the semiclassical high b-decay radial point estimates at the zero scattering section $R(0)$ (see \cite[\S 4--5]{Vas21a}), there exist $B'_0, G'_0\in\Psi^{0,0,0}_{\scop,\bop,h}(\CX)$ microlocally supported near $R(0)$ (in fact near the blown up $R(0)$ in the second microlocalized space introduced in \S\ref{subsubsec:secondmic}) with $R(0)\subset \mathrm{ell}_{\scop,\bop,h}(B'_0)$, and satisfying all the backward null characteristics from $\mathrm{WF}_{\scop,\bop,h}(B'_0)$ tend to $R(0)$, while remaining in the second microlocalized scattering elliptic set of $G'_0$, such that for any $N\in\BR$
\begin{equation}\label{eq:highbdecayradialesdual}
	\norm{B'_0 u}_{H^{1-s,-r-1,-\ell-1}_{\scop,\bop,h}(\CX)}\leq C h^{-1}\norm{G'_0P_h(z)^*u}_{H^{-s, -r,-\ell}_{\scop,\bop,h}(\CX)}+Ch^{N}\norm{u}_{\dot{H}^{-N,-N,-\ell-1}_{\scop,\bop,h}(\CX)}
\end{equation}
where the differential order $s$ is actually irrelevant here.

We again calculate
\[
\sigma_h(h^{-1}\IM P_{h}(z)^*)=\sigma_h(\frac{P_h(\bar{z})-P_h(z)}{2ih})<0<\frac{\nu_{\min}}{2}\quad \mbox{at}\quad K.
\] 
Then using the hyperbolic trapping estimates (see \cite{D16} and \cite[Theorem 4.7]{HV16}), there exist $B'_1, G'_1,\in\Psi^{0}_{h}(\CX)$ microlocally supported near $K$ with $K\subset \mathrm{ell}_{h}(B'_1)$ and $B'_2\in\Psi_{h}^{0}(\CX), \mathrm{WF}_{h}(B'_2)\cap \Gamma^+=\emptyset$, and satisfying that $
\mathrm{WF}_h(B'_1)\cup\mathrm{WF}_h(B'_2)$ is contained in the elliptic set of $G'_1$, one has for any $N, s',r',\ell'\in\BR$
\begin{equation}\label{eq:normallytrapes2}
	\begin{split}
	&\norm{B'_1 u}_{H^{1-s,-r-1,-\ell-1}_{\scop,\bop,h}(\CX)}\\
	&\quad\leq C h^{-2}\norm{G'_1P_h(z)^*u}_{H^{-s, -r,-\ell}_{\scop,\bop,h}(\CX)}+C\norm{B'_2 u}_{H^{1-s,-r-1,-\ell-1}_{\scop,\bop,h}(\CX)}+Ch^{N}\norm{u}_{\dot{H}^{s',r',\ell'}_{\scop,\bop,h}(\CX)}
	\end{split}
\end{equation}
where the differential order $s'$, the sc-decay order $r'$ and b-decay order $\ell'$ are actually irrelevant here.

By the same reasoning as in the proof of the estimate \eqref{eq:highenergysecondmic}, we have
\begin{equation}
	\begin{split}
		&\norm{\chi u}_{H^{1-s,-r-1,-\ell-1}_{\scop,\bop,h}(\CX)}\\
		&\quad\leq C h^{-1}\norm{P_h(z)^*u}_{\dot{H}^{-s, -r,-\ell}_{\scop,\bop,h}(\CX)}+C\norm{B_+ u}_{H^{1-s,-r-1,-\ell-1}_{\scop,\bop,h}(\CX)}\\
		&\quad\quad+C\norm{B'_- u}_{H^{1-s,-r-1,-\ell-1}_{\scop,\bop,h}(\CX)}
		+C\norm{B'_1 u}_{H^{1-s,-r-1,-\ell-1}_{\scop,\bop,h}(\CX)}\\
		&\quad\quad+C\norm{B'_0 u}_{H^{1-s,-r-1,-\ell-1}_{\scop,\bop,h}(\CX)}+C\norm{B'_z u}_{H^{1-s,-r-1,-\ell-1}_{\scop,\bop,h}(\CX)}+Ch^{N}\norm{u}_{\dot{H}^{-N,-N,-\ell-1}_{\scop,\bop,h}(\CX)},
	\end{split}
\end{equation}
\begin{equation}
	\begin{split}
		&\norm{E'_\pm u}_{H^{1-s,-r-1,-\ell-1}_{\scop,\bop,h}(\CX)}+\norm{E'_z u}_{H^{1-s,-r-1,-\ell-1}_{\scop,\bop,h}(\CX)}\\
		&\quad\leq C h^{-1}\norm{P_h(z)^*u}_{\dot{H}^{-s, -r,-\ell}_{\scop,\bop,h}(\CX)}+C\norm{(1-\chi) u}_{\dot{H}^{1-s,-r-1,-\ell-1}_{\scop,\bop,h}(\CX)}\\&\quad\quad+C\norm{B'_1 u}_{H^{1-s,-r-1,-\ell-1}_{\scop,\bop,h}(\CX)}+C\norm{B'_0 u}_{H^{1-s,-r-1,-\ell-1}_{\scop,\bop,h}(\CX)}+Ch^{N}\norm{u}_{\dot{H}^{-N,-N,-\ell-1}_{\scop,\bop,h}(\CX)}
	\end{split}
\end{equation}
and
\begin{equation}
	\begin{split}
	\norm{B'_2 u}_{H^{1-s,-r-1,-\ell-1}_{\scop,\bop,h}(\CX)}&\leq C h^{-1}\norm{P_h(z)^*u}_{\dot{H}^{-s, -r,-\ell}_{\scop,\bop,h}(\CX)}+C\norm{(1-\chi) u}_{\dot{H}^{1-s,-r-1,-\ell-1}_{\scop,\bop,h}(\CX)}\\
	&\quad+C\norm{B'_0 u}_{H^{1-s,-r-1,-\ell-1}_{\scop,\bop,h}(\CX)}+Ch^{N}\norm{u}_{\dot{H}^{-N,-N,-\ell-1}_{\scop,\bop,h}(\CX)}.
\end{split}
\end{equation}
Putting all the above estimates together yields for $s>\frac 12, r>-\frac 12, \ell<-\frac 12$
\begin{equation}\label{eq:semiglobales2}
\begin{split}
	\norm{\chi u}_{H^{1-s,-r-1,-\ell-1}_{\scop,\bop,h}(\CX)}&\leq C h^{-2}\norm{P_h(z)^*u}_{\dot{H}^{-s, -r,-\ell}_{\scop,\bop,h}(\CX)}+C\norm{(1-\chi) u}_{\dot{H}^{1-s,-r-1,-\ell-1}_{\scop,\bop,h}(\CX)}\\
	&\quad+Ch^{N}\norm{u}_{\dot{H}^{-N,-N,-\ell-1}_{\scop,\bop,h}(\CX)},
\end{split}
\end{equation}
Again using the semiclassical hyperbolic estimates (see \cite[Theorem E.57]{DZ19}), we have for all $s,r,\ell\in \BR$
\begin{equation}\label{eq:semihyperessupp}
	\norm{(1-\chi) u}_{\dot{H}^{1-s,-r-1,-\ell-1}_{\scop,\bop,h}(\CX)}\leq C h^{-1}\norm{P_h(z)^*u}_{\dot{H}^{-s, -r,-\ell}_{\scop,\bop}(\CX)}
\end{equation}
where the sc-decay order $r$ and b-decay order $\ell$ are actually relevant here. Combining \eqref{eq:semiglobales2} with \eqref{eq:semihyperessupp} yields for $s>\frac 12, r>-\frac 12, \ell<-\frac12$
\begin{equation}
	\norm{u}_{\dot{H}^{1-s,-r-1,-\ell-1}_{\scop,\bop,h}(\CX)}\leq C h^{-2}\norm{P_h(z)^*u}_{\dot{H}^{-s, -r,-\ell}_{\scop,\bop,h}(\CX)}+Ch^{N}\norm{u}_{\dot{H}^{-N,-N,-\ell-1}_{\scop,\bop,h}(\CX)}.
\end{equation}
Taking $h$ small enough, the term $Ch^{N}\norm{u}_{\dot{H}^{-N,-N,\ell}_{h,\scop,\bop}(\CX)}$ can be absorbed into the left-hand side and this proves the estimate \eqref{eq:highenergysecondmicdual}.
\item \underline{Invertibility of $\widehat{\Box_{g_{b}}}(\sigma)$ for $0\leq\IM\sigma\leq C_1, \RE\sigma\geq C_2$.} We only discuss the proof of the invertibility of the operator $\widehat{\Box_{g_{b}}}:\{u\in\bar{H}_{\bop,h}^{s,\ell}(\CX)\mid\bar{H}_{\bop,h}^{s,\ell+1}(\CX)\}\to\bar{H}_{\bop,h}^{s,\ell+1}(\CX)$ in detail as the operator $
\widehat{\Box_{g_{b}}}:\{u\in\bar{H}_{\bop,h}^{s,\ell}(\CX)\mid\bar{H}_{\bop,h}^{s-1,\ell+2}(\CX)\}\to\bar{H}_{\bop,h}^{s-1,\ell+2}(\CX)$ can be handled in a completely analogous manner. We define the space \[
\mathcal{X}(\sigma):=\{u\in\bar{H}_\bop^{s,\ell}(\CX)\mid \widehat{\Box_{g_{b}}}(\sigma)u\in\bar{H}_\bop^{s,\ell+1}(\CX)\},
\]
which is a Hilbert space endowed with the norm $\norm{u}_{\mathcal{X}(\sigma)}:=\norm{u}_{\bar{H}_\bop^{s,\ell}}+\norm{\widehat{\Box_{g_{b}}}(\sigma)u}_{\bar{H}_\bop^{s,\ell+1}}$. First, the injectivity of $\widehat{\Box_{g_{b}}}(\sigma):\mathcal{X}(\sigma)\to\bar{H}^{s,\ell}_\bop$ immediately follows from \eqref{eq:highenergy1}. As for the surjectivity, we need to show that for any $f\in\bar{H}_\bop^{s,\ell+1}$, there exists $u\in\bar{H}^{s,\ell}_\bop$ such that $\widehat{\Box_{g_{b}}}(\sigma)u=f$. We define 
\[
\mathcal{Y}(\sigma):=\{v\in\dot{H}_\bop^{-s,-\ell-1}(\CX)\mid \widehat{\Box_{g_{b}}}(\sigma)^*v\in\dot{H}_\bop^{-s,-\ell}(\CX)\}.
\]
According to \eqref{eq:highenergy1}, we have
\[
\abs{\angles{f}{v}}\leq C\norm{f}_{\bar{H}_\bop^{s,\ell+1}}\norm{\widehat{\Box_{g_{b}}}(\sigma)^*v}_{\dot{H}_\bop^{-s,-\ell}},\quad v\in\mathcal{Y}(\sigma).
\]
By Hahn-Banach Theorem, the anti-linear form $\widehat{\Box_{g_{b}}}(\sigma)^*v\to \angles{f}{v}$ with $v\in\mathcal{Y}(\sigma)$ can be extended to a continuous anti-linear form on $\dot{H}^{-s,-\ell}_\bop$. Since $(\dot{H}^{-s,-\ell}_\bop)^*=\bar{H}^{s,\ell}_\bop$, there exists $u\in\bar{H}^{s,\ell}_\bop$ such that $\angles{u}{\widehat{\Box_{g_{b}}}(\sigma)^*v}=\angles{f}{v}$ for $v\in\mathcal{Y}(\sigma)$. In particular, for $v\in C_c^\infty(X^\circ)$
\[
\angles{\widehat{\Box_{g_{b}}}(\sigma)u}{v}=\angles{u}{\widehat{\Box_{g_{b}}}(\sigma)^*v}=\angles{f}{v}.
\]
This proves that $\widehat{\Box_{g_{b}}}(\sigma)u=f$.
 \end{itemize}
\end{proof}

Now we further determine the index of the Fredholm operators $\widehat{\Box_{g_{b}}}(\sigma)$ on suitable function spaces for $\sigma\in\BC, \IM\sigma\geq 0$.
\begin{lem}\label{lem:indexofFredholmoperator}
	Let $s>\frac12, \ell<-\frac12, s+\ell>-\frac12$. Then the following holds.
		\begin{enumerate}
		\item For $\sigma\in\BC, \IM\sigma\geq 0, \sigma\neq 0$, the operators
		\begin{align}\label{eq:Fredholmindex1}
			&\widehat{\Box_{g_{b}}}(\sigma):\{u\in\bar{H}_{\bop}^{s,\ell}(\CX)\mid	\widehat{\Box_{g_{b}}}(\sigma)u\in\bar{H}_{\bop}^{s-1,\ell+2}(\CX)\}\to\bar{H}_{\bop}^{s-1,\ell+2}(\CX)\\\label{eq:Fredholmindex2}
			&\widehat{\Box_{g_{b}}}(\sigma):\{u\in\bar{H}_{\bop}^{s,\ell}(\CX)\mid	\widehat{\Box_{g_{b}}}(\sigma)u\in\bar{H}_{\bop}^{s,\ell+1}(\CX)\}\to\bar{H}_{\bop}^{s,\ell+1}(\CX)
		\end{align}
		are Fredholm of index $0$.
		\item Let $-\frac32<\ell<-\frac12$. Then
		\begin{equation}\label{eq:Fredholmindex0}
			\widehat{\Box_{g_{b}}}(0):\{u\in\bar{H}_{\bop}^{s,\ell}(\CX)\mid	\widehat{\Box_{g_{b}}}(0)u\in\bar{H}_{\bop}^{s-1,\ell+2}(\CX)\}\to\bar{H}_{\bop}^{s-1,\ell+2}(\CX)
		\end{equation}
		is Fredholm of index $0$.
	\end{enumerate}	
\end{lem}
\begin{proof}
 We now prove that \eqref{eq:Fredholmindex1} and \eqref{eq:Fredholmindex0} have index $0$, and the index $0$ property of \eqref{eq:Fredholmindex2} follows in a similar manner. 
 \begin{itemize}
 	\item\underline{Proof of the index $0$ property of \eqref{eq:Fredholmindex1}.} In view of the \textit{high energy estimates} in Theorem \ref{thm:highenergy}, \[\widehat{\Box_{g_b}}(\sigma):\{u\in\bar{H}_{\bop}^{s,\ell}(\CX)\mid	\widehat{\Box_{g_{b}}}(\sigma)u\in\bar{H}_{\bop}^{s-1,\ell+2}(\CX)\}\to\bar{H}_{\bop}^{s-1,\ell+2}(\CX)
 \]
  is invertible and thus has index $0$ for fixed $\IM\sigma=C\geq 0$ when $\abs{\RE\sigma}\gg1$. We claim that the index of $\widehat{\Box_{g_b}}(\sigma)$ with $\IM \sigma\geq 0,\sigma\neq 0$ is a constant on the horizontal line $\IM\sigma=C$ and thus is $0$ for all $\sigma$ with $\IM\sigma\geq 0,\sigma\neq0$. Let 
\[
\mathcal{X}(\sigma):=
\{u\in\eHb^{s,\ell}(\CX):\widehat{\Box_{g_b}}(\sigma)u\in \eHb^{s-1,\ell+2}(\CX)\}.
\]
For any fixed $\sigma_0\neq 0$ with $\IM\sigma_0=C\geq0$, without loss of generality, we may assume $\RE \sigma_0\geq 0$. First, there exists a $C'$ such that $\widehat{\Box_{g_b}}(\sigma)$ is invertible and thus has index $0$ for $\sigma\in[C'+iC, \infty+iC)$. We next prove that for each $\sigma'\in[\sigma_0=\RE\sigma_0+iC, C'+iC]$, there exists $\epsilon>0$ such that the index of $\widehat{\Box_{g_b}}(\sigma):\mathcal{X}(\sigma)\to\bar{H}^{s-1,\ell+2}_\bop(\CX)$ is a constant for $\sigma$ with $\IM\sigma\geq 0,\sigma\neq 0,\abs{\sigma-\sigma'}<\epsilon$. Then by compactness of $[\sigma_0=\RE\sigma_0+iC, C'+iC]$ we conclude that the index of $\widehat{\Box_{g_b}}(\sigma_0)$ is the same as that of $\widehat{\Box_{g_b}}(C'+iC)$ and thus is $0$. Suppose $\widehat{\Box_{g_b}}(\sigma')$ has kernel and cokernel of dimension $k_+$ and $k_-$, respectively, then we establish a Grushin problem \cite[Appendix C.1]{DZ19}. More specifically, we define the following operator
\[
L(\sigma)=\begin{pmatrix}
	\widehat{\Box_{g_b}}(\sigma)&L_-\\
	L_+&0
\end{pmatrix}: \mathcal{X}(\sigma)\oplus\BC^{k_-}\to \eHb^{s-1,\ell+2}(\CX)\oplus \BC^{k_+}
\]
where $L_+, L_-$ are defined as follows. If $\ker\widehat{\Box_{g_b}}(\sigma')$ is spanned by $x_j\in \mathcal{X}(\sigma'), j=1,\cdots, k_+$, then by Hahn-Banach Theorem  we can find $x_j^*\in \sHb^{-s,-\ell}(\CX)$ with $\norm{x_j^*}\leq 1$ such that $\angles{x_i}{x_j^*}=\delta_{ij}$. It follows that 
\[
L_+: \eHb^{s,\ell}(\CX)\to C^{k_+},\quad L_+(f):=(\angles{f}{x_1^*},\cdots,\angles{f}{x_{k_+}^*})
\]
restricts to an isomorphism $\ker\widehat{\Box_{g_b}}(\sigma')\to \BC^{k_+}$. As for the construction of $L_-$, we choose $y_j\in \eHb^{s-1,\ell+2}(\CX)$ with $\norm{y_j}\leq 1$, $ j=1, \cdots, k_-$ so that $y_j+\Ran\widehat{\Box_{g_b}}(\sigma')$ form a basis of $\eHb^{s-1,\ell+2}/\Ran\widehat{\Box_{g_b}}(\sigma')$. Then we define 
\[
L_-:\BC^{k_-}\to \eHb^{s-1,\ell+2}(\CX),\quad L_-((a_1, \cdots,a_{k_-})):=\sum_{j=1}^{k_-}a_jy_j.
\]
The operator $L_-$ has maximal rank and $\Ran L_-\cap \Ran\widehat{\Box_{g_b}}(\sigma')=\emptyset$. The above construction of $L_+,L_-$ implies $L(\sigma')$ has a trivial kernel and onto, and thus is invertible. Since $\widehat{\Box_{g_b}}(\sigma)$ satisfy the uniform Fredholm estimates \eqref{eq:Fredholmenergy2} for all $\sigma\in\BC, \IM\sigma\in[0,2C]$ satisfying $\abs{\sigma_0}/2\leq \abs{\sigma}\leq 2(C+C') $, so do $L(\sigma)$. That is, there exists a constant $\tilde{C}$ (independent of $b$ and $\sigma$) such that 
\[
\norm{(u, u^-)}_{\eHb^{s,\ell}(\CX)\oplus \BC^{k_-}}\leq \tilde{C}\Bigl(\norm{L(\sigma)(u, u^-)}_{\eHb^{s-1,\ell+2}(\CX)\oplus \BC^{k_+}}+\norm{(u,u^-)}_{\eHb^{s_0,\ell_0}(\CX)\oplus \BC^{k_-}}\Bigr)
\]
for $\sigma\in\BC, \IM\sigma\in[0,2C]$ satisfying $\abs{\sigma_0}/2\leq \abs{\sigma}\leq 2(C+C')$, and $s_0<s, \ell_0<\ell$. Then following the perturbation arguments in \cite[\S2.7]{Vas13}, we prove the invertibility of $L(\sigma)$ when $\sigma$ is close to $\sigma'$. Concretely, suppose there exists a sequence $\delta_j\to\delta'$ such that $L(\delta_j)$ is not invertible, so either $\ker L(\delta_j) $ on $\eHb^{s,\ell}(\CX)\oplus \BC^{k_-}$ or $\ker L(\delta_j)^* $ on $\sHb^{-s+1,-\ell-2}(\CX)\oplus \BC^{k_+}$ is non-trivial. By passing to a subsequence, we may assume that the former possibility holds for all $j$, as the case of the adjoint is completely analogous. Now, if $\norm{(u_j, u^-_j)}_{\eHb^{s,\ell}(\CX)\oplus \BC^{k_-}}=1$ and $L(\delta_j)(u_j, u^-_j)=0$, then the above uniform Fredholm estimate gives $\tilde{C}^{-1}\leq \norm{(u_j,u_j^-)}_{\eHb^{s_0,\ell_0}(\CX)\oplus \BC^{k_-}}$. Using the sequential compactness of the unit ball in $\eHb^{s,\ell}(\CX)\oplus \BC^{k_-}$ in the weak topology, and the compactness of the inclusion $\eHb^{s,\ell}(\CX)\oplus \BC^{k_-}\to\eHb^{s_0,\ell_0}(\CX)\oplus \BC^{k_-}, s_0<s,\ell_0<\ell$,  it follows that $(u_j, u^-_j)$ has a weakly convergent subsequence in $\eHb^{s,\ell}(\CX)\oplus \BC^{k_-}$ to some $(u_0, u^+_0) \in \eHb^{s,\ell}(\CX)\oplus \BC^{k_-}$ which is  norm-convergent in $\eHb^{s_0,\ell_0}(\CX)\oplus \BC^{k_-}$. We also have
\[
\qquad \qquad L(\delta')(u_0, u^+_0)=L(\delta')\left((u_0, u^+_0)-(u_j, u^-_j)\right)+(L(\delta')-L(\delta_j))(u_j, u^-_j)\to 0\quad\mbox{in}\quad \eHb^{s_0-2, \ell_0}(\CX)\oplus\BC^{k_-}
\]
since $L(\delta_j)\to L(\delta')$ as bounded operators in $\mathcal{L}(\eHb^{s_0, \ell_0}(\CX)\oplus\BC^{k_-}, \eHb^{s_0-2, \ell_0}(\CX)\oplus\BC^{k_-})$ and $(u_j, u^-_j)\to(u_0, u^-_0)$ in $\eHb^{s_0, \ell_0}(\CX)\oplus\BC^{k_-}$. Therefore, $L(\delta')(u_0, u^-_0)=0$. Since $\tilde{C}^{-1}\leq \norm{(u,u^-)}_{\eHb^{s_0,\ell_0}(\CX)\oplus \BC^{k_-}}$, $(u_0, u^-_0)\neq 0$ and thus $\ker L(\sigma')$ is non-trivial, which leads to a contradiction. This proves that $L(\sigma)$ is invertible for $\sigma$ close to $\sigma'$, which implies $\widehat{\Box_{g_b}}(\sigma)$ has index $k_+-k_-$ (see \cite[Theorem C.4]{DZ19}) for for $\sigma$ close to $\sigma'$.

\item\underline{Proof of $0$ index property of \eqref{eq:Fredholmindex0}.} Suppose that $-\frac32<\ell<-\frac 12$. Then using the uniform Fredholm estimates \eqref{eq:uniformFredholmenergy} (up to $\sigma=0$) and the argument as above yields the index $0$ property of \eqref{eq:Fredholmindex0}.
\end{itemize}
\end{proof}

\subsubsection{High energy estimates for tensor wave operators}
Now we prove an analogy to Theorem \ref{thm:highenergy} for the semicalssical operators  $h^2\widehat{\mathcal{P}_{b,\gamma}}(h^{-1}z),\  h^2\widehat{\mathcal{W}_{b,\gamma}}(h^{-1}z)$ and $h^2\widehat{L_{b, \gamma}}(h^{-1}z)$ acting on bundles.

\begin{thm}\label{thm:highenergyBundle}
	Let $b_0=(\Bm_0, \Ba_0, \BQ_0)$ with $\abs{\BQ_0}+\abs{\Ba_0}<\Bm_0$ and $\abs{\Ba_0}\ll\abs{\Bm_0}+\abs{\BQ_0}$. Then there exists $\epsilon>0$ such that for $b=(\Bm, \Ba,\BQ)$ with $\abs{b-b_0}<\epsilon$ and for $s>s_0>2$ (resp. $s>s_0>3$), $\ell<-\frac{1}{2}$ with $s+\ell>-\frac12$, the conclusions in Theorem \ref{thm:highenergy} hold for $h^2\widehat{\mathcal{P}_{b,\gamma}}(h^{-1}z)$ and $h^2\widehat{\mathcal{W}_{b,\gamma}}(h^{-1}z)$ (resp. $h^2\widehat{L_{b,\gamma}}(h^{-1}z)$).	\end{thm}
\begin{proof}
	The proof is analogous to that of Theorem \ref{thm:highenergy}, except for the computation of threshold regularity in the radial point estimate at event horizon and the threshold decay rate in the radial point estimate at spatial infinity $\pa_+\CX$, and the verification of 
	\begin{equation}
		\label{eq:verificationatK}
	\sigma_{1,h}\Big(\frac{1}{2ih}\big(h^2\widehat{\bullet}(h^{-1}z)-(h^2\widehat{\bullet}(h^{-1}z))^*\big)\Big)<\frac{\nu_{\min}}{2},\quad \bullet=\mathcal{P}_{b,\gamma},\ \mathcal{W}_{b,\gamma},\ L_{b,\gamma}
	\end{equation}
	at trapping $K$.

The computation of threshold regularity in the radial point estimate at event horizon and the threshold decay rate in the radial point estimate at spatial infinity $\pa_+\CX$ is the same as that in the fixed $\sigma$ case in Theorem \ref{thm:FredholmestimateBundle}. The verification of \eqref{eq:verificationatK} at trapping $K$ follows from the discussion in appendix \ref{app:trap} (see Theorems \ref{ThmSDSNilpotentAtTrapping}, \ref{ThmSDSLNilpotentAtTrapping} and Remark \ref{RemTrap}) and the relevant calculation in the proof of Theorem \ref{thm:highenergy}. 
\end{proof}
 
 Correspondingly, we have the following analogy to Lemma \ref{lem:indexofFredholmoperator} for $\widehat{\mathcal{P}_{b,\gamma}},\ \widehat{\mathcal{W}_{b,\gamma}}$ and $\widehat{L_{b,\gamma}}$.
 \begin{lem}\label{lem:indexofFredholmoperatorBundle}
 		Let $b=(\Bm, \Ba, \BQ)$ with $\abs{\BQ}+\abs{\Ba}<\Bm$ and $\abs{\Ba}\ll\abs{\Bm}+\abs{\BQ}$. Let $s>2$(resp. $s>3$), $\ell<-\frac12$ and $s+\ell>-\frac12$. Then the conclusions in Lemma \ref{lem:indexofFredholmoperator} also hold for $\widehat{\mathcal{P}_{b,\gamma}}(\sigma)$ and $\widehat{\mathcal{W}_{b,\gamma}}(\sigma)$ (resp. $\widehat{L_{b,\gamma}}(\sigma)$).	
 \end{lem}
\begin{proof}
	The proof is completely analogous to that of Lemma \ref{lem:indexofFredholmoperator}.
\end{proof}

\subsection{Energy estimates}
\label{subsec:energyestimates}
In this section, we shall establish the energy estimates for the solutions to various wave type equations on slowly rotating Kerr-Newman metrics.

 First, we introduce another modification of the Boyer-Lindquist coordinates, which is defined by 
\[
\bar{t}=t+F_t(r),\quad \bar{\varphi}=\varphi+F_\varphi(r).
\]
In the new coordinates $(\bar{t}, r, \theta, \bar{\varphi})$, the Kerr-Newman metric takes the following form
\begin{equation}\label{eq:extendedmetric}
	\begin{split}
	g_b&=-\frac{\Delta_b}{\rho_b^2}\Big( d\bar{t}-a\sin^2\theta d\bar{\varphi}+\big(a\sin^2\theta F_\varphi'(r)-F_t'(r)\big)dr\Big)^2\\
	&\quad+\frac{\sin^2\theta}{\rho_b^2}\Big(a d\bar{t}-(r^2+a^2)d\bar{\varphi}+\big((r^2+a^2)F_\varphi'(r)-a F_t'(r)\big)dr\Big)^2+\rho_b^2(\frac{dr^2}{\Delta_b}+d\theta^2)
	\end{split}
\end{equation}
and its inverse is
\begin{equation}\label{eq:extendedinversemetric}
	\begin{split}
	g_b^{-1}&=-\frac{1}{\rho_b^2\Delta_b}\Big((r^2+a^2)\pa_{\bar{t}}+a\pa_{\bar{\varphi}}\Big)^2+\frac{1}{\rho^2_b\sin^2\theta}\Big(\pa_{\bar{\varphi}}+a\sin^2\theta\pa_{\bar{t}}\Big)^2\\
	&\quad+\frac{1}{\rho_b^2}\pa_\theta^2+\frac{\Delta_b}{\rho_b^2}\Big(\pa_r+F_t'(r)\pa_{\bar{t}}+F_\varphi'(r)\pa_{\bar{\varphi}}\Big)^2.
\end{split}
\end{equation}
In order for $g_b, g_b^{-1}$ to be smooth at event horizon, $F_t(r)$ and $F_\varphi(r)$ are required to satisfy that $F'_t(r)=\pm(r^2+a^2)/\Delta_b, F'_\varphi(r)=\pm a/\Delta_b$ (up to a smooth modification) at event horizon. Now we shall prove
\begin{lem}
There exist $F_t(r), F_\varphi(r)$ such that they satisfy the following conditions
	\begin{enumerate}
			\item $F'_\varphi(r)=\frac{a}{\Delta_b}+\tilde{F}_\varphi(r)$ where $\tilde{F}_\varphi(r)$ is smooth on $[r_-, \infty)_r$, and $F_\varphi(r)=0$ near $r=\infty$.
			\item $F'_t(r)=\frac{r^2+a^2}{\Delta_b}+\tilde{F}_t(r)$ where $\tilde{F}_t(r)$ is smooth on $[r_-,\infty)$, and $F_t(r)=-r_{b,*}+\mathcal{O}(r^{-1})$ near $r=\infty$ where $\frac{dr_{b,*}}{dr}=\frac{1}{\Delta_b}$.
		\item The hypersurfaces $\bar{t}=const$ are spacelike, i.e., $d\bar{t}$ is timelike. More precisely, \[g_b^{-1}(d\bar{t}, d\bar{t})\leq-\Bm^2\rho_b^{-2}<0.
		\]
	\end{enumerate}
With this choice of $F_t(r), F_\varphi(r)$, $g_b$ in the coordinates $(\bar{t}, r, \theta, \bar{\varphi})$ is smooth and non degenerate on the extended manifold $\BR_{\bar{t}}\times [r_-, \infty)_r\times\BS^2$.
\end{lem}
\begin{proof}
	Let $\chi(r)\in C_c^\infty(\BR)$ such that $\chi(r)=1$ when $r\leq 3\Bm$ and $\chi=0$ when $r\geq 4\Bm$. We first define $F_\varphi(r)=\chi \int a/\Delta_b\,dr$ which satisfies statement ($1$). Next, we expect
	\begin{equation}\label{eq:requireforhyperblodal}
	F'_t(r)=\frac{r^2+a^2}{\Delta_b}+\mu_1(r)\quad \mbox{near}\quad r=\ehKN;\quad 	F'_t(r)=-\frac{r^2+a^2}{\Delta_b}+\mu_2(r)\quad \mbox{near}\quad r=\infty
	\end{equation}
	where $\mu_1(r)$ is smooth near event horizon and $\mu_2(r)\sim r^{-2}$ near $r=\infty$. We let
	\[
	g_b^{-1}(d\bar{t}, d\bar{t})=-\frac{1}{\rho_b^2}\Big(\frac{(r^2+a^2)^2}{\Delta_b}-\Delta_bF_t'(r)^2-a^2\sin^2\theta\Big)=-\frac{\Bm^2}{\rho_b^2}<0,
	\] 
	In order to arrange \eqref{eq:requireforhyperblodal}, we need
	\[
F_t'(r)=\begin{cases}
\sqrt{\Big(\frac{r^2+a^2}{\Delta_b}\Big)^2-\frac{\Bm^2+a^2\sin^2\theta}{\Delta_b}}=\frac{r^2+a^2}{\Delta_b}\sqrt{1-\frac{\Delta_b(\Bm^2+a^2\sin^2\theta)}{(r^2+a^2)^2}},\quad & r \mbox{ is near } \ehKN\\
-\sqrt{\Big(\frac{r^2+a^2}{\Delta_b}\Big)^2-\frac{1+a^2\sin^2\theta}{\Delta_b}}=-\frac{r^2+a^2}{\Delta_b}\sqrt{1-\frac{\Delta_b(\Bm^2+a^2\sin^2\theta)}{(r^2+a^2)^2}},\quad & r \mbox{ is near } \infty
\end{cases}.
	\]
	Therefore, we define 
	\[
	F_t'(r)=\chi(r)\bigg(\frac{r^2+a^2}{\Delta_b}\sqrt{1-\frac{\Delta_b(\Bm^2+a^2\sin^2\theta)}{(r^2+a^2)^2}}\bigg)+(1-\chi(r))\bigg(-\frac{r^2+a^2}{\Delta_b}\sqrt{1-\frac{\Delta_b(\Bm^2+a^2\sin^2\theta)}{(r^2+a^2)^2}}\bigg).
	\]
	With this choice of $F_t'(r)$, we have $
	g_b^{-1}(d\bar{t},d\bar{t})=-\Bm^2\rho_b^{-2}$ for $r\leq 3\Bm, r\geq 4\Bm$ and $
	g_b^{-1}(d\bar{t},d\bar{t})\leq -\Bm^2\rho_b^{-2}$ for $3\Bm\leq r\leq  4\Bm$, and thus the requirement statement ($3$) is satisfied. By choosing a proper integration constant we have $\bar{t}=r_{b,*}+\mathcal{O}(r^{-1})$ near $r=\infty$, and thus the requirement statement ($2$) is satisfied
\end{proof}

We now establish the energy estimates for the solutions to the equation $\Box_{g_b} u=f$ on the extended Kerr-Newman manifold $M=\BR_{\bar{t}}\times[r_-,\infty)_r\times \BS^2$. We denote by $\Sigma_{\bar{t}_0}$ the spacelike hypersurfaces $\{\bar{t}=\bar{t}_0\}
\cap M$. Here, the integral on $M$ is taken with respect to the volume form $dV=dvol_{g_b}=\rho_b^2\sin\theta d\bar{t}drd\theta d\varphi$.
% the integral on $\Sigma_{\bar{t}}$ is defined with respect to the induced volume form on $\Sigma_{\bar{t}}$, which is $\sim r^3\sin\theta drd\theta d\varphi$.

We begin with the
energy-momentum tensor
\[
T_{\alpha\beta}[g]=\partial_\alpha u \partial_\beta u -
\frac{1}{2}g_{\alpha\beta}\partial^\gamma u \partial_\gamma u
\]
Its contraction with respect to a vector field $X$ is denoted by
\[
P_\alpha[g,X]=T_{\alpha\beta}[g]X^\beta
\]
and its divergence is
\[
\nabla^\alpha P_\alpha[g,X] = \Box_g u \cdot Xu + \frac{1}{2}T_{\alpha\beta}[g]\pi_X^{\alpha \beta}
\]
and $\pi_X^{\alpha \beta} $ is the deformation tensor of $X$, which is given in terms of the Lie derivative by
\[
\pi_{\alpha\beta}^X=\nabla_\alpha X_\beta + \nabla_\beta X_\alpha=({\mathcal L}_X g)_{\alpha\beta}.
\]
Since $\Sigma_{\bar{t}}$ is ``asymptotically null'', we shall define the (degenerate) energy at $\Sigma_{\bar{t}_0}$ as
	\[
	E[u](\bar{t}_0):=\int_{r_-}^\infty\int_{\BS^2}\frac{1}{r^2}\abs{\pa_{\bar{t}}u(\bar{t}_0,\cdot)}^2+\abs{\pa_ru(\bar{t}_0,\cdot)}^2+\frac{1}{r^2}\Big(\abs{\pa_\theta u(\bar{t}_0, \cdot)}^2+\frac{1}{\sin^2\theta}\abs{\pa_\varphi u(\bar{t}_0, \cdot)}^2\Big) r^2\sin\theta\,dr\,d\theta\,d\varphi.
		\]
%Moreover, we define the higher order energy 
%\[
%E^k[u](\bar{t}_0):=\Sigma_{i=0}^k \Sigma_{\abs{\alpha}=0}E[\pa^\alpha u](\bar{t})
%\]
%where $\pa^\alpha=\pa^{\alpha_0}_{\bar{t}}\pa_{x^j}^{\alpha_j}, 1\leq j\leq 3$ and $(x^1,x^2,x^3)$ is the standard coordinates in $\BR^3\supset [r_-,\infty)_r\times\BS^2$.
Now we establish the energy estimate.
\begin{prop}\label{prop:energyestimate}
 Let $b=(\Bm,\Ba, \BQ)$ with $\abs{\Ba}\ll\abs{\BQ}+\Bm$. Let $u$ be the solution to the following initial value problem  
	\begin{equation}\label{eq:ivp}
		\begin{split}
			\Box_{g_b}u=f, \quad u|_{\bar{t}=0}=u_0,\quad \pa_{\bar{t}}u|_{\bar{t}=0}=u_1
		\end{split}	
	\end{equation}
	where $rf\in L^1([0, \infty)_{\bar{t}}; L^2(\Sigma_{\bar{t}}))$ and 
	$\frac{u_0}{r}, \frac{u_1}{r}, \pa u_0\in L^2(\Sigma_0; r^2\sin\theta\,dr\,d\theta\,d\varphi)$. Then there exists a constant $C(b)$ which depends on the parameter $b=(\Bm, \Ba,\BQ)$ such that
	\begin{equation}\label{eq:energyestimate}
E[u](\bar{t})\leq C(b)\Big(\norm{\frac{u_0}{r}}_{L^2(\Sigma_0)}+\norm{\pa u_0}_{L^2(\Sigma_0)}+\norm{\frac{u_1}{r}}_{L^2(\Sigma_0)}+\norm{rf}^2_{L^1([0, \bar{t}); L^2(\Sigma_{\bar{t}}))}\Big)e^{C(b)\bar{t}}.
	\end{equation}
where the integral is taken with respect the volume form $r^2\,dr\,d\theta\,d\varphi$ and $\pa\in\{\pa_r,\frac{1}{r}\pa_\theta, \frac{1}{r\sin\theta}\pa_\varphi\}$.

Moreover, the above statements also hold for the wave type operators $\mathcal{P}_{b,\gamma}, \mathcal{W}_{b,\gamma}$ and $L_{b,\gamma}$.
\end{prop}
\begin{proof}
%The existence and uniqueness of the solution $u$ to the initial value problem follows from the theory of linear hyperbolic equations (see \cite[theorem 23.1.2]{H07}). Now we turn to the discussion of the energy estimate. 
%Since $\pa_{\bar{t}}, d\bar{t}$ are timelike when $r$ is away from and outside event horizon, 
%We only need more careful analysis near event horizon, which is in a compact set. Then 
 We only do the analysis for the case $\Ba=0$ here and the same argument will also work for small $\Ba$ since all the calculation below near event horizon depends continuously on $b=(\Bm, \Ba,\BQ)$ (away from the event the horizon, we make use of the timelike property of $d\bar{t}$ and $\pa_{\bar{t}}$). Therefore, we let $b=(\Bm, 0,\BQ)$ and then we have
\begin{align*}
	g=g_b&=-\frac{\Delta_b}{r^2} d\bar{t}^2+2\frac{\Delta_b}{r^2}F_t'(r)d\bar{t}dr+\Big(\frac{r^2}{\Delta_b}-\frac{\Delta_b}{r^2}F'_t(r)^2\Big)dr^2+r^2d\theta^2+r^2\sin^2\theta d\varphi^2,\\
	g^{-1}=g_b^{-1}&=-\Big(\frac{r^2}{\Delta_b}-\frac{\Delta_b}{r^2}F'_t(r)^2\Big)\pa_{\bar{t}}^2+2\frac{\Delta_b}{r^2}F_t'(r)\pa_{\bar{t}}\pa_r+\frac{\Delta_b}{r^2}\pa_r^2+\frac{1}{r^2}\pa_\theta^2+\frac{1}{r^2\sin^2\theta}\pa_\varphi^2.
\end{align*}
By a density statement, we may assume that $u$ is supported away from spatial infinity.
\begin{itemize}
	\item \underline{Killing multiplier: $X=\pa_{\bar{t}}$.}
We compute
\begin{equation}\label{eq:boundaryterm1}
	\begin{split}
		g^{-1}(d\bar{t}, P[g,\pa_{\bar{t}}])&=\Big(\frac{\Delta_b}{r^2}F_t'(r)^2-\frac{r^2}{\Delta_b}\Big)T(\partial_{\bar{t}}, \partial_{\bar{t}})+\frac{\Delta_b}{r^2}F'_t(r)T(\partial_r, \partial_{\bar{t}})\\
		&=-\frac 12\bigg(\Big(\frac{r^2}{\Delta_b}-\frac{\Delta_b}{r^2}F_t'(r)^2\Big)\lvert\partial_{\bar{t}}u\rvert^2+\frac{\Delta_b}{r^2}\lvert\partial_r u\rvert^2+\abs{\pa_\theta u}^2+\frac{1}{\sin^2\theta}\abs{\pa_\varphi}^2\bigg)\\
		&\sim  -\Big(\frac{1}{r^2}\lvert\partial_{\bar{t}}u\rvert^2+\lvert\partial_r u\rvert^2+\frac{1}{r^2}\abs{\pa_\theta u}^2+\frac{1}{r^2\sin^2\theta}\abs{\pa_\varphi u}^2\Big)
		\quad \mbox{on}
		\quad [r_-,\infty),\\
		g^{-1}(dr, P[g,\pa_{\bar{t}}])&=\frac{\Delta_b}{r^2}F_t'(r)T(\partial_{\bar{t}}, \partial_{\bar{t}})+\frac{\Delta_b}{r^2}T(\partial_r, \partial_{\bar{t}})\\
		&=\frac{\Delta_b(r_-)}{r_-^2}F_t'(r_-)\lvert\partial_{\bar{t}}u\rvert^2+\frac{\Delta_b(r_-)}{r_-^2}\partial_{\bar{t}}u\partial_r u	\quad\mbox{at}\quad r=r_-.
		\end{split}
\end{equation}
We see that $g^{-1}(d\bar{t}, P[g,\pa_{\bar{t}}])$ gives control of $\abs{\frac{1}{r}\partial_{\bar{t}}u}$ and angular derivative but not $\partial_ru$ near event horizon because $\Delta_b(r)$ is nonpositive at and beyond event horizon. We also note that the first term of $g^{-1}(dr, P[g,\pa_{\bar{t}}])$ is positive if $r_-$ is sufficiently close to event horizon $r=\ehKN$ while the second term does not have a definite sign. So we need to introduce a suitable multiplier which allows us to deal with the above issues at event horizon.

\item \underline{Energy estimates near event horizon.} Following \cite{MMTT10}, We  introduce $X=b(r)\pa_r$ which is expressed in terms of the coordinates $(\bar{t}, r,\theta,\varphi)$ with $b(r)=-\chi(r)$ where $\chi(r)=1$ when $r\leq 3\Bm$ and $\chi=0$ when $r\geq 4\Bm$.
%In the $(t,r, \theta,\varphi)$ coordinates this means that
%\[
%X=b(r)\pa_r -b(r)F_t'(r)\pa_t= b_1(r)\big( \frac{1}{F'(r)}\pa_r -\pa_t\big) .
%\]
First we compute the boundary terms at the spacelike hypersurfaces $\bar{t}=const$ 
\begin{align*}
	g^{-1}(d\bar{t}, P[g,X])&=\Big(\frac{\Delta_b}{r^2}F_t'(r)^2-\frac{r^2}{\Delta_b}\Big)T(\partial_{\bar{t}}, b(r)\partial_r)+\frac{\Delta_b}{r^2}F_t'(r)T(\partial_r, b(r)\partial_r)\\
	&=-\chi(r)\bigg(-\Big(\frac{\Delta_b}{r^2}F_t'(r)^2-\frac{r^2}{\Delta_b}\Big)\partial_{\bar{t}}u\partial_r u+\frac{\Delta_b}{r^2}F_t'(r)\lvert\partial_r u\rvert^2\bigg).
\end{align*}
Since $\frac{\Delta_b}{r^2}F_t'(r)\sim 1$ near event horizon $r=\ehKN$, it follows that on $r\in[r_-,\infty)$
\begin{equation}\label{eq:boundaryt}
	g^{-1}(d\bar{t}, P[g,C\pa_{\bar{t}}+X])\sim
	 - E[u](\bar{t})
	%\Big(\lvert\partial_{\bar{t}}u\rvert^2+\lvert\partial_r u\rvert^2+\frac{1}{r^2}\abs{\pa_\theta^2 u}^2+\frac{1}{r^2\sin^2\theta}\abs{\pa_\varphi u}^2\Big)\
\end{equation}
provided $C$ is large enough.
At the hypersurface $r=r_-$, we have
\begin{align*}
	g^{-1}(dr, P[g,X])&=\frac{\Delta_b}{r^2}F_t'(r)T(\partial_{\bar{t}}, b(r)\partial_r)+\frac{\Delta_b}{r^2}T(\partial_r, b(r)\partial_r)\\
	&=-\frac 12\chi(r)\bigg(\Big(\frac{r^2}{\Delta_b}-\frac{\Delta_b}{r^2}F_t'(r)^2\Big)\lvert\partial_{\bar{t}}u\rvert^2+\frac{\Delta_b}{r^2}\lvert\partial_r u\rvert^2-\frac{1}{r^2}\abs{\pa_\theta^2 u}^2-\frac{1}{r^2\sin^2\theta}\abs{\pa_\varphi u}^2\bigg).
\end{align*}
Therefore, combining the calculation of $	g^{-1}(dr, P[g,X])$ with \eqref{eq:boundaryterm1} yields
\begin{equation}\label{eq:boundaryr}
	\begin{split}
	&g^{-1}(dr, C\pa_{\bar{t}}+X)\\
	&\quad=
	\Big(C\frac{\Delta_b(r_-)}{r_-^2}F_t'(r_-)+\frac12\big(\frac{r^2}{\Delta_b}-\frac{\Delta_b}{r^2}F_t'(r)^2\big)-\epsilon\Big)\lvert\partial_{\bar{t}}u\rvert^2+\epsilon(\partial_{\bar{t}}u+\frac{C\Delta_b(r_-)}{2\epsilon r_-^2}\partial_ru\Big)^2\\
		&\quad+\Big(-\frac{\Delta_b(r_-)}{2r_-^2}-\frac{C^2\Delta_b^2(r_-)}{4\epsilon r_-^4}\Big)\lvert\partial_r u\rvert^2+\frac{1}{2r_-^2}\abs{\pa_\theta^2 u}^2+\frac{1}{2r_-^2\sin^2\theta}\abs{\pa_\varphi u}^2\geq 0\quad\mbox{at}\quad r=r_-.
\end{split}
\end{equation}
 provided $C$ is sufficiently large, $\epsilon$ is small enough and $r_-$ is sufficiently close to $\ehKN$. 
 
 Then we integrate $\nabla^\alpha P_\alpha[g,	C\pa_{\bar{t}}+X]=\Box_gu\cdot (\pa_{\bar{t}}u+u)+\frac 12 T_{\alpha\beta}[g]\pi_X^{\alpha \beta}$ over the region \[\mathcal{D}=\{0\leq\bar{t}\leq\bar{t}_1, r\geq r_-\}
 \]
  with respect to the volume form $dV_{g_b}=r^2\sin\theta d\bar{t}drd\theta d\varphi$ and use Divergence Theorem 
 \[
 \int_\mathcal{D}\nabla^\alpha P_\alpha\,dV=\int_{\pa\mathcal{D}}\iota_{P}dV
 \]
 to obtain
\begin{align*}
	&\int_{\mathcal{D}}\Box_gu\cdot (C\pa_{\bar{t}}u+Xu) \,r^2\sin\theta d\bar{t}drd\theta d\varphi+\frac 12\int_{\mathcal{D}}T_{\alpha\beta}[g]\pi_X^{\alpha \beta} \,r^2\sin\theta d\bar{t}drd\theta d\varphi\\
	&\quad=\int_{r_-}^\infty\int_{\BS^2} g^{-1}(d\bar{t}, P[g,C\pa_{\bar{t}}+X])\, r^2\sin\theta drd\theta d\varphi|_{\bar{t}=0}^{\bar{t}=\bar{t}_1}\\
	&\quad\quad-\int_{0}^{\bar{t}_1}\int_{\BS^2}g^{-1}(dr, P[g,C\pa_{\bar{t}}+X]) r^2\sin\theta d\bar{t}d\theta d\varphi|_{r=r_-}.
\end{align*}
Since $g^{-1}(dr, P[g,C\pa_{\bar{t}}+X])|_{r=r_-}\geq0$, and $T_{\alpha\beta}[g]\pi_X^{\alpha \beta}$ is quadratic in $\partial u$,  we find that
\begin{align*}
	\sup_{0\leq\bar{t}\leq\bar{t}_1}E[u](\bar{t})
	%&\lesssim E[u](\bar{t}_1)+\int_{\mathcal{D}}\lvert \Box_{g}u\rvert\cdot\lvert C\pa_{\bar{t}}u+Xu\rvert\,dV+\frac 12\int_{\mathcal{D}}\lvert Q[g, X]\rvert\,dV\\
	&\leq C'\Big(\norm{\pa u_0}_{L^2(\Sigma_0)}+\norm{\frac{u_1}{r}}_{L^2(\Sigma_0)}+\norm{rf}^2_{L^1([0, \bar{t}); L^2(\Sigma_{\bar{t}}))}\Big)\\
	&\quad +C'\int_{0}^{\bar{t}_1}E[u](\bar{t})\,d\bar{t}.
\end{align*}
where $\pa\in\{\pa_r, \frac{1}{r}\pa_\theta, \frac{1}{r\sin\theta}\pa_\varphi\}$ and $L^2(\Sigma_0)=L^2(\Sigma_0;r^2\sin\theta\,dr\,d\theta\,d\varphi)$.

As for the $L^2$ norm of $u$, we have
\[
\norm{\frac{u}{r}}_{L^2(\Sigma_{\bar{t}})}\lesssim\norm{\frac{u_0}{r}}_{L^2(\Sigma_{0})}+\int_0^{\bar{t}_1}\norm{\frac{\pa_{\bar{t}}u}{r}}_{L^2(\Sigma_{\bar{t}})}\,d\bar{t}
\]
where $L^2(\Sigma_{\bar{t}})=L^2(\Sigma_{\bar{t}};r^2\sin\theta\,dr\,d\theta\,d\varphi)$.
Therefore, we have 
\begin{align*}
	\sup_{0\leq\bar{t}\leq \bar{t}_1} E[u](\bar{t})	&\leq C\Big(\norm{\frac{ u_0}{r}}_{L^2(\Sigma_0)}+\norm{\pa u_0}_{L^2(\Sigma_0)}+\norm{\frac{u_1}{r}}_{L^2(\Sigma_0)}+\norm{rf}^2_{L^1([0, \bar{t}); L^2(\Sigma_{\bar{t}}))}\Big)\\
	&\quad +C\int_{0}^{\bar{t}_1}E[u](\bar{t})\,d\bar{t}.
\end{align*}
 Finally by Gr\"{o}nwall inequality we obtain the exponential growth of the energy
\[
E[u](\bar{t})\leq C\Big(\norm{\frac{u_0}{r}}_{H^1(\Sigma_0)}+\norm{\pa u_0}_{L^2(\Sigma_0)}+\norm{\frac{u_1}{r}}_{L^2(\Sigma_0)}+\norm{rf}^2_{L^1([0, \bar{t}); L^2(\Sigma_{\bar{t}}))}\Big)e^{C\bar{t}}.
\]
%where $C(b)$ depends on the parameter $b=(\Bm, \Ba, \BQ)$ continuously.
\item\underline{Proof for $\mathcal{P}_{b,\gamma}, \mathcal{W}_{b,\gamma}$ and $L_{b,\gamma}$.} According to Lemma \ref{LemPWLtStar}, $-2\mathcal{P}_{b,\gamma}, -2\mathcal{W}_{b,\gamma}$ (resp. $-L_{b,\gamma}$) are $\Box_{g_b}$ tensored with $4\times 4$ (resp. $14\times 14$) identity matrix plus a a matrix whose entries are first order differential operators with coefficients decaying like $r^{-2}$, and then the above arguments in the proof for $\Box_{g_b}$ still apply here.
\end{itemize}
\end{proof}

We next show that if $\IM\sigma$ is sufficiently large, then $\widehat{\Box_{g_{b}}}(\sigma), \widehat{\mathcal{P}_{b, \gamma}}(\sigma),\widehat{\mathcal{W}_{b,\gamma}}(\sigma), \widehat{L_{b,\gamma}}(\sigma)$ are invertible on suitable function spaces.

\begin{thm}\label{thm:energyestimate}
	Let $b_0=(\Bm_0, 0,\BQ_0)$ with $\abs{\BQ_0}<\Bm_0$. Then there exists $\epsilon>0$ such that for $b=(\Bm, \BQ,\Ba)$ with $\abs{b-b_0}<\epsilon$ and for $s>\frac{1}{2}, \ell<-\frac{1}{2}$ with $s+\ell>-\frac12$, the operators
		\begin{align}\label{eq:largeiminvertible1}
			&\widehat{\Box_{g_{b}}}(\sigma):\{u\in\bar{H}_{\bop}^{s,\ell}(\CX)\mid	\widehat{\Box_{g_{b}}}(\sigma)u\in\bar{H}_{\bop}^{s,\ell+1}(\CX)\}\to\bar{H}_{\bop}^{s,\ell+1}(\CX)\\\label{eq:largeiminvertible2}
			&\widehat{\Box_{g_{b}}}(\sigma):\{u\in\bar{H}_{\bop}^{s,\ell}(\CX)\mid	\widehat{\Box_{g_{b}}}(\sigma)u\in\bar{H}_{\bop}^{s-1,\ell+2}(\CX)\}\to\bar{H}_{\bop}^{s-1,\ell+2}(\CX)
		\end{align}
		are invertible for $\IM\sigma\geq C'(b_0)$ where $C'(b_0)>0$ is a sufficiently large constant only depending on $b_0$.
		
		Moreover, the above statements also hold for the wave type operators $\mathcal{P}_{b,\gamma}, \mathcal{W}_{b,\gamma}$ and $L_{b,\gamma}$.
\end{thm}
\begin{proof}
According to Proposition \ref{prop:energyestimate}, there exist $\epsilon>0, \tilde{C}(b_0)>0$ such that for $\abs{b-b_0}<\epsilon$, the solution $u$ to the initial value problem \eqref{eq:ivp} satisfies
\[
	E[u](\bar{t})\leq \Big(\tilde{C}(b_0)E[u](0)+\tilde{C}(b_0)\big(\int_{0}^{\bar{t}}\lVert rf\rVert_{L^2(\Sigma_{\bar{t}})}\,d\bar{t}\big)^2\Big)e^{\tilde{C}(b_0)\bar{t}}.
\]
We now show that if $\IM\sigma>\tilde{C}(b_0)$, then \eqref{eq:largeiminvertible1} and \eqref{eq:largeiminvertible2} are invertible. According to Lemma \ref{lem:indexofFredholmoperator}, it suffices to prove that \eqref{eq:largeiminvertible1} and \eqref{eq:largeiminvertible2} are injective. Assume the contrary, by Proposition \ref{prop:desofkernel}, there exists a nontrivial solution $w=\rho C^\infty(\pa_+\CX)+\mathcal{A}^{2-}(\CX)$ to $\widehat{\Box_{g_{b}}}(\sigma)w=0$. Then, $u(t_{b,*}, r,\theta, \varphi)=e^{-it_{b,*}\sigma}w$ is a solution to $\Box_{g_b}u=0$. Let $\chi(\bar{t})\in C^\infty(\BR)$ be such that $\mbox{supp}\chi\subset[1,\infty)$ and $\mbox{supp}(1-\chi)\subset(-\infty, 2]$. Then $\chi u$ is the unique solution which is supported in $(1,\infty)$ to the inhomogeneous equation $\Box\tilde{u}=[\Box_{g_b},\chi]u$. According to Lemma \ref{LemPWL} and $\bar{t}-t_{b,*}=\mathcal{O}(\frac{1}{r})$, we have
\[
\Box\tilde{u}=[\Box_{g_b},\chi]u\sim\frac{1}{r^3-}
\]
and thus $\chi(\bar{t})u$ satisfies $E(\chi u)\lesssim e^{\tilde{C}(b_0)\bar{t}}$. But a direct calculation implies $E(\chi u)\sim e^{\IM\sigma t_{b,*}}\sim e^{\IM\sigma\bar{t}}$, which leads to a contradiction.

The proof for $\widehat{\mathcal{P}_{b, \gamma}}(\sigma), \widehat{\mathcal{W}_{b, \gamma}}(\sigma), \widehat{L_{b,\gamma}}(\sigma)$ is completely analogous.
\end{proof}

%%%%%%%%%%%%%%%%%%%%%%%%%%%%%%%%%%%%%%%%%%%%%%%%%%%%%%%%%%%%%%%%

\section{Spherical harmonic decompositions }\label{sec:basiccalculation}
In this section, we shall introduce the spherical harmonic decompositions, which will be used in the proof of the geometric mode stability of Reissner-Nordstr\"{o}m spacetime (see \S\ref{sec:gmodestability}).

%The first step in the proof of Theorem \ref{thm:modestability} is to decompose $(\dot{g}, \dot{A})$ into scalar and vector type spherical harmonics.

Let $\slashed{g}$ be the standard metric on the unit sphere $\BS^2$. We denote the operators on $\BS^2$ using a slash, e.g. $\slashed{\tr}=\tr_{\sg}, \slashed{\delta}=\delta_{\slashed{g}}, \slashed{\delta}^*=\delta_{\slashed{g}}^*$, etc. We denote by $Y_l^m$ with $l\in\BN, m\in\BZ, \abs{m}\leq l$ the spherical harmonic function of degree $l$ and order $m$ which satisfies $\slashed{\Delta}_{H}Y_l^m=l(l+1)Y_l^m$ where $\slashed{\Delta}_{H}=\slashed{d}\slashed{\delta}+\slashed{\delta}\slashed{d}$ is the Hodge Laplacian. We denote by $\BFS_l=\{Y_l^m: \abs{m}\leq l\}$ the space of degree $l$ spherical harmonic functions. Then one can conclude that $\{\BFS_l: l\in\BN\}$ form an orthogonal basis of $L^2(\BS^2)$.

According to Hodge decomposition Theorem and the fact that the de Rham cohomology group $H^1_{dR}(\BS^2)=0$, we see that any $1$-form $\omega\in C^{\infty}(\BS^2, T^*\BS^2)$ can be uniquely decomposed as $\omega=\slashed{d}\phi+\slashed{\delta} \eta$ where $\phi\in C^{\infty}(\BS^2), \eta\in C^\infty(\BS^2; \Lambda^2T^*\BS^2)$. Since $\sdelta=-\slashed{*}\sd \sstar$ where $\sstar$ is the Hodge star operator, we can rewrite $\omega=\sd\phi+\sstar\sd\psi$ where $\phi, \psi\in C^\infty(\BS^2)$. Therefore, one can conclude that an orthogonal basis of $L^2(\BS^2, T^*\BS^2)$ is given by $\{\sd\BFS_l, \BFV_l=\sstar\sd\BFS_l: l\in\BN\}$ where $\BFS_l, \BFV_l$ are the eigen-$1$-forms of the Hodge Laplacian $\Delta_H$ with eigenvalue $l(l+1)$. We note that $\sdelta\BFV_l=0$ and $\sd\BFS_0=\BFV_0=0$. In other words, a spectral decomposition of the negative tensor Laplacian $-\slashed{\Delta}=-\str\slashed{\nabla}_a\slashed{\nabla}_b$ satisfying $-\slashed{\Delta}=\slashed{\Delta}_H-1$ is given by the scalar part 
\begin{equation}
	\sd\BFS_l\in\ker(-\slashed{\Delta}-\left(l(l+1)-1\right))\quad\mbox{with}\quad l\geq 1
\end{equation}
and the vector part
\begin{equation}
	\BFV_l\in\ker(-\slashed{\Delta}-\left(l(l+1)-1\right))\quad\mbox{with}\quad l\geq 1.
\end{equation}

As for the symmetric $2$-tensors $h\in C^\infty(\BS^2; S^2T^*\BS^2)$, we have the transverse-traceless decomposition \cite{Y74} $h=\frac12(\str h)\sg+\sdelta_0^*\omega+h_{TT}$ where $\sdelta_0^*=\sdelta^*+\frac 12\sg\sdelta$ is the traceless symmetric gradient, $\omega\in C^\infty(\BS^2; T^*\BS^2)$ and $h_{TT}$ is both traceless and divergence free. According to \cite{H87}, we know that $h_{TT}=0$. We then decompose $\omega$ further into eigen-$1$-forms (we also call them spherical harmonic $1$-forms) of $-\slashed{\Delta}$ of scalar type $\sd\BFS_l$ and vector type $\BFV_l$, and $\str h$ into scalar spherical harmonic functions $\BFS_l$. In the spherical coordinates $(\theta, \varphi)$, since \[\sd Y_1^0=-\sin\theta d\theta, \quad\sd Y_1^1=\cos\theta \cos\varphi d\theta-\sin\theta\sin\varphi d\varphi, \quad\sd Y_1^{-1}=\cos\theta\sin\varphi d\theta+\sin\theta\cos\varphi d\varphi,
\]
we see that $\sd\BFS_1$ is conformal killing and thus $\sdelta^*_0\sd\BFS_1=0$. We also have \begin{equation}\label{eq:Vkilling}
	\sstar\sd Y_1^0=-\sin^2\theta d\varphi, \quad\sstar\sd Y_1^1=\sin\theta\cos\theta \cos\varphi d\varphi+\sin\varphi d\theta, \quad\sstar\sd Y_1^{-1}=\sin\theta\cos\theta \sin\varphi d\varphi-\cos\varphi d\theta.
\end{equation}
and find that $\BFV_1^{\sharp}$ are rotations, and thus killing, i.e. $\sdelta^*\BFV_1=0$. Then any symmetric $2$-tensors on sphere can be decomposed into the following spherical harmonic symmetric $2$-tensors of the scalar type
\begin{equation}
	\BFS_l\sg~(l\geq0),\quad \sdelta^*_0\sd\BFS_l~(l\geq 2)
\end{equation}
and the vector type
\begin{equation}
	\sdelta_0^*\BFV_l=	\sdelta^*\BFV_l~(l\geq2).
\end{equation}

We claim that the following geometric operators on sphere $\BS^2$ 
\[\sg, \str, \sd, \sdelta, \sdelta^*, -\slashed{\Delta} \mbox{ and their compositions}
\]
preserve the the scalar and vector type of the spherical harmonics. More precisely, these geometric operators map the scalar type functions/$1$-forms/symmetric $2$-tensors built out of a particular $\mathsf{S}\in \BFS_l$ into another scalar type tensor with the same $\mathsf{S}$, likewise for the vector type $1$-forms/symmetric $2$-tensors. This is obviously true for $\sg$ acting on functions, $\str$ on symmetric $2$-tensors, $\sd$ on functions, $\sdelta$ on $1$-forms, and $-\slashed{\Delta}$ on functions and $1$-forms. For the verification of $\sdelta, \sdelta^*, -\slashed{\Delta}$, we make use of the identities $\sdelta\sdelta^*\omega=-\frac 12 \slashed{\Delta}\omega+\frac 12\sd\sdelta\omega-\frac 12\omega$ and $-\slashed{\Delta}\sdelta^*\omega=\sdelta^*(-\slashed{\Delta}\omega)-3\sdelta^*\omega-2\sg\sdelta\omega$ to obtain
\begin{equation}
	\begin{split}
		\sdelta(\mathsf{S}\sg)&=-\sd \mathsf{S},~\sdelta \sdelta_0^*\sd \mathsf{S}=\frac{l(l+1)-2}{2}\sd\mathsf{S},~ \sdelta\sdelta^*\!\!\sstar\sd \mathsf{S}=\frac{l(l+1)-2}{2}\sstar\sd \mathsf{S}; ~\sdelta^*\!\!\sd \mathsf{S}=\sdelta_0^*\sd \mathsf{S}\!-\!\!\frac{l(l+1)}{2}\mathsf{S}\sg;\\
		-\slashed{\Delta}(\mathsf{S}\sg)&=l(l+1)\mathsf{S}\sg,\quad-\slashed{\Delta}(\sdelta_0^*\sd \mathsf{S})=\left(l(l+1)-4\right)(\sdelta_0^*\sd \mathsf{S}),\quad -\slashed{\Delta}(\delta^*\sstar\sd \mathsf{S})=\left(l(l+1)-4\right)	(\delta^*\sstar\sd \mathsf{S}).
	\end{split}
\end{equation}

We now decompose the spacetime $M$ into the aspherical $\hX=\BR_{t_*}\times [r_-, \infty)_r$ and spherical part $\BS^2$, i.e.
\begin{equation}\label{eq:as_sdecom}
	M=\hX\times \BS^2, \quad \hpi: M\to\hX,\quad \slashed{\pi}: M\to \BS^2.
\end{equation}
We then call the functions \textit{aspherical} if they are only functions of $(t_*, r)$, and \textit{spherical} if they are only functions on $\BS^2$. Alternatively, we can identify the aspherical and spherical functions as subspaces of $C^\infty(M)$ as follows:
\begin{equation}
	C^\infty(\hX)\cong\hpi^*C^\infty(\hX)\subset C^\infty(M),\quad C^\infty(\BS^2)\cong\slashed{\pi}^*C^\infty(\BS^2)\subset C^\infty(M).
\end{equation}
Then functions on $M$, say $L^2(M)$, can be decomposed into an infinite sum of products of aspherical functions and spherical harmonic functions. 

We can also split the cotangent bundle $T^*M$ into aspherical and spherical part as follows:
\begin{equation}\label{eq:splitof1}
	T^*M=T^*_{AS}\oplus T^*_{S}\quad \mbox{where}\quad T^*_{AS}=\hpi^* T^*\hX\quad\mbox{and}\quad T^*_{S}=\slashed{\pi}^*T^*\BS^2.
\end{equation}
We can further write $\omega\in T^*_S$ as an infinite sum of the spherical harmonic $1$-forms $\sd\BFS_l, \BFV_l$ of $-\slashed{\Delta}$. Therefore, a $1$-form $\omega\in C^\infty(M; T^*M)$ can be expressed as an infinite sum of products of aspherical functions/$1$-forms on $\hX$ and spherical harmonic functions/$1$-forms.

The above splitting of $T^*M$ induces a natural splitting of the symmetric $2$-tensors,into aspherical part $S^2T^*_{AS}$, mixed part $T^*_{AS}\otimes T^*_S$ and spherical part $S^2T^*_S$ as follows:
\begin{equation}\label{eq:splitofsym2}
	S^2T^*M=S^2T^*_{AS}\oplus(T^*_{AS}\otimes T^*_S)\oplus S^2T^*_S
\end{equation}
where we identify $a\otimes b\in T^*_{AS}\otimes T^*_S$ as $2a\otimes_s b=a\otimes b+b\otimes a$.
Therefore, a symmetric $2$-tensor $h\in C^\infty(M; S^2T^*M)$ can be expressed as an infinite sum of products of aspherical functions/$1$-forms/symmetric $2$-tensors on $\hX$ and spherical harmonic functions/$1$-forms/symmetric $2$-tensors.

The above decomposition for functions/$1$-form/symmetric $2$-tensors on $M$ implies that the perturbation $(\dot{g}, \dot{A})$ can be divided into two categories: the first consists of \textit{scalar perturbations} (also called even parity modes in \cite{RW57} and closed portions in \cite{HKW20}):
\begin{equation}\label{eq:scalarper}
	\begin{split}
		&	\mbox{scalar}~l\geq 2:\quad\begin{cases}
			\dot{g}=\widetilde{f}\sfS+(f\otimes_s\sd\sfS)+(H_L\sfS\sg+H_T\sdelta_0^*\sd\sfS)\\
			\dot{A}=\widetilde{K}\sfS+K\sd\sfS
		\end{cases}\quad\mbox{with}\quad \sfS\in\BFS_l\\
		&	\mbox{scalar}~l=1:\quad\begin{cases}
			\dot{g}=\widetilde{f}\sfS+(f\otimes_s\sd\sfS)+H_L\sfS\sg\\
			\dot{A}=\widetilde{K}\sfS+K\sd\sfS
		\end{cases}\quad\mbox{with}\quad \sfS\in\BFS_1\\
		&	\mbox{scalar}~l=0:\quad\begin{cases}
			\dot{g}=\widetilde{f}+H_L\sg\\
			\dot{A}=\widetilde{K}
		\end{cases}
	\end{split}
\end{equation}
where \[
H_L, H_T, K\in C^\infty(\hX), \quad f, \widetilde{K}\in C^\infty(\hX; T^*\hX), \quad\widetilde{f}\in C^\infty(\hX; S^2T^*\hX).
\]
The second comprises of the \textit{vector perturbations} (also called odd parity modes in \cite{RW57} and co-closed portions in \cite{HKW20}): 
\begin{equation}\label{eq:vectorper}
	\begin{split}
		&	\mbox{vector}~l\geq 2:\quad\begin{cases}
			\dot{g}=f\otimes_s\sfV+H_T\sdelta^*\sfV\\
			\dot{A}=K\sfV
		\end{cases}\quad\mbox{with}\quad \sfV\in\BFV_l\\
		&	\mbox{vector}~l=1:\quad\begin{cases}
			\dot{g}=f\otimes_s\sfV\\
			\dot{A}=K\sfV
		\end{cases}\quad\mbox{with}\quad \sfV\in\BFV_1
	\end{split}
\end{equation}
where $
H_T,K\in C^\infty(\hX)$ and $f\in C^\infty(\hX; T^*\hX)$.

We decompose the Reissner-Nordstr\"{o}m metric $g$ into aspherical and spherical part as in \eqref{eq:as_sdecom}, which takes the form in static coordinates $(t,r, \omega)$ with $\omega\in\BS^2$ as follows
\begin{equation}
	g=\hg+r^2\sg
\end{equation}
where $\hg$ is a Lorentzian metric on $\hX=\BR_{t_*}\times [r_-, \infty)$ and $\sg$ denote the standard metric on the unit sphere $\BS^2$. In static coordinates $(t,r)$, $\hg$ takes the form 
\begin{equation}
	\hg=-\weight dt^2+\weight^{-1}dr^2\quad\mbox{with}\quad\weight=1-\frac{2\Bm}{r}+\frac{\BQ^2}{r^2},
\end{equation}
and it can be extended beyond the event horizon by introducing $(t_*, r)$ coordinates. We denote the operators associated with $\hg$ and $\sg$ by rings and slashes respectively. Here we use the Einstein summation convention with Greek indices $\mu, \nu,\dots$ for coordinates on $M$, lower case Roman indices $i,j,k,m,n$ for coordinates on $\hX$ and $a,b,c,d,e$ for coordinates on $\BS^2$. Indices are raised and lowered using the Reissner-Nordstr\"{o}m metric $g$, except that we write $\sg^{ab}=(\sg^{-1})_{ab}$ for the inverse metric to $\sg$ on $\BS^2$.

%%%%%%%%%%%%%%%%%%%%%%%%%%%%%%%%%%%%%%%%%%%%%%%%%%%%%%%%%%%%%%%%%%%%%%%%%%%
\section{Mode stability of the Reissner-Nordstr\"{o}m spacetime}\label{sec:gmodestability}
In this section we shall characterize the non-decaying (generalized) modes of the linearized Einstein-Maxwell system around the Reissner-Nordstr\"{o}m spacetime $(M, g_{b_0}, A_{b_0})$ with subextremal charge, following the method of Kodama-Ishibashi \cite{KI03, KI04} for the study of perturbations of black holes without (or with) charge in high dimensions. Throughout this section, we drop the subscript $b_0$, i.e., we take $(g, A)=(g_{b_0}, A_{b_0})$ and $\Bm=\Bm_0, \BQ=\BQ_0$. Here we adopt some notations from \cite{KI04, H18, HHV21}.

\begin{thm}\label{thm:modestability}
	Let $\sigma\in \BC$ with $\IM\sigma\geq0$, suppose $(\dot{g}, \dot{A})$ is an outgoing mode solution to the linearized Einstein-Maxwell system, i.e., $(\dot{g}, \dot{A})=(e^{-i\sigma t_*}\dot{g}_0, e^{-i\sigma t_*}\dot{A}_0)$ with $(\dot{g}_0, \dot{A}_0)\in \eHb^{\infty, 
	\ell}(\CX; \scsym\oplus \scform)$ for some $\ell\in\BR$ solves
	\begin{equation}
		\begin{split}
			\mathscr{L}_1(\dot{g}, \dot{A})&=-\frac 12\Box_{g}\dot{g}-\delta^*_g\delta_gG_g\dot{g}+\mathscr{R}_g\dot{g}-2D_{(g, dA)}T(\dot{g}, d\dot{A})=0\\
				\mathscr{L}_2(\dot{g}, \dot{A})&=D_{(g, A)}\left(\delta_{g}dA\right)(\dot{g}, \dot{A})=0.
			\end{split}
	\end{equation}
Then there exist parameters $\dot{\Bm}\in \BR, \dot{\Ba}\in \BR^3, \dot{\BQ}\in\BR$ and an outgoing $1$-form $\omega$ and a function $\phi$ on $\CM$, i.e. $(\omega, \phi)\in e^{-i\sigma t_*}\eHb^{\infty, 
	\ell'}(\CX; \scform \oplus \BC)$ for some $\ell'\in\BR$, such that 
\begin{equation}\label{eq:modesoln}
	(\dot{g}, \dot{A})=\left(\dot{g}_{(\Bm, 0,\BQ)}(\dot{\Bm}, \dot{\Ba},\dot{\BQ})+2\delta_g^*\omega, ~\dot{A}_{(\Bm, 0,\BQ)}(\dot{\Bm}, \dot{\Ba},\dot{\BQ})+\mathcal{L}_{\omega^\#}A+d\phi\right).
\end{equation}
More specifically
\begin{enumerate}
	\item If $\sigma\neq 0$, then 
		\[
	(\dot{g}, \dot{A})=\left(2\delta_g^*\omega, ~\mathcal{L}_{\omega^\sharp}A+d\phi\right)
	\]
where $(\omega, \phi)\in e^{-i\sigma t_*}\eHb^{\infty, 
		\ell'}(\CX; \scform\oplus \BC)$ for some $\ell'\in\BR$.
	\item If $\sigma=0$ and $-\frac 32<\ell<-\frac 12$, we decompose the perturbation $(\dot{g}, \dot{A})$ into spherical harmonics whose detail is given in \S \ref{sec:basiccalculation}. Then
	\begin{itemize}
		\item [(\romannumeral 1)] If $(\dot{g}, \dot{A})$ is of scalar type with $l\geq 1$ or vector type $l\geq 2$, then 
		\[
			(\dot{g}, \dot{A})=\left(2\delta_g^*\omega, ~\mathcal{L}_{\omega^\sharp}A+d\phi\right)
		\]
		where $(\omega, \phi)\in \eHb^{\infty, 
			\ell-1}(\CX; \scform\oplus \BC)$ is of the same type as $(\dot{g}, \dot{A})$.
		\item[(\romannumeral 2)] If $(\dot{g}, \dot{A})$ is of scalar type with $l=0$, i.e. spherically symmetric, then 
		\[
		(\dot{g}, \dot{A})=\left(\dot{g}_{(\Bm, 0,\BQ)}(\dot{\Bm}, 0, \dot{\BQ})+2\delta_g^*\omega, ~\dot{A}_{(\Bm, 0,\BQ)}(0, 0,\dot{\BQ})+\mathcal{L}_{\omega^\sharp}A+d\phi\right)
		\]
		where $(\omega, \phi)\in\eHb^{\infty, 
			\ell-1}(\CX; \scform\oplus \BC)$ is spherically symmetric.
		\item[(\romannumeral 3)] If $(\dot{g}, \dot{A})$ is of vector type $l=1$, then 
		\[
		(\dot{g}, \dot{A})=\left(\dot{g}_{(\Bm, 0,\BQ)}(0, \dot{\Ba},0)+2\delta_g^*\omega, ~\dot{A}_{(\Bm, 0,\BQ)}(0, \dot{\Ba},0)+\mathcal{L}_{\omega^\sharp}A+d\phi\right)
		\]
		where $(\omega, \phi)\in\eHb^{\infty, 
			\ell-1}(\CX; \scform\oplus \BC)$ is of vector type with $l=1$.
\end{itemize}
The statement of stationary perturbation $(\dot{g}, \dot{A})$ ($\sigma=0$) of scalar type with $l=0,1$ can be extended to the following cases:
\begin{itemize}
	\item [(\romannumeral 4)] If $(\dot{g}, \dot{A})\in \eHb^{\infty, 
		\ell}(\CX; \scsym\oplus \scform)$ with $-\frac 52<\ell<-\frac 32$ is a stationary perturbation of scalar type $l=1$, then 
		\[
	(\dot{g}, \dot{A})=\left(2\delta_g^*\omega, ~\mathcal{L}_{\omega^\sharp}A+d\phi\right)
	\]
	where $(\omega, \phi)\in\eHb^{\infty, 
		\ell-1}(\CX; \scform\oplus \BC)$ is of the scalar type $l=1$.
	\item[(\romannumeral 5)] If $(\dot{g}, \dot{A})\in \mbox{Poly}(t_*)^k\eHb^{\infty, 
		\ell}(\CX; \scsym\oplus \scform)$ is of scalar type $l=0$, then 
		\[
	(\dot{g}, \dot{A})=\left(\dot{g}_{(\Bm, 0,\BQ)}(\dot{\Bm}, 0, \dot{\BQ})+2\delta_g^*\omega, ~\dot{A}_{(\Bm, 0,\BQ)}(0, 0,\dot{\BQ})+\mathcal{L}_{\omega^\sharp}A+d\phi\right)
	\]
	where $(\omega, \phi)\in\mbox{Poly}(t_*)^{k+1}\eHb^{\infty, 
		\ell'}(\CX; \scform\oplus \BC)$ is of scalar type $l=0$ for some $\ell'<\ell$.
\end{itemize}
\end{enumerate}
\end{thm}

\begin{rem}
	Our discussion corresponds to the case $n=2,K=1, \kappa^2=2, E_0=r^{-2}\BQ, Q^2=\BQ^2, \Lambda=0$ (and thus $\lambda=0$), in \cite{KI04}. We note that \cite{KI04} discusses the scalar and vector type perturbations for non-stationary modes ($\sigma\neq0$) with $l\geq2$ in detail, but the corresponding stationary perturbations ($\sigma=0$) are only treated in the uncharged case in\cite{KI03}. \cite{KI04} also studies the scalar (Appendix D) and vector (\S 4) perturbations for modes $l=1$ (which is called exceptional modes there), however, they do not give a full description of the scalar $l=0$ (spherically symmetric) perturbations.
\end{rem}

\subsection{Scalar type perturbations}
We shall consider the scalar type perturbations of modes $l\geq 2, l=1$ and $l=0$ separately.  Recall that we can write the perturbations under the splitting \eqref{eq:splitof1} and \eqref{eq:splitofsym2} as in \eqref{eq:scalarper}. We further introduce a rescaled version of the traceless part of the spherical harmonic symmetric $2$-tensor $\sdelta_0^*\sd \mathsf{S}_l$ which is built from the spherical harmonic function $\mathsf{S}\in \BFS_l$ with eigenvalues $k^2=l(l+1)$:
\[
\rsH_k\mathsf{S}=k^{-2}\sdelta_0^*\sd\mathsf{S}\quad \mbox{where}\quad \mathsf{S}\in\BFS_l\mbox{ with }l\neq 0.
\]

\subsubsection{Modes with $l\geq 2$}
Suppose $(\dg, \dA)$ is the scalar perturbations of the following form
\begin{equation}\label{eq:perturbation_l2}
\dg=\begin{pmatrix}
	\widetilde{f}\sfS\\
	-\frac{r}{k}f\otimes\sd\sfS\\
	2r^2(H_L\sfS\sg+H_T\rsH_k\sfS)
\end{pmatrix},\quad \dA=\begin{pmatrix}
\widetilde{K}\sfS\\
-\frac{r}{k}K\sd\sfS
\end{pmatrix}\quad\mbox{with}\quad \sfS\in\BFS_l ~(l\geq2).
\end{equation}
We notice that the pure gauge solutions take the form
\begin{equation}
\bdelta\dg=2\delta_g^*\omega=\begin{pmatrix}
2\hdelta^*T	\\
-\frac{r}{k}(-\frac{k}{r}T+r\hd r^{-1}L)\sd\sfS\\
2r^2(r^{-1}\iota_{dr}T+\frac{k}{2r}L)\sfS\sg-\frac{k}{r}L\rsH_k\sfS
\end{pmatrix},\quad\!\! \bdelta\dA=\mathcal{L}_{\omega^\sharp}A+d\phi=\begin{pmatrix}
(\hd\iota_AT+\iota_{T}\hd A+\hd P)\sfS\\
-\frac{r}{k}(-\frac{k}{r}\iota_AT-\frac{k}{r}P)\sd\sfS
\end{pmatrix} 
	\end{equation}
with 
\begin{equation}
\omega=\begin{pmatrix}
	T\sfS\\
	-\frac{r}{k}L\sd\sfS
\end{pmatrix},\quad\!\! \phi=P\sfS
\end{equation}
where $T\in C^\infty(\hX; T^*\hX), L, P\in C^\infty({\hX})$. When adding $(\bdelta\dg, \bdelta\dA)$ to $(\dg, \dA)$, the quantities $\tilde{f}, f, H_L, H_T, \tilde{K}, K$ change by
\begin{equation}
	\begin{gathered}
	\bdelta\widetilde{f}=2\hdelta^*T,\quad\bdelta f=-\frac{k}{r}T+r\hd (r^{-1}L),\quad \bdelta H_L=r^{-1}\iota_{dr}T+\frac{k}{2r}L,\quad \bdelta H_T=-\frac{k}{r}L;\\
	\bdelta\widetilde{K}=\hd\iota_AT+\iota_T\hd A+\hd P,\quad \bdelta K=-\frac{k}{r}(\iota_A T+P).
	\end{gathered}
\end{equation}
We define $\mathbf{X}:=\frac{r}{k}(f+\frac{r}{k}\hd H_T)$, then $\bdelta\mathbf{X}=\frac{r}{k}(\bdelta f+\frac{r}{k}\hd \bdelta H_T)=-T$ because $\bdelta$ commutes with any geometric operators we use in this manuscript. As a consequence, the following quantities
\begin{equation}
	\begin{split}
		\widetilde{F}&:=\widetilde{f}+2\hdelta^*\mathbf{X}\in C^{\infty}(\hX; S^2T^*\hX),\\
		J&:=H_L+\frac{1}{2}H_T+r^{-1}\iota_{dr}\mathbf{X}\in C^\infty(\hX;T^*\hX),\\
	N&:=\widetilde{K}+\hd\left(\frac{r}{k}K\right)+\iota_{\mathbf{X}}\hd A\in C^\infty(\hX; T^*\hX)
	\end{split}
\end{equation}
are gauge-invariant, that is, $\bdelta\widetilde{F}=0, \bdelta J=\bdelta N=0$. Conversely, if $\widetilde{F}=0,  J= N=0$, one can verify $(\dg, \dA)$ is a pure gauge solution
\begin{equation}
	(\dg, \dA)=(2\delta^*\omega, \mathcal{L}_{\omega^\sharp}A+d\phi) \quad\!\! \mbox{with}\quad\!\! \omega=\begin{pmatrix}
		-\mathbf{X}\sfS\\
		\frac{r^2}{k^2}H_T\sd\sfS
	\end{pmatrix},\quad \!\! \phi=-\frac{r}{k}K+\iota_A\mathbf{X}.
\end{equation}
	If $(\dg, \dA)$ are outgoing mode solutions with frequency $\sigma\neq 0$, then $\omega$ and $ \phi$ are modes of the same type, with an extra factor $r^2$ and $r$ respectively relative to $\dg$ and $\dA$.  If $\sigma=0$, $(\omega, \phi)$ grow at most by a factor $r$ more than $(\dg, \dA)$.
	
	Since the linearized Einstein-Maxwell system is gauge-invariant, we can express it in terms of the gauge-invariant quantities defined above. Concretely, one can choose a gauge, i.e., add a pure gauge solution $(\bdelta\dg, \bdelta\dA)$ to $(\dg, \dA)$ for suitable $\omega=\omega(T,L), \phi=\phi(P)$ such that the non gauge-invariant quantities $\tilde{f}$ etc. take a simple form in the new gauge. More specifically, we adds $(\bdelta\dg, \bdelta\dA)$ built from $T=\mathbf{X}, L=\frac{r}{k}H_T, P=\frac{r}{k}K-\iota_A\mathbf{X}$, then $\mathbf{X}+\bdelta\mathbf{X}=T-T=0, H_T+\bdelta H_T=\frac{k}{r}L-\frac{k}{r}L=0$ which implies $f+\bdelta f=0, \widetilde{f}+\bdelta\widetilde{f}=\tilde{F}, H_L+\bdelta H_L=J$, and $K+\bdelta K=K-\frac{k}{r}\iota_A\mathbf{X}-\frac{k}{r}(\frac{r}{k}K-\iota_A\mathbf{X})=0$ which implies $\widetilde{K}+\bdelta \widetilde{K}=N$. To summarize, if we replace $(\dg, \dA)$ by $(\dg+\bdelta\dg, \dA+\bdelta\dA)$, in the new gauge we have $(\widetilde{f}, f, H_L, H_T, \widetilde{K}, K)=(\widetilde{F}, 0, J, 0, N, 0)$. 
	Therefore, using the detailed calculation in \S \ref{sec:basiccalculation} and \S \ref{subsec:detailcal} we can write the linearized 	Einstein-Maxwell system, acting on the new $(\dg, \dA)$
	\[
	\dg=\begin{pmatrix}
	&	\widetilde{F}\sfS\\
		&0\\
		&2r^2 J\sfS\sg
	\end{pmatrix}, \quad \dA=\begin{pmatrix}
	&N\sfS\\
	&0
\end{pmatrix}
	\] in terms of the gauge-invariant quantities $\widetilde{F}, J,N$. 
	
	Now we express $2\mathcal{L}_1(\dg, \dA)=0$ and $\mathcal{L}_2(\dg, \dA)=0$ in the form of \eqref{eq:perturbation_l2} in terms of $\widetilde{f}^E, f^E, H_L^E, H_T^E$ and $\widetilde{K}^E, K^E$ respectively. Then the linearized Einstein-Maxwell system reads
\begin{subequations}
		\begin{align}
				\widetilde{f}^E&=\left(-\hBox-2\hdelta^*\hdelta-\hdelta^*\hd \htr\right)\widetilde{F}+2r^{-1}\left(2\hdelta^*\iota_{dr}\widetilde{F}-(\pa^i r)\hnabla_i\widetilde{F}\right)-4\hdelta^*\hd J-8r^{-1}(dr)\otimes_s \hd J\notag \\
				&\quad+(-\weight''+k^2r^{-2})\widetilde{F}+(-2\BQ^2 r^{-4}+\weight'')\hg\htr\widetilde{F}-4\BQ r^{-2}\hg\hstar\hd N=0,\label{eq:Etildef}\\
				-\frac{r}{k}f^E&=-\hdelta\widetilde{F}-2\hd J-r\hd(r^{-1}\htr\widetilde{F})+4\BQ r^{-2}\hstar N=0,\label{eq:Ef}\\
			2r^2H_L^E&=-\hBox(2r^2J)-2r\iota_{dr}\hdelta\widetilde{F}+2\iota_{dr}\iota_{dr}\widetilde{F}-r\iota_{dr}\hd\htr\widetilde{F}+(r\weight'+\frac{k^2}{2}+2\BQ^2 r^{-2})\htr \widetilde{F}\notag\\
			&\quad+	(2k^2-4\BQ^2r^{-2})J+4\BQ\hstar\hd N=0,\label{eq:EH_L}\\
		2r^2H_T^E&=-k^2\htr\widetilde{F}=0,\label{eq:EH_T}\\
			\widetilde K^E&=r^{-2}\hdelta(r^2\hd N)+k^2r^{-2}N+\frac{1}{2}\BQ r^{-2}\hstar\hd\htr\widetilde{F}-2\BQ r^{-2}\hstar\hd J=0,\label{eq:EtildeK}\\
	-\frac{r}{k}K^E&=-\hdelta N=0.\label{eq:EK}
\end{align}
	\end{subequations}
By \eqref{eq:EH_T}, all terms containing $\htr \widetilde{F}$ vanish, then plugging $ 
\hdelta\widetilde{F}=-2\hd J+4\BQ r^{-2}\hstar N$ into \eqref{eq:Etildef} yields the cancellation of the term $4\hdelta^*\hd J$ and thus one obtains a wave equation of $\widetilde{F}$ (coupled to $J, N$ only via subprincipal terms). Moreover, by \eqref{eq:EK}, one adds the zero term $\hd\hdelta N$ to \eqref{eq:EtildeK} and then can obtain a wave equation for $N$ (coupled to $J$ via subprincipal terms). In conclusion, we can rewrite equation \eqref{eq:Etildef}, \eqref{eq:EH_L} and \eqref{eq:EtildeK} as a principally scalar system of wave equations for $(\widetilde{F}, J, N)$
\begin{equation}
	-\hBox B-\mathscr{D}B=0\quad \mbox{where}\quad
	B=
	\begin{pmatrix}
		\widetilde{F}\\
		J\\
		N
	\end{pmatrix}
\end{equation}
where $\mathscr{D}$ is a first order stationary differential operator acting on $C^{\infty}(\hX; S^2T^*\hX\oplus \BR\oplus T^*\hX)$. When $(\dg, \dA)$ and thus $B$ are smooth modes, this system of equations become a system of ODEs on the one dimensional space $t_*^{-1}(0)$ with a regular singular point at the event horizon $r=\ehRN$ whose solution is given by Frobenius series, which means that the vanishing of $B$ in the static region $r>\ehRN$ implies the vanishing of $B$ on $\hX$. As a result, it suffices to prove $B=0$ in the static region $r>\ehRN$. To achieve this, we can work in the static coordinates $(t,r)$.

First, since $\hdelta N=\hstar\hd\hstar N=0$, we have $\hd\hstar N=0$ and then  $\hstar N=\hd\mathcal{N}$ for some $\mathcal{N}\in C^\infty(\hX)$ according to $H_{dR}^1(\hX)=0$, which implies $N=\hstar\hd\mathcal{N}$. Returning to the equation \eqref{eq:EtildeK}, we find
\begin{equation}
r^2\hstar\widetilde{K}^E=\hd(-r^2\hBox\mathcal{N})+k^2\hd\mathcal{N}-2\BQ
\hd J=\hd\left(-r^2\hBox \mathcal{N}+k^2\mathcal{N}-2\BQ J\right)=0
\end{equation}
and thus
\[
-r^2\hBox \mathcal{N}+k^2\mathcal{N}-2\BQ J=C\quad\mbox{for some}\quad\!\!C\in\BR.
\]
Replacing $\mathcal{N}$ by $\mathcal{N}-\frac{C}{k^2}$, we still have $N=\hstar\hd\mathcal{N}$ and meanwhile
\begin{equation}\label{eq:eqnforN}
	-\hBox \mathcal{N}+k^2r^{-2}\mathcal{N}-2\BQ r^{-2}J=0.
\end{equation}
We note that since $N$, and thus $\hstar N$ is a mode, $\mathcal{N}$ can also be chosen as a mode. More specifically, suppose $\hstar 	N=e^{-it_*\sigma}(N_{0}dt_*+N_1dr)$ where $N_0, N_1$ are smooth functions of $r$ and $\sigma\neq 0$, we let $\mathcal{N}=e^{-it_*\sigma}\mathcal{N}_0$ with $\mathcal{N}_0=i\sigma^{-1}N_0$. Then $N_1=\pa_r\mathcal{N}_0=i\sigma^{-1}\pa_r N_0$ will be satisfied automatically because $\hd\hstar N=e^{-it_*\sigma}(-\pa_rN_0-i\sigma N_1)dt_*\wedge dr=0$.

Next, according to the second Bianchi identity $\delta_gG_g\Ric(g)=0$ (which holds for any metric $g$), one concludes for all $(g, F)$
\[
\delta_g G_g(\Ric(g)-2T(g,F))=-2\delta_g T(g,F)=-2g^{\al\be}(\delta_gF)_\al F_{\n\be}-g^{\al\kappa}g^{\be\lambda}F_{\al\be}(\nabla_\n F_{\kappa\lambda}+2\nabla_{\kappa}F_{\lambda\n}).
\]
Since $dF_{\n\kappa\lambda}=\nabla_{\n}F_{\kappa\lambda}
-\nabla_{\lambda}F_{\n\kappa}+\nabla_{\kappa}F_{\lambda\n}$ and $g^{\al\kappa}g^{\be\lambda}F_{\al\be}(\nabla_{\lambda}F_{\n\kappa}+\nabla_{\kappa}F_{\lambda\n})=0$, it follows that
\begin{equation}\label{eq:Lsecondbianchi}
\delta_g G_g(\Ric(g)-2T(g,F))=-2\delta_g T(g,F)=-2g^{\al\be}(\delta_gF)_\al F_{\n\be}-g^{\al\kappa}g^{\be\lambda}F_{\al\be}(dF)_{\n\kappa\lambda}.
\end{equation}
Linearizing the above equation around $(g, F)=(g_{b_0}, dA_{b_0})$, if $(\dg, \dA)$ is a solution to the Maxwell part of the linearized Einstein-Maxwell system, i.e., $\mathcal{L}_2(\dg, \dA)=0$, we find by using $\Ric(g)-2T(g,dA)=0, \delta_g dA=0, d^2A=0$ that
\begin{equation}\label{eq:LsecondBianchiM}
\delta_g G_g\mathcal{L}_1(\dg, d\dA)=0,\quad\mbox{i.e., }\quad\!\! \delta_g G_g\begin{pmatrix}
	\widetilde{f}^E\sfS\\
	-\frac{r}{k}f^E\otimes \sd\sfS\\
	2r^2(H_L^E\sfS\sg+H_T^E\rsH_k\sfS)
\end{pmatrix}=0,
\end{equation}
which gives the following system of equations for $(\widetilde{f}^E, f^E, H_L^E, H_T^E)$
\begin{gather*}
	r^{-2}\hdelta(r^2\widetilde{f}^E)+\frac{1}{2}\hd\htr\widetilde{f}^E-\frac{k}{r}f^E+2r^{-2}\hd(r^2H_L^E)=0,\\
	\frac{1}{2}\htr\widetilde{f}^E-\frac{1}{kr^2}\hdelta(r^3f^E)+\frac{k^2-2}{k^2}H_T^E=0.
	\end{gather*}
We define 
\begin{equation}\label{eq:defoftildeE}
	\widetilde{E}=G_g\widetilde{f}^E-2H_L^E\hg,
\end{equation}
it is clear that $\widetilde{E}=0$ implies $(\widetilde{f}^E, H_L^E)=(h\hg, 0)$ for some $h\in C^\infty(\hX)$. If in addition equations \eqref{eq:Ef} and \eqref{eq:EH_T} hold (i.e. $f^E=0$ and $H_T^E=0$), one obtains $\htr\widetilde{f}^E=0, \hdelta(r^2\widetilde{E})=0$, and thus $(\widetilde{f}^E, H_L^E)=(0, 0)$. We further write $\hdelta(r^2\widetilde{E})=0$ in $(t,r)$ coordinates and find its $dr$ component
\[
-\frac {\weight'}{2\weight^2}\widetilde{E}_{tt}+\weight'\pa_t\widetilde{E}_{tr}+\weight^{-1/2}\pa_r(\weight^{3/2}\widetilde{E}_{rr})=0.
\]
Then the vanishing of $\widetilde{E}_{tr}$ and $\widetilde{E}_{rr}$ implies $\widetilde{E}_{tt}=0$ because $\weight^{-2}\weight'=\weight^{-2}(2\Bm r^{-2}-2\BQ^2r^{-3})\neq0$ in the static region $r>\ehRN=\Bm+\sqrt{\Bm^2-\BQ^2}>\BQ^2/\Bm$ in the subextremal charge case $\Bm>\BQ$. In summary, $(\widetilde{f}^E, f^E, H_L^E, H_T^E)=0$ is equivalent to $(f^E, H_T^E, \widetilde{E}_{tr}, \widetilde{E}_{rr})=0$. So our goal is to prove $(\widetilde{F}, J, \mathcal{N})=0$, provided that $(f^E, H_T^E, \widetilde{E}_{tr}, \widetilde{E}_{rr}, \widetilde{K}^E,K^E)=0$.

Following \cite[equation (5.27)]{KI04} or \cite[equation (5.37)]{H18}, we write
\begin{equation}\label{eq:defofXYZ}
	\widetilde{F}-2J\hg=\begin{pmatrix}
		\weight X\\
		-\weight^{-1} Z\\
		-\weight^{-1} Y
	\end{pmatrix}
\end{equation}
in the splitting \eqref{eq:splitofX}. From now on, we will use the fact $\htr\widetilde{F}=0$ which is implied by $H_T^E=0$ without specified till the end of this subsubsection. One can recover $\widetilde{F}, J$ as 
\begin{equation}\label{eq:recoverofFJ}
	\widetilde{F}=\begin{pmatrix}
		\frac{\weight}{2}(X-Y)\\
		-\weight^{-1}Z\\
		\frac{1}{2\weight}(X-Y)
	\end{pmatrix},\quad J=\frac{X+Y}{4}.
	\end{equation}
Then the equation \eqref{eq:Ef} $f^E=0$ implies $\hdelta(\widetilde{F}-2J\hg)=4\BQ r^{-2}\hg\hd\mathcal{N}$, which we express in terms of $X,Y,Z, \mathcal{N}$ as 
\begin{equation}\label{eq:XYZ1}
	\pa_tX+\pa_r Z=4\BQ r^{-2}\pa_t\mathcal{N},\quad -\frac{\weight'}{2\weight}(X-Y)-\weight^{-2}\pa_tZ+\pa_rY=4\BQ r^{-2}\pa_r\mathcal{N}.
\end{equation}
The equation $\widetilde{E}_{tr}=\widetilde{f}^E_{tr}=0$ reads (we use the equation \eqref{eq:XYZ1} during the calculation)
\begin{equation}\label{eq:XYZ2}
	-\pa_t\pa_r X-\pa_t\pa_r Y+\frac{\weight'}{2\weight}
\pa_t X+(\frac{\weight'}{2\weight}-\frac{2}{r})\pa_t Y-\frac{k^2}{r^2\weight}Z=0.
\end{equation}
Finally the equation $\weight\widetilde{E}_{rr}=\frac{1}{2}(\weight^{-1}\widetilde{f}^E_{tt}+\weight\widetilde{f}^E_{rr})-2H_L^E=0$ becomes (we use equation \eqref{eq:XYZ1} during the calculation)
\begin{equation}\label{eq:XYZ3}
	-\weight^{-1}\pa_t^2X-\weight^{-1}\pa_t^2 Y+\frac{\weight'}{2}\pa_rX+(\frac{\weight'}{2}+\frac{2\weight}{r})\pa_r Y-\frac{4}{r\weight}\pa_t Z-\frac{\BQ^2}{r^4}X-\left(\frac{k^2-2}{r^2}+\frac{3\BQ^2}{r^2}\right)Y+\frac{4k^2\BQ}{r^4}\mathcal{N}=0.
\end{equation}
Therefore, it suffices to prove $(X, Y, Z, \mathcal{N})=0$ provided that they satisfy the equations \eqref{eq:eqnforN} and \eqref{eq:XYZ1}--\eqref{eq:XYZ3}.

\underline{We now discuss the case that $(\dg,\dA)$ is a mode solution with $\sigma\neq 0$}. After taking Fourier transform with respect to $t$, setting $\pa_t (X, Y, Z,\mathcal{N})=-i\sigma(X, Y, Z, \mathcal{N})$ and solving for $(X', Y', Z')=\pa_r(X, Y,Z)$, we obtain from equations \eqref{eq:XYZ1} and \eqref{eq:XYZ2}
\[
\begin{pmatrix}
	X'\\
	Y'\\
	\frac{Z'}{i\sigma}
\end{pmatrix}=T\begin{pmatrix}
X\\
Y\\
\frac{Z}{i\sigma}
\end{pmatrix}+f \quad\mbox{where}\quad T=\begin{pmatrix}
0&\frac{\weight'}{\weight}-\frac{2}{r}&\frac{k^2}{r^2\weight}-\frac{\sigma^2}{\weight^2}\\
\frac{\weight'}{2\weight}&-\frac{\weight'}{2\weight}&\frac{\sigma^2}{\weight^2}\\
1&0&0
\end{pmatrix},\quad f=\frac{4\BQ}{r^2}\begin{pmatrix}
-\mathcal{N}'\\
\mathcal{N}'\\
-\mathcal{N}
\end{pmatrix},
\]
while the equation \eqref{eq:XYZ3} implies the following linear constraint on $X,Y,Z$
\begin{equation}\label{eq:constraintonXYZ}
\gamma\begin{pmatrix}
	X\\
	Y\\
	\frac{Z}{i\sigma}
\end{pmatrix}=h
\end{equation}
where $h=-\frac{4\BQ}{r^2}(\frac{2\weight}{r}\mathcal{N}'+\frac{k^2}{r^2}\mathcal{N})$ and 
\[
\gamma=(\frac{\sigma^2}{\weight}-\frac{\BQ^2}{r^4}+\frac{(\weight')^2}{4\weight}+\frac{\weight'}{r},\  \frac{\sigma^2}{\weight}-\frac{k^2-2}{r^2}-\frac{3\BQ^2}{r^4}+\frac{(\weight')^2}{4\weight}-\frac{2\weight'}{r},\  -\frac{2\sigma^2}{r\weight}+\frac{k^2\weight'}{2r^2\weight}).
\]

Following \cite{KI04} and \cite{H18}, one can find a linear combination $\Phi$ of $X, Y, \frac{Z}{i\sigma}$ which satisfies a second order ODE. Concretely, let
\begin{equation}
	\Phi:=\frac{\frac{2Z}{i\sigma}-r(X+Y)}{H}\quad\mbox{with}\quad m:=k^2-2,\quad x:=\frac{2\Bm}{r},\quad z:=\frac{\BQ^2}{r^2},\quad H:=m+3x-4z.
	\end{equation}
Then the second order ODE for $\Phi$ is (see \cite[equation (5.41)]{KI04}, \cite[equation (5.49)]{H18})
\begin{equation}\label{eq:eqforPhi}
	(\weight\pa_r)^2\Phi+(\sigma^2-V_{\Phi})\Phi=F_{\Phi}\mathcal{N}
\end{equation}
where $V_\Phi$ and $F_\Phi$ are (see \cite[equations (5.43), (5.44), (C.1)]{KI04}, \cite[equation  (B.1)]{H18})
\begin{equation}\label{eq:VFPhi}
	\begin{split}
	V_\Phi&=\frac{\weight}{r^2H^2}\Bigl(9x^3-9(6z-m)x^2+(72z^2-8(4m-3)z+3m^2)x-32z^3+24mz(z+1)+m^2(m+2)\Bigr)\\
	F_\Phi&=\frac{8\BQ\weight}{r^3H^2}\Bigl(-3x^2+(2z+6)x+m(m+4)\Bigr)
	\end{split}.
\end{equation}
Conversely, $X, Y, \frac{Z}{i\sigma}$ can be expressed in terms of $\Phi$ and $\mathcal{N}$ as
\begin{equation}\label{eq:eqforXYZs}
	\begin{split}
		X&=(\frac{\sigma^2r}{\weight}-\frac{P_{X_0}}{2rH^2})\Phi+\frac{P_{X_1}}{2H}\Phi'-\frac{2\BQ P_{X\mathcal{N}}}{r^2H^2}\mathcal{N}-\frac{8\BQ\weight}{rH}\mathcal{N}'\\
		Y&=(-\frac{\sigma^2r}{\weight}-\frac{P_{Y_0}}{2rH^2})\Phi+\frac{P_{Y_1}}{2H}\Phi'-\frac{2\BQ P_{Y\mathcal{N}}}{r^2H^2}\mathcal{N}+\frac{8\BQ\weight}{rH}\mathcal{N}'\\
		\frac{Z}{i\sigma}&=\frac{P_Z}{2H}\Phi-r\weight\Phi'-\frac{8\BQ\weight}{rH}\mathcal{N}'
	\end{split}
\end{equation}
where $P_{X_0}, P_{X_1},P_{X\mathcal{N}},P_{Y_0}, P_{Y_1},P_{Y\mathcal{N}}, P_Z$ are functions of $r$ (see \cite[equations (5.46), (C.4)--(C.8)]{KI04} and \cite[equations (B.2)]{H18})
\begin{equation}\label{eq:eqforP_XYZ}
	\begin{split}
		P_{X_0}&=27x^3-24(5z-m)x^2+\Bigl(152z^2-2(35m-12)z+3m(3m+2)\Bigr)x\\
		&\qquad -64z^3+48mz^2-8m(m-2)z+2m^2(m+2),\\
		P_{X_1}&=9x^2-(16z-5m+6)x+8z^2-6mz-4m,\\
		P_{X\mathcal{N}}&=-18x^2+4(8z-m+6)x-16z^2+4(m-4)z+2m(m+6),\\
		P_{Y_0}&=9x^3-6(10z-m)x^2+\Bigl(120z^2-2(11m-12)z+3m(m+2)\Bigr)x\\
		&\qquad -64z^3+16(m-4)z^2-8m(m+2)z,\\
		P_{Y_1}&=3x^2-(12z+m+6)x+8z^2+2(m+8)z,\\
		P_{Y\mathcal{N}}&=-6x^2+4(6z-m)x-16z^2+4(m-4)z-2m(m+2),\\
		P_Z&=3x^2+(-2z+3m)x-(4m+8)z-2m.
	\end{split}
\end{equation}

Therefore it suffices to prove $(\Phi, \mathcal{N})=0$. Recall that $\mathcal{N}$ satisfies the wave equation \eqref{eq:eqnforN} and since $\mathcal{N}$ is also a mode, we can rewrite it as 
\begin{equation}
(\weight\pa_r)^2\mathcal{N}+(\sigma^2-\frac{\weight k^2}{r^2})\mathcal{N}=-\frac{2\BQ J}{r^2}=-\frac{\weight\BQ(X+Y)}{2r^2}.
\end{equation}
Using \eqref{eq:eqforXYZs} and \eqref{eq:eqforP_XYZ}, we find
\begin{align*}
X+Y&=-(\frac{P_{X_0}}{2rH^2}+\frac{P_{Y_0}}{2rH^2	})\Phi+(\frac{P_{X_1}}{2H}+\frac{P_{Y_1}}{2	H})\Phi'-(\frac{2\BQ P_{X\mathcal{N}}}{r^2H^2}+\frac{2\BQ P_{Y\mathcal{N}}}{r^2H^2})\mathcal{N}\\
&=-(\frac{H}{r}-\frac{P_Z}{rH})\Phi-2\weight\Phi'-\frac{16\BQ \weight}{r^2H}\mathcal{N}
\end{align*}
and thus (see \cite[equation (5.49)]{KI04} and \cite[eqaution (5.52)]{H18})
\begin{equation}\label{eq:eqforN2}
	(\weight\pa_r)^2\mathcal{N}+(\sigma^2-V_{\mathcal{N}})\mathcal{N}=F_{\mathcal{N}0}\Phi+F_{\mathcal{N}1}\Phi'
\end{equation}
with
\begin{equation}\label{eq:VFN}
	V_{\mathcal{N}}=\frac{\weight k^2}{r^2}+\frac{8\BQ^2\weight^2}{r^4 H},\quad F_{\mathcal{N}0}=\frac{\BQ \weight}{2r^2}(\frac{H}{r}-\frac{P_Z}{rH}),\quad F_{\mathcal{N}1}=\frac{\BQ \weight^2}{r^2}.
\end{equation}
Now we have two coupled second order ODEs \eqref{eq:eqforPhi} and \eqref{eq:eqforN2}for $\Phi$ and $\mathcal{N}$ and in fact we can transform them into two decoupled second order ODEs by introducing suitable linear combinations of $\Phi$ and $\mathcal{N}$ (see \cite[equations (5.56)--(5.64)]{KI04} and \cite[equations (5.55)--(5.59)]{H18}). Concretely, we set 
\begin{equation}\label{eq:defofPsi+-}
\Psi_{\pm}=a_{\pm}\Phi+b_{\pm}\mathcal{N}
\end{equation}
where $a_{\pm}, b_{\pm}$ are smooth on $\hX$ (which extends a little bit beyond the event horizon $r=\ehRN$)
\begin{equation}\label{eq:a+-b+-}
	(a_+, b_+)=(\frac{\BQ m}{2\tilde{c}}+\frac{\BQ}{2r}, 1),\quad (a_-, b_-)=(\frac{\tilde{c}}{6\Bm}-\frac{2\BQ^2}{3\Bm r}, -\frac{4\BQ}{3\Bm}),\quad \tilde{c}=3\Bm+\sqrt{9\Bm^2+4\BQ^2 m}.
	\end{equation}
Since $\begin{vmatrix}
	\frac{\BQ m}{2\tilde{c}}+\frac{\BQ}{2r}&1\\
	\frac{\tilde{c}}{6\Bm}-\frac{2\BQ^2}{3\Bm r}& -\frac{4\BQ}{3\Bm}
	\end{vmatrix}=-\frac{2\BQ^2m}{3\Bm \tilde{c}}-\frac{\tilde{c}}{6\Bm}<0$, we can recover $\Phi, \mathcal{N}$ from $\Psi_{\pm}$. Next the decoupled equations for $\Phi_{\pm}$ are given as
\begin{equation}\label{eq:eqforPsi+-}
	(\weight\pa_r)^2\Psi_{\pm}+(\sigma^2-V_{\pm})\Psi_{\pm}=0
\end{equation}
where
\begin{equation}\label{eq:defofV+-}
	V_{\pm}=\frac{a_{\pm}F_{\Phi}+b_{\pm}V_{\mathcal{N}}}{b_{\pm}}.
\end{equation}
Introducing the tortoise coordinate $r_*=\int \weight^{-1}\,dr$, we see that the equations for $\Psi_{\pm}$ are reduced to an eigenvalue problem of the type $A\Psi=\sigma^2\Psi$ on the region $r^*\in(-\infty,\infty)$ where $A=-\pa_{r^*}^2+V(r)$ is a self-adjoint operator. As explained in \cite[\S 6]{KI04}, we expect $V(r)$ to be non-negative on $r\in[\ehRN, \infty)$ and then naturally so is $A$.
\begin{lem}
	Given $r\in [\ehRN, \infty)$, we have $V_{+}\geq0$ for $l\geq 1$, i.e., $k^2\geq 2$ and $m\geq0$.
	\end{lem}
	\begin{proof}
		Since $\ehRN=\Bm+\sqrt{\Bm^2-\BQ^2}$ and $H=m+\frac{2}{r}(3\Bm-\frac{2\BQ^2}{r})$, in the subextremal case $\BQ<\Bm$, we have $r\geq \ehRN>\Bm>\BQ>\frac{2\BQ^2}{3\Bm}$ and thus $H(r)>0$. We first analyze
		\[
		F_\Phi=\frac{8\BQ\weight}{r^3H^2}\Bigl(-3x^2+(2z+6)x+m(m+4)\Bigr),
		\]
		since $x=\frac{2\Bm}{r}\in[0, 2), z\geq 0, m\geq0$, we have $-3x^2+(2z+6)x+m(m+4)\geq m(m+4)\geq 0$ and thus $F_{\Phi}\geq0$. Next it is clear from \eqref{eq:VFN} that $V_{\mathcal{N}}\geq 0$. Therefore $V_+\geq0$.
	\end{proof}
However, the positivity of $V_-$ fails. To deal with this issue, we follow \cite[\S 6]{KI04} and introduce the procedure \textit{S-deformation} of $V_-$. More specifically, suppose $\Psi$ is compactly supported in $r_*$, we have \[(\Psi, A\Psi)_{L^2(dr_*)}=\int \abs{\pa_{r_*}\Psi}^2+V(r)\abs{\Psi}^2\,dr_*=\int \abs{\wt{D}\Psi}^2+\wt{V}(r)\abs{\Psi}^2\,dr_*\]
where $\wt{D}=\pa_{r_*}+S=\weight\pa_r+S$ and $\wt{V}=V+\weight\pa_{r}S-S^2$. We apply the $S$-deformation to $V_-$ with 
\[
S=S_-:=\frac{\weight\pa_r H_-}{H_-}=\weight\pa_r(\log H_-)\!\quad\!\mbox{where}\!\quad\! H_-=k^2-2+\frac{\tilde{c}}{r}>0 \!\quad\!\mbox{on}\!\quad\![\ehRN, \infty)\!\quad\!\mbox{for}\!\quad\! k^2\geq 2(l\geq1)
\]
and then
\begin{equation}\label{eq:Sdedor}
	\wt{V}_-=\frac{k^2(k^2-2)\weight}{r^2H_-}\geq 0\!\quad \!\mbox{on}\!\quad \![\ehRN, \infty)\quad\mbox{for}\quad k^2\geq 2(l\geq 1).
	\end{equation}
Namely, with $H_-$ and $V_-$ (which are also smooth near $r=\ehRN$) defined above, we write $\wt{D}=\pa_{r_*}+S_-=e^{-\int\! S_-dr_*}\pa_{r_*}e^{\int\!S_-dr^*}=H_-^{-1}\pa_{r_*}H_-=H_-\weight\pa_rH_-$ and find that $D^*$ which is the adjoint of $D$ with respect to $dr_*=\weight^{-1}dr$ can be expressed as $-H_-\pa_{r_*}H_-^{-1}=-H_-\weight\pa_rH_-^{-1}$. As a consequence, we have
\begin{equation}\label{eq:Psi-}
	\begin{split}
\wt{D}^*\wt{D}\Psi_-+\wt{V}_-\Psi_-&=-H_-\weight\pa_r\Bigl(H_-^{-2}\weight\pa_r(H_-\Psi)\Bigr)+\wt{V}_-\Psi\\
&=-(\weight\pa_r)^2\Psi-\Psi H_-\weight\pa_r(H_-^{-2}\weight\pa_rH_-)+(V+H_-^{-1}(\weight\pa_r)^2H_--2H_-^{-2}(\weight\pa_rH_-)^2)\Psi\\
&=-(\weight\pa_r)^2\Psi+V\Psi.
\end{split}
\end{equation}

Now we are at the position to show that $\Psi_{\pm}=0$ on $r>\ehRN$, and thus from the previous discussion the mode solution $(\dg, \dA)$ with $\sigma\neq0$ is a pure gauge solution. To achieve this, we first need to discuss the behavior of $\Psi_{\pm}$ near $r=\ehRN$ and $r=\infty$, i.e., $r_*=\mp\infty$. Since $H, a_{\pm}, b_{\pm}$ are bounded away from zero, we only need to analyze the contributions of $X, Y,Z,\mathcal{N}$. Note that $\wt{F}, J$ are smooth on $\hX$ (which extends a little bit beyond the event horizon). Therefore, the contribution of $J=(X+Y)/4$ to $\Phi$ is smooth on $\hX$. As for the contribution from $Z$, we notice that $\pa_t, \pa_{r_*}$, which are expressed as $\pa_t, \weight\pa_r$ in the static coordinates $(t,r)$, are smooth vector fields on $\hX$, and thus $Z=\weight\wt{F}_{tr}=-\wt{F}(\pa_t, \pa_{r^*})$ is smooth on $\hX$ as well. As a result, $\Phi=e^{-i\sigma t_*}C^{\infty}([\ehRN, \infty)_r)$, which is written in the static coordinates as
 \[
 \Phi=(t, r)=e^{-i\sigma t}\Theta(r)\quad\mbox{where}\quad \Theta(r)\in e^{\pm i\sigma r_*}C^\infty([\ehRN, \infty)) \quad \mbox{when}\quad r^*\to\pm\infty.
 \]
 which implies the exponential decay of $\Phi$ as $r_*\to-\infty$ when $\IM\sigma>0$. Also, when $r_*\to\infty$, since $\dg=e^{-i\sigma t_*}\dg_0$ with $\dg_0\in\eHb^{\infty, \ell}(\CX;\scsym)$, we still obtain the rapid decay of $\Phi$ when $r^*\to\infty$. As for $\mathcal{N}$, from the previous discussion, we know $\mathcal{N}$ is a mode, that is, $\mathcal{N}=e^{-i\sigma t_*}\mathcal{N}_0$ with $\mathcal{N}_0\in C^\infty([\ehRN, \infty)_r)\cap \eHb^{\infty, \ell}(\CX)$ as well, and thus decays rapidly when $r_*\to\pm\infty$. Consequently, $\Psi_{\pm}$ decays rapidly as well when $r_*\to\pm\infty$ when $\IM \sigma>0$.
 
 Then when $\IM \sigma>0$, by pairing $(-\pa_{r_*}^2+V_+-\sigma^2)\Psi_+=0$ and $(-\wt{D}^*\wt{D}+\wt{V}_--\sigma^2)\Psi_-=0$ with $\overline{\Psi}_+$ and $\overline{\Psi}_-$ respectively with respect to the $L^2(\BR_{r_*}; dr_*)$ and then integrating by parts (whose validity is guaranteed by the rapid decay of $\Psi_{\pm}$ when $r_*\to\pm\infty$ ), one obtains
 \[
 0=\norm{\pa_{r_*}\Psi_+}_{L^2}+\norm{V_+^{1/2}\Psi_+}_{L^2}-\sigma^2\norm{\Psi_+}_{L^2},\quad 0=\norm{\wt{D}\Psi_-}_{L^2}+\norm{\wt{V}_-^{1/2}\Psi_-}_{L^2}-\sigma^2\norm{\Psi_-}_{L^2}.
 \]
 When $\RE\sigma=0$ and thus $\sigma^2<0$, we have a sum of squares and thus $\Psi_{\pm}=0$. When $\RE \sigma\neq0$, the imaginary part $-2i\RE\sigma\IM\sigma\norm{\Psi_{\pm}}_{L^2}=0$ and thus $\Psi_{\pm}=0$ on $(\ehRN, \infty)$.
 
 When $\sigma\in \BR\setminus\{0\}$, we still have $e^{i\sigma t}\Phi, e^{i\sigma t}\mathcal{N}\in e^{\pm i\sigma r_*}\Bigl(C^\infty([\ehRN, \infty))\cap\eHb^{\infty, \ell}(\CX)\Bigr)$ when  
 $r_*\to\pm\infty$ for some $\ell\in\BR$, and thus so do $\Psi_{\pm}$. First, by normal operator arguments as in the proof of Proposition \ref{prop:desofkernel}, we find $\Psi_+=e^{-i\sigma t}e^{i\sigma r_*}(a+\mathcal{A}^{1-})$ with $a\in\BC$ when $r_*\to\infty$. Next, a standard boundary pairing argument, see the proof of Theorem \ref{thm:modesforscalarwave} below for the detail (starting at \eqref{eq:boundarypair}), allows us to conclude $\lim _{r_*\to\pm\infty}e^{\mp i\sigma r_*}\Psi_{+}=0$. Then an indicial root argument implies $\Psi_+\in e^{-i\sigma t}e^{i\sigma r_*}\mathcal{A}^\infty$, i.e., $\Psi_+$ vanishes to infinite order at $r_*=\infty$. Lastly, the unique continuation at infinity (\cite[Theorem 17.2.8]{H05}) implies $\Psi_+=0$ on $(\ehRN, \infty)$. The proof for $\Psi_-$ proceeds similarly.
 
\underline{Next we treat the stationary mode solutions, i.e., $\sigma=0$}. We still assume $l\geq2$. Now $(\dg, \dA)$ and thus $\wt{F}, J, N$ and $X, Y,Z$ are stationary. 

We first analyze the nature of $\mathcal{N}$ defined as $\hstar N=\hd\mathcal{N}$. Let $\hstar N=N_0dt_*+N_1dr$ where $N_0, N_1$ are smooth functions of $r$ only, and then $\hd\hstar N=0$ gives $\pa_rN_0=0$ and thus $N_0$ is a constant. Therefore, we have the unique (up to additive constants) $\mathcal{N}=N_0t_*+\int_{r_-}^r N_1(s)ds$, which is a generalized mode with frequency $\sigma=0$ and grows at most linearly in $t_*$. Since $X,Y,Z$ are stationary, equation \eqref{eq:XYZ2} implies
\begin{equation}\label{eq:Z=0}
	Z=0,
	\end{equation}
which in turn implies $\pa_t\mathcal{N}=0$ by using equation \eqref{eq:XYZ1}. Therefore, we conclude $N_0=0$ and $\mathcal{N}$ is stationary as well.
\begin{rem}
	Alternatively, using $N\in\eHb^{\ell, \infty}(\CX; \scform)\subset \mathcal{A}^{\ell+3/2}\subset\mathcal{A}^{0+}$ (because $-3/2<\ell<-1/2$) when $\sigma=0$, it also follows that $N_0=c=0$.
\end{rem}
As for $\Phi$, we notice that $Z/(i\sigma)$ is not well-defined when $\sigma=0$ and we need to do some remedies. Using \eqref{eq:constraintonXYZ}, $Z/(i\sigma)$ can be expressed as a linear combination of $X, Y,\mathcal{N}, \mathcal{N}'$ and thus we can rewrite $\Psi_{\pm}$ as linear combinations of $X,Y,\mathcal{N},\mathcal{N}'$ whose coefficients depend on $\sigma$
\begin{equation}\label{eq:newexpressionforPsi+-l2}
	\Psi_{\pm}(\sigma)=C_{X\pm}(\sigma)X+C_{Y\pm}(\sigma)Y+C_{\mathcal{N}\pm}(\sigma)\mathcal{N}+C_{\mathcal{N}'\pm}(\sigma)\mathcal{N}'
\end{equation}
where $C_{\bullet \pm}(
\sigma)$ are rational functions of $r$ depending on $\sigma\in\BC$. To make the dependence on $\sigma$ explicit, we write (see \cite[equations (5.65), (B.3)]{H18})
\begin{equation}\label{eq:CXYNN'l2}
	\begin{split}
		C_{X\pm}&=a_{\pm}\frac{P_X}{3\wt{H}},\quad
C_{Y\pm}=a_{\pm}\frac{P_Y}{3\wt{H}},\\
C_{\mathcal{N}+}&=1+\frac{\weight(m+2)P_{\mathcal{N}}}{\tilde{c}r\wt{H}},\quad C_{\mathcal{N}-}=b_-\Bigl(1+\frac{6a_-\weight(m+2)x}{r\wt{H}}\Bigr)\\
C_{\mathcal{N}'+}&=\frac{2\weight^2P_{\mathcal{N}}}{\tilde{c}\wt{H}},\quad C_{\mathcal{N}'-}=\frac{4b_-\weight^2(\tilde{c}-4rz)}{r\tilde{H}}
\end{split}
\end{equation}
where $\wt{H}(\sigma)=H(k^2\weight'-4r\sigma^2)$ and $P_X, P_Y, P_{\mathcal{N}}$ are smooth on $\hX$
\begin{equation}\label{eq:eqforP_XYNl2}
	\begin{split}
	P_X&=(9x-36z-3m-18)x+6(4z+m+8)z,\\
	P_Y&=-3(9x-16z+5m-6)x-6(4z-3m)z+12m,\\
	P_{\mathcal{N}}&=-8(\tilde{c}+mr)z,
	\end{split}
	\end{equation}
Since $\weight'=2 r^{-2}(\Bm-r^{-1}\BQ^2)>0$ in the  region $r\geq \ehRN>\Bm>\Bm^{-1}\BQ^2$ and thus $\wt{H}(0)=Hk^2\weight'>0$ there as well, we conclude that $\Psi_{\pm}(\sigma)$ exist down to $\sigma=0$. We define $\Psi_{\pm}:=\Psi_{\pm}(0)$. Conversely, using the first two equations in \eqref{eq:eqforXYZs} with $\sigma=0$, one can recover $X, Y$ in terms of $\Phi, \mathcal{N},\mathcal{N}'$ and thus $\Psi_{\pm}, \mathcal{N},\mathcal{N}'$. One can also check that $\Psi_{\pm}$ satisfy the equation \eqref{eq:eqforPsi+-} with $\sigma=0$ (because $\Psi_{\pm}(\sigma)$ is holomorphic near $\sigma=0$), i.e.,
\begin{equation}\label{eq:eqforPsi+-sigma0l2}
	(\weight\pa_r)^2\Psi_{\pm}-V_{\pm}\Psi_{\pm}=0
\end{equation}
with $V_{\pm}$ defined as before in \eqref{eq:defofV+-}, \eqref{eq:a+-b+-}, \eqref{eq:VFN} and \eqref{eq:VFPhi}. 

Since $(\dg, \dA)\in\eHb^{\infty, \ell}(\CX; \scsym\oplus\scform)\subset \mathcal{A}^{\ell+3/2}$, we have $X, Y,\mathcal{N}'\in\mathcal{A}^{\ell+3/2}(\CX)$ and $\mathcal{N}\in \mathcal{A}^{\ell+1/2}$. Let $\rho=1/r$, we notice that $H=m+6\Bm\rho-4\BQ^2\rho^2$ and $k^2\weight'=2k^2\Bm\rho^2-2k^2\BQ^2\rho^3$, so $\wt{H}(0)\in 2k^2m\Bm\rho^2+\rho^3C^\infty(\CX)$. In view of the expression in \eqref{eq:eqforP_XYNl2}, we find $P_Y\in C^\infty(\CX)$ and $P_X, P_{\mathcal{N}}\in \rho C^\infty(\CX)$, and then $C_{Y\pm}\in \rho^{-2}C^\infty(\CX)$, $C_{X\pm},C_{\mathcal{N}'\pm}\in \rho^{-1}C^\infty(\CX)$ and $C_{\mathcal{N}\pm}\in C^\infty(\CX)$. As a consequence, we have $\Psi_{\pm}\in\mathcal{A}^{\ell-1/2}\subset \mathcal{A}^{-2+}$ a priori because $-3/2<\ell<-1/2$. Now we are ready to describe the asymptotic behavior of $\Psi_\pm$ near $r=\ehRN$ and $r=\infty$. We only present the proof of $\Psi_+$ because the proof of $\Psi_-$ follows analogously. Since $F_{\Phi}, V_{\mathcal{N}}, a_{\pm}, b_{\pm}$ are smooth across the event horizon, we have $\weight(\pa_r\Psi_+)\overline{\Psi}_+|_{r=\ehRN}=0$. This implies that boundary term at $r=\ehRN$ arising in the integration by parts we will do later on vanishes. Near $r=\infty$, owing to \eqref{eq:VFPhi}, \eqref{eq:VFN} \eqref{eq:a+-b+-} and \eqref{eq:defofV+-}, we conclude $F_{\Phi}\in\rho^3C^\infty(\CX)$, $V_{\mathcal{N}}\in \frac{k^2}{r^2}+\rho^4C^\infty(\CX)$ and then $V_{\pm}\in\frac{k^2}{r^2}+\rho^3C^\infty(\CX)$, which implies $V_{\pm}=\frac{k^2}{r^2_*}+\mathcal{A}^{3-}$ as a function of $r_*$ near $r_*=\infty$. We set $x=1/r_*$ for the moment, \[
(\weight\pa_r)^2\Psi_+-V_+\Psi_+=\pa_{r_*}^2\Psi_+-V_+(r_*)\Psi=x^4\pa_x\Psi_++2x^3\pa_x\Psi_+-V_+(1/x)=0,
\]
which implies $x=0$(i.e. $r_*=\infty$) is a regular singular point and its indicial equation is $\lambda(\lambda-1) +2\lambda-k^2=0$. The roots to the indicial equation are $\lambda_{\pm}=(-1\pm\sqrt{1+4k^2})/2$. Since $k^2\geq 6$, we see that $\lambda_+\geq 2$  corresponds to one solution $\Psi_+$ with decay $\sim r_*^{-\lambda_+}$ near $r_*=\infty$, while $\lambda_-\leq -3$ implies another linearly independent solution $\Psi_+$ with growth $\sim r_*^{-\lambda_-}$ near $r_*=\infty$. Since we have $\Psi_+\in \mathcal{A}^{-2+}$ a priori, this excludes the second solution with growth. Namely, $\Psi_+\sim r^{-\lambda_+}$ which means that $\Psi_+\sim r_*^{-2}$ near $r_*=\infty$. With this decay rate near $r_*=\infty$, we can justify the $L^2(\BR_{r_*};dr_*)$ pairing between $((\weight\pa_r)^2-V_+)\Psi_+$ and $\overline{\Psi}_+$ and also find that the boundary term at $r_*=\infty$ vanishes when integrating by parts. Therefore
we proceed in the same way as in the proof of the case $\IM \sigma>0$ (use the $L^2(\BR_{r_*}; dr_*)$ pairing and then integrate by parts) and conclude that $\Psi_{\pm}=0$ on $(\ehRN, \infty)$. Therefore, $(\dg, \dA)$ is a pure gauge solution.

 \subsubsection{Modes with $l=1$}
 We now let $k^2=l(l+1)=2$. Our goal is again to prove that an outgoing mode solution $(\dg, \dA)$ is a pure gauge solution, with the gauge potential an outgoing mode as well. The scalar perturbations with $l=1$ take the form
 \begin{equation}\label{eq:perturbation_l1}
 	\dg=\begin{pmatrix}
 		\widetilde{f}\sfS\\
 		-\frac{r}{k}f\otimes\sd\sfS\\
 		2r^2H_L\sfS\sg
 	\end{pmatrix},\quad \dA=\begin{pmatrix}
 		\widetilde{K}\sfS\\
 		-\frac{r}{k}K\sd\sfS
 	\end{pmatrix}\quad\mbox{with}\quad \sfS\in\BFS_l ~(l=1).
 \end{equation}
 where the term $H_T$ is absent because $\rsH_1\sfS=0$. Now we follow the notations and expressions in the case $l\geq 2$ with $H_T$ set to be $0$. First, the pure gauge solutions take the form
 \begin{equation}
 	\bdelta\dg=2\delta_g^*\omega=\begin{pmatrix}
 		2\hdelta^*T	\\
 		-\frac{r}{k}(-\frac{k}{r}T+r\hd r^{-1}L)\sd\sfS\\
 		2r^2(r^{-1}\iota_{dr}T+\frac{k}{2r}L)\sfS\sg
 	\end{pmatrix},\quad\!\! \bdelta\dA=\mathcal{L}_{\omega^\sharp}A+d\phi=\begin{pmatrix}
 		(\hd\iota_AT+\iota_{T}\hd A+\hd P)\sfS\\
 		-\frac{r}{k}(-\frac{k}{r}\iota_AT-\frac{k}{r}P)\sd\sfS
 	\end{pmatrix} 
 \end{equation}
 with 
 \begin{equation}
 	\omega=\begin{pmatrix}
 		T\sfS\\
 		-\frac{r}{k}L\sd\sfS
 	\end{pmatrix},\quad\!\! \phi=P\sfS
 \end{equation}
 where $T\in C^\infty(\hX; T^*\hX), L, P\in C^\infty({\hX})$.  When adding $(\bdelta\dg, \bdelta\dA)$ to $(\dg, \dA)$, the quantities $\tilde{f}, f, H_L, \tilde{K}, K$ change by
 \begin{equation}
 	\begin{gathered}
 		\bdelta\widetilde{f}=2\hdelta^*T,\quad\bdelta f=-\frac{k}{r}T+r\hd (r^{-1}L),\quad \bdelta H_L=r^{-1}\iota_{dr}T+\frac{k}{2r}L,\quad %\bdelta H_T=-\frac{k}{r}L;\\
 		\\
 		\bdelta\widetilde{K}=\hd\iota_AT+\iota_T\hd A+\hd P,\quad \bdelta K=-\frac{k}{r}(\iota_A T+P).
 	\end{gathered}
 \end{equation}
 We define $\mathbf{X}:=\frac{r}{k}f$, then $\bdelta\mathbf{X}=\frac{r}{k}\bdelta f=-T+\frac{r^2}{k}\hd(r^{-1}L)$. We observe that for any given $L$, one can choose 
 \begin{equation}\label{eq:puregaugen=1}
 T=\mathbf{X}+\frac{r^2}{k}\hd(r^{-1}L),\quad P=-\iota_AT+\frac{r}{k}K
 \end{equation}
 which implies $\mathbf{X}+\bdelta\mathbf{X}=0, K+\bdelta K=0$.  If $(\dg, \dA)$ are outgoing mode solutions and $L$ is a mode, then $T, P$ are modes of the same type which grows at most by the factor $r$ relative to $\dg, \dA$ and $L$.  We again define
 \begin{equation}
 	\begin{split}
 		\widetilde{F}&:=\widetilde{f}+2\hdelta^*\mathbf{X}\in C^{\infty}(\hX; S^2T^*\hX),\\
 		J&:=H_L+r^{-1}\iota_{dr}\mathbf{X}\in C^\infty(\hX;T^*\hX),\\
 		N&:=\widetilde{K}+\hd\left(\frac{r}{k}K\right)+\iota_{\mathbf{X}}\hd A\in C^\infty(\hX; T^*\hX).	\end{split}
 \end{equation}
 However, $\wt{F}, J, N$ are \textit{not} gauge-invariant in this case. In fact
 \begin{equation}
 	\bdelta\wt{F}=\frac{2}{k}\hdelta^*\Bigl(r^2\hd(r^{-1}L)\Bigr),\quad\bdelta J=\frac{k}{2r}L+\frac{r}{k}\iota_{dr}\hd(r^{-1}L),\quad \bdelta N=\frac{r^2}{k}\iota_{\hd(r^{-1}L)}\hd A.
 	\end{equation}
 For any fixed $L$, we pick $T, P$ as explained above and then can work in the gauge $\mathbf{X}=0, K=0$, which implies
 \[
 (\wt{f}, f,H_L, \wt{K}, K)=(\wt{F},0,J,N,0),
 \]
 and thus the linearized Einstein-Maxwell system $2\mathcal{L}_1(\dg, \dA)=0, \mathcal{L}_2(\dg, \dA)=0$ is again given by \eqref{eq:Etildef}--\eqref{eq:EK} with the equation \eqref{eq:EH_T} for $H^E_T$ absent. Conversely, in the gauge $\mathbf{X}=0, K=0$, if $(\wt{F}, J, N)=0$, we have $(\dg, \dA)=0$, which implies the original perturbations $(\dg, \dA)$ take the form 
 \begin{equation}\label{eq:gaugesolnl=1}
 	(\dg, \dA)=(2\delta_g^*\omega, \mathcal{L}_{\omega^\sharp}A+d\phi) \quad\!\! \mbox{with}\quad\!\! \omega=-\begin{pmatrix}
 		\Bigl(\mathbf{X}+\frac{r^2}{k}\hd(r^{-1}L)\Bigr)\sfS\\
 		-\frac{r}{k}L\hd\sfS
 	\end{pmatrix},\quad \!\! \phi=\iota_A(\mathbf{X}+\frac{r^2}{k}\hd(r^{-1}L))-\frac{r}{k}K.
 \end{equation}
	If $(\dg, \dA)$ are outgoing mode solutions and $L$ is a mode , then $\omega$ and $ \phi$ are modes of the same type which grow at most by a factor $r$ more than $(\dg, \dA)$ and $L$.

Now we exploit this extra gauge freedom resulting from $L$ to simplify the structure of the linearized Einstein-Maxwell system. Concretely, we \textit{define} $H_T^E:=-\frac{k^2}{2r^2}\htr\wt{F}$ and we shall arrange $H^E_T+\bdelta H_T^E=0$ by choosing $L$ properly and thus the absent equation \eqref{eq:EH_T} still holds in the case $l=1$. In order to achieve this, we should choose $L$ such that
\begin{equation}\label{eq:eqnforL}
	\begin{split}
	\bdelta H_T^E&=-\frac{k}{r^2}\htr\hdelta^*\Bigl(r^2\hd(r^{-1}L)\Bigr)=\frac{k}{r^2}\hdelta\Bigl(r^2\hd(r^{-1}L)\Bigr)=-k\hBox_{\hg}(r^{-1}L)-2kr^{-1}\pa^i(r)\pa_i(r^{-1}L)\\
	&=-k\hBox_{\hg}(r^{-1}L)+kg^{ab}\cs_{ab}^i\pa_i(r^{-1}L)=-k\Box_g(r^{-1}L)=-H_T^E.
	\end{split}
\end{equation}

\underline{We first consider the non-stationary mode solutions with $\IM \sigma\geq0,  \sigma\neq0$}. We again define $\mathcal{N}$ as $\hstar N=\hd\mathcal{N}$ and $X, Y,Z$ as in \eqref{eq:defofXYZ}. As explained in the case $l\geq 2$, $\mathcal{N}, X, Y,Z$ are modes with the same frequency $\sigma$. We shall find a mode $L$ satisfying the above equation \eqref{eq:eqnforL}. Writing $H_T^E=e^{-i\sigma t_*}H_{T,0}^E(r)$, then the equation \eqref{eq:eqnforL} for $L=e^{-i\sigma t_*}\hat{L}(r)$ becomes
\begin{equation}
	\widehat{\Box_g}(\sigma)(r^{-1}\hat{L})=\frac{H_{T,0}^E}{k}
\end{equation}
which has a spherically symmetric solution since by Theorem \ref{thm:modesforscalarwave} $\widehat{\Box_g}(\sigma): \eHb^{\infty, \ell}(\CX)\to\eHb^{\infty, \ell+1}(\CX)$ is invertible for any $\ell<-1/2$. Therefore, as discussed around \eqref{eq:puregaugen=1}, we add a pure gauge solution using this $L$ and the corresponding $T, P$ as in \eqref{eq:puregaugen=1} , the linearized Einstein-Maxwell system is again given by \eqref{eq:Etildef}--\eqref{eq:EtildeK}. This allows us to repeat the proof of the case of non-stationary mode solutions with $l\geq 2$ to conclude that in the case $l=1$, $(\dg, \dA)$ is a gauge solution as well. More precisely, $(\dg, \dA)$ take the form as in \eqref{eq:gaugesolnl=1}.

\underline{Next we discuss the stationary case $\sigma=0$}. Suppose $(\dg, \dA)\in\eHb^{\infty, \ell}(\CX; \scsym\oplus\scform)$ where we now assume $-5/2<\ell<-1/2$, then $\mathbf{X}=rf/k\in \eHb^{\infty, \ell-1}(\CX; \scform)$ and thus $\hdelta^*\mathbf{X}\in\eHb^{\infty, \ell}(\CX; \scsym)$ (because $\hdelta^*$ acts as $\rho\mbox{Diff}^1_b$). Therefore, the right-handed side of \eqref{eq:eqnforL} is in $\eHb^{\infty, \ell+2}(\CX)$. According to Theorem \ref{thm:modesforscalarwave},%and Proposition \ref{prop:gscalarzeromode},  
 $\widehat{\Box_g}(0): \eHb^{\infty, \ell}(\CX)\to \eHb^{\infty, \ell+2}(\CX)$ is surjective for $-5/2<\ell<-1/2$,%when restricted in function spaces of scalar type $l=1$ 
 and then we can solve \eqref{eq:eqnforL} for $r^{-1}L\in \eHb^{\infty, \ell}(\CX)$ and thus $L\in \eHb^{\infty, \ell-1}(\CX)$. Again, according to \eqref{eq:puregaugen=1}, by adding a pure gauge solution $(2\hdelta_g^*\omega, \mathcal{L}_{\omega^\sharp}A+d\phi)$ with $(\omega, \phi)\in \eHb^{\infty, \ell-1}(\CX; \scform\oplus\BR)$, the linearized Einstein-Maxwell system is again given by \eqref{eq:Etildef}--\eqref{eq:EtildeK}.

Next, we need to modify the argument in the case $l\geq 2, \sigma=0$ a little bit. We again write
\begin{equation}\label{eq:newexpressionforPsi+-l=1}
	\Psi_{\pm}(\sigma)=C_{X\pm}(\sigma)X+C_{Y\pm}(\sigma)Y+C_{\mathcal{N}\pm}(\sigma)\mathcal{N}+C_{\mathcal{N}'\pm}(\sigma)\mathcal{N}'
\end{equation}
where $C_{\bullet \pm}(
\sigma)$ are rational functions of $r$ depending on $\sigma\in\BC$ (obtained by letting $k^2=2, m=0$ in \eqref{eq:CXYNN'l2})
\begin{equation}\label{eq:CXYNN'l1}
	\begin{split}
		C_{X\pm}&=a_{\pm}\frac{P_X}{3\wt{H}},\quad
		C_{Y\pm}=a_{\pm}\frac{P_Y}{3\wt{H}},\\
		C_{\mathcal{N}+}&=1+\frac{2\weight P_{\mathcal{N}}}{\tilde{c}r\wt{H}},\quad C_{\mathcal{N}-}=b_-\Bigl(1+\frac{12a_-\weight x}{r\wt{H}}\Bigr)\\
		C_{\mathcal{N}'+}&=\frac{2\weight^2P_{\mathcal{N}}}{\tilde{c}\wt{H}},\quad C_{\mathcal{N}'-}=\frac{4b_-\weight^2(\tilde{c}-4rz)}{r\tilde{H}}
	\end{split}
\end{equation}
where $\wt{H}(\sigma)=H(2\weight'-4r\sigma^2)$ with $H(r)=6\Bm r^{-1}-4\BQ^2 r^{-2}$, $(a_+, b_+)=(\frac{\BQ}{2r}, 1)$, $(a_-, b_-)=(1-\frac{2\BQ^2}{3\Bm r}, -\frac{4\BQ}{3\Bm})$, $\tilde{c}=6\Bm$ and $P_X, P_Y, P_{\mathcal{N}}$ are smooth on $\hX$(we again let $k^2=2, m=0$ in \eqref{eq:eqforP_XYNl2})
\begin{equation}\label{eq:eqforP_XYNl1}
	\begin{split}
		P_X&=(9x-36z-18)x+6(4z+8)z,\\
		P_Y&=-3(9x-16z-6)x-24z^2,\\
		P_{\mathcal{N}}&=-8\tilde{c}z,
	\end{split}
\end{equation}
Since $H(r)=2r^{-1}(3\Bm-2\BQ^2 r^{-1})>0$ and $\weight'=2 r^{-2}(\Bm-r^{-1}\BQ^2)>0$ in the  region $r\geq\ehRN>\Bm>\Bm^{-1}\BQ^2>(2/3)\Bm^{-1}\BQ^2$, we have $\wt{H}(0)=2H\weight'>0$ there as well. We again define $\Psi_{\pm}:=\Psi_{\pm}(0)$. Conversely, using the the first two equations in \eqref{eq:eqforXYZs} with $\sigma=0$ and $m=0$, one can recover $X, Y$ in terms of $\Phi, \mathcal{N},\mathcal{N}'$ and thus $\Psi_{\pm}, \mathcal{N},\mathcal{N}'$ as well. Again, $\Psi_{\pm}(0)$ satisfy
\begin{equation}\label{eq:eqforPsi+-sigma0l1}
	(\weight\pa_r)^2\Psi_{\pm}-V_{\pm}\Psi_{\pm}=0
\end{equation}
with $V_{\pm}$ defined as before in \eqref{eq:defofV+-} \eqref{eq:a+-b+-}, \eqref{eq:VFN} and \eqref{eq:VFPhi} with $m=0, k^2=2$.

Since $(\dg, \dA)\in\eHb^{\infty, \ell}(\CX; \scsym\oplus\scform)\subset \mathcal{A}^{\ell+3/2}(\CX)$, we have $X, Y, \mathcal{N}'\in\mathcal{A}^{\ell+3/2}(\CX)$ and $\mathcal{N}\in \mathcal{A}^{\ell+1/2}(\CX)$. Let $\rho=1/r$, we notice that $H=6\Bm\rho-4\BQ^2\rho^2$ and $k^2\weight'=4\Bm\rho^2-4\BQ^2\rho^3$, so $\wt{H}(0)\in 24\Bm^2\rho^3+\rho^4C^\infty(\CX)$. In view of the expression in \eqref{eq:eqforP_XYNl1}, we find $P_X,  P_Y\in \rho C^\infty(\CX)$ and $P_{\mathcal{N}}\in \rho^2 C^\infty(\CX)$, and then $C_{X+}, C_{Y+}, C_{\mathcal{N}'+}\in \rho^{-1}C^\infty(\CX)$ and $C_{\mathcal{N}+}\in C^\infty(\CX)$. As a consequence, we have
 \[
 \Psi_{+}\in\mathcal{A}^{\ell+1/2}(\CX)\subset \mathcal{A}^{-2+}(\CX)\quad\!\!\mbox{a priori if}\!\!\quad -5/2<\ell<-1/2.
 \] 
Now we are ready to describe the asymptotic behavior of $\Psi_+$ near $r=\ehRN$ and $r=\infty$. Since $F_{\Phi}, V_{\mathcal{N}}, a_{\pm}, b_{\pm}$ are smooth across the event horizon, we have $\weight(\pa_r\Psi_+)\overline{\Psi}_+|_{r=\ehRN}=0$. This implies that boundary term at $r=\ehRN$ arising in the integration by parts we will do later on vanishes. Near $r=\infty$, owing to \eqref{eq:VFPhi}, \eqref{eq:VFN} \eqref{eq:a+-b+-} and \eqref{eq:defofV+-} with $m=0$, we conclude $F_{\Phi}\in \frac{8\BQ}{3\Bm}\rho^2+\rho^3C^\infty(\CX)$, $V_{\mathcal{N}}\in \frac{2}{r^2}+\rho^3C^\infty(\CX)$ and then $V_{+}\in\frac{2}{r^2}+\rho^3C^\infty(\CX)$, which implies $V_{+}=\frac{2}{r^2_*}+\mathcal{A}^{3-}$ as a function of $r_*$ near $r_*=\infty$. We set $x=1/r_*$ for the moment, \[
(\weight\pa_r)^2\Psi_+-V_+\Psi_+=\pa_{r_*}^2\Psi_+-V_+(r_*)\Psi=x^4\pa_x\Psi_++2x^3\pa_x\Psi_+-V_+(1/x)=0,
\]
which implies $x=0$(i.e. $r_*=\infty$) is a regular singular point and its indicial equation is $\lambda(\lambda-1) +2\lambda-2=0$. The roots to the indicial equation are $\lambda_+=1, \lambda_-=-2$. Again we see that  $\lambda_+=1$  corresponds to one solution $\Psi_+$ with decay $\sim r_*^{-1}$ near $r_*=\infty$, while $\lambda_-\leq -2$ implies another linearly independent solution $\Psi_+$ with growth $\sim r_*^{2}$ near $r_*=\infty$. Since we have $\Psi_+\in \mathcal{A}^{-2+}$ a priori, this excludes the second solution with growth. Namely,  $\Psi_+\sim r_*^{-1}$ near $r_*=\infty$. With this decay rate near $r_*=\infty$, we can justify the $L^2(\BR_{r_*};dr_*)$ pairing between $((\weight\pa_r)^2-V_+)\Psi_+$ and $\overline{\Psi}_+$ and also find that the boundary term at $r_*=\infty$ vanishes when integrating by parts. Therefore
we proceed in the same way as in the proof of the case $\IM \sigma>0$ (use the $L^2(\BR_{r_*}; dr_*)$ pairing and then integrate by parts) and conclude that $\Psi_+=0$ on $(\ehRN, \infty)$ and thus on $\hX$. 

As for $\Psi_-$, recall the \textit{S-deformation} we discussed around \eqref{eq:Psi-} and we can rewrite the equation $((\weight\pa_r)^2-V_-)\Psi_-=0$ as
\[
\wt{D}^*\wt{D}\Psi_-=-H_-\pa_r\Bigl(H_-^{-2}\weight\pa_r(H_-\Psi_-)\Bigr)=0\quad \mbox{with}\quad H_-(r)=\frac{\tilde{c}}{r}=\frac{6\Bm}{r}>0\ \mbox{on}\ r\in[\ehRN, \infty)
\]
because $\wt{V}_-=\frac{k^2(k^2-2)\weight}{r^2H_-}=0$ if $l=1$. Solving the above equation for $\Psi_-$ yields
\begin{equation}\label{eq:SPsi_-}
	\Psi_-=\frac{c_1}{H_-}\int\frac{H_-^2}{\weight}dr+\frac{c_2}{H_-}=d_1\ln\left\lvert\frac{r-r_H}{r-(\Bm-\sqrt{\Bm^2-\BQ^2})}\right\rvert+d_2r.
\end{equation}
where $c_1, c_2, d_1, d_2\in \BR$. Since $\Psi_-$ is smooth across the event horizon $r=\ehRN$, we conclude $d_1=0$ and thus $\Psi_-=d_2r$. Using $(\Psi_+, \Psi_-)=(0, d_2 r)$,  \eqref{eq:defofPsi+-} and \eqref{eq:a+-b+-}, we have $(\Psi, \mathcal{N})=(d_2r, -d_2\BQ/2)$. Exploiting the first two equations in \eqref{eq:eqforXYZs} we obtain $(X, Y)=(-d_2,-d_2)$. As discussed previously, the equation \eqref{eq:XYZ2} implies $Z=0$ in the stationary perturbation. Finally using \eqref{eq:recoverofFJ} and $\hstar N=\hd\mathcal{N}$ we see
\[
\wt{F}=0, \quad J=-\frac{d_2}{2},\quad  N=0
\]
Since $J\in\eHb^{\infty, \ell}(\CX)$ and $0\neq d_2\in\eHb^{\infty, -\frac{3}{2}-}(\CX)$, we have $d_2=0$ if $\ell\geq-3/2$ and thus we are done. If $-5/2<\ell<-3/2$, $d_2$ may be non-zero and then in the gauge $\mathbf{X}=0$
\[
\dg=\begin{pmatrix}
	0\\
	0\\
	-2r^2\frac{d_2}{2}\sfS\sg
\end{pmatrix}, \quad \dA=0.
\]
This implies in the gauge $\mathbf{X}=0$, $(\dg, \dA)$ is a gauge solution with $T=0, L=-d_2r/k, P=0$. Therefore, according to the discussion around \eqref{eq:gaugesolnl=1}, the original perturbations $(\dg, \dA)=(2\delta_g^*\omega, \mathcal{L}_{\omega^\sharp}A+d\phi)$ with $(\omega, \phi)\in \eHb^{\infty, \ell-1}(\CX; \scform\oplus\BR)$ for $-5/2<\ell<-1/2$.

\subsubsection{Modes with $l=0$}
As stated in \cite{H18, HHV21}, the linearization of the simple proof of the Birkhoff' theorem \cite{SW10} can be extended to the spherically symmetric Einstein-Maxwell system. Following the argument in \cite{SW10}, we work in the incoming Eddington-Finkelstein coordinates $(t_0=t+r^*, r)$ where the Reissner-Nordstr\"{o}m metric and the electromagnetic $4$-potential take the form
\begin{equation}\label{eq:incomingEF}
g=-\weight dt_0^2+2dt_0dr+r^2\sg,\quad A=\frac{\BQ}{r}dt_0
\end{equation}
\begin{rem}
	The electromagnetic $4$-potential defined above differs from \eqref{eq:4potentialt*} by an exact $1$-form $f(r)dr$. We note that adding an exact $1$-form to $A$ changes neither the electromagnetic field nor the linearized Einstein-Maxwell system because the linearized system only involves $dA$.
\end{rem}
Representing the Reissner-Nordstr\"{o}m metric and the electromagnetic $4$-potential in the way \eqref{eq:incomingEF}, we can linearize $(g_b,A_b)$ with respect to the parameters $(\Bm, 0,\BQ)$ in the direction $(\dot{\Bm},0, \dot{\BQ})$ to obtain the following linearized metric and  electromagnetic $4$-potential
\begin{equation}\label{eq:linearizedRNv}
g'=(\frac{2\dot{\Bm}}{r}-\frac{2\BQ\dot{\BQ}}{r^2})dt_0^2,\quad A'=\frac{\dot{\BQ}}{r}dt_0.
\end{equation}
We note that $(g', A')$ differs from $(\dg_{(\Bm, 0, \BQ)}(\dot{\Bm}, 0, \dot{\BQ}), \dA_{(\Bm, 0, \BQ)}(\dot{\Bm}, 0, \dot{\BQ}))$ by a pure gauge term.

With this we now turn to the analysis of the spherically symmetric modes. Since there is no spherical component in $1$-forms and no aspherical-spherical mixed component in symmetric $2$-tensors for spherically symmetric perturbations, we use the following splitting of $1$-forms and symmetric $2$-tensors respectively:$ \langle dt_0\rangle\oplus \langle dr\rangle$ and $\langle dt_0^2\rangle\oplus \langle 2dt_0dr\rangle\oplus\langle dr^2\rangle\oplus \langle\sg\rangle$. Under this splitting,  we write the scalar perturbations of modes $l=0$ (spherically symmetric modes) in the following form
\[
\dg=\begin{pmatrix}
	\dot{X}\\
	\dot{Y}\\
	\dot{Z}\\
	2r^2H_L
\end{pmatrix},\quad \dA=\begin{pmatrix}
\dA_0\\
\dA_1
\end{pmatrix}
\]
where $\dot{X}, \dot{Y}, \dot{Z}, H_L$ are smooth function on $\BR_{t_0}\times [r_-, \infty)$. To describe the pure gauge perturbations, we use the following lemma
\begin{lem}
Let $(g, A)	$ be as defined in \eqref{eq:incomingEF} , under the splitting $\langle\pa_{t_0}\rangle\oplus\langle\pa_r\rangle$ of $TM$ and $\langle dv^2\rangle\oplus \langle 2dt_0dr\rangle\oplus\langle dr^2\rangle$ of $S^2T^*M$, the map $\mathcal{L}_{(\bullet)}g: TM\to S^2T^*M$ takes the form
	\begin{equation}
		\mathcal{L}_{(\bullet)}g=\begin{pmatrix}
			-2\weight\pa_{t_0}&2\pa_{t_0}-\weight'\\
			\pa_{t_0}-\weight\pa_r&\pa_r\\
			2\pa_r&0\\
			0&2r
		\end{pmatrix};
	\end{equation}
the map $\mathcal{L}_{(\bullet)}A$ between sections of $\langle \pa_{t_0}\rangle\oplus \langle \pa_r\rangle$ and $\langle dt_0\rangle\oplus \langle dr\rangle$ is given by
\begin{equation}
	\mathcal{L}_{(\bullet)}A=\begin{pmatrix}
	\BQ r^{-1}\pa_{t_0}&-\BQ r^{-2}\\
		\BQ r^{-1}\pa_r& 0
	\end{pmatrix},
\end{equation}
and  as a mapping between the sections of the trivial line bundle $\BR$ and $\langle dt_0\rangle\oplus \langle dr\rangle$ , $d=\begin{pmatrix}
	\pa_{t_0}\\
	\pa_r
\end{pmatrix}$.
\end{lem}
\begin{proof}
It follows from the formula $(\mathcal{L}_Wg)_{\alpha\beta}=W^\gamma\pa_\gamma g_{\al\be}+(\pa_{\al}W^\gamma)g_{\gamma\be}+(\pa_\be 	W^\gamma)g_{\al\gamma}$, $(\mathcal{L}_W A)_\al=W^\gamma\pa_\gamma A_{\al}+(\pa_{\al}W^\gamma)A_{\gamma}$.
\end{proof}
Therefore, adding a pure gauge solution $(\mathcal{L}_Wg, \mathcal{L}_W A)$ to $(\dg, \dA)$ with $W:=Z_1\pa_{t_0}-rH_L\pa_r$ and $Z_1:=(-\int_{3\Bm}^r \dot{Z}(t_0, s)ds)/2$, the $dr^2$ and spherical components of $\dg$ vanish. Next, by adding another pure gauge solution $(0, d\phi)$ with $\phi=-\int_{3\Bm}^r(A_1 +\BQ s^{-1}\pa_r Z_1)(t_0, s)ds$, the $dr$-component of $\dA$ can be eliminated. 

Suppose the original perturbations $(\dg, \dA)\in e^{-i\sigma t_*}\eHb^{\infty, \ell}(\CX;\scsym\oplus\scform)$ are outgoing modes, then so are $\dot{X}, \dot{Y}, \dot{Z}, H_L$. If $\sigma=0$, it is obvious that $(\omega=W^\flat, \phi)\in \eHb^{\infty, \ell-1}(\CM; \scform\oplus \BR)$ are still stationary modes. If $\sigma\neq 0$, since $t_*=t_0+f(r)$ where $f$ is smooth on $[r_-, \infty)$ and $\sim -2 r^*$ near infinity, we find that $(\omega=W^\flat, \phi)\in e^{-i\sigma t_*}\eHb^{\infty, \ell'}(\CX;\scform\oplus\BR)$ for some $\ell'<\ell$ is also outgoing modes. Further, if $(\dg, \dA)\in \mbox{Poly}(t_*)^k\eHb^{\infty, \ell}(\CX;\scsym\oplus\scform)$ are the generalized stationary modes, the integration in the definition of $W$ and $\phi$ implies $(\omega=W^\flat, \phi)\in \mbox{Poly}(t_*)^k\eHb^{\infty, \ell'}(\CX;\scform\oplus\BR)$ for some $\ell'<\ell$. To summarize, by adding pure gauge solutions which are outgoing modes(or generalized stationary modes), the perturbations are reduced to the following form
\[
\dg=\dot{X}dt_0^2+2\dot{Y}dt_0dr,\quad \dA=\dA_0dt_0.
\] 

Next, we analyze the linearized Einstein-Maxwell system for spherically symmetric modes. Concretely, for $(\check{g}, \check{A})$ of the general form $\check{g}=Xdt_0^2+2Ydt_0dr+r^2\sg$ and $\check{A}=A_0dt_0$ where $X, Y,A_0$ are functions of $(t_0,r)$, the $dr^2$-component of $\Ric(\check{g})-2T(\check{g}, d\check{A})=0$ implies $2\pa_r Y/(rY)=0$. Linearizing it around $(g, A)$ in \eqref{eq:incomingEF}, we obtain the corresponding equation for $(\dg, \dA)$: $\pa_r \dot{Y}=0$, so $\dot{Y}=\dot{Y}(t_0)$. If $\dot{Y}$ is an outgoing mode with $\sigma\neq 0$, we must have $\dot{Y}=\dot{Y}(t_0)=ce^{-i\sigma t_0}$ which contradicts the definition of outgoing modes unless $\dot{Y}=0$. If $\dot{Y}$ is a stationary mode, then $\dot{Y}=c$. However the assumption $\dot{Y}\in \eHb^{\infty, \ell}(\CX)\subset \mathcal{A}^{0+}(\CX)$ for $-3/2<\ell<-1/2$ require $\dot{Y}=0$. Lastly, if $\dot{Y}$ is a generalized zero mode, we then have $\dot{Y}=\dot{Y}(t_0)=\mbox{Poly}(t_0)^k$. Letting $W_1=f\pa_{t_0}$ with $f(t_0)\in\mbox{Poly}(t_0)^{k+1}$ satisfying $\pa_{t_0} f(t_0)=-\dot{Y}(t_0)$, and adding the pure gauge solution $(\mathcal{L}_{W_1}g, \mathcal{L}_{W_1}A=0)$ to $(\dg, \dA)$, $dvdr$-component of $\dg$ again vanishes. Therefore, we now have
\begin{equation}
	\dg=\dot{X}dt_0^2,\quad \dA=\dA_0dt_0.
\end{equation}
Again, for $(\check{g}, \check{A})$ the Maxwell equation becomes
\begin{equation}
	\begin{split}
	\delta_{\check{g}}d\check{A}&=\Bigl(-\frac{1}{Y}\pa_{t_0}\pa_r A_0+\frac{X}{Y^2}\pa^2_r A_0+\frac{2X}{rY^2}\pa_r A_0-\frac{X}{Y^3}\pa_r Y\pa_r A_0+\frac{1}{Y^2}\pa_r A_0\pa_{t_0} Y\Bigr)dv\\
	&\quad+\Bigl(\frac{1}{Y}\pa_r^2 A_0+\frac{2}{rY}\pa_r A_0-\frac{1}{Y^2}\pa_r Y\pa_rA_0\Bigr) dr.
	\end{split}
\end{equation} 
Linearizing around $(g, A)$ (i.e. $X=-\weight, Y=1, A_0=r^{-1}\BQ$) and using the previously obtained fact $\dot{Y}=0$, the $dr$ component tells $\pa_r^2\dA_0+2r^{-1}\pa_r\dA_0=0$, so $\dA_0=c_1(t_0)r^{-1}+c_2(t_0)$. Since $c_2(t_0)dt_0$ is an exact $1$-form and thus a pure gauge solution, we may assume $\dA_0=c_1(t_0)r^{-1}$. Then the equation for $dt_0$-component gives $-\pa_{t_0}\pa_r\dA_0-\weight\pa_r^2\dA_0-2\weight r^{-1}\pa_r\dA_0=r^{-2}\pa_{t_0}c_1(t_0)=0$ and thus $c_1(t_0)$ is a constant that we denote by $\dot{\BQ}$. That is, $\dA_0=r^{-1}\dot{\BQ}$.

Next, the spherical component of $\Ric(\check{g})-2T(\check{g}, d\check{A})=0$ is given by 
\[
1+\frac{X}{Y^2}+\frac{r}{Y^2}\pa_r X-\frac{rX}{Y^3}\pa_r Y-\frac{r^2}{Y^2}(\pa_r A_0)^2=0.
\]
Linearizing around $(g, A)$ (i.e. $X=-\weight, Y=1, A_0=r^{-1}\BQ$) and using the previously obtained facts $\dot{Y}=0, \dA_0=r^{-1}\dot{\BQ}$, we find $\dot{X}+r\pa_r \dot{X}-2r^{-2}\BQ\dot{\BQ}=0$ and thus
\[
\dot{X}=-\frac{2\BQ\dot{\BQ}}{r^2}+\frac{c(t_0)}{r}.
\]

Lastly, the $dt_0^2$-component of  $\Ric(\check{g})-2T(\check{g}, d\check{A})=0$ is 
\[
-\frac{X}{Y^2}\pa_{t_0}\pa_r Y+\frac{X}{2Y^2}\pa_r^2 X+\frac{X}{Y^3}\pa_{t_0}Y\pa_r Y-\frac{X}{2Y^3}\pa_r X\pa_r Y-\frac{2X}{rY^2}\pa_{t_0} Y+\frac{X}{rY^2}\pa_r X+\frac{1}{rY}\pa_{t_0} X+\frac{X}{Y^2}(\pa_r A_0)^2=0.
\]
Again linearizing around $(g, A)$ (i.e. $X=-\weight, Y=1, A_0=r^{-1}\BQ$) and using the previously obtained facts $\dot{Y}=0, \dA_0=r^{-1}\dot{\BQ}, \dot{X}=-\frac{2\BQ\dot{\BQ}}{r^2}+\frac{c(v)}{r}$, we obtain $r^{-2}\pa_vc(v)=0$ and thus $c(v)$ is a constant which we denote by $2\dot{\Bm}$. Therefore, we conclude that up to pure gauge solutions
\begin{equation}
	\dg=(\frac{2\dot{\Bm}}{r}-\frac{2\BQ\dot{\BQ}}{r^2})dt_0^2=g',\quad \dA_0=\frac{\dot{\BQ}}{r}dt_0=A'.
\end{equation}

\subsection{Vector type perturbations}
We shall consider the vector type perturbations of modes $l\geq 2$ and $l=1$ separately.  Recall that we can write the perturbations under the splitting \eqref{eq:splitof1} and \eqref{eq:splitofsym2} as in \eqref{eq:vectorper}. We denote by $\sfV\in\BFV_l$ the spherical harmonic $1$-form with eigenvalues $k^2=l(l+1)-1$.
\subsubsection{Modes with $l\geq 2$}
Suppose $(\dg, \dA)$ is the vector perturbations of the following form
\begin{equation}\label{eq:vperturbation_l2}
	\dg=\begin{pmatrix}
		0\\
		rf\otimes\sfV\\
		-\frac{2}{k}r^2H_T\sdelta^*\sfV
	\end{pmatrix},\quad \dA=\begin{pmatrix}
		0\\
	rK\sfV
	\end{pmatrix}\quad\mbox{with}\quad \sfV\in\BFV_l ~(l\geq2).
\end{equation}
We notice that the pure gauge solutions take the form
\begin{equation}
	\bdelta\dg=2\delta_g^*\omega=\begin{pmatrix}
		0	\\
		r^2\hd(r^{-1}L)\sd\sfS\\
		-\frac{2}{k}r^2(-kr^{-1}L)\sdelta^*\sfV
	\end{pmatrix},\quad\!\! \bdelta\dA=\mathcal{L}_{\omega^\sharp}A=0
\end{equation}
with 
\begin{equation}
	\omega=\begin{pmatrix}
		0\\
		rL\sfV
	\end{pmatrix}
\end{equation}
where $L\in C^\infty({\hX})$. When adding $(\bdelta\dg, \bdelta\dA)$ to $(\dg, \dA)$, the quantities $ f,  H_T, K$ change by
\begin{equation}
 \bdelta f=r\hd (r^{-1}L),\quad \bdelta H_T=-kr^{-1}L, \quad \bdelta K=0.
\end{equation}
Then we have the following gauge-invariant quantities
\begin{equation}
	J:=f+\frac{r}{k}\hd H_T\in C^\infty(\hX; T^*\hX),\quad K\in C^\infty(\hX)
\end{equation}
We also note that if $J=0, K=0$, $(\dg, \dA)$ is a pure gauge solution
\begin{equation}
	(\dg, \dA)=(2\delta^*\omega, \mathcal{L}_{\omega^\sharp}A) \quad\!\! \mbox{with}\quad\!\! \omega=\begin{pmatrix}
	0\\
		-\frac{r^2}{k}H_T\sfV
	\end{pmatrix}
\end{equation}
If $(\dg, \dA)$ are outgoing mode solutions, then $\omega$ is a mode of the same type, which grows by a factor $r$ more than $(\dg, \dA)$.

Again we can express the linearized Einstein-Maxwell system in terms of the gauge-invariant quantities $J, K$ defined above. %Concretely, one can choose a gauge, i.e., add a pure gauge solution $(\bdelta\dg, \bdelta\dA)$ to $(\dg, \dA)$ for suitable $\omega=\omega(T,L), \phi=\phi(P)$ such that the non gauge-invariant quantities $\tilde{f}$ etc. take a simple form in the new gauge. 
More specifically, we adds $(\bdelta\dg, \bdelta\dA)$ built from $ L=\frac{r}{k}H_T$, then $ H_T+\bdelta H_T=0$, and thus $(f, H_T,K)=(J, 0, K)$. 
Therefore, using the detailed calculation in \S \ref{sec:basiccalculation} and \S \ref{subsec:detailcal} we can write the linearized 	Einstein-Maxwell system, acting on the new $(\dg, \dA)$
\[
\dg=\begin{pmatrix}
	0\\
	rJ\otimes\sfV\\
0
\end{pmatrix}, \quad \dA=\begin{pmatrix}
	0\\
	rK\sfV
\end{pmatrix}
\] in terms of the gauge-invariant quantities $J, K$. 

Now we express $2\mathcal{L}_1(\dg, \dA)=0$ and $\mathcal{L}_2(\dg, \dA)=0$ in the form of \eqref{eq:perturbation_l2} in terms of $f^E,  H_T^E$ and $K^E$ respectively. Then the linearized Einstein-Maxwell system reads
\begin{subequations}
	\begin{align}
	rf^E&=r^{-2}\hdelta(r^4\hd(r^{-1}J))+\frac{k^2-1}{r}J-4\BQ r^{-2}\hstar\hd(rK)=0,\label{eq:EfV}\\		
		-\frac{2r^2}{k}H_T^E&=-2\hdelta(rJ)=0,\label{eq:EH_TV}\\
	rK^E&=-\hBox(rK)+\frac{k^2+1}{r}K+\BQ \hstar\hd(r^{-1}J)=0.\label{eq:EKV}
	\end{align}
\end{subequations}
By \eqref{eq:EH_TV}, one adds the zero term $\hd\hdelta (rJ)$ to \eqref{eq:EfV} and then can obtain a wave equation for $J$ (coupled to $K$ via subprincipal terms) as $(\hdelta\hd+\hd\hdelta) J=(-\hBox
-\frac 12\weight'')J$. In conclusion, we can rewrite equation \eqref{eq:EfV} and \eqref{eq:EKV} as a principally scalar system of wave equations for $( J, K)$
\begin{equation}
	-\hBox B-\mathscr{D}B=0\quad \mbox{where}\quad
	B=
	\begin{pmatrix}
		J\\
		N
	\end{pmatrix}
\end{equation}
where $\mathscr{D}$ is a first order stationary differential operator acting on $C^{\infty}(\hX;T^*\hX \oplus T^*\hX)$. %{\color{red} When $(\dg, \dA)$ and thus $B$ are smooth modes, this system of equations become an ODE on the one dimensional space $t_*^{-1}(0)$ with a regular singular point at the event horizon $r=r_	H$ whose solution is given by Frobenius series, which means that the vanishing of $B$ in the static region $r>r_H$ implies the vanishing of $B$ on $\hX$}. 
As discussed in the scalar perturbations, it suffices to prove $B=0$ in the static region $r>\ehRN$. To achieve this, we again work in the static coordinates $(t,r)$.

First, by \eqref{eq:EH_TV} we have $\hd\hstar(rJ)=0$ (because $\hdelta=\hstar\hd\hstar$), and then $\hstar rJ=\hd(r\Phi)$ for some function $\Phi$ on $\hX$. Applying $\hstar r^2$ on both sides of equation \eqref{eq:EfV} yields
\[
\hd \Bigl(r^4\hdelta(r^{-2}\hd(r\Phi))+(k^2-1)r\Phi-4\BQ rK\Bigr)=0
\]
which implies $r^4\hdelta(r^{-2}\hd(r\Phi))+(k^2-1)r\Phi-4\BQ rK=c$ for some constant $c$. Replacing $\Phi$ by $\Phi-\frac{c}{(k^2-1)r}$, we obtain $
r^4\hdelta(r^{-2}\hd(r\Phi))+(k^2-1)r\Phi-4\BQ rK=0.
$ and thus 
\begin{equation}\label{eq:eqnforPhiversion1}
r\hdelta(r^{-2}\hd(r\Phi))+\frac{(k^2-1)}{r^2}\Phi-\frac{4\BQ}{r^2}K=0.
\end{equation}
We can also rewrite it as
\begin{equation}\label{eq:eqnforPhiVl2}
	-\hBox\Phi+\frac{1}{r^2}\Bigl(k^2+1-\frac{6\Bm}{r}+\frac{4\BQ^2}{r^2}\Bigr)\Phi-\frac{4\BQ}{r^2}K=0.
\end{equation}
Next, equation \eqref{eq:EKV} can be rewritten as (where we use equation \eqref{eq:eqnforPhiversion1})
\begin{equation}\label{eq:eqnforrK}
	-\hBox(rK)+\frac{1}{r^2}\Bigl(k^2+1+\frac{4\BQ^2}{r^2}\Bigr)(rK)-(k^2-1)\frac{\BQ}{r^3}\Phi=0
\end{equation}
We note that \eqref{eq:eqnforPhiVl2} and \eqref{eq:eqnforrK} recover \cite[equations (4.28) and (4.30)]{KI04} with $\varOmega=r\Phi, \mathcal{A}=-rK, m_V=k^2-1, \kappa^2=2, q=\BQ, n=2, K=1, \bar{\tau}=0, J=0$ there. Following \cite[equations(4.34) and (4.35)]{KI04}, we introduce the following combinations
\begin{equation}\label{eq:defforPsi+-Vl2}
	\Psi_{\pm}=a_{\pm}\Phi+b_{\pm}rK
\end{equation}
with 
\begin{equation}\label{eq:defofab+-Vl2}
	(a_+, b_+)=(\frac{\BQ(k^2-1)}{3\Bm+\sqrt{9\Bm^2+4(k^2-1)\BQ^2}}, -1),\quad (a_-, b_-)=(1, \frac{4\BQ}{3\Bm+\sqrt{9\Bm^2+4(k^2-1)\BQ^2}}).
\end{equation}
When expressed in terms of $\Psi_\pm$, equations \eqref{eq:eqnforPhiVl2} and \eqref{eq:eqnforrK} are transformed into two decoupled wave equations (see \cite[equations (4.37)--(4.39)]{KI04})
\begin{equation}\label{eq:eqnforPsi+-Vl2}
	\hBox\Psi_\pm-\frac{V_{\pm}}{\weight}\Psi_\pm=0
\end{equation}
where 
\begin{equation}
	V_\pm=\frac{\weight}{r^2}\Bigl(k^2+1+\frac{4\BQ^2}{r^2}-\frac{3\Bm}{r}\pm\frac{\sqrt{9\Bm^2+4(k^2-1)\BQ^2}}{r}\Bigr).
\end{equation}
Suppose $(\dg, \dA)\in e^{-i\sigma t_*}\eHb^{ \infty,\ell}(\CX; \scsym\oplus\scform)$ are modes, then $(J, K)\in e^{-i\sigma t_*}\eHb^{ \infty,\ell-1}(\CX; \scform\oplus\BR)$ for $\sigma\neq 0$, while $(J, K)\in\eHb^{\infty,\ell}(\CX; \scform\oplus\BR)$ for $\sigma=0$. As discussed around \eqref{eq:eqnforN} and \eqref{eq:Z=0} and using \eqref{eq:eqnforrK} in the $\sigma=0$ case, we conclude $\Phi, rK\in e^{-i\sigma t_*}\eHb^{ \infty,\ell-1}(\CX)$ for $\IM\sigma\geq0$, and so are $\Psi_\pm$. Therefore in the static coordinates $(t,r)$, equations \eqref{eq:eqnforPsi+-Vl2} take the form
\[
(\weight\pa_r)^2\Psi_\pm-(V_\pm-\sigma^2)\Psi_\pm=0.
\]
We observe that when $r>\ehRN=\Bm+\sqrt{\Bm^2-\BQ^2}$, $V_+>0$ and
 \begin{align*}
r^2\weight^{-1}V_-&>(r^2\weight^{-1}V_-)(\ehRN)=k^2-1+(\Bm-\sqrt{9\Bm^2+4(k^2-1)\BQ^2})/\ehRN\\
&=\ehRN^{-1}\Bigl(k^2\Bm+(k^2-1)\sqrt{\Bm^2-\BQ^2}-\sqrt{(4k^2+5)\BQ^2+9(\Bm^2-\BQ^2)}\Bigr)\\
&>\ehRN^{-1}\Bigl(k^2\Bm-\sqrt{4k^2+5}\BQ+(k^2-4)\sqrt{\Bm^2-\BQ^2}\Bigr)>0
\end{align*}
where we use $k^2\geq 5, \Bm>\BQ$ in the last inequality, so $V_->0$ in the static region as well. We also notice the asymptotic behavior of $V_\pm$ is $V_\pm=(k^2+1)\rho^2+\rho^3C^\infty(\CX)$ where $\rho=1/r$. Then we can proceed as in the scalar perturbations $l\geq 2$ case to conclude that $\Psi_\pm=0$ and thus $(J,K)=0$. More specifically, when $\IM\sigma>0$, we pair $(\weight\pa_r)^2\Psi_\pm-(V_\pm-\sigma^2)\Psi_\pm$ with $\overline{\Psi}_\pm$ with respect to $L^2(\BR_{r_*}; dr_*)$ and integrate by parts, then the positivity of $V_\pm$ allows us to obtain $\Psi_\pm=0$. When $\sigma\in\BR\setminus\{0\}$, we use the boundary paring argument. Lastly, when $\sigma=0$, we first obtain the a priori decay rate $\Psi_\pm\in\eHb^{ \infty,\ell-1}(\CX)\subset\mathcal{A}^{\ell+1/2}(\CX)\subset\mathcal{A}^{-1+}(\CX)$ because $-3/2<\ell<-1/2$. Next we use the Frobenius method (i.e. analyze the indicial equation) and the decay rate of $V_\pm=(k^2+1)\rho^2+\rho^3C^\infty(\CX)$ to conclude that $\Psi_\pm\sim r^{-2}$ near $r=\infty$. So the conclusion $\Psi_\pm=0$ again follows from the $L^2$ pairing and integration by parts.

\subsubsection{Modes with $l=1$}
We now let $k^2=l(l+1)-1=1$. Suppose $(\dg, \dA)$ is the vector perturbations of the following form
\begin{equation}\label{eq:vperturbation_l1}
	\dg=\begin{pmatrix}
		0\\
		rf\otimes\sfV\\
	0
	\end{pmatrix},\quad \dA=\begin{pmatrix}
		0\\
		rK\sfV
	\end{pmatrix}\quad\mbox{with}\quad \sfV\in\BFV_l ~(l=1).
\end{equation}
We notice that the pure gauge solutions take the form
\begin{equation}
	\bdelta\dg=2\delta_g^*\omega=\begin{pmatrix}
		0	\\
		r^2\hd(r^{-1}L)\sd\sfS\\
	0
	\end{pmatrix},\quad\!\! \bdelta\dA=\mathcal{L}_{\omega^\sharp}A=0
\end{equation}
with 
\begin{equation}
	\omega=\begin{pmatrix}
		0\\
		rL\sfV
	\end{pmatrix}
\end{equation}
where $L\in C^\infty({\hX})$. When adding $(\bdelta\dg, \bdelta\dA)$ to $(\dg, \dA)$, the quantities $ f,  H_T, K$ change by
\begin{equation}
	\bdelta f=r\hd (r^{-1}L), \quad \bdelta K=0.
\end{equation}
Then we define the following gauge-invariant quantities
\begin{equation}
\hd(r^{-1}f)\in C^\infty(\hX; \Lambda^2T^*\hX),\quad K\in C^\infty(\hX)
\end{equation}
If $\hd(r^{-1}f)=0, K=0$, since $\hX$ is contractible, we can find $L\in C^\infty(\hX)$ such that $r^{-1}f=\hd(r^{-1}L)$, which implies 
\begin{equation}
	(\dg, \dA)=(2\delta^*\omega, \mathcal{L}_{\omega^\sharp}A) \quad\!\! \mbox{with}\quad\!\! \omega=\begin{pmatrix}
		0\\
	rL\sfV
	\end{pmatrix}
\end{equation}
is a pure gauge solution. As explained around \eqref{eq:eqnforN} and \eqref{eq:Z=0} again, we have $L\in e^{-i\sigma t_*}\eHb^{\infty, \ell}(\CX)$ for $\sigma\neq 0$, while $L\in \eHb^{\infty, \ell-1}(\CX)$ for $\sigma=0$.

Again we can express the linearized Einstein-Maxwell system in terms of the gauge-invariant quantities $\hd(r^{-1}f), K$ as follows
\begin{subequations}
	\begin{align}
		rf^E&=r^{-2}\hdelta(r^4\hd(r^{-1}f))-4\BQ r^{-2}\hstar\hd(rK)=0,\label{eq:EfVl1}\\		
		rK^E&=-\hBox(rK)+\frac{2}{r}K+\BQ \hstar\hd(r^{-1}f)=0.\label{eq:EKVl1}
	\end{align}
\end{subequations}

Before solving the above system, we first describe the linearized Kerr-Newman solution in which we linearize the Kerr-Newman solution family $(g_b, A_b)$ around Reissner-Nordstr\"{o}m solution $(g_{b_0}, A_{b_0})$ with respect to the angular momentum parameter $\Ba$ in the direction $\dot{\Ba}$, which  is parallel to the rotation axis of $\sfV$. We see that in the static coordinates
\begin{equation}\label{eq:linearizedKNa}
	g'_{(\Bm, 0, \BQ)}(0, \dot{\Ba}, 0)=2(\weight-1)\sin^2\theta dtd\varphi,\quad 
	A'_{(\Bm, 0, \BQ)}(0, \dot{\Ba}, 0)=-\frac{\BQ}{r}\sin^2\theta d\varphi
\end{equation}
where $\abs{\dot{\Ba}}=1$ and $(\theta, \varphi)$ are the spherical coordinates adapted to $\dot{\Ba}$, i.e, $\dot{\Ba}$ is defined by $\theta=0$. We note that \eqref{eq:linearizedKNa} is of the form \eqref{eq:vperturbation_l1} with $f=(-2\Bm r^{-2}+\BQ^2r^{-3})dt, K=-\BQ r^{-2}$ and $\sfV=\sin^2\theta d\phi$. Then the associated gauge-invariant quantities are $\hstar\hd(r^{-1}f)=6\Bm r^{-4}-4\BQ^2r^{-5}, K=-\BQ r^{-2}$.

Now we turn to solving the equations \eqref{eq:EfVl1} and \eqref{eq:EKVl1}. First \eqref{eq:EfVl1} implies $\hd\Bigl(r^4\hstar\hd(r^{-1}f)-4\BQ rK\Bigr)=0$ and thus
\begin{equation}\label{eq:solnoff}
r^4\hstar\hd(r^{-1}f)-4\BQ rK=c
\end{equation}
for some constant $c$. Plugging this into the equation \eqref{eq:EKVl1} yields
\[
(\hBox-V)(rK)=\frac{c\BQ}{r^4}\quad \mbox{where}\quad V=\frac{2}{r^2}+\frac{4\BQ^2}{r^4}.
\]
Since $V>0$ and the right-handed side is stationary, as discussed previously, $rK=0, c=0$ is the only mode solution with frequency $\sigma\neq 0$ to the above equation, and thus $\hd(r^{-1}f)=0$ as well. As for the case $\sigma=0$, since $V=2\rho^2+4\BQ^2\rho^4$ where $\rho=1/r$, using the Frobenius method and analyzing the indicial equation, it follows that for each $c\in\BR$, there is at most one solution $K\in\eHb^{\infty, \ell}(\CX)$ for $-3/2<\ell<-1/2$. On the other hand, one can verify \[K=-\frac{c\BQ}{6\Bm r^2}\in\eHb^{\infty, \ell}(\CX) 
\] is indeed a solution. Returning to \eqref{eq:solnoff} we find \[\hstar\hd(r^{-1}f)=\frac{c}{6\Bm}(\frac{6\Bm}{r^4}-\frac{4\BQ^2}{ r^5}).
\]
Therefore, replacing $(\dg, \dA)$ by $(\dg, \dA)-\frac{c}{6\Bm}(g', A')$, we may assume $c=0$ and thus $K=0, \hd(r^{-1}f)=0$, which implies $(\dg, \dA)$ is a pure gauge solution.

This finishes the proof of Theorem \ref{thm:modestability}.

%%%%%%%%%%%%%%%%%%%%%%%%%%%%%%%%%%%%%%%%%%%%%%%%%%%%%%%%%%%%%%%%%%%%%%%%%%%
\section{Mode analysis of the scalar wave operator}
\label{sec:modeanalysisofscalar}
In this section we shall discuss the modes of the scalar wave operator on slowly rotating KN spacetimes with subextremal charge.  It not only occurs as the gauge propagation operator acting on the gauge function $D_{(g_b,A_b)}\widetilde{\Upsilon}_M(\dg,\dA;g_b,A_b)=-\delta_{g_b}\dA$, but an operator on a scalar function which generate the pure gauge solutions to the Maxwell equation. We define the Fourier-transformed scalar wave operator on slowly rotating Kerr-Newman spacetime $g_b$ with parameters $b=(\Bm, \Ba, \BQ)$ near $b_0=(\Bm_0, 0,\BQ_0)$ as \[
\widehat{\Box_{g_b}}(\sigma):=e^{i\sigma t_{b,*}}\Box_{g_b}e^{-i\sigma t_{b,*}}
\]
where $ t_{b,*}=\chi_0(r)(t+r_{b_0,*})+(1-\chi_0(r))(t-r_{(\Bm,0,\BQ),*})$ is defined as in \eqref{EqKNTimeFn}.

For ease of notation, we drop $\BC$ in the notations of the functions space $\eHb^{s,\ell}
(\CX;\BC), \eHb^{s,\ell}
(\CX;\BC),\mathcal{A}(\CX;\BC)$ and $C^\infty(\CX;\BC)$ if it is clear from the context.

\begin{thm}\label{thm:modesforscalarwave}
		Let $b_0=(\Bm_0, 0,\BQ_0)$ and $g=g_{b_0}$ be the RN metric with subextremal charge. There exists $\epsilon>0$ such that for $b=(\Bm,\Ba,\BQ)$ with $\abs{b-b_0}<\epsilon$, the following holds
	\begin{enumerate}
	\item If $\IM \sigma\geq0$ and $\sigma\neq 0$, 
	\begin{equation}\label{eq:scalarmodenonzero}
	\widehat{\Box_{g_b}}(\sigma):\{u\in\eHb^{s, \ell}(\CX;\BC): \widehat{\Box_g}(\sigma)u\in\eHb^{s,\ell+1}(\CX; \BC)\}\to \eHb^{s,\ell+1}(\CX;\BC)
	\end{equation}
	is invertible for $s>\frac 12, \ell<-\frac12, s+\ell>-\frac 12$.\\
	\item If $s>\frac 12$ and $-\frac 32<\ell<-\frac 12$, then stationary operator
	\begin{equation}\label{eq:scalarmodezero}
	\widehat{\Box_{g_b}}(0):\{u\in\eHb^{s, \ell}(\CX;\BC): \widehat{\Box_g}(0)u\in\eHb^{s-1,\ell+2}(\CX; \BC)\}\to \eHb^{s-1,\ell+2}(\CX;\BC)
	\end{equation}
is invertible.
\end{enumerate}		
\end{thm}

\begin{proof}
	The proof closely follows \cite[Theorem 6.1]{HHV21}. We first prove \eqref{eq:scalarmodenonzero} and \eqref{eq:scalarmodezero} for the RN metric $g$ and then use a perturbation argument to extend them to the slowly rotating KN metric $g_b$ with $b$ near $b_0$.

	\begin{itemize}
		\item \underline{Injectivity of $\widehat{\Box_g}(0)$.} Suppose $u\in\eHb^{s,\ell}(\CX)$ with $s>\frac 12,\ell\in(-\frac 32, -\frac 12)$ and 
	\[
	\widehat{\Box_g}(0)u=\Box_gu=\frac{1}{r^2}\pa_r\weight r^2\pa_ru+\frac{1}{r^2}\sL u=0,\quad \weight=1-\frac{2\Bm_0}{r}+\frac{\BQ^2_0}{r^2}.	\]
Then by Proposition \ref{prop:desofkernel}, $u\in\rho C^\infty(\CX)+\mathcal{A}^{2-}(\CX)$ where $\rho=r^{-1}$, which implies $\abs{u},\abs{r\pa_r u},\abs{\snabla u}\lesssim r^{-1}$. As a consequence, the boundary term at $r=\infty$ in the following integration by parts
\begin{equation}\label{eq:ibpzero}
0=-\int_{\BS^2}\!\int_{\ehRN}^\infty\!\Box_gu\cdot\overline{u}\,r^2drd\sg=\int_{\BS^2}\!\int_{\ehRN}^\infty\!\Bigl(\weight(r)\abs{r\pa_r u}^2+\abs{\snabla u}^2\Bigr)\,drd\sg
\end{equation}
vanishes, and the boundary term at $r=\ehRN$ vanishes since $u$ is smooth there and $\weight(\ehRN)=0$. Then $u$ is a constant in $r\geq \ehRN$ and thus vanishes there since $\lim_{r\to\infty}u=0$. Since $u$ vanishes to infinite order at $r=\ehRN$ and smooth in $r\leq \ehRN$, we shall prove that $u=0$ in $r\leq \ehRN$ as well by following the arguments in \cite[Lemma 1]{Zwo16}. Concretely, we first compute the following energy identity in $r_-\leq r<\ehRN$
\begin{equation}\label{eq:energyidentity}
\begin{split}
	&\qquad\qquad\pa_r\Bigl(\abs{r-\ehRN}^{-N}\bigl(-\weight(r)\abs{\pa_r u}^2+\frac{1}{r^2}\abs{\sd u}_{\sg}^2+\abs{u}^2\bigr)\Bigr)+2\abs{r-\ehRN}^{-N}\frac{1}{r^2}\sdelta\big(\RE(\overline{\pa_r u}\sd u)\big)\\
	&\qquad\qquad=N\abs{r-\ehRN}^{-N-1}\bigl(-\weight(r)\abs{\pa_r u}^2+\frac{1}{r^2}\abs{\sd u}_{\sg}^2+\abs{u}^2\bigr)-\abs{r-\ehRN}^{-N}\bigl(2\RE(\overline{\pa_r u}\cdot\widehat{\Box_g}(0)u)-R(u, du)\bigr)
\end{split}
\end{equation}
where 
\[
R(u, du)=\bigl(4r^{-1}\weight(r)-\weight'(r)\bigr)\abs{\pa_r u}^2+2\RE(\overline{u}\pa_r u)-2r^{-3}\abs{\sd u}_{\sg}^2
\]
is a quadratic form in $u$ and $du$ independent of $N$. As a result, $R(u, du)$ can be controlled by the first term on the right-hand side of the above energy identity \eqref{eq:energyidentity} when $N$ is sufficiently large. Then for any fixed $r_0\in[r_-, \ehRN)$ we apply Stokes' Theorem in $[r_0, \ehRN-\epsilon]\times\BS^2$. For $N$ large enough we have
\begin{equation}\label{eq:Stokesthm}
	\begin{split}
&\int_{\BS^2}\!(-\weight(r)\abs{\pa_r u}^2+\abs{\sd u}^2+\abs{u}^2)|_{r=r_0}\,d\sg\\
&\quad\leq (\ehRN-r_0)^N\epsilon^{-N}\int_{\BS^2}\!(-\weight(r)\abs{\pa_r u}^2+\abs{\sd u}^2+\abs{u}^2)|_{r=\ehRN-\epsilon}\,d\sg\leq C_K\epsilon^{-N+K}
\end{split}
\end{equation}
for any $K$ as $\epsilon\to 0^+$ since $u$ is smooth and vanishes to infinite order at $r=\ehRN$. By choosing $K>N$ and letting $\epsilon\to0^+$ we conclude that $u=0$ at $r=r_0$.

\item\underline{Surjectivity of $\widehat{\Box_g}(0)$.} The index $0$ property of $\widehat{\Box_{g}}(0)$ established in Lemma \ref{lem:indexofFredholmoperator} directly gives the surjectivity. But here we present an alternative proof of the surjectivity. To prove the surjectivity of $\widehat{\Box_g}(0)$, we note that it is equivalent to proving the injectivity of the adjoint $\widehat{\Box_g}(0)^*$. Suppose $v\in \sHb^{-s+1, -\ell-2}(\CX)$ with $s>\frac 12,\ell\in(-\frac 32, -\frac 12)$ satisfies $\widehat{\Box_g}(0)^* v=\widehat{\Box_g}(0)v=0$. By Proposition \ref{prop:desofkernel}, $v$ is smooth in $r_-\leq r<\ehRN$ and $r>\ehRN$, and $u\in\rho C^\infty(\CX)+\mathcal{A}^{2-}(\CX)$ near $r=\infty$. First we have $v=0$ in $r<\ehRN$ since $\widehat{\Box_g}(0)^*$ is a hyperbolic operator there with $r$ being the timelike function, see \cite[Theorem E.56]{DZ19}. Next, according to \cite[Theorem 6.3]{HaV15}, $v$ is conormal at $r=\ehRN$. Since the normal operator of $\weight\widehat{\Box_g}(0)^*$ is $(\weight\pa_r)^2$ whose boundary spectrum consists of $\{(0,0), (0,1)\}$,
\begin{comment}Once we know that $v$ is conormal at $r=\ehRN$, say $u\in\mathcal{A}^l([\ehRN,\infty)\times\BS^2)$ for some $l\in\BR$, which can be identified with $C^\infty(\BS^2;\mathcal{A}^l([\ehRN,\infty))$. Then we can treat $\omega\in\BS^2$ as a parameter and solve the PDE $\widehat{\Box_g}(0)^* v=0$ as an ODE with respect to $r$ as follows
\begin{equation}
(\weight r^2\pa_r)u=-\weight r^2\sL u\in \mathcal{A}^{l+1}.
\end{equation}
This ODE has regular singular point at $r=\ehRN$ with indicial root being $0$ with multiplicity $2$. If $l+1<0$, we conclude that $v\in\mathcal{A}^{l+1}$ which improves the decay as $r\to \ehRN^+$ by $(r-\ehRN)$. 
\end{comment}
it follows that $v=H(r-\ehRN)(v_0+v_1\log\weight+\tilde{v})$ where $v_0, v_1\in C^\infty(\BS^2)$ and $\tilde{v}\in\mathcal{A}^{1-}([\ehRN,\infty))$ which means that  $\tilde{v}$ is conormal at $r=\ehRN$ and $\tilde{v}\sim (r-\ehRN)^{1-}$ as $r\to\ehRN^+$. Since $\widehat{\Box_g}(0)H(r-\ehRN)=r^{-2}\pa_r(r^2\weight(r)\delta(r-\ehRN))=0$ and $\widehat{\Box_g}(0)(H(r-\ehRN)\log\weight)=\weight'(r)\delta(r-\ehRN)+2\BQ_0^2r^{-4}H(r-\ehRN)$, we have
\begin{align*}
0=\widehat{\Box_g}(0)v=\weight'(r)\delta(r-\ehRN)v_1+R
\end{align*}
where
\[
R=\frac{H(r-\ehRN)}{r^2}\bigl(\sL v_0+\log(r-\ehRN)\sL v_1+\frac{2\BQ_0^2}{r^2}v_1\bigr)+H(r-\ehRN)\widehat{\Box_g}(0)\tilde{v}\in\mathcal{A}^{0-}.
\]
However, $\delta(r-\ehRN)\notin \mathcal{A}^{0-}$, and this implies $v_1=0$. Since $v_1=0$ implies $\weight\pa_r v\to 0$ as $r\to \ehRN^+$, we find that the integration by parts \eqref{eq:ibpzero} still works here and thus gives $v=0$ in $r>\ehRN$. Then we can conclude that $v$ is supported at $r=\ehRN$, and consequently $v\notin L^2$ \cite[Theorem 7.1.27]{H03} unless $v=0$. However, the radial point estimates at the event horizon imply that $v\in H^{1/2-}$ near $r=\ehRN$. Therefore, $v$ must be $0$.

\item\underline{Invertibility of $\widehat{\Box_g}(\sigma)$ for $\sigma\in\BR\setminus\{0\}$.} According to Lemma \ref{lem:indexofFredholmoperator}, it suffices to prove the injectivity of $\widehat{\Box_g}(\sigma)$ for non-zero real $\sigma$. We first illustrate the relationship between $\widehat{\Box_g}(\sigma)$ and $\wt{\Box_g}(\sigma)=e^{i\sigma t}\Box_g e^{-i\sigma t}$, that is, $u\in\ker\widehat{\Box_g}(\sigma)$ gives rise to an outgoing solution 
\begin{equation}\label{eq:differentconjBox}
\wt{\Box_g}\tilde{u}=0\quad \mbox{where}\quad  \tilde{u}=e^{i\sigma(t-t_{b_0,*})}u. 
\end{equation}
We next prove that $\tilde{u}=0$ and thus $u=0$ in $r\geq \ehRN$ by a boundary pairing argument (see\cite[\S 2.3]{Mel95}\cite[\S3.2]{HV18_2}\cite[\S7]{HHV21}), and a unique continuation theorem at infinity. Concretely, let $f(r)\in C^\infty([\ehRN, \infty))$ be a nonnegative function with $f(r)=r-\ehRN$ for $\ehRN\leq r<3\Bm_0$ and $f(r)=r^{-1}$ for $r>4\Bm_0$. Fix a cutoff $\chi\in C^\infty([0,\infty))$ which is $0$ on $[0,1/2]$ and $1$ on $[1,\infty)$. For small $\epsilon>0$, we define
\begin{equation}\label{eq:cutoff}
	\chi_\epsilon(r):=\chi(\frac{f}{\epsilon})\in C^\infty_c((\ehRN, \infty)).
\end{equation}
We see that $\chi_\epsilon=1$ when $r-\ehRN\geq \epsilon, r\leq\epsilon^{-1}$, and $r-\ehRN\geq \epsilon/2, r\leq 2\epsilon^{-1}$ on supp$\chi_\epsilon$. We then calculate
\begin{equation}\label{eq:boundarypair}
	\begin{split}
		0&=\lim_{\epsilon\to0}\Big(\int_{\BS^2}\!\int_{\ehRN}^\infty\!\chi_\epsilon\tilde{u}\cdot\overline{\wt{\Box_g}(\sigma)\tilde{u}}\,r^2drd\sg-\int_{\BS^2}\!\int_{\ehRN}^\infty\!\chi_\epsilon\wt{\Box_g}(\sigma)\tilde{u}\cdot\overline{\tilde{u}}\,r^2drd\sg\Big)\\
		&=\lim_{\epsilon\to 0}\int_{\BS^2}\!\int_{\ehRN}^\infty\![\wt{\Box_g}(\sigma),\chi_\epsilon]\tilde{u}\cdot\overline{\tilde{u}}\,r^2drd\sg\\
		&=\lim_{\epsilon\to 0}\Bigl(\int_{\BS^2}\!\int_{\ehRN}^\infty\!2\weight\pa_r\chi_{\epsilon}\cdot\pa_r \tilde{u}\cdot\overline{\tilde{u}}\,r^2drd\sg-\int_{\BS^2}\!\int_{\ehRN}^\infty\!\weight\pa_r\chi_{\epsilon}\cdot\pa_r \abs{\tilde{u}}^2\,r^2drd\sg\Bigr).
	\end{split}
\end{equation}
where $\pa_r\chi_\epsilon$ may be nonzero only near the event horizon $r=\ehRN$ and near infinity $r=\infty$. Near $r=\ehRN$, we have $\tilde{u}=e^{-i\sigma r_*}u$ where we write $r_*=r_{b_0, *}$ for brevity of notation,  $\pa_r\chi_{\epsilon}=\epsilon^{-1}\chi'(\frac{r-\ehRN}{\epsilon})\to\delta(r-\ehRN)$ as $\epsilon\to 0$, and thus
\begin{equation}\label{eq:commutatoratEH}
	\begin{split}
		&\lim_{\epsilon\to 0}\Bigl(\int_{\BS^2}\!\int_{\ehRN}^\infty\!2\weight\pa_r\chi_{\epsilon}\cdot\pa_r \tilde{u}\cdot\overline{\tilde{u}}\,r^2drd\sg-\int_{\BS^2}\!\int_{\ehRN}^\infty\!\weight\pa_r\chi_{\epsilon}\cdot\pa_r \abs{\tilde{u}}^2\,r^2drd\sg\Bigr)\\
		&\quad=\int_{\BS^2}\!\Big(-2i\sigma r^2\abs{u}^2+2\weight r^2\pa_r u\cdot \overline{u}-r^2\weight\pa_r\abs{u}^2\Big)|_{r=\ehRN}\,d\sg\\
		&\quad=\int_{\BS^2}\!\big(-2i\sigma r^2\abs{u}^2\big)|_{r=\ehRN}\,d\sg\\
	\end{split}
	\end{equation}
where we use the smoothness of $u$ at $r=\ehRN$ in the last step. While near $r=\infty$, we have $r^2\pa_r\chi_{\epsilon}=-\epsilon^{-1}\chi'(\frac{\rho}{\epsilon})\to-\delta(\rho)$ with $\rho=r^{-1}$ as $\epsilon\to 0$, $\tilde{u}=e^{i\sigma r_*}u$ where $u=\rho u_+(\omega)+e(\rho,\omega)$ with $u_+(\omega)\in C^\infty(\BS^2)$ and $e(\rho, \omega)\in\mathcal{A}^{2-}(\CX)$ (see Proposition \ref{prop:desofkernel}), and thus 
\begin{equation}\label{eq:commutatoratinfty}
	\begin{split}
		&\lim_{\epsilon\to 0}\Bigl(\int_{\BS^2}\!\int_{\ehRN}^\infty\!2\weight\pa_r\chi_{\epsilon}\cdot\pa_r \tilde{u}\cdot\overline{\tilde{u}}\,r^2drd\sg-\int_{\BS^2}\!\int_{\ehRN}^\infty\!\weight\pa_r\chi_{\epsilon}\cdot\pa_r \abs{\tilde{u}}^2\,r^2drd\sg\Bigr)\\
		&\quad=\int_{\BS^2}\!\Big(-2i\sigma r^2\abs{u}^2-2\weight r^2\pa_r u\cdot \overline{u}+r^2\weight\pa_r\abs{u}^2\Big)|_{r=\infty}\,d\sg\\
		&\quad=\int_{\BS^2}\!\big(-2i\sigma r^2\abs{u}^2\big)|_{r=\infty}\,d\sg=\int_{\BS^2}\!-2i\sigma \abs{u_+}^2\,d\sg\\
	\end{split}
\end{equation}
Putting \eqref{eq:commutatoratEH} and \eqref{eq:commutatoratinfty} together yields
\begin{equation}
0=-2i\sigma\Bigl(\int_{\BS^2}\! r^2\abs{u}^2|_{r=\ehRN}\,d\sg+\int_{\BS^2}\!\abs{u_+}^2\,d\sg\Bigr),
\end{equation}
and this implies $u|_{r=\ehRN}=0$ and $u_+(\omega)=0$. This proves that the leading order term of $u$ at infinity vanishes and thus $u\in\mathcal{A}^{2-}(\CX)$. For the next step, we rewrite $\widehat{\Box_g}(\sigma)u=0$ near infinity as follows
\[
2i\sigma\pa_r (ru)=-\frac{1}{r}\pa_rr^2\weight\pa_ru-\frac{1}{r}\sL u=\rho\mbox{Diff}_b^2u.
\]
Then integrating both sides from infinity to $r$ and using $ru|_{r=\infty}=0$ yield $u\in\mathcal{A}^{3-}(\CX)$. Proceeding iteratively, we find that $u\in\mathcal{A}^{\infty}$, which means that $u$ vanishes to infinite order at $r=\infty$. Then using the unique continuation theorem at infinity \cite[Theorem 17.2.8]{H05} for $\wt{\Box_g}(\sigma)\tilde{u}=0$ gives $\tilde{u}=0$ and thus $u=0$ in $r>\ehRN$. By Proposition \ref{prop:desofkernel} again, $u$ is smooth and vanishes to infinite order at $r=\ehRN$. Therefore, we follow the arguments in \cite[Lemma 1]{Zwo16} again to prove $u=0$ in $r<\ehRN$. Since we are working in the region near the $r=\ehRN$ where $t_{b_0,*}=t_0$, it would be more convenient to use the coordinates $(t_{b_0}, r,\omega)$ with $\omega\in\BS^2$, and then we write
\begin{equation}
	\widehat{\Box_g}(\sigma)u=\frac{1}{r^2}\pa_rr^2\weight\pa_ru+\frac{1}{r^2}\sL u-\frac{2i\sigma}{r}\pa_rr u.
\end{equation}
We can again obtain the energy identity as in \eqref{eq:energyidentity} with $R(u, du)$ replaced by \[R(u,du)-2\RE\bigl(2i\sigma r^{-1}\overline{\pa_ru}\cdot\pa_r(ru)\bigr)
\] which can still be controlled when $N$ is sufficiently large. Proceeding in the same way as in \eqref{eq:Stokesthm}, it follows that $u=0$ in $r<\ehRN$ as well.

\item\underline{Invertibility of $\widehat{\Box_g}(\sigma)$ for $\IM\sigma>0$.} Again according to Lemma \ref{lem:indexofFredholmoperator}, it suffices to prove the injectivity of $\widehat{\Box_g}(\sigma)$ for $\sigma\!\in\!\BC, \IM\sigma\!\!>0$. We again use relationship between $\widehat{\Box_g}(\sigma)$ and $\wt{\Box_g}(\sigma)$ as illustrated in \eqref{eq:differentconjBox}. Now $\tilde{u}$ decays exponentially when $r_{b_0, *}\to\pm\infty$. So we can apply integration by parts to conclude
\[
0=-\int_{\BS^2}\!\int_{\ehRN}^\infty\!\wt{\Box_g}(\sigma)\tilde{u}\cdot\overline{\tilde{u}}\,r^2dr\sg=\int_{\BS^2}\!\int_{\ehRN}^\infty\!\Bigl(\weight(r)\abs{r\pa_r \tilde{u}}^2+\abs{\snabla \tilde{u}}^2-\sigma^2\weight^{-1}(r)\abs{\tilde{u}}^2\Bigr)\,drd\sg.
\]
Then by taking imaginary part for $\RE\sigma\neq 0$, while using $-\sigma^2>0$ for $\RE\sigma=0$, it follows that $\tilde{u}=0$ and thus $u=0$ in $r>\ehRN$. The proof of $u=0$ in $r<\ehRN$ is the same as that in the non-zero real $\sigma$ case.

\item\underline{Proof for $\widehat{\Box_{g_b}}(\sigma)$.} Finally, we consider the slowly rotating KN metric $g_b$, with $b$ near $b_0$. By Theorem \ref{thm:energyestimate} and high energy estimate \ref{thm:highenergy}, $\widehat{\Box_{g_b}}(\sigma)$ is invertible for $\abs{\sigma}$ large, say $\abs{\sigma}\geq C$, when $\abs{b-b_0}<\epsilon_1$, so we only need to consider the bounded region$\abs{\sigma}\leq C$. Near $\sigma=0$, the invertibility of $\widehat{\Box_{g_b}}(\sigma)$ for $b$ close to $b_0$ follows from the perturbation argument as above and the fact that $\widehat{\Box_{g}}(0)$ is invertible. For $\sigma\in\BC, \IM\sigma\geq 0$, bounded away from $0$ and infinity, we prove the theorem by perturbation arguments (starting with the invertibility of $\widehat{\Box_{g}}(\sigma) $) in a compact set of $\IM\sigma\geq0$. Concretely, suppose there exist sequences $\delta_j\to 0, b_j\to b_0$ such that $\ker\widehat{\Box_{g_{b_j}}}(\sigma_j)$ is non-trivial. Then we can find $u_j\in\eHb^{s,\ell}(\CX)\cap\ker\widehat{\Box_{g_{b_j}}}(\sigma_j)$ with $\norm{u_j}_{\eHb^{s,\ell}(\CX)}=1$. By the uniform Fredholm estimate \ref{thm:Fredholmestimate}, we have $1\leq C'\norm{u_j}_{\eHb^{s_0, l_0}(\CX)}$. Therefore, there exists a subsequence $u_j\to u$ weakly in $\eHb^{s, \ell}(\CX)$ which is norm convergent in $\eHb^{s_0, \ell_0}(\CX)$ with the limit $u$ being non-zero and satisfying $\widehat{\Box_{g}}(0)u=0$. But this contradicts the invertibility of $\widehat{\Box_{g}}(0)$. So this proves the injectivity. The surjectivity which is equivalent to the injectivity of the adjoint $\widehat{\Box_{g_{b}}}(\sigma)^*$ can be proved in a similar manner. This proves that for $s>\frac 12,\ell\in (-\frac 32, \frac 12)$ and $\IM\sigma\geq 0,\abs{\sigma}<c, \abs{b-b_0}<\epsilon_2$ for some $c, \epsilon_2>0$,
\[\widehat{\Box_{g_{b}}}(\sigma):\{u\in\eHb^{s,\ell}(\CX):\widehat{\Box_{g_{b}}}(\sigma)u\in\eHb^{s-1,\ell+2}(\CX)\}\to \eHb^{s-1,
\ell+2}(\CX)
\]
 is invertible. By Proposition \ref{prop:desofkernel}, this implies that $\widehat{\Box_{g_{b}}}(\sigma)$ acting on the function spaces in \eqref{eq:scalarmodenonzero} is injective for $\IM\sigma\geq0, 0<\abs{\sigma}<c$, and the invertibility follows from the fact that $\widehat{\Box_{g_{b}}}(\sigma)$ has index $0$. This proves the theorem for $\IM\sigma\geq 0,0\leq\abs{\sigma}<c$ and $\abs{b-b_0}<\epsilon_2$. For $\IM\sigma\geq 0, c\leq\abs{\sigma}\leq C$, the fact that $\widehat{\Box_g}(\sigma)$ is invertible and a perturbation argument as above imply the invertibility of $\widehat{\Box_{g_b}}(\sigma)$ for $(b, \sigma)$ in an open neighborhood of $(b_0, \sigma)$. Then the compactness of the region $c\leq\abs{\sigma}\leq C$ implies that there exists an $\epsilon_3>0$ (depending on $c, C$) such that $\widehat{\Box_{g_b}}(\sigma)$ in \eqref{eq:scalarmodenonzero} is invertible for $\IM\sigma\geq 0, c\leq\abs{\sigma}\leq C$ and $\abs{b-b_0}<\epsilon_3$. Therefore, the theorem holds for $\abs{b-b_0}<\epsilon$ where $\epsilon=\min\{\epsilon_1, \epsilon_2, \epsilon_3\}$.
 \end{itemize}
\end{proof}

\subsection{Growing zero modes}
For later use, we now discuss the explicit form of the scalar functions in $\ker\widehat{\Box_{g_{b}}}(0)$ which are allowed to have more growth at infinity.
\begin{prop}\label{prop:gscalarzeromode}
For $b=(\Bm, \Ba, \BQ)$ near $b_0=(\Bm_0, 0,\BQ_0)$, we have
\begin{align}\label{eq:gscalar3/2}
	\ker\widehat{\Box_{g_b}}(0)\cap\eHb^{\infty, -3/2-}(\CX)&=\langle u_{b, s_0}\rangle, \quad u_{b,s_0}=1.\\\label{eq:gscalardual3/2}
		\ker\widehat{\Box_{g_b}}(0)^*\cap\sHb^{-\infty, -3/2-}(\CX)&=\langle u^*_{b, s_0}\rangle,\quad u^*_{b,s_0}=H(r-\ehKN).
	\end{align}
The following spaces
\begin{align}\label{eq:gscalar5/2}
		\ker\widehat{\Box_{g_b}}(0)\cap\eHb^{\infty, -5/2-}(\CX)&=\langle u_{b, s_0}\rangle\oplus\{u_{b, s_1}(\sfS): \sfS\in\BFS_1\},\\\label{eq:gscalardual5/2}
	\ker\widehat{\Box_{g_b}}(0)^*\cap\sHb^{-\infty, -5/2-}(\CX)&=\langle u^*_{b, s_0}\rangle\oplus\{u^*_{b, s_1}(\sfS): \sfS\in\BFS_1\},
\end{align}
are $4$-dimensional. Moreover, the maps $b\mapsto u_{b, s_1}(\sfS)$ and $b\mapsto u^*_{b,s_1}(\sfS)$ can be chosen to be continuous in $b$ with values in the respective spaces. More explicitly, we have
\begin{equation}\label{eq:gscalar5/2RN}
	u_{b=(\Bm, 0,\BQ),s_1}(\sfS)=(r-\Bm)\sfS,\quad u_{b=(\Bm, 0,\BQ),s_1}(\sfS)=(r-\Bm)H(r-\ehKN)\sfS.
	\end{equation}
\end{prop}
\begin{rem}
	We use the notation $u_{b, s_0}, u_{b,s_1}$ because they are of scalar type $l=0$ and $l=1$ respectively when $b=(\Bm, 0,\BQ)$ is a parameter of RN metric.
\end{rem}
\begin{proof}[Proof of Proposition \ref{prop:gscalarzeromode}]
We first prove \eqref{eq:gscalar3/2}. Let $u\in\ker\widehat{\Box_{g_b}}(0)\cap \eHb^{\infty,-3/2-}(\CX)$. By the normal operator argument as in  Proposition \ref{prop:desofkernel}, %(with the modification that we add $\{(0,0)\}$ in the boundary spectrum corresponding to $\lambda=0$ and $l=0$ there to include $\eHb^{\infty, -3/2-}(\CX)$ since $\eHb^{\infty, -3/2-}(\CX)\subset\mathcal{A}^{0-}(\CX)$), 
$u$ must have the form $u=u_0+\tilde{u}$ where $u_0\in\BC$ and $\tilde{u}\in\mathcal{A}^{1-}(\CX)\subset\eHb^{\infty, -1/2-}(\CX)$. Then we have 
\[
0=\widehat{\Box_{g_b}}(0)u=\widehat{\Box_{g_b}}(0)u_0+\widehat{\Box_{g_b}}(0)\tilde{u}=\widehat{\Box_{g_b}}(0)\tilde{u}.
\]
In view of Theorem \ref{thm:modesforscalarwave}, $\tilde{u}$ must be $0$, and thus $u=u_0$ is a constant. Conversely, we certainly have $\widehat{\Box_{g_b}}(0)1=0$. This proves \eqref{eq:gscalar3/2}.

The proof of \eqref{eq:gscalardual3/2} is analogous. Let $u\in\ker\widehat{\Box_{g_b}}(0)\cap \sHb^{-\infty,-3/2-}(\CX)$. By the normal operator argument again, we conclude that $u=\chi u_0+\tilde{u}$ where $\chi$ is a smooth cutoff with $\chi=0$ for $r\leq 3\Bm_0$, $\chi=1$ with $r\geq 4\Bm_0$, and $u_0\in\BC, \tilde{u}\in\sHb^{\infty, -1/2-}(\CX)$. We calculate
\[
-\widehat{\Box_{g_b}}(0)^*\chi u_0\in\sHb^{-\infty,\infty}(\CX)\subset\sHb^{-\infty, 3/2-}(\CX).
\]
Owing to Theorem \ref{thm:modesforscalarwave}, there exists a unique $\tilde{u}\in\sHb^{-\infty, -1/2-}(\CX)$ satisfying $\widehat{\Box_{g_b}}(0)^*\tilde{u}=-\widehat{\Box_{g_b}}(0)^*\chi u_0$. This proves that the space in \eqref{eq:gscalardual3/2} is at most $1$-dimensional. Conversely, a direct calculation implies \[
\widehat{\Box_{g_b}}(0)H(r-\ehKN)=\rho_b^{-2}\pa_r(\Delta_b\delta(r-\ehKN))=0
\] since $\Delta_b|_{r=\ehKN}=0$. This proves \eqref{eq:gscalardual3/2}.

As for \eqref{eq:gscalar5/2}, since $\eHb^{\infty, -5/2-}(\CX)\subset\mathcal{A}^{-1-}(\CX)$, in the normal operator argument as in Proposition \ref{prop:desofkernel}, we shift the contour of integration through the pole $i$ with the space of resonant states given by $r\sfS$ with $\sfS\in\BFS_1$. That is to say, $u$ must take the form $u=\chi r\sfS+\tilde{u}$ where $\chi$ is the smooth cutoff as defined above and $\sfS\in\BFS_1, \tilde{u}\in\eHb^{\infty, -3/2-}(\CX)$. Since $\BFS_1$ is $3$-dimensional, we find that the space in \eqref{eq:gscalar5/2} is at most $4$-dimensional. Now we prove that it is indeed $4$-dimensional. We compute
\begin{equation}\label{eq:eqnforerror}
-\widehat{\Box_{g_b}}(0)\chi r\sfS=-\chi\widehat{\Box_{\underline{g}}}(0)r\sfS+[\chi, \widehat{\Box_{\underline{g}}}(0)]r\sfS+(\widehat{\Box_{\underline{g}}}(0)-\widehat{\Box_{g_b}}(0))\chi r\sfS=0+\eHb^{\infty, \infty}(\CX)+\eHb^{\infty, 1/2-}(\CX)
\end{equation}
where we use $\widehat{\Box_{\underline{g}}}(0)-\widehat{\Box_{g_b}}(0)\in\rho^3\mbox{Diff}_b^2$ for the third summand. Since $\ker\widehat{\Box_{g_b}}(0)\cap \sHb^{-\infty, -1/2-}(\CX)=0$, there exists $\tilde{u}\in\eHb^{\infty, -3/2-}(\CX)$, which is unique modulo a multiple of $u_{b, s_0}$, satisfying the above equation \eqref{eq:eqnforerror}. Concretely, let $v\in C^\infty_c(X^\circ)$ such that $\langle u_{b,s_0},v\rangle=1$. Then there exists a unique $\tilde{u}\in \ann(v)$ satisfying the equation \eqref{eq:eqnforerror}. Therefore, $\chi r\sfS+\tilde{u}$, where $\tilde{u}$ is uniquely determined by $\sfS$ and $b$, indeed gives rise to an element in the space \eqref{eq:gscalar5/2} with leading order term $r\sfS$ at infinity. This proves \eqref{eq:gscalar5/2}. The proof of $\eqref{eq:gscalardual5/2}$ is completely analogous.

Now we turn to discussing the continuous dependence of $u_{b, s_1}(\sfS), u^*_{b, s_1}(\sfS)$ on $b$.
Suppose $b_j\to b$ and $u_{b_j}(\sfS)=\chi r\sfS+\tilde{u}_{b_j}(\sfS)\in\ker\widehat{\Box_{g_{b_j}}}\cap \eHb^{\infty, -5/2-}(\CX)$ where $\tilde{u}_{b_j}(\sfS)$ is in the annihilator of $v\in C^\infty_c(X^\circ)$ which satisfies $\langle u_{b,s_0}, v\rangle \neq 0$. Let $u=\chi r\sfS+\tilde{u}(\sfS)\in\ker\widehat{\Box_{g_b}}(0)$ and $e_j(\sfS)=u_{b_j}(\sfS)-u(\sfS)=\tilde{u}_{b_j}(\sfS)-\tilde{u}(\sfS)$, and we find
\[
\widehat{\Box_{g_b}}(0)e_j(\sfS)=(\widehat{\Box_{g_b}}(0)-\widehat{\Box_{g_{b_j}}}(0))(\tilde{u}_{b_j}(\sfS)+\chi r\sfS).
\]
Since $\{\tilde{u}_{b_j}(\sfS)\}$ is uniformly bounded in $\ann(v)$ and $\widehat{\Box_{g_b}}(0)-\widehat{\Box_{g_{b_j}}}(0)\in (b-b_j)\rho^3\mbox{Diff}_b^2$, we see that $\widehat{\Box_{g_b}}(0)e_j(\sfS)\in (b-b_j)\eHb^{\infty, 1/2-}(\CX)$. As $\widehat{\Box_{g_b}}(0)|_{\ann(v)}: \ann{v}\to\eHb^{\infty,1/2-}(\CX)$ is invertible, it follows that $e_j(\sfS)$ is of size $\mathcal{O}_{\eHb^{\infty, -3/2-}(\CX)}(b-b_j)$. This completes the proof of the continuity.

It remains to derive the explicit expressions in \eqref{eq:gscalar5/2RN}. We first consider the case $b=b_0$, i.e. $g_b=g$. Using $\sL\sfS=-2\sfS$, it follow that 
\begin{equation}
	\begin{split}
	\widehat{\Box_{g}}(0)((r-\Bm_0)\sfS)&=\frac{1}{r^2}\Bigl(\pa_r(r^2-2\Bm_0r+\BQ_0^2)\sfS-2(r-\Bm_0)\sfS\Bigr)=0.\\
	\widehat{\Box_{g}}(0)^*((r-\Bm_0)H(r-\ehRN)\sfS)&=H(r-\ehRN)\widehat{\Box_{g}}(0)((r-\Bm_0)\sfS)+2\weight\delta(r-\ehRN)\sfS\\
	&\quad+r^{-2}\pa_r(r^2\weight\delta(r-\ehRN))(r-\Bm_0)\sfS=0
	\end{split}
	\end{equation}
For the general RN metric $g_b$ with parameter $b=(\Bm, 0,\BQ)$, we notice that the expression of $g_b$ is the same as that of $g$ with $(\Bm_0, 0,\BQ_0)$ replaced by $(\Bm, 0,\BQ)$. Therefore, the above calculation still follows and gives the explicit form of $u_{(\Bm, 0,\BQ),s_1}(\sfS)$ and $u^*_{(\Bm, 0,\BQ),s_1}(\sfS)$.
\end{proof}

%%%%%%%%%%%%%%%%%%%%%%%%%%%%%%%%%%%%%%%%%%%%%%%%%%%%%%%%%%%%%%%%%%%%%%%%%%%
\section{Mode analysis of the $1$-form wave-type operator}
\label{sec:modeanalysisof1form}
In this section we shall analyze the modes of the gauge propagation operator $\mathcal{P}_{b,\gamma}=\delta_{g_b} G_{g_b}\wt{\delta}^*_{g_b,\gamma}$ and the gauge potential wave operator $\mathcal{W}_{b,\gamma}=\wt{\delta}_{g_b,\gamma} G_{g_b}\delta_{g_b}^*$, both of which are wave-type operators acting on $1$-form. The former one occurs as the (Einstein part) of gauge propagation operator acting on the gauge $1$-form $D_{g_b}\widetilde{\Upsilon}^E(\dg;g_b)=\tilde{\delta}_{g_b,\gamma}G_{g_b}\dg$, while the latter one acts on the gauge potentials (which generate the pure gauge solutions). We also note that from this section on, the smallness of the charge $\BQ$ is required, i.e., we are restricted in the weakly charged RN and KN (slowly rotating) metric. Throughout this section, the bundle we mostly deal with is scattering $1$-form, so we shall drop ``$\scform$'' in $\eHb^{s,\ell}(\CX;\scform), \sHb^{s,\ell}(\CX;\scform)$ and $\mathcal{A}(\CX,\scform)$ when it is clear from the context.

We first recall the \textit{ linearized gauge-fixed Einstein-Maxwell system} \[L_{b,\gamma}=(2L^E_{b,\gamma}(\dg,\dA),\ L^M_{b,\gamma}(\dg,\dA))=0
\] where
\begin{gather}\label{eq:modifiedlinearizedEinstein}
L^E_{b,\gamma}(\dg,\dA)=D_{g_b}\Ric(\dg)-2D_{(g_b, dA_b)}T(\dg,d\dA)+\wt{\delta}^*_{g_b,\gamma}\wt{\delta}_{g_b,\gamma}G_{g_b}\dg=0,\\\label{eq:modifiedlinearizedMaxwell}
L^M_{b,\gamma}(\dg,\dA)=D_{(g_b, A_b)}(\delta_gdA)(\dg, \dA)+d\delta_{g_b}\dA=0.
\end{gather}

We note that for $(g_b, A_b)$ satisfying the Einstein-Maxwell system, $(2\delta_{g_b}^*\omega,\ \mathcal{L}_{\omega^\sharp}A_b+d\phi)$ solves the linearized gauge-fixed Einstein-Maxwell system \eqref{eq:modifiedlinearizedEinstein} and \eqref{eq:modifiedlinearizedMaxwell} provided that the $1$-form $\omega$ and the scalar function $\phi$ satisfy 
\begin{equation}\label{eq:eqnforgaugepotential}
	\wt{\delta}_{g_b,\gamma}G_{g_b}\delta^*_{g_b}\omega=0,\quad \delta_{g_b}d\phi=-\delta_{g_b}(\mathcal{L}_{\omega^\sharp}A_b).
\end{equation}
 Therefore, zero mode solutions to the above wave-type system of equations \eqref{eq:eqnforgaugepotential} give rise to pure gauge zero mode solutions to $L_{(g_b,A_b),\gamma}$.

Dually,  using \eqref{eq:DT}, \eqref{eq:L_1}, \eqref{eq:L_2} and the following calculation for $	(D_{g_b}\Ric)^*:\scsym\to\scsym$, $(D_{(g_b,dA_b)}T)^*:\scsym\to\scform\oplus\scsym$ and $(D_{(g_b, A_b)}(\delta_gdA))^*:\scform\to\scform\oplus\scsym$
\begin{gather*}
	(D_{g_b}\Ric)^*(\dg)=G_{g_b}D_{g_b}\Ric(G_{g_b}\dg)\\
	 (D_{(g_b,dA_b)}T)^*(\dg)=(-D_{(g_b, A_b)}(\delta_gdA)(\dg,0),\ D_{(g_b,dA_b)}T(\dg,0)-\frac12(g_{b})_{\mu\nu}\dg^{\al\be}F_{\al}^{\ \gamma}F_{\be\gamma}+\frac12\tr_{g_b}\dg F_{\mu\al}F_{\nu}^{\ \al}),\\ 
	 =(-D_{(g_b, A_b)}(\delta_gdA)(G_{g_b}\dg,0),\ G_{g_b}D_{(g_b,dA_b)}T(G_{g_b}\dg,0))\\
	(D_{(g_b, A_b)}(\delta_gdA))^*(\dA)=\Bigl(\delta_{g_b}d\dA,\ -D_{(g_b,dA_b)}T(0,d\dA)\Bigr)=\Bigl(d\delta_{g_b}\dA,\ -G_{g_b}D_{(g_b,dA_b)}T(0,d\dA)\Bigr),
\end{gather*}
we conclude
\[L^*_{(g_b,A_b),\gamma}(\dg, \dA)=(2L^{*E}_{(g_b,A_b),\gamma}(\dg,\dA),\  4L^{*M}_{(g_b,A_b),\gamma}(\dg,\dA))
\]
with
\begin{gather}\label{eq:modifiedlinearizedEinsteinDual}
	L^{*E}_{(g_b,A_b),\gamma}(\dg,\dA)=G_{g_b}\Bigl(D_{g_b}\Ric-2D_{(g_b, dA_b)}T+\wt{\delta}^*_{g_b,\gamma}\wt{\delta}_{g_b,\gamma}G_{g_b}\Bigr)(G_{g_b}\dg,\frac14d\dA),\\\label{eq:modifiedlinearizedMaxwellDual}
	L^{*M}_{(g_b,A_b),\gamma}(\dg,\dA)=\Bigl(D_{(g_b, A_b)}(\delta_gdA)+d\delta_{g_b}\Bigr)(G_{g_b}\dg, \frac14\dA).
\end{gather}
Therefore, we have $L^*_{(g_b, A_b),\gamma}(2G_{g_b}\delta_{g_b}^*\omega^*, 4\mathcal{L}_{\omega^{*\sharp}}A_b+d\phi^*)=0$ provided that
\begin{equation}\label{eq:eqfordualpotential}
\wt{\delta}_{g_b,\gamma}G_{g_b}\delta_{g_b}^*\omega^*=0,\quad \delta_{g_b}d\phi^*=-4\delta_{g_b}(\mathcal{L}_{\omega^{*\sharp}}A_b).
\end{equation}
Such \textit{dual pure gauge} zero modes $(2G_{g_b}\delta_{g_b}^*\omega^*, 4\mathcal{L}_{\omega^{*\sharp}}A_b+d\phi^*)$ give rise to zero mode solutions to \[L^*_{(g_b, A_b),\gamma}(\dg,\dA)=0.
\]

\begin{rem}
	The reason we introduce the constraint damping, i.e., replace $\delta_{g_b}^*$ by $\delta_{g_b, \gamma}^*$ in this manuscript, is that we need to use the invertibility of the corresponding gauge propagation operator $\delta_{g_b}G_{g_b}\tilde{\delta}^*_{g_b}$ to exclude the generalized modes, which grows linearly in $t_{b,*}$ and whose leading term is given by linearized (in $(\dot{\Bm}, 0,\dot{\BQ})$) KN solutions, to the linearized gauge-fixed Einstein-Maxwell equations. We also note that we use a perturbation argument to %extend \cite[Proposition 10.3 and 10.12]{HHV21} to the 
	prove the invertibility of the gauge propagation operator $\delta_{g_b}G_{g_b}\tilde{\delta}^*_{g_b}$ on the weakly charged and slowly rotating KN metrics $g_b$. Moreover, we use the generalized linearized gauge condition for the Einstein part $\wt{\delta}_{g_b, \gamma}G_{g_b}\dg=0$ in this paper to exclude the zero mode of no geometric significance, see \cite[Remark 10.13]{HHV21}.
\end{rem}

Let $\tilde{\chi}(r)\in C_c^\infty((r_-, 3\Bm_0))$ such that $\tilde{\chi}=1$ near $r=2\Bm_0$ (and thus near the event horizon of $g_b$ with $b$ near $(\Bm_0, 0,\BQ_0)$ where $\abs{\BQ_0}\ll\Bm_0$). We use the following
\begin{equation}\label{eq:defofc}
	\mathfrak{c}=\tilde{\chi}(r)(dt_0-\mathfrak{b}dr), \quad t_0=t+r_{(\Bm_0, 0,\BQ_0),*}
\end{equation}
with $\mathfrak{b}$ to be determined later, to define $\wt{\delta}^*_{g_b, \gamma}, \wt{\delta}_{g_b, \gamma}, L_{(g_b, A_b),\gamma}, L^*_{(g_b, A_b),\gamma}$. As discussed in \eqref{eq:eqnforgaugepotential}, we then define \textit{gauge propagation operator}
\begin{equation}
	\mathcal{P}_{b,\gamma}:=\delta_{g_b}G_{g_b}\wt{\delta}_{g_b, \gamma}^*
\end{equation}
and \textit{gauge potential wave operator} 
\begin{equation}
	\mathcal{W}_{b,\gamma}:=\wt{\delta}_{g_b,\gamma}G_{g_b}\delta_{g_b}^*.
\end{equation}
\begin{rem}
We notice that $\mathcal{P}_{b,\gamma}$ and $\mathcal{W}_{b,\gamma}$ are formally dual to each other. However, on the level of function spaces, they are not because we need to work with the extendible $b$-Sobolev spaces for both when analyzing the modes of $L_{(g_b, A_b),\gamma}$. We also note that the \textit{dual pure gauge} zero mode solutions to $L^*_{(g_b, A_b),\gamma}(\dg,\dA)=0$ (as discussed in \eqref{eq:eqfordualpotential}) are closely related to the zero mode solutions to $\mathcal{W}_{b,\gamma}\omega^*=0$.
\end{rem}

In order to study the modes of $\mathcal{P}_{b,\gamma}$ and $\mathcal{W}_{b, \gamma}$ for $b=(\Bm,\Ba,\BQ)$ near $b_0=(\Bm_0,0,\BQ_0)$, we define
\begin{equation}\label{eq:fourierCPW}
	\widehat{\mathcal{P}_{b,\gamma}}(\sigma):=e^{i\sigma t_{b,*}}\mathcal{P}_{b,\gamma}e^{-i\sigma t_{b,*}},\quad \widehat{\mathcal{W}_{b,\gamma}}(\sigma):=e^{i\sigma t_{b,*}}\mathcal{W}_{b,\gamma}e^{-i\sigma t_{b,*}}
\end{equation}
where $t_{b,*}=(t+r_{b_0,*})\chi_0(r)+(t-r_{(\Bm,0,\BQ),*})(1-\chi_0(r))$ is defined as in \eqref{EqKNTimeFn}.

We first record the following result about Schwarzschild metric which we rely on in the analysis of $\widehat{\mathcal{P}_{b,\gamma}}(\sigma)$ and $\widehat{\mathcal{W}_{b, \gamma}}(\sigma)$ on weakly charged and slowly rotating KN metrics.

\begin{prop}[{\cite[Theorem 7.1]{HHV21}}]\label{prop:modeofCPSchw}
	Let $(g,M_{(\Bm_0,0,0)})$ be a Schwarzschild spacetime with mass $\Bm_0$ and $\mathcal{P}_0=\delta_gG_g\delta_{g}^*$ be the wave operator acting on $1$-form. Define $\widehat{\mathcal{P}_0}(\sigma):=e^{i\sigma t_{0,*}}\mathcal{P}_0e^{-i\sigma t_{0,*}}$ with $t_{0,*}=(t+r_{0,*})\chi_0(r)+(t-r_{0,*})(1-\chi_0(r))$ and $ r_{0,*}=r+2\Bm_0\log(r-2\Bm_0)$. Then we have
	\begin{enumerate}
		\item If $\IM \sigma\geq0$ and $\sigma\neq 0$, then
		\begin{equation}
	\widehat{\mathcal{P}_0}(\sigma):\{\omega\in\eHb^{s, \ell}(\CX;\scform): \widehat{\mathcal{P}_0}(\sigma)\omega\in\eHb^{s,\ell+1}(\CX; \scform)\}\to \eHb^{s,\ell+1}(\CX;\scform)
		\end{equation}
		is invertible for $s>\frac 32, \ell<-\frac12, s+\ell>-\frac 12$.\\
		\item If $s>\frac 32$ and $-\frac 32<\ell<-\frac 12$, then stationary operator
		\begin{equation}
		\widehat{\mathcal{P}_0}(0):\{\omega\in\eHb^{s, \ell}(\CX;\scform): 	\widehat{\mathcal{P}_0}(0)\omega\in\eHb^{s-1,\ell+2}(\CX; \scform)\}\to \eHb^{s-1,\ell+2}(\CX;\scform)
		\end{equation}
		has $1$-dimensional kernel and cokernel.
	\end{enumerate}
	More explicitly, with $v=t+r_{0,*}=t+r+2\Bm_0\log(r-2\Bm_0)$ we have
	\begin{equation}\label{eq:g1formmodeSchw}
		\begin{split}
			\ker\widehat{\mathcal{P}_0}(0)\cap \eHb^{\infty, -1/2-}(\CX; \scform)&=\langle\omega_{ s_0}\rangle,\quad 	\omega_{ s_0}=r^{-1}(dv-dr)\\
			\ker\widehat{\mathcal{P}_0}(0)^*\cap \sHb^{-\infty, -1/2-}(\CX; \scform)&=\langle\omega^*_{ s_0}\rangle,\quad \omega^*_{ s_0}=\delta(r-2\Bm_0)dr
		\end{split}.
	\end{equation} 
\end{prop}

\begin{rem}\label{rem:invertibilityofscalar1}
According to \eqref{eq:g1formmodeSchw}, $\widehat{\mathcal{P}_0}(0)$ restricted to $1$-forms of scalar type $l=1$ (and vector type $l=1$ respectively)
 is invertible. Since  $\widehat{\mathcal{W}_{b_0,\gamma}}(0)$ also maps $1$-forms of scalar type $l=1$ (and vector type $l=1$ respectively) to the same type and is a small perturbation of $\widehat{P_0}(0)$ for  $b_0=(\Bm_0, 0, \BQ_0)$ with $\abs{\BQ_0}\ll\Bm_0$ and $\gamma\ll1$, it follows that $\widehat{\mathcal{W}_{b_0,\gamma}}(0)^{-1}$ restricted to $1$-forms of scalar type $l=1$ (and $1$-forms of vector type $l=1$ respectively) exists with norm $\mathcal{O}(1)$.
\end{rem}

\subsection{Mode stability of the gauge propagation operator}
We first record the following two propositions about Schwarzschild metric which are needed in the course of the proof of mode stability of the gauge propagation operator $\mathcal{P}_{b,\gamma}$ for the weakly charged RN metric and also slowly rotating and weakly charged KN metric.
\begin{prop}[{\cite[Proposition 10.3]{HHV21}}]\label{prop:PzeroinSchw}
	Let $(g,M_{(\Bm_0,0,0)})$ be the Schwarzschild spacetime with mass $\Bm_0$ and $\mathcal{P}_{0,\gamma}=\delta_gG_g\tilde{\delta}_{g,\gamma}^*$ where $\tilde{\delta}^*_{g,\gamma}$ is defined as in \eqref{eq:modifieddeltaFirst} and \eqref{eq:defofBFirst} with $\mathfrak{c}=\tilde{\chi}(r)(dv-\mathfrak{b}dr), v=t+r_{0,*}=t+r+2\Bm_0\log(r-2\Bm_0)$ and $\mathfrak{b}\neq 1$. Then there exists $\gamma_0>0$ such that for fixed $\gamma$ with $0<\abs{\gamma}<\gamma_0$, the following holds for $\widehat{\mathcal{P}_{0,\gamma}}(\sigma):=e^{i\sigma t_{0,*}}\mathcal{P}_{0,\gamma}e^{-i\sigma t_{0,*}}$ with $t_{0,*}=(t+r_{0,*})\chi_0(r)+(t-r_{0,*})(1-\chi_0(r))$.
\begin{equation}
		\widehat{\mathcal{P}_{0,\gamma}}(0):\{\omega\in\eHb^{s, \ell}(\CX;\scform): 	\widehat{\mathcal{P}_{0,\gamma}}(0)\omega\in\eHb^{s-1,\ell+2}(\CX; \scform)\}\to \eHb^{s-1,\ell+2}(\CX;\scform)
\end{equation}
with $s>2, -\frac 32<\ell<-\frac 12$ is invertible.
\end{prop}
\begin{prop}[{\cite[Proposition 10.12]{HHV21}}]\label{prop:PnonzeroinSchw}
Let $\widehat{\mathcal{P}_{0,\gamma}}(\sigma)$ be defined as in Proposition \ref{prop:PzeroinSchw} with $\mathfrak{b}>1$, then we have for $\IM\sigma\geq 0, \sigma\neq 0$ and $s>2, \ell<-\frac 12, s+\ell>-\frac 12$
\begin{equation}
	\widehat{\mathcal{P}_{0,\gamma}}(\sigma):\{\omega\in\eHb^{s, \ell}(\CX;\scform): 	\widehat{\mathcal{P}_{0,\gamma}}(\sigma)\omega\in\eHb^{s,\ell+1}(\CX; \scform)\}\to \eHb^{s,\ell+1}(\CX;\scform)
\end{equation}
is invertible.
\end{prop}

\begin{rem}
	We note that $\delta^*_{g,\gamma}$ defined in \cite[\S10]{HHV21} is slightly different from that in this manuscript (\eqref{eq:modifieddeltaFirst} and \eqref{eq:defofBFirst}) in the way that the coefficient of the second term in $B$ in \cite{HHV21} (which is denoted by $E$ there) is $-1$ instead of $-1/2$. However, this difference has no influence on the conclusions stated in the above two propositions. In particular, it does not affect the calculation in \cite[equation (10.5)]{HHV21}.
\end{rem}
Next we shall prove the analogues of the above two propositions for RN and slowly rotating KN metric with small charge by a perturbation argument.

We define $\mathfrak{c}$ as in \eqref{eq:defofc} with $\mathfrak{b}>1$. With $\gamma_0$ as in Propositions \ref{prop:PzeroinSchw} and \ref{prop:PnonzeroinSchw}, we fix $\gamma\in(0,\gamma_0)$.
\begin{thm}\label{thm:modesforCP}
	Let $b_0=(\Bm_0, 0,\BQ_0)$ and $\mathfrak{b}>1$. Fix $\gamma\in(0,\gamma_0)$, there exists a small constant $C(\gamma)>0$ such that for the weakly charged RN metric $g_{b_0}$ with $\abs{\BQ_0}<C(\gamma)$, the following holds.
	\begin{enumerate}
		\item If $\IM \sigma\geq0$ and $\sigma\neq 0$, 
		\begin{equation}\label{eq:CPmodenonzero}
			\widehat{\mathcal{P}_{b_0,\gamma}}(\sigma):\{\omega\in\eHb^{s, \ell}(\CX;\scform): 	\widehat{\mathcal{P}_{b_0,\gamma}}(\sigma)\omega\in\eHb^{s,\ell+1}(\CX; \scform)\}\to \eHb^{s,\ell+1}(\CX;\scform)
		\end{equation}
		is invertible when $s>2, \ell<-\frac12, s+\ell>-\frac 12$.\\
		\item  If $s>2$ and $-\frac 32<\ell<-\frac 12$, then stationary operator
		\begin{equation}\label{eq:CPmodezero}
			\widehat{\mathcal{P}_{b_0,\gamma}}(0):\{\omega\in\eHb^{s, \ell}(\CX;\scform): 	\widehat{\mathcal{P}_{b_0,\gamma}}(0)\omega\in\eHb^{s-1,\ell+2}(\CX; \scform)\}\to \eHb^{s-1,\ell+2}(\CX;\scform)
		\end{equation}
		is invertible.
	\end{enumerate}
Both statements also hold for the weakly charged and slowly rotating KN metric $g_b$ with $b=(\Bm, \Ba,\BQ)$ near $b_0=(\Bm_0, 0,\BQ_0)$.		
\end{thm}
\begin{proof}
	We note that a perturbation argument as in the final step in the proof of Theorem \ref{thm:modesforscalarwave} allows us to extend Propositions \ref{prop:PzeroinSchw} and \ref{prop:PnonzeroinSchw} to weakly charged RN metrics, which however is a smooth family of Lorentzian metric on the fixed manifold $M_{(\Bm_0, 0,0)}$ and is a pull back of the RN metric $g_{b_0}$ defined in \eqref{EqRNMetricNull}. We need to prove that Propositions \ref{prop:PzeroinSchw} and \ref{prop:PnonzeroinSchw} in fact hold for $g_{b_0}$ with sufficiently small charge as well. Specifically, let $(g, M_{(\Bm_0, 0,0)})$ be the Schwarzschild spacetime
	\[
	g=-(1-\frac{2\Bm_0}{r})dv^2+2dv dr+r^2\sg,\quad
	M_{(\Bm_0, 0,0)}=\BR_{v}\times[r_-, \infty)\times\BS^2
	\]
	where $r_-\in(0,2\Bm_0)$ is a fixed number and $v=t+r_{0,*}=t+r+2\Bm_0\log(r-2\Bm_0)$. We introduce another coordinates for $(g, M_{(\Bm_0, 0,0)})$: $(\tilde{t}_{\chi_0}, r,\theta,\phi)$ where $\tilde{t}_{\chi_0}=t+\int_{4\Bm_0}^r s^{-2}(s^2-2\Bm_0 s)\chi_0(s)\,ds$. Let $g_{b_0}$ be the RN metric as defined in \eqref{EqRNMetricNull}. We then define $\tilde{g}_{b_0}=(\Psi_{b_0})^*g_{b_0}$ as the pull back of $g_{b_0}$ under a diffeomorphism $\Psi_{b_0}$
	\[
	\Psi_{b_0}: M_{(\Bm, 0,0)}\to M, \quad (t_{\chi_0}, r, \theta ,\phi)(\Psi_{b_0}(p))=(\tilde{t}_{\chi_0}, r,\theta, \phi)(p).
	\]
	Then we have
	\begin{align}\label{eq:defofg_b}
		g_{b_0}&=-\weight dt_{\chi_0}^2+2\chi_0 dt_{\chi_0} dr+\frac{1-\chi_0^2}{\weight}dr^2+r^2\sg,\\\label{eq:defoftildeg_b}
		\tilde{g}_{b_0}&=-\weight d\tilde{t}_{\chi_0}^2+2\chi_0 d\tilde{t}_{\chi_0} dr+\frac{1-\chi_0^2}{\weight}dr^2+r^2\sg.
	\end{align}
We now define 
\[
\mathcal{P}_{\tilde{g}_{b_0},\gamma}:=\delta_{\tilde{g}_{b_0}}G_{\tilde{g}_{b_0}}\delta^*_{\tilde{g}_{b_0},\gamma}
\]
with $\mathfrak{c}=\tilde{\chi}(r)(dv-\mathfrak{b}dr)$ and $v=t+r_*$ in the definition of $\delta^*_{\tilde{g}_{b_0},\gamma}$. In view of \eqref{eq:defofg_b} and \eqref{eq:defoftildeg_b}, we find that $\mathcal{P}_{b_0,\gamma}$ can be obtained by replacing $\pa_{\tilde{t}_\chi}$ in $\mathcal{P}_{\tilde{g}_{b_0},\gamma}$ by $\pa_{t_\chi}$. 
Since 
\begin{align*}
\widehat{\mathcal{P}_{b_0,\gamma}}(\sigma)&=e^{i\sigma t_{b_0,*}}\mathcal{P}_{b_0,\gamma}e^{-i\sigma t_{b_0,*}},\quad t_{b_0,*}=(t+r_{b_0,*})\chi_0(r)+(t-r_{b_0,*})(1-\chi_0(r)), \\
\widehat{\mathcal{P}_{\tilde{g}_{b_0},\gamma}}(\sigma)&=e^{i\sigma \tilde{t}_{b_0,*}}\mathcal{P}_{\tilde{g}_{b_0},\gamma}e^{-i\sigma \tilde{t}_{b_0,*}},\quad \tilde{t}_{b_0,*}=(t+r_{0,*})\chi_0(r)+(t-r_{b_0,*})(1-\chi_0(r)), 
\end{align*}
we rewrite
\begin{align*}
\widehat{\mathcal{P}_{b_0,\gamma}}(\sigma)&=	e^{i\sigma (t_{\chi_0}+F_1(r))}\mathcal{P}_{b_0,\gamma}e^{-i\sigma (t_{\chi_0}+F_1(r))},\\
\widehat{\mathcal{P}_{\tilde{g}_{b_0},\gamma}}(\sigma)&=	e^{i\sigma (\tilde{t}_{\chi_0}+F_2(r))}\mathcal{P}_{\tilde{g}_{b_0},\gamma}e^{-i\sigma (\tilde{t}_{\chi_0}+F_2(r))}.
	\end{align*}
Therefore \[
\widehat{\mathcal{P}_{b_0,\gamma}}(\sigma)=e^{i\sigma(F_1(r)-F_2(r))}\widehat{\mathcal{P}_{\tilde{g}_{b_0},\gamma}}(\sigma)e^{-i\sigma(F_1(r)-F_2(r))}
\]
where $F_1(r)-F_2(r)\in C_c^\infty((2\Bm_0, \infty))$. As discussed at the beginning, Propositions \ref{prop:PzeroinSchw} and \ref{prop:PnonzeroinSchw} hold for $\widehat{\mathcal{P}_{\tilde{g}_{b_0},\gamma}}(\sigma)$, and thus for $\widehat{\mathcal{P}_{b_0,\gamma}}(\sigma)$ as well with $\abs{\BQ_0}<C(\gamma)$ where $C(\gamma)$ is a sufficiently small constant. 

Finally  it remains to analyze the weakly charged and slowly rotating KN case.  It follows from a perturbation argument as in the final step in the proof of Theorem \ref{thm:modesforscalarwave}.
\end{proof}

\subsection{Mode stability of the gauge potential wave operator}
Now we turn to the analysis of the gauge potential wave operator $\mathcal{W}_{b,\gamma}=\wt{\delta}_{g_b,\gamma}G_{g_b}\delta_{g_b}^*$. We will first prove the analogues of Propositions \ref{prop:PzeroinSchw} and \ref{prop:PnonzeroinSchw} for $\mathcal{W}_{0,\gamma}$. We note that the proofs closely follow \cite[Proposition 10.3, 10.12]{HHV21}. However, we present the proofs in detail for completeness here. Then proceeding as in Theorem \ref{thm:modesforCP}, we can obtain its counterpart for $\mathcal{W}_{b,\gamma}$.

\begin{prop}\label{prop:WzeroinSchw}
 Let $(g,M_{(\Bm_0,0,0)})$ be the Schwarzschild spacetime with mass $\Bm_0$ and $\mathcal{W}_{0,\gamma}=\tilde{\delta}_{g,\gamma}G_g\delta_{g}^*$ where $\tilde{\delta}_{g,\gamma}$ is defined as in \eqref{eq:modifieddeltaFirst} and \eqref{eq:defofFFirst} with $\mathfrak{c}=\tilde{\chi}(r)(dv-\mathfrak{b}dr), v=t+r_{0,*}=t+r+2\Bm_0\log(r-2\Bm_0)$ and $\mathfrak{b}\neq 1$. Then there exists $\gamma_0>0$ such that for fixed $\gamma$ with $0<\abs{\gamma}<\gamma_0$, the following holds for $\widehat{\mathcal{W}_{0,\gamma}}(\sigma):=e^{i\sigma t_{0,*}}\mathcal{W}_{0,\gamma}e^{-i\sigma t_{0,*}}$ with $t_{0,*}=(t+r_{0,*})\chi_0(r)+(t-r_{0,*})(1-\chi_0(r))$.
	\begin{equation}
		\widehat{\mathcal{W}_{0,\gamma}}(0):\{\omega\in\eHb^{s, \ell}(\CX;\scform): 	\widehat{\mathcal{W}_{0,\gamma}}(0)\omega\in\eHb^{s-1,\ell+2}(\CX)\}\to \eHb^{s-1,\ell+2}(\CX)
	\end{equation}
	with $s>2, -\frac 32<\ell<-\frac 12$ is invertible.
\end{prop}
\begin{proof}
Fix $-\frac 32<\ell<-\frac 12, s>2$ and let
\[
\mathcal{X}^{s,\ell}:=\{u\in\eHb^{s,\ell}(\CX):\widehat{\mathcal{W}_{0,\gamma}}(0)u\in\eHb^{s-1,\ell+2}(\CX)\}.
\]	
In view of Lemma \ref{lem:indexofFredholmoperatorBundle}, we know that $\widehat{\mathcal{W}_{0,\gamma}}(0):\mathcal{X}^{s,\ell}\to\eHb^{s-1,\ell+2}(\CX)$ is Fredholm of index $0$ when $\gamma$ is sufficiently small. As $\widehat{\mathcal{W}_{0,\gamma_1}}(0)-\widehat{\mathcal{W}_{0,\gamma_2}}(0)$ is a first-order differential operator with compactly supported coefficients, the space $\mathcal{X}^{s,\ell}$ does not depend on $\gamma$.

We split the domain $\mathcal{X}^{s,\ell}$ and the target space $\eHb^{s,\ell}(\CX)$ as follows
\begin{align*}
	\mathcal{X}^{s,\ell}&=\mathcal{K}^{\perp}\oplus \mathcal{K},\quad \mathcal{K}=\ker\widehat{\mathcal{P}_0}(0)\cap\mathcal{X}^{s,\ell}=\langle \omega_{s_0}\rangle\\
	\eHb^{s,\ell}(\CX)&=\mathcal{R}\oplus \mathcal{R}^\perp,\quad \mathcal{R}=\Ran_{\mathcal{X}^{s,\ell}}\widehat{\mathcal{P}_0}(0)=\mbox{ann }\omega^*_{s_0}
	\end{align*}
where ann means annihilator, $\mathcal{K}^\perp$ is a complementary subspace to $\mathcal{K}$ inside $\mathcal{X}^{s,\ell}$ while $\mathcal{R}^\perp=\langle \eta\rangle$ is complementary to $\mathcal{R}$ inside $\eHb^{s-1,\ell+2}$ (in other word, $\mathcal{R}^\perp$ is the cokernel of $\widehat{\mathcal{P}_0}(0)$), and $ \omega_0, \omega_0^*$ are defined as in \eqref{eq:g1formmodeSchw}. We may choose $\eta\in C_c^\infty(X;\scform)$ satisfying $\angles{\eta}{\omega_{s_0}^*}=1$. Then the operator $\widehat{\mathcal{W}_0}(0)$ can be rewritten as
\begin{equation}\label{decomofW}
	\widehat{\mathcal{W}_{0,\gamma}}(0)=\begin{pmatrix}
		\mathcal{W}_{00}+\gamma\mathcal{W}^\flat_{00}&\gamma\mathcal{W}^\flat_{01}\\
		\gamma\mathcal{W}^\flat_{10}&\gamma\mathcal{W}^\flat_{11}
	\end{pmatrix}.
\end{equation}
We notice that $\mathcal{W}_{00}=\widehat{\mathcal{P}_0}(0)|_{\mathcal{K}^\perp}:\mathcal{K}^\perp\to \mathcal{R}$ is invertible, and so is $\mathcal{W}_{00}+\gamma\mathcal{W}^\flat_{00}$ provided that $\gamma$ is small enough. Moreover, if we identify $\mathcal{K}\cong\BC$ via $c\omega_{s_0}\mapsto c$ and $\mathcal{R}^\perp\cong\BC$ via $c\eta\mapsto \angles{c\eta}{\omega_{s_0}^*}=c$, correspondingly, $\mathcal{W}^\flat_{11}$ can be identified as an endmorphism of $\BC$, and thus is simply a number which can be computed as follows.
\begin{equation}\label{eq:calofW_11flat}
	\begin{split}
	\mathcal{W}_{11}^\flat&=\gamma^{-1}\angles{\bigl(\widehat{\mathcal{W}_{0,\gamma}}(0)-\widehat{\mathcal{P}_{0}}(0)\bigr)\omega_{s_0}}{\omega_{s_0}^*}\\
	&=\gamma^{-1}\angles{\bigl(F_gG_{g}\delta_g^*\bigr)\omega_{s_0}}{\omega_{s_0}^*}=\gamma^{-1}\angles{\delta_g^*\omega_{s_0}}{G_gB_g\omega_{s_0}^*}\\
	&=\angles{\delta_g^*\omega_{s_0}}{\big(2\mathfrak{c}\otimes_s\omega_{s_0}^*-\frac 12 g^{-1}(\mathfrak{c},\omega_{s_0}^*)g\big)}\\
	&=4\pi(\mathfrak{b}-1)
	\end{split}
\end{equation}
where we use $g=-(1-\frac{2\Bm_0}{r})dv^2+2dvdr+r^2\sg$, $2\mathfrak{c}\otimes_s\omega^*_{s_0}=\delta(r-2\Bm_0)\tilde{\chi}(r)(2dvdr-2\mathfrak{b}dr^2)$, $g^{-1}(\mathfrak{c},\omega_{s_0}^*)=\delta(r-2\Bm_0)\tilde{\chi}(r)(1-\mathfrak{b}(1-\frac{2\Bm_0}{r}))$ and $\delta_g^*\omega_{s_0}=-\frac{2\Bm_0^2}{r^4}dv^2-2(\frac{1}{2r^2}+\frac{\Bm_0}{r^3})dvdr+\frac{1}{r^2}dr^2-(1-\frac{2\Bm_0}{r})\sg$ in the last step.

Now suppose $(\omega_0,\omega_1)\in(\mathcal{K}^\perp\oplus\mathcal{K})\cap\ker	\widehat{\mathcal{W}_{0,\gamma}}(0)$, then we have 
\[
\omega_0=-\gamma(\mathcal{W}_{00}+\gamma\mathcal{W}_{00}^\flat)^{-1}\mathcal{W}_{01}^\flat\omega_1
\]
and  thus
\[
\Big(\mathcal{W}_{11}^\flat-\gamma\mathcal{W}_{10}^\flat(\mathcal{W}_{00}+\gamma\mathcal{W}_{00}^\flat)^{-1}\mathcal{W}_{01}^\flat\Big)\omega_1=0.
\]
if $\gamma$ is nonzero. When $\mathfrak{b}\neq 1$, $\mathcal{W}_{11}^\flat$ is invertible, and thus so is $\mathcal{W}_{11}^\flat-\gamma\mathcal{W}_{10}^\flat(\mathcal{W}_{00}+\gamma\mathcal{W}_{00}^\flat)^{-1}\mathcal{W}_{01}^\flat$ if $\gamma$ is sufficiently small. So we must have $\omega_1=0$ and thus $\omega_0=0$. This proves the injectivity of $\widehat{\mathcal{W}_{0,\gamma}}(0)$. The surjectivity follows from the fact that $\widehat{\mathcal{W}_{0,\gamma}}(0)$ is Fredholm of index $0$.
\end{proof}

As explained in \eqref{eq:eqnforgaugepotential} and \eqref{eq:eqfordualpotential}, $\mathcal{W}_{b,\gamma}$ acts as the gauge potential wave operator. In order to obtain the mode stability of $L_{g_b,\gamma}(\dg,\dA)=0$ for $\IM\sigma\geq 0, \sigma\neq 0$, we need to  prove that for the operator $\mathcal{W}_{b,\gamma}$. In other words, we need to show that for suitable choices of $\mathfrak{b}$ and small $\gamma$, $\mathcal{W}_{b,\gamma}$ has no modes with $\sigma\neq 0$ in the closed upper half plane. We first prove this for Schwarzschild metric, and then the RN and slowly rotating KN metric with small charge follows from a perturbation argument as in Theorem \ref{thm:modesforCP}.

\begin{prop}\label{prop:WnonzeroinSchw}
	Let $\widehat{\mathcal{W}_{0,\gamma}}(\sigma)$ be defined as in Proposition \ref{prop:WzeroinSchw} with $\mathfrak{b}>1$, then there exists $\tilde{\gamma}_0>0$ such that for $0<\gamma<\tilde{\gamma}_0$, $\IM\sigma\geq 0, \sigma\neq 0$ and $s>2, \ell<-\frac 12, s+\ell>-\frac 12$
	\begin{equation}\label{eq:WnonzeroinSchw}
		\widehat{\mathcal{W}_{0,\gamma}}(\sigma):\{\omega\in\eHb^{s, \ell}(\CX;\scform): 	\widehat{\mathcal{W}_{0,\gamma}}(\sigma)\omega\in\eHb^{s,\ell+1}(\CX; \scform)\}\to \eHb^{s,\ell+1}(\CX;\scform)
	\end{equation}
	is invertible.
\end{prop}
In the course of the proof of Proposition \ref{prop:WnonzeroinSchw}, we closely follow the arguments in \cite[\S10.2]{HHV21}. As explained there, the proof consists of two steps:
\begin{enumerate}
	\item We show in Lemma \ref{lem:step1} that for $\mathfrak{b}>1$ and for \textit{all} sufficiently small $\gamma$, $\widehat{\mathcal{W}_{0,\gamma}}(\sigma)$ has no modes in a \textit{fixed} neighborhood of $\sigma=0$ in the closed upper half plane.
	\item In the proof of Proposition \ref{prop:WnonzeroinSchw}, we proceed as in the final step in the proof of Theorem \ref{thm:modesforscalarwave} to prove the mode stability of $\widehat{\mathcal{W}_{0,\gamma}}(\sigma)$ in the closed upper half plane.
\end{enumerate}

Fix $s>2, -\frac 32<\ell<-\frac12$, the domains
\begin{equation}
	\mathcal{X}^{s,\ell}(\sigma):=\{\omega\in\eHb^{s,\ell}(\CX): \widehat{\mathcal{W}_{0,\gamma}}(\sigma)\omega\in\eHb^{s-1,\ell+2}(\CX)\}
\end{equation}
of the operators $\widehat{\mathcal{W}_{0,\gamma}}(\sigma):\mathcal{X}^{s,\ell}(\sigma)\to\eHb^{s-1,\ell+2}$ now \textit{depend} on $\sigma$ (but still independent of $\gamma$). As a consequence, we need the following technical lemma which allows us to pass to operators with fixed domain and target space.

\begin{lem}\label{lem:prestep1}
Let $s>2, -\frac 32<\ell<-\frac12$ and $\mathfrak{b}\neq 1$ and fix $\gamma_1$ satisfying $0<\abs{\gamma_1}<\gamma_0$ where $\gamma_0$ is as in Proposition \ref{prop:WzeroinSchw}. Then $\widehat{\mathcal{W}_{0,\gamma_1}}(\sigma):\mathcal{X}^{s,\ell}(\sigma)\to\eHb^{s-1,\ell+2}(\CX)$ is invertible for $\sigma\in\BC$ with $\IM\sigma\geq 0, \abs{\sigma}<c$ where $c$ is some small constant. Moreover, $\widehat{\mathcal{W}_{0,\gamma_1}}(\sigma)^{-1}$ is continuous in $\sigma$ with values in $\mathcal{L}_{\mathrm{weak}}(\eHb^{s-1,\ell+2}(\CX), \eHb^{s,\ell}(\CX))$ (the space of linear bounded operators equipped with weak operator topology), while continuous in $\sigma$ with values in $\mathcal{L}_{\mathrm{op}}(\eHb^{s-1+\epsilon,\ell+2+\epsilon}(\CX), \eHb^{s-\epsilon,\ell-\epsilon}(\CX))$ (norm topology) for any $\epsilon>0$.
\end{lem}

\begin{proof}
	By Theorem \ref{thm:FredholmestimateBundle}, we have the uniform Fredholm estimates for $\omega\in\mathcal{X}^{s,\ell}(\sigma)$
	\begin{equation}
		\norm{\omega}_{\eHb^{s,\ell}(\CX)}\leq C\Bigl(\norm{\widehat{\mathcal{W}_{0,\gamma_1}}(\sigma)\omega}_{\eHb^{s-1,\ell+2}(\CX)}+\norm{\omega}_{\eHb^{s_0,\ell_0}(\CX)}\Bigr)
	\end{equation}
for $\IM\sigma\geq 0$ with $\abs{\sigma}$ being small and $s_0<s,\ell_0<\ell$. We now follow the arguments in \cite[Propsition 4.4]{Vas21a} to show  that the compact error term $\norm{\omega}_{\eHb^{s_0,\ell_0}(\CX)}$ can be dropped (while we may need to take larger constant $C$) if $\abs{\sigma}$ is sufficiently small. Concretely, suppose for the sake of contradiction that there exists a sequence $\delta_j\to0$ and a sequence $\omega_j$ with $\norm{\omega_j}_{\eHb^{s,\ell}(\CX)}=1$ and $\widehat{\mathcal{W}_{0,\gamma_1}}(\sigma_j)\omega_j\in\eHb^{s-1,\ell+2}(\CX)$ such that $1=\norm{\omega_j}_{\eHb^{s,\ell}(\CX)}\geq j\norm{\widehat{\mathcal{W}_{0,\gamma_1}}(\sigma_j)\omega_j}_{\eHb^{s-1,\ell+2}(\CX)}$, then $\widehat{\mathcal{W}_{0,\gamma_1}}(\sigma_j)\omega_j\to 0$ in $\eHb^{s-1,\ell+2}(\CX)$ and $\liminf_{j\to\infty}\norm{\omega_j}_{\eHb^{s_0,\ell_0}(\CX)}\geq C^{-1}>0$. As a consequence, there exists a subsequence $\omega_j\to \omega$ weakly in $\eHb^{s, \ell}(\CX)$ and strongly in $\eHb^{s_0, \ell_0}(\CX)$ with the limit $\omega$ being non-zero. Moreover,  
\[
\widehat{\mathcal{W}_{0,\gamma_1}}(\sigma_j)\omega_j-\widehat{\mathcal{W}_{0,\gamma_1}}(0)\omega=\bigl(\widehat{\mathcal{W}_{0,\gamma_1}}(\sigma_j)-\widehat{\mathcal{W}_{0,\gamma_1}}(0)\bigr)\omega_j+\widehat{\mathcal{W}_{0,\gamma_1}}(0)\bigl(\omega_j-\omega\bigr)\to 0\quad\mbox{in}\quad \eHb^{s_0-2, \ell_0}(\CX)
\]
since $\widehat{\mathcal{W}_{0,\gamma_1}}(\sigma_j)\to\widehat{\mathcal{W}_{0,\gamma_1}}(0)$ as bounded operators in $\mathcal{L}(\eHb^{s_0, \ell_0}(\CX), \eHb^{s_0-2, \ell_0}(\CX))$ and $\omega_j\to\omega$ in $\eHb^{s_0, \ell_0}(\CX)$. Therefore, $\widehat{\mathcal{W}_{0,\gamma_1}}(0)\omega=0$. But this contradicts Proposition \ref{prop:WzeroinSchw}, and thus proves the injectivity of $\widehat{\mathcal{W}_{0,\gamma_1}}(\sigma)$ for $\sigma$ with small $\abs{\sigma}$ in the closed upper half plane. Since $\widehat{\mathcal{W}_{0,\gamma_1}}(\sigma)$ has index $0$, it follows that $\widehat{\mathcal{W}_{0,\gamma_1}}(\sigma)^{-1}$ exists with a uniform bound in a small neighborhood of $\sigma=0$ in the closed upper half plane.

Next we prove the sequential continuity of $\widehat{\mathcal{W}_{0,\gamma_1}}(\sigma)^{-1}$ in $\sigma$ (which is equivalent to continuity). Suppose $\delta_j\to\delta$, one need to show $\widehat{\mathcal{W}_{0,\gamma_1}}(\sigma_j)^{-1}\to\widehat{\mathcal{W}_{0,\gamma_1}}(\sigma)^{-1}$ in weak operator topology of $\mathcal{L}(\eHb^{s-1,\ell+2}(\CX), \eHb^{s,\ell}(\CX))$. Suppose $\{f_j\}$ is a sequence in $\eHb^{s-1,\ell+2}$ which is norm converging to $f$, let $\omega_j=\widehat{\mathcal{W}_{0,\gamma_1}}(\sigma_j)^{-1}f_j$ and we shall show that $\omega_j$ converges weakly to $\widehat{\mathcal{W}_{0,\gamma_1}}(\sigma)^{-1}f$ in $\eHb^{s,\ell}(\CX)$. In view of the uniform bound of $\widehat{\mathcal{W}_{0,\gamma_1}}(\sigma_j)^{-1}$ proved in the first step above, we see that $\omega_j$ is uniformly bounded in $\eHb^{s,\ell}(\CX)$. Therefore, there exists a subsequence $\omega_{j_k}\to \omega$ weakly in $\eHb^{s,\ell}(\CX)$, then $f_{j_k}= \widehat{\mathcal{W}_{0,\gamma_1}}(\sigma_{j_k})\omega_{j_k}\to \widehat{\mathcal{W}_{0,\gamma_1}}(\sigma)\omega$ weakly in $\eHb^{s-2,\ell}(\CX)$. Meanwhile by assumption $f_{j_k}\to f$ strongly in $\eHb^{s-1,\ell+2}(\CX)$ and thus $\widehat{\mathcal{W}_{0,\gamma_1}}(\sigma)\omega=f$. This implies that $\omega$ is uniquely determined by $f$ and independent of the subsequence. Namely, every subsequence of $\omega_j$ has a subsequence converging weakly to the same $\omega$. This means that the full sequence $\omega_j\to\omega=\widehat{\mathcal{W}_{0,\gamma_1}}(\sigma_j)^{-1}f$ weakly in $\eHb^{s,\ell}(\CX)$ and thus proves the continuity in the weak operator topology (Otherwise, one can find $g\in\sHb^{-s,-\ell}(\CX)), \delta>0$ and a subsequence $\{\omega_{j_k}\}\subset\{w_j\}$ such that $\abs{g(\omega_{j_k})-g(\omega)}\geq \delta$. However, by assumption $\{\omega_{j_k}\}$ should have a subsequence converging weakly to $\omega$, which is impossible).

As for the continuity in the norm topology, we shall prove it by contradiction. Suppose there exists a $\delta>0$ and a sequence $\sigma_j\to\sigma$ such that $\norm{\widehat{\mathcal{W}_{0,\gamma_1}}(\sigma_j)^{-1}-\widehat{\mathcal{W}_{0,\gamma_1}}(\sigma)^{-1}}_{\mathcal{L}(\eHb^{s-1+\epsilon,\ell+2+\epsilon}(\CX), \eHb^{s-\epsilon,\ell-\epsilon}(\CX))}\geq \delta$, then one can find a bounded sequence $f_j\in\eHb^{s-1+\epsilon,\ell+2+\epsilon}$ such that
\begin{equation}\label{eq:nonormconti}
\norm{\widehat{\mathcal{W}_{0,\gamma_1}}(\sigma_j)^{-1}f_j-\widehat{\mathcal{W}_{0,\gamma_1}}(\sigma)^{-1}f_j}_{ \eHb^{s-\epsilon,\ell-\epsilon}(\CX)}\geq \delta.
\end{equation}
Passing to a subsequence, we may assume that $f_j\to f$ weakly in $\eHb^{s-1+\epsilon,\ell+2+\epsilon}(\CX)$ and strongly in $\eHb^{s-1,\ell+2}(\CX)$. Then it follows that  $\widehat{\mathcal{W}_{0,\gamma_1}}(\sigma)^{-1}f_j\to \widehat{\mathcal{W}_{0,\gamma_1}}(\sigma)^{-1}f$ in $\eHb^{s,\ell}(\CX)$. However, by the continuity in the weak operator topology, we see that $\widehat{\mathcal{W}_{0,\gamma_1}}(\sigma_j)^{-1}f_j\to \widehat{\mathcal{W}_{0,\gamma_1}}(\sigma)^{-1}f$ weakly in $\eHb^{s,\ell}(\CX)$ and thus (passing to a subsequence) strongly in $\eHb^{s-\epsilon,\ell-\epsilon}(\CX)$, which contradicts \eqref{eq:nonormconti}. This finishes the proof of the continuity in the norm topology.
\end{proof}

With the help of the invertibility of $\widehat{\mathcal{W}_{0,\gamma_1}}(\sigma)$, the analysis of $\widehat{\mathcal{W}_{0,\gamma}}(\sigma)$ can be converted to that of another operator $ \widehat{\mathcal{W}_{0,\gamma}}(\sigma)\widehat{\mathcal{W}_{0,\gamma_1}}(\sigma)^{-1}:\eHb^{s-1,\ell+2}(\CX)\to\eHb^{s-1,\ell+2}(\CX)$ with fixed domain and target space independent of $\sigma$. This will enable us to prove the invertibility of $\widehat{\mathcal{W}_{0,\gamma}}(\sigma)$  for \textit{all} sufficiently small  $\gamma>0$ in a fixed small neighborhood of $\sigma=0$ in the closed upper half plane.

\begin{lem}\label{lem:step1}
	Let $s>2, -\frac 32<\ell<-\frac 12$ and $\mathfrak{b}>1$. Then there exists $C_0>0$ such that for all $\gamma>0$ and $\sigma\in\BC, \IM\sigma\geq 0$ with $\gamma+\abs{\sigma}<C_0$, the operator $\widehat{\mathcal{W}_{0,\gamma_1}}(\sigma):\mathcal{X}^{s,\ell}(\sigma)\to\eHb^{s-1,\ell+2}(\CX)$ is invertible.
\end{lem}
\begin{proof}
Let
	\[
	\widehat{\mathcal{W}_{0,\gamma}}(\sigma)\widehat{\mathcal{W}_{0,\gamma_1}}(\sigma)^{-1}:\eHb^{s-1,\ell+2}(\CX)\to\eHb^{s-1,\ell+2}(\CX),
	\]
	then it suffices to prove the injectivity of $	\widehat{\mathcal{W}_{0,\gamma}}(\sigma)\widehat{\mathcal{W}_{0,\gamma_1}}(\sigma)^{-1}$ provided that $\mathfrak{b}>1$ and $\gamma>0, \IM\sigma\geq 0$ with $\gamma+\abs{\sigma}<C_0$ where $C_0$ is a small constant. To this end, we split the domains and the target spaces as follows
	\begin{align*}
	\mbox{domain: }\eHb^{s-1,\ell+2}(\CX)&=\mathcal{K}^\perp\oplus\mathcal{K},\quad \mathcal{K}=\ker	\widehat{\mathcal{P}_{0}}(0)\widehat{\mathcal{W}_{0,\gamma_1}}(0)^{-1}\cap\eHb^{s-1,\ell+2}(\CX)=\langle\tilde{\omega}\rangle\quad\!\!\!\mbox{with}\!\!\quad\! \tilde{\omega}=\widehat{\mathcal{W}_{0,\gamma_1}}(0)\omega_{s_0}\\
		\mbox{target: }\eHb^{s-1,\ell+2}(\CX)&=\mathcal{R}\oplus\mathcal{R}^\perp,\quad \mathcal{R}=\Ran_{\mathcal{X}^{s,\ell}(0)}\widehat{\mathcal{P}_{0}}(0)
	\end{align*}
	where $\mathcal{K}^\perp$ and $\mathcal{R}^\perp=\langle\eta\rangle$ are the complementary subspaces to $\mathcal{K}$ and $\mathcal{R}$ respectively inside $\eHb^{s-1,\ell+2}(\CX)$, and $\omega_{s_0}$ is defined as in \eqref{eq:g1formmodeSchw}. We notice that 
	\[
	\tilde{\omega}=\widehat{\mathcal{W}_{0,\gamma_1}}(0)\omega_{s_0}=\bigl(\widehat{\mathcal{W}_{0,\gamma_1}}(0)-\widehat{\mathcal{P}_{0}}(0)\bigr)\omega_{s_0}=\widehat{F_gG_{g}\delta_g}\omega_{s_0}\in C^\infty_c(X^\circ)
	\]
	and we can identify $\mathcal{R}^\perp\cong\BC$ via $c\eta\mapsto\angles{c\eta}{\omega_{s_0}^*}$ where $\omega_{s_0}^*$ is defined as in \eqref{eq:g1formmodeSchw}. Under these splittings, $	\widehat{\mathcal{W}_{0,\gamma}}(\sigma)\widehat{\mathcal{W}_{0,\gamma_1}}(\sigma)^{-1}$ takes the form
	\begin{equation}
		\widehat{\mathcal{W}_{0,\gamma}}(\sigma)\widehat{\mathcal{W}_{0,\gamma_1}}(\sigma)^{-1}=\begin{pmatrix}
			\mathcal{W}_{00}(\gamma, \sigma)&	\mathcal{W}_{01}(\gamma, \sigma)\\
				\mathcal{W}_{10}(\gamma, \sigma)&	\mathcal{W}_{11}(\gamma, \sigma)
		\end{pmatrix}.
		\end{equation}
	We first list several basic facts: $\mathcal{W}_{01}(0,0), \mathcal{W}_{10}(0,0), \mathcal{W}_{11}(0,0)$ are $0$, and $\mathcal{W}_{00}(0,0)$ is invertible. Since $	\widehat{\mathcal{W}_{0,\gamma}}(\sigma)\widehat{\mathcal{W}_{0,\gamma_1}}(\sigma)^{-1}\in\mathcal{L}(\eHb^{s-1,\ell+2}(\CX),\eHb^{s-1,\ell+2}(\CX))$ is Fredholm of index $0$ (see Lemma \ref{lem:indexofFredholmoperatorBundle}) for $\gamma+\abs{\sigma}$ small and the three components $(0,1),(1,0),(1,1)$ are compact operators (because they all have finite rank), we can conclude that $\mathcal{W}_{00}(\gamma, \sigma)$ is Fredholm of index $0$ as well for $\gamma+\abs{\sigma}$ small.
	
	Suppose $\omega=(\omega_0, \omega_1)\in(\mathcal{K}^\perp\oplus\mathcal{K})\cap\ker\widehat{\mathcal{W}_{0,\gamma}}(\sigma)\widehat{\mathcal{W}_{0,\gamma_1}}(\sigma)^{-1}$, we have $\mathcal{W}_{00}(\gamma, \sigma)\omega_0+\mathcal{W}_{01}(\gamma,\sigma)\omega_1=0$. Assume for the moment that $\mathcal{W}_{00}(\gamma, \sigma)$ is invertible for $\gamma+\abs{\sigma}$ small, then 
	\begin{equation}\label{eq:roughdis1}
	\omega_0=-\mathcal{W}_{00}(\gamma,\sigma)^{-1}\mathcal{W}_{01}(\gamma,\sigma)\omega_1
	\end{equation}
	and thus
	\begin{equation}\label{eq:roughdis2}
	\Bigl(\mathcal{W}_{11}(\gamma,\sigma)-\mathcal{W}_{10}(\gamma, \sigma)\mathcal{W}_{00}(\gamma,\sigma)^{-1}\mathcal{W}_{01}(\gamma,\sigma)\Bigr)\omega_1=0.	\end{equation}
	We want to show that $\omega_1=0$ and thus $\omega_0=0$ for $\gamma+\abs{\sigma}$.  small. Formally speaking, since $\mathcal{W}_{01}(0,0), \mathcal{W}_{10}(0,0)$ are $0$,  $\mathcal{W}_{01}(\gamma,\sigma), \mathcal{W}_{10}(\gamma, \sigma)$ are of size $\mathcal{O}(\gamma+\abs{\sigma})$. Therefore, it suffices to show that $\mathcal{W}_{11}(\gamma,\sigma)$ is nonzero modulo $\mathcal{O}(\gamma^2+\abs{\sigma}^2)$ for $\gamma+\abs{\sigma}$ small. To make these arguments rigorous, we need to prove the injectivity of $\mathcal{W}_{00}(\gamma,\sigma)$ (and thus the invertibility in view of its $0$ index property) for $\gamma+\abs{\sigma}$ small and the differentiability of $\mathcal{W}_{01}, \mathcal{W}_{10}$ at $(\gamma, \sigma)=(0,0)$, and calculate the Taylor expansion of $\mathcal{W}_{11}(\gamma, \sigma)$ at $(0,0)$ up to the first order.	\begin{itemize}
		\item \underline{Invertibility of $\mathcal{W}_{00}$.} We shall prove that for $\gamma+\abs{\sigma}$ small, there exists a uniform constant $C>0$ such that 
		\begin{equation}\label{eq:uniformnound}
			\norm{\mathcal{W}_{00}(\gamma, \sigma)^{-1}}_{\mathcal{R}\to\mathcal{K}^\perp}\leq C.
		\end{equation}
	The proof is similar to that in the first step of the proof of Lemma \ref{lem:prestep1}. Specifically, we define $\Pi:\eHb^{s-1,\ell+2}(\CX)\to\mathcal{R}$ to be the projection onto $\mathcal{R}$. Suppose \eqref{eq:uniformnound} is not true, one can find a sequence $(\gamma_j, \delta_j)\to(0,0)$ and a sequence $\{f_j\}\subset\mathcal{K}^\perp$ with $\norm{f_j}_{\eHb^{s-1,\ell+2}(\CX)}=1$ such that for $\omega_j=\widehat{\mathcal{W}_{0,\gamma_1}}(\sigma_j)^{-1}f_j$
	\[
	1=\norm{f_j}_{\eHb^{s-1,\ell+2}(\CX)}\geq j\norm{\Pi	\widehat{\mathcal{W}_{0,\gamma_j}}(\sigma_j)\omega_j}_{\eHb^{s-1,\ell+2}(\CX)}
	\]
	which implies $\Pi	\widehat{\mathcal{W}_{0,\gamma_j}}(\sigma_j)\omega_j\to 0$ in $\eHb^{s-1,\ell+2}(\CX)$. In view of Theorem \ref{thm:FredholmestimateBundle}, we have the following uniform Fredholm estimates for $\gamma+\abs{\sigma}$ small 
	\begin{equation}
		\begin{split}
		\norm{\omega_j}_{\eHb^{s,\ell}(\CX)}&\leq \tilde{C}\Bigl(\norm{\Pi	\widehat{\mathcal{W}_{0,\gamma_j}}(\sigma_j)\omega_j}_{\eHb^{s-1,\ell+2}(\CX)}+\norm{(I-\Pi)	\widehat{\mathcal{W}_{0,\gamma_j}}(\sigma_j)\omega_j}_{\eHb^{s-1,\ell+2}(\CX)}\\
		&\qquad\qquad+\norm{\omega_j}_{\eHb^{s_0,\ell_0}(\CX)}\Bigr).
		\end{split}
	\end{equation}
Using the identification $\mathcal{R}^\perp\cong \BC$ introduced previously, we calculate
\begin{equation}
	\begin{split}
&\norm{(I-\Pi)	\widehat{\mathcal{W}_{0,\gamma_j}}(\sigma_j)\omega_j}_{\eHb^{s-1,\ell+2}(\CX)}\sim\abs{\angles{	\widehat{\mathcal{W}_{0,\gamma_j}}(\sigma_j)\omega_j}{\omega_{s_0}^*}}=\abs{\angles{	\omega_j}{\widehat{\mathcal{W}_{0,\gamma_j}}(\sigma_j)^*\omega_{s_0}^*}}\\
	&\qquad=\abs{\angles{	\omega_j}{\Bigl(\widehat{\mathcal{W}_{0,\gamma_j}}(\sigma_j)^*-\widehat{\mathcal{P}_{0}}(0)^*\Bigr)\omega_{s_0}^*}}\to 0\quad\mbox{as}\quad (\gamma_j, \sigma_j)\to(0,0)
	\end{split}
	\end{equation}
where we use $\omega_{s_0}^*\in\sHb^{-\frac 12-,\infty}(\CX)$ in the first equality and $\widehat{\mathcal{P}_{0,\gamma_j}}(\sigma_j)^*-\widehat{\mathcal{P}_{0}}(0)^*\to 0$ as bounded operators $\mathcal{L}(\sHb^{1-s,-\ell-2}(\CX),\sHb^{-s,-\ell-1}(\CX))$ in the last step. Since $\omega_j=\widehat{\mathcal{W}_{0,\gamma_1}}(\sigma_j)^{-1}f_j$, we have $\norm{\omega_j}_{\eHb^{s,\ell}(\CX)}\sim 1$ and thus $\liminf_{j\to\infty}\norm{\omega_j}_{\eHb^{s_0,\ell_0}(\CX)}\geq \tilde{C}^{-1}\liminf_{j\to\infty}\norm{\omega_j}_{\eHb^{s,\ell}(\CX)}>0$. As a result, there exists a subsequence (not shown in the notation) $\omega_j\to\omega\neq 0$ weakly in $\eHb^{s,\ell}(\CX)$. Then $\widehat{\mathcal{W}_{0,\gamma_j}}(\sigma_j)\omega_j\to\widehat{\mathcal{W}_{0}}(0)\omega=\widehat{\mathcal{P}_{0}}(0)\omega$ weakly in $\eHb^{s-2,\ell}(\CX)$. Since $\widehat{\mathcal{W}_{0,\gamma_j}}(\sigma_j)\omega_j\to 0$ strongly in $\eHb^{s-1,\ell+2}(\CX)$, we obtain $\widehat{\mathcal{P}_{0}}(0)\omega=0$. 

Meanwhile, since $\norm{f_j}_{\eHb^{s-1,\ell+2}(\CX)}=1$, passing to a subsequence, we see that $f_j\to f$ weakly in $\eHb^{s-1,\ell+2}(\CX)$ and strongly in $\eHb^{s-1-\epsilon, \ell+2-\epsilon}(\CX)$. Owing to Lemma \ref{lem:prestep1} (the continuity in weak operator topology), $\omega_j=\widehat{\mathcal{W}_{0,\gamma_1}}(\sigma_j)^{-1}f_j\to \widehat{\mathcal{W}_{0,\gamma_1}}(0)^{-1}f$ weakly in $\eHb^{s-\epsilon, \ell-\epsilon}$. Therefore, we obtain $0\neq \omega=\widehat{\mathcal{W}_{0,\gamma_1}}(0)^{-1}f$, i.e., $0\neq f=\widehat{\mathcal{W}_{0,\gamma_1}}(0)\omega\in\mathcal{K}$. It is clear that one can find $g\in\sHb^{1-s,\ell-2}(\CX)$ such that $g(\mathcal{K}^\perp)=0$ while $g(\tilde{w})=1$ ($\mathcal{K}=\langle \tilde{\omega}\rangle$), but this contradicts the fact that $f_j\to f$ weakly, and thus proves \eqref{eq:uniformnound}. From \eqref{eq:uniformnound}, we conclude that for $\gamma+\abs{\sigma}$ small,  $\mathcal{W}_{00}$ is injective and then the invertibility follows from the fact that $\mathcal{W}_{00}$ has index $0$.

		\item \underline{Differentiability of $\mathcal{W}_{01}, \mathcal{W}_{10}$.} We define
		\begin{equation}
			\mathcal{W}_1(\gamma,\sigma):=\mathcal{W}_{01}(\gamma, \sigma)\oplus\mathcal{W}_{11}(\gamma, \sigma):\mathcal{K}=\langle \tilde{\omega}\rangle\to\eHb^{s-1,\ell+2}(\CX).
		\end{equation}
	We will show that $\mathcal{W}_1(\gamma, \sigma)$ is continuous for $\gamma+\abs{\sigma}$ small and differentiable at $(0,0)$.
	
We write
\begin{equation}\label{eq:splitofW1}
	\begin{split}
	\widehat{\mathcal{W}_{0,\gamma}}(\sigma)\widehat{\mathcal{W}_{0,\gamma_1}}(\sigma)^{-1}&=\Big(\widehat{\mathcal{W}_{0,\gamma}}(\sigma)-\widehat{\mathcal{W}_{0,\gamma_1}}(\sigma)+\widehat{\mathcal{W}_{0,\gamma_1}}(\sigma)\Big)\widehat{\mathcal{W}_{0,\gamma_1}}(\sigma)^{-1}\\
	&=\Big(\widehat{\mathcal{W}_{0,\gamma}}(\sigma)-\widehat{\mathcal{W}_{0,\gamma_1}}(\sigma)\Big)\widehat{\mathcal{W}_{0,\gamma_1}}(\sigma)^{-1}+I.
	\end{split}
\end{equation}
Then it suffices to analyze the first term on the right hand side of \eqref{eq:splitofW1}. For the continuity, since $\widehat{\mathcal{W}_{0,\gamma_1}}(\sigma)^{-1}\tilde{\omega}\in\eHb^{s,\ell}(\CX)$ depending on $\sigma$ continuously for any $s>2, -\frac 32<\ell<-\frac 12$ (because $\tilde{\omega}\in C^\infty_c(X^\circ)$ and $\widehat{\mathcal{W}_{0,\gamma_1}}(\sigma)^{-1}$ is continuous in $\sigma$ with value in $\mathcal{L}_{\mathrm{op}}(\eHb^{s-1+\epsilon, \ell+2+\epsilon}(\CX),\eHb^{s-\epsilon, \ell-\epsilon}(\CX))$) and  $\widehat{\mathcal{W}_{0,\gamma}}(\sigma)-\widehat{\mathcal{W}_{0,\gamma_1}}(\sigma)\in\mbox{Diff}^1_b$ with $C^\infty_c(X)$ coefficients and depends on $\gamma, \sigma$ smoothly, the continuity of $	\widehat{\mathcal{W}_{0,\gamma}}(\sigma)\widehat{\mathcal{W}_{0,\gamma_1}}(\sigma)^{-1}\tilde{\omega}\in C^\infty_c(X)$ in $(\gamma, \sigma)$ follows

Now we turn to the analysis of the differentiability at $(0,0)$. Since $\widehat{\mathcal{W}_{0,\gamma}}(\sigma)-\widehat{\mathcal{W}_{0,\gamma_1}}(\sigma)$ are polynomials in $\gamma, \sigma$ with coefficients being first order differential operators, it is certainly differentiable at $(0,0)$. So it suffices to prove that $\widehat{\mathcal{W}_{0,\gamma_1}}(\sigma)^{-1}\tilde{\omega}$ is differentiable at $\sigma=0$. Now we claim that if $\widehat{\mathcal{W}_{0,\gamma_1}}(0)^{-1}\omega\in\rho C^\infty(\CX)+\eHb^{\infty, 1/2-}(\CX)$, then the following resolvent identity holds
\begin{equation}\label{eq:resolventidentity}
	\Bigl(\widehat{\mathcal{W}_{0,\gamma_1}}(\sigma)^{-1}-\widehat{\mathcal{W}_{0,\gamma_1}}(0)^{-1}\Bigr)\omega=\widehat{\mathcal{W}_{0,\gamma_1}}(\sigma)^{-1}\Bigl(\widehat{\mathcal{W}_{0,\gamma_1}}(0)-\widehat{\mathcal{W}_{0,\gamma_1}}(\sigma)\Bigr)\widehat{\mathcal{W}_{0,\gamma_1}}(0)^{-1}\omega.
\end{equation}
The proof of the above resolvent identity closely follows \cite[\S 6]{Vas21b} and uses a regularization argument. Concretely, let $\chi_\epsilon(\rho)=\chi(\epsilon^{-1}\rho)$ where $\chi$ is a nonnegative smooth cutoff such that $\chi=1$ on $[1,\infty)$ and $\chi=0$ on $(-\infty, 1/2]$. With the limit being strong operator limits, we have
\begin{align*}	&\widehat{\mathcal{W}_{0,\gamma_1}}(\sigma)^{-1}-\widehat{\mathcal{W}_{0,\gamma_1}}(0)^{-1}\\
	&\quad=\lim_{\epsilon\to 0}\Big(\widehat{\mathcal{W}_{0,\gamma_1}}(\sigma)^{-1}\chi_\epsilon(\rho)\widehat{\mathcal{W}_{0,\gamma_1}}(0)\widehat{\mathcal{W}_{0,\gamma_1}}(0)^{-1}-\widehat{\mathcal{W}_{0,\gamma_1}}(\sigma)^{-1}\widehat{\mathcal{W}_{0,\gamma_1}}(\sigma)\chi_{\epsilon}(\rho)\widehat{\mathcal{W}_{0,\gamma_1}}(0)^{-1}\Big)\\
	&\quad=\lim_{\epsilon\to 0}\widehat{\mathcal{W}_{0,\gamma_1}}(\sigma)^{-1}\Big([\chi_\epsilon(\rho),\widehat{\mathcal{W}_{0,\gamma_1}}(0)]+\widehat{\mathcal{W}_{0,\gamma_1}}(0)\chi_{\epsilon}(\rho)-\widehat{\mathcal{W}_{0,\gamma_1}}(\sigma)\chi_{\epsilon}(\rho)\Big)\widehat{\mathcal{W}_{0,\gamma_1}}(0)^{-1}\\
		&\quad=\widehat{\mathcal{W}_{0,\gamma_1}}(\sigma)^{-1}\lim_{\epsilon\to 0}\bigg(\Big(\widehat{\mathcal{W}_{0,\gamma_1}}(0)-\widehat{\mathcal{W}_{0,\gamma_1}}(\sigma)\Big)\chi_{\epsilon}(\rho)\bigg)\widehat{\mathcal{W}_{0,\gamma_1}}(0)^{-1}\\
		&\quad\quad+\widehat{\mathcal{W}_{0,\gamma_1}}(\sigma)^{-1}\lim_{\epsilon\to 0}\Big([\chi_\epsilon(\rho),\widehat{\mathcal{W}_{0,\gamma_1}}(0)]\Big)\widehat{\mathcal{W}_{0,\gamma_1}}(0)^{-1}:=\Romanupper{1}+\Romanupper{2}.
\end{align*}
As for $\Romanupper{1}$, since $\widehat{\mathcal{W}_{0,\gamma_1}}(0)^{-1}\omega\in\rho C^\infty(\CX)+\eHb^{\infty, 1/2-}(\CX)$ which is annihilated modulo $\eHb^{\infty, 3/2-}(\CX)$ by the normal operator $-i\sigma\rho(\rho \pa_\rho-1)$ of $\widehat{\mathcal{W}_{0,\gamma_1}}(0)-\widehat{\mathcal{W}_{0,\gamma_1}}(\sigma)$, it follows that
\begin{align*}
	&\bigg(\Big(\widehat{\mathcal{W}_{0,\gamma_1}}(0)-\widehat{\mathcal{W}_{0,\gamma_1}}(\sigma)\Big)\chi_{\epsilon}(\rho)\bigg)\widehat{\mathcal{W}_{0,\gamma_1}}(0)^{-1}\omega\\
	&\quad=\chi_{\epsilon}(\rho)\Big(\widehat{\mathcal{W}_{0,\gamma_1}}(0)-\widehat{\mathcal{W}_{0,\gamma_1}}(\sigma)\Big)\widehat{\mathcal{W}_{0,\gamma_1}}(0)^{-1}\omega+\epsilon^{-1}\chi'(\epsilon^{-1}\rho)\Big(\rho^3 C^\infty(\CX)+\eHb^{\infty, 5/2-}(\CX)\Big)\\
	&\quad\in \chi_{\epsilon}(\rho)\eHb^{\infty, 1/2-}(\CX)+\epsilon^{-1}\chi'(\epsilon^{-1}\rho)\Big(\rho^3 C^\infty(\CX)+\eHb^{\infty, 5/2-}(\CX)\Big).
\end{align*}
Since the $\eHb^{\infty, 1/2-}(\CX)$ norm of the second term $\epsilon\chi'(\epsilon^{-1}\rho)\Big(\rho C^\infty(\CX)+\eHb^{\infty, 1/2-}(\CX)\Big)$ goes to $0$ as $\epsilon\to0$, we arrive at
\[
\Romanupper{1}=\widehat{\mathcal{W}_{0,\gamma_1}}(\sigma)^{-1}\Big(\widehat{\mathcal{W}_{0,\gamma_1}}(0)-\widehat{\mathcal{W}_{0,\gamma_1}}(\sigma)\Big)\widehat{\mathcal{W}_{0,\gamma_1}}(0)^{-1}\omega.
\]
For $\Romanupper{2}$, since $\widehat{\mathcal{W}_{0,\gamma_1}}(0)\in\rho^2\mbox{Diff}_b^2$, it follows that $[\chi_\epsilon(\rho), \widehat{\mathcal{W}_{0,\gamma_1}}(0)]$ is uniform (in $\epsilon$) in $\rho^2\mathrm{Diff}_\bop^1$ and $[\chi_\epsilon(\rho), \widehat{\mathcal{W}_{0,\gamma_1}}(0)]$ goes to $0$ in the strong operator topology $\mathcal{L}(\eHb^{s,\ell}(\CX),\eHb^{s-1,\ell+2}(\CX))$, and thus $\Romanupper{2}=0$. This proves \eqref{eq:resolventidentity}.

Since $\widehat{\mathcal{W}_{0,\gamma_1}}(0)^{-1}\tilde{\omega}=\omega_{s_0}=r^{-1}(dv-dr)\in\rho C^\infty(\CX)+\eHb^{\infty, 1/2-}(\CX)$, applying \eqref{eq:resolventidentity} yields
\begin{equation}
	\Bigl(\widehat{\mathcal{W}_{0,\gamma_1}}(\sigma)^{-1}-\widehat{\mathcal{W}_{0,\gamma_1}}(0)^{-1}\Bigr)\tilde{\omega}=\widehat{\mathcal{W}_{0,\gamma_1}}(\sigma)^{-1}\Bigl(\widehat{\mathcal{W}_{0,\gamma_1}}(0)-\widehat{\mathcal{W}_{0,\gamma_1}}(\sigma)\Bigr)\widehat{\mathcal{W}_{0,\gamma_1}}(0)^{-1}\tilde{\omega}.
\end{equation}
Let
\begin{equation}
	\tilde{\omega}(\sigma):=\Bigl(\widehat{\mathcal{W}_{0,\gamma_1}}(0)-\widehat{\mathcal{W}_{0,\gamma_1}}(\sigma)\Bigr)\widehat{\mathcal{W}_{0,\gamma_1}}(0)^{-1}\tilde{\omega}\in\eHb^{\infty,3/2-}(\CX),
	\end{equation}
and write
\begin{equation}
	\tilde{\omega}(\sigma)=-\sigma(\pa_{\sigma}\widehat{\mathcal{W}_{0,\gamma_1}}(0)\omega_{s_0})+\mathcal{O}_{\eHb^{\infty,3/2-}(\CX)}(\abs{\sigma}^2)
\end{equation}
where the notation $\mathcal{O}_{\eHb^{\infty,3/2-}(\CX)}(\abs{\sigma}^2)$ means that the term has norm in $\eHb^{\infty, 3/2-}(\CX)$ of size $\mathcal{O}(\abs{\sigma}^2)$. Using the fact that $\widehat{\mathcal{W}_{0,\gamma_1}}(\sigma)^{-1}$ is continuous in $\sigma$ with value in $\mathcal{L}_{\mathrm{op}}(\eHb^{s-1+\epsilon, \ell+2+\epsilon}(\CX),\eHb^{s-\epsilon, \ell-\epsilon}(\CX))$, it follows that 
\begin{equation}
	\Bigl(\widehat{\mathcal{W}_{0,\gamma_1}}(\sigma)^{-1}-\widehat{\mathcal{W}_{0,\gamma_1}}(0)^{-1}\Bigr)\tilde{\omega}=-\sigma\widehat{\mathcal{W}_{0,\gamma_1}}(0)^{-1}\pa_{\sigma}\widehat{\mathcal{W}_{0,\gamma_1}}(0)\omega_{s_0}+o_{\eHb^{s,\ell}(\CX)}(\abs{\sigma})
	\end{equation}
for any $s>2, -\frac 32<\ell<-\frac 12$. This proves the differentiability of $\mathcal{W}_1$ and thus $\mathcal{W}_{01}$ at $(\gamma, \sigma)=(0,0)$.

As for $\mathcal{W}_{10}$, we proceed in a similar manner. We define 
	\begin{equation}
		\begin{split}
	\mathcal{W}_2(\gamma,\sigma):=\mathcal{W}_{10}(\gamma, \sigma)\oplus\mathcal{W}_{11}(\gamma, \sigma):\eHb^{s-1,\ell+2}(\CX)&\to \mathcal{R}^\perp\cong\BC\\
	\omega&\mapsto\angles{\widehat{\mathcal{W}_{0,\gamma}}(\sigma)\widehat{\mathcal{W}_{0,\gamma_1}}(\sigma)^{-1}\omega}{\omega^*_{s_0}}
	\end{split}
\end{equation}
We will show that $\mathcal{W}_2(\gamma, \sigma)$ is continuous for $\gamma+\abs{\sigma}$ small and differentiable at $(0,0)$.

As $\omega_{s_0}^*=\delta(r-2\Bm_0)dr\in\sHb^{-1/2-,\infty}(\CX)$, we can rewrite
\begin{equation}
	\begin{split}
	\angles{\widehat{\mathcal{W}_{0,\gamma}}(\sigma)\widehat{\mathcal{W}_{0,\gamma_1}}(\sigma)^{-1}\omega}{\omega^*_{s_0}}&=	\angles{\widehat{\mathcal{W}_{0,\gamma_1}}(\sigma)^{-1}\omega}{\widehat{\mathcal{W}_{0,\gamma}}(\sigma)^*\omega^*_{s_0}}\\
	&=\angles{\widehat{\mathcal{W}_{0,\gamma_1}}(\sigma)^{-1}\omega}{\bigl(\widehat{\mathcal{W}_{0,\gamma}}(\sigma)^*-\widehat{\mathcal{W}_{0}}(0)^*\bigr)\omega^*_{s_0}}.
	\end{split}
\end{equation}
Since $\widehat{\mathcal{W}_{0,\gamma_1}}(\sigma)^{-1}$ is continuous in $\sigma$ with value in $\mathcal{L}_{\mathrm{op}}(\eHb^{s-1, \ell+2}(\CX),\eHb^{s-\epsilon, \ell-\epsilon}(\CX))$, i.e.,
\[
\widehat{\mathcal{W}_{0,\gamma_1}}(\sigma)^{-1}=
\widehat{\mathcal{W}_{0,\gamma_1}}(0)^{-1}+o_{\mathcal{L}(\eHb^{s-1,\ell+2}(\CX), \eHb^{s-\epsilon,\ell-\epsilon}(\CX))}(1),
\] 
and $\widehat{\mathcal{W}_{0,\gamma}}(\sigma)^*-\widehat{\mathcal{W}_{0}}(0)^*$ is a polynomial in $\gamma, \overline{\sigma}$ with coefficients in $\mathcal{L}(\sHb^{1-s,-\ell-2}(\CX),\sHb^{-s,-\ell-1}(\CX))$
\begin{align*}
&\qquad\widehat{\mathcal{W}_{0,\gamma}}(\sigma)^*-\widehat{\mathcal{W}_{0}}(0)^*\\
&\qquad\quad=\overline{\sigma}\Big(\pa_{\overline{\sigma}}(\widehat{\mathcal{W}_{0,\gamma}}(\sigma)^*)|_{(0,0)}\Big)+\gamma\Big(\pa_\gamma(\widehat{\mathcal{W}_{0,\gamma}}(\sigma)^*)|_{(0,0)}\Big)+\mathcal{O}_{\mathcal{L}(\sHb^{1-s,-\ell-2}(\CX),\sHb^{-s,-\ell-1}(\CX))}((\gamma+\abs{\sigma})^2),
\end{align*}
it follows that
\begin{equation}\label{eq:diffofW_10}
		\begin{split}
	&\angles{\widehat{\mathcal{W}_{0,\gamma}}(\sigma)\widehat{\mathcal{W}_{0,\gamma_1}}(\sigma)^{-1}\omega}{\omega^*_{s_0}}\\
		&\quad=\sigma\angles{\widehat{\mathcal{W}_{0,\gamma_1}}(0)^{-1}\omega}{\pa_{\overline{\sigma}}(\widehat{\mathcal{W}_{0,\gamma}}(\sigma)^*)|_{(0,0)}\omega_{s_0}^*}+\gamma\angles{\widehat{\mathcal{W}_{0,\gamma_1}}(0)^{-1}\omega}{\pa_\gamma(\widehat{\mathcal{W}_{0,\gamma}}(\sigma)^*)|_{(0,0)}\omega_{s_0}^*}\\
		&\qquad+o(\gamma+\abs{\sigma})\cdot\norm{\omega}_{\eHb^{s-1,\ell+2}(\CX)}.
	\end{split}
\end{equation}
This proves the differentiability of $\mathcal{W}_{10}$ at $(0,0)$.

		\item\underline{Taylor expansion of $\mathcal{W}_{11}$.} Since $\mathcal{W}_{11}:\mathcal{K}=\langle\tilde{\omega}\rangle\to\mathcal{R}^\perp\cong\BC$ is given by $\angles{\widehat{\mathcal{W}_{0,\gamma}}(\sigma)\widehat{\mathcal{W}_{0,\gamma_1}}(\sigma)^{-1}\tilde{\omega}}{\omega^*_{s_0}}$, by \eqref{eq:diffofW_10} we find
		\begin{equation}
				\mathcal{W}_{11}(\gamma,\sigma)=\sigma\angles{\pa_\sigma\widehat{\mathcal{W}_{0,\gamma}}(\sigma)|_{(0,0)}\omega_{s_0}}{\omega_{s_0}^*}+\gamma\angles{\pa_\gamma\widehat{\mathcal{W}_{0,\gamma}}(\sigma)|_{(0,0)}\omega_{s_0}}{\omega_{s_0}^*}+o(\gamma+\abs{\sigma}).
		\end{equation}
	Since $\pa_\gamma\widehat{\mathcal{W}_{0,\gamma}}(\sigma)|_{(0,0)}=\gamma^{-1}\bigl(\widehat{\mathcal{W}_{0,\gamma}}(0)-\widehat{\mathcal{P}_{0}}(0)\bigr)$, using the calculation of $\mathcal{W}_{11}^\flat$ in \eqref{eq:calofW_11flat} gives
	\begin{equation}
		\angles{\pa_\gamma\widehat{\mathcal{W}_{0,\gamma}}(\sigma)|_{(0,0)}\omega_{s_0}}{\omega_{s_0}^*}=4\pi(\mathfrak{b}-1).
	\end{equation}
We notice that $\pa_\sigma\widehat{\mathcal{W}_{0,\gamma}}(\sigma)|_{(0,0)}=\pa_\sigma\widehat{\mathcal{P}_{0}}(\sigma)|_{\sigma=0}=-i[\mathcal{P}_0, t_{0,*}]$, so we have
\begin{equation}
	\begin{split}
	\angles{\pa_\sigma\widehat{\mathcal{P}_{0,\gamma}}(\sigma)|_{(0,0)}\omega_{s_0}}{\omega_{s_0}^*}&=-i\angles{[\mathcal{P}_0, t_{0,*}]\omega_{s_0} }{\omega_{s_0}^*}=-i\angles{[\mathcal{P}_0, v]\omega_{s_0} }{\omega_{s_0}^*}=-2\pi i
	\end{split}
\end{equation}
where we use 
\begin{align*}
	[\mathcal{P}_0, v]\omega_{s_0}&=-(G_g\delta_g^*\omega_{s_0})_{\al\be}(\nabla^\al v)+\delta_gG_g(\nabla_\al v\otimes_s\omega_{s_0})\\
	&=\Big((-\frac{1}{2r^2}+\frac{\Bm_0}{r^3})dv-\frac{1}{r^2}dr\Big)+\Big(\frac{1}{2r^2}dv+\frac{1}{r^2}dr\Big)=\frac{\Bm_0}{r^3}dv.
	\end{align*}
Therefore, we arrive at 
\begin{equation}\label{eq:valueofW_11}
	\mathcal{W}_{11}(\gamma,\sigma)=-2\pi i\sigma+4\pi (\mathfrak{b}-1)\gamma+o(\gamma+\abs{\sigma}).
\end{equation}
\item\underline{Final step.} 
Owing to the previous discussion starting at \eqref{eq:roughdis1}, in order to prove that $\widehat{\mathcal{W}_{0,\gamma}}(\sigma)\widehat{\mathcal{W}_{0,\gamma_1}}(\sigma)^{-1}$ is injective, it suffices to prove the injectivity of the following operator
\begin{equation}
\mathcal{W}_{11}(\gamma,\sigma)-\mathcal{W}_{10}(\gamma, \sigma)\mathcal{W}_{00}(\gamma,\sigma)^{-1}\mathcal{W}_{01}(\gamma,\sigma):\mathcal{K}\to\mathcal{R}^\perp.
\end{equation}
Combining the differentiability of $\mathcal{W}_{01}, \mathcal{W}_{10}$ at $(0,0)$ and \eqref{eq:valueofW_11} together yields
\[
\mathcal{W}_{11}(\gamma,\sigma)-\mathcal{W}_{10}(\gamma, \sigma)\mathcal{W}_{00}(\gamma,\sigma)^{-1}\mathcal{W}_{01}(\gamma,\sigma)=-2\pi i\sigma+4\pi (\mathfrak{b}-1)\gamma+o(\gamma+\abs{\sigma})
\]
which is nonzero for $\gamma>0, \IM\sigma\geq 0$ with $\gamma+\abs{\sigma}$ small if we set $\mathfrak{b}>1$, and thus is injective. Finally, the invertibility follows from the fact that $\widehat{\mathcal{W}_{0,\gamma}}(\sigma)\widehat{\mathcal{W}_{0,\gamma_1}}(\sigma)^{-1}$ has index $0$ for $\gamma+\abs{\sigma}$ small.
	\end{itemize}
\end{proof}

Now we are at the position to prove Proposition \ref{prop:WnonzeroinSchw}.

\begin{proof}[Proof of Proposition \ref{prop:WnonzeroinSchw}]
	The proof proceed in a similar fashion to that in the last step in the proof of Theorem \ref{thm:modesforscalarwave}. Specifically, by Lemma \ref{lem:step1}, there exists $\gamma_0'>0, C'>0$ such that for $0<\gamma<\gamma_0'$, $\IM\sigma\geq 0,\abs{\sigma}<C'$ and $s>2,\ell\in (-\frac 32, \frac 12)$,
	\[\widehat{\mathcal{W}_{0,\gamma}}(\sigma):\{\omega\in\eHb^{s,\ell}(\CX):\widehat{\mathcal{W}_{0,\gamma}}(\sigma)u\in\eHb^{s-1,\ell+2}(\CX)\}\to \eHb^{s-1,
		\ell+2}(\CX)
	\]
	is invertible. By Proposition \ref{prop:desofkernel}, this implies for $0<\gamma<\gamma_0'$, $\IM\sigma\geq 0,\abs{\sigma}<C'$ and $s>2,\ell<-\frac12, s+\ell>-\frac 12$,
	\[\widehat{\mathcal{W}_{0,\gamma}}(\sigma):\{\omega\in\eHb^{s,\ell}(\CX):\widehat{\mathcal{W}_{0,\gamma}}(\sigma)u\in\eHb^{s,\ell+1}(\CX)\}\to \eHb^{s,
		\ell+1}(\CX)
	\]
	is injective as well, and the invertibility follows from the fact that $\widehat{\mathcal{P}_{0,\gamma}}(\sigma)$ has index $0$ (see Lemma \ref{lem:indexofFredholmoperatorBundle}).
	
	Next, according to the high energy estimates (Theorem \ref{thm:highenergyBundle}) and energy estimate (Proposition \ref{prop:energyestimate} and Theorem \ref{thm:energyestimate}), there exists $\gamma_0''>0, C''>0$ such that for $0<\gamma<\gamma_0''$ and $\IM\sigma\geq 0,\abs{\sigma}>C''$, $\widehat{\mathcal{W}_{0,\gamma}}(\sigma)$ defined as in \eqref{eq:WnonzeroinSchw} is invertible. 
	
	For $\sigma$ with $\IM\sigma\geq 0, C'\leq\abs{\sigma}\leq C''$, the fact that $\widehat{\mathcal{W}_{0,0}}(\sigma)=\widehat{\mathcal{P}_{0}}(\sigma)$ (see Proposition \ref{prop:modeofCPSchw}) is invertible and a perturbation argument (as in the final step in the proof of Theorem \ref{thm:modesforscalarwave}) give the invertibility of $\widehat{\mathcal{W}_{0,\gamma}}(\sigma)$ for $(\gamma, \sigma)$ in an open neighborhood of $(0, \sigma)$. Then the compactness of the region $C'\leq\abs{\sigma}\leq C''$ implies that there exists an $\gamma_0'''>0$ (depending on $C', C''$) such that $\widehat{\mathcal{W}_{0,\gamma}}(\sigma)$ in \eqref{eq:WnonzeroinSchw} is invertible for $\IM\sigma\geq 0, C'\leq\abs{\sigma}\leq C''$ and $0<\gamma<\gamma_0'''$. Therefore, the proposition holds for $0<\gamma<\tilde{\gamma}_0$ where $\tilde{\gamma}_0=\min\{\gamma_0',\gamma_0'',\gamma_0'''\}$.
\end{proof}

Finally, having Propositions \ref{prop:WzeroinSchw} and \ref{prop:WnonzeroinSchw} at our disposal, we are able to establish the following theorem for $\widehat{\mathcal{W}_{b,\gamma}}(\sigma)$.
\begin{thm}\label{thm:modesforW}
	Let $b_0=(\Bm_0, 0,\BQ_0)$ and $\mathfrak{b}>2$. Fix $\gamma\in(0,\tilde{\gamma}_0)$, there exists a small constant $C(\gamma)>0$ such that for the weakly charged 	RN metric $g_{b_0}$ with $\abs{\BQ_0}<C(\gamma)$, the following holds for 
	\begin{enumerate}
		\item If $\IM \sigma\geq0$ and $\sigma\neq 0$, 
		\begin{equation}\label{eq:Wmodenonzero}
			\widehat{\mathcal{W}_{b_0,\gamma}}(\sigma):\{\omega\in\eHb^{s, \ell}(\CX;\scform): 	\widehat{\mathcal{W}_{b_0,\gamma}}(\sigma)\omega\in\eHb^{s,\ell+1}(\CX; \scform)\}\to \eHb^{s,\ell+1}(\CX;\scform)
		\end{equation}
		is invertible when $s>2, \ell<-\frac12, s+\ell>-\frac 12$.
		\item If $s>2$ and $-\frac 32<\ell<-\frac 12$, then stationary operator
		\begin{equation}\label{eq:Wmodezero}
			\widehat{\mathcal{W}_{b_0,\gamma}}(0):\{\omega\in\eHb^{s, \ell}(\CX;\scform): 	\widehat{\mathcal{W}_{b_0,\gamma}}(0)\omega\in\eHb^{s-1,\ell+2}(\CX; \scform)\}\to \eHb^{s-1,\ell+2}(\CX;\scform)
		\end{equation}
		is invertible.
	\end{enumerate}
	Both statements also hold for the weakly charged and slowly rotating Kerr Newman metric $g_b$ with $b=(\Bm, \Ba,\BQ)$ near $b_0=(\Bm_0, 0,\BQ_0)$.		
\end{thm}
\begin{proof}
	The proof is completely analogous to that of Theorem \ref{thm:modesforCP}.
\end{proof}

\subsection{Growing  zero modes of the gauge potential wave operator}\label{subsec:growingzeromodesof1form}
As discussed around \eqref{eq:eqnforgaugepotential} and \eqref{eq:eqfordualpotential}, the zero mode solutions to $\mathcal{W}_{b,\gamma}\omega=0$ on extendible and supported $b$-Sobolev spaces are related to the pure gauge zero mode solutions of $L_{g_b,\gamma}(\dg, \dA)=0$ and dual pure gauge zero mode solutions of $L_{g_b, \gamma}^*(\dg, \dA)=0$, respectively. In this section we will discuss the zero mode solutions to $\mathcal{W}_{b,\gamma}\omega=0$ which are allowed to grow at infinity and are asymptotic to translations and rotations. We will see in the next section that these growing zero modes may generate (dual) pure gauge zero modes solutions to $L^{(*)}_{g_b, \gamma}(\dg,\dA)=0$, which have the expected decay rate at infinity (more precisely, lie in the space $\eHb^{\infty, \ell}(\CX)$ or $\sHb^{-\infty, \ell}(\CX)$ with $-\frac32<\ell<-\frac12$). Throughout this subsection, we focus on the RN metric and slowly rotating metric with sufficiently small charge. Namely, for the parameter $b_0=(\Bm_0, 0, \BQ_0)$ of the RN metric, we assume that \[\BQ_0\ll\Bm_0.
\]

We first discuss the growing zero modes which are asymptotic to translations.
\begin{prop}\label{prop:asym_trans}
	For $b=(\Bm, \Ba, \BQ)$ near $b_0=(\Bm_0, 0,\BQ_0)$, we obtain
	\begin{align}\label{eq:g1form3/2}
	\ker \widehat{\mathcal{W}_{b,\gamma}}(0)\cap\eHb^{\infty,-3/2-}(\CX)&=\langle\omega_{b,s_0}\rangle\oplus\{ \omega_{b,s_1}(\sfS):\sfS\in\BFS_1\}, \quad \omega_{b,s_0}=\pa_t^\flat,\\\label{eq:g1formdual3/2}
		\ker \widehat{\mathcal{W}_{b,\gamma}}(0)\cap\sHb^{-\infty,-3/2-}(\CX)&=\langle\omega^*_{b,s_0}\rangle\oplus\{ \omega^*_{b,s_1}(\sfS):\sfS\in\BFS_1\},
	\end{align}
where $\flat$ denotes the musical isomorphism $V^\flat:=g_b(V,\bullet)$ and $\omega_{b,s_1}(\sfS)$ and $\omega_{b,s_1}^*(\sfS)$ depend on $\sfS\in\BFS_1$ linearly. They satisfy
\begin{equation}\label{eq:symgra3/2}
	\delta^*_{g_b}\omega_{b,s_1}\in\eHb^{\infty,1/2-}(\CX;\scsym),\quad G_{g_b}\delta_{g_b}^*\omega^*_{b,\bullet}\in\sHb^{-\infty,1/2-}(\CX;\scsym)
\end{equation}
Moreover, the maps $b\mapsto \omega_{b,s_0}, \omega_{b,s_1}(\sfS)$ and $b\mapsto \omega^*_{b,s_0},\omega^*_{b,s_1}(\sfS)$ can be chosen to be continuous in $b$ with values in the respective spaces. 

For later use, we further determine the leading term of $\omega_{b_0,s_1}(\sfS)$ and $\omega^*_{b_0,s_1}(\sfS)$
\begin{equation}\label{eq:refinedg1formmode}
	\omega_{b_0,s_1}(\sfS)=du_{b_0,s_1}(\sfS)+\mathcal{O}_{\eHb^{\infty,-1/2-}(\CX)}(\gamma+\abs{\BQ_0}^2),\quad \omega^*_{b_0,s_1}(\sfS)=du^*_{b_0,s_1}(\sfS)+\mathcal{O}_{\sHb^{-\infty,-1/2-}(\CX)}(\gamma+\abs{\BQ_0}^2)
\end{equation}
where $u_{b_0,s_1}(\sfS)=(r-\Bm_0)\sfS$ and  $u_{b_0,s_1}^*(\sfS)=(r-\Bm_0)H(r-\ehRN)\sfS$ are defined as in \eqref{eq:gscalar5/2RN}.
\end{prop}

\begin{proof}
	The proof exploits a normal operator argument as in Propositions \ref{prop:desofkernel} and \ref{prop:gscalarzeromode}. In particular, the normal operator of $-2\widehat{\mathcal{W}_{b,\gamma}}(0)$ is equal to $\widehat{\Box_{\underline{g},1}}(0)$ which is the Euclidean Laplacian $\Delta_{\BR^3}=\rho^2(\rho\pa_\rho(\rho\pa_\rho-1)+\sL)$ tensored with $4\times 4$ identity matrix when working in the standard coordinate trivialization of $\scform$ and annihilates the following $1$-forms
	\begin{equation}
		dt,\ dx^1,\ dx^2,\ dx^3.
	\end{equation}
We note that $dt$ is of scalar type $l=0$ while $dx^i=\sfS dr+rd\sfS$ with $\sfS=x^i/r$ ($i=1,2,3$)  is of scalar type $l=1$.

We first prove \eqref{eq:g1form3/2}. Let $u\in\ker\widehat{\mathcal{W}_{b, \gamma}}(0)\cap \sHb^{-\infty,-3/2-}(\CX)$. By the normal operator argument as in Proposition \ref{prop:desofkernel} ( where we shift the contour of integration through the pole $0$ with the space of resonant states given by $\BFS_0$), $\omega$ must take the form $\omega=\chi \omega_0+\tilde{\omega}$ where $\omega_0$ is in the $4$-dimensional space span$\{dt, dx^1, dx^2,dx^3\}$, $\tilde{\omega}\in\eHb^{\infty, -1/2-}(\CX)$ and $\chi$ is a smooth cutoff with $\chi=0$ for $r\leq 3\Bm_0$, $\chi=1$ with $r\geq 4\Bm_0$. This proves that the space in \eqref{eq:g1form3/2} is at most $4$-dimensional. Now we prove that it is indeed $4$-dimensional. We compute for $\omega_0\in\{dt, dx^1, dx^2,dx^3\}$
\[
-\widehat{\mathcal{W}_{b, \gamma}}(0)\chi\omega_0=-(\widehat{\mathcal{W}_{b,\gamma}}(0)-\Box_{\underline{g},1})\chi\omega_0-[\Box_{\underline{g},1}, \chi]\omega_0\in\eHb^{\infty, 3/2-}(\CX)+\eHb^{\infty,\infty}(\CX)=\eHb^{\infty, 3/2-}(\CX).
\]
In view of Theorem \ref{thm:modesforW}, there exists a unique $\tilde{\omega}\in\eHb^{\infty, -1/2-}(\CX)$ satisfying $\widehat{\mathcal{W}_{b, \gamma}}(0)\tilde{\omega}=-\widehat{\mathcal{W}_{b, \gamma}}(0)\chi \omega_0$. Therefore, $\chi\omega_0+\tilde{\omega}$ gives rise to an element in the space \eqref{eq:g1form3/2} which we denote by $\omega_{b,s_0}$ if $\omega_0=dt$ and $\omega_{b,s_1}(\sfS)$ if $\omega_0=d(r\sfS)\in\{dx^1, dx^2,dx^3\}$. The explicit expression for $\omega_{b,s_0}$ follows from the direction calculation $\widehat{\mathcal{W}_{b,\gamma}}\pa_t^\flat=\delta_{g_b,\gamma}G_g\delta_g^*\pa_t^\flat=0$ since $\pa_t $ is killing (i.e. $\delta_g^*\pa_t^\flat=0$). As for the symmetric gradient, we calculate
\[
\delta_{g_b}^*\omega_{b,s_1}=\delta_{g_b}\tilde{\omega}+(\delta_{g_b}^*-\delta_{\underline{g}}^*)\chi\omega_0+[\delta^*_{\underline{g}},\chi]\omega_0\in\eHb^{\infty,1/2-}(\CX)+\eHb^{\infty,\infty}(\CX)=\eHb^{\infty,1/2-}(\CX)
\]
where we use $\delta_{g_b}^*-\delta_{\underline{g}}^*\in\rho^2\mbox{Diff}_\bop^1, \delta_{g_b}^*\in\rho\mbox{Diff}_\bop^1$ and $\delta_{\underline{g}}^*\omega_0=0$.

Now let us turn to the continuous dependence of $\omega_{b, s_1}(\sfS)$ on $b$.
Suppose $b_j\to b$ and $\omega_{b_j,s_1}(\sfS)=\chi\omega_0+\tilde{\omega}_{b_j}(\sfS)$ where $\omega_0=d(r\sfS)\in\{ dx^1,dx^2,dx^3\}$ and $\tilde{\omega}_{b_j}(\sfS)$ is uniquely determined by $b_j$ and $\sfS$. Let $\omega_{b,s_1}(\sfS)=\chi\omega_0+\tilde{\omega}_b(\sfS)$ and $e_{j,s_1}(\sfS)=\omega_{b_j,s_1}(\sfS)-\omega_{b,s_1}(\sfS)=\tilde{\omega}_{b_j,s_1}(\sfS)-\tilde{\omega}_{b,s_1}(\sfS)$, we find
\[
\widehat{\mathcal{W}_{b, \gamma}}(0)e_{j,s_1}(\sfS)=(\widehat{\mathcal{W}_{b, \gamma}}(0)-\widehat{\mathcal{W}_{b_j, \gamma}}(0))(\tilde{\omega}_{b_j,s_1}(\sfS)+\chi \omega_0).
\]
Since $\{\tilde{\omega}_{b_j,s_1}(\sfS)\}$ is uniformly bounded in $\eHb^{\infty,-1/2-}(\CX)$ and $\widehat{\mathcal{W}_{b, \gamma}}(0)-\widehat{\mathcal{W}_{b_j, \gamma}}(0)\in (b-b_j)\rho^3\mbox{Diff}_b^2$, we see that $\widehat{\mathcal{W}_{b, \gamma}}(0)e_{j,s_1}(\sfS)\in (b-b_j)\eHb^{\infty, 3/2-}$, and thus $e_{j,s_1}(\sfS)\in\eHb^{\infty,-1/2-}(\CX)$ is of size $O(b-b_j)$. This proves the continuous dependence of $\omega_{b,s_1}(\sfS)$ on $b$.

The proof for the dual zero modes $\omega^*_{b,s_0}$ and $\omega^*_{b,s_1}(\sfS$ )proceeds in an analogous manner (we use Theorem \ref{thm:modesforCP} because $\widehat{\mathcal{W}_{b, \gamma}}(0)$ acting on supported $b$-Sobolev space is identical to $\widehat{\mathcal{P}_{b,\gamma}}(0)^*$).

It remains to derive the expressions in \eqref{eq:refinedg1formmode}. Since
\[
\widehat{\mathcal{W}_{b_0,\gamma}}(0)=\delta_{g_{b_0}}G_{g_{b_0}}\delta^*_{g_{b_0}}+F_{g_{b_0}}G_{g_{b_0}}\delta^*_{g_{b_0}}=\frac 12(d\delta_{g_{b_0}}+\delta_{g_{b_0}}d)-\Ric(g_{b_0})+F_{g_{b_0}}G_{g_{b_0}}\delta^*_{g_{b_0}},
\]
we have
\begin{equation}
	\begin{split}
\widehat{\mathcal{W}_{b_0,\gamma}}(0)(\omega_{b_0, s_1}(\sfS)-du_{b,s_1}(\sfS))&=\Ric(g)_\al^{\ \be}(du_{b_0,s_1}(\sfS))_\be-F_{g_{b_0}}G_{g_{b_0}}\delta^*_{g_{b_0}}du_{b_0,s_1}(\sfS)\\
&=\mathcal{O}_{\eHb^{\infty,5/2-}}(\abs{\BQ_0}^2)+\mathcal{O}_{\eHb^{\infty,\infty}(\CX)}(\gamma).
\end{split}
\end{equation}
Since $\widehat{\mathcal{W}_{b_0,\gamma}}(0)(\omega_{b_0, s_1}(\sfS)-du_{b,s_1}(\sfS))$ is of scalar type $l=1$ and $\widehat{\mathcal{W}_{b_0,\gamma}}(0)^{-1}$ restricted in scalar type $l=1$ $1$-form spaces exists with norm of size $\mathcal{O}(1)$ (see Remark \ref{rem:invertibilityofscalar1}), it follows that $\omega_{b_0, s_1}(\sfS)-du_{b,s_1}(\sfS)=\mathcal{O}_{\eHb^{\infty,-1/2-}(\CX)}(\gamma+\abs{\BQ_0}^2)$.
\end{proof} 

Next we turn to the growing zero modes asymptotic to rotations.

\begin{prop}\label{prop:asym_rota}
	For $b=(\Bm, \Ba, \BQ)$ near $b_0=(\Bm_0, 0,\BQ_0)$, there exists continuous (in $b$) families
	\begin{align}\label{eq:g1form5/2}
	b\mapsto \omega_{b,v_1}(\sfV)\in\ker \widehat{\mathcal{W}_{b,\gamma}}(0)\cap\eHb^{\infty,-5/2-}(\CX),\\\label{eq:g1formdual5/2}
		b\mapsto \omega^*_{b,v_1}(\sfV)\in\ker \widehat{\mathcal{W}_{b,\gamma}}(0)\cap\sHb^{-\infty,-5/2-}(\CX),
	\end{align} 
depending linearly on $\sfV\in\BFV_1$, such that 
\begin{equation}\label{eq:symmgra5/2}
	\delta_{g_b}^*\omega_{b,v_1}(\sfV)\in\eHb^{\infty,1/2-}(\CX;\scsym),\quad 	G_{g_b}\delta_{g_b}^*\omega^*_{b,v_1}(\sfV)\in\sHb^{-\infty,1/2-}(\CX;\scsym).
	\end{equation}
More explicitly, we have 
\begin{equation}\label{eq:refinedg1formzeromode_rota}
	\omega_{b_0,v_1}(\sfV)=r^2\sfV,\quad\omega^*_{b_0,v_1}(\sfV)=r^2H(r-\ehRN)\sfV+\mathcal{O}_{\sHb^{-\infty,-1/2-}(\CX)}(\gamma)
\end{equation}
\end{prop}

\begin{proof}
	We first notice that, with respect to RN metric, $r^2\sfV\in\eHb^{\infty,-5/2-}(\CX)$ is dual to the rotation vector field (see the calculation in \eqref{eq:Vkilling}), and thus killing. Namely, we have $\delta_{g_{b_0}}^*r^2\sfV=0$ and thus $\widehat{\mathcal{W}_{b_0,\gamma}}(0)r^2\sfV=\delta_{g_{b_0,\gamma}}G_{g_{b_0}}\delta^*_{g_{b_0}}r^2\sfV=0$. As for KN metric $g_b$ with $b=(\Bm,\Ba,\BQ)$ near $b_0=(\Bm_0,0,\BQ_0)$, a direct calculation implies
	\[
-\widehat{\mathcal{W}_{b, \gamma}}(0)\chi r^2\sfV=-\Bigl(\widehat{\mathcal{W}_{b,\gamma}}(0)-\widehat{\mathcal{W}_{b',\gamma}}(0)\Bigr)\chi r^2\sfV-[\widehat{\mathcal{W}_{b',\gamma}}(0), \chi]r^2\sfV\in\eHb^{\infty, 3/2-}(\CX)+\eHb^{\infty, \infty}(\CX)=\eHb^{\infty, 3/2-}(\CX).
	\]
where $b'=(\Bm,0,\BQ)$ and we use $\widehat{\mathcal{W}_{b,\gamma}}(0)-\widehat{\mathcal{W}_{b',\gamma}}(0)\in\rho^4\mbox{Diff}_\bop^2$. In view of Theorem \ref{thm:modesforW}, there exists a unique $\tilde{\omega}_b(\sfV)\in\eHb^{\infty, -1/2-}(\CX)$ satisfying $\widehat{\mathcal{W}_{b, \gamma}}(0)\tilde{\omega}_b(\sfV)=-\widehat{\mathcal{W}_{b, \gamma}}(0)\chi r^2\sfV$. We then define $\omega_{b,v_1}(\sfV):=\chi r^2\sfV+\tilde{\omega}_b(\sfV)$. For $b=(\Bm,\Ba,\BQ)$, the symmetric gradient $\delta_{g_b}\omega_{b,v_1}(\sfV)$ is given by
\[
\delta_{g_b}^*\omega_{b,v_1}(\sfV)=\delta_{g_b}^*\tilde{\omega}_b(\sfV)+(\delta_{g_b}^*\!\!-\!\delta_{g_{(\Bm,0,\BQ)}}^*)\chi r^2\sfV+[\delta^*_{g_{(\Bm,0,\BQ)}},\chi]r^2\sfV\in\eHb^{\infty,1/2-}(\CX)+\eHb^{\infty,\infty}(\CX)=\eHb^{\infty,1/2-}(\CX)
\]
where we use $\delta_{g_b}^*-\delta_{g_{(\Bm,0,\BQ)}}^*\in\rho^3\mbox{Diff}_\bop^1, \delta_{g_b}^*\in\rho\mbox{Diff}_\bop^1$ and $\delta_{g_{(\Bm,0,\BQ)}}^*r^2\sfV=0$.

The proof of the continuous dependence of $\omega_{b, v_1}$ on $b$ proceeds as in Proposition \ref{prop:asym_trans}. Suppose $b_j=(\Bm_j,\Ba_j,\BQ_j)\to b=(\Bm,\Ba,\BQ)$ and let $\omega_{b_j,v_1}(\sfV)=\chi r^2\sfV+\tilde{\omega}_{b_j}(\sfV)$, $\omega_{b, v_1}(\sfV)=\chi r^2\sfV+\tilde{\omega}_b(\sfV)$ and $e_{j}(\sfV)=\omega_{b_j,v_1}(\sfV)-\omega_{b,v_1}(\sfV)=\tilde{\omega}_{b_j}(\sfV)-\tilde{\omega}_{b}(\sfV)$, we find
\begin{align*}
\widehat{\mathcal{W}_{b, \gamma}}(0)e_{j}(\sfV)&=(\widehat{\mathcal{W}_{b, \gamma}}(0)-\widehat{\mathcal{W}_{b_j, \gamma}}(0))(\tilde{\omega}_{b_j}(\sfV)+\chi r^2\sfV)\\
&=\abs{b-b_j}\rho^3\mbox{Diff}_b^2\tilde{\omega}_{b_j}(\sfV)+\Big(\big(\widehat{\mathcal{W}_{b, \gamma}}(0)-\widehat{\mathcal{W}_{b', \gamma}}(0)\big)-\big(\widehat{\mathcal{W}_{b_j, \gamma}}(0)-\widehat{\mathcal{W}_{b'_j, \gamma}}(0)\big)\Big)\chi r^2\sfV\\
&\qquad+\Big(\widehat{[\mathcal{W}_{b', \gamma}}(0)-\widehat{\mathcal{W}_{b'_j, \gamma}}(0),\chi]r^2\sfV\Big)\\
&=\abs{b-b_j}\rho^3\mbox{Diff}_b^2\tilde{\omega}_{b_j}(\sfV)+\abs{b-b_j}\rho^4\mbox{Diff}_b^2\chi r^2\sfV+\abs{b-b_j}\rho^\infty\mbox{Diff}_b^1r^2\sfV\\
&\in\abs{b-b_j}\Bigl(\eHb^{\infty,5/2-}(\CX)+\eHb^{\infty,3/2-}(\CX)+\eHb^{\infty,\infty}(\CX)\Bigr)\subset\abs{b-b_j}\eHb^{\infty,3/2-}(\CX)
\end{align*}
where $b_j'=(\Bm_j,0,\BQ_j)$ and $b'=(\Bm,0,\BQ)$. As a consequence, $e_j(\sfV)\in\eHb^{\infty,-1/2-}(\CX)$ is of size $O(b-b_j)$. This proves the continuous dependence of $\omega_{b,v_1}(\sfV)$ on $b$.

The statements for dual zero modes $\omega^*_{b,v_1}(\sfV)$ can be proved in a similar way (we use Theorem \ref{thm:modesforCP} because $\widehat{\mathcal{W}_{b, \gamma}}(0)$ acting on supported $b$-Sobolev space is identical to $\widehat{\mathcal{P}_{b,\gamma}}(0)^*$). Here, we present the detail of the derivation of the expression for $\omega^*_{b_0,v_1}(\sfV)$ in \eqref{eq:refinedg1formzeromode_rota}. First, we calculate
\[
\delta_{g_{b_0}}G_{g_{b_0}}\delta^*_{g_{b_0}} r^2H(r-\ehRN)\sfV=\delta_{g_{b_0}}\Big((r^2\delta(r-\ehRN)dr)\otimes_s\sfV\Big)=-\frac{1}{2r^2}\pa_r(r^4\delta(r-\ehRN)\weight)\sfV=0.
\]
It follows that
\[
	\begin{split}
	\widehat{\mathcal{W}_{b_0,\gamma}}(0)(\omega^*_{b_0, v_1}(\sfV)-r^2H(r-\ehRN)\sfV)&=-F_{g_{b_0}}\Big((r^2\delta(r-\ehRN)dr)\otimes_s\sfV\Big)=\mathcal{O}_{\eHb^{-\infty,\infty}(\CX)}(\gamma).
\end{split}
\]
Since $\widehat{\mathcal{W}_{b_0,\gamma}}(0)(\omega_{b_0, v_1}(\sfV)-r^2H(r-\ehRN)\sfV)$ is of vector type $l=1$ and $\widehat{\mathcal{W}_{b_0,\gamma}}(0)^{-1}$ restricted in vector type $l=1$ $1$-form spaces exists with norm of size $\mathcal{O}(1)$ (see Remark \ref{rem:invertibilityofscalar1}), we conclude that \[\widehat{\mathcal{W}_{b_0,\gamma}}(0)(\omega_{b_0, v_1}(\sfV)-r^2H(r-\ehRN)\sfV)=\mathcal{O}_{\eHb^{-\infty,-1/2-}(\CX)}(\gamma).
\]

\end{proof}

\subsection{Generalized  zero modes of the gauge potential wave operator}
Apart from the growing modes, we also need to take the generalized zero modes  into consideration (the zero modes which allow for polynomial growth in $t_{b,*}$). Before the discussion about the generalized zero modes, we establish the following technical lemma about another family of growing zero modes.
\begin{lem}\label{lem:pre_asym_lor}
		For $b=(\Bm, \Ba, \BQ)$ near $b_0=(\Bm_0, 0,\BQ_0)$, there exists continuous (in $b$) families
	\begin{align}
		b\mapsto \omega^{(1)}_{b,s_1}(\sfS)\in\ker \widehat{\mathcal{W}_{b,\gamma}}(0)\cap\eHb^{\infty,-5/2-}(\CX),\\
		b\mapsto \omega^{(1)*}_{b,s_1}(\sfS)\in\ker \widehat{\mathcal{W}_{b,\gamma}}(0)\cap\sHb^{-\infty,-5/2-}(\CX),
	\end{align} 
	depending linearly on $\sfS\in\BFS_1$, such that 
	\begin{equation}\label{eq:symgra5/2}
		\begin{split}
		\delta_{g_b}^*\omega^{(1)}_{b,s_1}(\sfS)&=\chi dt\otimes_s d(r\sfS)+\eHb^{\infty,-1/2-}(\CX;\scsym)\\
		\delta_{g_b}^*\omega^{(1)*}_{b,s_1}(\sfS)&=\chi dt\otimes_s d(r\sfS)+\sHb^{\infty,-1/2-}(\CX;\scsym)
		\end{split}
	\end{equation}
Moreover, the leading terms of $\omega^{(1)}_{b,s_1}(\sfS), \omega^{(1)*}_{b,s_1}(\sfS)$are given by
\begin{equation}
	\omega_{b,s_1}^{(1)}(\sfS)-\chi r\sfS dt\in\eHb^{\infty,-3/2-}(\CX),\quad \omega_{b,s_1}^{(1)*}(\sfS)-\chi r\sfS dt\in\sHb^{-\infty,-3/2-}(\CX)
\end{equation}
where $\chi$ is a smooth cutoff satisfying $\chi=1$ for $r\geq 4\Bm_0$ and $\chi=0$ for $r\leq 3\Bm_0$. 

 For later use, we further determine the leading term of $\omega^{(1)}_{b_0,s_1}(\sfS)$
\begin{equation}\label{eq:refinedg1formmode5/2}
	\omega^{(1)}_{b_0,s_1}(\sfS)=r\sfS(\weight dt_0-dr)+\mathcal{O}_{\eHb^{\infty,-1/2-}(\CX)}(\gamma+\abs{\BQ_0}^2).
\end{equation}
\end{lem}

\begin{proof}
		The proof uses the fact that the normal operator of $-2\widehat{\mathcal{W}_{b,\gamma}}(0)$ is equal to $\widehat{\Box_{\underline{g},1}}(0)$, which is the Euclidean Laplacian $\Delta_{\BR^3}=\rho^2(\rho\pa_\rho(\rho\pa_\rho-1)+\sL)$ tensored with $4\times 4$ identity matrix when working in the standard coordinate trivialization of $\scform$ and annihilates the following $1$-form $x^idt=r\sfS dt$ ($i=1,2.3$) where $\sfS=x^i/r\in\BFS_1$. Concretely, we have 	
\[
-\widehat{\mathcal{W}_{b, \gamma}}(0)\chi r\sfS dt=-\Bigl(\widehat{\mathcal{W}_{b,\gamma}}(0)-\Box_{\underline{g},1}\Bigr)\chi r\sfS dt-[\Box_{\underline{g},1}, \chi]r\sfS dt\in\eHb^{\infty, 1/2-}(\CX)+\eHb^{\infty, \infty}(\CX)=\eHb^{\infty, 1/2-}(\CX).
\]
Since $\ker\widehat{\mathcal{W}_{b, \gamma}}(0)^*\cap\sHb^{-\infty,-1/2+}(\CX)=\{0\}$ by Theorem \ref{thm:modesforW}, the above equation is solvable. More precisely, there exists a unique $\tilde{\omega}^{(1)}_b(\sfS)\in\eHb^{\infty, -3/2-}(\CX)$ which is in $\ann \{f_1,\cdots, f_4\}$, where $\{f_1,\cdots,f_4\}\subset C^\infty_c(X^\circ)$ is a set of linearly independent functionals on $\widehat{\mathcal{W}_{b_0, \gamma}}(0)\cap\eHb^{\infty,-3/2-}(\CX)$, such that $\widehat{\mathcal{W}_{b, \gamma}}(0)\tilde{\omega}^{(1)}_b(\sfS)=-\widehat{\mathcal{W}_{b, \gamma}}(0)\chi r\sfS dt$. We then define $\omega^{(1)}_{b,s_1}(\sfS):=\chi r\sfS dt+\tilde{\omega}^{(1)}_b(\sfS)$ . As for the symmetric gradient $\delta_{g_b}\omega^{(1)}_{b,s_1}(\sfS)$, we find
\begin{align*}
\delta_{g_b}^*\omega^{(1)}_{b,s_1}(\sfS)&=\chi\delta_{\underline{g}}^*r\sfS dt+[\delta^*_{\underline{g}},\chi]r\sfS dt+(\delta_{g_b}^*\!\!-\!\delta_{\underline{g}}^*)\chi r\sfS dt+\delta_{g_b}^*\tilde{\omega}^{(1)}_b(\sfS)\\
&=\chi dt\otimes_s d(r\sfS)+\eHb^{\infty,\infty}(\CX)+\eHb^{\infty,-1/2-}(\CX)+\eHb^{\infty,-1/2-}(\CX)\\
&=\chi dt\otimes_s d(r\sfS)+\eHb^{\infty,-1/2-}(\CX)
\end{align*}
where we use $\delta_{g_b}^*-\delta_{\underline{g}}^*\in\rho^2\mbox{Diff}_\bop^1$ and $\delta_{g_b}^*\in\rho\mbox{Diff}_\bop^1$.

As for the continuous dependence of $\omega^{(1)}_{b, s_1}(\sfS)$ on $b$, suppose $b_j\to b$ and let $\omega^{(1)}_{b_j,s_1}(\sfS)=\chi r\sfS dt+\tilde{\omega}^{(1)}_{b_j}(\sfS)$, $\omega^{(1)}_{b,s_1}(\sfS)=\chi r\sfS dt+\tilde{\omega}^{(1)}_{b}(\sfS)$ and $e_{j}(\sfS)=\omega^{(1)}_{b_j,s_1}(\sfS)-\omega^{(1)}_{b,s_1}(\sfS)=\tilde{\omega}^{(1)}_{b_j}(\sfS)-\tilde{\omega}^{(1)}_{b}(\sfS)$, we find
\begin{align*}
	\widehat{\mathcal{W}_{b, \gamma}}(0)e_{j}(\sfS)&=(\widehat{\mathcal{W}_{b, \gamma}}(0)-\widehat{\mathcal{W}_{b_j, \gamma}}(0))(\tilde{\omega}^{(1)}_{b_j}(\sfS)+\chi r\sfS dt)\in\abs{b-b_j}\eHb^{\infty,1/2-}(\CX).
\end{align*}
As $\widehat{\mathcal{W}_{b,\gamma}}(0)|_{\ann\{f_1,\cdots,f_4\}}:\ann\{f_1,\cdots,f_4\}\to\eHb^{\infty,1/2-}(\CX)$ is invertible, it follows that $e_j(\sfS)$ is of size $\mathcal{O}_{\eHb^{\infty, -3/2-}(\CX)}(b-b_j)$. This completes the proof of the continuity.

The proofs for dual zero modes $\hat{\omega}^{(1)*}_{b,s_1}(\sfS)$ are analogous where we use \[\ker\widehat{\mathcal{W}_{b, \gamma}}(0)^*\cap\eHb^{\infty, -1/2+}(\CX)=\ker\widehat{\mathcal{P}_{b, \gamma}}(0)\cap\eHb^{\infty, -1/2+}(\CX)=\{0\}.
\]

It remains to derive the expressions in \eqref{eq:refinedg1formmode5/2}. Since
\[
\delta_{g_{b_0}}G_{g_{b_0}}\delta_{g_{b_0}}\big(r\sfS(\weight dt_0-dr)\big)=\frac{\BQ_0^2}{r^3}\sfS(\weight dt_0-dr)\in\BQ_0^2\eHb^{\infty,3/2-}(\CX),
\]
we have
\[
		\widehat{\mathcal{W}_{b_0,\gamma}}(0)\Big(\omega^{(1)}_{b_0, s_1}(\sfS)-r\sfS(\weight dt_0-dr)\Big)=\mathcal{O}_{\eHb^{\infty,3/2-}}(\abs{\BQ_0}^2)+\mathcal{O}_{\eHb^{\infty,\infty}(\CX)}(\gamma).
\]
Since the right-hand side of the above equation is of scalar type $l=1$ and $\widehat{\mathcal{W}_{b_0,\gamma}}(0)^{-1}$ restricted in scalar type $l=1$ $1$-form spaces exists with norm of size $\mathcal{O}(1)$ (see Remark \ref{rem:invertibilityofscalar1}), it follows that $\omega^{(1)}_{b_0, s_1}(\sfS)-r\sfS(\weight dt_0-dr)=\mathcal{O}_{\eHb^{\infty,-1/2-}(\CX)}(\gamma+\abs{\BQ_0}^2)$.
\end{proof}

Now we are ready to discuss (dual) generalized zero modes which have less decay rate in $r$ at infinity for fixed $t_{b,*}$ and grow linearly in $t_{b,*}$. In the course of the proof, we will find that they are asymptotic to Lorentz boosts.

\begin{prop}\label{prop:asym_lor}
		For $b=(\Bm, \Ba, \BQ)$ near $b_0=(\Bm_0, 0,\BQ_0)$, there exists continuous (in $b$) families
	\begin{align}\label{eq:ge1formzeromode5/2}
		b\mapsto \hat{\omega}_{b,s_1}(\sfS)\in\ker \mathcal{W}_{b,\gamma}\cap\mbox{Poly}^1(t_{b,*})\eHb^{\infty,-5/2-}(\CX),\\\label{eq:ge1formdualzeromode5/2}
		b\mapsto \hat{\omega}^*_{b,s_1}(\sfS)\in\ker \mathcal{W}_{b,\gamma}\cap\mbox{Poly}^1(t_{b,*})\sHb^{-\infty,-5/2-}(\CX),
	\end{align} 
	depending linearly on $\sfS\in\BFS_1$, such that 
	\begin{align}\label{eq:gesymgra}
		\delta_{g_b}^*\hat{\omega}_{b,s_1}(\sfS)\in\mbox{Poly}^1(t_{b,*})\eHb^{\infty,-1/2-}(\CX;\scsym),\\\label{eq:gesymgradual}	G_{g_b}\delta_{g_b}^*\hat{\omega}^*_{b,s_1}(\sfS)\in\mbox{Poly}^1(t_{b,*})\sHb^{-\infty,-1/2-}(\CX;\scsym).
	\end{align}
Moreover, the leading terms of $\hat{\omega}_{b,s_1}(\sfS)$ are given by
\begin{equation}
	\hat{\omega}_{b,s_1}^{(1)}(\sfS)-(tdx^i-x^idt) \in\mbox{Poly}^1(t_{b,*})\eHb^{\infty,-3/2-}(\CX),\quad r\gg 1
\end{equation}
where $\sfS=x^i/r$.

For later use, we further determine the leading term of $\hat{\omega}_{b_0,s_1}(\sfS)$
\begin{equation}\label{eq:refinedge1formmode}
	\begin{split}
	\hat{\omega}_{b_0,s_1}(\sfS)&=(t_0-r)\omega_{b_0,s_1}(\sfS)-r\sfS(\weight dt_0-dr)+\Bm_0\sfS(dt_0+dr)\\
	&\quad+d\bigg(2\Bm_0\Big(2\Bm_0+(\Bm_0-r)\log(\frac{r}{\Bm_0})\Big)\sfS\bigg)+\mathcal{O}_{\eHb^{\infty,-1/2-}(\CX)}(\gamma+\abs{\BQ_0}^2).
	\end{split}
\end{equation}
\end{prop}

\begin{proof}
%Let $\mathfrak{t}=t_{\chi}$ and 
We make the ansatz
\begin{equation}
	\hat{\omega}_{b,s_1}(\sfS)=t_{\chi_0}\omega_{b,s_1}(\sfS)+\check{\omega}_{b,s_1}(\sfS)
\end{equation}	
with $\check{\omega}_{b,s_1}(\sfS)\in\eHb^{\infty,-5/2-}(\CX)$ to be determined. Our aim is to find \[\hat{\omega}_{b,s_1}\in\ker\mathcal{W}_{b, \gamma}\cap\!\mbox{Poly}^1(t_{b,*})\eHb^{\infty,5/2-}\!(\CX)
\]
 such that $\delta_{g_b}^*	\hat{\omega}_{b,s_1}(\sfS)\in\mbox{Poly}^1(t_{b,*})\eHb^{\infty,-1/2-}(\CX)$. To this end, since \[t_{\chi_0}-t_{b_*}\in\mathcal{A}^{-1}(\CX)\quad \mbox{and} \quad\delta_{g_b}^*\omega_{b,s_1}\in\eHb^{\infty,1/2-}(\CX),
 \]
 it suffices to solve the following two equations for $\check{\omega}_{b,s_1}(\sfS)\in\eHb^{\infty,-5/2-}(\CX)$
\begin{align}\label{eq:eqnfromsgcondi}
	[\delta^*_{g_b},t_{\chi_0}]\omega_{b,s_1}(\sfS)+\delta_{g_b}^*\check{\omega}_{b,s_1}(\sfS)\in\eHb^{\infty,-1/2-}(\CX),\\\label{eq:eqnfromW=0}
	\widehat{\mathcal{W}_{b, \gamma}}(0)\check{\omega}_{b,s_1}(\sfS)=-[\mathcal{W}_{b, \gamma},t_{\chi_0}]\omega_{b,s_1}(\sfS).
\end{align}
We first analyze \eqref{eq:eqnfromsgcondi}. Since
\[
[\delta^*_{g_b},t_{\chi_0}]\omega_{b,s_1}(\sfS)=dt_{\chi_0}\otimes_s\omega_{b,s_1}(\sfS),
\]
we further make the ansatz
\begin{equation}\label{eq:defofcheckomega}
	\check{\omega}_{b,s_1}(\sfS)=-\omega^{(1)}_{b,s_1}(\sfS)+\tilde{\omega}_{b,s_1}(\sfS)\in\eHb^{\infty,-5/2-}(\CX).
	\end{equation}
with $\tilde{\omega}_{b,s_1}(\sfS)\in\eHb^{\infty,-3/2-}(\CX)$ to be determined. Therefore, using $\omega_{b,s_1}(\sfS)=\chi d(r\sfS)+\eHb^{\infty,-1/2-}(\CX)$ and \eqref{eq:symgra5/2}, it follows that
\[
[\delta^*_{g_b},t_{\chi_0}]\omega_{b,s_1}(\sfS)+\delta_{g_b}^*\check{\omega}_{b,s_1}\in\eHb^{\infty,-1/2-}(\CX)
\]
and the requirement \eqref{eq:eqnfromsgcondi} is satisfied with the choice of $\check{\omega}_{b,s_1}(\sfS)$ defined as in \eqref{eq:defofcheckomega}. 

Next we solve \eqref{eq:eqnfromW=0} to determine $\tilde{\omega}_{b,s_1}(\sfS)$ in \eqref{eq:defofcheckomega}. We rewrite \eqref{eq:eqnfromW=0} as
\begin{equation}\label{eq:eqnfortildeomega}
	\begin{split}
	&	\widehat{\mathcal{W}_{b, \gamma}}(0)\tilde{\omega}_{b,s_1}(\sfS)=\widehat{\mathcal{W}_{b, \gamma}}(0)\omega^{(1)}_{b,s_1}(\sfS)-[\mathcal{W}_{b, \gamma},t_{\chi_0}]\omega_{b,s_1}(\sfS)\\
	&\quad=	\tilde{\delta}_{g_b,\gamma}G_{g_b}\Bigl(\delta_{g_b}^*\omega^{(1)}_{b,s_1}(\sfS)-[\delta_{g_b}^*, t_{\chi_0}]\omega_{b,s_1}(\sfS)\Bigr)-[\tilde{\delta}_{g_b,\gamma},t_{\chi_0}]G_{g_b}\delta_{g_b}^*\omega_{b,s_1}(\sfS)\in\eHb^{\infty,1/2-}(\CX)
		\end{split}
	\end{equation}
where we use $ \tilde{\delta}_{g_b,\gamma}\in\rho\mbox{Diff}_\bop^1$, $\delta_{g_b}^*\omega^{(1)}_{b,s_1}(\sfS)-[\delta_{g_b}^*,t_{|chi_0}]\omega_{b,s_1}(\sfS)\in\eHb^{\infty,-1/2-}(\CX)$ and \[\delta_{g_b}^*\omega_{b,s_1}(\sfS)\in\eHb^{\infty,1/2-}(\CX),\quad [\tilde{\delta}_{g_b,\gamma},t_{\chi_0}]\in\mathcal{A}^0(\CX).
\]Since $\ker\widehat{\mathcal{W}_{b,\gamma}}(0)^*\cap\sHb^{-\infty,-1/2+}(\CX)=\{0\}$, we can solve the above equation \eqref{eq:eqnfortildeomega} for $\tilde{\omega}_{b,s_1}(\sfS)\in\eHb^{\infty,-3/2-}(\CX)$, which is unique modulo $\ker\widehat{\mathcal{W}_{b,\gamma}}(0)\cap\eHb^{\infty,-3/2-}(\CX)$. We then define \[\hat{\omega}^{(1)}_{b,s_1}(\sfS):=t_{\chi_0}\omega_{b,s_1}(\sfS)-\omega_{b,s_1}^{(1)}(\sfS)+\tilde{\omega}_{b, s_1}(\sfS)
\]
where $\tilde{\omega}_{b,s_1}(\sfS)$ is in the orthogonal complement $\big(\ker\widehat{\mathcal{W}_{b, \gamma}}(0)\cap\eHb^{\infty,-3/2-}(\CX)\big)^\perp$ and thus uniquely determined by $\sfS$ and $b$. As a result, we have for $r\gg1$
\[
\hat{\omega}_{b,s_1}(\sfS)=t\omega_{b,s_1}(\sfS)-\chi r\sfS dt+\eHb^{\infty,-3/2}(\CX)=t d(r\sfS)-r\sfS dt+\mbox{Poly}^1(t_{b,*})\eHb^{\infty,-3/2-}(\CX).
\]
Finally, the continuity in $b$ follows in the same way as in the proof of Lemma \ref{lem:pre_asym_lor}.

The proof for $\hat{\omega}^*_{b,s_1}(\sfS)$ is completely analogous.

It remains to derive the expressions in \eqref{eq:refinedge1formmode}. For the simplicity of calculation, now we let $\tilde{t}=t_0-r$ and make the ansatz
\[
\hat{\omega}_{b_0,s_1}(\sfS)=\tilde{t}\omega_{b_0,s_1}(\sfS)-\omega^{(1)}_{b_0,s_1}+\tilde{\omega}_{b_0,s_1}(\sfS).
\]
Our aim is to solve the following equation for $\tilde{\omega}_{b_0,s_1}(\sfS)\in\eHb^{-3/2-}(\CX)$ 
\begin{equation}\label{eq:eqfortildeomega}
	\begin{split}
\widehat{\mathcal{W}_{b_0, \gamma}}(0)\tilde{\omega}_{b_0,s_1}(\sfS)&=-[\mathcal{W}_{b_0,\gamma}, \tilde{t}\,]\omega_{b_0,s_1}(\sfS)=\frac 12[\Box_{g_{b_0}}, \tilde{t}\,]\omega_{b_0,s_1}(\sfS)+\mathcal{O}_{\eHb^{\infty,\infty}(\CX)}(\gamma)\\
&=\sfS\Big((\frac{\Bm_0}{r^2}-\frac{\BQ_0^2}{r^3})dt_0+\frac{\BQ_0^2}{r^3}dr\Big)+\frac{\Bm_0}{r}(1+\frac{\Bm_0}{r}-\frac{\BQ_0^2}{r^2})\sd\sfS+\mathcal{O}_{\eHb^{\infty,3/2-}(\CX)}(\gamma+\abs{\BQ_0}^2)
\end{split}
\end{equation}
where we use $[\Box_{g_{b_0}}, \tilde{t}\,]=\Box_{g_{b_0},0}\tilde{t}+2\nabla^\al\tilde{t}\nabla_\al\in\rho^2\mbox{Diff}_\bop^1$ and \eqref{eq:refinedg1formmode} in the last step. A direct calculation implies 
\begin{align*}
	&\widehat{\mathcal{W}_{b_0, \gamma}}(0)\Bm_0(dt_0+dr)=\delta_{g_{b_0}}G_{g_{b_0}}\delta^*_{g_{b_0}}\Bm_0(dt_0+dr)+\mathcal{O}_{\eHb^{\infty,\infty}(\CX)}(\gamma)\\
	&\quad=\sfS\Big((\frac{\Bm_0}{r^2}+\frac{\Bm_0\BQ_0^2}{r^4})dt_0\!+\!(\frac{3\Bm_0}{r^2}-\frac{2\Bm_0^2}{r^3}\!+\!\frac{\Bm_0\BQ_0^2}{r^4})dr\Big)\!+\!\frac{\Bm_0}{r}(-2\!+\!\frac{2\Bm_0}{r}\!-\!\frac{2\BQ_0^2}{r^2})\sd\sfS+\!\mathcal{O}_{\eHb^{\infty,\infty}(\CX)}(\gamma).
\end{align*}
Then
\begin{align*}
	&\widehat{\mathcal{W}_{b_0, \gamma}}(0)\Big(\tilde{\omega}_{b_0,s_1}(\sfS)-\Bm_0(dt_0+dr)\Big)\\
	&\qquad=\sfS(-\frac{3\Bm_0}{r^2}+\frac{2\Bm_0^2}{r^3})dr+\!\frac{\Bm_0}{r}(3-\frac{\Bm_0}{r})\sd\sfS+\mathcal{O}_{\eHb^{\infty,3/2-}(\CX)}(\gamma+\abs{\BQ_0}^2)\\
	&\qquad=d\Big((\frac{3\Bm_0}{r}-\frac{\Bm_0^2}{r^2})\sfS\Big)+\mathcal{O}_{\eHb^{\infty,3/2-}(\CX)}(\gamma+\abs{\BQ_0}^2)
\end{align*}
Meanwhile, with $u=2\Bm_0(2\Bm_0+(\Bm_0-r)\log(r/\Bm_0))$, we have
 \begin{align*}
 	\widehat{\mathcal{W}_{b_0, \gamma}}(0)du&=\frac 12d\delta_{g_{b_0}}du-\Ric(g_0)_\al^{\ \beta}(du)_\beta+F_{g_{b_0}}G_{g_{b_0}}du\\
 	&=d\Big((\frac{3\Bm_0}{r}-\frac{\Bm_0^2}{r^2})\sfS\Big)+\mathcal{O}_{\eHb^{\infty,5/2-}(\CX)}(\gamma+\abs{\BQ_0}^2).
 \end{align*}
Therefore, we obtain
\[
\widehat{\mathcal{W}_{b_0,\gamma}}(0)\Big(\tilde{\omega}_{b_0, s_1}(\sfS)-\Bm_0(dt_0+dr)-du\Big)=\mathcal{O}_{\eHb^{\infty,3/2-}(\CX)}(\abs{\BQ_0}^2+\gamma).
\]
Since the right-hand side of the above equation is of scalar type $l=1$ and $\widehat{\mathcal{W}_{b_0,\gamma}}(0)^{-1}$ restricted in scalar type $l=1$ $1$-form spaces exists with norm of size $\mathcal{O}(1)$ (see Remark \ref{rem:invertibilityofscalar1}), it follows that $\tilde{\omega}_{b_0, s_1}(\sfS)-\Bm_0(dt_0+dr)-du=\mathcal{O}_{\eHb^{\infty,-1/2-}(\CX)}(\gamma+\abs{\BQ_0}^2)$. Therefore, \eqref{eq:refinedge1formmode} follows from the combination of this fact and \eqref{eq:refinedg1formmode5/2}.
\end{proof}

%%%%%%%%%%%%%%%%%%%%%%%%%%%%%%%%%%%%%%%%%%%%%%%%%%%%%%%%%%%%%%%%%%%%%%%%%%%
\section{Modes of the linearized gauge-fixed Einstein-Maxwell system}
\label{sec:modeanalsysiofGFEM}
In this section, we will discuss the modes of the linearized gauge-fixed Einstein-Maxwell system 
\begin{equation}\label{eq:mlgEM}
	L_{(g_b,A_b),\gamma}(\dg,\dA)=(2L^E_{(g_b,A_b),\gamma}(\dg,\dA),\ L^M_{(g_b,A_b),\gamma}(\dg,\dA))=0
\end{equation}
where
\begin{gather}\label{eq:mlgE}
	L^E_{(g_b,A_b),\gamma}(\dg,\dA)=D_{g_b}\Ric(\dg)-2D_{(g_b, dA_b)}T(\dg,d\dA)+\wt{\delta}^*_{g_b,\gamma}\wt{\delta}_{g_b,\gamma}G_{g_b}\dg=0,\\\label{eq:mlgM}
	L^M_{(g_b,A_b),\gamma}(\dg,\dA)=D_{(g_b, A_b)}(\delta_gdA)(\dg, \dA)+d\delta_{g_b}\dA=0.
\end{gather}
on the RN metric and $4$-electromagnetic potential, i.e., the case $b=b_0=(\Bm_0,0,\BQ_0), \abs{\BQ_0}\ll\Bm_0$.

For the brevity of notation, we rewrite the Lie derivative of a $1$-form $\omega$ with respect to the vector field $X$ as
\[
\wt{\mathcal{L}}_{\omega}X:=\mathcal{L}_{X}\omega
\]
and put
\begin{equation}\label{eq:simplifiedmlgEM}
	L_{b,\gamma}:=L_{(g_b,A_b),\gamma},\quad 	L^E_{b,\gamma}:=L^E_{(g_b,A_b),\gamma},\quad	L^M_{b,\gamma}:=L^M_{(g_b,A_b),\gamma}.
\end{equation}
For the various spaces in this section, for instance, $\eHb^{s,\ell}(\CX;E), \sHb^{s,\ell}(\CX;E),\mathcal{A}(\CX;E)$, etc., where $E\to \CX$ is a vector bundle over $\CX$, we will drop the notation $E$ of the bundle if it it clear from the context. 

We shall prove that $L_{b_0,\gamma}$ has no non-zero modes in the closed upper half plane for RN metric and $4$-electromagnetic potential $(g_{b_0},A_{b_0})$ (we note that the same result for slowly rotating KN is nontrivial and will be discussed in detail in next section). Moreover, we will describe the space of zero modes and generalized zero modes (with linearly growth in $t_{b,*}$) of $L_{b,\gamma}$ for both RN case and slowly rotating KN case.

\subsection{Modes in the closed upper half plane} 
Let $(g_b,A_b)$ be the KN metric and $4$-electromagnetic potential with $b=(\Bm,\Ba,\BQ)$ near $b_0=(\Bm_0, 0,\BQ_0)$ and the linearized gauge-fixed Einstein-Maxwell operator $L_{b,\gamma}:\eHb^{s,\ell}(\CX;\scsym\oplus\scform)\to\eHb^{s,\ell}(\CX;\scsym\oplus\scform)$ be defined as in \eqref{eq:simplifiedmlgEM}, \eqref{eq:mlgEM}, \eqref{eq:mlgE} and \eqref{eq:mlgM}. In the definition of $\tilde{\delta}_{g_b,\gamma}^*,\tilde{\delta}_{g_b,\gamma}$(see \eqref{eq:modifieddeltaFirst}, \eqref{eq:defofBFirst} and \eqref{eq:defofFFirst}), $\mathfrak{c}$ is defined as in \eqref{eq:defofc} and $\gamma\in(0, \min\{\gamma_0, \tilde{\gamma}_0\})$ where $\gamma_0,\tilde{\gamma}_0$ are as in Theorem \ref{thm:modesforCP} and \ref{thm:modesforW} respectively. We also note that we consider the weakly charged case, namely, $\abs{\BQ_0}\ll\Bm_0$ (more precisely, $\abs{\BQ_0}\leq C(\gamma)$ for some small constant $C(\gamma)$). As in the previous sections, we define the Fourier-transformed operator 
\begin{equation}
	\widehat{L_{b,\gamma}}(\sigma):=e^{i\sigma t_{b,*}}L_{b,\gamma}e^{-i\sigma t_{b,*}}
\end{equation} 
using the function $t_{b,*}=\chi_0(r)(t+r_{b_0,*})+(1-\chi_0(r))(t-r_{(\Bm,0,\BQ),*})$ defined in \eqref{EqKNTimeFn}.

\begin{thm}\label{thm:modestabilityofmlgEM}
	Let $(g_{b_0}, A_{b_0})$ be the RN metric and $4$-electromagnetic potential where $b_0=(\Bm_0, 0,\BQ_0)$ with $\abs{\BQ_0}\ll\Bm_0$. Then the Fourier-transformed operator $	\widehat{L_{b_0,\gamma}}(\sigma)$ satisfies the following statements:
	\begin{enumerate}
		\item If $\IM \sigma\geq0$ and $\sigma\neq 0$, 
	\begin{equation}\label{eq:mlgEMmodenonzero}
		\begin{split}
&\widehat{L_{b_0,\gamma}}(\sigma):\{(\dg,\dA)\in\eHb^{s, \ell}(\CX;\scsym\oplus\scform): 	\widehat{L_{b_0,\gamma}}(\sigma)(\dg,\dA)\in\eHb^{s,\ell+1}(\CX; \scsym\oplus\scform)\}\\
 &\qquad\qquad\qquad\to\eHb^{s,\ell+1}(\CX;\scsym\oplus\scform)
\end{split}
	\end{equation}
	is invertible for $s>3, \ell<-\frac12$ and $s+\ell>-\frac 12$.
	\item If $s>3$ and $-\frac 32<\ell<-\frac 12$, then stationary operator
	\begin{equation}\label{eq:mlgEMmodezero}
		\begin{split}
&\widehat{L_{b_0,\gamma}}(0):\{(\dg,\dA)\in\eHb^{s, \ell}(\CX;\scsym\oplus\scform): 	\widehat{L_{b_0,\gamma}}(0)(\dg,\dA)\in\eHb^{s-1,\ell+2}(\CX; \scsym\oplus\scform)\}\\
&\qquad\qquad\qquad\to\eHb^{s-1,\ell+2}(\CX;\scsym\oplus\scform)
		\end{split}
	\end{equation}
	has $8$-dimensional kernel and cokernel.
\end{enumerate}
The second statement also holds for the weakly charged and slowly rotating KN metric and $4$-electromagnetic potential $(g_b, A_b)$ with $b=(\Bm, \Ba,\BQ)$ near $b_0=(\Bm_0, 0,\BQ_0)$. Concretely, we have
	\begin{align}\label{eq:mlgEMzeromode}
\mathcal{K}_b:=	\ker \widehat{L_{b,\gamma}}(0)\cap\eHb^{\infty,-1/2-}(\CX)&=\{(\dg_{b,s_0},\dA_{b,s_0})\}\oplus\{ (\dg_{b,s_1}(\sfS),\dA_{b,s_1}(\sfS)):\sfS\in\BFS_1\}\\\notag
	&\qquad\oplus\{ (\dg_{b,v_1}(\sfV),\dA_{b,v_1}(\sfV)):\sfV\in\BFV_1\},\\\label{eq:mlgEMdualzeromode}
\mathcal{K}_b^*:=	\ker \widehat{L_{b,\gamma}}(0)^*\cap\sHb^{-\infty,-1/2-}(\CX)&=\{(\dg^*_{b,s_0},\dA^*_{b,s_0})\}\oplus\{ (\dg^*_{b,s_1}(\sfS),\dA^*_{b,s_1}(\sfS)):\sfS\in\BFS_1\}\\\notag
	&\qquad\oplus\{ (\dg^*_{b,v_1}(\sfV),\dA^*_{b,v_1}(\sfV)):\sfV\in\BFV_1\},
\end{align}
where 
\begin{align}\label{eq:mlgzeromodes0}
	(\dg_{b,s_0},\dA_{b,s_0})&=(\dg_{b}(\dot{\Bm},0,\dot{\BQ})+2\delta_{g_b}\tilde{\omega}_{b,s_0},\  \dA_b(0,0,\dot{\BQ})+\wt{\mathcal{L}}_{A_b}\tilde{\omega}_{b,s_0}^{\sharp}+d\phi_{b,s_0}),\quad\dot{\Bm},\dot{\BQ}\in\BR,\\\label{eq:mlgzeromodes1}
	(\dg_{b,s_1}(\sfS),\dA_{b,s_1}(\sfS))&=(2\delta_{g_b}^*\omega_{b,s_1}(\sfS),\  \wt{\mathcal{L}}_{A_b}(\omega_{b,s_1}(\sfS))^{\sharp}+d\phi_{b,s_1}(\sfS)),\\\label{eq:mlgzeromodev1}
	(\dg_{b,v_1}(\sfV),\dA^*_{b,v_1}(\sfV))&=(\dg_{b}(0,\dot{\Ba},0)+2\delta_{g_b}^*\tilde{\omega}_{b,v_1}(\sfV),\  \dA_b(0,\dot{\Ba},0)+\wt{\mathcal{L}}_{A_b}(\tilde{\omega}_{b,v_1}(\sfV))^{\sharp}+d\phi_{b,v_1}(\sfV)),
\end{align}
and
\begin{align}\label{eq:mlgdualzeromodes0}
(\dg^*_{b,s_0},\dA^*_{b,s_0})&=c_1(2G_{g_b}\delta_{g_b}^*\omega^*_{b,s_0},\  4\wt{\mathcal{L}}_{A_b}(\omega^*_{b,s_0})^{\sharp}+d\phi_{b,s_0}^*)+c_2(0, \delta(r-\ehKN)dr),\quad c_1,c_2\in\BR,\\\label{eq:mlgdualzeromodes1}
(\dg^*_{b,s_1}(\sfS),\dA^*_{b,s_1}(\sfS))&=(2G_{g_b}\delta_{g_b}^*\omega^*_{b,s_1}(\sfS),\  4\wt{\mathcal{L}}_{A_b}(\omega^*_{b,s_1}(\sfS))^{\sharp}+d\phi_{b,s_1}^*(\sfS)),\\\label{eq:mlgdualzeromodev1}
(\dg^*_{b,v_1}(\sfV),\dA^*_{b,v_1}(\sfV))&=(2G_{g_b}\delta_{g_b}^*\omega^*_{b,v_1}(\sfV),\  4\wt{\mathcal{L}}_{A_b}(\omega^*_{b,v_1}(\sfV))^{\sharp}+d\phi_{b,v_1}^*(\sfV))
\end{align}
with $\omega_{b,s_1}(\sfS)$, $\omega^*_{b,s_0}$ and $\omega^*_{b,s_1}(\sfS)$, $\omega^*_{b,v_1}(\sfV)$ defined as in \eqref{eq:g1form3/2}, \eqref{eq:g1formdual3/2} and \eqref{eq:g1formdual5/2}. 

Moreover, we have \[(\dg_{b,s_1}(\sfS),\ \dA_{b,s_1}(\sfS)),\quad(\dg_{b,v_1}(\sfV),\ \dA_{b,v_1}(\sfV))\in\eHb^{\infty,1/2-}(\CX),
\]
and the maps $b\mapsto (\dg_{b, \bullet}(\bullet),\ \dA_{b,\bullet}(\bullet))$ and $b\mapsto (\dg^*_{b, \bullet}(\bullet),\ \dA^*_{b,\bullet}(\bullet))$ can be chosen to be continuous in $b$ with values in the respective spaces. 

For later use, we further determine the leading term of $(\dg_{b_0,v_1}(\sfV),\ \dA_{b_0,v_1}(\sfV))$
\begin{equation}\label{eq:refinedhv1}
	\begin{split}
	\dg_{b_0,v_1}(\sfV)&=\Big((-\frac{4\Bm_0}{r}+\frac{2\BQ_0^2}{r^2})dt_0+\frac{4\Bm_0}{r-(\Bm_0-\sqrt{\Bm_0^2-\BQ_0^2})}dr\Big)\otimes_s\sfV+\mathcal{O}_{\eHb^{\infty,1/2-}(\CX)}(\gamma),\\
	\dA_{b_0,v_1}(\sfV)&=\mathcal{O}_{\eHb^{\infty,1/2-}(\CX)}(\abs{\BQ_0}).
	\end{split}
	\end{equation}
\end{thm}

\begin{rem}\label{rem:gaugecondition}
	In the course of the proof of the above theorem, we will find that the zero modes in $\mathcal{K}_b$ satisfy both the linearized Einstein-Maxwell system and the linearized generalized wave map gauge and Lorenz gauge conditions.
\end{rem}

\begin{proof}[Proof of Theorem \ref{thm:modestabilityofmlgEM}]
	We first analyze the RN case $b=b_0$ and write $(g, A)=(g_{b_0}, A_{b_0})$. Suppose $\widehat{L_{b_0,\gamma}}(\sigma)(\dg,\dA)=0$. Then by Proposition \ref{prop:desofkernel}, $(\dg,\dA)\in\eHb^{\infty,-1/2-}(\CX)$. In view of \eqref{eq:mlgM}, applying $\delta_g$ to $L^ M_{b_0,\gamma}e^{-i\sigma t_{b_0,*}}(\dg,\dA)=0$ yields
	\[
	\delta_g d(\delta_ge^{-i\sigma t_{b_0,*}}\dA)=0.
	\]
	Since $\dA\in\eHb^{\infty,-1/2-}(\CX;\scform)$,  we obtain that  $\delta_g(e^{-i\sigma t_{b_0,*}}\dA)$ is a mode. More precisely, we have \[
	\delta_g(e^{-i\sigma t_{b_0,*}}\dA)\in\begin{cases} e^{-i\sigma t_{b_0,*}}\eHb^{\infty,-1/2-}(\CX;\BC)&\quad\mbox{if}\quad \sigma\neq0,\\
		\eHb^{\infty,1/2-}(\CX;\BC)&\quad\mbox{if}\quad \sigma=0.
		\end{cases}
	\]
	By Theorem \ref{thm:modesforscalarwave}, it follows that
	\begin{equation}\label{eq:wavepotemtialeqnM}
		\delta_g(e^{-i\sigma t_{b_0,*}}\dA)=0,
	\end{equation}
	and thus $e^{-i\sigma t_{b_0,*}}(\dg, \dA)$ is a mode solution to the Maxwell part of the linearized Einstein-Maxwell system. Applying $\delta_gG_g$ to $L^ E_{b_0,\gamma}e^{-i\sigma t_{b_0,*}}(\dg,\dA)=0$ and using the linearized second Bianchi identity (see the discussion around \eqref{eq:Lsecondbianchi} and \eqref{eq:LsecondBianchiM}) give
	\[
	\delta_gG_g\tilde{\delta}_{g,\gamma}^*(\tilde{\delta}_{g,\gamma}G_ge^{-i\sigma t_{b_0,*}}\dg)=\mathcal{P}_{b_0,\gamma}(\tilde{\delta}_{g,\gamma}G_ge^{-i\sigma t_{b_0,*}}\dg)=0.
	\]
	Again, since $\tilde{\delta}_{g,\gamma}G_ge^{-i\sigma t_{b_0,*}}\dg$ is a mode in the same way as $\delta_g(e^{-i\sigma t_{b_0,*}}\dA)$, it follows from Theorem \ref{thm:modesforCP} that 
	\begin{equation}\label{eq:wavepotemtialeqnE}
		\tilde{\delta}_{g,\gamma}G_ge^{-i\sigma t_{b_0,*}}\dg=0,
	\end{equation}
	and thus $e^{-i\sigma t_{b_0,*}}(\dg, \dA)$ is a mode solution to the linearized Einstein-Maxwell system.

	\begin{itemize}
		\item \underline{The case $\IM\sigma\geq 0, \sigma\neq0$.} Then we apply the mode stability result Theorem \ref{thm:modestability} and find that
	\begin{equation}\label{eq:exfornonzeromodesoltoEM}
e^{-i\sigma t_{b_0,*}}(\dg, \dA)=(2\delta^*_g\omega, \mathcal{L}_{\omega^\sharp}A+d\phi)
	\end{equation}
	where $(\omega,\phi)=e^{-i\sigma t_{b_0,*}}(\omega_0,\phi_0)\in e^{-i\sigma t_{b_0,*}}\eHb^{\infty,\ell'}(\CX;\scform\oplus\BC)$ with $\ell'<-1/2$. Plugging \eqref{eq:exfornonzeromodesoltoEM} into the equation \eqref{eq:wavepotemtialeqnE}, we find
	\[
	\widehat{\mathcal{W}_{b_0,\gamma}}(\sigma)\omega_0=0,
	\]
	and thus $\omega_0=0$ by Theorem \ref{thm:modesforW}. We then plug \eqref{eq:exfornonzeromodesoltoEM} with $\omega=0$ into the equation \eqref{eq:wavepotemtialeqnM} and obtain
	\[
	\widehat{\Box_g}(\sigma)\phi_0=0.
	\]
	Therefore, we have $\phi_0=0$ by Theorem \ref{thm:modesforscalarwave}. This proves the injectivity of $\widehat{L_{b_0,\gamma}}(\sigma)$ for $\IM\sigma\geq 0, \sigma\neq0$, and the surjectivity follows from the fact that $\widehat{L_{b_0,\gamma}}(\sigma)$ is Fredholm of index $0$.

	\item\underline{Scalar type $l\geq2$ or vector type $l\geq2$ zero modes.} We apply the mode stability result Theorem \ref{thm:modestability} and find that
	\begin{equation}\label{eq:exforl2zeromodesoltoEM}
	(\dg, \dA)=(2\delta^*_g\omega, \mathcal{L}_{\omega^\sharp}A+d\phi)
	\end{equation}
	where $(\omega,\phi)\in \eHb^{\infty,-3/2-}(\CX;\scform\oplus\BC)$. Plugging \eqref{eq:exforl2zeromodesoltoEM} into the equation \eqref{eq:wavepotemtialeqnE}, we find
	\[
	\widehat{\mathcal{W}_{b_0,\gamma}}(0)\omega=0.
	\]
	Since $\omega\in\eHb^{\infty, -3/2-}(\CX)$ is of scalar or vector type $l\geq 2$, it follows that $\omega=0$ by Theorem \ref{thm:modesforW} and Proposition \ref{prop:asym_trans}. We then plug \eqref{eq:exforl2zeromodesoltoEM} with $\omega=0$ into the equation \eqref{eq:wavepotemtialeqnM} and obtain
	\[
	\widehat{\Box_g}(0)\phi=0.
	\]
	Again, since $\phi\in\eHb^{\infty,-3/2-}(\CX)$ with vector or scalar type $l\geq2$, we conclude that  $\phi=0$ by Theorem \ref{thm:modesforscalarwave} and Proposition \ref{prop:gscalarzeromode}. This proves the injectivity of $\widehat{L_{b_0,\gamma}}(0)$ when restricted on scalar or vector type $l\geq 2$ modes.

	\item\underline{Scalar type $l=1$ zero modes.} As in the previous case (scalar or vector $l\geq 2$ case), we find that $(\dg, \dA)=(2\delta^*_g\omega, \mathcal{L}_{\omega^\sharp}A+d\phi)$
	where $(\omega,\phi)\in \eHb^{\infty,-3/2-}(\CX;\scform\oplus\BC)$ and satisfies first
	\[
	\widehat{\mathcal{W}_{b_0,\gamma}}(0)\omega=0.
	\]
	Since $\omega\in\eHb^{\infty, -3/2-}(\CX)$  is of scalar type $l=1$, in view of Proposition \ref{prop:asym_trans} we see that 
	\begin{equation}\label{eq:exforscalarl1gpE}
		\omega=\omega_{b_0,s_1}(\sfS)\quad\mbox{for some}\quad \sfS\in\BFS_1
	\end{equation}
where $\omega_{b_0,s_1}(\sfS)$ is defined as in \eqref{eq:g1form3/2}. We then plug $\dA=\mathcal{L}_{\omega^\sharp}A+d\phi$ and \eqref{eq:exforscalarl1gpE} into the equation \eqref{eq:wavepotemtialeqnM} and obtain
	\[
-\widehat{\Box_g}(0)\phi=-\delta_g\wt{\mathcal{L}}_A(\omega_{b_0,s_1}(\sfS))^\sharp\in\BQ_0\eHb^{\infty,3/2-}(\CX)
	\]
	where we use $\delta_g\in\rho\mbox{Diff}_\bop^1$ and $\mathcal{L}_{(\cdot)^\sharp} A\in\rho^2\mbox{Diff}_\bop^1$ (see \eqref{eq:liederiofA}). Owing to Theorem \ref{thm:modesforscalarwave} and the fact that $\ker \widehat{\Box_g}(0)\cap\eHb^{\infty,-3/2-}(\CX)$ restricted to scalar type $l=1$ spaces is $0$, there exists a unique $\phi\in\eHb^{\infty,-1/2-}(\CX)$, which we denote by $\phi_{b_0,s_1}(\sfS)$, such that the above equation holds. Therefore, the scalar type $l=1$ nullspace of $\widehat{L_{b_0,\gamma}}(0)$ is $3$-dimensional.  We also note that $\phi_{b_0,s_1}(\sfS)$ is of size $\mathcal{O}_{\eHb^{\infty,-1/2-}(\CX)}(\abs{\BQ_0})$ and thus $\dA_{b_0,s_1}(\sfS)=\mathcal{O}_{\eHb^{\infty,1/2-}(\CX)}(\abs{\BQ_0})$. In view of \eqref{eq:symgra3/2}, we see that $\dg_{b_0,s_1}(\sfS)=2\delta_g^*\omega_{b_0,s_1}(\sfS)\in\eHb^{\infty,1/2-}(\CX;\scsym)$.

	\item\underline{Scalar type $l=0$ zero modes.} If $(\dg,\dA)$ is of scalar type $l=0$, by Theorem \ref{thm:modestability} we have
	\begin{equation}\label{eq:exforl0zeromode}
		(\dg,\dA)=(\dg_{b_0}(\dot{\Bm},0,\dot{\BQ})+2\delta^*_{g}\omega,\  \dA_{b_0}(0,0,\dot{\BQ})+\mathcal{L}_{\omega^\sharp}A+d\phi)
	\end{equation}
where $(\omega,\phi)\in \eHb^{\infty,-3/2-}(\CX;\scform\oplus\BC)$. Plugging \eqref{eq:exforl0zeromode} into \eqref{eq:wavepotemtialeqnE}, it follows that
\[
2\widehat{\mathcal{W}_{b_0, \gamma}}(0)\omega=-\tilde{\delta}_{g,\gamma}G_g\dg_{b_0}(\dot{\Bm},0,\dot{\BQ})\in\eHb^{\infty,1/2-}(\CX).
\]
In view of Theorem \ref{thm:modesforW}, $\ker\widehat{\mathcal{W}_{b_0,\gamma}}(0)^*\cap\sHb^{-\infty,-1/2+}(\CX;\scform)=\{0\}$. Therefore, we can solve the above equation for $\omega\in\eHb^{\infty,-3/2-}(\CX)$ which is of scalar type $l=0$ and unique modulo $\langle\pa_t^\flat\rangle$. However, since $\delta^*_g\pa_t^\flat=0$ and $\mathcal{L}_{\pa_t}A=0$, a multiple of $\pa_t^\flat$ does not change $\dg$ and the equation for $\phi$ below. This implies $\omega$ is unique and we denote it by $\tilde{\omega}_{b_0,s_0}$. Then we plug $\dA=\dA_{b_0}(0,0,\dot{\BQ})+\mathcal{L}_{\omega^\sharp}A+d\phi$ and $\omega=\tilde{\omega}_{b_0,s_0}$ into the equation \eqref{eq:wavepotemtialeqnM} and obtain
\[
-\widehat{\Box_g}(0)\phi=-\delta_g\dA_{b_0}(0,0,\dot{\BQ})-\delta_g\mathcal{L}_{\tilde{\omega}_{b_0,s_0}^\sharp}A\in\eHb^{\infty,1/2-}(\CX)
\]
By Theorem \ref{thm:modesforscalarwave}, we have $\ker\widehat{\Box_g}(0)^*\cap\sHb^{-\infty,-1/2+}(\CX;\BC)=\{0\}$. Therefore, we can solve the above equation for $\phi\in\eHb^{\infty,-3/2-}(\CX;\BC)$ which is of scalar type $l=0$ and unique modulo $\langle1\rangle$. Again, adding a constant to $\phi$ make no change to $\dA$, which means that $\phi$ is unique. We then denote this unique $\phi$ by $\phi_{b_0,s_0}$. This proves that the scalar type $l=0$ kernel of $\widehat{L_{b_0,\gamma}}(0)$ is $2$-dimensional.

	\item\underline{Vector type $l=1$ zero modes.} If $(\dg,\dA)$ is of vector type $l=1$, by Theorem \ref{thm:modestability} we have
	\begin{equation}\label{eq:exforvectorl1zeromode}
		(\dg,\dA)=(\dg_{b_0}(0,\dot{\Ba},0)+2\delta^*_{g}\omega,\  \dA_{b_0}(0,\dot{\Ba},0)+\mathcal{L}_{\omega^\sharp}A+d\phi)
	\end{equation}
	where $(\omega,\phi)\in \eHb^{\infty,-3/2-}(\CX;\scform\oplus\BC)$. Plugging \eqref{eq:exforvectorl1zeromode} into \eqref{eq:wavepotemtialeqnE}, it follows that
	\[
	2\widehat{\mathcal{W}_{b_0, \gamma}}(0)\omega=-\tilde{\delta}_{g,\gamma}G_g\dg_{b_0}(0,\dot{\Ba},0)\in\eHb^{\infty,3/2-}(\CX).
	\]
	In view of Theorem \ref{thm:modesforW},
	\begin{comment} $\ker\widehat{\mathcal{W}_{b_0,\gamma}}(0)^*\cap\sHb^{-\infty,-1/2+}(\CX;\scform)=\{0\}$, and thus the above equation is solvable for $\omega\in\eHb^{\infty,-3/2-}(\CX;\scform)$. Meanwhile, since $\omega$ is of vector type $l=1$, by Proposition \ref{prop:asym_trans},
	\end{comment}
	 there exists a unique $\omega\in\eHb^{\infty,-1/2-}(\CX;\scform)$ of vector type $l=1$ satisfying the above equation. We then denote this unique $\omega$ by $\tilde{\omega}_{b_0,v_1}(\sfV)$. Therefore, we have $\dg_{b_0, v}(\sfV)\in\eHb^{\infty,1/2-}(\CX)$.  Plugging $\dA=\dA_{b_0}(0,\dot{\Ba},0)+\mathcal{L}_{\omega^\sharp}A+d\phi$ and $\omega=\tilde{\omega}_{b_0,v_1}$ into the equation \eqref{eq:wavepotemtialeqnM}, we obtain that
	\[
	-\widehat{\Box_g}(0)\phi=-\delta_g\dA_{b_0}(0,\dot{\Ba},0)-\delta_g\wt{\mathcal{L}}_A(\tilde{\omega}_{b_0,v_1}(\sfV))^\sharp\in\BQ_0\eHb^{\infty,3/2-}(\CX)
	\]
	By Theorem \ref{thm:modesforscalarwave}, there exists a unique $\phi\in\eHb^{\infty,-1/2-}$ of vector type $l=1$ satisfying the above equation. We then denote this unique $\phi$ by $\phi_{b_0,v_1}(\sfV)$. Therefore, the vector $l=1$ kernel of $\widehat{L_{b_0,\gamma}}(0)$ is $3$-dimensional. We also note that $\phi_{b_0,v_1}$ is of size $\mathcal{O}_{\eHb^{\infty,-1/2-}(\CX;\BC)}(\abs{\BQ_0})$, and thus $\dA_{b_0,v_1}=\mathcal{O}_{\eHb^{\infty,1/2-}(\CX;\BC)}(\abs{\BQ_0})$.
	
	Now we shall derive the expression for $(\dg_{b_0,v_1}(\sfV),\dA_{b_0,v_1}(\sfV))$ in \eqref{eq:refinedhv1}. We write
	\[
	\dg_{b_0,v_1}(\sfV)=\dg^0_{b_0}(0,\dot{\Ba},0)+2\delta_{g}^*\omega^0.
	\]
	When rescaling to $\abs{\dot{\Ba}}=1$ and letting $(\theta,\varphi)$ be the spherical coordinates adapted to $\dot{\Ba}$, we can rewrite
	\[
		\dg_{b_0,v_1}(\sfV)=(2(\weight-1)dt_0-2dr)\otimes_s\sfV+2\delta_{g}^*\omega^0\quad\mbox{with}\quad\sfV=\sin^2\theta d\varphi.
	\]
	Plugging $\dg_{b_0,v_1}(\sfV)$ into \eqref{eq:wavepotemtialeqnE} yields
	\[
2\widehat{\mathcal{W}_{b_0, \gamma}}(0)\omega^0=-\tilde{\delta}_{g,\gamma}G_g\Big((2(\weight-1)dt_0-2dr)\otimes_s\sfV\Big)=-2r^{-1}\sfV+\mathcal{O}_{\eHb^{\infty,\infty}(\CX)}(\gamma).
	\]
	We also find that
	\[
\delta_g^*f(r)\sfV=\Big(1+\frac{2\Bm_0}{r-(\Bm_0-\sqrt{\Bm_0^2-\BQ_0^2})}\Big)dr\otimes_s\sfV,\quad	\delta_gG_g\delta_g^* f(r)\sfV=-r^{-1}\sfV
	\]
	with
	\[
	f(r)=(\frac{2\Bm_0}{\Bm_0-\sqrt{\Bm_0^2-\BQ_0^2}}-1)r+\frac{2\Bm_0}{(\Bm_0-\sqrt{\Bm_0^2-\BQ_0^2})^2}r^2\ln(1-\frac{\Bm_0-\sqrt{\Bm_0^2-\BQ_0^2}}{r}).
	\]
Then it follows that
\[
2\widehat{\mathcal{W}_{b_0, \gamma}}(0)(\omega^0-f(r)\sfV)=\mathcal{O}_{\eHb^{\infty,\infty}(\CX)}(\gamma),
\]
and thus $\omega^0-f(r)\sfV=\mathcal{O}_{\eHb^{\infty,-1/2-}(\CX)}(\gamma)$ (according to Remark \ref{rem:invertibilityofscalar1}, $\widehat{\mathcal{W}_{b_0,\gamma}}(0)^{-1}$ restricted in vector type $l=1$ $1$-form spaces exists with norm of size $\mathcal{O}(1)$ ). This implies
\begin{equation}
	\begin{split}
	\quad \qquad\dg_{b_0,v_1}(\sfV)&\!\!=(2(\weight-1)dt_0-2dr)\otimes_s\sfV+2\delta_g^*f(r)\sfV\!+\!\mathcal{O}_{\eHb^{\infty,1/2-}(\CX)}(\gamma)\\
	&=\Big((-\frac{4\Bm_0}{r}+\frac{2\BQ_0^2}{r^2})dt_0+\frac{4\Bm_0}{r-(\Bm_0-\sqrt{\Bm_0^2-\BQ_0^2})}dr\Big)\otimes_s\sfV+\mathcal{O}_{\eHb^{\infty,1/2-}(\CX)}(\gamma).
	\end{split}
\end{equation}
Therefore, $\dg_{b_0,v_1}(\sfV)\in\eHb^{\infty,1/2-}(\CX;\scsym)$.

	\item\underline{Dual zero modes for RN and KN case.} Combining the discussion around \eqref{eq:modifiedlinearizedEinsteinDual}, \eqref{eq:modifiedlinearizedMaxwellDual} and \eqref{eq:eqfordualpotential} with the growing dual zero modes established in \S\ref{subsec:growingzeromodesof1form}, we see that \[(2G_{g_b}\delta_{g_b}^*\omega_{b,\bullet}^*(\bullet),\  4\wt{\mathcal{L}}_{A_b}(\omega^*_{b,\bullet}(\bullet))^{\sharp}+d\phi_{b,\bullet}^*(\bullet))\in\ker \widehat{L_{b,\gamma}}(0)^*
	\] provided that
	\begin{equation}\label{eq:eqfordualpotentialM}
	-\widehat{\Box_{g_b}}(0)\phi_{b,\bullet}^*(\bullet)=-4\delta_{g_b}(\wt{\mathcal{L}}_{A_b}(\omega^*_{b,\bullet}(\bullet))^{\sharp}).
	\end{equation}
For $\omega^*_{b,s_\bullet}(\bullet)$, in view of \eqref{eq:g1formdual3/2}, $\delta_{g_b}\in\rho\mbox{Diff}_\bop^1$ and $\mathcal{L}_{(\bullet)^\sharp}A_b\in\rho^2\mbox{Diff}_\bop^1$, the right-hand side of \eqref{eq:eqfordualpotentialM} is in $\sHb^{-\infty,3/2-}(\CX)$. For the case $\phi^*_{b,s_0}$, by Theorem \ref{thm:modesforscalarwave}, there exists a unique $\phi^*_{b,s_0}\in\sHb^{-\infty,-1/2-}(\CX)$ such that \eqref{eq:eqfordualpotentialM} holds. We note that adding a multiple of $H(r-\ehRN)$ to $\phi^*_{b,s_0}$ still gives rise to $A^*_{b,s_0}\in\sHb^{-\infty,1/2-}(\CX)$. For the case $\phi^*_{b,s_1}(\sfS), \sfS\in\BFS_1$, by Theorem \ref{thm:modesforscalarwave} again, there exists a unique $\phi^*_{b,s_1}(\sfS)\in\sHb^{-\infty,-1/2-}(\CX)$ such that \eqref{eq:eqfordualpotentialM} holds. Therefore, with the above choice of $\phi_{b,\bullet}^*(\bullet)$, we have $\dA^*_{b,s_\bullet}(\bullet)\in\sHb^{-\infty,1/2-}(\CX)$. As for $\omega_{b,v_1}(\sfV)$, according to \eqref{eq:g1formdual5/2}, now the right-hand side of \eqref{eq:eqfordualpotentialM} is in $\sHb^{-\infty,1/2-}(\CX)$. Since $\ker\widehat{\Box_{g_b}}(0)\cap\eHb^{\infty,-1/2+}(\CX)=\{0\}$ by Theorem \ref{thm:modesforscalarwave}, \eqref{eq:eqfordualpotentialM} is solvable for $\phi_{b,v_1}^*(\sfV)$. More precisely,  there exists a unique $\phi^*_{b,v_1}(\sfV)\in\sHb^{-\infty,-3/2-}(\CX)$, which lies in $\ann(v)$ with $v\in C^\infty_c(X)$ and satisfying $\langle v,H(r-r_b)\rangle\neq 0$, such that \eqref{eq:eqfordualpotentialM} holds, and thus $\dA^*_{b,v_1}\in\sHb^{-\infty,-1/2-}(\CX;\scform)$. We also note that by \eqref{eq:symgra3/2} and \eqref{eq:symmgra5/2}, $G_{g_b}\delta_{g_b}^*\omega^*_{b,\bullet}(\bullet)\in\sHb^{-\infty,1/2-}(\CX;\scsym)$. As a consequence, \[(2G_{g_b}\delta_{g_b}^*\omega_{b,\bullet}^*(\bullet),\  4\wt{\mathcal{L}}_{A_b}(\omega^*_{b,\bullet}(\bullet))^{\sharp}+d\phi_{b,\bullet}^*(\bullet))\quad \mbox{and}\quad(0, \delta(r-\ehKN)dr)
\]
indeed span a $8$-dimension subspace of $\mathcal{K}^*_b$.

	\item\underline{Zero modes for KN case.} First, the above construction of the dual zero modes of $\widehat{L_{b,\gamma}}(0)^*$ and the index $0$ property of $\widehat{L_{b,\gamma}}(0)$ imply that $\mathcal{K}_b$ is at least $8$-dimensional. Next, we shall prove by a contradiction argument that it is at most $8$-dimensional for $b$ near $b_0$, and thus must be $8$-dimensional. Suppose $\{h_1,\cdots,h_8\}$ is a basis of $\mathcal{K}_{b_0}$ and choose $h_1^\flat,\cdots, h_8^\flat\in C^\infty_c(X^\circ;\scsym\oplus \scform)$ such that $\angles{h_i}{h_j^\flat}=\delta_{ij}$. (Consider the map $\Phi: C^\infty_c(X^\circ)\to\BC^{8}$ defined by $\Phi(h^\flat)=(\langle h_{1},h^\flat \rangle,\cdots,\langle h_{8},h^\flat\rangle)$, and note that $\Phi$ is surjective. Otherwise, there would exist some $a=(a_1, \cdots,a_{8})\neq 0\in\BC^{8}$ such that $a\cdot\Phi(h^\flat)=\langle\sum_{i=1}^{8}a_i h_{i},h^\flat\rangle=0$ for all $h^\flat\in C^\infty_c(X^\circ)$, which implies $\langle \sum_{i=1}^{8}a_i h_{i}, h^\flat\rangle=0$ all $h^\flat\in\sHb^{-\infty,-\ell}(\CX)$ since $C^\infty_c(X^\circ) $ is dense in $\sHb^{-\infty,-\ell}(\CX)$. However, this is impossible because $\sum_{i=1}^{8}a_i h_{i}\neq 0$). Assume for the sake of contradiction that there exists a sequence $b_j\to 0$ such that for all $j$, the dimension of $\mathcal{K}_b$ is greater than $8$. Then one can find $u_j\in\mathcal{K}_{b_j}$ such that $u_j\in\mbox{ann}\{h^\flat_1,\cdots,h^\flat_8\}$ with $\norm{u_j}_{\eHb^{\infty,-1/2-}(\CX)}=1$. As in the proof of the first step of Lemma \ref{lem:step1} (using the Fredholm estimates for $\widehat{L_{b,\gamma}}(0)$), there exists a subsequence $u_j\to u\neq0$ weakly in $\eHb^{\infty,-1/2-}(\CX)$ and then $u\in\ker\widehat{L_{b_0,\gamma}}(0)$. Therefore, one can pick $h^\flat\in\mbox{span}\{h_1^\flat,\cdots,h_8^\flat\}$ such that $\angles{u}{h^\flat}=1$, but $0=\angles{u_j}{h^\flat}\to\angles{u}{h^\flat}=1$, which leads to a contradiction.
	
	Getting the explicit expressions for $(\dg_{b,\bullet}(\bullet),\dA_{b,\bullet}(\bullet))$ as in \eqref{eq:mlgzeromodes0}, \eqref{eq:mlgzeromodes1} and \eqref{eq:mlgzeromodev1} requires a direct argument. First,  by \eqref{eq:symgra3/2}, we indeed have $(\dg_{b,s_1}(\sfS),\dA_{b,s_1}(\sfS))=(2\delta_{g_b}\omega_{b,s_1}(\sfS),\ \wt{\mathcal{L}}_{	A_b}(\omega_{b,s_1}(\sfS))^\sharp+d\phi_{b,s_1}(\sfS) )\in\ker\widehat{L_{b,\gamma}}\cap\eHb^{\infty,1/2-}(\CX)$ for suitable $\phi_{b,s_1}$. It remains to construct $(\dg_{b,s_0},\dA_{b,s_0})$ and $(\dg_{b,v_1}(\sfV),\dA_{b,v_1}(\sfV))$ which extend $(\dg_{b_0,s_0},\dA_{b_0,s_0})$ and $(\dg_{b_0,v_1}(\sfV),\dA_{b_0,v_1}(\sfV))$, respectively. We make the ansatz
	\begin{equation}
		(\dg_{b,s_0},\dA_{b,s_0})=(\dot{g}_b(\dot{\Bm},0,\dot{\BQ})+2\delta^*_{g_b}\omega,\  \dA_b(0,0,\dot{\BQ})+\mathcal{L}_{\omega^\sharp}A_b+d\phi)
	\end{equation}
with $\omega, \phi$ to be found. Since 	$(\dg_{b,s_0},\dA_{b,s_0})$ of the form in the above ansatz satisfies the linearized Einstein-Maxwell system, we have $	(\dg_{b,s_0},\dA_{b,s_0})\in\ker\widehat{L_{b,\gamma}}(0)$ provided that the generalized linearized gauge conditions are satisfied, namely, 
\begin{equation}\label{eq:mgaugecondieqn}
\tilde{\delta}_{g_b,\gamma}G_{g_b}\dg_{b,s_0}=0\quad\mbox{and}\quad \delta_{g_b}d\dA_{b,s_0}=0.
\end{equation}
Then as in the proof of scalar type $l=0$ zero modes of $\widehat{L_{b_0,\gamma}}(0)$, we can find a unique $\omega_{b,s_0}\in\eHb^{\infty,-3/2-}(\CX;\scform)$, which lies in $\ann\{f_1,\cdots,f_4\}$ with $\{f_1,\cdots,f_4\}\subset C^\infty_c(X)$ and being a set of linearly independent functionals on $\ker\widehat{\mathcal{W}_{b_0, \gamma}}(0)\cap\eHb^{\infty,-3/2-}(\CX;\scform)$, and $\phi_{b,s_0}\in\eHb^{\infty,-3/2-}(\CX;\BC)$, which lies in $\ann(v)$ with $v\in C^\infty_c(X)$ and satisfying $\langle 1,v\rangle\neq 0$, such that \eqref{eq:mgaugecondieqn} holds. The construction of $(\dg_{b,v_1}(\sfV),\dA_{b,v_1}(\sfV))$ is analogous. We make the ansatz
\begin{equation}
	(\dg_{b,v_1}(\sfV),\dA_{b,v_1}(\sfV))=(\dot{g}_b(0,\dot{\Ba},0)+2\delta^*_{g_b}\omega,\  \dA_b(0,\dot{\Ba},0)+\mathcal{L}_{\omega^\sharp}A_b+d\phi)
\end{equation}
where $\sfV\in\BFV_1$ is dual to the rotation around the axis $\dot{\Ba}$ and $\omega,\phi$ are to be found. Then by using the discussion above and proceeding as in the proof of vector type $l=1$ zero modes of $\widehat{L_{b_0,\gamma}}(0)$, the conclusion \eqref{eq:mlgzeromodev1} follows and $(\dg_{b,v_1}(\sfV), \dA_{b,v_1}(\sfV))\in\eHb^{\infty, 1/2-}(\CX)$. %In fact, we can further obtain that $\dg_{b,v_1}(\sfV)\in\eHb^{\infty, 1/2-}(\CX)$ as follows. Let $\dg_{b,v_1}(\sfV)=\dg_{b_0,v_1}(\sfV)+e_{b,v_1}$ where $e_{b,v_1}=(\dot{g}_b(0,\dot{\Ba},0)-\dot{g}_{b_0}(0,\dot{\Ba},0))+2\delta^*_{g_b}\omega_{b}-2\delta^*_{g_{b_0}}\omega_{b_0}$
	  \item\underline{Continuity.} The continuity (in $b$) of the zero modes and dual zero modes follows from the explicit construction above (see the proof of continuity in \S\ref{subsec:growingzeromodesof1form}) and that of $\omega_{b,s_1}(\sfS)$ and $\omega^*_{b, \bullet}(\bullet)$.
 \end{itemize}
\end{proof}

\subsection{Generalized zero modes: linear growth}
Now we turn to discussing the generalized zero mode solutions of $L_{b,\gamma}(\dg,\dA)=0$. Suppose $(\dg,\dA)=(\sum^k_{j=1}t^j_{b,*}\dg_j,\sum^k_{j=1}t^j_{b,*}\dA_j)$ and $L_{b,\gamma}(\dg,\dA)=0$. Due to the fact that $[L_{b,\gamma},\pa_{t_{b,*}}]=0$, we find that $\pa_{t_{b,*}}^j(\dg,\dA)\in\ker L_{b,\gamma}$ for $0\leq j\leq k$. Taking $j=k$, we obtain that the leading order term $(\dg_k,\dA_k)$ of $(\dg,\dA)$ lies in the $\ker \widehat{L_{b,\gamma}}(0)$. Next we further determine what the leading order term $(\dg_k,\dA_k)$ should be.

\begin{lem}\label{lem:gezeromodesv1}
	Let $k\geq 1$. Then there does not exist  $(\dg,\dA)=(\sum^k_{j=1}t^j_{b,*}\dg_j,\sum^k_{j=1}t^j_{b,*}\dA_j)$, with $0\neq(\dg_k,\dA_k)\in\{(\dg_{b,v_1}(\sfV),\dA_{b,v_1}(\sfV)):\sfV\in\BFV_1\}$ and $(\dg_j,\dA_j)\in\eHb^{\infty,-1/2-}(\CX;\scsym\oplus\scform)$, such that $L_{b,\gamma}(\dg,\dA)=0$.
\end{lem}

\begin{proof}
	According to the discussion above, it suffices to consider the case $k=1$. Assume \[(\dg,\dA)=(t_{b,*}\dg_{b,v_1}(\sfV)+\dg_0,\  t_{b,*}\dA_{b,v_1}(\sfV)+\dA_0)
	\]
	 with $\sfV\neq 0$. Then we need to show that there is no $(\dg_0,\dA_0)\in\eHb^{\infty,-1/2-}(\CX)$ such that the following equation holds
	 \[
	 \widehat{L_{b,\gamma}}(0)(\dg_0,\dA_0)=-[L_{b,\gamma},\ t_{b,*}](\dg_{b,v_1}(\sfV),\dA_{b,v_1}(\sfV))\in\eHb^{\infty,3/2-}(\CX)
	 \]
	 where we use $[L_{b,\gamma},\ t_{b,*}]\in\rho\mbox{Diff}_\bop^1$ and $(\dg_{b,v_1}(\sfV),\dA_{b,v_1}(\sfV)\in\eHb^{\infty,1/2-}(\CX)$. To this end, we need to verify that the pairing of the right-hand side with $(\dg^*_{b,v_1}(\sfV),\dA_{b,v_1}^*(\sfV))$ is non-zero. We note that it suffices to check the RN case $b=b_0$ because the general KN case with $b$ near $b_0$ follows directly by the continuity (in $b$). Using the expression \eqref{eq:refinedhv1} for $(\dg_{b_0,v_1}(\sfV),\dA_{b_0,v_1}(\sfV))$, we find that
	 \begin{equation}
	 	\begin{split}
	 &\quad\angles{[L_{b_0,\gamma},\ t_{b_0,*}](\dg_{b_0,v_1}(\sfV),\dA_{b_0,v_1}(\sfV))}{(\dg^*_{b_0,v_1}(\sfV'),\dA^*_{b_0,v_1}(\sfV'))}\\
	% &=\angles{[L_{b_0,\gamma},\ t_{b_0,*}](\dg_{b_0,v_1}(\sfV),0)}{(\dg^*_{b_0,v_1}(\sfV'),\dA^*_{b_0,v_1}(\sfV'))}+\mathcal{O}(\abs{\BQ_0})\\
	 &=\angles{[L^E_{b_0,\gamma},\ t_{b_0,*}](\dg_{b_0,v_1}(\sfV),0)}{(\dg^*_{b_0,v_1}(\sfV'),0)}+\mathcal{O}(\abs{\BQ_0})\\
	 &=\angles{[L^E_{b_0,\gamma},\ t_{b_0,*}](\dg_{b_0,v_1}(\sfV),0)}{(2G_g\delta^*_g(r^2H(r-\ehRN)\sfV'),0)}+\mathcal{O}(\abs{\BQ_0}+\gamma)
	 \end{split}
	 \end{equation}
 where we use %$[L^M_{b,\gamma},\ t_{b,*}](\dg_{b,v_1}(\sfV),0)=\mathcal{O}_{\eHb^{\infty,3/2-}(\CX)}(\abs{\BQ_0})$ in the second step and 
 \eqref{eq:refinedg1formzeromode_rota} in the last step. Since $G_{g_{b_0}}\delta_{g_{b_0}}(r^2H(r-\ehRN)\sfV')=r^2\delta(r-\ehRN)dr\otimes_s\sfV'$ is supported at $r=\ehRN$ where $t_{b_0,*}=t_0$, we can replace $t_{b_0,*}$ by $t_0$ in the subsequent calculation. We then compute
 \begin{equation}
 	\begin{split}
 	&\quad	\angles{[L^E_{b_0,\gamma},\ t_0](\dg_{b_0,v_1}(\sfV),0)}{\ 2r^2\delta(r-\ehRN)dr\otimes_s\sfV'}\\
 		&=\angles{[L^E_{b_0,\gamma},\ t_0](4\Bm_0r^{-1}(dr-dt_0)\otimes_s\sfV,0)}{\ 2r^2\delta(r-\ehRN)dr\otimes_s\sfV'}+\mathcal{O}(\abs{\BQ_0}^2+\gamma)\\
 		&=-\angles{[\Box_{g_{b_0},2},\ t_0](4\Bm_0r^{-1}(dr-dt_0)\otimes_s\sfV)}{\ 2r^2\delta(r-\ehRN)dr\otimes_s\sfV'}+\mathcal{O}(\abs{\BQ_0}^2+\gamma)\\
 		&=\angles{8\Bm_0r^{-2}\big((dr-(1+\frac{\Bm_0}{r})dt_0)\big)\otimes_s\sfV}{\ 2r^2\delta(r-\ehRN)dr\otimes_s\sfV'}+\mathcal{O}(\abs{\BQ_0}^2+\gamma)\\
 		&=-12\Bm_0\angles{V}{V'}_{L^2(\BS^2;T^*\BS^2)}+\mathcal{O}(\abs{\BQ_0}^2+\gamma).
 	\end{split}
 \end{equation}
where we use \eqref{eq:refinedhv1} in the first step, \eqref{eq:mlgE} and \eqref{eq:L_1} in the second step and $[\Box_{g_{b_0},2},\ t_0]h=h\Box_{g_{b_0}}t_0+2(\nabla^\al t_0) \nabla_\al h$ in the third step. Therefore, the pairing is indeed non-zero if $\sfV=\sfV'$ and $\abs{\BQ_0}+\gamma$ is sufficiently small.
\end{proof}

We now exclude the generalized zero modes whose leading order term is $(\dg_{b,s_0},\dA_{b,s_0})$.

\begin{lem}\label{lem:gezeromodess0}
		Let $k\geq 1$. Then there does not exist  $(\dg,\dA)=(\sum^k_{j=1}t^j_{b,*}\dg_j,\sum^k_{j=1}t^j_{b,*}\dA_j)$, with $0\neq(\dg_k,\dA_k)=(\dg_{b,s_0},\dA_{b,s_0})$ and $(\dg_j,\dA_j)\in\eHb^{\infty,-1/2-}(\CX;\scsym\oplus\scform)$, such that $L_{b,\gamma}(\dg,\dA)=0$.
\end{lem}

\begin{proof}
	As discussed in the proof of Lemma \ref{lem:gezeromodesv1}, it suffices to prove the statement for the case $k=1$. We first consider the case $b=b_0$. Suppose 
	\[(\dg,\dA)=(t_{b_0,*}\dg_{b_0,s_0}+\dg_0,\  t_{b_0,*}\dA_{b_0,s_0}+\dA_0)\in\ker L_{b_0,\gamma}
	\]
	with $(\dg_0,\dA_0)\in\eHb^{\infty,-1/2-}(\CX)$. Applying $\delta_{g_{b_0}}$ to $L^M_{b_0,\gamma}(\dg,\dA)=0$ gives $\delta_{g_{b_0}}\dA\in\ker\Box_{g_{b_0}}$. As pointed out in Remark \ref{rem:gaugecondition},  we have $\delta_{g_{b_0}}\dA_{b_0,s_0}=0$. This implies
	\[
\ker\widehat{\Box_{g_{b_0}}}(0)\ni\delta_{g_{b_0}}\dA=[\delta_{g_{b_0}},t_{b_0,*}]\dA_{b_0,s_0}+\delta_{g_{b_0}}\dA_0\in\eHb^{\infty,-1/2-}(\CX;\BC).	\] 
Therefore, $\delta_{g_{b_0}}\dA=0$ by Theorem \ref{thm:modesforscalarwave}. That is, $(\dg,\dA)$ satisfies the Maxwell part of the linearized Einstein-Maxwell system. Then as in the proof of Theorem \ref{thm:modestabilityofmlgEM}, we apply $\delta_{g_{b_0}}G_{g_{b_0}}$ to $L^E_{b_0,\gamma}(\dg,\dA)=0$ and use the linearized second Bianchi identity to obtain $\tilde{\delta}_{g_{b_0},\gamma}G_{g_{b_0}}\dg\in\ker\mathcal{P}_{b_0,\gamma}$. Since $\tilde{\delta}_{g_{b_0},\gamma}G_{g_{b_0}}\dg_{b_0,s_0}=0$ (see Remark \ref{rem:gaugecondition}), it follows that
\[
\ker\widehat{\mathcal{P}_{b_0,\gamma}}(0)\ni\tilde{\delta}_{g_{b_0},\gamma}G_{g_{b_0}}\dg=[\tilde{\delta}_{g_{b_0},\gamma}G_{g_{b_0}},t_{b_0,*}]\dg_{b_0,s_0}+\tilde{\delta}_{g_{b_0},\gamma}G_{g_{b_0}}\dg_0\in\eHb^{\infty,-1/2-}(\CX;\scform),
\]
 and thus $\tilde{\delta}_{g_{b_0},\gamma}G_{g_{b_0}}\dg=0$ by Theorem \ref{thm:modesforCP}. As a consequence, $(\dg,\dA)$ satisfies the linearized Einstein-Maxwell system. By the statement (\romannumeral 5) in mode stability Theorem \ref{thm:modestability}, we have
 \[
 	(\dg,\dA)=(\dg_{b_0}(\dot{\Bm},0,\dot{\BQ})+2\delta^*_{g_{b_0}}\omega,\ \dA_{b_0}(0,0,\dot{\BQ})+\mathcal{L}_{\omega^\sharp}A_{b_0}+d\phi)
 \]
 where $(\omega,\phi)\in\mbox{Poly}^2(t_{b_0,*})\eHb^{\infty,-3/2-}(\CX;\scform\oplus\BC)$. To make the calculation simpler, we work in $(t_0=t+r_{b_0,*}, r,\theta,\varphi)$ coordinates instead.  Since $\dg$ is linear in $t_0$, it follows that the coefficient of $t_0^2$ in $\omega$ and $\phi$ must be a multiple of $\pa_{t_0}^\flat$ and $1$, respectively. Then equating the coefficient of $t_0$ of the equality $\dg=\dg_{b_0}(\dot{\Bm},0,\dot{\BQ})+2\delta^*_{g_{b_0}}\omega$ yields
 \begin{equation}\label{eq:RNnotpuregauge}
 (\frac{2\dot{\Bm}}{r}-\frac{2\BQ_0\dot{\BQ}}{r^2})dt_0^2=\dg^0_{b_0}(\dot{\Bm},0,\dot{\BQ})=c(dt_0)\otimes_s\pa_{t_0}^\flat+\delta^*_{g_{b_0}}\tilde{\omega}
 \end{equation}
 where $c\in\BR$ and $\tilde{\omega}\in\eHb^{\infty, \ell'}(\CX;\scform)$ for some $\ell'\in\BR$ is of scalar type $l=0$. Assume $\tilde{\omega}=f(r)dt_0+g(r)dr$ and let $S=c(dt_0)\otimes_s\pa_{t_0}^\flat+\delta^*_{g_{b_0}}\tilde{\omega}$. A direct calculation implies $0=S_{rr}=g'(r)$ and thus $g(r)=m$ for some $m\in\BR$. Then $0=S_{t_0r}=\frac c2+\frac 12 f'(r)+\frac m2 \weight'$ implies $f(r)=-cr+\frac{2m\Bm_0}{r}-\frac{m\BQ_0^2}{r^2}+n$ for some $n\in\BR$. Next $0=S_{\theta\theta}=r(f(r)+m\weight)=-cr^2+(m+n)r$ gives $c=0,m+n=0$, and thus $\tilde{\omega}=-m\weight dt_0+mdr$. Finally, $ \frac{2\dot{\Bm}}{r}-\frac{2\BQ_0\dot{\BQ}}{r^2}=S_{t_0t_0}=\frac{m}{2}\weight\weight'-\frac{m}{2}\weight\weight'=0$ implies that $\dot{\Bm}=\BQ_0\dot{\BQ}=0$ and thus the leading term $g_{b_0,s_0}$ of $\dg$ is $0$. If $\dot{\BQ}\neq 0, \BQ_0=0$, equating the coefficient of $t_0$ of the equality $\dA=\dA_{b_0}(0,0,\dot{\BQ})+d\phi$ yields
 \begin{equation}
 	\frac{\dot{\BQ}}{r}dt_0=cdt_0+d\tilde{\phi}
 \end{equation} 
where $c\in\BR$ and $\tilde{\phi}\in\eHb^{\infty,\ell'}(\CX;\BC)$ for some $\ell'\in\BR$ is of scalar type $l=0$. This implies that $\dot{\BQ}=0$ and thus  the leading term $A_{b_0,s_0}$ of $\dA$ is $0$. %Therefore, there does not exist $c\in\BR$ and $\tilde{\omega}$ of scalar type $l=0$ such that \eqref{eq:RNnotpuregauge} holds. 
This proves the lemma for the RN case $b=b_0$. 
 
 This non-existence result for $b=b_0$ in turn means that the following equation
 \[
  \widehat{L_{b_0,\gamma}}(0)(\dg_0,\dA_0)=-[L_{b_0,\gamma},\ t_{b_0,*}](\dg_{b_0,s_0},\dA_{b_0,s_0})\in\eHb^{\infty,3/2-}(\CX)
 \]
 cannot be solved for $(\dg_0,\dA_0)\in\eHb^{\infty,-1/2-}(\CX;\scsym\oplus\scform)$ (The right-hand side is in the stated Sobolev space because by Proposition \ref{prop:desofkernel}, $(\dg_{b_0,s_0},\dA_{b_0,s_0})=\rho C^\infty(\CX)+\eHb^{\infty,1/2-}(\CX)$ whose leading term $\rho C^\infty(\CX)$ is annihilated by the normal operator $2\rho(\rho\pa_\rho-1)$ of $-[L_{b_0,\gamma},\ t_{b_0,*}]$). Suppose $\{(\dg^1_{b, s_0},\dA^1_{b,s_0}),\,(\dg^2_{b, s_0},\dA^2_{b,s_0})\}$ is a basis of $\{(\dg_{b, s_0},\dA_{b,s_0})\}$ and  $\{(\dg^{*1}_{b, s_0},\dA^{*1}_{b,s_0}),\,(\dg^{*2}_{b, s_0},\dA^{*2}_{b,s_0})\}$ is a basis of $\{(\dg^*_{b, s_0},\dA^*_{b,s_0})\}$.  In terms of pairing,  we have
 \[
 \angles{[L_{b_0,\gamma},\ t_{b_0,*}](\dg^i_{b_0,s_0},\dA^i_{b_0,s_0})}{(\dg^{*j}_{b_0,s_0},\dA^{*j}_{b_0,s_0})},\quad 1\leq i,j\leq 2
 \]
 is non-degenerate. 
 Due to the continuity of $(\dg_{b,s_0},\dA_{b,s_0})$ and $(\dg^*_{b,s_0},\dA^*_{b,s_0})$ in $b$, this pairing is still non-zero for KN case with $b$ near $b_0$. This finishes the proof of the lemma. 
\end{proof}

However, there do exist generalized zero modes (growing linearly in $t_{b,*}$) whose leading order term is given by $(\dg_{b,s_1}(\sfS), \dA_{b,s_1}(\sfS))$ for $\sfS\in\BFS_1$.

\begin{prop}\label{prop:gezeromodess1}
Let $b_0=(\Bm_0,0,\BQ_0)$ with $\abs{\BQ_0}\ll\Bm_0$. Then the following space
\begin{align}
	\widehat{\mathcal{K}}_{b_0}&:=\ker L_{b_0,\gamma}\cap\mbox{Poly}^1(t_{b_0,*})\eHb^{\infty,-1/2-}(\CX;\scsym\oplus\scform).	
\end{align}	
is $11$-dimensional. The $3$-dimensional quotient space	$\widehat{\mathcal{K}}_{b_0}/\mathcal{K}_{b_0}$ is spanned by the following continuous (in $b$) families
 \begin{equation}
\{(\hat{g}_{b_0,s_1}(\sfS),\hat{A}_{b_0,s_1}(\sfS)):\ \sfS\in\BFS_1\}.
	\end{equation}
with
\begin{equation}
	(\hat{g}_{b_0,s_1}(\sfS),\hat{A}_{b_0,s_1}(\sfS))=(2\delta_{g_{b_0}}^*\hat{\omega}_{b_0,s_1}(\sfS),\ \wt{\mathcal{L}}_{A_{b_0}}(\hat{\omega}_{b_0,s_1}(\sfS))^{\sharp}+d\hat{\phi}_{b_0,s_1}(\sfS))
\end{equation}
 where $\hat{\omega}_{b_0,s_1}(\sfS)$ is defined as in \eqref{eq:ge1formzeromode5/2} and $\hat{\phi}_{b_0,s_0}(\sfS)\in\mbox{Poly}^1(t_{b_0,*})\eHb^{\infty,-3/2-}(\CX;\BC)$ depends linearly in $\sfS\in\BFS_1$. 
 
 The generalized zero modes $(\hat{g}_{b_0,s_1}(\sfS),\hat{A}_{b_0,s_1}(\sfS))$ extend to continuous families in $b$
 \begin{equation}
 	b\to(\hat{g}_{b,s_1}(\sfS),\hat{A}_{b,s_1}(\sfS))\in \ker L_{b,\gamma}\cap\mbox{Poly}^1(t_{b,*})\eHb^{\infty,-1/2-}(\CX;\scsym\oplus\scform)
 \end{equation} 
 with
 \begin{equation}
 	(\hat{g}_{b,s_1}(\sfS),\hat{A}_{b,s_1}(\sfS))=(2\delta_{g_{b}}^*\hat{\omega}_{b,s_1}(\sfS),\ \wt{\mathcal{L}}_{A_{b}}(\hat{\omega}_{b,s_1}(\sfS))^{\sharp}+d\hat{\phi}_{b,s_1}(\sfS))
 \end{equation}
 where $\hat{\omega}_{b,s_1}(\sfS)$ is defined as in \eqref{eq:ge1formzeromode5/2} and $\hat{\phi}_{b,s_0}(\sfS)\in\mbox{Poly}^1(t_{b,*})\eHb^{\infty,-3/2-}(\CX;\BC)$ depends linearly in $\sfS\in\BFS_1$. Moreover, $(\hat{g}_{b,s_1}(\sfS),\hat{A}_{b,s_1}(\sfS))$ satisfies both the linearized Einstein-Maxwell system and the linearized generalized wave map gauge and generalized Lorenz gauge conditions.
 
 Meanwhile, there exists continuous families (in $b$)
 \begin{equation}
 	b\to(\hat{g}_{b,s_1}^*(\sfS),\hat{A}_{b,s_1}^*(\sfS))\in\ker L_{b,\gamma}^*\cap\mbox{Poly}^1(t_{b,*})\sHb^{-\infty,-1/2-}(\CX;\scsym\oplus\scform)
 \end{equation}
with
\begin{equation}
	(\hat{g}_{b,s_1}^*(\sfS),\hat{A}_{b,s_1}^*(\sfS))=(2G_{g_b}\delta_{g_b}^*\hat{\omega}_{b,s_1}^*(\sfS),\ 4\wt{\mathcal{L}}_{A_{b_0}}(\hat{\omega}^*_{b,s_1}(\sfS))^{\sharp}+d\hat{\phi}^*_{b,s_1}(\sfS))
\end{equation}
where $\hat{\omega}_{b,s_1}^*(\sfS)$ is defined as in \eqref{eq:ge1formzeromode5/2} and $\hat{\phi}^*_{b,s_0}(\sfS)\in\mbox{Poly}^1(t_{b,*})\sHb^{\infty,-3/2-}(\CX;\BC)$.
\end{prop}

\begin{proof}
Let $(\dg,\dA)=(t_{b,*}\dg_1+\dg_0, t_{b,*}\dA_1+\dA_0)\in\ker L_{b,\gamma}$ with $(\dg_{j},\dA_{j})\in\eHb^{\infty,-1/2-}(\CX)$.
\begin{itemize}
	\item \underline{Generalized zero modes in the RN case $b=b_0$.}  In view of Lemma \ref{lem:gezeromodesv1} and \ref{lem:gezeromodess0}, $(\dg_1,\dA_1)$ must take the form $(2\delta_{g_{b_0}}\omega_{b_0,s_1}(\sfS), \wt{\mathcal{L}}_{A_{b_0}}(\omega_{b_0,s_1}(\sfS))^\sharp+d\phi_{b_0,s_1}(\sfS))$ for some $\sfS\in\BFS_1$. By the argument at the beginning of the proof of Lemma  \ref{lem:gezeromodess0}, we see that $(\dg,\dA)$ satisfies the generalized linearized gauge condition, that is, 
\begin{equation}\label{eq:temmlgc}
	\tilde{\delta}_{g_{b_0}}G_{g_{b_0}}\dg=0,\quad \delta_{g_{b_0}}\dA=0.
\end{equation}
Therefore, $(\dg,\dA)$ solves the linearized Einstein-Maxwell system. Since
\begin{equation}\label{eq:exforgA}
	\begin{split}
	\dg&=2\delta^*_{g_{b_0}}(t_{b_0,*}\omega_{b_0,s_1}(\sfS))-2[\delta^*_{g_{b_0}},t_{b_0,*}]\omega_{b_0,s_1}(\sfS)+\dg_0=2\delta^*_{g_{b_0}}(t_{b_0,*}\omega_{b_0,s_1}(\sfS))+\dg_0' , \\
	\dA&=\wt{\mathcal{L}}_{A_{b_0}}(t_{b_0,*}\omega_{b_0,s_1}(\sfS))^\sharp+d\big(t_{b_0,*}\phi_{b_0,s_1}(\sfS)\big)-\Big(\iota_{\omega_{b_0,s_1}(\sfS)^\sharp}A_{b_0}+\phi_{b_0,s_1}(\sfS)\Big)dt_{b_0,*}+\dA_0\\
	&=\wt{\mathcal{L}}_{A_{b_0}}(t_{b_0,*}\omega_{b_0,s_1}(\sfS))^\sharp+d\big(t_{b_0,*}\phi_{b_0,s_1}(\sfS)\big)+\dA_0',
	\end{split}
\end{equation}
with
\begin{align*}
\dg_0'&=\dg_0-2[\delta^*_{g_{b_0}},t_{b_0,*}]\omega_{b_0,s_1}(\sfS)\in\eHb^{\infty,-3/2-}(\CX),\\
		\dA_0'&=\dA_0-\Big(\iota_{\omega_{b_0,s_1}(\sfS)^\sharp}A_{b_0}+\phi_{b_0,s_1}(\sfS)\Big)dt_{b_0,*}\in\eHb^{\infty,-1/2-}(\CX).
\end{align*}
where we use $[\delta^*_{g_{b_0}},t_{b_0,*}]\in\mathcal{A}^0(\CX), \omega_{b_0,s_1}(\sfS)\in\eHb^{\infty,-3/2-}(\CX)$ and $\phi_{b_0,s_1}(\sfS)\in\eHb^{\infty,-1/2-}(\CX)$, it follows that $(\dg_0',\dA_0')$ solves the linearized Einstein-Maxwell system as well. By the statement (\romannumeral 4) in Theorem \ref{thm:modestability}, we have
\begin{equation}\label{eq:exforg'A'}
(\dg_0',\dA_0')=(2\delta_{g_{b_0}}^*\omega,\  \mathcal{L}_{\omega^{\sharp}}A_{b_0}+d\phi)
\end{equation}
with $(\omega,\phi)\in\eHb^{\infty,-5/2-}(\CX;\scform\oplus\BC)$. Plugging \eqref{eq:exforgA} and \eqref{eq:exforg'A'} into \eqref{eq:temmlgc} yields
\begin{align}\label{eq:eqnforEpotential}
		\widehat{\mathcal{W}_{b_0, \gamma}}(0)\omega&=-[\mathcal{W}_{b_0, \gamma},\  t_{b_0,*}]\omega_{b_0,s_1}(\sfS)\in\eHb^{\infty,-1/2-}(\CX),\\\label{eq:eqnforMpotential}
		-\widehat{\Box_{g_{b_0}}}(0)\phi&=-\delta_{g_{b_0}}\Big(\mathcal{L}_{\omega^{\sharp}}A_{b_0}+(\iota_{\omega_{b_0,s_1}(\sfS)^\sharp}A_{b_0})dt_{b_0,*}\Big)\\\notag
		&\quad+[\Box_{g_{b_0}},\  t_{b_0,*}]\phi_{b_0,s_1}(\sfS)-[\delta_{g_{b_0}},t_{b_0,*}]\wt{\mathcal{L}}_{A_{b_0}}(\omega_{b_0,s_1}(\sfS))^\sharp
\end{align}
We first consider \eqref{eq:eqnforEpotential}. Since $\ker\widehat{\mathcal{W}_{b_0, \gamma}}(0)^*\cap\sHb^{-\infty,1/2+}(\CX)=\{0\}$ by Theorem \ref{thm:modesforW}, it follows that \eqref{eq:eqnforEpotential} can be solved for $\omega\in\eHb^{\infty,-5/2-}(\CX)$. According to Proposition \ref{prop:asym_lor}, we know that 
\[
\hat{\omega}_{b_0,s_1}(\sfS)=t_{b_0,*}\omega_{b_0,s_1}(\sfS)+\check{\omega}_{b_0,s_1}(\sfS)\in\ker\mathcal{W}_{b_0, \gamma}
\]
with $\check{\omega}_{b_0,s_1}(\sfS)\in\eHb^{\infty,-5/2-}(\CX)$. Therefore,
\[
	\widehat{\mathcal{W}_{b_0, \gamma}}(0)\big(\omega-\check{\omega}_{b_0,s_1}(\sfS)\big)=-\mathcal{W}_{b_0, \gamma}\hat{\omega}_{b_0,s_1}(\sfS)=0
\]
and $\omega-\check{\omega}_{b_0,s_1}(\sfS)\in \mbox{span}\{\omega^{(1)}_{b_0,s_1}(\sfS):\  \sfS\in\BFS_1\}$ (In fact $\ker\widehat{\mathcal{W}_{b_0, \gamma}}(0)\cap\eHb^{\infty,-5/2-}(\CX)$ restricted to the scalar type $l=1$ $1$-form is equal to $\mbox{span}\{\omega_{b_0,s_1}(\sfS),\omega^{(1)}_{b_0,s_1}(\sfS):\  \sfS\in\BFS_1\}$. Since $(\dg,\dA)$ is in the quotient space $\widehat{\mathcal{K}}_{b_0}/\mathcal{K}_{b_0}$, we can exclude $\omega_{b_0,s_1}(\sfS)$). Using \eqref{eq:gesymgra}, we rewrite
\begin{align*}
\dg&=2\delta^*_{g_{b_0}}\hat{\omega}_{b_0,s_1}(\sfS)+2\delta^*_{g_{b_0}}\Big(\omega-\check{\omega}_{b_0,s_1}(\sfS)\Big)\\
&=\mbox{Poly}^1(t_{b_0,*})\eHb^{\infty,-1/2-}(\CX)+2\delta^*_{g_{b_0}}\Big(\omega-\check{\omega}_{b_0,s_1}(\sfS)\Big).
\end{align*}
Since we require $\dg\in\mbox{Poly}^1(t_{b_0,*})\eHb^{\infty,-1/2-}(\CX)$, it follows that $\delta^*_{g_{b_0}}\big(\omega-\check{\omega}_{b_0,s_1}(\sfS)\big)$ must lie in $\eHb^{\infty,-1/2-}(\CX)$. Owing to \eqref{eq:symgra5/2}, $\omega-\check{\omega}_{b_0,s_1}(\sfS)$ must be $0$. That it, $\dg=2\delta_{g_{b_0}}\hat{\omega}_{b_0,s_1}(\sfS)$. 

Next we turn to \eqref{eq:eqnforMpotential}. Since $\omega=\check{\omega}_{b_0,s_1}(\sfS)\in\eHb^{\infty,-5/2}(\CX)$ and $\phi_{b_0,s_1}\in\eHb^{\infty,-1/2-}(\CX)$, the right-hand side of \eqref{eq:eqnforMpotential} lies in $\eHb^{\infty,1/2-}(\CX)$. By Theorem \ref{thm:modesforscalarwave} ($\ker\widehat{\Box_{g_{b_0}}}(0)^*\cap\sHb^{-\infty,-1/2+}(\CX)=\{0\}$) and Proposition \ref{prop:gscalarzeromode} ($\ker\widehat{\Box_{g_{b_0}}}(0)\cap\eHb^{\infty,-3/2-}(\CX)$ restricted to scalar type $l=1$ is $0$), there exists a unique $\phi\in\eHb^{\infty,-3/2-}(\CX)$ which we denote by $\check{\phi}_{b_0,s_1}(\sfS)$ satisfying \eqref{eq:eqnforMpotential}. This proves that $\widehat{\mathcal{K}}_{b_0}/\mathcal{K}_{b_0}$ is at most $3$-dimensional.

Using $\dg=2\delta_{g_{b_0}}^*\hat{\omega}_{b_0,s_1}(\sfS)\in\mbox{Poly}^1(t_{b_0,*})\eHb^{\infty,-1/2}(\CX)$ (see \eqref{eq:gesymgradual}) and rewriting \begin{equation}\label{eq:vefiofdecayofA}
	\begin{split}
		\dA&=\wt{\mathcal{L}}_{A_{b_0}}(\hat{\omega}_{b_0,s_1}(\sfS))^\sharp+d\big(t_{b_0,*}\phi_{b_0,s_1}(\sfS)+\check{\phi}_{b_0,s_1}(\sfS)\big)\\
	&=\Big(\wt{\mathcal{L}}_{A_{b_0}}(\check{\omega}_{b_0,s_1}(\sfS))^\sharp+\phi_{b_0,s_1}dt_{b_0,*}+\iota_{({\omega}_{b_0,s_1}(\sfS))^\sharp}A_{b_0}dt_{b_0,*}+d\check{\phi}_{b_0,s_1}(\sfS)\Big)\\
	&\quad+t_{b_0,*}\Big(\wt{\mathcal{L}}_{A_{b_0}}({\omega}_{b_0,s_1}(\sfS))^\sharp+d\phi_{b_0,s_1}(\sfS)\Big)\in\mbox{Poly}^1(t_{b_0,*})\eHb^{\infty,-1/2}(\CX),
	\end{split}
	\end{equation}
we conclude that $(\dg,\dA)=(\hat{g}_{b_0,s_1}(\sfS),\hat{A}_{b_0,s_1}(\sfS))=(2\delta_{g_{b_0}}^*\hat{\omega}_{b_0,s_1}(\sfS), \wt{\mathcal{L}}_{A_{b_0}}(\hat{\omega}_{b_0,s_1}(\sfS))^\sharp+d(\hat{\omega}_{b_0,s_1}(\sfS)))$ with $\hat{\phi}_{b_0,s_1}(\sfS)=t_{b_0,*}\phi_{b_0,s_1}(\sfS)+\check{\omega}_{b_0,s_1}(\sfS)\in\mbox{Poly}^1(t_{b_0,*})\eHb^{\infty,-3/2}(\CX)$ indeed lies in $\widehat{\mathcal{K}}_{b_0}$.

\item\underline{Generalized zero modes in the KN case.} For $b$ near $b_0$, we make the ansatz
\[
(\dg,\dA)=(2\delta_{g_{b}}^*\hat{\omega}_{b,s_1}(\sfS), \wt{\mathcal{L}}_{A_{b}}(\hat{\omega}_{b,s_1}(\sfS))^\sharp+d(t_{b,*}\phi_{b,s_1}(\sfS))+d\phi).
\]
with $\phi\in\eHb^{\infty,-3/2}(\CX)$ to be determined. By the arguments in the proof of RN case above (the last two paragraphs), there exists $\check{\phi}_{b,s_1}(\sfS)\in\eHb^{\infty,-3/2}(\CX)$ such that with $\hat{\phi}_{b,s_1}(\sfS)=t_{b,*}\phi_{b,s_1}(\sfS)+\check{\phi}_{b,s_1}(\sfS)\in\mbox{Poly}^1(t_{b,*})\eHb^{\infty,-3/2}(\CX)$,
\begin{align*}
&\quad(\hat{g}_{b,s_1}(\sfS),\hat{A}_{b,s_1}(\sfS))\\
&=(2\delta_{g_{b}}^*\hat{\omega}_{b,s_1}(\sfS), \wt{\mathcal{L}}_{A_{b}}(\hat{\omega}_{b,s_1}(\sfS))^\sharp+d(\hat{\phi}_{b,s_1}(\sfS)))\in\ker L_{b,\gamma}\cap\mbox{Poly}^1(t_{b,*})\eHb^{\infty,-1/2-}(\CX)
\end{align*}
extends $(\hat{g}_{b_0,s_1}(\sfS), \hat{A}_{b_0,s_1}(\sfS))$ and is continuous in $b$.

\item\underline{Generalized dual zero modes for RN and KN cases.} Combining the discussion around \eqref{eq:modifiedlinearizedEinsteinDual}, \eqref{eq:modifiedlinearizedMaxwellDual} and \eqref{eq:eqfordualpotential} with the generalized dual zero mode $\hat{\omega}_{b,s_1}^*(\sfS)=t_{b,*}\omega_{b,s_1}^*(\sfS)+\check{\omega}_{b,s_1}^*(\sfS)$ of $\mathcal{W}_{b,\gamma}$ established in Proposition \ref{prop:asym_lor}, we see that \[(\hat{g}^*_{b,s_1}(\sfS),\hat{A}^*_{b,s_1}(\sfS))=(2G_{g_b}\delta_{g_b}^*\hat{\omega}_{b,s_1}^*(\sfS),\  4\wt{\mathcal{L}}_{A_b}(\hat{\omega}^*_{b,s_1}(\sfS))^{\sharp}+d(t_{b,*}\phi^*_{b,s_1}(\sfS))+d\check{\phi}_{b,s_1}^*(\sfS))\in\ker L_{b,\gamma}^*
\] provided that
\begin{equation}\label{eq:eqnforgedualpotentialM}
	\begin{split}
-\widehat{\Box_{g_{b}}}(0)\check{\phi}^*_{b,s_1}(\sfS)&=-4\delta_{g_{b}}\Big(\wt{\mathcal{L}}_{A_b}(\check{\omega}_{b,s_1}^*(\sfS))^{\sharp}+(\iota_{\omega^*_{b,s_1}(\sfS)^\sharp}A_{b})dt_{b,*}\Big)\\\notag
&\quad+[\Box_{g_{b}},\  t_{b,*}]\phi^*_{b,s_1}(\sfS)-4[\delta_{g_{b}},t_{b,*}]\wt{\mathcal{L}}_{A_{b}}(\omega^*_{b,s_1}(\sfS))^\sharp\in\sHb^{-\infty,1/2-}(\CX).
\end{split}
\end{equation}
In view of Theorem \ref{thm:modesforscalarwave} and Proposition \ref{prop:gscalarzeromode}, $\ker\widehat{\Box_{g_{b}}}(0)\cap\eHb^{\infty,-1/2+}(\CX)=\{0\}$, and thus there exists a unique $\check{\phi}^*_{b,s_1}(\sfS)\in\sHb^{-\infty,-3/2-}(\CX)$ (up to addition of a multiple of $H(r-\ehKN)$)) such that \eqref{eq:eqnforgedualpotentialM} holds. We put $\hat{\phi}^*_{b,s_1}(\sfS)=t_{b,*}\phi^*_{b,s_1}+\check{\phi}^*_{b,s_1}(\sfS)$ and find that $
\hat{A}_{b,s_1}(\sfS)\in\mbox{poly}^1(t_{b,*})\sHb^{-\infty,-1/2-}(\CX)$ by the argument in \eqref{eq:vefiofdecayofA}. We also recall that $G_{g_b}\delta_{g_b}^*\hat{\omega}^*_{b,s_1}(\sfS)\in\mbox{Poly}^1(t_{b,*})\sHb^{-\infty,-1/2-}(\CX)$ (see \eqref{eq:gesymgradual}). As a consequence,  the continuous (in $b$) families $(\hat{g}^*_{b,s_1}(\sfS),\hat{A}^*_{b,s_1}(\sfS))$ indeed lie in $\ker L_{b,\gamma}^*\cap\mbox{poly}^1(t_{b,*})\sHb^{-\infty,-1/2-}(\CX)$.
\end{itemize}
\end{proof}

Let
\begin{align}
	(\check{g}_{b,s_1}(\sfS),\ \check{A}_{b,s_1}(\sfS))&:=(\hat{g}_{b,s_1}(\sfS)-t_{b,*}\dg_{b,s_1}(\sfS),\  \hat{A}_{b,s_1}(\sfS)-t_{b,*}\dA_{b,s_1}(\sfS))\in\eHb^{\infty,-1/2-}(\CX),\\
		(\check{g}^*_{b,s_1}(\sfS),\ \check{A}^*_{b,s_1}(\sfS))&:=(\hat{g}^*_{b,s_1}(\sfS)-t_{b,*}\dg^*_{b,s_1}(\sfS),\  \hat{A}^*_{b,s_1}(\sfS)-t_{b,*}\dA^*_{b,s_1}(\sfS))\in\sHb^{-\infty,-1/2-}(\CX).
\end{align}
be the coefficients of $t_{b,*}^0$ of the generalized (dual) zero modes constructed above. For later use, we determine the leading order behavior of them.

\begin{lem}\label{lem:leadingtermofgAhat}
For $\sfS\in\BFS_1$, we have
\begin{equation}\label{eq:leadingtermofgAhat}
	\begin{split}
	(\check{g}_{b,s_1}(\sfS),\ \check{A}_{b,s_1}(\sfS))&\in\rho C^\infty(\CX)+\eHb^{\infty,1/2-}(\CX),\\ (\check{g}^*_{b,s_1}(\sfS),\ \check{A}^*_{b,s_1}(\sfS))&\in\rho C^\infty(\CX)+\sHb^{-3/2-C(a,\gamma),1/2-}(\CX)
\end{split}
\end{equation}	
where $C(a,\gamma)>0$ is a sufficiently small constant depending on $a,\gamma$.
\end{lem}

\begin{proof}	
	Arguing as in the proof of Proposition \ref{prop:desofkernel}, we again exploit the normal operator argument. First, since $(\dg_{b,s_1}(\sfS),\ \dA_{b,s_1}(\sfS))\in\eHb^{\infty,1/2-}(\CX)$, according to Proposition \ref{prop:desofkernel} (where we shift the contour of integration through the pole $\xi=-2i$ and thus the space of resonant states is $\BFS_1$), it follows that
	\[
		(\dg_{b,s_1}(\sfS),\ \dA_{b,s_1}(\sfS))\in \rho^2\Omega_1+\eHb^{\infty,3/2-}(\CX)
	\]
	where 
	\[
	\Omega_1\in\mbox{span}\{\sfS dt^2,\ \sfS dt\otimes_s dx^i,\ \sfS dx^i\otimes_s dx^j,\ \sfS dt,\ \sfS dx^i\mid \sfS\in\BFS_1,\ 1\leq i,j\leq 3\}.
	\]
Since
	\[
	-\widehat{L_{b,\gamma}}(0)	(\check{g}_{b,s_1}(\sfS),\ \check{A}_{b,s_1}(\sfS))=[L_{b,\gamma}, t_{b,*}](\dg_{b,s_1}(\sfS),\ \dA_{b,s_1}(\sfS))\in-2\rho^3\Omega_1+\eHb^{\infty, 5/2-}(\CX)
	\]
	where we use the fact $[L_{b,\gamma}, t_{b,*}]\in-2\rho(\rho\pa_\rho-1)+\rho^2\mbox{Diff}_\bop^1$, it follows that we can solve the above equation for $(\check{g}_{b,s_1}(\sfS),\ \check{A}_{b,s_1}(\sfS))$ by first solving away the leading term $-2\rho^3\Omega_1$ and then the error term which lies in $\eHb^{\infty, 5/2-}(\CX)$. 
	
	For the leading term $2\rho^3\Omega_1$, proceeding as in the proof of Proposition \ref{prop:desofkernel} (here we have $k=-1, l=1$), we need to solve the following equation 
	\[
	\Big(\rho\pa_\rho(\rho\pa_\rho -1)+\sL\Big)(\dg_1,\dA_1)=-2\rho\Omega_1.
	\]
	Since now $k(k+1)=(-1)\cdot 0\neq 1\cdot 2=l(l+1)$, we conclude that $(\dg_1, \dA_2)\in(-1/2)\cdot(-2\rho\Omega_1)=\rho\Omega_1$.
	
	For the error term, we need to solve the following equation 
	\[
	\Big(\rho\pa_\rho(\rho\pa_\rho -1)+\sL\Big)(\dg_2,\dA_2)=f\in \eHb^{\infty, 5/2-}(\CX),\quad (\dg_2,\dA_2)\in\eHb^{\infty, -1/2}(\CX).
	\]
	Arguing as in the proof of Proposition \ref{prop:desofkernel}, we see that $(\dg_2,\dA_2)\in\eHb^{\infty, -1/2}(\CX)\in \rho C^\infty(\pa_+\CX)+\eHb^{\infty, 1/2-}(\CX)$. This proves \eqref{eq:leadingtermofgAhat} for $(\check{g}_{b,s_1}(\sfS),\ \check{A}_{b,s_1}(\sfS))$.
	
	For $(\check{g}^*_{b,s_1}(\sfS),\ \check{A}^*_{b,s_1}(\sfS))$, the above proof also applies. As for the regularity of $(\check{g}^*_{b,s_1}(\sfS),\ \check{A}^*_{b,s_1}(\sfS))$ near the radial point at event horizon (by the same reasoning as in the proof of statement (2) of Proposition \ref{prop:desofkernel}, $(\check{g}^*_{b,s_1}(\sfS),\ \check{A}^*_{b,s_1}(\sfS))$ is smooth away from the radial point at even horizon), we notice that $[L_{b,\gamma}, t_{b,*}](\dg^*_{b,s_1}(\sfS),\ \dA^*_{b,s_1}(\sfS))$ is one derivative less regular than $(\dg^*_{b,s_1}(\sfS),\ \dA^*_{b,s_1}(\sfS))$ and $(\widehat{L_{b,\gamma}}(0)^*)^{-1}$ gains one derivative. Therefore, $(\check{g}^*_{b,s_1}(\sfS),\ \check{A}^*_{b,s_1}(\sfS))$ has at least the same regularity as $(\dg^*_{b,s_1}(\sfS),\ \dA^*_{b,s_1}(\sfS))$, which is $-\frac32-C(a,\gamma)$ in view of Proposition \ref{prop:desofkernel}.
\end{proof}

\subsection{Generalized zero modes: quadratic growth}
Finally, we exclude the generalized zero mode solutions with quadratic growth in $t_{b,*}$ of $L_{b,\gamma}$.

\begin{lem}\label{lem:qgezeromodess1}
	Let $k\geq 2$. Then there does not exist  $(\dg,\dA)=(\sum^k_{j=1}t^j_{b,*}\dg_j,\sum^k_{j=1}t^j_{b,*}\dA_j)$, with $0\neq(\dg_k,\dA_k)=\{(\dg_{b,s_1}(\sfS),\dA_{b,s_1}(\sfS)): \sfS\in\BFS_1\}$ and $(\dg_j,\dA_j)\in\eHb^{\infty,-1/2-}(\CX;\scsym\oplus\scform)$, such that $L_{b,\gamma}(\dg,\dA)=0$.
\end{lem}

\begin{proof}
	According to the discussion at the beginning of the previous subsection, it suffices to consider the case $k=2$. For the simplicity of calculation, we replace $t_{b,*}$ by another smooth function $t_1:=t_0-2r$. Since $t_{b,*}-t_1\in\mathcal{A}^{0-}(\CX)$, we assume \[(\dg,\dA)=(t_1^2\dg_{b,s_1}(\sfS)+t_1\dg_1+\dg_0,\  t_1^2\dA_{b,s_1}(\sfS)+t_1\dA_1+\dA_0)
	\]
	with $\sfS\neq 0$ and $(\dg_{j},\dA_{j})\in\eHb^{\infty,-1/2-}(\CX)$ for $j=0,1$. We write $L_{b,\gamma}(\dg,\dA)$ as a polynomial of degree $1$ in $t_1$ as follows
	\begin{align*}
		L_{b,\gamma}(\dg,\dA)&=t_1\Big(2[L_{b,\gamma},\  t_1](\dg_{b,s_1}(\sfS),\dA_{b,s_1}(\sfS))+L_{b,\gamma}(\dg_1,\dA_1)\Big)\\
		&\quad+\Big([[L_{b,\gamma},\  t_1],\ t_1](\dg_{b,s_1}(\sfS),\dA_{b,s_1}(\sfS))+[L_{b,\gamma},\  t_1](\dg_1,\dA_1)+L_{b,\gamma}(\dg_0,\dA_0)\Big).
	\end{align*}
Then $L_{b,\gamma}(\dg,\dA)=0$ is equivalent to the vanishing of the coefficients of the above polynomial. Namely,
\begin{align}\label{eq:cooft1}
	2[L_{b,\gamma},\  t_1](\dg_{b,s_1}(\sfS),\dA_{b,s_1}(\sfS))+L_{b,\gamma}(\dg_1,\dA_1)&=0,\\\label{eq:cooft0}
	[[L_{b,\gamma},\  t_1],\ t_1](\dg_{b,s_1}(\sfS),\dA_{b,s_1}(\sfS))+[L_{b,\gamma},\  t_1](\dg_1,\dA_1)+L_{b,\gamma}(\dg_0,\dA_0)&=0.
\end{align}
First, \eqref{eq:cooft1} implies $L_{b,\gamma}\big(2t_1(\dg_{b,s_1}(\sfS),\dA_{b,s_1}(\sfS))+(\dg_1,\dA_1)\big)=0$, and thus by Proposition \ref{prop:gezeromodess1} 
\[
2t_1(\dg_{b,s_1}(\sfS),\dA_{b,s_1}(\sfS))+(\dg_1,\dA_1)=2(\hat{g}_{b_,s_1}(\sfS),\ \hat{A}_{b,s_1}(\sfS))+c(\dg_{b,s_1}(\sfS'),\ \dA_{b,s_1}(\sfS'))
\]
for some $c\in\BR$ and $\sfS'\in\BFS_1$. Subtracting $c(\dg_{b,s_1}(\sfS'),\ \dA_{b,s_1}(\sfS'))$ from $(\dg,\dA)$, we may assume $c=0$, and this implies
\[
(\dg_1,\dA_1)=2(\hat{g}_{b_,s_1}(\sfS),\ \hat{A}_{b,s_1}(\sfS))-2t_1(\dg_{b,s_1}(\sfS),\dA_{b,s_1}(\sfS)).
\]
Plugging the above expression for $(\dg_1,\dA_1)$	into \eqref{eq:cooft0} yields
\begin{equation}\label{eq:eqnforh0s1}
	\begin{split}
L_{b,\gamma}(\dg_0,\dA_0)&=-[[L_{b,\gamma},\ t_1],\ t_1](\dg_{b,s_1}(\sfS),\dA_{b,s_1}(\sfS))\\
&\quad-[L_{b,\gamma},\  t_1]\Big(2(\hat{g}_{b_,s_1}(\sfS),\ \hat{A}_{b,s_1}(\sfS))-2t_1(\dg_{b,s_1}(\sfS),\dA_{b,s_1}(\sfS))\Big) \in\eHb^{\infty,3/2-}(\CX)
\end{split}
\end{equation}
Our aim is to prove that there is no solution $(\dg_0,\dA_0)\in\eHb^{\infty,-1/2-}(\CX)$ to the above equation. To this end, we need to verify that the pairing of the right-hand side of \eqref{eq:eqnforh0s1} with $(\dg^*_{b,s_1}(\sfS),\dA_{b,s_1}^*(\sfS))$ is non-zero. We note that it suffices to check the RN case $b=b_0$ because the general KN case with $b$ near $b_0$ follows directly by the continuity (in $b$). Using $\dA_{b_0,s_1}(\sfS)=\mathcal{O}_{\eHb^{\infty,1/2-}(\CX)}(\abs{\BQ_0})$ and the same argument as in the proof of Lemma \ref{lem:gezeromodesv1}, we obtain that the the pairing of the right-hand side of \eqref{eq:eqnforh0s1} with $(\dg^*_{b_0,s_1}(\sfS),\dA_{b_0,s_1}^*(\sfS))$ is equal to 
	\begin{align*}
		&\langle[[\Box_{g_{b_0},2},\  t_1],\ t_1]\dg_{b_0,s_1}(\sfS),\ \dg^*_{b_0,s_1}(\sfS)\rangle+2\langle[\Box_{g_{b_0},2},\  t_1](\hat{g}_{b_0,s_1}(\sfS)-t_1\dg_{b_0,s_1}(\sfS)),\  \dg^*_{b_0,s_1}(\sfS)\rangle+\mathcal{O}(\abs{\BQ_0}+\gamma)\\
	&\quad:=P_1+P_2+\mathcal{O}(\abs{\BQ_0}+\gamma).
	\end{align*}
We first calculate $P_1$. Using $\dg_{b_0,s_1}(\sfS)=2\delta^*_{g_{b_0}}\omega_{b_0,s_1}(\sfS)$, $\dg^*_{b_0,s_1}(\sfS)=2G_{g_{b_0}}\delta^*_{g_{b_0}}\omega^*_{b_0,s_1}(\sfS)$ and \eqref{eq:refinedg1formmode}, we find that 
\begin{align*}
	P_1&=2\langle (\nabla^\al t_1\nabla_\al t_1)\dg_{b_0,s_1}(\sfS),\ \dg^*_{b_0,s_1}(\sfS)\rangle=2\langle 4(\weight-1)\dg_{b_0,s_1}(\sfS),\ \dg^*_{b_0,s_1}(\sfS)\rangle\\
	&=2\langle16(\weight-1) \delta^*_{g_{b_0}}d\big((r-\Bm_0)\sfS\big),G_{g_{b_0}}\delta^*_{g_{b_0}}d\big((r-\Bm_0)H(r-\ehRN)\sfS\big)\rangle+\mathcal{O}(\abs{\BQ_0}^2+\gamma)\\
&=-\frac{512}{9}\pi\Bm_0+\mathcal{O}({\abs{\BQ_0}^2}+\gamma).
	\end{align*}
Next, using $\hat{g}_{b_0,s_1}(\sfS)=2\delta_{g_{b_0}}^*\hat{\omega}_{b_0,s_1}(\sfS)$, \eqref{eq:refinedge1formmode} and \eqref{eq:refinedg1formmode}, we compute
\begin{align*}
	P_2&=2\langle\big(2\nabla^\al t_1\nabla_\al+\Box_{g_{b_0},0}t_1\big)(\hat{g}_{b_,s_1}(\sfS)-t_1\dg_{b,s_1}(\sfS)),\ \dg^*_{b_0,s_1}(\sfS)\rangle\\
	&=-\frac{64}{9}(7-\log 8)\pi\Bm_0+\mathcal{O}(\abs{\BQ_0}^2+\gamma).
\end{align*}
	Therefore, the pairing is indeed non-zero if $\abs{\BQ_0}+\gamma$ is sufficiently small.
\end{proof}

%%%%%%%%%%%%%%%%%%%%%%%%%%%%%%%%%%%%%%%%%%%%%%%%%%%%%%%%%%%%%%%%%%%%%%%%%%%
\section{Structure of the resolvent of the linearized gauge-fixed Einstein-Maxwell operator}\label{sec:structureofmlEM}
In this section, we shall prove that $\widehat{L_{b,\gamma}}(\sigma)$ is invertible for $b=(\Bm,\Ba,\BQ)$ near $b_0=(\Bm_0,0,\BQ_0),\abs{\BQ_0}\ll\Bm_0$ and $\IM\sigma\geq 0, \sigma\neq 0$. This is a generalization of the first statement in Theorem \ref{thm:modestabilityofmlgEM} from the RN case to the KN case. The proof uses the relevant conclusions of the RN case and the arguments in the proof of Proposition \ref{prop:WnonzeroinSchw}.

Let $\gamma>0$ be a fixed sufficiently small constant such that all the conclusions in the previous section hold. We recall the linearized gauge-fixed Einstein-Maxwell operator
\[
	L_{b,\gamma}:=L_{(g_b,A_b),\gamma},\quad 	L^E_{b,\gamma}:=L^E_{(g_b,A_b),\gamma},\quad	L^M_{b,\gamma}:=L^M_{(g_b,A_b),\gamma},
\]
and its Fourier transformed version
\[
	\widehat{L_{b,\gamma}}(\sigma):=e^{i\sigma t_{b,*}}L_{b,\gamma}e^{-i\sigma t_{b,*}}
	\] where $t_{b,*}=\chi_0(r)(t+r_{b_0,*})+(1-\chi_0(r))(t-r_{(\Bm,0,\BQ),*})$ is defined in \eqref{EqKNTimeFn}. Since $\gamma$ is fixed, we will drop the subscript $\gamma$ from the notations from now on.

Let $s>3$ and $-3/2<\ell<-1/2$. We redefine
\[
\mathcal{X}^{s,\ell}_{b}(\sigma):=\{(\dg,\dA)\in\eHb^{s,
\ell}(\CX;\scsym\oplus\scform): \widehat{L_{b}}(\sigma)(\dg,\dA)\in\eHb^{s-1,
\ell+2}(\CX;\scsym\oplus\scform)\}.
\]
As in the previous section, from now on, we will drop the notation of the bundle if it it clear from the context. 

We also decompose zero energy nullspace $\mathcal{K}_b$ and the space of zero energy dual states $\mathcal{K}_b^*$ of $L_{b}$ as follows
\begin{equation}
	\mathcal{K}_b=\mathcal{K}_{b,1}\oplus\mathcal{K}_{b,2},\quad \mathcal{K}^*_b=\mathcal{K}^*_{b,1}\oplus\mathcal{K}^*_{b,2}
\end{equation}
where
\begin{align}
	\mathcal{K}_{b,1}:=\{(\dg_{b,s_1}(\BFS_1),\dA_{b,s_1}(\BFS_1))\},\quad \mathcal{K}_{b,2}:=\{(\dg_{b,s_0},\dA_{b,s_0})\}\oplus\{(\dg_{b,v_1}(\BFV_1),\dA_{b,v_1}(\BFV_1))\},\\
	\mathcal{K}^*_{b,1}:=\{(\dg^*_{b,s_1}(\BFS_1),\dA^*_{b,s_1}(\BFS_1))\},\quad \mathcal{K}_{b,2}:=\{(\dg^*_{b,s_0},\dA^*_{b,s_0})\}\oplus\{(\dg^*_{b,v_1}(\BFV_1),\dA^*_{b,v_1}(\BFV_1))\}.
\end{align}
The motivation for doing such a decomposition is that $\widehat{L_{b}}(\sigma)$ is more singular when acting on $\mathcal{K}_{b,1}$ because of the existence of the generalized solutions to $L_{b}(\dg,\dA)=0$ with linear growth in $t_{b,*}$ and leading order terms in $\mathcal{K}_{b,1}$. Moreover, we recall the generalized zero energy states $(\hat{g}_{b,s_1}(\BFS_1),\hat{A}_{b,s_1}(\BFS_1))$ and dual states $(\hat{g}^*_{b,s_1}(\BFS_1),\hat{A}^*_{b,s_1}(\BFS_1))$ which is constructed in Proposition \ref{prop:gezeromodess1} and define
\begin{equation}
	\label{eq:stationaryofgezeromode}
	\check{\mathcal{K}}_b:=\{(\check{g}_{b,s_1}(\sfS),\check{A}_{b,s_1}(\sfS)):\sfS\in\BFS_1\},\quad\check{\mathcal{K}}^*_b:=\{(\check{g}^*_{b,s_1}(\sfS),\check{A}^*_{b,s_1}(\sfS)):\sfS\in\BFS_1\}
	\end{equation}
where
\begin{align}
	(\check{g}_{b,s_1}(\sfS),\check{A}_{b,s_1}(\sfS))=(\hat{g}_{b,s_1}(\sfS),\hat{A}_{b,s_1}(\sfS))-t_{b,*}(\dg_{b,s_1}(\sfS),\dA_{b,s_1}(\sfS)),\\
	(\check{g}^*_{b,s_1}(\sfS),\check{A}^*_{b,s_1}(\sfS))=(\hat{g}^*_{b,s_1}(\sfS),\hat{A}^*_{b,s_1}(\sfS))-t_{b,*}(\dg^*_{b,s_1}(\sfS),\dA^*_{b,s_1}(\sfS)).
\end{align}

\subsection{Construction of the reference operator $\check{L}_{b}(\sigma)$.}
We first construct a reference operator $\check{L}_{b}(\sigma)$ which is invertible near $\sigma=0$ by perturbing $\widehat{L_{b}}(\sigma)$ by a smoothing operator.

\begin{lem}\label{lem:checkL_b}
Let $b_0=(\Bm_0,0,\BQ_0)$ with $\abs{\BQ_0}\ll\Bm_0$.	There exist $V_b\in\Psi^{-\infty}(X;\scsym\oplus \scform)$ which is continuous in $b$ with uniformly compactly supported Schwartz kernel, and a constant $C_0>0$, such that the following statements hold.
	\begin{enumerate}
		\item The operator \[\check{L}_{b}(\sigma):=\widehat{L_{b}}(\sigma)+V_b:\mathcal{X}_b^{s,\ell}(\sigma)\to\eHb^{s-1,\ell+2}(\CX;\scsym\oplus\scform)
		\]
		is invertible for $\sigma\in\BC, \IM\sigma\geq 0$ with $\abs{\sigma}+\abs{b-b_0}\leq C_0$.
		\item  The inverse operator $\check{L}_{b}(\sigma)^{-1}$ is continuous in $\sigma$ with values in $\mathcal{L}_{\mbox{weak}}(\eHb^{s-1,\ell+2}(\CX),\eHb^{s,\ell}(\CX))$ (the space of linear bounded operators equipped with weak operator topology), and  continuous in $\sigma$ with values in $\mathcal{L}_{\mbox{op}}(\eHb^{s-1+\epsilon,\ell+2+\epsilon}(\CX),\eHb^{s-\epsilon,\ell-\epsilon}(\CX))$ (norm topology) for any $\epsilon>0$.
		\item %For a suitable chosen complementary subspace $\wt{\mathcal{K}}_b^\perp$ of $\wt{\mathcal{K}}_b=\check{L}_b(0)\mathcal{K}_b=V_b(\mathcal{K}_b)$, 
		One has
		\begin{align}\label{eq:restriction1}
			V_b(\check{g}_b,\check{A}_b)&=0\quad\mbox{for}\quad (\check{g}_b,\check{A}_b)\in\check{\mathcal{K}}_b.%\\\label{eq:restriction2}
			%\langle(\dg,\dA),\left(I-(\check{L}_b(0)^{-1})^*V_b^*\right)(\check{g}^*_b,\check{A}^*_b)\rangle&=0\quad \mbox{for}\quad (\dg,\dA)\in\wt{\mathcal{K}}_b^\perp \quad  (\check{g}^*_b,\check{A}^*_b)\in\check{\mathcal{K}}^*_b.
		\end{align}
		\end{enumerate}
	\end{lem}

\begin{proof}
The proof closely follows \cite[Lemma 11.1 and Lemma 11.7]{HHV21}. Concretely, we shall construct \[V_b:\mathcal{D}'(X;\scsym\oplus\scform)\to C^\infty_c(X;S^2T^*X\oplus T^*X)
\]
where $V_{b}$ satisfy
\begin{align}\label{eq:requirement1}
\check{\mathcal{K}}_b\subset\ker V_{b},\quad
\Ran V_{b}\subset\mathcal{R}_b^\perp\quad\mbox{and}\quad	V_{b}|_{\mathcal{K}_b}:\mathcal{K}_b\to\mathcal{R}_b^\perp\quad\mbox{is invertible}.
%\\\label{eq:requirement2}
%\check{\mathcal{K}}_b\in\ker V_{b,2},\quad\left(I-(\check{L}_b(0)^{-1})^*V_b^*\right)\check{\mathcal{K}}_b^*\cap \mbox{ann}(\wt{\mathcal{K}}_b)=0.
\end{align}
where $\mathcal{R}_b:=\left(\Ran_{\mathcal{X}_{b}^{s,\ell}(0)}\widehat{L_{b}}(0)\right)$. We now prove that once the above requirements are satisfied, all the three statements in the lemma hold. As for the proof of statements (1) and (2), it suffices to prove the invertiblity of $\check{L}_b(\sigma)=\widehat{L_{b}}(\sigma)+V_b$ for the case $b=b_0, \sigma=0$ since the invertibility and continuity for $(b,\sigma)$ near $(b_0,0)$ follow from the same arguments as in the proof of Lemma \ref{lem:prestep1}. We split the domain $\mathcal{X}_{b_0}^{s,\ell}(0)=\mathcal{K}_{b_0}^{\perp}\oplus\mathcal{K}_{b_0}$ and the target space $\eHb^{s-1,\ell+2}(\CX;\scsym\oplus\scform)=\mathcal{R}_{b_0}\oplus\mathcal{R}_{b_0}^\perp$. Under these splittings, $\check{L}_{b_0}(0)=\widehat{L_{b_0}}(0)+V_b$ takes the form
\[
\check{L}_{b_0}(0)=\begin{pmatrix}
	L_{00}&0\\
	V_{10}&V_{11}
\end{pmatrix}
\]
where $L_{00}=\widehat{L_{b_0}}(0)|_{\mathcal{K}_{b_0}^{\perp}}:\mathcal{K}_{b_0}^{\perp}\to\mathcal{R}_{b_0}$ is invertible. Then the invertibility of $\check{L}_{b_0}(0)$ follows 
%from that of $\widehat{L_{b_0}}(0)+V_{b_0}$if $c$ is sufficiently small (see the arguments in the proof of Proposition \ref{prop:WzeroinSchw}), and thus 
from that of  $V_{11}:\mathcal{K}_{b_0}\to\mathcal{R}_{b_0}^\perp$. We notice that the invertibility of $V_{11}$ holds provided that the above conditions in \eqref{eq:requirement1} are satisfied. 
For the statement (3), it is clear that the conclusion \eqref{eq:restriction1} directly follows from the condition $\check{\mathcal{K}}_b\subset \ker V_{b}$.
% It remains to prove the second conclusion \eqref{eq:restriction2} in the statement (3). Since $\left(I-(\check{L}_b(0)^{-1})^*V_b^*\right)\check{\mathcal{K}}_b^*\cap \mbox{ann}(\wt{\mathcal{K}}_b)=0$, for any $0\neq f\in\left(I-(\check{L}_b(0)^{-1})^*V_b^*\right)\check{\mathcal{K}}_b^*$, $f|_{\wt{\mathcal{K}}_b}$ is a nonzero linear functional on $\wt{\mathcal{K}}_b$. Therefore, $\left(I-(\check{L}_b(0)^{-1})^*V_b^*\right)\check{\mathcal{K}}_b^*$ can be extended the $7$-dimensional space of linear functionals on $\wt{\mathcal{K}}_b$, which is spanned by $\{f_1^*,\cdots,f_7^*\}$. By Hahn-Banach Theorem, each $f_i^*$ can be extended by a continuous functional $\tilde{f}_i^*$ on $\mathcal{X}_b^{s,\ell}(0)$ (if $f_i$ is obtained by restricting some $f\in\left(I-(\check{L}_b(0)^{-1})^*V_b^*\right)\check{\mathcal{K}}_b^*$ on $\wt{\mathcal{K}}_b$, we take $\tilde{f}_i^*=f$). We define $\wt{\mathcal{K}}_b^\perp:=\cap_{i=1}^7(\tilde{f}_i^*)^{-1}(0)$ and then it is clear that \eqref{eq:restriction2} holds for this choice of $\wt{\mathcal{K}}_b$.

 First, since $(\dg_{b_0,s_1}(\BFS_1), \dA_{b_0,s_1}(\BFS_1))\in\eHb^{\infty,1/2-}(\CX)$ (see Theorem \ref{thm:modestabilityofmlgEM}) and $(\check{g}_{b_0,s_1}(\BFS_1), \check{A}_{b_0,s_1}(\BFS_1))$ have nonzero $r^{-1}$ leading terms (see Lemma \ref{lem:leadingtermofgAhat}), we conclude that $\mathcal{K}_{b_0,1}\cap\check{\mathcal{K}}_{b_0}=0$, and thus $\mathcal{K}_{b_0}\cap\check{\mathcal{K}}_{b_0}=0$. By the continuity of $\mathcal{K}_b$ and $\check{\mathcal{K}}_b$ in $b$, this holds for $b$ near $b_0$ as well. This implies that fixing the basis $\{h_{b,1},\cdots, h_{b,8}\}$ of $\mathcal{K}_b$ and $\{\check{h}_{b,1},\check{h}_{b,2},\check{h}_{b,3}\}$ of $\check{\mathcal{K}}_b$, both of which depend continuously on $b$, by the same reasoning as in the beginning of the seventh step in the proof of Theorem \ref{thm:modestabilityofmlgEM}, there exists $\{h^\flat_{b,1},\cdots, h^\flat_{b,8}\}\subset C^\infty_c(X^\circ;\scsym\oplus \scform)$ such that $\langle h_{b,i}, h^\flat_{b,j} \rangle=\delta_{ij}$ for $1\leq i,j \leq 8$ and $\langle\check{h}_{b,i}, h^\flat_{b,j}\rangle=0$ for $1\leq i\leq 3,1\leq j\leq 8$. (One can verify that $\{h^\flat_{b,1},\cdots, h^\flat_{b,8}\}\subset C^\infty_c(X^\circ)$ can be chosen to be continuous in $b$.) Next, we pick a basis of $\mathcal{R}_b^\perp$ in the following way. We note that $\mathcal{K}^*_b\cap\check{\mathcal{K}}^*_b=0$, which can be proved in a similar manner to the statement $\mathcal{K}_b\cap\check{\mathcal{K}}_b=0$. Fixing the continuous (in $b$) basis $\{h^*_{b,1},\cdots, h^*_{b,8}\}$ of $\mathcal{K}^*_b=\mbox{ann}(\mathcal{R}_b)$ and $\{\check{h}^*_{b,1}, \check{h}^*_{b,2}, \check{h}^*_{b,3}\}$ of $\check{\mathcal{K}}^*_b$, by the same reasoning mentioned above, there exists $\{h^\sharp_{b,1},\cdots, h^\sharp_{b,8}\}\subset C^\infty_c(X;\scsym\oplus \scform)$ (which can be chosen continuous in $b$) such that $\langle h^\sharp_{b,i}, h^*_{b,j}\rangle=\delta_{ij}$ for $1\leq i,j\leq 8$ and $\langle h^\sharp_{b,i}, \check{h}^*_{b,j}\rangle=0$ for $1\leq i\leq 8, 1\leq j\leq 3$. It is easy to check that $\{h^\sharp_{b,1},\cdots, h^\sharp_{b,8}\}$ are linearly independent and the space generated by $h^\sharp_{b,j}, 1\leq j\leq 8$ is a complement of $\mathcal{R}_b$. Therefore, we define
	\[
	V_{b}(h):=\sum_{i=1}^8h_{b,i}^\sharp\langle h, h_{b,i}^\flat\rangle,
	\]
	which satisfies the requirement \eqref{eq:requirement1}.

\end{proof}

Using $\check{L}_{b}(\sigma)$ constructed in Lemma \ref{lem:checkL_b}, we define the following spaces
\begin{equation}
	\wt{\mathcal{K}}_{b,1}:=\check{L}_b(0)\mathcal{K}_{b,1},\quad	\wt{\mathcal{K}}_{b,2}:=\check{L}_b(0)\mathcal{K}_{b,2},\quad	\wt{\mathcal{K}}_{b}:=\check{L}_b(0)\mathcal{K}_b=	\wt{\mathcal{K}}_{b,1}\oplus	\wt{\mathcal{K}}_{b,2}.
\end{equation}
Since $\check{L}_{b}(0)\mathcal{K}_b=V_{b}(\mathcal{K}_b)$, we have $\wt{\mathcal{K}}_{b,1}, 	\wt{\mathcal{K}}_{b,2}, \wt{\mathcal{K}}_{b}\subset C^\infty_c(X;\scsym\oplus \scform)$.

Next, we decompose the target space $\eHb^{s-1,\ell+2}(\CX;\scsym\oplus\scform)$ into the range of $\widehat{L_{b}}(0)$ and its complement in a continuous (in $b$) way.
\begin{lem}
	Let $b_0=(\Bm_0,0,\BQ_0)$ with $\abs{\BQ_0}\ll\Bm_0$. There exists a linear projection map \[
	\Pi_b^\perp:\eHb^{s-1,\ell+2}(\CX;\scsym\oplus\scform)\to\eHb^{s-1,\ell+2}(\CX;\scsym\oplus\scform),
	\]
	which is of rank $8$ and depends continuously on $b$ near $b_0$ in the norm topology, such that
	\[
	\langle (I-\Pi_b^\perp)h,\  h^*\rangle=0\quad \mbox{for all}\quad h^*\in\mathcal{K}_b^*.
	\]
	Moreover, the $\Pi_{b}^\perp$ can be chosen such that it maps $\eHb^{s-1,\ell+2}(\CX;\scsym\oplus\scform)$ to $\C^\infty_c (X; \scsym\oplus \scform)$.
\end{lem}
\begin{proof}
	Let $\{h^*_{b,1}, \cdots,h^*_{b,8}\}$ be a basis which is continuous in $b$ of $\mathcal{K}^*_b$. There exists $\{h_{b,1}^\sharp, \cdots,h^\sharp_{b,8}\}\subset C^\infty_c(X;\scsym\oplus \scform)$ which is continuous in $b$, such that $\langle h_{b,i}^\sharp, h_{b,j}^*\rangle=\delta_{ij}$. We then define
	\[
	\Pi_b^\perp h=\sum_{i=1}^8h^\sharp_{b,i}\langle h, h^*_{b,i}\rangle.
	\]
	It is easy to check that $\Pi_b^\perp$ has rank $8$ and satisfies $	\langle (I-\Pi_b^\perp)h,\  h^*\rangle=0$ for all $h^*\in\mathcal{K}_b^*$.
\end{proof}
We then define the projection onto $\mathcal{R}_b$ which is the range of $\widehat{L_{b}}(0)$
\[
\Pi_{b}=I-\Pi_{b}^\perp:\eHb^{s-1,\ell+2}(\CX;\scsym\oplus\scform)\to\mathcal{R}_b=\mbox{ann }\mathcal{K}_b^*.
\]
Now we are able to split the domain and target space of the operator $\widehat{L_{b}}(\sigma)\check{L}_b(\sigma)^{-1}:\eHb^{s-1,\ell+2}\to\eHb^{s-1,\ell+2}$ as follows.
\begin{equation}\label{eq:splitofdoandran}
	\begin{split}
	\mbox{domain:}\quad \eHb^{s-1,\ell+2}(\CX;\scsym\oplus\scform)=\wt{\mathcal{K}}_b^\perp\oplus\wt{\mathcal{K}}_{b,1}\oplus\wt{\mathcal{K}}_{b,2},\\
	\mbox{target:}\quad \eHb^{s-1,\ell+2}(\CX;\scsym\oplus\scform)=\mathcal{R}_b\oplus\mathcal{R}^\perp_{b,1}\oplus\mathcal{R}^\perp_{b,2}
	\end{split}
\end{equation}
where $\mathcal{R}_{b,1}^\perp$, resp. $\mathcal{R}_{b,2}^\perp$ is a subspace of dimension $3$, resp. $5$ and defined as follows. Fixing the continuous (in $b$) basis $\{h^{*(1)}_{b,1},\cdots, h^{*(1)}_{b,3}\}$ of $\mathcal{K}^*_{b,1}$ and $\{h^{*(2)}_{b,1}, \cdots, h^{*(2)}_{b,5}\}$ of $\mathcal{K}^*_{b,2}$, there exist \[\{v^{(1)}_{b,1},v^{(1)}_{b,2},v^{(1)}_{b,3}\},\   \{v^{(2)}_{b,1},v^{(2)}_{b,2}, v^{(2)}_{b,3},v^{(2)}_{b,4},v^{(2)}_{b,5}\}\subset C^\infty_c(X; \scsym\oplus \scform)
\]
such that $\langle v_{b,i}^{(1)}, h^{*(1)}_{b,j}\rangle=\delta_{ij}$ and $\langle v_{b,i}^{(2)}, h^{*(2)}_{b,j}\rangle=\delta_{ij}$. We then define 
\[
\mathcal{R}_{b,1}^\perp=\{v^{(1)}_{b,1}, v^{(1)}_{b,2}, v^{(1)}_{b,3}\}\quad \mbox{and}\quad \mathcal{R}_{b,2}^\perp=\{v^{(2)}_{b,1}, v^{(2)}_{b,2}, v^{(2)}_{b,3} ,v^{(2)}_{b,4},v^{(2)}_{b,5}\}.
\]

\subsection{Structure of the resolvent near zero energy}\label{subsec:structureof rel}

Now we are at the position of establishing the mode stability of the operator $\widehat{L_{b}}(\sigma)$ for $b$ near $b_0$ at $\IM\sigma\geq 0, \sigma\neq0$, and we also provide a description of the structure of the resolvent $\widehat{L_{b}}(\sigma)^{-1}$ near $\sigma=0$ along the way.

\begin{thm}\label{thm:modestabilityofmlgEMofKN}
	Let $(g_{b}, A_{b})$ be the RN metric and $4$-electromagnetic potential where $b=(\Bm, \Ba, \BQ)$ is near $b_0=(\Bm_0, 0,\BQ_0)$ with $\abs{\BQ_0}\ll\Bm_0$. Then the Fourier-transformed operator 
		\begin{equation}\label{eq:mlgEMmodenonzeroforRN}
			\begin{split}
				&\widehat{L_{b}}(\sigma):\{(\dg,\dA)\in\eHb^{s, \ell}(\CX;\scsym\oplus\scform): 	\widehat{L_{b}}(\sigma)(\dg,\dA)\in\eHb^{s,\ell+1}(\CX; \scsym\oplus\scform)\}\\
				&\qquad\qquad\qquad\to\eHb^{s,\ell+1}(\CX;\scsym\oplus\scform)
			\end{split}
		\end{equation}
		is invertible for $\sigma\in\BC, \IM\sigma\geq 0, \sigma\neq 0$ and $s>3, \ell<-\frac12, s+\ell>-\frac 12$.
		\end{thm}
	
	\begin{proof}
		By the arguments in the last step of the proof of Theorem \ref{thm:modesforscalarwave} and Proposition \ref{prop:WnonzeroinSchw}, it suffices to prove that $\mathcal{X}^{s,\ell}_b(\sigma)\to\eHb^{s,\ell+1}(\CX;\scsym\oplus\scform)$ is invertible for $s>3, -3/2<\ell<-1/2$ and $(b,\sigma)$ with $\IM\sigma\geq 0, \sigma\neq 0$ near $(b_0,0)$. To this end, we shall prove the invertibility of the operator
		\[
		\widehat{L_{b}}(\sigma)\check{L}_b(\sigma)^{-1}:\eHb^{s-1,\ell+2}(\CX;\scsym\oplus\scform)\to\eHb^{s-1,\ell+2}(\CX;\scsym\oplus\scform)
		\]
		for $s>3, -3/2<\ell<-1/2$ and $(b,\sigma)$ with $\IM\sigma\geq 0, \sigma\neq 0$ near $(b_0,0)$.
		
		Under the splitting \eqref{eq:splitofdoandran}, we express
		\begin{equation}\label{eq:exofLL-1}
				\widehat{L_{b}}(\sigma)\check{L}_b(\sigma)^{-1}=
			\begin{pmatrix}
				L_{00}(b,\sigma)&L_{01}(b,\sigma)&L_{02}(b,\sigma)\\
				L_{10}(b,\sigma)&L_{11}(b,\sigma)&L_{12}(b,\sigma)\\
				L_{20}(b,\sigma)&L_{21}(b,\sigma)&L_{22}(b,\sigma)\\
			\end{pmatrix}.
		\end{equation}
	It is clear that $L_{ij}(b,0)=0$ when $(i,j)\neq (0,0)$ and $L_{00}(b,0)$ is invertible. We will use the following resolvent identity (whose proof is similar to that of the resolvent identity \eqref{eq:resolventidentity} in the proof of Lemma \ref{lem:step1}) frequently in the subsequent analysis
	\begin{align}\label{eq:resolventidentity1}
		\big(\check{L}_b(\sigma)^{-1}-\check{L}_b(0)^{-1}\big)(\dg,\dA)&=\check{L}_b(\sigma)^{-1}\big(\check{L}_b(0)-\check{L}_b(\sigma)\big)\check{L}_b(0)^{-1}(\dg,\dA)\\\label{eq:dualresolventidentity}			\big((\check{L}_b(\sigma)^{-1})^*-(\check{L}_b(0)^{-1})^*\big)(\dg^*,\dA^*)&=(\check{L}_b(\sigma)^{-1})^*\big(\check{L}_b(0)^*-\check{L}_b(\sigma)^*\big)(\check{L}_b(0)^{-1})^*(\dg^*,\dA^*)
	\end{align}
for $\check{L}_b(0)^{-1}(\dg,\dA)\in \rho C^\infty(\CX)+\eHb^{s, 1/2-}(\CX)$ and $(\check{L}_b(0)^{-1})^*(\dg^*,\dA^*)\in \rho C^\infty(\CX)+\sHb^{-s+1, 1/2-}(\CX)$.
\begin{itemize}
	\item\underline{Analysis of $L_{01}$ and $L_{02}$.} For $(\tilde{g}_2, \tilde{A}_2)=\check{L}_b(0)(\dg_2,\dA_2)\in\wt{\mathcal{K}}_{b,2}$, we have 
	\begin{equation}\label{eq:calofL02}
		\begin{split}
			\widehat{L_{b}}(\sigma)\check{L}_b(\sigma)^{-1}(\tilde{g}_2, \tilde{A}_2)&=\big(\check{L}_b(\sigma)-V_b\big)\check{L}_b(\sigma)^{-1}\check{L}_b(0)(\dg_2,\dA_2)\\
			&=-V_b\big(\check{L}_b(\sigma)^{-1}-\check{L}_b(0)^{-1}\big)\check{L}_b(0)(\dg_2,\dA_2)\\
			&=V_b\check{L}_b(\sigma)^{-1}\big(\check{L}_b(\sigma)-\check{L}_b(0)\big)\check{L}_b(0)^{-1}\check{L}_b(0)(\dg_2,\dA_2)\\
			&=\sigma V_b\check{L}_b(\sigma)^{-1}\pa_\sigma\widehat{L_{b}}(0)(\dg_2,\dA_2)+\frac{\sigma^2}{2}V_b\check{L}_b(\sigma)^{-1}\pa^2_\sigma\widehat{L_{b}}(0)(\dg_2,\dA_2)
		\end{split}
	\end{equation}
where we use $\check{L}_b(0)^{-1}\check{L}_b(0)(\dg_2,\dA_2)=(\dg_2,\dA_2)\in \rho C^\infty(\CX)+\eHb^{\infty, 1/2-}(\CX)$ by Proposition \ref{prop:desofkernel} and the resolvent identity \eqref{eq:resolventidentity1} in the third step. Since $(\dg_2,\dA_2)\in\rho C^\infty(\CX)+\eHb^{\infty,1/2-}(\CX)$ which is annihilated by normal operator of $\pa_\sigma\widehat{L_{b}}(0)$ modulo $\eHb^{\infty,3/2-}(\CX)$, it follows that \[
\pa_\sigma\widehat{L_{b}}(0)(\dg_2,\dA_2)\in\eHb^{\infty,3/2-}(\CX).
\]
 As $\pa^2_\sigma\widehat{L_{b}}(0)\in\rho^2C^\infty(\CX;\mbox{End}(\scsym\oplus\scform))$, we have \[\pa^2_\sigma\widehat{L_{b}}(0)(\dg_2,\dA_2)\in\eHb^{\infty,3/2-}(\CX)
 \]
  as well. By using the continuity of $\check{L}_b(\sigma)^{-1}$ in the norm operator topology \[\mathcal{L}_{\mbox{op}}(\eHb^{s-1+\epsilon, \ell+2+\epsilon}(\CX), \eHb^{s-\epsilon, \ell-\epsilon}(\CX)),
  \]
   we conclude that
\[
\check{L}_b(\sigma)^{-1}\pa_\sigma\widehat{L_{b}}(0)(\dg_2,\dA_2),\ \check{L}_b(\sigma)^{-1}\pa^2_\sigma\widehat{L_{b}}(0)(\dg_2,\dA_2)
\] 
are continuous in $(b,\sigma)$, and thus \[
L_{02}(b,\sigma)=\sigma\wt{L}_{02}(b,\sigma)
\]
 where $\wt{L}_{02}(b,\sigma):\wt{\mathcal{K}}_{b,2}\to\mathcal{R}_b$ is continuous in $(b,\sigma)$  in the norm operator topology.
 
For $(\tilde{g}_1, \tilde{A}_1)=\check{L}_b(0)(\dg_1,\dA_1)\in\wt{\mathcal{K}}_{b,1}$, the calculation in \eqref{eq:calofL02} still applies. We further calculate 
\begin{align*}
	&\widehat{L_{b}}(\sigma)\check{L}_b(\sigma)^{-1}(\tilde{g}_1, \tilde{A}_1)\\
	&\quad=\sigma V_b\check{L}_b(\sigma)^{-1}\pa_\sigma\widehat{L_{b}}(0)(\dg_1,\dA_1)+\frac{\sigma^2}{2}V_b\check{L}_b(\sigma)^{-1}\pa^2_\sigma\widehat{L_{b}}(0)(\dg_1,\dA_1)\\
	&\quad=i\sigma V_b\check{L}_b(\sigma)^{-1}\widehat{L_{b}}(0)(\check{g}_1,\check{A}_1)+\frac{\sigma^2}{2}V_b\check{L}_b(\sigma)^{-1}\pa^2_\sigma\widehat{L_{b}}(0)(\dg_1,\dA_1)\\
	&\quad=i\sigma\big(I-V_b\check{L}_b(0)^{-1}\big)V_b(\check{g}_1,\check{A}_1)+i\sigma V_b(\check{L}_b(\sigma)^{-1}-\check{L}_b(0)^{-1})\widehat{L_{b}}(0)(\check{g}_1,\check{A}_1)\\
	&\quad\quad+\frac{\sigma^2}{2}V_b\check{L}_b(\sigma)^{-1}\pa^2_\sigma\widehat{L_{b}}(0)(\dg_1,\dA_1)\\
	&\quad =i\sigma V_b(\check{L}_b(\sigma)^{-1}-\check{L}_b(0)^{-1})\widehat{L_{b}}(0)(\check{g}_1,\check{A}_1)+\frac{\sigma^2}{2}V_b\check{L}_b(\sigma)^{-1}\pa^2_\sigma\widehat{L_{b}}(0)(\dg_1,\dA_1)
\end{align*}
where we use Lemma \ref{lem:checkL_b} in the last step. Since by Lemma \ref{lem:checkL_b} and Lemma \ref{lem:leadingtermofgAhat}, we have \[\check{L}_b(0)^{-1}\widehat{L_{b}}(0)(\check{g}_1,\check{A}_1)=(\check{g}_1,\check{A}_1)\in\rho C^\infty(\CX)+\eHb^{\infty,1/2-}(\CX),
\]
and it follows that we can apply the resolvent identity \eqref{eq:resolventidentity1} again and obtain
\begin{equation}\label{eq:calofL01}
	\begin{split}
	&\widehat{L_{b}}(\sigma)\check{L}_b(\sigma)^{-1}(\tilde{g}_1, \tilde{A}_1)\\
	&\quad=i\sigma V_b\check{L}_b(\sigma)^{-1}\big(\check{L}_b(0)-\check{L}_b(\sigma)\big)\check{L}_b(0)^{-1}\widehat{L_{b}}(0)(\check{g}_1,\check{A}_1)+\frac{\sigma^2}{2}V_b\check{L}_b(\sigma)^{-1}\pa^2_\sigma\widehat{L_{b}}(0)(\dg_1,\dA_1)\\
	&\quad=\sigma^2\Big(-iV_b\check{L}_b(\sigma)^{-1}\big(\pa_\sigma\widehat{L_{b}}(0)+\frac{\sigma}{2}\pa^2_\sigma\widehat{L_{b}}(0)\big)(\check{g}_1,\check{A}_1)+\frac{1}{2}V_b\check{L}_b(\sigma)^{-1}\pa^2_\sigma\widehat{L_{b}}(0)(\dg_1,\dA_1)\Big).
	\end{split}
\end{equation}
By the same reasoning as in the component $L_{02}(b,\sigma)$, we find that 
\[
L_{01}(b,\sigma)=\sigma^2\wt{L}_{01}(b,\sigma)
\]
where $\wt{L}_{01}(b,\sigma):\wt{\mathcal{K}}_{b,1}\to\mathcal{R}_b$ is continuous in $(b,\sigma)$ in the norm operator topology.

\item\underline{Analysis of $L_{10}$ and $L_{20}$.} For $(\tilde{g}_0, \tilde{A}_0)\in\wt{\mathcal{K}}_{b}^\perp$ and $(\dg^*_{2}, \dA^*_2)\in\mathcal{K}^*_{b,2}$, we have 
\begin{equation}\label{eq:calofL20}
	\begin{split}
		&\langle\widehat{L_{b}}(\sigma)\check{L}_b(\sigma)^{-1}(\tilde{g}_0, \tilde{A}_0),\ (\dg^*_{2}, \dA^*_2)\rangle\\
		&\quad=\langle(\tilde{g}_0, \tilde{A}_0),\ (\check{L}_b(\sigma)^{-1})^*\widehat{L_{b}}(\sigma)^*(\dg^*_{2}, \dA^*_2)\rangle\\
		&\quad=\langle(\tilde{g}_0, \tilde{A}_0),(\dg^*_{2}, \dA^*_2)\rangle-\langle(\tilde{g}_0, \tilde{A}_0),(\check{L}_b(\sigma)^{-1})^*V_b^*(\dg^*_{2}, \dA^*_2)\rangle\\
		&\quad=-\Big\langle(\tilde{g}_0, \tilde{A}_0),\big((\check{L}_b(\sigma)^{-1})^*-(\check{L}_b(0)^{-1})^*\big)V_b^*(\dg^*_{2}, \dA^*_2)\Big\rangle\\
		&\quad=\Big\langle(\tilde{g}_0, \tilde{A}_0),(\check{L}_b(\sigma)^{-1})^*\big(\check{L}_b(\sigma)^*-\check{L}_b(0)^*\big)(\check{L}_b(0)^{-1})^*V_b^*(\dg^*_{2}, \dA^*_2)\Big\rangle\\
		&\quad=\sigma\Big\langle (\tilde{g}_0, \tilde{A}_0), (\check{L}_b(\sigma)^{-1})^*\big(\pa_\sigma\widehat{L_{b}}(0)+\frac{\sigma}{2}\pa^2_\sigma\widehat{L_{b}}(0)\big)^*(\dg^*_2,\dA^*_2)\Big\rangle
	\end{split}
\end{equation}
where we use $(\check{L}_b(0)^{-1})^*V_b^*(\dg^*_2,\dA^*_2)=(\dg^*_2,\dA^*_2)\in \rho C^\infty(\CX)+\sHb^{-3/2-C(a,\gamma), 1/2-}(\CX)$ where $C(a,\gamma)$ is a small constant by Proposition \ref{prop:desofkernel} and the resolvent identity \eqref{eq:dualresolventidentity} in the fourth step. Since $(\dg^*_2,\dA^*_2)$ is annihilated by normal operator of $(\pa_\sigma\widehat{L_{b}}(0))^*$ modulo $\sHb^{-5/2-C(a,\gamma),3/2-}(\CX)$, it follows that $(\pa_\sigma\widehat{L_{b}}(0))^*(\dg^*_2,\dA^*_2)\in\sHb^{-5/2-C(a,\gamma),3/2-}(\CX)$. As $(\pa^2_\sigma\widehat{L_{b}}(0))^*\in\rho^2C^\infty(\CX)$, we have \[(\pa^2_\sigma\widehat{L_{b}}(0))^*(\dg^*_2,\dA^*_2)\in\sHb^{-3/2-\nu-,3/2-}(\CX).
\]
 Since $(\check{L}_b(\sigma)^{-1})^*$ is continuous with values in $\mathcal{L}_{\mathrm{op}}(\sHb^{-s+\epsilon, -\ell+\epsilon}(\CX), \sHb^{-s+1-\epsilon, -\ell-2-\epsilon}(\CX))$, we conclude that
\[
(\check{L}_b(\sigma)^{-1})^*(\pa_\sigma\widehat{L_{b}}(0))^*(\dg^*_2,\dA^*_2),\ (\check{L}_b(\sigma)^{-1})^*(\pa^2_\sigma\widehat{L_{b}}(0))^*(\dg^*_2,\dA^*_2)
\] 
are continuous in $(b,\sigma)$, and thus \[
L_{20}(b,\sigma)=\sigma\wt{L}_{20}(b,\sigma)
\]
where $\wt{L}_{20}(b,\sigma):\wt{\mathcal{K}}^\perp_{b}\to\mathcal{R}^\perp_{b,2}$ is continuous in $(b,\sigma)$ in the norm operator topology.

For $(\tilde{g}_0, \tilde{A}_0)\in\wt{\mathcal{K}}_{b}^\perp$ and $(\dg^*_{1}, \dA^*_1)\in\mathcal{K}^*_{b,1}$, the calculation in \eqref{eq:calofL20} still applies. We further calculate
	\begin{align*}
		&\langle\widehat{L_{b}}(\sigma)\check{L}_b(\sigma)^{-1}(\tilde{g}_0, \tilde{A}_0),\ (\dg^*_1, \dA^*_1)\rangle\\
		&\quad=\sigma\Big\langle (\tilde{g}_0, \tilde{A}_0), (\check{L}_b(\sigma)^{-1})^*\big(\pa_\sigma\widehat{L_{b}}(0)+\frac{\sigma}{2}\pa_\sigma^2\widehat{L_{b}}(0)\big)^*(\dg^*_1,\dA^*_1)\Big\rangle\\
		&\quad=\sigma\Big\langle (\tilde{g}_0, \tilde{A}_0), i(\check{L}_b(\sigma)^{-1})^*\widehat{L_{b}}(0)^*(\check{g}^*_1,\check{A}^*_1)\Big\rangle\\
		&\quad\quad+\sigma^2\Big\langle(\tilde{g}_0, \tilde{A}_0),\frac{1}{2}(\check{L}_b(\sigma)^{-1})^*(\pa^2_\sigma\widehat{L_{b}}(0))^*(\dg^*_1,\dA^*_1)\Big\rangle\\
		&\quad=\sigma\Big\langle (\tilde{g}_0, \tilde{A}_0), i\big(I-(\check{L}_b(0)^{-1})^*V_b^*\big)(\check{g}^*_1,\check{A}^*_1)\Big\rangle\\
		&\quad\quad+\sigma\Big\langle (\tilde{g}_0, \tilde{A}_0), i\big((\check{L}_b(\sigma)^{-1})^*-(\check{L}_b(0)^{-1})^*\big)\widehat{L_{b}}(0)^*(\check{g}^*_1,\check{A}^*_1)\Big\rangle\\
		&\quad\quad\quad+\sigma^2\Big\langle(\tilde{g}_0, \tilde{A}_0),\frac{1}{2}(\check{L}_b(\sigma)^{-1})^*(\pa^2_\sigma\widehat{L_{b}}(0))^*(\dg^*_1,\dA^*_1)\Big\rangle
	\end{align*}
By a normal operator argument as in the proof of Proposition \ref{prop:desofkernel}, we have
\[
(\check{L}_b(0)^{-1})^*: C^\infty_c(X)\to \rho C^\infty(\CX)+\sHb^{-3/2-C(a,\gamma),1/2-}(\CX).
\]
Therefore, by Lemma \ref{lem:leadingtermofgAhat}, we conclude \[(\check{L}_b(0)^{-1})^*\widehat{L_{b}}(0)^*(\check{g}_1,\check{A}_1)=(\check{g}_1,\check{A}_1)-(\check{L}_b(0)^{-1})^*V^*_b(\check{g}_1,\check{A}_1)\in\rho C^\infty(\CX)+\sHb^{-3/2-C(a,\gamma),1/2-}(\CX).
\]
This implies that we can apply the resolvent identity \eqref{eq:dualresolventidentity} again and obtain
\begin{equation}\label{eq:calofL10}
	\begin{split}
		&\langle\widehat{L_{b}}(\sigma)\check{L}_b(\sigma)^{-1}(\tilde{g}_0, \tilde{A}_0),\ (\dg^*_1, \dA^*_1)\rangle\\
		&\quad=\sigma\Big\langle (\tilde{g}_0, \tilde{A}_0), i\big(I-(\check{L}_b(0)^{-1})^*V_b^*\big)(\check{g}^*_2,\check{A}^*_2)\Big\rangle\\
		&\quad\quad-\sigma^2\Big\langle (\tilde{g}_0, \tilde{A}_0), i(\check{L}_b(\sigma)^{-1})^*(\pa_\sigma\widehat{L_{b}}(0)+\frac{\sigma}{2}\pa^2_\sigma\widehat{L_{b}}(0))^*\big(I-(\check{L}_b(0)^{-1})^*V_b^*\big)(\check{g}^*_2,\check{A}^*_2)\Big\rangle\\
		&\quad\quad\quad+\sigma^2\Big\langle(\tilde{g}_0, \tilde{A}_0),\frac{1}{2}(\check{L}_b(\sigma)^{-1})^*(\pa^2_\sigma\widehat{L_{b}}(0))^*(\dg^*_2,\dA^*_2)\Big\rangle.
	\end{split}
\end{equation}
By the same reasoning as in the component $L_{20}(b,\sigma)$, we find that 
\[
L_{10}(b,\sigma)=\sigma \wt{L}_{10}(b,\sigma)
\]
where $\wt{L}_{10}(b,\sigma)=\wt{L}^0_{10}(b)+\sigma\wt{L}^e_{10}(b,\sigma)$ with $\wt{L}^e_{10}(b,\sigma):\wt{\mathcal{K}}_{b}^\perp\to \mathcal{R}_{b,1}^\perp$ continuous in $(b,\sigma)$ in the norm operator topology.

\item\underline{Analysis of $L_{22}$.} Using \eqref{eq:calofL02}, we find that 
\begin{equation}\label{eq:calofL22}
	\begin{split}
		&\langle\widehat{L_{b}}(\sigma)\check{L}_b(\sigma)^{-1}(\tilde{g}_2,\tilde{A}_2),\  (\dg_2^*, \dA_2^*)\rangle\\
		&\quad=\langle\sigma\pa_\sigma\widehat{L_{b}}(0)(\dg_2,\dA_2)+\frac{\sigma^2}{2}\pa^2_\sigma\widehat{L_{b}}(0)(\dg_2,\dA_2),\  (\check{L}_b(\sigma)^{-1})^*V_b^*(\dg_2^*, \dA_2^*)\rangle\\
		&\quad=\langle\sigma\pa_\sigma\widehat{L_{b}}(0)(\dg_2,\dA_2)+\frac{\sigma^2}{2}\pa^2_\sigma\widehat{L_{b}}(0)(\dg_2,\dA_2),\  (\dg_2^*, \dA_2^*)\rangle\\
		&\quad\quad+\langle\sigma\pa_\sigma\widehat{L_{b}}(0)(\dg_2,\dA_2)+\frac{\sigma^2}{2}\pa^2_\sigma\widehat{L_{b}}(0)(\dg_2,\dA_2),\ \big((\check{L}_b(\sigma)^{-1})^*-(\check{L}_b(0)^{-1})^*\big)V_b^*(\dg_2^*, \dA_2^*)\rangle\\
			&\quad=\langle\sigma\pa_\sigma\widehat{L_{b}}(0)(\dg_2,\dA_2)+\frac{\sigma^2}{2}\pa^2_\sigma\widehat{L_{b}}(0)(\dg_2,\dA_2),\  (\dg_2^*, \dA_2^*)\rangle\\
		&\quad\quad+\langle\sigma\pa_\sigma\widehat{L_{b}}(0)(\dg_2,\dA_2)+\frac{\sigma^2}{2}\pa^2_\sigma\widehat{L_{b}}(0)(\dg_2,\dA_2),\ (\check{L}_b(\sigma)^{-1})^*\big(\check{L}_b(0)^*-\check{L}_b(\sigma)^*\big)(\dg_2^*, \dA_2^*)\rangle\\
		&\quad=\sigma\langle-i[L_b, t_{b,*}](\dg_2, \dA_2)\,\  (\dg_2^*, \dA_2^*)\rangle+\sigma^2\Big(\langle\frac{1}{2}\pa^2_\sigma\widehat{L_{b}}(0)(\dg_2,\dA_2),\  (\dg_2^*, \dA_2^*)\rangle\\
		&\quad\quad-\big\langle\pa_\sigma\widehat{L_{b}}(0)(\dg_2,\dA_2)+\frac{\sigma}{2}\pa^2_\sigma\widehat{L_{b}}(0)(\dg_2,\dA_2), (\check{L}_b(\sigma)^{-1})^*(\pa_\sigma\widehat{L_{b}}(0)+\frac{\sigma}{2}\pa^2_\sigma\widehat{L_{b}}(0))^*(\dg_2^*, \dA_2^*)\big\rangle\Big)
	\end{split}
\end{equation}
where we use $(\check{L}_b(0)^{-1})^*V_b^*(\dg_2^*, \dA_2^*)=(\dg_2^*, \dA_2^*)$ and the resolvent identity \eqref{eq:dualresolventidentity} in the third step. According to the proof of the non-existence of the linearly growing generalized zero modes with leading order term $(\dg_2, \dA_2)\in\mathcal{K}_{b,2}$ (see Lemma \ref{lem:gezeromodesv1} and \ref{lem:gezeromodess0}), the pairing on $\mathcal{K}_{b,2}\times\mathcal{K}_{b,2}^*$
\[
\langle-i[L_b, t_{b,*}](\dg_2, \dA_2),\  (\dg_2^*, \dA_2^*)\rangle
\]
is non-degenerate for $b$ near $b_0$. Therefore, we find that 
\[
L_{22}(b,\sigma)=\sigma\wt{L}_{22}(b,\sigma)\quad\mbox{with}\quad \wt{L}_{22}(b,\sigma)=\wt{L}^0_{22}(b)+\sigma\wt{L}^e_{22}(b,\sigma)
\]
where $\wt{L}^0_{22}(b)$ is invertible and $\wt{L}^e_{22}(b,\sigma):\wt{\mathcal{K}}_{b,2}\to\mathcal{R}_{b,2}^\perp$ is continuous in $(b,\sigma)$ in the norm operator topology.

\item\underline{Analysis of $L_{11}$.} Using the calculation in \eqref{eq:calofL01} and \eqref{eq:calofL10}, we see that 
\begin{equation}\label{eq:calofL11}
	\begin{split}
		&\langle\widehat{L_{b}}(\sigma)\check{L}_b(\sigma)^{-1}(\tilde{g}_1,\tilde{A}_1),\  (\dg_1^*, \dA_1^*)\rangle\\
			&\ =\sigma^2\langle-i\big(\pa_\sigma\widehat{L_{b}}(0)+\frac{\sigma}{2}\pa^2_\sigma\widehat{L_{b}}(0)\big)(\check{g}_1,\check{A}_1)+\frac{1}{2}\pa_\sigma^2\widehat{L_{b}}(0)(\dg_1,\dA_1),\  (\check{L}_b(\sigma)^{-1})^*V_b^*(\dg_1^*, \dA_1^*)\rangle\\
			&\ =\sigma^2\langle-i\big(\pa_\sigma\widehat{L_{b}}(0)+\frac{\sigma}{2}\pa^2_\sigma\widehat{L_{b}}(0)\big)(\check{g}_1,\check{A}_1)+\frac{1}{2}\pa_\sigma^2\widehat{L_{b}}(0)(\dg_1,\dA_1),\  (\dg_1^*, \dA_1^*)\rangle\\	
			&\ \ +\sigma^2\langle-i\big(\pa_\sigma\widehat{L_{b}}(0)+\frac{\sigma}{2}\pa^2_\sigma\widehat{L_{b}}(0)\big)(\check{g}_1,\check{A}_1)+\frac{1}{2}\pa_\sigma^2\widehat{L_{b}}(0)(\dg_1,\dA_1),\  (\check{L}_b(\sigma)^{-1})^*(\check{L}_{b}(0)-\check{L}_{b}(\sigma))^*(\dg_1^*, \dA_1^*)\rangle\\
			&\ =\sigma^2\langle-i\big(\pa_\sigma\widehat{L_{b}}(0)+\frac{\sigma}{2}\pa^2_\sigma\widehat{L_{b}}(0)\big)(\check{g}_1,\check{A}_1)+\frac{1}{2}\pa_\sigma^2\widehat{L_{b}}(0)(\dg_1,\dA_1),\  (\dg_1^*, \dA_1^*)\rangle\\
			&\ \ -\sigma^3\langle-i\big(\pa_\sigma\widehat{L_{b}}(0)+\frac{\sigma}{2}\pa^2_\sigma\widehat{L_{b}}(0)\big)(\check{g}_1,\check{A}_1)+\frac{1}{2}\pa_\sigma^2\widehat{L_{b}}(0)(\dg_1,\dA_1),\  i(\check{L}_b(\sigma)^{-1})^*\widehat{L_{b}}(0)^*(\check{g}_1^*, \check{A}_1^*)\rangle\\	
			&\ \ \  -\sigma^4\langle-i\big(\pa_\sigma\widehat{L_{b}}(0)+\frac{\sigma}{2}\pa^2_\sigma\widehat{L_{b}}(0)\big)(\check{g}_1,\check{A}_1)+\frac{1}{2}\pa_\sigma^2\widehat{L_{b}}(0)(\dg_1,\dA_1),\  (\check{L}_b(\sigma)^{-1})^*(\frac{1}{2}\pa_\sigma^2\widehat{L_{b}}(0))^*(\dg_1^*, \dA_1^*)\rangle	\\	
			&\ =\sigma^2\langle-i\pa_\sigma\widehat{L_{b}}(0)(\check{g}_1,\check{A}_1)+\frac{1}{2}\pa_\sigma^2\widehat{L_{b}}(0)(\dg_1,\dA_1),\  (\dg_2^*, \dA_2^*)\rangle+\sigma^3\langle\frac{-i}{2}\pa^2_\sigma\widehat{L_{b}}(0)(\check{g}_1,\check{A}_1), (\dg_1^*,\dA_1^*)\rangle\\
			&\ \ -\sigma^3\langle-i\pa_\sigma\widehat{L_{b}}(0)(\check{g}_1,\check{A}_1)+\frac{1}{2}\pa_\sigma^2\widehat{L_{b}}(0)(\dg_1,\dA_1),\  i(I-(\check{L}_b(0)^{-1})^*V_b^*)(\check{g}_1^*, \check{A}_1^*)\rangle\\
			&\ \ \ -\sigma^4\langle\frac{-i}{2}\pa_\sigma^2\widehat{L_{b}}(0)(\check{g}_1,\check{A}_1),\  i(I-(\check{L}_b(0)^{-1})^*V_b^*)(\check{g}_1^*, \check{A}_1^*)\rangle\\
				&\ \ \ \  +\sigma^4\Big\langle-i\big(\pa_\sigma\widehat{L_{b}}(0)+\frac{\sigma}{2}\pa^2_\sigma\widehat{L_{b}}(0)\big)(\check{g}_1,\check{A}_1)+\frac{1}{2}\pa_\sigma^2\widehat{L_{b}}(0)(\dg_1,\dA_1),\\
				&\quad\quad\quad\quad\quad\quad (\check{L}_b(\sigma)^{-1})^*(\pa_\sigma\widehat{L_{b}}(0)+\frac{\sigma}{2}\pa_\sigma^2\widehat{L_{b}}(0))^*(I-(\check{L}_b(0)^{-1})^*V_b^*)(\check{g}_1^*, \check{A}_1^*)\Big\rangle	\\	
			&\ \ \ \ \ \  -\sigma^4\langle-i\big(\pa_\sigma\widehat{L_{b}}(0)+\frac{\sigma}{2}\pa^2_\sigma\widehat{L_{b}}(0)\big)(\check{g}_1,\check{A}_1)+\frac{1}{2}\pa_\sigma^2\widehat{L_{b}}(0)(\dg_1,\dA_1),\  (\check{L}_b(\sigma)^{-1})^*(\frac{1}{2}\pa_\sigma^2\widehat{L_{b}}(0))^*(\dg_1^*, \dA_1^*)\rangle	.
	\end{split}
\end{equation}
According to the proof of the non-existence of the quadratically growing generalized zero modes with leading order term $(\dg_1, \dA_1)\in\mathcal{K}_{b,1}$ (see Lemma \ref{lem:qgezeromodess1}), the pairing on $\mathcal{K}_{b,1}\times\mathcal{K}_{b,1}^*$
\begin{align*}
&\langle-i\pa_\sigma\widehat{L_{b}}(0)(\check{g}_1,\check{A}_1)+\frac{1}{2}\pa_\sigma^2\widehat{L_{b}}(0)(\dg_1,\dA_1),\  (\dg_2^*, \dA_2^*)\rangle\\
&\quad=-\langle[L_b, t_{b,*}](\check{g}_1,\check{A}_1)+\frac{1}{2}[[L_b, t_{b,*}],t_{b,*}](\dg_1,\dA_1),\  (\dg_2^*, \dA_2^*)\rangle
\end{align*}
is non-degenerate for $b$ near $b_0$. Therefore, we find that 
\[
L_{11}(b,\sigma)=\sigma^2\wt{L}_{11}(b,\sigma)\quad\mbox{with}\quad \wt{L}_{11}(b,\sigma)=\wt{L}^0_{11}(b)+\sigma\wt{L}^1_{22}(b)+\sigma^2\wt{L}^e_{11}(b,\sigma)
\]
where $\wt{L}^0_{11}(b)$ is invertible and $\wt{L}^e_{11}(b,\sigma):\wt{\mathcal{K}}_{b,1}\to\mathcal{R}_{b,1}^\perp$ is continuous in $(b,\sigma)$ in the norm operator topology.

\item\underline{Analysis of $L_{12}$ and $L_{21}$.} For $(\tilde{g}_1,\tilde{A}_1)\in\wt{\mathcal{K}}_{b,1}$ and $(\dg_2^*, \dA_2^*)\in\mathcal{K}_{b,2}^*$, using \eqref{eq:calofL01} and \eqref{eq:calofL22}, we compute
\begin{equation}\label{eq:calofL21}
	\begin{split}
		&\langle\widehat{L_{b}}(\sigma)\check{L}_b(\sigma)^{-1}(\tilde{g}_1,\tilde{A}_1),\  (\dg_2^*, \dA_2^*)\rangle\\
		&\ =\sigma^2\langle-i\pa_\sigma\widehat{L_{b}}(0)(\check{g}_1,\check{A}_1)+\frac{1}{2}\pa_\sigma^2\widehat{L_{b}}(0)(\dg_1,\dA_1),\ (\dg_2^*, \dA_2^*)\rangle+\sigma^3\langle\frac{-i}{2}\pa^2_\sigma\widehat{L_{b}}(0)(\check{g}_1,\check{A}_1), (\dg_2^*,\dA_2^*)\rangle\\
		&\ \ -\sigma^3\Big\langle-i\big(\pa_\sigma\widehat{L_{b}}(0)+\frac{\sigma}{2}\pa^2_\sigma\widehat{L_{b}}(0)\big)(\check{g}_1,\check{A}_1)+\frac{1}{2}\pa_\sigma^2\widehat{L_{b}}(0)(\dg_1,\dA_1),\\
		&\qquad\qquad\quad\  (\check{L}_b(\sigma)^{-1})^*\big(\pa_\sigma\widehat{L_{b}}(0)+\frac{\sigma}{2}\pa^2_\sigma\widehat{L_{b}}(0)\big)^*(\dg_2^*, \dA_2^*)\Big\rangle.
	\end{split}
\end{equation}
Therefore, 
\[
L_{21}(b,\sigma)=\sigma^2\wt{L}_{21}(b,\sigma)\quad\mbox{with}\quad \wt{L}_{21}(b,\sigma)=\wt{L}^0_{21}(b)+\sigma\wt{L}^e_{21}(b,\sigma)
\]
where $\wt{L}^e_{21}(b,\sigma):\wt{\mathcal{K}}_{b,1}\to\mathcal{R}^\perp_{b,2}$ is continuous in $(b,\sigma)$ in the norm operator topology.

For $(\tilde{g}_2,\tilde{A}_2)=\check{L}_b(0)(\dg_2, \dA_2)\in\wt{\mathcal{K}}_{b,2}=\check{L}_b(0)\mathcal{K}_{b,2}$ and $(\dg_1^*, \dA_1^*)\in\mathcal{K}^*_{b,1}$, using \eqref{eq:calofL10}, we calculate
\begin{equation}\label{eq:calofL12}
	\begin{split}
		&\langle\widehat{L_{b}}(\sigma)\check{L}_b(\sigma)^{-1}(\tilde{g}_2, \tilde{A}_2),\ (\dg^*_1, \dA^*_1)\rangle\\
		&\quad=\sigma\Big\langle \big(I-V_b(\check{L}_b(0)^{-1})\big)\check{L}_b(0)(\dg_2, \dA_2), i(\check{g}^*_1,\check{A}^*_1)\Big\rangle\\
		&\quad\quad-\sigma^2\Big\langle \check{L}_b(\sigma)^{-1}\check{L}_b(0)(\dg_2, \dA_2), i(\pa_\sigma\widehat{L_{b}}(0))^*\big(I-(\check{L}_b(0)^{-1})^*V_b^*\big)(\check{g}^*_1,\check{A}^*_1)\Big\rangle\\
		&\quad\quad\quad+\sigma^2\Big\langle\check{L}_b(\sigma)^{-1}\check{L}_b(0)(\dg_2, \dA_2),\frac{1}{2}(\pa^2_\sigma\widehat{L_{b}}(0))^*(\dg^*_1,\dA^*_1)\Big\rangle\\
		&\quad\quad\quad\quad-\sigma^3\Big\langle \check{L}_b(\sigma)^{-1}\check{L}_b(0)(\dg_2, \dA_2), \frac{i}{2}(\pa_\sigma\widehat{L_{b}}(0))^*\big(I-(\check{L}_b(0)^{-1})^*V_b^*\big)(\check{g}^*_1,\check{A}^*_1)\Big\rangle\\
		&\quad=-\sigma^2\Big\langle (\dg_2, \dA_2), i(\pa_\sigma\widehat{L_{b}}(0))^*\big(I-(\check{L}_b(0)^{-1})^*V_b^*\big)(\check{g}^*_2,\check{A}^*_2)-\frac{1}{2}(\pa^2_\sigma\widehat{L_{b}}(0))^*(\dg^*_1,\dA^*_1)\Big\rangle\\
			&\quad\quad+\sigma^3\Big\langle \check{L}_b(\sigma)^{-1}(\pa_\sigma\widehat{L_{b}}(0)+\frac{\sigma}{2}\pa_\sigma^2\widehat{L_{b}}(0))(\dg_2, \dA_2), i(\pa_\sigma\widehat{L_{b}}(0))^*\big(I-(\check{L}_b(0)^{-1})^*V_b^*\big)(\check{g}^*_1,\check{A}^*_1)\Big\rangle\\
			&\quad\quad\quad+\sigma^3\Big\langle \check{L}_b(\sigma)^{-1}(\pa_\sigma\widehat{L_{b}}(0)+\frac{\sigma}{2}\pa_\sigma^2\widehat{L_{b}}(0))(\dg_2, \dA_2), -\frac{1}{2}(\pa^2_\sigma\widehat{L_{b}}(0))^*(\dg^*_1,\dA^*_1)\Big\rangle\\
			&\quad\quad\quad\quad-\sigma^3\Big\langle \check{L}_b(\sigma)^{-1}\check{L}_b(0)(\dg_2, \dA_2), \frac{i}{2}(\pa^2_\sigma\widehat{L_{b}}(0))^*\big(I-(\check{L}_b(0)^{-1})^*V_b^*\big)(\check{g}^*_1,\check{A}^*_1)\Big\rangle.
	\end{split}
\end{equation}
Therefore, 
\[
L_{12}(b,\sigma)=\sigma^2\wt{L}_{12}(b,\sigma)\quad\mbox{with}\quad \wt{L}_{12}(b,\sigma)=\wt{L}^0_{12}(b)+\sigma\wt{L}^e_{12}(b,\sigma)
\]
where $\wt{L}^e_{12}(b,\sigma):\wt{\mathcal{K}}_{b,2}\to\mathcal{R}^\perp_{b,1}$ is continuous in $(b,\sigma)$ in the norm operator topology.

\item\underline{Analysis of $L_{00}$.}
Following the arguments in the first step in the proof of Lemma \ref{lem:step1}, we can prove that for $(b,\sigma)$ near $(b,0)$, $L_{00}(b,\sigma)$ is invertible and there exists a uniform constant $C>0$ such that 
\begin{equation}
	\norm{L_{00}(b,\sigma)^{-1}}_{\mathcal{R}_b\to\wt{\mathcal{K}}_{b}^\perp}\leq C.
\end{equation}
Moreover, for $(\tilde{g}_0, \tilde{A}_0)\in\wt{\mathcal{K}}_{b}^\perp$, we write
\begin{equation}\label{eq:calofL00}
	\begin{split}
	L_{00}(b,\sigma)(\tilde{g}_0, \tilde{A}_0)&=\widehat{L_{b}}(\sigma)\check{L}_b(\sigma)^{-1}(\tilde{g}_0, \tilde{A}_0)-\big(L_{10}(b,\sigma)+L_{20}(b,\sigma)\big)(\tilde{g}_0, \tilde{A}_0)\\
	&=(\tilde{g}_0, \tilde{A}_0)-V_b\check{L}_b(\sigma)^{-1}(\tilde{g}_0, \tilde{A}_0)-\big(L_{10}(b,\sigma)+L_{20}(b,\sigma)\big)(\tilde{g}_0, \tilde{A}_0).
	\end{split}
\end{equation}
Since $\check{L}_b(\sigma)^{-1}$ is continuous in the norm operator topology $\mathcal{L}_{\mathrm{op}}(\eHb^{s-1+\epsilon, \ell+2+\epsilon}(\CX), \eHb^{s-\epsilon, \ell-\epsilon}(\CX))$ and $V_b$ maps $\mathcal{D}'(X^\circ)$ to $C^\infty_c(X)$, it follows that $L_{00}(b,\sigma):\wt{\mathcal{K}}_b^\perp\to\mathcal{R}_b$ is continuous in $(b,\sigma)$ in the norm operator topology.

\item\underline{The inverse of $\widehat{L_{b}}(\sigma)\check{L}_b(\sigma)^{-1}$.} Putting all the above analysis of the components $L_{ij}(b,\sigma)$ together, we conclude that 
\begin{equation}\label{eq:detailedexofLL-1}
	\begin{split}
	&\widehat{L_{b}}(\sigma)\check{L}_b(\sigma)^{-1}=\begin{pmatrix}
		L_{00}(b,\sigma)& \sigma^2\wt{L}_{01}(b,\sigma)&\sigma\wt{L}_{02}(b,\sigma)\\
			\sigma\wt{L}_{10}(b,\sigma)& \sigma^2\wt{L}_{11}(b,\sigma)&\sigma^2\wt{L}_{12}(b,\sigma)\\
				\sigma\wt{L}_{20}(b,\sigma)& \sigma^2\wt{L}_{21}(b,\sigma)&\sigma\wt{L}_{22}(b,\sigma)\\
	\end{pmatrix}\\
	&\quad=\begin{pmatrix}
		L_{00}(b,\sigma)& \sigma^2\wt{L}_{01}(b,\sigma)&\sigma\wt{L}_{02}(b,\sigma)\\
		\sigma\big(\wt{L}^0_{10}(b)+\sigma\wt{L}^e_{10}(b,\sigma)\big)&\sigma^2\big(\wt{L}^0_{11}(b)+\sigma\wt{L}^1_{11}(b)+\sigma^2\wt{L}^e_{11}(b,\sigma)\big)&\sigma^2\big(\wt{L}^0_{12}(b)+\sigma\wt{L}^e_{12}(b,\sigma)\big)\\
		\sigma\wt{L}_{20}(b,\sigma)&\sigma^2\big(\wt{L}^0_{21}(b)+\sigma\wt{L}^e_{21}(b)\big)&\sigma\big(\wt{L}^0_{22}(b)+\sigma\wt{L}^e_{22}(b,\sigma)\big)
	\end{pmatrix}
\end{split}
\end{equation}
where $L_{00}(b,\sigma), \wt{L}^0_{11}(b), \wt{L}^0_{22}(b)$ are invertible, and $L_{00}(b,\sigma)$, $\wt{L}_{ij}(b,\sigma)$ with $(i,j)=(0,1), (0,2), (2,0)$, $\wt{L}^e_{ij}(b,\sigma)$ with $(i,j)=(1,0), (1,1), (1,2), (2,1), (2,2)$ are continuous in $(b,\sigma)$ in the norm operator topology. 

Solving the system 
\[
\widehat{L_{b}}(\sigma)\check{L}_b(\sigma)^{-1}\big((\tilde{g}_0,\tilde{A}_0), (\tilde{g}_1,\tilde{A}_1),(\tilde{g}_2,\tilde{A}_2)\big)=(f_0, f_1,f_2)
\]
yields
\begin{equation}\label{eq:exofinverseR}
	R(b,\sigma)=(\widehat{L_{b}}(\sigma)\check{L}_b(\sigma)^{-1})^{-1}=
	\begin{pmatrix}
		\wt{R}_{00}(b,\sigma)&\wt{R}_{01}(b,\sigma)&\wt{R}_{02}(b,\sigma)\\
		\sigma^{-1}\wt{R}_{10}(b,\sigma)&\sigma^{-2}\wt{R}_{11}(b,\sigma)&\sigma^{-1}\wt{R}_{12}(b,\sigma)\\
		\wt{R}_{20}(b,\sigma)&\sigma^{-1}\wt{R}_{21}(b,\sigma)&\sigma^{-1}\wt{R}_{22}(b,\sigma)
	\end{pmatrix}
\end{equation}
where
\begin{equation}\label{eq:exofcompoofR}
	\begin{split}
 \wt{R}_{11}=\big(\wt{L}^\sharp_{11}-\sigma\wt{L}^\sharp_{12}(\wt{L}^\flat_{22})^{-1}\wt{L}^\sharp_{21}\big)^{-1},\quad \wt{R}_{10}=\wt{R}_{11}\big(-\wt{L}_{10}+\sigma\wt{L}_{12}^\sharp(\wt{L}^\flat_{22})^{-1}\wt{L}_{20}\big)L_{00}^{-1},\\
 \wt{R}_{12}=-\wt{R}_{11}\wt{L}_{12}^\sharp(\wt{L}^\flat_{22})^{-1},\quad \wt{R}_{21}=-(\wt{L}^\flat_{22})^{-1}\wt{L}_{21}^\sharp\wt{R}_{11},\quad\wt{R}_{22}=(\wt{L}^\flat_{22})^{-1}-\sigma(\wt{L}^\flat_{22})^{-1}\wt{L}_{21}^\sharp \wt{R}_{12},\\
\wt{R}_{20}=-(\wt{L}^\flat_{22})^{-1}\big(\wt{L}_{20}L_{00}^{-1}+\wt{L}_{21}^\sharp\wt{R}_{10}\big),\quad \wt{R}_{01}=-L_{00}^{-1}\big(\wt{L}_{01}\wt{R}_{11}+\wt{L}_{02}\wt{R}_{21}\big),\\
\wt{R}_{02}=-L_{00}^{-1}\big(\sigma\wt{L}_{01}\wt{R}_{12}+\wt{L}_{02}\wt{R}_{22}\big),\quad \wt{R}_{00}=L_{00}^{-1}\big(I-\sigma\wt{L}_{01}\wt{R}_{10}-\sigma\wt{L}_{02}\wt{R}_{20}\big)
\end{split}
	\end{equation}
with
\begin{equation}\label{eq:exofauxofR}
	\begin{split}
	&\wt{L}^\sharp_{i1}=\wt{L}_{i1}-\sigma\wt{L}_{i0}L_{00}^{-1}\wt{L}_{01}\quad\mbox{for}\quad i=1,2, \quad \wt{L}^\sharp_{12}=\wt{L}_{12}-\wt{L}_{10}L_{00}^{-1}\wt{L}_{02},\\
&\qquad\qquad\qquad\qquad	\wt{L}_{22}^\flat=\wt{L}_{22}-\sigma\wt{L}_{20}L_{00}^{-1}\wt{L}_{02}.
	\end{split}
\end{equation}
Since $\check{L}_b(\sigma)$ is invertible for $(b,\sigma)$ near $(b_0,0)$, it follows that $\widehat{L_{b}}(\sigma)^{-1}=\check{L}_b(\sigma)^{-1} R(b,\sigma)$, which implies the invertibility of $\widehat{L_{b}}(\sigma):\mathcal{X}_b^{s,\ell}\to\eHb^{s-1,\ell+1}(\CX)$ for $(b,\sigma)$ with $\IM\sigma\geq 0, \sigma\neq 0$ near $(b_0, 0)$.
This finishes the proof of the theorem.
	\end{itemize}
	\end{proof}
We now give a more detailed description of the structure of the singular part of $\widehat{L_{b}}(\sigma)^{-1}$.

\begin{cor}\label{cor:singandreg}
	For $(b,\sigma)$ with $\IM\sigma\geq 0$ near $(b_0,0)$, we have
	\[
	\widehat{L_{b}}(\sigma)^{-1}=P(b,\sigma)+L_b^-(\sigma):\eHb^{s-1,\ell+2}(\CX;\scsym\oplus\scform)\to\eHb^{s,\ell}(\CX;\scsym\oplus\scform).
	\]
	The regular part $L^-_b(\sigma)$ is uniformly bounded, and is continuous in $\sigma$ with values in  \[\mathcal{L}_{\mathrm{weak}}(\eHb^{s-1,\ell+2}(\CX),\eHb^{s,\ell}(\CX))\cap\mathcal{L}_{\mathrm{op}}(\eHb^{s-1+\epsilon,\ell+2+\epsilon}(\CX),\eHb^{s-\epsilon,\ell-\epsilon}(\CX))
	\]
	 for $\epsilon>0$. 
	 
	 The singular part $P(b,\sigma)$ satisfies
	\begin{equation}
		P(b,\sigma)f=(\sigma^{-2}(\dg_1,\dA_1)-i\sigma(\check{g}_1, \check{A}_1))+i\sigma^{-1}\big((\dg_2, \dA_2)
+(\dg'_1, \dA'_1)\big)+i\sigma^{-1}(\dg_1''(\sigma), \dA_1''(\sigma))
	\end{equation}
where 
\[
(\dg_1,\dA_1),\  (\dg_1',\dA_1'),\ (\dg_1''(\sigma),\dA_1''(\sigma))\in\mathcal{K}_{b,1},\quad (\check{g}_1, \check{A}_1)\in \check{\mathcal{K}}_b
\]
 with $(\dg_1''(\sigma),\dA_1''(\sigma))=o(1)$ in $\mathcal{K}_{b,1}$ as $\sigma\to 0$, and \[
 (\dg_2,\dA_2)\in\mathcal{K}_{b,2}.
 \]
  Moreover, $(\dg_1,\dA_1), (\dg_1',\dA_1')$ and $(\dg_2,\dA_2)$ are determined by the conditions
\begin{align}\notag
	&-\Big\langle[L_b, t_{b,*}](\check{g}_1, \check{A}_1)+\frac{1}{2}[[L_b, t_{b,*}], t_{b,*}](\dg_1,\dA_1), (\dg^*, \dA^*)\Big\rangle\\
\label{eq:condition1}&\quad\quad+\Big\langle[L_b, t_{b,*}](\dg_2, \dA_2), (\dg^*, \dA^*)\Big\rangle=\langle f, (\dg^*, \dA^*)\rangle\quad\mbox{for all}\quad(\dg^*,\dA^*)\in\mathcal{K}_b^*,\\\notag
&\Big\langle[L_b, t_{b,*}](\check{g}'_1, \check{A}'_1)+\frac{1}{2}[[L_b, t_{b,*}], t_{b,*}](\dg'_1,\dA'_1), (\dg_1^*, \dA_1^*)\Big\rangle=-\langle f, (\check{g}_1^*, \check{A}_1^*)\rangle\\\notag
&\quad -\Big\langle\frac{1}{2}[[L_b, t_{b,*}], t_{b,*}](\dg_1,\dA_1)+[L_b, t_{b,*}](\dg_2-\check{g}_1, \dA_2-\check{A}_1), (\check{g}_1^*, \check{A}_1^*)\Big\rangle\\
\label{eq:condition2}
&\quad\quad+\Big\langle\frac{1}{2}[[L_b, t_{b,*}], t_{b,*}](\check{g}_1-\dg_2,\check{A}_1-\dA_2), (\dg_1^*, \dA_1^*)\Big\rangle\quad\mbox{for all}\quad (\dg_1^*, \dA_1^*)\in\mathcal{K}_{b,1}^*.
\end{align}
\end{cor}

\begin{proof}
	For the simplicity of notation, we define
	\begin{gather*} \wt{\mathcal{K}}_0:=\wt{\mathcal{K}}_b^\perp,\quad \wt{\mathcal{K}}_1:=\wt{\mathcal{K}}_{b,1},\quad \wt{\mathcal{K}}_2:=\wt{\mathcal{K}}_{b,2},\\
		\mathcal{R}_0:=\mathcal{R}_b,\quad \mathcal{R}_1:=\mathcal{R}^\perp_{b,1},\quad \mathcal{R}_2:=\mathcal{R}^\perp_{b,2}.
	 \end{gather*}
 First, we prove that if an operator $A(\sigma):\wt{\mathcal{K}}_i\to\mathcal{R}_j$ is invertible, and continuous in $\sigma$ (in norm operator topology) in a small neighborhood of $\sigma=0$, then so is $A(\sigma)^{-1}$. We write
 \[
 A(\sigma)=A(0)+A^e(\sigma)=A(0)(I+A(0)^{-1}A^e(\sigma))
 \]
 with
 \[ A^e(\sigma)=A(\sigma)-A(0)=o(1)\quad\mbox{in}\quad \mathcal{L}_{\mathrm{op}}(\wt{\mathcal{K}}_i,\mathcal{R}_j)\quad \mbox{as}\quad \sigma\to 0.
 \]
 Since $\norm{A(0)^{-1}A^e(\sigma)}_{\wt{\mathcal{K}}_i\to\wt{\mathcal{K}}_i}<1/2$ when $\sigma$ is near $0$, the inverse of $A(\sigma)$ is given by the Neumann series
 \[
 A(\sigma)^{-1}=\Big(I+\sum_{k=1}^\infty(-1)^k\big(A(0)^{-1}A^e(\sigma)\big)^k\Big)A(0)^{-1},
 \]
 from which we find that
 \begin{align*}
 \norm{A(\sigma)^{-1}-A(0)^{-1}}_{\mathcal{R}_j\to\wt{\mathcal{K}}_i}&\leq\Big(\sum_{k=1}^\infty\norm{A(0)^{-1}A^e(\sigma)}^k_{\wt{\mathcal{K}}_i\to\wt{\mathcal{K}}_i}\Big)\norm{A(0)^{-1}}_{\mathcal{R}_j\to\wt{\mathcal{K}}_i}\\
 &\leq2\norm{A(0)^{-1}}^2_{\mathcal{R}_j\to\wt{\mathcal{K}}_i}\norm{A^e(\sigma)}_{\wt{\mathcal{K}}_i\to\mathcal{R}_j}=o(1)\quad \mbox{in}\quad \mathcal{L}_{\mathrm{op}}(\mathcal{R}_j,\wt{\mathcal{K}}_i)\quad \mbox{as}\quad \sigma\to 0.
 \end{align*}
This proves the continuity of $A(\sigma)^{-1}$ in the norm operator topology in a small neighborhood of $\sigma=0$. Applying this result to $L_{00}$ and $\wt{L}_{22}^\flat$ (defined in \eqref{eq:exofauxofR}) yields the continuity (in the norm operator topology) of $L_{00}^{-1}$ and $(\wt{L}_{22}^\flat)^{-1}$. Then the continuity of $\wt{R}_{ij}$ follows from the explicit expressions in \eqref{eq:exofcompoofR}. Furthermore, we find that 
\begin{equation}\label{eq:expansionofcomofR}
	\begin{split}
\wt{R}_{21}(b,\sigma)&=\wt{R}^0_{21}(b)+\sigma\wt{R}^e_{21}(b,\sigma)\quad \mbox{with}\quad \wt{R}^0_{21}(b)=-(\wt{L}^0_{22}(b))^{-1}\wt{L}^0_{21}(b)(\wt{L}^0_{11}(b))^{-1},\\
\wt{R}_{22}(b,\sigma)&=\wt{R}^0_{22}(b)+\sigma\wt{R}^e_{22}(b,\sigma)\quad \mbox{with}\quad \wt{R}^0_{22}(b)=(\wt{L}^0_{22}(b))^{-1},\\
\wt{R}_{11}(b,\sigma)&=\wt{R}^0_{11}(b)+\sigma\wt{R}^e_{11}(b,\sigma)\quad \mbox{with}\quad \wt{R}^0_{11}(b)=(\wt{L}^0_{11}(b))^{-1}
\end{split}
\end{equation}
where $\wt{R}^e_{21}, \wt{R}^e_{22}, \wt{R}^e_{11}$ are continuous in $(b,\sigma)$ for $(b,\sigma)$ near $(b_0, 0)$ in the norm operator topology. Therefore, we rewrite
\begin{equation}\label{eq:simplifiedR}
	R(b,\sigma)=R_{\mathrm{sing}}(b,\sigma)+R_{\mathrm{reg}}(b,\sigma)=
	\begin{pmatrix}
		0&0&0\\
		\frac{\wt{R}_{10}(b,\sigma)}{\sigma}&	\frac{\wt{R}^0_{11}(b)}{\sigma^2}+	\frac{\wt{R}^e_{11}(b,\sigma)}{\sigma}&	\frac{\wt{R}_{12}(b,\sigma)}{\sigma}\\
		0&	\frac{\wt{R}^0_{21}(b)}{\sigma}&	\frac{\wt{R}^0_{22}(b)}{\sigma}
	\end{pmatrix}+R_{\mathrm{reg}}(b,\sigma)
\end{equation}
where all the entries in $R_{\mathrm{reg}}(b,\sigma)$ are continuous in $(b,\sigma)$ in the norm operator topology.

For $(\tilde{g}, \tilde{A})=\check{L}_b(0)(\dg, \dA)\in\wt{\mathcal{K}}_{b,1}$ with $(\dg,\dA)\in\mathcal{K}_{b,1}$, we have 
\begin{equation}\label{eq:exofcheckLon1}
	\begin{split}
	\check{L}_b(\sigma)^{-1}(\tilde{g}, \tilde{A})&=\check{L}_b(\sigma)^{-1}\check{L}_b(0)(\dg, \dA)=(\dg,\dA)-i\sigma\check{L}_b(\sigma)^{-1}\widehat{L_{b}}(0)(\check{g}, \check{A})-\frac{\sigma^2}{2}\check{L}_b(\sigma)^{-1}\pa_\sigma^2\widehat{L_{b}}(0)(\dg,\dA)\\
	&=(\dg,\dA)-i\sigma\check{L}_b(\sigma)^{-1}\big(\check{L}_b(\sigma)+\widehat{L_{b}}(0)-\widehat{L_{b}}(\sigma)-V_b\big)(\check{g}, \check{A})-\frac{\sigma^2}{2}\check{L}_b(\sigma)^{-1}\pa_\sigma^2\widehat{L_{b}}(0)(\dg,\dA)\\
	&=(\dg,\dA)-i\sigma(\check{g}, \check{A})+\sigma^2\check{L}_b(\sigma)^{-1}\Big(i(\pa_\sigma\widehat{L_{b}}(0)+\frac{\sigma}{2}\pa_\sigma^2\widehat{L_{b}}(0))(\check{g},\check{A})-\frac{1}{2}\pa_\sigma^2\widehat{L_{b}}(0)(\dg,\dA)\Big)
	\end{split}
	\end{equation}
where the coefficient of $\sigma^2$ is continuous in $(b,\sigma)$ in the norm operator topology.

For $(\tilde{g}, \tilde{A})=\check{L}_b(0)(\dg, \dA)\in\wt{\mathcal{K}}_{b,2}$ with $(\dg,\dA)\in\mathcal{K}_{b,2}$, we have 
\begin{equation}\label{eq:exofcheckLon2}
	\begin{split}
		\check{L}_b(\sigma)^{-1}(\tilde{g}, \tilde{A})&=\check{L}_b(\sigma)^{-1}\check{L}_b(0)(\dg, \dA)=(\dg,\dA)-\sigma\check{L}_b(\sigma)^{-1}\Big(\big(\pa_\sigma\widehat{L_{b}}(0)+\frac{\sigma}{2}\pa_\sigma^2\widehat{L_{b}}(0)\big)(\dg,\dA)\Big)
	\end{split}
\end{equation}
where the coefficient of $\sigma$ is continuous in $(b,\sigma)$ in the norm operator topology.

Since $R_{\mathrm{reg}}(b,\sigma)$ is continuous in $\sigma$ in norm operator topology and $\check{L}_b(\sigma)^{-1}$ is continuous in $\sigma$ with values in $\mathcal{L}_{\mathrm{weak}}(\eHb^{s-1,\ell+2}(\CX),\eHb^{s,\ell}(\CX))\cap\mathcal{L}_{\mathrm{op}}(\eHb^{s-1+\epsilon,\ell+2+\epsilon}(\CX),\eHb^{s-\epsilon,\ell-\epsilon}(\CX))$ for $\epsilon>0$, it follows that $\check{L}_b(\sigma)^{-1}R_{\mathrm{reg}}(b,\sigma)$ is uniformly bounded and has the same continuity as $\check{L}_b(\sigma)^{-1}$. According to \eqref{eq:exofcheckLon1} and \eqref{eq:exofcheckLon2}, we conclude that $\check{L}_b(\sigma)^{-1}R_{\mathrm{sing}}(b,\sigma)$ is equal to a singular part $P(b,\sigma)$ plus a uniformly bounded and continuous operator with value in $\mathcal{L}_{\mathrm{op}}(\eHb^{s-1,\ell+2}(\CX),\eHb^{s,\ell}(\CX))$. Moreover, $P(b,\sigma)$ satisfies
\begin{equation}\label{eq:exofPbsigma}
	P(b,\sigma)f=(\sigma^{-2}(\dg_1,\dA_1)-i\sigma^{-1}(\check{g}_1, \check{A}_1))+i\sigma^{-1}\big((\dg_2, \dA_2)
	+(\dg'_1, \dA'_1)\big)+i\sigma^{-1}(\dg_1''(\sigma), \dA_1''(\sigma))
\end{equation}
where $(\dg_1,\dA_1), (\dg_1',\dA_1'),(\dg_1''(\sigma),\dA_1''(\sigma))\in\mathcal{K}_{b,1}$ with $(\dg_1''(\sigma),\dA_1''(\sigma))=o(1)$ in $\mathcal{K}_{b,1}$ as $\sigma\to 0$, and $(\dg_2,\dA_2)\in\mathcal{K}_{b,2}$.

It remains to prove that $(\dg_1,\dA_1), (\dg_1',\dA_1'),(\dg_2,\dA_2)$ are determined by conditions \eqref{eq:condition1} and \eqref{eq:condition2}. Let $f\in\eHb^{s-1,\ell+2}(\CX)$ and $(\dg(\sigma), \dA(\sigma)):=\widehat{L_{b}}(\sigma)^{-1}f$, and then we have
\begin{align*}
(\dg(\sigma), \dA(\sigma))&=(\sigma^{-2}(\dg_1,\dA_1)-i\sigma^{-1}(\check{g}_1, \check{A}_1))+i\sigma^{-1}\big((\dg_2, \dA_2)
+(\dg'_1, \dA'_1)\big)+i\sigma^{-1}(\dg_1''(\sigma), \dA_1''(\sigma))\\
&\qquad+(\tilde{\dg},\tilde{\dA})+(\tilde{\dg}(\sigma), \tilde{\dA}(\sigma))
\end{align*}
where $(\tilde{\dg},\tilde{\dA})\in\eHb^{s,\ell}(\CX)$ and $(\tilde{\dg}(\sigma), \tilde{\dA}(\sigma))=o(1)$ in $\eHb^{s-\epsilon, \ell-\epsilon}(\CX)$ as $\sigma\to 0$. Applying $\widehat{L_{b}}(\sigma)$ on both sides gives
\begin{equation}\label{eq:exoff}
	\begin{split}
		f&=\widehat{L_{b}}(\sigma)(\dg(\sigma), \dA(\sigma))=\widehat{L_{b}}(0)(\dg(\sigma), \dA(\sigma))+\big(\sigma\pa_\sigma\widehat{L_{b}}(0)+\frac{\sigma^2}{2}\pa_\sigma^2\widehat{L_{b}}(0)\big)(\dg(\sigma), \dA(\sigma))\\
		&=\Big(\frac{1}{2}\pa_\sigma^2\widehat{L_{b}}(0)(\dg_1,\dA_1)-i\pa_\sigma\widehat{L_{b}}(0)(\check{g}_1, \check{A}_1)-\widehat{L_{b}}(0)(\check{g}_1',\check{A}_1')+i\pa_\sigma\widehat{L_{b}}(0)(\dg_2, \dA_2)+\widehat{L_{b}}(0)(\tilde{\dg},\tilde{\dA})\Big)\\
		&\quad+i\pa_\sigma\widehat{L_{b}}(0)(\dg_{1}''(\sigma),\dA_1''(\sigma))+\widehat{L_{b}}(0)(\tilde{\dg}(\sigma), \tilde{\dA}(\sigma))\\
		&\quad\quad+\frac{i\sigma}{2}\Big(-\pa_\sigma^2\widehat{L_{b}}(0)(\check{g}_1, \check{A}_1)+\pa_\sigma^2\widehat{L_{b}}(0)(\dg_1', \dA_1')+\pa_\sigma^2\widehat{L_{b}}(0)(\dg_2, \dA_2))\Big)\\
		&\quad\quad\quad+\frac{i\sigma}{2}\pa_\sigma^2\widehat{L_{b}}(0)(\dg_1''(\sigma), \dA_1''(\sigma)+\big(\sigma\pa_\sigma\widehat{L_{b}}(0)+\frac{\sigma^2}{2}\pa_\sigma^2\widehat{L_{b}}(0)\big)(\tilde{\dg}+\tilde{\dg}(\sigma), \tilde{\dA}+\tilde{\dA}(\sigma)).
	\end{split}
\end{equation}
Letting $\sigma\to0$, we arrive at
\begin{equation}\label{eq:limitofexoff}
		f=\frac{1}{2}\pa_\sigma^2\widehat{L_{b}}(0)(\dg_1,\dA_1)-i\pa_\sigma\widehat{L_{b}}(0)(\check{g}_1, \check{A}_1)-\widehat{L_{b}}(0)(\check{g}_1',\check{A}_1')+i\pa_\sigma\widehat{L_{b}}(0)(\dg_2, \dA_2)+\widehat{L_{b}}(0)(\tilde{\dg},\tilde{\dA}).
\end{equation}
Pairing both sides of \eqref{eq:limitofexoff} with $(\dg^*, \dA^*)\in\mathcal{K}_b^*$, we obtain
\begin{equation}\label{eq:combinedpairing}
	\begin{split}
	\langle f, (\dg^*, \dA^*)\rangle&=\Big\langle\frac{1}{2}\pa_\sigma^2\widehat{L_{b}}(0)(\dg_1,\dA_1)-i\pa_\sigma\widehat{L_{b}}(0)(\check{g}_1, \check{A}_1)+i\pa_\sigma\widehat{L_{b}}(0)(\dg_2, \dA_2), \ (\dg^*, \dA^*)\Big\rangle\\
	&=-\Big\langle[L_b, t_{b,*}](\check{g}_1, \check{A}_1)+\frac{1}{2}[[L_b, t_{b,*}], t_{b,*}](\dg_1,\dA_1)-[L_b, t_{b,*}](\dg_2, \dA_2), (\dg^*, \dA^*)\Big\rangle
	\end{split}
\end{equation}
Since the pairing on $\mathcal{K}_b\times\mathcal{K}_b^*$ on the right-hand side of \eqref{eq:combinedpairing} is non-degenerate, $(\dg_1,\dA_1)$ (thus $(\check{g}_1, \check{A}_1)$) and $(\dg_2, \dA_2)$ are uniquely determined by \eqref{eq:combinedpairing}. This proves the condition \eqref{eq:condition1}.

Rewriting \eqref{eq:limitofexoff} gives rise to the following PDE
\begin{equation}\label{eq:PDEfoftwoterms}
	\widehat{L_{b}}(0)\big((\check{g}_1',\check{A}_1')-(\tilde{\dg},\tilde{\dA})\big)=-f+\frac{1}{2}\pa_\sigma^2\widehat{L_{b}}(0)(\dg_1,\dA_1)-i\pa_\sigma\widehat{L_{b}}(0)(\check{g}_1, \check{A}_1)+i\pa_\sigma\widehat{L_{b}}(0)(\dg_2, \dA_2)
\end{equation}
whose right-hand side is uniquely determined by $f$. Due to \eqref{eq:combinedpairing}, the pairings of the right-hand side of \eqref{eq:PDEfoftwoterms} with any $(\dg^*, \dA^*)\in\mathcal{K}_b^*$ is $0$, which implies that \eqref{eq:PDEfoftwoterms} can be solved for $(\check{g}_1',\check{A}_1')-(\tilde{\dg},\tilde{\dA})$ and $(\check{g}_1',\check{A}_1')-(\tilde{\dg},\tilde{\dA})$ is uniquely determined by $f$ modulo $\mathcal{K}_b$. Namely, 
\begin{equation}\label{eq:exoftwoterms}
	(\check{g}_1',\check{A}_1')-(\tilde{\dg},\tilde{\dA})=(\bar{\dg}, \bar{\dA})+(\dg''', \dA''')
\end{equation}
where $(\bar{\dg}, \bar{\dA})$ is uniquely determined by $f$ and $(\dg''', \dA''')\in\mathcal{K}_b$. Pairing both sides of \eqref{eq:exoff} with $(\dg_1^*, \dA_1^*)\in\mathcal{K}^*_{b,1}$ and using the identity \eqref{eq:combinedpairing}, we obtain, upon multiplying $\sigma^{-1}$ and letting $\sigma\to0$,
\begin{align*}
	0&=\frac{i}{2}\Big\langle-\pa_\sigma^2\widehat{L_{b}}(0)(\check{g}_1, \check{A}_1)+\pa_\sigma^2\widehat{L_{b}}(0)(\dg_1', \dA_1')+\pa_\sigma^2\widehat{L_{b}}(0)(\dg_2, \dA_2)-2i\pa_\sigma\widehat{L_{b}}(0)(\tilde{\dg},\tilde{\dA}),\  (\dg_1^*, \dA_1^*)\Big\rangle\\
	&=\frac{i}{2}\Big\langle-\pa_\sigma^2\widehat{L_{b}}(0)(\check{g}_1, \check{A}_1)+\pa_\sigma^2\widehat{L_{b}}(0)(\dg_2, \dA_2)+2i\pa_\sigma\widehat{L_{b}}(0)(\bar{\dg},\bar{\dA}),\  (\dg_1^*, \dA_1^*)\Big\rangle\\
	&\quad+\frac{i}{2}\Big\langle\pa_\sigma^2\widehat{L_{b}}(0)(\dg_1', \dA_1')-2i\pa_\sigma\widehat{L_{b}}(0)(\check{g}_1', \check{A}_1'),\  (\dg_1^*, \dA_1^*)\Big\rangle+i\langle(\dg''', \dA'''),\widehat{L_{b}}(0)^*(\check{g}^*_1, \check{A}_1^*)\rangle
	\end{align*}
where we use \eqref{eq:exoftwoterms} and $-i(\pa_\sigma\widehat{L_{b}}(0))^*(\dg^*_1, \dA_1^*)=\widehat{L_{b}}(0)^*(\check{g}^*_1, \check{A}_1^*)$ in the second step. Since \[\langle(\dg''', \dA'''),\widehat{L_{b}}(0)^*(\check{g}^*_1, \check{A}_1^*)\rangle=\langle\widehat{L_{b}}(0)(\dg''', \dA'''),(\check{g}^*_1, \check{A}_1^*)\rangle=0,
\]
 we arrive at
\begin{equation}\label{eq:determinationofh1'}
	\begin{split}
	&\Big\langle[[L_b, t_{b,*}], t_{b,*}](\dg_1', \dA_1')+2[L_b, t_{b,*}](\check{g}_1', \check{A}_1'),\  (\dg_1^*, \dA_1^*)\Big\rangle\\
	&\quad=\Big\langle[[L_b, t_{b,*}], t_{b,*}](\check{g}_1, \check{A}_1)-[[L_b, t_{b,*}], t_{b,*}](\dg_2, \dA_2)+2[L_b, t_{b,*}](\bar{\dg},\bar{\dA}),\  (\dg_1^*, \dA_1^*)\Big\rangle.
	\end{split}
\end{equation}
Since the pairing on $\mathcal{K}_{b,1}\times\mathcal{K}_{b,1}^*$ on the left-hand side of \eqref{eq:determinationofh1'} is non-degenerate and the right-hand side is uniquely determined by $f$, so is $(\dg_1', \dA_1')$. Actually, we can further calculate the right-hand side
\begin{align*}
	\Big\langle[L_b, t_{b,*}](\bar{\dg},\bar{\dA}),\  (\dg_1^*, \dA_1^*)\Big\rangle&=\Big\langle(\bar{\dg},\bar{\dA}),\  \widehat{L_{b}}(0)^*(\check{g}_1^*, \check{A}_1^*)\Big\rangle=\Big\langle\widehat{L_{b}}(0)(\bar{\dg},\bar{\dA}),\  (\check{g}_1^*, \check{A}_1^*)\Big\rangle\\
	&=-\langle f+\frac{1}{2}[[L_b, t_{b,*}], t_{b,*}](\dg_1, \dA_1)+[L_b, t_{b,*}](\dg_2-\check{g}_1, \dA_2-\check{A}_1), (\check{g}_1^*, \check{A}_1^*)\rangle
	\end{align*}
where we use \eqref{eq:PDEfoftwoterms} and \eqref{eq:exoftwoterms} in the third step. Plugging this into \eqref{eq:determinationofh1'} proves the condition \eqref{eq:condition2}.
\end{proof}
\section{Regularity of the resolvent of the linearized gauge-fixed Einstein-Maxwell operator}\label{sec:regularityofmlEM}
In this section, we continue using the notations from \S\ref{sec:structureofmlEM}. In \S\ref{subsec:regularityoflowfre}, we give a refined description of the singular part $P(b,\sigma)$ and the regular part $L^-_b(\sigma)$ of the operator $\widehat{L_{b}}(0)^{-1}$ defined in Corollary \ref{cor:singandreg}. 	Concretely, we prove that $(\dg_1''(\sigma), \dA_1''(\sigma))$ defined in Corollary \ref{cor:singandreg} is H\"{o}lder regular at $\sigma=0$, while $L^-_b(\sigma)$ is H\"{o}lder regular at $\sigma=0$ only when one relaxes the target space. In \S\ref{subsec:regularityofhighfre}, we provide estimates of the bounds of any derivatives (in $\sigma$) of $L^-_b(\sigma)$ for $\sigma$ in the region $0\leq\IM\sigma\leq C, \abs{\sigma}\geq c_0$ for some $c_0,C>0$. Finally, in \S\ref{subsec:conormalregularity}, we discuss the conormal regularity of $L^-_b(\sigma)$ at $\sigma=0$. Specifically, after applying any number of derivatives $\sigma\pa_\sigma$ to $L^-_b(\sigma)$, it satisfies the same properties as $L^-_b(\sigma)$.

\subsection{Regularity at low frequencies}\label{subsec:regularityoflowfre}
In \S\ref{subsubsec:sing}, we study in detail the structure of $\widehat{L_{b}}(\sigma)\check{L}_b(\sigma)^{-1}$, i.e., prove the $\epsilon$-regularity of $\widehat{L_{b}}(\sigma)\check{L}_b(\sigma)^{-1}$. In \S\ref{subsubsec:reg}, we establish the $\epsilon$-regularity of $R(b,\sigma)$, which allows us to give a refined description of the singular part $P(b,\sigma)$ of $\widehat{L_{b}}(\sigma)^{-1}$ near $\sigma=0$. We also discuss a certain H\"{o}lder regularity of the regular part $L^-_b(\sigma)$ at $\sigma=0$.

\subsubsection{$\epsilon$-regularity of $\widehat{L_{b}}(\sigma)\check{L}_b(\sigma)^{-1}$.}\label{subsubsec:sing}
First, we introduce several definitions.
\begin{defn}[{\cite[Definition 12.3]{HHV21}}]
	Let $X,Y$ be two Banach spaces and $\al, \be\in\BR$. Suppose $A(\sigma):X\to Y$ is a family of operators depending on parameter $\sigma\in\BC\setminus\{0\}$. Then we write
	\[
	A(\sigma): \abs{\sigma}^\al X\to\abs{\sigma}^\be Y
	\]
	if and only if there exists a constant $C>0$ such that $\norm{A(\sigma)}_{X\to Y}\leq  C\abs{\sigma}^{\be-\al}$ for all $\sigma\in\BC\setminus\{0\}$.
\end{defn}

Let $\wt{\mathcal{K}}_i, \mathcal{R}_i$ ($i=0,1,2$) be defined as in \S\ref{sec:structureofmlEM}.
\begin{defn}[{\cite[Definition 12.6]{HHV21}}]
\begin{enumerate}
	\item An operator $L(\sigma):\wt{\mathcal{K}}_i\to\mathcal{R}_j$ is \textit{$\epsilon$-regular at $\sigma=0$} if for $\sigma\in \BC, \IM\sigma\geq 0$ and $\abs{\sigma}$ small, there exists a constant $C>0$ such that 
\begin{subequations}
	\begin{align}
	\norm{L(\sigma)}_{\wt{\mathcal{K}}_i\to\mathcal{R}_j}&\leq C,\\
	\norm{\pa_\sigma L(\sigma)}_{\wt{\mathcal{K}}_i\to\mathcal{R}_j}&\leq C\abs{\sigma}^{-\epsilon},\\
		\norm{\pa^2_\sigma L(\sigma)}_{\wt{\mathcal{K}}_i\to\mathcal{R}_j}&\leq C\abs{\sigma}^{-\epsilon-1}.
	\end{align}
\end{subequations}
\item  An operator $L(\sigma):\wt{\mathcal{K}}_i\to\mathcal{R}_j$ has an \textit{$\epsilon$-regular expansion at $\sigma=0$} in $\sigma$ up to order one if 
\begin{equation}
	L(\sigma)=L^0+\sigma L^e(\sigma)
	\end{equation}
where $L^0:\wt{\mathcal{K}}_i\to\mathcal{R}_j$ is independent of $\sigma$ and $L^e(\sigma):\wt{\mathcal{K}}_i\to\mathcal{R}_j$ is $\epsilon$-regular at $\sigma=0$.
\item An operator $L(\sigma):\wt{\mathcal{K}}_i\to\mathcal{R}_j$ has an \textit{$\epsilon$-regular expansion at $\sigma=0$} in $\sigma$ up to order two if 
\begin{equation}
	L(\sigma)=L^0+\sigma L^1+\sigma^2L^e(\sigma)
\end{equation}
where $L^0, L^1:\wt{\mathcal{K}}_i\to\mathcal{R}_j$ are independent of $\sigma$ and $L^e(\sigma):\wt{\mathcal{K}}_i\to\mathcal{R}_j$ is $\epsilon$-regular at $\sigma=0$.
\end{enumerate}	
\end{defn}

We recall the analysis of the entries $L_{ij}$ of the operator in the proof of Theorem \ref{thm:modestabilityofmlgEMofKN}, and note that each entry is a sum of polynomials in $\sigma$ whose coefficients are either independent of $\sigma$, terms involving $\check{L}_b(\sigma)^{-1}$ acting on an element in $\eHb^{\infty, 3/2-}(\CX)$, or terms involving $(\check{L}_b(\sigma)^{-1})^*$ acting on an element in $\sHb^{-5/2-C(a,\gamma), 3/2-}(\CX)$ where $C(a,\gamma_0)$ is a sufficiently small constant depending on $a,\gamma$. Therefore, we first discuss the properties of the operator $\check{L}_b(\sigma)^{-1}$.
\begin{prop}\label{prop:epsilonregofcheckL-1} 
Let $-\frac 32<\ell<-\frac 12$, $0<\epsilon<1$ with $-\frac 12<\ell+\epsilon<\frac 12$, and $s-\epsilon>4$. For $\IM\sigma\geq 0$, $0<\abs{\sigma}\leq c_0$ with $c_0$ small, we have
\begin{align}\label{eq:firstderiofcheckL-1}
	&\pa_\sigma\check{L}_b(\sigma)^{-1}: \eHb^{s-1,\ell+2}(\CX)\to \abs{\sigma}^{-\epsilon}\eHb^{s-\epsilon, \ell+\epsilon-1}(\CX),\\\label{eq:secondderiofcheckL-1}
	&\pa^2_\sigma\check{L}_b(\sigma)^{-1}: \eHb^{s-1,\ell+2}(\CX)\to \abs{\sigma}^{-\epsilon-1}\eHb^{s-1-\epsilon, \ell+\epsilon-1}(\CX).
\end{align}
\end{prop}

\begin{proof}
Formally, we have 
\[
\pa_\sigma\check{L}_b(\sigma)^{-1}=-\check{L}_b(\sigma)^{-1}(\pa_\sigma\check{L}_b(\sigma))\check{L}_b(\sigma)^{-1}
\]
where by Proposition \ref{PropFourierPWL}
\begin{equation}\label{eq:deriofcheckL}
\pa_\sigma\check{L}_b(\sigma)=\pa_\sigma\widehat{L_{b}}(\sigma)\in2i\rho(\rho\pa_\rho-1)+\rho^3\mbox{Diff}_\bop^1+\rho^2C^\infty+\sigma\rho^2C^\infty.
\end{equation}
Motivated by this, we first analyze $\rho(\rho\pa_\rho-1)\check{L}_b(\sigma)^{-1}$ and $\check{L}_b(\sigma)^{-1}\rho(\rho\pa_\rho-1)$. Since $\rho(\rho\pa_\rho-1): \eHb^{s,\ell}(\CX)\to\eHb^{s-1,\ell+1}(\CX)$, it is clear that 
\[
\rho(\rho\pa_\rho-1)\check{L}_b(\sigma)^{-1}: \eHb^{s-1,\ell+2}(\CX)\to\eHb^{s-1, \ell+1}(\CX).
\]
Meanwhile, in view of Proposition \ref{PropFourierPWL}
\[
2i\rho(\rho\pa_\rho-1)=\sigma^{-1}\big(\check{L}_b(\sigma)-\check{L}_b(0)+Q\big)\quad\mbox{with}\quad Q\in\sigma\rho^3\mbox{Diff}_\bop^1+\sigma\rho^2C^\infty+\sigma\rho^2C^\infty,
\]
then it follows that
\[
2i\rho(\rho\pa_\rho-1)\check{L}_b(\sigma)^{-1}=\sigma^{-1}\Big(I-(\widehat{L_{b}}(0)+V_b+Q)\check{L}_b(\sigma)^{-1}\Big): \eHb^{s-1,\ell+2}(\CX)\to\abs{\sigma}^{-1}\eHb^{s-2, \ell+2}(\CX).
\]
Then an interpolation gives 
\begin{equation}
\rho(\rho\pa_\rho-1)\check{L}_b(\sigma)^{-1}: \eHb^{s-1,\ell+2}(\CX)\to\abs{\sigma}^{-\epsilon}\eHb^{s-1-\epsilon, \ell+1+\epsilon}(\CX),\quad 0\leq \epsilon\leq 1.	
\end{equation}
\begin{comment}
As for $\check{L}_b(\sigma)^{-1}\rho(\rho\pa_\rho-1)$, interpolation between 
\[
\check{L}_b(\sigma)^{-1}\rho(\rho\pa_\rho-1):\eHb^{s,\ell+1}(\CX)\to\eHb^{s,\ell}(\CX)
\]
and
\[
\check{L}_b(\sigma)^{-1}\big(-2i\rho(\rho\pa_\rho-1)\big)=\sigma^{-1}\Big(I-\check{L}_b(\sigma)^{-1}(\widehat{L_{b}}(0)+V_b-Q)\Big):\eHb^{s+1, \ell}(\CX)\to\abs{\sigma}^{-1}\eHb^{s,\ell}(\CX)
\]
yields
\begin{equation}
\check{L}_b(\sigma)^{-1}\rho(\rho\pa_\rho-1):\eHb^{s+\epsilon,\ell+1-\epsilon}(\CX)\to\abs{\sigma}^{-\epsilon}\eHb^{s,\ell}(\CX), \quad 0\leq \epsilon\leq 1.
\end{equation}
\end{comment}
Since $s-1-\epsilon>2$ and $1/2<\ell+1+\epsilon<3/2$, we have
\begin{equation}\label{eq:maincal}
\eHb^{s-1, \ell+2}(\CX)\xrightarrow{2i\rho(\rho\pa_\rho-1)\check{L}_b(\sigma)^{-1}}\abs{\sigma}^{-\epsilon}\eHb^{s-1-\epsilon, \ell+\epsilon+1}(\CX)\xrightarrow{\check{L}_b(\sigma)^{-1}}\abs{\sigma}^{-\epsilon}\eHb^{s-\epsilon, \ell+\epsilon-1}(\CX).
\end{equation}
We also have
\begin{equation}\label{eq:easycal}
\eHb^{s-1, \ell+2}(\CX)\xrightarrow{\check{L}_b(\sigma)^{-1}\!\!}\eHb^{s, \ell}(\CX)\xrightarrow{\rho^2\mbox{Diff}^1_\bop+\sigma\rho^2C^\infty\!\!}\eHb^{s-1,\ell+2}(\CX)\xrightarrow{\check{L}_b(\sigma)^{-1}\!\!\!}\eHb^{s,\ell}(\CX)\subset\abs{\sigma}^{-\epsilon}\eHb^{s-\epsilon, \ell+\epsilon-1}(\CX).
\end{equation}
By \eqref{eq:deriofcheckL}, we finishes the proof of \eqref{eq:firstderiofcheckL-1}.

For $\pa_\sigma^2\check{L}_b(\sigma)^{-1}$, we note that 
\begin{align*}
\pa_\sigma^2\check{L}_b(\sigma)^{-1}&=2\check{L}_b(\sigma)^{-1}(\pa_\sigma\check{L}_b(\sigma))\check{L}_b(\sigma)^{-1}(\pa_\sigma\check{L}_b(\sigma))\check{L}_b(\sigma)^{-1}-\check{L}_b(\sigma)^{-1}(\pa^2_\sigma\check{L}_b(\sigma))\check{L}_b(\sigma)^{-1}\\
&:=\Romanupper{1}+\Romanupper{2}.
\end{align*}
Since $\pa_\sigma^2\check{L}_b(\sigma)\in\rho^2C^\infty$, according to \eqref{eq:easycal}, we obtain 
\[
\Romanupper{2}: \eHb^{s-1,\ell+2}(\CX)\to\eHb^{s,\ell}(\CX)\subset\abs{\sigma}^{-\epsilon-1}\eHb^{s-1-\epsilon, \ell+\epsilon-1}(\CX).
\]
It remains to analyze $\Romanupper{1}$. By the assumption on $s,\ell, \epsilon$, we find that $1/2<(1+\epsilon)/2<1$, and thus using the reasoning in \eqref{eq:maincal} and \eqref{eq:easycal}, we obtain
\begin{align*}
	\eHb^{s-1,\ell+2}(\CX)&\xrightarrow{\pa_\sigma\check{L}_b(\sigma)\check{L}_b(\sigma)^{-1}}\abs{\sigma}^{-\frac{1+\epsilon}{2}}\eHb^{s-1-\frac{1+\epsilon}{2}, \ell+\frac{1+\epsilon}{2}+1}(\CX)\xrightarrow{\pa_\sigma\check{L}_b(\sigma)\check{L}_b(\sigma)^{-1}}\abs{\sigma}^{-1-\epsilon}\eHb^{s-2-\epsilon, \ell+\epsilon+1}(\CX)\\
	&\xrightarrow{\check{L}_b(\sigma)^{-1}}\abs{\sigma}^{-1-\epsilon}\eHb^{s-1-\epsilon, \ell+\epsilon-1}(\CX)
	\end{align*}
where we use $s-\frac{3}{2}-\frac{\epsilon}{2}>2, \frac{1}{2}<\ell+\frac{3}{2}+\frac{\epsilon}{2}<\frac{3}{2}$ in the second step and $s-2-\epsilon>2,\frac{1}{2}< \ell+\epsilon+1<\frac{3}{2}$ in the last step. This finishes the proof of \eqref{eq:secondderiofcheckL-1}.
\end{proof}

Recall the expression \eqref{eq:detailedexofLL-1} of $\widehat{L_{b}}(\sigma)\check{L}(\sigma)^{-1}$
\[
\begin{split}
	&\widehat{L_{b}}(\sigma)\check{L}_b(\sigma)^{-1}=\begin{pmatrix}
		L_{00}(b,\sigma)& \sigma^2\wt{L}_{01}(b,\sigma)&\sigma\wt{L}_{02}(b,\sigma)\\
		\sigma\wt{L}_{10}(b,\sigma)& \sigma^2\wt{L}_{11}(b,\sigma)&\sigma^2\wt{L}_{12}(b,\sigma)\\
		\sigma\wt{L}_{20}(b,\sigma)& \sigma^2\wt{L}_{21}(b,\sigma)&\sigma\wt{L}_{22}(b,\sigma)\\
	\end{pmatrix}\\
	&\quad=\begin{pmatrix}
		L_{00}(b,\sigma)& \sigma^2\wt{L}_{01}(b,\sigma)&\sigma\wt{L}_{02}(b,\sigma)\\
		\sigma\big(\wt{L}^0_{10}(b)+\sigma\wt{L}^e_{10}(b,\sigma)\big)&\sigma^2\big(\wt{L}^0_{11}(b)+\sigma\wt{L}^1_{11}(b)+\sigma^2\wt{L}^e_{11}(b,\sigma)\big)&\sigma^2\big(\wt{L}^0_{12}(b)+\sigma\wt{L}^e_{12}(b,\sigma)\big)\\
		\sigma\wt{L}_{20}(b,\sigma)&\sigma^2\big(\wt{L}^0_{21}(b)+\sigma\wt{L}^e_{21}(b)\big)&\sigma\big(\wt{L}^0_{22}(b)+\sigma\wt{L}^e_{22}(b,\sigma)\big).
	\end{pmatrix}
\end{split}
\]

\begin{prop}\label{prop:epsilonregofLL-1}
	Let $s,\ell,\epsilon$ be defined as in Proposition \ref{prop:epsilonregofcheckL-1}. The entries of $\widehat{L_{b}}(\sigma)\check{L}_b(\sigma)^{-1}$ satisfy the following properties.
\begin{enumerate}
	\item $L_{00}, \wt{L}_{01}, \wt{L}_{02}, \wt{L}_{20}$ are $\epsilon$-regular at $\sigma=0$.
	\item $\wt{L}_{10}, \wt{L}_{12}, \wt{L}_{21}, \wt{L}_{22}$ have an $\epsilon$-regular expansion up to order one, that is, $\wt{L}^e_{10}, \wt{L}^e_{12}, \wt{L}^e_{21}, \wt{L}^e_{22}$ are $\epsilon$-regular at $\sigma=0$.
	\item $\wt{L}_{11}$ has an $\epsilon$-regular expansion up to order two, that is, $\wt{L}^e_{11}$ is $\epsilon$-regular at $\sigma=0$.
\end{enumerate}
\end{prop}
\begin{proof}
	The key point in the proof was already mentioned before Proposition \ref{prop:epsilonregofcheckL-1}.
	\begin{itemize}

		\item\underline{Analysis of $\wt{L}^e_{10}$ and $\wt{L}_{20}$.}
		Here, we only discuss $\wt{L}^e_{10}$ in detail because the proof of $\wt{L}_{02}$ follows in a similar manner. Recall the calculation in \eqref{eq:calofL10}
		\[
			\begin{split}
			&\wt{L}^e_{10}(\tilde{g}_0, \tilde{A}_0),\ (\dg^*_1, \dA^*_1)\rangle\\
			&\quad=-\Big\langle (\tilde{g}_0, \tilde{A}_0), i(\check{L}_b(\sigma)^{-1})^*(\pa_\sigma\widehat{L_{b}}(0)+\frac{\sigma}{2}\pa_\sigma^2\widehat{L_{b}}(0))^*\big(I-(\check{L}_b(0)^{-1})^*V_b^*\big)(\check{g}^*_2,\check{A}^*_2)\Big\rangle\\
			&\quad\quad+\Big\langle(\tilde{g}_0, \tilde{A}_0),\frac{1}{2}(\check{L}_b(\sigma)^{-1})^*(\pa^2_\sigma\widehat{L_{b}}(0))^*(\dg^*_2,\dA^*_2)\Big\rangle
			\end{split}
		\]
		where
		\[
		(\pa_\sigma\widehat{L_{b}}(0)+\frac{\sigma}{2}\pa_\sigma^2\widehat{L_{b}}(0))^*\big(I-(\check{L}_b(0)^{-1})^*V_b^*\big)(\check{g}^*_2,\check{A}^*_2),\quad(\pa^2_\sigma\widehat{L_{b}}(0))^*(\dg^*_2,\dA^*_2)\in\sHb^{-5/2-C(a,\gamma),3/2-}(\CX).
		\]
		Since $(\tilde{g}_0, \tilde{A}_0)\in\eHb^{s-1,\ell+2}(\CX)$, and by Proposition \eqref{prop:epsilonregofcheckL-1}
		\begin{gather*}
		(\check{L}_b^{-1}(\sigma))^*\sHb^{-5/2-C(a,\gamma),\, 3/2-}(\CX)\in\sHb^{-3/2-C(a,\gamma), -1/2-}(\CX),\\	(\pa^j_\sigma\check{L}_b^{-1})^*\sHb^{-5/2-C(a,\gamma),\, 3/2-}(\CX)\in\abs{\sigma}^{-\epsilon-j+1}\sHb^{-s+1, -\ell-2}(\CX),\quad j=1,2
		\end{gather*}
	where we use the facts that $-5/2-C(a,\gamma)>-3>-s+\epsilon+1$ and $-\ell-\epsilon+1<3/2$, the pairings above (and their up to second order derivatives) are well-defined. This proves the $\epsilon$-regularity of $\wt{L}^e_{10}$.
	  
	  	\item\underline{Analysis of $\wt{L}^e_{11},\wt{L}^e_{12},\wt{L}^e_{22}$ and $\wt{L}^e_{22}$.} The statements for these entries can be proved in the same way as in the proof of $\wt{L}^e_{10}$.

		\item \underline{Analysis of $\wt{L}_{01}$ and $\wt{L}_{02}$.} 	Here, we only prove the statement for $\wt{L}_{01}$ as the proof of $\wt{L}_{02}$ follows in an analogous manner. Using the calculation in \eqref{eq:calofL01}, we find that 
	\begin{align*}
\wt{L}_{01}(\tilde{g}_1, \tilde{A}_1)\!&=\!\!\Big(\!\!-iV_b\check{L}_b(\sigma)^{-1}\big(\pa_\sigma\widehat{L_{b}}(0)+\frac{\sigma}{2}\pa^2_\sigma\widehat{L_{b}}(0)\big)(\check{g}_1,\check{A}_1)+\frac{1}{2}V_b\check{L}_b(\sigma)^{-1}\pa^2_\sigma\widehat{L_{b}}(0)(\dg_1,\dA_1)\Big)\\
&\quad-(\wt{L}_{11}+\wt{L}_{21})(\tilde{g}_1, \tilde{A}_1).
	\end{align*}
	where
	\[
	\pa_\sigma\widehat{L_{b}}(0)(\check{g}_1,\check{A}_1),\quad \pa^2_\sigma\widehat{L_{b}}(0)(\check{g}_1,\check{A}_1),\quad\pa^2_\sigma\widehat{L_{b}}(0)(\dg_1,\dA_1)\in\eHb^{\infty, 3/2-}(\CX).
	\]
	Since by Proposition \ref{prop:epsilonregofLL-1} 
	\[
	  \check{L}_b^{-1}(\sigma)\eHb^{\infty, 3/2-}(\CX)\in\eHb^{\infty, -1/2-}(\CX),\  	\pa^j_\sigma\check{L}_b^{-1}(\sigma)\in\abs{\sigma}^{-\epsilon-j+1}\eHb^{\infty, \ell+\epsilon-1}(\CX),\  j=1,2,
	\]
	and $V_b$ maps $\mathcal{D}'(X^\circ)$ to $C_c^\infty(X)$, it follows that the first term on the right-hand side is $\epsilon$-regular. Meanwhile, it is clear that the second term is $\epsilon$-regular. This proves the $\epsilon$-regularity for $\wt{L}_{01}$.
	
	 \item\underline{Analysis of $L_{00}$.} By the calculation in \eqref{eq:calofL00}, we have
	\[
	L_{00}(b,\sigma)(\tilde{g}_0, \tilde{A}_0)=(\tilde{g}_0, \tilde{A}_0)-V_b\check{L}_b(\sigma)^{-1}(\tilde{g}_0, \tilde{A}_0)-\big(\sigma\wt{L}^0_{10}(b)+\sigma^2\wt{L}^e_{10}(b,\sigma)+\sigma\wt{L}_{20}(b,\sigma)\big)(\tilde{g}_0, \tilde{A}_0).
	\]
	Again, it is clear that the third term on the right-hand side is $\epsilon$-regular.	Since by Proposition \ref{prop:epsilonregofLL-1} 
	\[
	\check{L}_b^{-1}(\sigma)(\tilde{g}_0, \tilde{A}_0)\in\eHb^{s,\ell}(\CX),\  	\pa^j_\sigma\check{L}_b^{-1}(\sigma)\in\abs{\sigma}^{-\epsilon-j+1}\eHb^{s-\epsilon-j+1, \ell+\epsilon-1}(\CX),\  j=1,2,
	\]
	and $V_b$ maps $\mathcal{D}'(X^\circ)$ to $C_c^\infty(X)$, it follows that the second term on the right-hand side is $\epsilon$-regular. This finishes the proof.
	\end{itemize}
\end{proof}

\subsubsection{H\"{o}lder regularity of $P(b,\sigma)$ and $L^-_b(\sigma)$ at $\sigma=0$}\label{subsubsec:reg}
Now we are ready to analyze the $\epsilon$-regularity of $R(b,\sigma)$. To this end, we need the following lemma.

\begin{lem}\label{lem:epsilonregofinverse}
	Let $s,\ell,\epsilon$ be defined as in Proposition \ref{prop:epsilonregofcheckL-1}. Suppose that the operator $A(\sigma):\wt{\mathcal{K}}_i\to\mathcal{R}_j$ with $i,j=0,1,2$ has an $\epsilon$-regular expansion up to order one, i.e.,
	\[
	A(\sigma)=A^0+\sigma A^e(\sigma)
	\]
	where $A^0:\wt{\mathcal{K}}_i\to\mathcal{R}_j$ is invertible and independent of $\sigma$, and $A^e(\sigma):\wt{\mathcal{K}}_i\to\mathcal{R}_j$ is $\epsilon$-regular. Then for $\abs{\sigma}$ sufficiently small, $A(\sigma)$ is invertible, and its inverse $B(\sigma)=A(\sigma)^{-1}$ also has an $\epsilon$-regular expansion up to order one, that is, 
	\[
	B(\sigma)=B^0+\sigma B^e(\sigma)
	\]
	where $B^0:\mathcal{R}_j\to\wt{\mathcal{K}}_i$ is independent of $\sigma$, and $B^e(\sigma):\mathcal{R}_j\to\wt{\mathcal{K}}_i$ is $\epsilon$-regular.
	
	Similarly, if $A(\sigma)$ is $\epsilon$-regular and has uniformly bounded inverse $B(\sigma)$, then $B(\sigma)$ is $\epsilon$-regular as well.
\end{lem}

\begin{proof}
 Since $A^e(\sigma)$ is $\epsilon$-regular (and thus uniformly bounded), $\norm{\sigma (A^0)^{-1}A^e(\sigma)}_{\wt{\mathcal{K}}_i\to\wt{\mathcal{K}}_i}<1/2$ when $\abs{\sigma}$ is sufficiently small, the inverse of $A(\sigma)$ is given by the Neumann series
\begin{align*}
A(\sigma)^{-1}\!\!&=\!\!\Big(I+\sum_{k=1}^\infty(-1)^k\big(\sigma (A^0)^{-1}A^e(\sigma)\big)^k\Big)(A^0)^{-1}\\
&=(A^0)^{-1}-\sigma\Big((A^0)^{-1}A^e(\sigma)\sum_{k=0}^\infty(-1)^k\big(\sigma (A^0)^{-1}A^e(\sigma)\big)^k(A^0)^{-1}\Big):=B^0+\sigma B^e(\sigma).
\end{align*}
First, we have 
\begin{align*}
	\norm{B^e(\sigma)}_{\mathcal{R}_j\to\wt{\mathcal{K}}_i}&\leq\Big(\sum_{k=0}^\infty\norm{\sigma (A^0)^{-1}A^e(\sigma)}^k_{\wt{\mathcal{K}}_i\to\wt{\mathcal{K}}_i}\Big)\norm{(A^0)^{-1}}^2_{\mathcal{R}_j\to\wt{\mathcal{K}}_i}\norm{A^e(\sigma)}_{\wt{\mathcal{K}}_i\to\mathcal{R}_j}\\
	&\leq2\norm{A(0)^{-1}}^2_{\mathcal{R}_j\to\wt{\mathcal{K}}_i}\norm{A^e(\sigma)}_{\wt{\mathcal{K}}_i\to\mathcal{R}_j},
\end{align*}
which implies the uniform boundedness of $B^e(\sigma)$. Secondly, the first derivative of $B^e(\sigma)$ is a sum of the terms of the types
\begin{align*}
&\sigma^{-2}\big(\sigma (A^0)^{-1}A^e(\sigma)\big)^k(A^0)^{-1}\lesssim 2^{-k+2},\quad k\geq 2\\
&\big(\sigma (A^0)^{-1}A^e(\sigma)\big)^{k_1}\circ\big((A^0)^{-1}\pa_\sigma A^e(\sigma)\big)\circ\big(\sigma (A^0)^{-1}A^e(\sigma)\big)^{k_2}(A^0)^{-1}\lesssim2^{-k_1-k_2}\abs{\sigma}^{-\epsilon},
\end{align*}
where we use $\norm{\pa_\sigma A^e(\sigma)}_{\wt{\mathcal{K}}_i\to\mathcal{R}_j}\lesssim \abs{\sigma}^{-\epsilon}$. As a consequence, $\norm{\pa_\sigma B^e(\sigma)}_{\mathcal{R}_j\to\wt{\mathcal{K}}_i}\lesssim \abs{\sigma}^{-\epsilon}$. Finally, the second derivative of $B^e(\sigma)$ is a sum of the terms of the types
\begin{align*}
	&\sigma^{-3}\big(\sigma (A^0)^{-1}A^e(\sigma)\big)^k(A^0)^{-1}\lesssim 2^{-k+3},\quad k\geq 3\\
	&\sigma^{-1}\big(\sigma (A^0)^{-1}A^e(\sigma)\big)^{k_1}\circ\big((A^0)^{-1}\pa_\sigma A^e(\sigma)\big)\circ\big(\sigma (A^0)^{-1}A^e(\sigma)\big)^{k_2}(A^0)^{-1}\lesssim2^{-k_1-k_2+1}\abs{\sigma}^{-\epsilon}, \quad k_1+k_2\geq 1,\\
	&\sigma\big(\sigma (A^0)^{-1}A^e(\sigma)\big)^{k_1}\circ\big((A^0)^{-1}\pa_\sigma A^e(\sigma)\big)\circ\big(\sigma (A^0)^{-1}A^e(\sigma)\big)^{k_2}\circ\big((A^0)^{-1}\pa_\sigma A^e(\sigma)\big)\circ\big(\sigma (A^0)^{-1}A^e(\sigma)\big)^{k_3}(A^0)^{-1}\\
	&\quad\lesssim2^{-k_1-k_2-k_3}\abs{\sigma}^{-2\epsilon+1},\\
	&\big(\sigma (A^0)^{-1}A^e(\sigma)\big)^{k_1}\circ\big((A^0)^{-1}\pa^2_\sigma A^e(\sigma)\big)\circ\big(\sigma (A^0)^{-1}A^e(\sigma)\big)^{k_2}(A^0)^{-1}\lesssim2^{-k_1-k_2}\abs{\sigma}^{-\epsilon-1},
\end{align*}
where we use $\pa_\sigma A^e(\sigma)\lesssim \abs{\sigma}^{-\epsilon}$ and $\pa^2_\sigma A^e(\sigma)\lesssim \abs{\sigma}^{-\epsilon-1}$. Therefore, $\pa^2_\sigma B^e(\sigma)\lesssim \abs{\sigma}^{-\epsilon-1}$. 
This proves that $B^e(\sigma)$ is $\epsilon$-regular.

The proof of the second statement is similar by using 
\begin{align*}
	\pa_\sigma B(\sigma)&=-B(\sigma)\circ\pa_\sigma A(\sigma) \circ B(\sigma),\\ \pa_\sigma^2B(\sigma)&=-B(\sigma)\circ \pa^2_\sigma A(\sigma) \circ B(\sigma)+2B(\sigma)\circ\pa_\sigma A(\sigma) \circ B(\sigma)\circ \pa_\sigma A(\sigma)\circ B(\sigma).
\end{align*}
\end{proof}

Recall the expressions \eqref{eq:exofinverseR} and \eqref{eq:expansionofcomofR} of $R_{b,\sigma}=(\widehat{L_{b}}(\sigma)\check{L}_b(\sigma)^{-1})^{-1}$
\[
\begin{split}
	R(b,\sigma)&=(\widehat{L_{b}}(\sigma)\check{L}_b(\sigma)^{-1})^{-1}=
	\begin{pmatrix}
		\wt{R}_{00}(b,\sigma)&\wt{R}_{01}(b,\sigma)&\wt{R}_{02}(b,\sigma)\\
		\sigma^{-1}\wt{R}_{10}(b,\sigma)&\sigma^{-2}\wt{R}_{11}(b,\sigma)&\sigma^{-1}\wt{R}_{12}(b,\sigma)\\
		\wt{R}_{20}(b,\sigma)&\sigma^{-1}\wt{R}_{21}(b,\sigma)&\sigma^{-1}\wt{R}_{22}(b,\sigma)
	\end{pmatrix}\\
&=
\begin{pmatrix}
	\wt{R}_{00}(b,\sigma)&\wt{R}_{01}(b,\sigma)&\wt{R}_{02}(b,\sigma)\\
	\sigma^{-1}\wt{R}_{10}(b,\sigma)&\sigma^{-2}\big(\wt{R}^0_{11}(b)+\sigma\wt{R}^e_{11}(b,\sigma)\big)&\sigma^{-1}\wt{R}_{12}(b,\sigma)\\
	\wt{R}_{20}(b,\sigma)&\sigma^{-1}\big(\wt{R}^0_{21}(b)+\sigma\wt{R}^e_{21}(b,\sigma)\big)&\sigma^{-1}\big(\wt{R}^0_{22}(b)+\sigma\wt{R}^e_{22}(b,\sigma)\big)
\end{pmatrix}
\end{split}
\]
where
\begin{equation}
	\begin{split}
		\wt{R}_{11}=\big(\wt{L}^\sharp_{11}-\sigma\wt{L}^\sharp_{12}(\wt{L}^\flat_{22})^{-1}\wt{L}^\sharp_{21}\big)^{-1},\quad \wt{R}_{10}=\wt{R}_{11}\big(-\wt{L}_{10}+\sigma\wt{L}_{12}^\sharp(\wt{L}^\flat_{22})^{-1}\wt{L}_{20}\big)L_{00}^{-1},\\
		\wt{R}_{12}=-\wt{R}_{11}\wt{L}_{12}^\sharp(\wt{L}^\flat_{22})^{-1},\quad \wt{R}_{21}=-(\wt{L}^\flat_{22})^{-1}\wt{L}_{21}^\sharp\wt{R}_{11},\quad\wt{R}_{22}=(\wt{L}^\flat_{22})^{-1}-\sigma(\wt{L}^\flat_{22})^{-1}\wt{L}_{21}^\sharp \wt{R}_{12},\\
		\wt{R}_{20}=-(\wt{L}^\flat_{22})^{-1}\big(\wt{L}_{20}L_{00}^{-1}+\wt{L}_{21}^\sharp\wt{R}_{10}\big),\quad \wt{R}_{01}=-L_{00}^{-1}\big(\wt{L}_{01}\wt{R}_{11}+\wt{L}_{02}\wt{R}_{21}\big),\\
		\wt{R}_{02}=-L_{00}^{-1}\big(\sigma\wt{L}_{01}\wt{R}_{12}+\wt{L}_{02}\wt{R}_{22}\big),\quad \wt{R}_{00}=L_{00}^{-1}\big(I-\sigma\wt{L}_{01}\wt{R}_{10}-\sigma\wt{L}_{02}\wt{R}_{20}\big)
	\end{split}
\end{equation}
with
\begin{equation}
	\begin{split}
		&\wt{L}^\sharp_{i1}=\wt{L}_{i1}-\sigma\wt{L}_{i0}L_{00}^{-1}\wt{L}_{01}\quad\mbox{for}\quad i=1,2, \quad \wt{L}^\sharp_{12}=\wt{L}_{12}-\wt{L}_{10}L_{00}^{-1}\wt{L}_{02},\\
		&\qquad\qquad\qquad\qquad	\wt{L}_{22}^\flat=\wt{L}_{22}-\sigma\wt{L}_{20}L_{00}^{-1}\wt{L}_{02}.
	\end{split}
\end{equation}

\begin{prop}\label{prop:epsilonregofR}
	Let $s,\ell,\epsilon$ be defined as in Proposition \ref{prop:epsilonregofcheckL-1}. The entries of $R(b,\sigma)$ satisfy the following properties.
\begin{enumerate}
	\item $\wt{R}_{00}, \wt{R}_{01}, \wt{R}_{02}, \wt{R}_{10}, \wt{R}_{12}, \wt{R}_{20}$ are $\epsilon$-regular at $\sigma=0$.
	\item $\wt{R}_{11}, \wt{R}_{21}, \wt{R}_{22}$ have an $\epsilon$-regular expansion up to order one, that is, $\wt{R}^e_{11}, \wt{R}^e_{21}, \wt{R}^e_{22}$ are $\epsilon$-regular at $\sigma=0$.
\end{enumerate}
\end{prop}

\begin{proof}
First, since $L_{00}$ is $\epsilon$-regular, by Lemma \ref{lem:epsilonregofinverse}, so is $L_{00}^{-1}$. Then by Proposition \ref{prop:epsilonregofLL-1}, we find that $\wt{L}^\sharp_{11}, \wt{L}^\sharp_{21}, \wt{L}^\flat_{22}$ (and thus $(\wt{L}^\flat_{22})^{-1}$ by Lemma \ref{lem:epsilonregofinverse}) have an $\epsilon$-regular expansion up to order one, while $\wt{L}^\sharp_{12}$ is $\epsilon$-regular. Therefore, we see that 
\[
\wt{L}^\sharp_{11}-\sigma\wt{L}^\sharp_{12}(\wt{L}^\flat_{22})^{-1}\wt{L}^\sharp_{21}
\]
has an $\epsilon$-regular expansion up to order one whose coefficient of $\sigma^0$ is $\wt{L}_{11}^0(b)$. Then the $\epsilon$-regular expansion property of $\wt{R}_{11}$ follows from Lemma \ref{lem:epsilonregofinverse}. Similarly, the $\epsilon$-regular expansion property of $\wt{R}_{21},\wt{R}_{22}$ follows from that of $\wt{R}_{11}, (\wt{L}^\flat_{22})^{-1},\wt{L}^\sharp_{21}$ and the above expressions of $\wt{R}_{ij}$. Finally, the $\epsilon$-regularity of the remaining components can be obtained by using the above expressions of $\wt{R}_{ij}$ and Proposition \ref{prop:epsilonregofLL-1}.
\end{proof}

Recall the singular part $P(b,\sigma)$ of $\widehat{L_{b}}(\sigma)^{-1}$ as defined in Corollary \ref{cor:singandreg}, which satisfies 
\begin{equation}\label{eq:recallofsingex}
	\begin{split}
	P(b,\sigma)f&=(\sigma^{-2}(\dg_1,\dA_1)-i\sigma^{-1}(\check{g}_1, \check{A}_1))+i\sigma^{-1}\big((\dg_2, \dA_2)
+(\dg'_1, \dA'_1)\big)+i\sigma^{-1}(\dg_1''(\sigma), \dA_1''(\sigma))\\
&:=\sigma^{-2}P^2(b)f+\sigma^{-1}P^1(b)f+\sigma^{-1}P^e(b,\sigma)f
\end{split}
\end{equation}
where $(\dg_1''(\sigma),\dA_1''(\sigma))=P^e(b,\sigma)f=o(1)$ in $\mathcal{K}_{b,1}$ as $\sigma\to 0$. With the $\epsilon$-regularity of $R(b,\sigma)$ at our disposal, now we are able to prove that $P^e(b,\sigma)$ is $\epsilon$-regular as well, and thus give a refined description of $\sigma^{-1}(\dg_1''(\sigma),\dA_1''(\sigma))$. \begin{comment}Concretely, we shall prove that $(\dg_1''(\sigma),\dA_1''(\sigma))$ is H\"{o}lder continuous in a small interval of $0$, and thus $(\dg_1''(\sigma),\dA_1''(\sigma))=\mathcal{O}(\abs{\sigma}^{1-\epsilon})$ in $\mathcal{K}_{b,1}$ for some $0<\epsilon<1$.
\end{comment}

\begin{rem}
	We now make a remark that there is a relationship between the Sobolev spave $\sH^k([0,c_0))$, which is the restriction to $[0, c_0)$ of elements in $H^k(\BR)$ with support in $[0, \infty)$, and the $\bop$-Sobolev space $H_{\bop}^k([0, c_0))$, which is a completion of $\{\phi|_{(0, c_0]}\}$ where $\phi\in C_c^\infty((0,\infty))$ with respect the following  norm 
	\[
	\norm{\phi}^2_{H_{\bop}^k([0, C_0))}:=\sum_{j=0}^k\norm{(\sigma\pa_\sigma)^j\phi}^2_{L^2(d\sigma)}.
	\]
	First, it is clear that $\sigma^kH_{\bop}^k([0, c_0))\subset \sH^k([0, c_0))$. Second, Hardy's inequality tells us that for $\phi\in C_c^\infty(0, \infty)$, we have 
	\[
	\int_0^\infty \frac{\abs{\phi}^2}{r^2}\,dr\leq 4\int_0^\infty \abs{\pa_rf}^2\,dr.
	\]
	Applying Hardy's inequality $k$ times, we find that 
	\[
	\int_0^\infty \frac{\abs{\phi}^2}{r^{2k}}\,dr\leq C_k\int_0^\infty \abs{\pa^k_rf}^2\,dr.
	\]
	for $\phi\in C_c^\infty(0, \infty)$. This implies $ \sH^k([0, c_0))\subset\sigma^kH_{\bop}^k([0, c_0))$. Therefore, $\sH^k([0, c_0))=\sigma^kH_{\bop}^k([0, c_0))$. Then an interpolation gives $\sH^\al([0, c_0))=\sigma^\al H_{\bop}^\al([0, c_0))$ for all $\al\geq 0$.
\end{rem}

As a preparation, we record the following technical lemma.

\begin{lem}\label{lem:technicallem}
	Suppose that $u_{\pm}\in H^s(\BR_{\pm})$ with $s\geq 0$ and $s\neq \frac12+\BN_0$ and let $u(x)=u_+(x)$ for $x>0$ and $u(x)=u_-(x)$ for $x<0$. If $\pa_ju_+(x)=\pa_j u_-(x)$ for all $j\in\BN_0, j<s-\frac 12$, then $u(x)\in H^s(\BR)$.
\end{lem}

\begin{proof}
	See \cite[Theorem 11.4 and 11.5]{LM72}.
\end{proof}

\begin{thm}\label{thm:holderregularityofsing}
Let $\sigma^{-1}(\dg_1''(\sigma),\dA_1''(\sigma))=\sigma^{-1}P^e(b,\sigma)f\in\mathcal{K}_{b,1}$ be defined as in \eqref{eq:recallofsingex} and let $s,\ell,\epsilon$ be defined as in Proposition \ref{prop:epsilonregofcheckL-1}. Then 
%for any $\delta>0$ satisfying $1-\epsilon-\delta>0$ and $\frac{3}{2}-\epsilon-\delta\neq \frac{1}{2}+\BN_0$, 
we have
\begin{equation}\label{eq:epsilonregofPe}
	P^e(b,\sigma)\in H^{\frac 32-\epsilon-}((-c_0,c_0);\mathcal{L}(\eHb^{s-1,\ell+2}(\CX), \mathcal{K}_{b,1}))
	\end{equation}
where $c_0>0$ is a sufficiently small constant. Moreover,
\begin{equation}\label{eq:conormalofPe}
	\sigma^{-1}(\dg_1''(\sigma),\dA_1''(\sigma))\in\sum_{\pm}\mathcal{A}^{-\epsilon}\big(\pm[0, c_0);\mathcal{K}_{b,1}\big).
	\end{equation}
\end{thm}

\begin{proof}
	Using the calculation \eqref{eq:exofcheckLon1} and \eqref{eq:exofcheckLon2} in the proof of Corollary \ref{cor:singandreg}, we find that for $f=(f_0, f_1, f_2)\in\eHb^{s-1,\ell+2}(\CX)=\mathcal{R}_0\oplus\mathcal{R}_{1}\oplus\mathcal{R}_{2}$,
	\begin{equation}\label{eq:anotherexofP}
		\begin{split}
			&\check{L}_b(0)P^e(b,\sigma)f\\
	&\quad=\big(\wt{R}_{10}(b,\sigma)-\wt{R}_{10}(b,0)\big)f_0+\big(\wt{R}^e_{11}(b,\sigma)-\wt{R}^e_{11}(b,0)\big)f_1+\big(\wt{R}_{12}(b,\sigma)-\wt{R}_{12}(b,0)\big)f_2.
	\end{split}
\end{equation}
	According to Proposition \ref{prop:epsilonregofR}, we have $\norm{\pa_\sigma\wt{R}^e_{11}(b,\sigma)}_{\mathcal{R}_1\to\wt{\mathcal{K}}_1}\lesssim \abs{\sigma}^{-\epsilon}$ and $\norm{\pa^2_\sigma\wt{R}^e_{11}(b,\sigma)}_{\mathcal{R}_1\to\wt{\mathcal{K}}_1}\lesssim \abs{\sigma}^{-\epsilon-1}$ for $0<\abs{\sigma}\leq c_0$ where $c_0$ is a sufficiently small constant, which implies that
	\[
	\norm{\pa_\sigma\sigma\pa_\sigma\wt{R}^e_{11}(b,\sigma)}_{\mathcal{R}_1\to\wt{\mathcal{K}}_1}\lesssim \abs{\sigma}^{-\epsilon}.\]
%	and
%	\begin{equation}\label{eq:secondderiofRe11}
%	(\sigma\pa_\sigma)^2\wt{R}^e_{11}(\sigma)\in\sigma^{1-\epsilon}L^\infty([0, c_0);\mathcal{L}(\mathcal{R}_1, \wt{\mathcal{K}}_1))\subset \sigma^{\frac 32-\epsilon-}H^0_b([0, c_0);\mathcal{L}(\mathcal{R}_1, \wt{\mathcal{K}}_1)).
	%\end{equation}
	Integrating $\pa_\sigma\sigma\pa_\sigma\wt{R}^e_{11}(b,\sigma)$ and $\pa_\sigma\wt{R}^e_{11}(b,\sigma)$ from $\sigma\in(0, c_0)$ to $0$ and using $\sigma\pa_\sigma\wt{R}^e_{11}(\sigma)|_{b,\sigma=0}=0$ yield
	\begin{align}\label{eq:firstderoofRe11}
	\sigma\pa_\sigma\wt{R}^e_{11}(b,\sigma)\in\sigma^{1-\epsilon}L^\infty([0, c_0);\mathcal{L}(\mathcal{R}_1, \wt{\mathcal{K}}_1))\subset \sigma^{\frac 32-\epsilon-}H^0_{\bop}([0, c_0);\mathcal{L}(\mathcal{R}_1, \wt{\mathcal{K}}_1))
	,\\\label{eq:zeroderiofFRe11}
	\wt{R}^e_{11}(b,\sigma)-\wt{R}^e_{11}(b,0)\in\sigma^{1-\epsilon}L^\infty([0, c_0);\mathcal{L}(\mathcal{R}_1, \wt{\mathcal{K}}_1))\subset \sigma^{\frac 32-\epsilon-}H^0_{\bop}([0, c_0);\mathcal{L}(\mathcal{R}_1, \wt{\mathcal{K}}_1))
	.
	\end{align}
This implies
\[
	(\wt{R}^e_{11}(\sigma)-\wt{R}^e_{11}(0))\in\sigma^{\frac32-\epsilon-}H^2_{\bop}([0, c_0);\mathcal{L}(\mathcal{R}_1, \wt{\mathcal{K}}_1))\subset\sH^{\frac 32-\epsilon-}([0, c_0);\mathcal{L}(\mathcal{R}_1, \wt{\mathcal{K}}_1)).
\]
Similarly, we also have 
\[
			(\wt{R}^e_{11}(\sigma)-\wt{R}^e_{11}(0))\in\sH^{\frac 32-\epsilon-}((-c_0,0];\mathcal{L}(\mathcal{R}_1, \wt{\mathcal{K}}_1)).
\]
Since $	(\wt{R}^e_{11}(\sigma)-\wt{R}^e_{11}(0))\to0$ as $\sigma\to0$ from both sides, by Lemma \ref{lem:technicallem}, we obtain
\[
(\wt{R}^e_{11}(\sigma)-\wt{R}^e_{11}(0))\in H^{\frac 32-\epsilon-}((-c_0,c_0);\mathcal{L}(\mathcal{R}_1, \wt{\mathcal{K}}_1)).
\] 
Similarly, $\wt{R}_{10}(b,\sigma)-\wt{R}_{10}(b,0)$ and $\wt{R}_{12}(b,\sigma)-\wt{R}_{12}(b,0)$ satisfy the same estimates as $\wt{R}^e_{11}(b,\sigma)-\wt{R}^e_{11}(b,0)$. This proves \eqref{eq:epsilonregofPe}.

Moreover, the estimate in \eqref{eq:zeroderiofFRe11} implies 
\begin{equation}\label{eq:LinfityestimateofR11}
	\wt{R}^e_{11}(b,\sigma)-\wt{R}^e_{11}(b,0)\in\sum_{\pm}\sigma^{1-\epsilon}L^\infty(\pm[0, c_0);\mathcal{L}(\mathcal{R}_1, \wt{\mathcal{K}}_1))
\end{equation}
and this also holds for $\wt{R}_{10}(b,\sigma)-\wt{R}_{10}(b,0), \wt{R}_{20}(b,\sigma)-\wt{R}_{20}(b,0)$.
Using Proposition \ref{prop:conormalregofR}, the estimate \eqref{eq:LinfityestimateofR11} is satisfied after applying any number of derivatives $\sigma\pa_\sigma$. This proves \eqref{eq:conormalofPe}.
%Since 
%\[
%\norm{\sigma^{-\delta}\big(\wt{R}^e_{11}(\sigma)-\wt{R}^e_{11}(0)\big)}\lesssim \abs{\sigma}^{1-\epsilon-\delta}\to0
%\] 
%as $\sigma\to0$ from both sides, by using Lemma \ref{lem:technicallem}, we arrives at
%\[
%\sigma^{-\delta}\big(\wt{R}^e_{11}(\sigma)-\wt{R}^e_{11}(0)\big)\in H^{\frac 32-(\epsilon+\delta)-}((-c_0, c_0);\mathcal{L}(\mathcal{R}_1, \wt{\mathcal{K}}_1)).
%\]
%The above statement also holds for $\sigma^{-\delta}\big(\wt{R}_{10}(\sigma)-\wt{R}_{10}(0)\big)$ and $\sigma^{-\delta}\big(\wt{R}_{12}(\sigma)-\wt{R}_{12}(0)\big)$ with $\mathcal{R}_1$ replaced by $\mathcal{R}_0$ and $\mathcal{R}_2$ respectively. This implies 
%\[
%\delta^{-1}(\dg_1''(\sigma),\dA_1''(\sigma))\in\sigma^{-1+\delta} H^{\frac 32-(\epsilon+\delta)-}((-c_0, c_0);\mathcal{K}_{b,1}).
%\]
\end{proof}

Now we turn to the analysis of the regular part $L^-_b(\sigma)$.
\begin{prop}\label{prop:epsilonregofregpart}
Let $s,\ell,\epsilon$ be defined as in Proposition \ref{prop:epsilonregofcheckL-1}.
	 For $ \IM\sigma\geq0$ and $0<\abs{\sigma}\leq c_0$ where $c_0>0$ is a small constant, the regular part $L^-_b(\sigma)$ of the resolvent $\widehat{L_{b}}(\sigma)^{-1}$ defined in Corollary \ref{cor:singandreg} satisfies
	\begin{equation}\label{eq:epsilonregofregpart}
		\pa_\sigma^jL^-_b(\sigma):\eHb^{s-1,\ell+2}(\CX)\to\abs{\sigma}^{-\epsilon-j+1}\eHb^{s-\epsilon-j+1,\ell+\epsilon+1}(\CX),\quad j=1,2.
		\end{equation}
\end{prop}

\begin{proof}
According to Proposition \ref{prop:epsilonregofR} and the expression \eqref{eq:anotherexofP}, we see that $P(b,\sigma)$ has the form $P(b,\sigma)=\sigma^{-2}P^{2}(b)+\sigma^{-1} P^1(b)+\sigma^{-1}P^e(b,\sigma)$ 
where 
\[
P^2(b), P^1(b):\eHb^{s-1,\ell+2}(\CX)\to\rho C^\infty(\CX)+\eHb^{\infty, \frac12-}(\CX)
\]
and $P^e(b,\sigma):\eHb^{s-1,\ell+2}(\CX)\to\mathcal{K}_{b,1}$ is $\epsilon$-regular as proved in Theorem \ref{thm:holderregularityofsing}. We also write $R(b,\sigma)=R_{\mathrm{sing}}(b,\sigma)+R_{\mathrm{reg}}(b,\sigma)$ where
\[
R_{\mathrm{sing}}(b,\sigma)=\begin{pmatrix}
	0&0&0\\
	\sigma^{-1}\wt{R}_{10}(b,\sigma)&\sigma^{-2}\wt{R}^0_{11}(b)+\sigma^{-1}\wt{R}^e_{11}(b,\sigma)&\sigma^{-1}\wt{R}_{12}(b,\sigma)\\
	0&\sigma^{-1}\wt{R}^0_{21}(b)&\sigma^{-1}\wt{R}^0_{22}(b)
\end{pmatrix}
\]
and
\[
R_{\mathrm{reg}}(b,\sigma)=\begin{pmatrix}
	\wt{R}_{00}(b,\sigma)&\wt{R}_{01}(b,\sigma)&\wt{R}_{02}(b,\sigma)\\
	0&0&0\\
	\wt{R}_{20}(b,\sigma)&\wt{R}^e_{21}(b,\sigma)&\wt{R}^e_{22}(b,\sigma)
\end{pmatrix}.
\]
Using the calculation in \eqref{eq:exofcheckLon1}, \eqref{eq:exofcheckLon2} and \eqref{eq:exofPbsigma}, we find that 
\begin{align*}
	L^-_b(\sigma)&=\check{L}_b(\sigma)^{-1}R_{\mathrm{reg}}(b,\sigma)-\check{L}_b(\sigma)^{-1}\Big(\big(\pa_\sigma\widehat{L_{b}}(0)+\frac{\sigma}{2}\pa_\sigma^2\widehat{L_{b}}(0)\big)(P^1(b)+P^e(b,\sigma))+\frac{1}{2}\pa_\sigma^2\widehat{L_{b}}(0)P^2(b)\Big)\\
	&:=\Romanupper{1}+\Romanupper{2}.
	\end{align*}
For the term $\Romanupper{1}$, we compute
\begin{align*}
	\pa_\sigma\Romanupper{1}&=\pa_\sigma\check{L}_b(\sigma)^{-1}R_{\mathrm{reg}}(b,\sigma)+\check{L}_b(\sigma)^{-1}\pa_\sigma R_{\mathrm{reg}}(b,\sigma),\\
	\pa^2_\sigma\Romanupper{1}&=\pa^2_\sigma\check{L}_b(\sigma)^{-1}R_{\mathrm{reg}}(b,\sigma)+2\pa_\sigma\check{L}_b(\sigma)^{-1}\pa_\sigma R_{\mathrm{reg}}(b,\sigma)+\check{L}_b(\sigma)^{-1}\pa^2_\sigma R_{\mathrm{reg}}(b,\sigma).
\end{align*}
Then by proposition \ref{prop:epsilonregofcheckL-1} and \ref{prop:epsilonregofR}, we conclude that 
\[
	\pa_\sigma^j\Romanupper{1}:\eHb^{s-1,\ell+2}(\CX)\to\abs{\sigma}^{-\epsilon-j+1}\eHb^{s-\epsilon-j+1,\ell+\epsilon+1}(\CX),\quad j=1,2.
\]
As for the term $\Romanupper{2}$, we first note that $P^2(b), P^1(b), P^e(b,\sigma)$ map to $\rho C(\CX)+\eHb^{\infty, 1/2-}(\CX)$, and thus \[
\pa_\sigma^k\widehat{L_{b}}(0)P^2(b),\  \pa_\sigma^k\widehat{L_{b}}(0)P^1(b),\ \pa_\sigma^k\widehat{L_{b}}(0)P^e(b,\sigma):\eHb^{s-1,\ell+2}(\CX)\to\eHb^{\infty, \frac 32-}(\CX),\quad k=1,2.
\]
Next, we calculate
\begin{align*}
\pa_\sigma\Romanupper{2}&=-\pa_\sigma\check{L}_b(\sigma)^{-1}\Big(\big(\pa_\sigma\widehat{L_{b}}(0)+\frac{\sigma}{2}\pa_\sigma^2\widehat{L_{b}}(0)\big)(P^1(b)+P^e(b,\sigma))+\frac{1}{2}\pa_\sigma^2\widehat{L_{b}}(0)P^2(b)\Big)\\
&\quad-\check{L}_b(\sigma)^{-1}\Big(\big(\pa_\sigma\widehat{L_{b}}(0)+\frac{\sigma}{2}\pa_\sigma^2\widehat{L_{b}}(0)\big)\pa_\sigma P^e(b,\sigma)+\frac{1}{2}\pa_\sigma^2\widehat{L_{b}}(0) P^e(b,\sigma)\Big),\\
\pa_\sigma^2\Romanupper{2}&=-\pa^2_\sigma\check{L}_b(\sigma)^{-1}\Big(\big(\pa_\sigma\widehat{L_{b}}(0)+\frac{\sigma}{2}\pa_\sigma^2\widehat{L_{b}}(0)\big)(P^1(b)+P^e(b,\sigma))+\frac{1}{2}\pa_\sigma^2\widehat{L_{b}}(0)P^2(b)\Big)\\
&\quad-2\pa_\sigma\check{L}_b(\sigma)^{-1}\Big(\big(\pa_\sigma\widehat{L_{b}}(0)+\frac{\sigma}{2}\pa_\sigma^2\widehat{L_{b}}(0)\big)\pa_\sigma P^e(b,\sigma)+\frac{1}{2}\pa_\sigma^2\widehat{L_{b}}(0) P^e(b,\sigma)\Big)\\
&\quad-\check{L}_b(\sigma)^{-1}\Big(\big(\pa_\sigma\widehat{L_{b}}(0)+\frac{\sigma}{2}\pa_\sigma^2\widehat{L_{b}}(0)\big)\pa^2_\sigma P^e(b,\sigma)+\pa_\sigma^2\widehat{L_{b}}(0) \pa_\sigma P^e(b,\sigma)\Big).
\end{align*}
Therefore, using Proposition \ref{prop:epsilonregofcheckL-1} and the fact that $P^e(b,\sigma):\eHb^{s-1,\ell+2}\to\mathcal{K}_{b,1}$ is $\epsilon$-regular, we arrive at
\[
	\pa_\sigma^j\Romanupper{2}:\eHb^{s-1,\ell+2}(\CX)\to\abs{\sigma}^{-\epsilon-j+1}\eHb^{s-\epsilon-j+1,\ell+\epsilon+1}(\CX),\quad j=1,2.
\]
This completes the proof of the estimates \eqref{eq:epsilonregofregpart}.
\end{proof}

\begin{thm}\label{thm:holderregularity}
Let $s,\ell,\epsilon$ be defined as in Proposition \ref{prop:epsilonregofcheckL-1}. Then we have
\begin{equation}\label{eq:holderregularity}
L^-_b(\sigma)\in H^{\frac{3}{2}-\epsilon-}\Big((-c_0, c_0);\mathcal{L}\big(\eHb^{s-1,\ell+2}(\CX), \eHb^{s-\max\{\epsilon, \frac 12\},\ell+\epsilon-1}(\CX)\big)\Big)
\end{equation}
where $c_0>0$ is a sufficiently small constant.
\end{thm}

\begin{proof}
Using the estimates in Proposition \ref{prop:epsilonregofcheckL-1} and following the proof at the beginning of Theorem \ref{thm:holderregularityofsing}, we find that 
\begin{equation}\label{eq:estimate1}
L_b^-(\sigma)-L_b^-(0)\in\sigma^{\frac32-\epsilon-}H_{\bop}^2\Big([0, c_0);\mathcal{L}\big(\eHb^{s-1,\ell+2}(\CX), \eHb^{s-1-\epsilon,\ell+\epsilon-1}(\CX)\big)\Big).
\end{equation}
Again integrating $\pa_\sigma L^-_b(\sigma)$ from $\sigma\in(0, c_0)$ to $0$ and using the estimate \eqref{eq:epsilonregofregpart} for $k=1$ yield
\begin{equation}\label{eq:estimate2}
L_b^-(\sigma)-L_b^-(0)\in\sigma^{\frac32-\epsilon-}H_{\bop}^1\Big([0, c_0);\mathcal{L}\big(\eHb^{s-1,\ell+2}(\CX), \eHb^{s-\epsilon,\ell+\epsilon-1}(\CX)\big)\Big).
\end{equation}
Interpolating between \eqref{eq:estimate1} and \eqref{eq:estimate2} gives
\begin{equation}\label{eq:interpo1and2}
	L_b^-(\sigma)-L_b^-(0)\in\sigma^{\frac32-\epsilon-}H_{\bop}^{1+\theta}\Big([0, c_0);\mathcal{L}\big(\eHb^{s-1,\ell+2}(\CX), \eHb^{s-\theta-\epsilon,\ell+\epsilon-1}(\CX)\big)\Big), \quad0\leq \theta\leq 1.
\end{equation}
If $\epsilon\in(0,1/2]$, we take $\theta=1/2-\epsilon$, and thus $3/2-\epsilon=1+\theta$. If $\epsilon\in(1/2, 1)$, we take $\theta=0$. This gives
\begin{equation}
		L_b^-(\sigma)-L_b^-(0)\in\begin{cases}
			\sigma^{\frac32-\epsilon-}H_{\bop}^{\frac32-\epsilon}\Big([0, c_0);\mathcal{L}\big(\eHb^{s-1,\ell+2}(\CX), \eHb^{s-\frac12,\ell+\epsilon-1}(\CX)\big)\Big),\quad &0<\epsilon\leq \frac12,\\
				\sigma^{\frac32-\epsilon-}H_{\bop}^{1}\Big([0, c_0);\mathcal{L}\big(\eHb^{s-1,\ell+2}(\CX), \eHb^{s-\epsilon,\ell+\epsilon-1}(\CX)\big)\Big),\quad &\frac 12<\epsilon<1,
		\end{cases}
	\end{equation}
which implies
\begin{equation}
	L_b^-(\sigma)-L_b^-(0)\in\begin{cases}
	\sH^{\frac32-\epsilon}\Big([0, c_0);\mathcal{L}\big(\eHb^{s-1,\ell+2}(\CX), \eHb^{s-\frac12,\ell+\epsilon-1}(\CX)\big)\Big),\quad &0<\epsilon\leq \frac12,\\
	\sH^{\frac 32-\epsilon-}\Big([0, c_0);\mathcal{L}\big(\eHb^{s-1,\ell+2}(\CX), \eHb^{s-\epsilon,\ell+\epsilon-1}(\CX)\big)\Big),\quad &\frac 12<\epsilon<1.
	\end{cases}
\end{equation}
Similarly, we also have the above statement for the other half interval $(-c_0, 0]$. Since $	L_b^-(\sigma)-L_b^-(0)\to0$ as $\sigma\to0$ from both sides, by Lemma \ref{lem:technicallem}, we obtain
\[
L^-_b(\sigma)-L^-_b(0)\in H^{\frac{3}{2}-\epsilon-}\Big((-c_0, c_0);\mathcal{L}\big(\eHb^{s-1,\ell+2}(\CX), \eHb^{s-\max\{\epsilon, \frac 12\},\ell+\epsilon-1}(\CX)\big)\Big),
\]
and thus the conclusion \eqref{eq:holderregularity}.
\end{proof}

\subsection{Regularity for frequencies away from $0$.}\label{subsec:regularityofhighfre}
Now we discuss the regularity of $\widehat{L_{b}}(\sigma)^{-1}$ for $ \abs{\sigma}\geq c_0,\ 0\leq\IM\sigma\leq C$ for some $c_0,\, C>0$.

\begin{thm}\label{thm:regularityaway0}
	Let $\ell<-\frac12, s>3+m$ and $s+\ell>-\frac 12+m$ where $m\in\BN_0$. Then for $0\leq \IM\sigma\leq C, \abs{\sigma}\geq c_0$ with $c_0,C>0$, the operator
	\begin{equation}\label{eq:regularityaway0}
		\pa_\sigma^m \widehat{L_{b}}(\sigma)^{-1}: \bar{H}_{\bop,h}^{s,\ell+1}(\CX)\to h^{-m}\bar{H}_{\bop,h}^{s-m,\ell}(\CX)
	\end{equation}
where $h=\abs{\sigma}^{-1}$, is uniformly bounded.
\end{thm}
  
  \begin{proof}
  	The estimate \eqref{eq:regularityaway0} for the case $m=0$ follows from high energy estimate \ref{thm:highenergyBundle}. For $m\geq 1$, $\pa_\sigma^m\widehat{L_{b}}(\sigma)^{-1}$ is a linear combination of the operators of the type
  	\begin{align*}
  \Big(	\widehat{L_{b}}(\sigma)^{-1}\circ\pa^{m_1}_\sigma \widehat{L_{b}}(\sigma)\Big)\circ \cdots\circ\Big(\widehat{L_{b}}(\sigma)^{-1}\circ\pa^{m_k}_\sigma \widehat{L_{b}}(\sigma)\Big)\circ \widehat{L_{b}}(\sigma)^{-1}
  		\end{align*}
  	where $1\leq m_1, \cdots, m_k\leq 2$ and $m_1+\cdots+m_k=m$. Since 
  	\[
  	\pa_\sigma \widehat{L_{b}}(\sigma)\in \rho\mbox{Diff}_\bop^1+\rho^2C^\infty\subset h^{-1}\rho\mbox{Diff}_{\bop,h}^1\quad \mbox{and}\quad \pa_\sigma^2\widehat{L_{b}}(\sigma)\in\rho^2C^\infty,
  \]
  it follows that
  \begin{align*}
&\bar{H}_{\bop,h}^{s,\ell+1} \xrightarrow{\widehat{L_{b}}(\sigma)^{-1}}\bar{H}^{s,\ell}_{\bop,h}\xrightarrow{\pa^{m_k}_\sigma\widehat{L_{b}}(\sigma)}h^{m_k-2}\bar{H}^{s+m_k-2,\ell+m_k}_{\bop,h}\subset h^{m_k-2}\bar{H}^{s+m_k-2,\ell+1}_{\bop,h}
 \\
  &\xrightarrow {\widehat{L_{b}}(\sigma)^{-1}}h^{m_k-2}\bar{H}^{s+m_k-2,\ell}_{\bop,h} \xrightarrow{\cdots}\cdots\xrightarrow{\cdots}h^{m-m_1-2(k-1)}\bar{H}^{s+m-m_1-2(k-1), \ell}_{\bop,h}\\
 &\xrightarrow{\pa_\sigma^{m_1}\widehat{L_{b}}(0)}h^{m-2k}\bar{H}^{s+m-2k,\ell+m_1}_{\bop,h}\xrightarrow{\widehat{L_{b}}(\sigma)^{-1}}h^{m-2k}\bar{H}^{s+m-2k,\ell}_{\bop,h}.
	\end{align*}
Since $1\leq k\leq m$, we have $h^{m-2k}\bar{H}^{s+m-2k,\ell+m-k}_{\bop,h}\subset h^{-m}\bar{H}^{s-m,\ell}_{\bop,h}$. This finishes the proof.
\end{proof}

\subsection{Conormal regularity}\label{subsec:conormalregularity}
Finally, we consider the conormal regularity of $L^-_b(\sigma)$ at $\sigma=0$. Specifically, after applying any number of derivatives $\sigma\pa_\sigma$ to $L^-_b(\sigma)$, it satisfies the same properties as $L^-_b(\sigma)$.

Recall that 
\[
L^-_b(\sigma)=\check{L}_b(\sigma)^{-1}R_{\mathrm{reg}}(b,\sigma)-\check{L}_b(\sigma)^{-1}\Big(\big(\pa_\sigma\widehat{L_{b}}(0)+\frac{\sigma}{2}\pa_\sigma^2\widehat{L_{b}}(0)\big)(P^1(b)+P^e(b,\sigma))+\frac{1}{2}\pa_\sigma^2\widehat{L_{b}}(0)P^2(b)\Big),
\]
so in order to study the conormal regularity, we need to analyze that of $\check{L}_b(\sigma)^{-1}$, $R_{\mathrm{reg}}(b,\sigma)$ and $P^e(b,\sigma)$.

We first discuss the conormal regularity of $\check{L}_b(\sigma)^{-1}$ and prove a analogous result of Proposition \ref{prop:epsilonregofcheckL-1} for $(\sigma\pa_\sigma)^m\check{L}_b(\sigma)^{-1}$.

\begin{prop}\label{prop:conormalregofcheckL-1}
Let $-\frac 32<\ell<-\frac 12$, $0<\epsilon<1$ with $-\frac 12<\ell+\epsilon<\frac 12$, and $s-\epsilon>4+m$ where $m\in\BN$. Define
\[
\check{\mathcal{L}}^{(m)}_b(\sigma):=(\sigma\pa_\sigma)^m\check{L}_b(\sigma)^{-1}.
\]
For $0<\abs{\sigma}\leq c_0, \IM\sigma\geq 0$ with $c_0$ small, we have
\begin{align}\label{eq:zeroderiofcocheckL-1}
	&\check{\mathcal{L}}^{(m)}_b(\sigma):\eHb^{s-1,\ell+2}(\CX)\to \eHb^{s-m, \ell}(\CX),\\\label{eq:firstderiofcocheckL-1}
	&\pa_\sigma\check{\mathcal{L}}^{(m)}_b(\sigma): \eHb^{s-1,\ell+2}(\CX)\to \abs{\sigma}^{-\epsilon}\eHb^{s-m-\epsilon, \ell+\epsilon-1}(\CX),\\\label{eq:secondderiofcocheckL-1}
	&\pa^2_\sigma\check{\mathcal{L}}^{(m)}_b(\sigma): \eHb^{s-1,\ell+2}(\CX)\to \abs{\sigma}^{-\epsilon-1}\eHb^{s-m-1-\epsilon, \ell+\epsilon-1}(\CX).
\end{align}
\end{prop}

\begin{proof}
Since \[
\pa_\sigma^2\check{\mathcal{L}}^{(m)}_b(\sigma)=\sigma^{-1}\pa_\sigma(\sigma\pa_\sigma-1)\check{\mathcal{L}}^{(m)}_b(\sigma)=\sigma^{-1}\pa_\sigma\check{\mathcal{L}}^{(m+1)}_b(\sigma)-\sigma^{-1}\pa_\sigma\check{\mathcal{L}}^{(m)}_b(\sigma),\]
it suffices to prove the statements \eqref{eq:zeroderiofcocheckL-1} and \eqref{eq:firstderiofcocheckL-1}, and then the conclusion \eqref{eq:secondderiofcocheckL-1} follows directly.

Now we prove \eqref{eq:zeroderiofcocheckL-1} and \eqref{eq:firstderiofcocheckL-1} by induction on $m$. For $m=1$, we calculate 
\[
\check{\mathcal{L}}_b^{(1)}=\sigma\pa_\sigma\check{L}_b(\sigma)^{-1}=-\check{L}_b(\sigma)^{-1}(\sigma\pa_\sigma\check{L}_b(\sigma))\check{L}_b(\sigma)^{-1}
\]
where 
\begin{equation}\label{eq:deriofcocheckL}
	\sigma\pa_\sigma\check{L}_b(\sigma)=\check{L}_b(\sigma)-\check{L}_b(0)+\frac{\sigma^2}{2}\pa^2_\sigma\check{L}_b(0)\in\check{L}_b(\sigma)+\rho^2\mbox{Diff}_\bop^2+\sigma^2\rho^2C^\infty.
\end{equation}
As a consequence,
\[
\check{\mathcal{L}}_b^{(1)}=-\check{L}_b(\sigma)^{-1}+\check{L}_b(\sigma)^{-1}\big(\rho^2\mbox{Diff}_\bop^2+\sigma^2\rho^2C^\infty\big)\check{L}_b(\sigma)^{-1}
\]
We then compute
\[
\eHb^{s-1,\ell+2}(\CX)\xrightarrow{\check{L}_b(\sigma)^{-1}}\eHb^{s,\ell}(\CX)\xrightarrow{\rho^2\mbox{Diff}_\bop^2+\sigma^2\rho^2C^\infty}\eHb^{s-2,\ell+2}(\CX)\xrightarrow{\check{L}_b(\sigma)^{-1}}\eHb^{s-1,\ell}(\CX).
\]
This proves \eqref{eq:zeroderiofcocheckL-1} for $m=1$. As for the first derivative $\pa_\sigma\check{\mathcal{L}}_b^{(1)}(\sigma)$, we calculate
\begin{align*}
\pa_\sigma\check{\mathcal{L}}_b^{(1)}(\sigma)&=-\pa_\sigma\check{L}_b(\sigma)^{-1}+\pa_\sigma\check{L}_b(\sigma)^{-1}\big(\rho^2\mbox{Diff}_b^2+\sigma^2\rho^2C^\infty\big)\check{L}_b(\sigma)^{-1}+\check{L}_b(\sigma)^{-1}\big(\sigma\rho^2C^\infty\big)\check{L}_b(\sigma)^{-1}\\
&\quad+\check{L}_b(\sigma)^{-1}\big(\rho^2\mbox{Diff}_\bop^2+\sigma^2\rho^2C^\infty\big)\pa_\sigma\check{L}_b(\sigma)^{-1}:=\Romanupper{1}+\Romanupper{2}+\Romanupper{3}+\Romanupper{4}.
\end{align*}
Using Proposition \ref{prop:epsilonregofcheckL-1}, we find that 
\begin{align*}
	\Romanupper{1}:&\eHb^{s-1,\ell+2}(\CX)\to\abs{\sigma}^{-\epsilon}\eHb^{s-\epsilon, \ell+\epsilon+1}(\CX),\\
	\Romanupper{2}:&\eHb^{s-1,\ell+2}(\CX)\xrightarrow{\check{L}_b(\sigma)^{-1}}\eHb^{s,\ell}(\CX)\xrightarrow{\rho^2\mbox{Diff}_\bop^2+\sigma^2\rho^2C^\infty}\eHb^{s-2,\ell+2}(\CX)\xrightarrow{\pa_\sigma\check{L}_b(\sigma)^{-1}}\abs{\sigma}^{-\epsilon}\eHb^{s-1-\epsilon, \ell+\epsilon-1}(\CX),\\
\Romanupper{3}:&\eHb^{s-1,\ell+2}(\CX)\xrightarrow{\check{L}_b(\sigma)^{-1}}\eHb^{s,\ell}(\CX)\xrightarrow{\sigma\rho^2C^\infty}\eHb^{s,\ell+2}(\CX)\xrightarrow{\check{L}_b(\sigma)^{-1}}\eHb^{s-1, \ell}(\CX)\subset \abs{\sigma}^{-\epsilon}\eHb^{s-1-\epsilon, \ell+\epsilon-1}(\CX),\\
\Romanupper{4}:&\eHb^{s-1,\ell+2}(\CX)\xrightarrow{\pa_\sigma\check{L}_b(\sigma)^{-1}}\abs{\sigma}^{-\epsilon}\eHb^{s-\epsilon,\ell+\epsilon-1}(\CX)\xrightarrow{\rho^2\mbox{Diff}_\bop^2+\sigma^2\rho^2C^\infty}\abs{\sigma}^{-\epsilon}\eHb^{s-\epsilon-2,\ell+\epsilon+1}(\CX)\\
&\xrightarrow{\check{L}_b(\sigma)^{-1}}\abs{\sigma}^{-\epsilon}\eHb^{s-1-\epsilon, \ell+\epsilon-1}(\CX)
\end{align*}
where we use $1/2<\ell+\epsilon+1<3/2$ in the last step of the analysis of $\Romanupper{4}$. This proves \eqref{eq:firstderiofcocheckL-1} for $m=1$.

Suppose that \eqref{eq:zeroderiofcocheckL-1} and \eqref{eq:firstderiofcocheckL-1} hold for $k\leq m$.
Then for $m+1$, by a direct calculation, we obtain that $\check{\mathcal{L}}_b^{(m+1)}(\sigma)$ is a linear combination of the operators of the type
\[
L_1=\check{\mathcal{L}}_b^{(m)}(\sigma),\quad L_2:=\check{\mathcal{L}}_b^{(m_1)}(\sigma)\circ(\rho^2\mbox{Diff}_\bop^2)\circ\check{\mathcal{L}}_b^{(m_2)}(\sigma)
\]
where $m_1, m_2\geq 0$ and $m_1+m_2\leq m$. Then by the induction hypothesis on $k\leq m$, it follows that
\begin{align*}
L_1:&\eHb^{s-1,\ell+2}(\CX)\xrightarrow{\check{\mathcal{L}}_b^{(m)}(\sigma)}\eHb^{s-m,\ell}(\CX)\subset\eHb^{s-m-1,\ell}(\CX),\\
L_2:&\eHb^{s-1,\ell+2}(\CX)\xrightarrow{\check{\mathcal{L}}_b^{(m_1)}(\sigma)}\eHb^{s-m_1,\ell}(\CX)\xrightarrow{\rho^2\mbox{Diff}_\bop^2}\eHb^{s-m_1-2,\ell+2}(\CX)\\
&\xrightarrow{\check{\mathcal{L}}_b^{(m_1)}(\sigma)}\eHb^{s-m_1-m_2-1,\ell}(\CX)=\eHb^{s-m-1,\ell}(\CX).
\end{align*}
This proves \eqref{eq:zeroderiofcocheckL-1} for $m+1$. As for the first derivative $\pa_\sigma\check{\mathcal{L}}_b^{(m+1)}$, it is a linear combination of the operators of the following types
\begin{align*}
D_1=\pa_\sigma\check{\mathcal{L}}_b^{(m)}(\sigma),\quad D_2:=\pa_\sigma\check{\mathcal{L}}_b^{(m_1)}(\sigma)\circ(\rho^2\mbox{Diff}_\bop^2)\circ\check{\mathcal{L}}_b^{(m_2)}(\sigma),\quad D_3:=\check{\mathcal{L}}_b^{(m_1)}(\sigma)\circ(\rho^2\mbox{Diff}_\bop^2)\circ\pa_\sigma\check{\mathcal{L}}_b^{(m_2)}(\sigma)
\end{align*}
where $m_1, m_2\geq 0$ and $m_1+m_2\leq m$. Then we have
\begin{align*}
	D_1:&\eHb^{s-1,\ell+2}(\CX)\xrightarrow{\pa_\sigma\check{\mathcal{L}}_b^{(m)}(\sigma)}\abs{\sigma}^{-\epsilon}\eHb^{s-m-\epsilon,\ell+\epsilon-1}(\CX)\subset\abs{\sigma}^{-\epsilon}\eHb^{s-m-1\epsilon,\ell+\epsilon-1}(\CX),\\
		D_2:&\eHb^{s-1,\ell+2}(\CX)\xrightarrow{\check{\mathcal{L}}_b^{(m_2)}(\sigma)}\eHb^{s-m_2,\ell}(\CX)\xrightarrow{\rho^2\mbox{Diff}_\bop^2}\eHb^{s-m_2-2,\ell+2}(\CX)\\
	&\xrightarrow{\pa_\sigma\check{\mathcal{L}}_b^{(m_1)}(\sigma)}\abs{\sigma}^{-\epsilon}\eHb^{s-m_1-1-m_2-\epsilon,\ell+\epsilon-1}(\CX)\subset\abs{\sigma}^{-\epsilon}\eHb^{s-m-1-\epsilon,\ell+\epsilon-1}(\CX),\\
	D_3:&\eHb^{s-1,\ell+2}(\CX)\xrightarrow{\pa_\sigma\check{\mathcal{L}}_b^{(m_2)}(\sigma)}\abs{\sigma}^{-\epsilon}\eHb^{s-m_2-\epsilon,\ell+\epsilon-1}(\CX)\xrightarrow{\rho^2\mbox{Diff}_\bop^2}\abs{\sigma}^{-\epsilon}\eHb^{s-m_2-\epsilon-2,\ell+\epsilon+1}(\CX)\\
	&\xrightarrow{\check{\mathcal{L}}_b^{(m_1)}(\sigma)}\abs{\sigma}^{-\epsilon}\eHb^{s-m_1-m_2-\epsilon-1,\ell+\epsilon-1}(\CX)\subset\abs{\sigma}^{-\epsilon}\eHb^{s-m-1-\epsilon,\ell+\epsilon-1}(\CX)
\end{align*}
where we use $1/2<\ell+\epsilon+1<3/2$ in the third step of the analysis of $D_3$. This proves \eqref{eq:firstderiofcocheckL-1} for $m+1$.

This finishes the proof of the proposition.
\end{proof}	

With the conormal regularity of $\check{L}_b(\sigma)^{-1}$ with our disposal, we are ready to discuss that of the entries $\wt{L}_{ij}$ of the operator $\widehat{L_{b}}(\sigma)\check{L}_b(\sigma)^{-1}$.

Recall the expression \eqref{eq:detailedexofLL-1} of $\widehat{L_{b}}(\sigma)\check{L}(\sigma)^{-1}$
\[
\begin{split}
	&\widehat{L_{b}}(\sigma)\check{L}_b(\sigma)^{-1}=\begin{pmatrix}
		L_{00}(b,\sigma)& \sigma^2\wt{L}_{01}(b,\sigma)&\sigma\wt{L}_{02}(b,\sigma)\\
		\sigma\wt{L}_{10}(b,\sigma)& \sigma^2\wt{L}_{11}(b,\sigma)&\sigma^2\wt{L}_{12}(b,\sigma)\\
		\sigma\wt{L}_{20}(b,\sigma)& \sigma^2\wt{L}_{21}(b,\sigma)&\sigma\wt{L}_{22}(b,\sigma)\\
	\end{pmatrix}\\
	&\quad=\begin{pmatrix}
		L_{00}(b,\sigma)& \sigma^2\wt{L}_{01}(b,\sigma)&\sigma\wt{L}_{02}(b,\sigma)\\
		\sigma\big(\wt{L}^0_{10}(b)+\sigma\wt{L}^e_{10}(b,\sigma)\big)&\sigma^2\big(\wt{L}^0_{11}(b)+\sigma\wt{L}^1_{11}(b)+\sigma^2\wt{L}^e_{11}(b,\sigma)\big)&\sigma^2\big(\wt{L}^0_{12}(b)+\sigma\wt{L}^e_{12}(b,\sigma)\big)\\
		\sigma\wt{L}_{20}(b,\sigma)&\sigma^2\big(\wt{L}^0_{21}(b)+\sigma\wt{L}^e_{21}(b)\big)&\sigma\big(\wt{L}^0_{22}(b)+\sigma\wt{L}^e_{22}(b,\sigma)\big).
	\end{pmatrix}
\end{split}
\]

\begin{prop}\label{prop:conormalregofLL-1}
	Let $s,\ell,\epsilon,m$ be defined as in Proposition \ref{prop:conormalregofcheckL-1}. Then  $(\sigma\pa_\sigma)^m\wt{L}_{ij}$ satisfy the following properties.
	\begin{itemize}
		\item [(1)] $(\sigma\pa_\sigma)^mL_{00}, (\sigma\pa_\sigma)^m\wt{L}_{01}, (\sigma\pa_\sigma)^m\wt{L}_{02}, (\sigma\pa_\sigma)^m\wt{L}_{20}$ are $\epsilon$-regular at $\sigma=0$.
		\item[(2)] $(\sigma\pa_\sigma)^m\wt{L}_{10}, (\sigma\pa_\sigma)^m\wt{L}_{12}, (\sigma\pa_\sigma)^m\wt{L}_{21}, (\sigma\pa_\sigma)^m\wt{L}_{22}$ have an $\epsilon$-regular expansion up to order one, that is, $(\sigma\pa_\sigma)^m\wt{L}^e_{10}, (\sigma\pa_\sigma)^m\wt{L}^e_{12}, (\sigma\pa_\sigma)^m\wt{L}^e_{21}, (\sigma\pa_\sigma)^m\wt{L}^e_{22}$ are $\epsilon$-regular at $\sigma=0$.
		\item[(3)]$(\sigma\pa_\sigma)^m\wt{L}_{11}$ has an $\epsilon$-regular expansion up to order two, that is, $(\sigma\pa_\sigma)^m\wt{L}^e_{11}$ is $\epsilon$-regular at $\sigma=0$.
	\end{itemize}
\end{prop}
\begin{proof}
		The proof is analogous to that of Proposition \ref{prop:epsilonregofLL-1}, now by using Proposition \ref{prop:conormalregofcheckL-1} instead of Proposition \ref{prop:epsilonregofcheckL-1}. Here, we only demonstrate the proof of $(\sigma\pa_\sigma)^m\wt{L}_{10}$ in detail because the proof of the other entries follows in a similar manner.
		
		Recall the calculation in \eqref{eq:calofL10}
		\[
		\begin{split}
			&(\sigma\pa_\sigma)^m\wt{L}^e_{10}(\tilde{g}_0, \tilde{A}_0),\ (\dg^*_1, \dA^*_1)\rangle\\
			&\quad=-\sum_{j+k=m}\Big\langle (\tilde{g}_0, \tilde{A}_0), i\big(\check{\mathcal{L}}_b^{(k)}(\sigma)\big)^*\Big((\sigma\pa_\sigma)^j\big(\pa_\sigma\widehat{L_{b}}(0)+\frac{\sigma}{2}\pa_\sigma^2\widehat{L_{b}}(0)\big)\Big)^*\big(I-(\check{L}_b(0)^{-1})^*V_b^*\big)(\check{g}^*_2,\check{A}^*_2)\Big\rangle\\
			&\quad\quad+\Big\langle(\tilde{g}_0, \tilde{A}_0),\frac{1}{2}(\check{\mathcal{L}}_b^{(m)}(\sigma))^*(\pa^2_\sigma\widehat{L_{b}}(0))^*(\dg^*_2,\dA^*_2)\Big\rangle
		\end{split}
		\]
		where
		\[
		\Big((\sigma\pa_\sigma)^j\big(\pa_\sigma\widehat{L_{b}}(0)+\frac{\sigma}{2}\pa_\sigma^2\widehat{L_{b}}(0)\big)\Big)^*\big(I-(\check{L}_b(0)^{-1})^*V_b^*\big)(\check{g}^*_2,\check{A}^*_2),\quad(\pa^2_\sigma\widehat{L_{b}}(0))^*(\dg^*_2,\dA^*_2)\in\sHb^{-5/2-C(a,\gamma),\,3/2-}(\CX).
		\]
		Since $(\tilde{g}_0, \tilde{A}_0)\in\eHb^{s-1,\ell+2}(\CX)$ with $s-m-\epsilon>4$, and by Proposition \ref{prop:conormalregofcheckL-1}
		\begin{gather*}
		(\check{\mathcal{L}}_b^{(k)}(\sigma))^*\sHb^{-5/2-C(a,\gamma), \,3/2-}(\CX)\in\sHb^{-3/2-k-C(a,\gamma), -1/2-}(\CX),\quad 0\leq k\leq m, \ k\in\BN_0,\\	(\pa^j_\sigma\check{\mathcal{L}}_b^{(k)}(\sigma))^*\sHb^{-5/2-C(a,\gamma),\, 3/2-}(\CX)\in\abs{\sigma}^{-\epsilon-j+1}\sHb^{-s+1, -\ell-2}(\CX),\quad j=1,2\quad\mbox{and}\quad 0\leq k\leq m, \ k\in\BN_0
		\end{gather*}
		where we use the facts that $-5/2-C(a,\gamma)>-3>-s+k+\epsilon+1$ for $0\leq k\leq m$ and $-\ell-\epsilon+1<3/2$, the pairings above (and their up to second order derivatives) are well-defined. This proves the $\epsilon$-regularity of $(\sigma\pa_\sigma)\wt{L}^e_{10}$.
\end{proof}

Next, we turn to discussing the conormal regularity of $R(b,\sigma)$. To this end, we need the following lemma, which is an analogy of Lemma \ref{lem:epsilonregofinverse}.

\begin{lem}\label{lem:conormalregofinverse}
	Let $s,\ell,\epsilon,m$ be defined as in Proposition \ref{prop:conormalregofcheckL-1}. Suppose that the operator $A(\sigma):\wt{\mathcal{K}}_i\to\mathcal{R}_j$ with $i,j=0,1,2$ has an $\epsilon$-regular expansion up to order one, i.e.,
	\[
	A(\sigma)=A^0+\sigma A^e(\sigma)
	\]
	where $A^0:\wt{\mathcal{K}}_i\to\mathcal{R}_j$ is invertible and independent of $\sigma$, and $(\sigma\pa_\sigma)^kA^e(\sigma):\wt{\mathcal{K}}_i\to\mathcal{R}_j$ is $\epsilon$-regular for any $0\leq k\leq m, k\in\BN_0$. Then for $\abs{\sigma}$ sufficiently small, $A(\sigma)$ is invertible with inverse $B(\sigma)=A(\sigma)^{-1}$, and $(\sigma\pa_\sigma)^kB(\sigma)$ has an $\epsilon$-regular expansion up to order one, that is, 
	\[
	(\sigma\pa_\sigma)^kB(\sigma)=B^0_k+\sigma B_k^e(\sigma), \quad 0\leq k\leq m,\  k\in\BN_0
	\]
	where $B^0_k:\mathcal{R}_j\to\wt{\mathcal{K}}_i$ is independent of $\sigma$, and $B_k^e(\sigma):\mathcal{R}_j\to\wt{\mathcal{K}}_i$ is $\epsilon$-regular.
	
	Similarly, if $A(\sigma)$ has uniformly bounded inverse $B(\sigma)$, and $(\sigma\pa_\sigma)^kA(\sigma)$ is $\epsilon$-regular for any $0\leq k\leq m, k\in\BN_0$, then $(\sigma\pa_\sigma)^kB(\sigma)$ is $\epsilon$-regular as well.
\end{lem}

\begin{proof}
	%According to Lemma \ref{lem:epsilonregofinverse}, it suffices to prove that $(\sigma\pa_\sigma)^kB(\sigma)$ has an $\epsilon$-regular expansion up to order one for any $1\leq k\leq m, k\in\BN_0$. 
	As established in the proof of Lemma \ref{lem:epsilonregofinverse}, we have
	\begin{align*}
		A(\sigma)^{-1}&=\Big(I+\sum_{k=1}^\infty(-1)^k\big(\sigma (A^0)^{-1}A^e(\sigma)\big)^k\Big)(A^0)^{-1}\\
		&=(A^0)^{-1}-\sigma\Big((A^0)^{-1}A^e(\sigma)\sum_{k=0}^\infty(-1)^k\big(\sigma (A^0)^{-1}A^e(\sigma)\big)^k(A^0)^{-1}\Big):=B^0+\sigma B^e(\sigma).
	\end{align*}
For $1\leq k\leq m$, we have
\[
(\sigma\pa_\sigma)^kB(\sigma)=\sigma\sum_{j=0}^k\binom{k}{j}(\sigma\pa_\sigma)^jB^e(\sigma).
\]
A direct calculation implies that  $(\sigma\pa_\sigma)^jB^e(\sigma)$ is a sum of the operators of the type $\sigma^{-1}L\circ(A^0)^{-1}$, where $L$ is a composition of the terms of the form $\sigma(A^0)^{-1}(\sigma\pa_\sigma)^lA^e(\sigma)$ with $0\leq l\leq j$. 
%We also note that there are at most $j$ components $\sigma(A^0)^{-1}(\sigma\pa_\sigma)^lA^e(\sigma)$ with $1\leq l\leq j$ in each composition $L$. 
Then proceeding as in the proof of Lemma \ref{lem:epsilonregofinverse}, we conclude that $(\sigma\pa_\sigma)^jB^e(\sigma)$ with $0\leq j\leq k$ is $\epsilon$-regular, and thus $(\sigma\pa_\sigma)^kB^e(\sigma)$ with $1\leq k\leq m$ has an $\epsilon$-regular expansion up to order one.

Since $(\sigma\pa_\sigma)^mB(\sigma)$ is a linear combination of the operators of the form
\[
(\sigma\pa_\sigma)^{m_1}B(\sigma)\circ(\sigma\pa_\sigma)^{m_2}A(\sigma)\circ(\sigma\pa_\sigma)^{m_3}B(\sigma)
\]
where $1\leq m_2\leq m, 0\leq m_1, m_2\leq m-1$ and $m_1+m_2+m_3=m$, it follows that the second statement follows by an induction argument on $m$.
\end{proof}

Recall the expressions \eqref{eq:exofinverseR} and \eqref{eq:expansionofcomofR} of $R_{b,\sigma}=(\widehat{L_{b}}(\sigma)\check{L}_b(\sigma)^{-1})^{-1}$
\[
\begin{split}
	R(b,\sigma)&=(\widehat{L_{b}}(\sigma)\check{L}_b(\sigma)^{-1})^{-1}=
	\begin{pmatrix}
		\wt{R}_{00}(b,\sigma)&\wt{R}_{01}(b,\sigma)&\wt{R}_{02}(b,\sigma)\\
		\sigma^{-1}\wt{R}_{10}(b,\sigma)&\sigma^{-2}\wt{R}_{11}(b,\sigma)&\sigma^{-1}\wt{R}_{12}(b,\sigma)\\
		\wt{R}_{20}(b,\sigma)&\sigma^{-1}\wt{R}_{21}(b,\sigma)&\sigma^{-1}\wt{R}_{22}(b,\sigma)
	\end{pmatrix}\\
	&=
	\begin{pmatrix}
		\wt{R}_{00}(b,\sigma)&\wt{R}_{01}(b,\sigma)&\wt{R}_{02}(b,\sigma)\\
		\sigma^{-1}\wt{R}_{10}(b,\sigma)&\sigma^{-2}\big(\wt{R}^0_{11}(b)+\sigma\wt{R}^e_{11}(b,\sigma)\big)&\sigma^{-1}\wt{R}_{12}(b,\sigma)\\
		\wt{R}_{20}(b,\sigma)&\sigma^{-1}\big(\wt{R}^0_{21}(b)+\sigma\wt{R}^e_{21}(b,\sigma)\big)&\sigma^{-1}\big(\wt{R}^0_{22}(b)+\sigma\wt{R}^e_{22}(b,\sigma)\big)
	\end{pmatrix}
\end{split}
\]
where
\begin{equation}
	\begin{split}
		\wt{R}_{11}=\big(\wt{L}^\sharp_{11}-\sigma\wt{L}^\sharp_{12}(\wt{L}^\flat_{22})^{-1}\wt{L}^\sharp_{21}\big)^{-1},\quad \wt{R}_{10}=\wt{R}_{11}\big(-\wt{L}_{10}+\sigma\wt{L}_{12}^\sharp(\wt{L}^\flat_{22})^{-1}\wt{L}_{20}\big)L_{00}^{-1},\\
		\wt{R}_{12}=-\wt{R}_{11}\wt{L}_{12}^\sharp(\wt{L}^\flat_{22})^{-1},\quad \wt{R}_{21}=-(\wt{L}^\flat_{22})^{-1}\wt{L}_{21}^\sharp\wt{R}_{11},\quad\wt{R}_{22}=(\wt{L}^\flat_{22})^{-1}-\sigma(\wt{L}^\flat_{22})^{-1}\wt{L}_{21}^\sharp \wt{R}_{12},\\
		\wt{R}_{20}=-(\wt{L}^\flat_{22})^{-1}\big(\wt{L}_{20}L_{00}^{-1}+\wt{L}_{21}^\sharp\wt{R}_{10}\big),\quad \wt{R}_{01}=-L_{00}^{-1}\big(\wt{L}_{01}\wt{R}_{11}+\wt{L}_{02}\wt{R}_{21}\big),\\
		\wt{R}_{02}=-L_{00}^{-1}\big(\sigma\wt{L}_{01}\wt{R}_{12}+\wt{L}_{02}\wt{R}_{22}\big),\quad \wt{R}_{00}=L_{00}^{-1}\big(I-\sigma\wt{L}_{01}\wt{R}_{10}-\sigma\wt{L}_{02}\wt{R}_{20}\big)
	\end{split}
\end{equation}
with
\begin{equation}
	\begin{split}
		&\wt{L}^\sharp_{i1}=\wt{L}_{i1}-\sigma\wt{L}_{i0}L_{00}^{-1}\wt{L}_{01}\quad\mbox{for}\quad i=1,2, \quad \wt{L}^\sharp_{12}=\wt{L}_{12}-\wt{L}_{10}L_{00}^{-1}\wt{L}_{02},\\
		&\qquad\qquad\qquad\qquad	\wt{L}_{22}^\flat=\wt{L}_{22}-\sigma\wt{L}_{20}L_{00}^{-1}\wt{L}_{02}.
	\end{split}
\end{equation}

\begin{prop}\label{prop:conormalregofR}
	Let $s,\ell,\epsilon,m$ be defined as in Proposition \ref{prop:conormalregofcheckL-1}. Then $(\sigma\pa_\sigma)^m\wt{R}_{ij}$ satisfy the following properties.
	\begin{itemize}
		\item [(1)] $(\sigma\pa_\sigma)^m\wt{R}_{00}, (\sigma\pa_\sigma)^m\wt{R}_{01}, (\sigma\pa_\sigma)^m\wt{R}_{02}, (\sigma\pa_\sigma)^m\wt{R}_{10}, (\sigma\pa_\sigma)^m\wt{R}_{12}, (\sigma\pa_\sigma)^m\wt{R}_{20}$ are $\epsilon$-regular at $\sigma=0$.
		\item[(2)] $(\sigma\pa_\sigma)^m\wt{R}_{11}, (\sigma\pa_\sigma)^m\wt{R}_{21}, (\sigma\pa_\sigma)^m\wt{R}_{22}$ have an $\epsilon$-regular expansion up to order one, that is, $(\sigma\pa_\sigma)^m\wt{R}^e_{11}$, $(\sigma\pa_\sigma)^m\wt{R}^e_{21}, (\sigma\pa_\sigma)^m\wt{R}^e_{22}$ are $\epsilon$-regular at $\sigma=0$.
	\end{itemize}
\end{prop}

\begin{proof}
	Applying Proposition \ref{prop:conormalregofLL-1} and Lemma \ref{lem:conormalregofinverse}, we conclude that for $0\leq k\leq m$, $(\sigma\pa_\sigma)^kL_{00}^{-1}$ is $\epsilon$-regular. Then by Proposition \ref{prop:conormalregofLL-1}, we find that $(\sigma\pa_\sigma)^k\wt{L}^\sharp_{11}, (\sigma\pa_\sigma)^k\wt{L}^\sharp_{21}, (\sigma\pa_\sigma)^k\wt{L}^\flat_{22}$ (and thus $(\sigma\pa_\sigma)^k(\wt{L}^\flat_{22})^{-1}$ by Lemma \ref{lem:conormalregofinverse}) have an $\epsilon$-regular expansion up to order one, while $(\sigma\pa_\sigma)^k\wt{L}^\sharp_{12}$ is $\epsilon$-regular. Therefore, we see that 
	\[
	(\sigma\pa_\sigma)^k\Big(\wt{L}^\sharp_{11}-\sigma\wt{L}^\sharp_{12}(\wt{L}^\flat_{22})^{-1}\wt{L}^\sharp_{21}\Big)
	\]
	has an $\epsilon$-regular expansion up to order one whose coefficient of $\sigma^0$ is $(\sigma\pa_\sigma)^k\wt{L}_{11}^0(b)$. Then the $\epsilon$-regular expansion property of $(\sigma\pa_\sigma)^m\wt{R}_{11}$ follows from Lemma \ref{lem:conormalregofinverse}. Similarly, the $\epsilon$-regular expansion property of $(\sigma\pa_\sigma)^m\wt{R}_{21},(\sigma\pa_\sigma)^m\wt{R}_{22}$ follows from that of $(\sigma\pa_\sigma)^k\wt{R}_{11}, (\sigma\pa_\sigma)^k(\wt{L}^\flat_{22})^{-1},(\sigma\pa_\sigma)^m\wt{L}^\sharp_{21}$ and the above expressions of $\wt{R}_{ij}$. Finally, the $\epsilon$-regularity of the remaining components can be obtained by using the above expressions of $\wt{R}_{ij}$ and Proposition \ref{prop:conormalregofLL-1}.
\end{proof}

\begin{prop}
	Let $L^-_b(\sigma)$ be the regular part of $\widehat{L_{b}}(\sigma)^{-1}$ defined in Corollary \ref{cor:singandreg} and define
	\[
	\mathcal{L}^{-(m)}_b(\sigma):=(\sigma\pa_\sigma)^mL^-_b(\sigma).
	\]
	Let $-\frac 32<\ell<-\frac12, 0<\epsilon<1, -\frac12<\ell+\epsilon< \frac12$ and $s-\epsilon>4+m$ where $m\in\BN_0$. For $ \IM\sigma\geq0$ and $0<\abs{\sigma}\leq c_0$ where $c_0>0$ is a small constant, $\mathcal{L}^{-(m)}_b(\sigma)$ satisfies
	\begin{equation}\label{eq:conormalbound}
		\mathcal{L}^{-(m)}_b(\sigma):\eHb^{s-1,\ell+2}(\CX)\to\eHb^{s-m,\ell+\epsilon-1}(\CX)
	\end{equation}
	and
	\begin{equation}\label{eq:conormalholderreg}
		\pa_\sigma^j\mathcal{L}^{-(m)}_b(\sigma):\eHb^{s-1,\ell+2}(\CX)\to\abs{\sigma}^{-\epsilon-j+1}\eHb^{s-m-\epsilon-j+1,\ell+\epsilon-1}(\CX),\quad j=1,2.
	\end{equation}
\end{prop}

\begin{proof}
	Recall that 
	\[
	L^-_b(\sigma)=\check{L}_b(\sigma)^{-1}R_{\mathrm{reg}}(b,\sigma)-\check{L}_b(\sigma)^{-1}\Big(\big(\pa_\sigma\widehat{L_{b}}(0)+\frac{\sigma}{2}\pa_\sigma^2\widehat{L_{b}}(0)\big)(P^1(b)+P^e(b,\sigma))+\frac{1}{2}\pa_\sigma^2\widehat{L_{b}}(0)P^2(b)\Big)
	\]
	where $
	P^2(b), P^1(b):\eHb^{s-1,\ell+2}(\CX)\to\rho C^\infty(\CX)+\eHb^{\infty, \frac12-}(\CX)$,  \[(\sigma\pa_\sigma)^kP^e(b,\sigma):\eHb^{s-1,\ell+2}(\CX)\to\mathcal{K}_{b,1}
	\]
	 is $\epsilon$-regular for $0\leq k\leq m$, and
	\[
	R_{\mathrm{reg}}(b,\sigma)=\begin{pmatrix}
		\wt{R}_{00}(b,\sigma)&\wt{R}_{01}(b,\sigma)&\wt{R}_{02}(b,\sigma)\\
		0&0&0\\
		\wt{R}_{20}(b,\sigma)&\wt{R}^e_{21}(b,\sigma)&\wt{R}^e_{22}(b,\sigma)
	\end{pmatrix}.
	\]
	Then a direct calculation implies $
	(\sigma\pa_\sigma)^mL^-_b(\sigma)$ is a linear combination of the operators of the forms
	\begin{gather*}
\check{\mathcal{L}}_b^{(m_1)}(\sigma)\circ(\sigma\pa_\sigma)^{m_2}R_{\mathrm{reg}}(b,\sigma),\quad\check{\mathcal{L}}_b^{(m_1)}(\sigma)\circ\pa^2_\sigma \widehat{L_{b}}(0)\circ P^2(b),\\
\check{\mathcal{L}}_b^{(m_1)}(\sigma)\circ(\sigma\pa_\sigma)^{m_2}\big(\pa_\sigma\widehat{L_{b}}(0)+\frac{\sigma}{2}\pa_\sigma^2\widehat{L_{b}}(0)\big)\circ(\sigma\pa_\sigma)^{m_3}\big(P^1(b)+P^e(b,\sigma)\big).
	\end{gather*}
Applying Proposition \ref{prop:conormalregofcheckL-1} and \ref{prop:conormalregofR}, we obtain \eqref{eq:conormalbound} and \eqref{eq:conormalholderreg}.	
\end{proof}

%%%%%%%%%%%%%%%%%%%%%%%%%%%%%%%%%%%%%%%%%%%%%%%%%%%%%%%%%%%%%%%%%%%%%%%%%%%
\section{Decay estimates}\label{sec:decayestimates}
In this section, we continue using the notation from \S\ref{sec:regularityofmlEM}. We will study the decay of the solution $(\dg, \dA)$ to the equations
\begin{equation}\label{eq:mlgEMeqn}
	L_b(\dg, \dA)=f,\quad f\in C_c^\infty((0,\infty)_{t_{b,*}};\eHb^{s,\ell+2}(\CX)).
\end{equation}
We define spacetime Sobolev spaces
\begin{equation}
	\wt{H}^{s,\ell,k}_{\bop,\cop},\quad k\in\BN_0,
\end{equation}
which is equal to $L^2(t_{b,*}^{-1}([0, \infty));\abs{dg_b})$ for $s,\ell,k=0$. Here, the index $s$ measures the regularity with respect to the derivatives $\pa_{t_{b,*}}$ and the stationary $b$-vector fields on $\CX$ (which is spanned by $r\pa_r$ and rotation vector fields). The index $\ell$ is the weight in $\rho=r^{-1}$, that is, $	\wt{H}^{s,\ell,k}_{\bop,\cop}=\rho^\ell	\wt{H}^{s,k}_{\bop,\cop}$. The index $k\in\BN_0$ measures the regularity with respect to $\jb{t_{b,*}}\pa_{t_{b,*}}$ (which is related to the conormal regularity of $\widehat{L_{b}}(\sigma)^{-1}$ on the Fourier transform side). Therefore, $u\in	\wt{H}^{s,\ell,k}_{\bop,\cop}$ if and only if $(\jb{t_{b,*}}\pa_{t_{b,*}})^ju\in	\wt{H}^{s-j,\ell}_{\bop,\cop}:=\wt{H}^{s-j,\ell,0}_{\bop,\cop}$ for any $0\leq j\leq k$.

\begin{thm}
	\label{thm:decayestimate}
	Let $-\frac 32<\ell<-\frac12, 0<\epsilon<1, -\frac12<\ell+\epsilon< \frac12$ and $s-\epsilon>4+m$ where $m\in\BN_0$. Let $(\dg, \dA)$ be the solution to the equation \eqref{eq:mlgEMeqn} with $f\in C_c^\infty((0,\infty)_{t_{b,*}};\eHb^{s,\ell+2}(\CX))$. Then there exists a generalized zero mode $(\hat{g}, \hat{A})\in\widehat{\mathcal{K}}_b$ (defined in Proposition \ref{prop:gezeromodess1}) such that 
	\begin{equation}
		(\dg, \dA)=(\hat{g}, \hat{A})+(\bar{g}, \bar{A})+(\tilde{g}, \tilde{A})
	\end{equation}
and  $(\hat{g}, \hat{A}), (\bar{g}, \bar{A})$ and $(\tilde{g}, \tilde{A})$ satisfy the following properties. 
\begin{enumerate}
		\item The term $(\bar{g}, \bar{A})$ can be written as  $
		(\bar{g}, \bar{A})=(\bar{g}_1, \bar{A}_1)+(\bar{g}_2, \bar{A}_2)$ where $(\bar{g}_1, \bar{A}_1)$ is a pure gauge solution. Moreover, for $m\geq 1$ and $j\geq 0$, we have the pointwise estimates
\begin{equation}\label{eq:estimatesforbar1}
	\abs{\big(\jb{t_{b,*}}\pa_{t_{b,*}}\big)^{m-1}V^j(\bar{g}_1, \bar{A}_1)}\lesssim
		\big(\jb{t_{b,*}+r}^{-2+\epsilon}+\jb{t_{b,*}}^{-1+\epsilon}\jb{r}^{-2}\big)\norm{f}_{\jb{t_{b,*}}^{-\frac 32-}	\wt{H}^{s-1,\ell+2}_{\bop,\cop}}
\end{equation}
and
\begin{equation}\label{eq:estimatesforgbar2}
	\abs{\big(\jb{t_{b,*}}\pa_{t_{b,*}}\big)^{m-1}V^j(\bar{g}_2,\bar{A}_2)}\lesssim
			\jb{t_{b,*}+r}^{-2+\epsilon}\norm{f}_{\jb{t_{b,*}}^{-\frac 32-}	\wt{H}^{s-1,\ell+2}_{\bop,\cop}}
			%\\\label{eq:estimatesforAbar2}
				%\abs{\big(\jb{t_{b,*}}\pa_{t_{b,*}}\big)^{m-1}V^j \bar{A}_2}&\lesssim
			%\Big(\jb{t_{b,*}+r}^{-2+\epsilon}\jb{r}^{-1}+	\jb{t_{b,*}+r}^{-2}\jb{t_{b,*}}^{-1+\epsilon}\Big)\norm{f}_{\jb{t_{b,*}}^{-\frac 32-}	\wt{H}^{s,\ell+2}_{\bop,\cop}}
\end{equation}
where $V^j$ denotes the product of $j$ vector fields $V\in\{\pa_{t_{b,*}}, r\pa_r, \mbox{ rotation fields }\}$.
\item $(\tilde{g}, \tilde{A})$ satisfies
\begin{equation}
	\norm{(\tilde{g}, \tilde{A})}_{\jb{t_{b,*}}^{-\frac 32+\epsilon}	\wt{H}^{s-2,\ell+\epsilon-1,m}_{\bop,\cop}}\lesssim\norm{f}_{\jb{t_{b,*}}^{-\frac 72+\epsilon}	\wt{H}^{s+m,\ell+2,m}_{\bop,\cop}}.
\end{equation}
For $m\geq 1$, we also have
\begin{equation}
		\norm{\big(\jb{t_{b,*}}\pa_{t_{b,*}}\big)^{m-1}(\tilde{g}, \tilde{A})}_{\jb{t_{b,*}}^{-2+\epsilon}	L^\infty(\BR_{t_{b,*}};\eHb^{s-2-m,\ell+\epsilon-1}(\CX))}\lesssim\norm{f}_{\jb{t_{b,*}}^{-\frac 72+\epsilon}	\wt{H}^{s+m,\ell+2,m}_{\bop,\cop}}.
\end{equation}
For $m\geq 1$ and $0\leq j<s-4-m$ with $j\in\BN_0$, we have the pointwise decay estimate
\begin{equation}
	\abs{\big(\jb{t_{b,*}}\pa_{t_{b,*}}\big)^{m-1}V^j(\tilde{g}, \tilde{A})}\lesssim\jb{t_{b,*}}^{-2+\epsilon}\jb{r}^{-\frac12-\ell-\epsilon}\norm{f}_{\jb{t_{b,*}}^{-\frac 72+\epsilon}	\wt{H}^{s+m,\ell+2,m}_{\bop,\cop}}.
\end{equation}
\item Recall the definition of $\mathcal{K}_{b,1}, \mathcal{K}_{b,2}$ and $\mathcal{K}^*_{b}$ at the beginning of \S\ref{sec:structureofmlEM}. The leading term $(\hat{g}, \hat{A})\in\widehat{K}_b$ is uniquely determined by $f$ and can be expressed as follows:
\begin{equation}
	(\hat{g}, \hat{A})=-\frac{1}{2}\big(t_{b,*}(\dg_1,\dA_1)+(\check{g}_1, \check{A}_1)\big)
	+\frac12\Big((\dg'_1, \dA'_1)+(\dg''_1, \dA''_1)+(\dg_2, \dA_2)\Big)
\end{equation}
where $(\dg_1,\dA_1), (\dg_1',\dA_1'),(\dg''_1, \dA''_1)\in\mathcal{K}_{b,1}$, and $(\dg_2,\dA_2)\in\mathcal{K}_{b,2}$. Moreover, $(\dg_1,\dA_1), (\dg_1',\dA_1')$ and $(\dg_2,\dA_2)$ are determined by the conditions \eqref{eq:condition1} and \eqref{eq:condition2} with $f$ there replaced by $\int_{\BR}f(t_{b,*})\,dt_{b,*}$, and $(\dg''_1, \dA''_1)\in\mathcal{K}_{b,1}$ is uniquely determined by the condition
\begin{align}\notag
	&-\Big\langle[L_b, t_{b,*}](\check{g}''_1, \check{A}''_1)+\frac{1}{2}[[L_b, t_{b,*}], t_{b,*}](\dg''_1,\dA''_1), (\dg^*, \dA^*)\Big\rangle+\Big\langle[L_b, t_{b,*}](\dg''_2, \dA''_2), (\dg^*, \dA^*)\Big\rangle\\
	&\quad\quad=\langle\int_{\BR} t_{b,*}f(t_{b,*})\,dt_{b,*}, \ (\dg^*, \dA^*)\rangle\quad\mbox{for all}\quad(\dg^*,\dA^*)\in\mathcal{K}_b^*.
	\end{align}
for some $(\dg''_2, \dA''_2)\in\mathcal{K}_{b,2}$.
\end{enumerate}
\end{thm}

\begin{proof}
	%We take the Fourier transform of $f$ as follows
	
The strategy of the proof is to take Fourier transform of the inhomogeneous equation $L_b(\dg,\dA)=f$ in $t_{b,*}$. We define the Fourier transform of $(\dg, \dA)$ as 
\begin{equation}
	(\hat{\dg}(\sigma), \hat{\dA}(\sigma)):=\int_{\BR}e^{it_{b,*}\sigma}(\dg, \dA)(t_{b,*})\,dt_{b,*}
	\end{equation}
and likewise for $f$ (here we drop the notation for the spatial variables.).
	
	First, according to the energy estimate in Proposition \ref{prop:energyestimate}, we see that the solution $(\dg,\dA)$ satisfies \[\norm{(\dg, \dA)(t_{b,*},\cdot)}_{\eHb^{1,-1}(\CX)}\lesssim e^{Mt_{b,*}}
	\]
	for some $M>0$. This implies that the Fourier transform $(\hat{\dg}(\sigma), \hat{\dA}(\sigma))$ of $(\dg, \dA)$ 
	is well defined for $\IM\sigma>M$.
	 Since $f(t_{b,*})$ has compact support in $t_{b,*}$, $\hat{f}(\sigma)$ is holomorphic in the full plane $\sigma\in\BC$ and decays superpolynomially  as $\abs{\RE\sigma}\to\infty$ with $\IM\sigma>0$ fixed.  Fourier transforming of the equation $L_b(\dg,\dA)=f$, we obtain
	 \begin{equation}
	 	\widehat{L_{b}}(\sigma)(\hat{\dg}(\sigma), \hat{\dA}(\sigma))=\hat{f}(\sigma),\quad 	\IM\sigma>M.
	 \end{equation}
 In view of Theorem \ref{thm:energyestimate}, we see that $\widehat{L_{b}}(\sigma)$ is invertible in $\IM\sigma>M$ and 
 \[
 	(\hat{\dg}(\sigma), \hat{\dA}(\sigma))=\widehat{L_{b}}(\sigma)^{-1}\hat{f}(\sigma),\quad 	\IM\sigma>M.
 \]
Therefore, we have the following integral representation 

%Since by {\color{red}energy estimate}, $\widehat{L_{b}}(\sigma)^{-1}$ exists with a uniform bound for $\IM\sigma>M$, it follows that $(\hat{\dg}(\sigma), \hat{\dA}(\sigma))=\widehat{L_{b}}(\sigma)^{-1}\hat{f}(\sigma)$ holds for such $\sigma$, and thus we have the integral representation
\begin{equation}\label{eq:integralrepsoln}
	(\dg(t_{b,*}), \ \dA(t_{b,*}))=
\frac{1}{2\pi}\int_{\IM\sigma=M+1}\!e^{-i\sigma t_{b,*}}\widehat{L_{b}}(\sigma)^{-1}\hat{f}(\sigma)\,d\sigma.
\end{equation}
Owing to Theorem \ref{thm:regularityaway0}, we see that the integrand in the above integral representation is holomorphic in $\sigma$ (with values in $\eHb^{s,\ell}(\CX)$) in upper half plane $\IM\sigma>0$, and decay superpolynomially in $\abs{\RE\sigma}$ as with $\IM\sigma$ bounded. Using Cauchy's Theorem, we obtain
\begin{equation}\label{eq:integralrepsoln2}
	(\dg(t_{b,*}), \ \dA(t_{b,*}))=
	\frac{1}{2\pi}\int_{\IM\sigma=C}\!e^{-i\sigma t_{b,*}}\widehat{L_{b}}(\sigma)^{-1}\hat{f}(\sigma)\,d\sigma.
\end{equation}
for any $C>0$.
%, and different values $C>0$ generate the same result. Moreover,according to Theorem {\color{red} energy estimate} and Paley-Wiener Theorem, we conclude that the integral \eqref{eq:integralrepsoln} is supported in $t_{b,*}\geq 0$ and thus gives rise to a forward solution to the equations $L_b(\dg,\dA)=f$.
 Now we let $C\to 0+$.

Fixing a frequency cutoff $\chi(s)\in C_c^\infty(\BR)$ such that $\chi(s)=1$ for $\abs{s}\leq c_0/2$ and $\chi(s)=0$ for $\abs{s}\geq c_0$ where $c_0$ is defined as in Theorem \ref{thm:holderregularity}. We now write
\[
\hat{f}(\sigma)=\chi(\RE\sigma)\hat{f}(\sigma)+(1-\chi(\RE\sigma))\hat{f}(\sigma):=\hat{f}_1(\sigma)+\hat{f}_2(\sigma).
\]
Correspondingly, by Theorem \ref{thm:modestabilityofmlgEMofKN} and Corollary \ref{cor:singandreg}, we split $\widehat{L_{b}}(\sigma)^{-1}\hat{f}(\sigma)$ into two parts
\[
\widehat{L_{b}}(\sigma)^{-1}\hat{f}(\sigma)=(\hat{\dg}_{\mathrm{reg}}(\sigma), \hat{\dA}_{\mathrm{reg}}(\sigma))+(\hat{\dg}_{\mathrm{sing}}(\sigma), \hat{\dA}_{\mathrm{sing}}(\sigma))
\]
where
\[
 (\hat{\dg}_{\mathrm{reg}}(\sigma), \hat{\dA}_{\mathrm{reg}}(\sigma))=L^-_b(\sigma)\hat{f}_1(\sigma)+\widehat{L_{b}}(\sigma)^{-1}\hat{f}_2(\sigma),\quad (\hat{\dg}_{\mathrm{sing}}(\sigma), \hat{\dA}_{\mathrm{sing}}(\sigma))=P(b,\sigma)\hat{f}_1(\sigma).
\]
As discussed above, we can shift the integration contour to $\IM\sigma=C$ for any $C>0$. Letting $C\to0+$, we obtain
\[
(\dg(t_{b,*}), \dA(t_{b,*}))=(\dg_{\mathrm{reg}}(t_{b,*}),\dA_{\mathrm{reg}}(t_{b,*}))+(\dg_{\mathrm{sing}}(t_{b,*}),\dA_{\mathrm{sing}}(t_{b,*}))
\]
where 
\[
(\dg_{\mathrm{reg}}(t_{b,*}),\dA_{\mathrm{reg}}(t_{b,*}))=\frac{1}{2\pi}\int_{\BR}\!e^{-i\sigma t_{b,*}} (\hat{\dg}_{\mathrm{reg}}(\sigma), \hat{\dA}_{\mathrm{reg}}(\sigma))\,d\sigma
\]
and 
\[
(\dg_{\mathrm{sing}}(t_{b,*}),\dA_{\mathrm{sing}}(t_{b,*}))=\lim_{C\to0+}\frac{1}{2\pi}\int_{\IM\sigma=C}\!e^{-i\sigma t_{b,*}} (\hat{\dg}_{\mathrm{sing}}(\sigma), \hat{\dA}_{\mathrm{sing}}(\sigma))\,d\sigma.
\]
\begin{itemize}
	\item \underline{Analysis of $(\dg_{\mathrm{reg}}(t_{b,*}),\dA_{\mathrm{reg}}(t_{b,*}))$.} Since $f\in C_c^\infty((0,\infty)_{t_{b,*}};\eHb^{s,\ell+2}(\CX))$, we see that $\hat{f}_1(\sigma), \hat{f}_2(\sigma)\in H^{s'}(\BR_{\sigma};\eHb^{s,\ell+2}(\CX))$ for any $s'\in\BR$. As $H^{3/2-\epsilon'}(\BR)$ is an algebra for any $0<\epsilon'<1$, by Theorem \ref{thm:holderregularity}, we have
	\[
	L^-_b(\sigma)\hat{f}_1(\sigma)\in H^{\frac32-\epsilon'}((-c_0,c_0);\eHb^{s+1-\max\{\epsilon, \frac 12\}, \ell+\epsilon-1}(\CX)),\quad 0<\epsilon<\epsilon'<1.
	\]
	Using Moser estimate, we bound norm of $	L^-_b(\sigma)\hat{f}_1(\sigma)$ as follows
	\begin{align*}
&	\norm{L^-_b(\sigma)\hat{f}_1(\sigma)}_{H^{\frac32-\epsilon'}(\BR_\sigma;\eHb^{s+1-\max\{\epsilon, \frac 12\}, \ell+\epsilon-1}(\CX))}\lesssim \norm{\hat{f}_1(\sigma)}_{H^{\frac32-\epsilon'}(\BR_\sigma;\eHb^{s, \ell+2}(\CX))}\\
	&\quad \lesssim\norm{\hat{f}(\sigma)}_{H^{\frac32-\epsilon'}(\BR_\sigma;\eHb^{s, \ell+2}(\CX))}=\norm{\jb{t_{b,*}}^{\frac32-\epsilon'}f}_{L^2(\BR_{t_{b,*}}; \eHb^{s,\ell+2}(\CX))}.
	\end{align*}
By exploiting the conormal regularity of $L^-_b(\sigma)$ proved in Proposition \ref{prop:conormalregofcheckL-1}, we find that for $0\leq k\leq m$, 
	\[
(\sigma\pa_\sigma)^k\big(L^-_b(\sigma)\hat{f}_1(\sigma)\big)\in H^{\frac32-\epsilon'}((-c_0,c_0);\eHb^{s+1-k-\max\{\epsilon, \frac 12\}, \ell+\epsilon-1}(\CX)), \quad 0<\epsilon<\epsilon'<1.
\]
In a similar manner, we arrive at 
\begin{equation}\label{eq:esofhatreg1}
	\begin{split}
&\norm{(\sigma\pa_\sigma)^k\big(L^-_b(\sigma)\hat{f}_1(\sigma)\big)}_{H^{\frac32-\epsilon'}(\BR_\sigma;\eHb^{s+1-k-\max\{\epsilon, \frac 12\}, \ell+\epsilon-1}(\CX))}\\
&\quad\lesssim\sum_{j=0}^k\norm{(\sigma\pa_\sigma)^j\hat{f}_1(\sigma)}_{H^{\frac32-\epsilon'}(\BR_\sigma;\eHb^{s-j, \ell+2}(\CX))}\lesssim\sum_{j=0}^k\norm{(\sigma\pa_\sigma)^j\hat{f}(\sigma)}_{H^{\frac32-\epsilon'}(\BR_\sigma;\eHb^{s-j, \ell+2}(\CX))}\\
&\quad=\sum_{j=0}^k\norm{(t_{b,*}\pa_{t_{b,*}})^j\jb{t_{b,*}}^{\frac32-\epsilon'}f}_{L^2(\BR_{t_{b,*}}; \eHb^{s-j,\ell+2}(\CX))}\lesssim \norm{f}_{\jb{t_{b,*}}^{-\frac 32+\epsilon'}	\wt{H}^{s,\ell+2,k}_{\bop,\cop}}.
\end{split}
\end{equation}

As for the second part $\widehat{L_{b}}(\sigma)^{-1}\hat{f}_2(\sigma)$ in 	 $(\hat{\dg}_{\mathrm{reg}}(\sigma), \hat{\dA}_{\mathrm{reg}}(\sigma))$, %repeating the proof of 
by Theorem \ref{thm:regularityaway0}, we have
%and using the high energy estimate of $\widehat{L_{b}}(\sigma)$ on $\BR_{\sigma}\setminus[-\frac{c_0}{2}, \frac{c_0}{2}]$ (i.e. $\norm{\widehat{L_{b}}(\sigma)}\sim\abs{\sigma}^{-1}$ there) yields the following improvement for $\widehat{L_{b}}(\sigma)^{-1}$
\begin{equation}\label{eq:improvementforfreawar0}
\widehat{L_{b}}(\sigma)^{-1}\in W^{j,\infty}\Big(\BR_{\sigma}\setminus[-\frac{c_0}{2}, \frac{c_0}{2}];\mathcal{L}\big(\bar{H}^{s,\ell+2}_{\bop,h}, h^{-j}\bar{H}^{s-j,\ell}_{\bop,h}\big)\Big), \quad j\in\BN_0
\end{equation}
where we take $h=\jb{\sigma}^{-1}$. Taking $j=2$ gives
\[
\widehat{L_{b}}(\sigma)^{-1}\hat{f}_2(\sigma)\in  H^{\frac32-\epsilon}(\BR; \jb{\sigma}^{-s+2}\bar{H}_{\bop, \jb{\sigma}^{-1}}^{s-2,\ell}(\CX)).
\]
whose norm is bounded by
\[
\norm{\hat{f}_2(\sigma)}_{H^{\frac32-\epsilon}(\BR; \jb{\sigma}^{-s}\bar{H}_{\bop, \jb{\sigma}^{-1}}^{s,\ell+2}(\CX))}\lesssim \norm{f}_{\jb{t_{b,*}}^{-\frac 32+\epsilon}	\wt{H}^{s,\ell+2}_{\bop,\cop}}.
\]
More generally, we have
\[
(\sigma\pa_\sigma)^k\big(\widehat{L_{b}}(\sigma)^{-1}\hat{f}_2(\sigma)\big)\in  H^{\frac32-\epsilon}(\BR; \jb{\sigma}^{-s+2+2k}\bar{H}_{\bop, \jb{\sigma}^{-1}}^{s-2-k,\ell}(\CX)).
\]
and
\begin{equation}\label{eq:esofhatreg2}
	\begin{split}
&\norm{(\sigma\pa_\sigma)^k\big(\widehat{L_{b}}(\sigma)^{-1}\hat{f}_2(\sigma)\big)}_{H^{\frac 32-\epsilon}(\BR; \jb{\sigma}^{-s+2+k}\bar{H}_{\bop, \jb{\sigma}^{-1}}^{s-2-k,\ell}(\CX))}\\
&\quad\lesssim\sum_{j=0}^k\norm{(\sigma\pa_\sigma)^j\hat{f}_2(\sigma)}_{H^{\frac32-\epsilon}(\BR; \jb{\sigma}^{-s-k+2j}\bar{H}_{\bop, \jb{\sigma}^{-1}}^{s+k-2j,\ell+2}(\CX))}\\
&\quad\lesssim\sum_{j=0}^k\norm{(\sigma\pa_\sigma)^j\hat{f}(\sigma)}_{H^{\frac32-\epsilon}(\BR; \jb{\sigma}^{-s-k+2j}\bar{H}_{\bop, \jb{\sigma}^{-1}}^{s+k-2j,\ell+2}(\CX))}\\
&\quad\lesssim\sum_{j=0}^k\norm{(t_{b,*}\pa_{t_{b,*}})^jf}_{\jb{t_{b,*}}^{-\frac 32+\epsilon}	\wt{H}^{s+k-2j,\ell+2}_{\bop,\cop}}
\lesssim \norm{f}_{\jb{t_{b,*}}^{-\frac 32+\epsilon}	\wt{H}^{s+k,\ell+2,k}_{\bop,\cop}}.
\end{split}
\end{equation}

Taking inverse Fourier transform of $(\hat{\dg}_{\mathrm{reg}}(\sigma), \hat{\dA}_{\mathrm{reg}}(\sigma))$ and using the estimates \eqref{eq:esofhatreg1} and \eqref{eq:esofhatreg2}, we obtain for $0\leq k\leq m$ and $0<\epsilon<\epsilon'<1$
\begin{equation}\label{eq:L2esofgAreg}
	\begin{split}
&	\norm{(\dg_{\mathrm{reg}}(t_{b,*}),\dA_{\mathrm{reg}}(t_{b,*}))}_{\jb{t_{b,*}}^{-\frac 32+\epsilon'}\wt{H}^{s-2,\ell+\epsilon-1,k}_{\bop,\cop}}\\
&\quad\sim\sum_{j=0}^k\norm{(\sigma\pa_\sigma)^j(\hat{\dg}_{\mathrm{reg}}(\sigma), \hat{\dA}_{\mathrm{reg}}(\sigma))}_{H^{\frac 32-\epsilon'}(\BR; \jb{\sigma}^{-s+2+j}\bar{H}_{\bop, \jb{\sigma}^{-1}}^{s-2-j,\ell+\epsilon-1}(\CX))}\\
	&\quad\lesssim\sum_{j=0}^k\norm{(\sigma\pa_\sigma)^j\big(L^-_b(\sigma)\hat{f}_1(\sigma)\big)}_{H^{\frac32-\epsilon'}(\BR_\sigma;\eHb^{s+1-j-\max\{\epsilon, \frac 12\}, \ell+\epsilon-1}(\CX))}\\
	&\quad\quad+\sum_{j=0}^k\norm{(\sigma\pa_\sigma)^j\big(\widehat{L_{b}}(\sigma)^{-1}\hat{f}_2(\sigma)\big)}_{H^{\frac 32-\epsilon'}(\BR; \jb{\sigma}^{-s+2+j}\bar{H}_{\bop, \jb{\sigma}^{-1}}^{s-2-j,\ell}(\CX))}\\
	&\quad
	\lesssim \norm{f}_{\jb{t_{b,*}}^{-\frac 32+\epsilon'}	\wt{H}^{s+k,\ell+2,k}_{\bop,\cop}}.
	\end{split}
\end{equation}
Now we integrate $\pa_{t_{b,*}}(\dg_{\mathrm{reg}}(t_{b,*}),\dA_{\mathrm{reg}}(t_{b,*}))=t_{b,*}^{-1}(t_{b,*}\pa_{t_{b,*}})(\dg_{\mathrm{reg}}(t_{b,*}),\dA_{\mathrm{reg}}(t_{b,*}))$ and exploit the above estimate \eqref{eq:L2esofgAreg} for $k=1$ to obtain the decay estimate of $(\dg_{\mathrm{reg}}(t_{b,*}),\dA_{\mathrm{reg}}(t_{b,*}))$ in $t_{b,*}$. Concretely, for $t_{b,*}\geq 0$ and $0<\epsilon<\epsilon'<1$, we have
\begin{align*}
&	\norm{(\dg_{\mathrm{reg}}(t_{b,*}),\dA_{\mathrm{reg}}(t_{b,*}))}_{\eHb^{s-3,\ell+\epsilon-1}(\CX)}=	\norm{\int_{t_{b,*}}^\infty\!\pa_s\big(\dg_{\mathrm{reg}}(s),\dA_{\mathrm{reg}}(s)\big)\,ds}_{\eHb^{s-3,\ell+\epsilon-1}(\CX)}\\
&\quad\lesssim\Big(\int_{t_{b,*}}^\infty\!\jb{s}^{-3+2\epsilon'}\jb{s}^{-2}\,ds\Big)^{\frac{1}{2}}	\norm{(\jb{t_{b,*}}\pa_{t_{b,*}})(\dg_{\mathrm{reg}}(t_{b,*}),\dA_{\mathrm{reg}}(t_{b,*}))}_{\jb{t_{b,*}}^{-\frac 32+\epsilon'}	\wt{H}^{s-3,\ell+\epsilon-1}_{\bop,\cop}}\\
&\quad\lesssim\jb{t_{b,*}}^{-2+\epsilon'}\norm{f}_{\jb{t_{b,*}}^{-\frac 32+\epsilon'}	\wt{H}^{s+1,\ell+2,1}_{\bop,\cop}}.
	\end{align*}
Since $(\dg_{\mathrm{reg}}(t_{b,*}),\dA_{\mathrm{reg}}(t_{b,*}))$ has support in $t_{b,*}\geq -C$ for some $C>0$, it follows that
\[
	\norm{(\dg_{\mathrm{reg}}(t_{b,*}),\dA_{\mathrm{reg}}(t_{b,*}))}_{\jb{t_{b,*}}^{-2+\epsilon}	L^\infty(\BR_{t_{b,*}};\eHb^{s-3,\ell+\epsilon-1}(\CX))}\lesssim\norm{f}_{\jb{t_{b,*}}^{-\frac 32+\epsilon'}	\wt{H}^{s+1,\ell+2,1}_{\bop,\cop}}.
\]
By the same reasoning, the above decay estimate can be generalized for all $0\leq k\leq m$,
\begin{equation}\label{eq:LinftyesofgAreg}
		\norm{\big(\jb{t_{b,*}}\pa_{t_{b,*}}\big)^{k-1}(\dg_{\mathrm{reg}}(t_{b,*}),\dA_{\mathrm{reg}}(t_{b,*}))}_{\jb{t_{b,*}}^{-2+\epsilon}	L^\infty(\BR_{t_{b,*}};\eHb^{s-2-k,\ell+\epsilon-1}(\CX))}\lesssim\norm{f}_{\jb{t_{b,*}}^{-\frac 32+\epsilon'}	\wt{H}^{s+k,\ell+2,k}_{\bop,\cop}}.
\end{equation}
The estimates \eqref{eq:L2esofgAreg} and \eqref{eq:LinftyesofgAreg} imply that $(\dg_{\mathrm{reg}}(t_{b,*}),\dA_{\mathrm{reg}}(t_{b,*}))$ occurs as a part of the remainder $(\tilde{g}, \tilde{A})$.

		\item \underline{Analysis of $(\dg_{\mathrm{sing}}(t_{b,*}),\dA_{\mathrm{sing}}(t_{b,*}))$.}
	We now turn to discussing the singular part 
	\[
	(\dg_{\mathrm{sing}}(t_{b,*}),\dA_{\mathrm{sing}}(t_{b,*}))=\lim_{C\to0+}\frac{1}{2\pi}\int_{\IM\sigma=C}\!e^{-i\sigma t_{b,*}} P(b,\sigma)\hat{f}_1(\sigma)\,d\sigma.
	\]
	Since $\hat{f}$ is holomorphic in the full plane $\sigma\in\BC$, we write
	\[
	\hat{f}_1(\sigma)=\chi(\RE\sigma)\hat{f}(\sigma)
=\chi(\RE\sigma)\big(f^{(0)}+i\sigma f^{(1)}+\sigma^2f^{(2)}(\sigma))=\chi(\RE\sigma)\big(f^{(0)}+i\sigma f^{(1)}\big)+\sigma^2\hat{f}_3(\sigma)
	\]
	where
	 \[
	f^{(0)}=\hat{f}(0)=\int_{\BR} f(t_{b,*})\,dt_{b,*},\quad f^{(1)}=-i\pa_\sigma\hat{f}(0)=\int_{\BR} t_{b,*}f(t_{b,*})\,dt_{b,*},
	\]
	and $f^{(2)}(\sigma)$ is holomorphic in the full plane $\sigma\in\BC$. Moreover, by Sobolev embedding theorem, we have for $j=0,1$
	\[
	\norm{f^{(j)}}_{\eHb^{s,\ell+2}(\CX)}\lesssim \norm{\hat{f}(\sigma)}_{H^{3/2+s'}(\BR_\sigma; \eHb^{s,\ell+2}(\CX))}\lesssim\norm{f}_{\jb{t_{b,*}}^{-\frac32-s'}\wt{H}^{s,\ell}_{\bop,\cop}}\quad\mbox{for any}\quad s'>0.
	\]
	As for $\hat{f}_3(\sigma)$, since $\hat{f}_3(\sigma)=\sigma^{-2}(\hat{f}_1(\sigma)-\chi(\RE\sigma)f^{(0)}-i\sigma \chi(\RE\sigma)f^{(1)})$, it follows that $\hat{f}_3(\sigma)\in C_c^\infty(\BR_\sigma)$ and 
	\[
	\norm{\hat{f}_3(\sigma)}_{H^{s'}(\BR_\sigma;\eHb^{s,\ell+2}(\CX))}\lesssim \norm{\hat{f}(\sigma)}_{H^{s'+2}(\BR_\sigma;\eHb^{s,\ell+2}(\CX))},\quad s'>-\frac12.
	\]
	
	Recall that $P(b,\sigma)=\sigma^{-2}P^2(b)+\sigma^{-1} P^1(b)+\sigma^{-1}P^e(\sigma)$ where $P^2(b), P^1(b):\eHb^{s,\ell+2}\to\eHb^{\infty, -1/2-}(\CX)$ and $P^e(b,\sigma):\eHb^{s,\ell+2}\to\mathcal{K}_{b,1}$ is $\epsilon$-regular. Therefore, following the arguments in the proof of \eqref{eq:esofhatreg1}, we find that for $0\leq k\leq m$, 
	\[
	(\sigma\pa_\sigma)^k\Big(P(b,\sigma)(\sigma^2\hat{f}_3(\sigma))\Big)\in H^{\frac32-\epsilon}((-c_0, c_0);\eHb^{\infty,-\frac12-}(\CX)),
	\]
	whose norm is bounded in the following manner
	\begin{equation}
		\begin{split}
		&\norm{(\sigma\pa_\sigma)^k\big(P(b,\sigma)(\sigma^2\hat{f}_3(\sigma))\big)}_{H^{\frac32-\epsilon}(\BR;\eHb^{\infty,-\frac12-}(\CX))}\\
	&\quad	\lesssim\sum_{j=0}^k\norm{(\sigma\pa_\sigma)^j\hat{f}_3(\sigma)}_{H^{\frac32-\epsilon}(\BR;\eHb^{s,\ell+2}(\CX))}\lesssim\sum_{j=0}^k\norm{(\sigma\pa_\sigma)^j\hat{f}(\sigma)}_{H^{\frac72-\epsilon}(\BR;\eHb^{s,\ell+2}(\CX))}\\
	&\quad\lesssim\norm{f}_{\jb{t_{b,*}}^{-\frac 72+\epsilon}	\wt{H}^{s,\ell+2,m}_{\bop,\cop}}.
\end{split}
\end{equation} 
Similarly, we have 
\[
(\sigma\pa_\sigma)^k\Big(\big(P^1(b)+P^e(b,\sigma)\big)(\chi(\sigma)f^{(1)})\Big)\in H^{\frac32-\epsilon}((-c_0, c_0);\eHb^{\infty,-\frac12-}(\CX)),
\]
and
\begin{equation}
\norm{(\sigma\pa_\sigma)^k\big(\big(P^1(b)+P^e(b,\sigma)\big)(\chi(\sigma)f^{(1)})\big)}_{H^{\frac32-\epsilon}(\BR;\eHb^{\infty,-\frac12-}(\CX))}\\
		\lesssim\norm{f}_{\jb{t_{b,*}}^{-\frac32-}	\wt{H}^{s,\ell+2}_{\bop,\cop}}.
\end{equation} 
As a result, it remains to analyze $P(b,\sigma)(\chi(\RE\sigma)f^{(0)})$ and $P^2(b)(i\sigma^{-1}\chi(\RE\sigma)f^{(1)})$. We write 
\begin{align*}
		&\Big(\sigma^{-2}P^2(b)f^{(0)}+\sigma^{-1}P^1(b)f^{(0)}+P^2(b)(i\sigma^{-1} f^{(1)})\Big)+\sigma^{-1}P^e(b,\sigma)(\chi(\RE\sigma)f^{(0)})\\
		&=\Big((\sigma^{-2}(\dg_1,\dA_1)-i\sigma^{-1}(\check{g}_1, \check{A}_1))+i\sigma^{-1}\big((\dg_2, \dA_2)
		+(\dg'_1, \dA'_1)+(\dg''_1, \dA''_1)\big)\Big)\\
		&\quad+i\sigma^{-1}(\dg_1'''(\sigma), \dA_1'''(\sigma)):=\Romanupper{1}+\Romanupper{2}
\end{align*}
	where $(\dg_1,\dA_1),(\dg_1',\dA_1'), (\dg''_1,\dA''_1),(\dg_1'''(\sigma),\dA_1'''(\sigma))\in\mathcal{K}_{b,1}$ and $(\dg_2,\dA_2)\in\mathcal{K}_{b,2}$. Moreover, $(\dg''_1, \dA''_1)$ is uniquely determined by the condition
\begin{align}\notag
	&-\Big\langle[L_b, t_{b,*}](\check{g}''_1, \check{A}''_1)+\frac{1}{2}[[L_b, t_{b,*}], t_{b,*}](\dg''_1,\dA''_1), (\dg^*, \dA^*)\Big\rangle+\Big\langle[L_b, t_{b,*}](\dg''_2, \dA''_2), (\dg^*, \dA^*)\Big\rangle\\
	\label{eq:condition3}
	&\quad\quad=\langle\int_{\BR} t_{b,*}f(t_{b,*})\,dt_{b,*}, \ (\dg^*, \dA^*)\rangle\quad\mbox{for all}\quad(\dg^*,\dA^*)\in\mathcal{K}_b^*
\end{align}
for some $(\dg''_2, \dA''_2)\in\mathcal{K}_{b,2}$.
	Since $\sigma^{-j}(1-\chi(\sigma))\in H^{s'}(\BR_\sigma)$ for $j=1,2$ and any $s'\in\BR$, we have
\[
(\sigma\pa_\sigma)^k\Big(\chi(\RE\sigma)\big(P(b,\sigma)f^{(0)}+P^2(b)(i\sigma^{-1}f^{(1)})\big)-(\Romanupper{1}+\Romanupper{2})\Big)\in H^{\frac32-\epsilon}((-c_0, c_0);\eHb^{\infty,-\frac12-}(\CX)),
\]
and
\begin{equation}
	\begin{split}
	&\norm{(\sigma\pa_\sigma)^k\big(\chi(\RE\sigma)\big(P(b,\sigma)f^{(0)}+P^2(b)(i\sigma^{-1}f^{(1)})\big)-(\Romanupper{1}+\Romanupper{2})\big)}_{H^{\frac32-\epsilon}(\BR;\eHb^{\infty,-\frac12-}(\CX))}\\
	&\quad\lesssim\norm{f}_{\jb{t_{b,*}}^{-\frac32-}	\wt{H}^{s,\ell+2}_{\bop,\cop}}.
	\end{split}
\end{equation} 
As a consequence, it suffices to discuss the inverse Fourier transform of $\Romanupper{1}$ and $\Romanupper{2}$.
		\begin{itemize}
			\item\underline{Analysis of the leading order term $(\hat{g}, \hat{A})$.}
			Since 
		\[
			\mathcal{F}^{-1}(\sigma^{-2})=\lim_{C\to0+}\frac{1}{2\pi}\int_{\IM\sigma=C}\!e^{-i\sigma t_{b,*}}\sigma^{-2}\,d\sigma=-\frac{1}{2}t_{b,*}H(t_{b,*})
			\]
			and \[ \mathcal{F}^{-1}(i \sigma^{-1})=\lim_{C\to0+}\frac{1}{2\pi}\int_{\IM\sigma=C}\!e^{-i\sigma t_{b,*}}i\sigma^{-1}\,d\sigma=\frac{1}{2}H(t_{b,*}),
			\]
			we obtain
			\begin{equation}
				\begin{split}
			\mathcal{F}^{-1}(\Romanupper{1})&=\lim_{C\to0+}\frac{1}{2\pi}\int_{\IM\sigma=C}\!e^{-i\sigma t_{b,*}}\Romanupper{1}\,d\sigma\\
			&=-\frac{1}{2}\big(t_{b,*}(\dg_1,\dA_1)+(\check{g}_1, \check{A}_1)\big)
			+\frac{1}{2}\Big((\dg'_1, \dA'_1)+(\dg''_1, \dA''_1)+(\dg_2, \dA_2)\Big):=(\hat{g}, \hat{A}),
			\end{split}
			\end{equation}
		where $(\dg_1,\dA_1), (\dg_1',\dA_1'),(\dg''_1, \dA''_1)\in\mathcal{K}_{b,1}$, and $(\dg_2,\dA_2)\in\mathcal{K}_{b,2}$. Moreover, $(\dg_1,\dA_1), (\dg_1',\dA_1')$ and $(\dg_2,\dA_2)$ are determined by the conditions \eqref{eq:condition1} and \eqref{eq:condition2} with $f$ there replaced by $f^{(0)}=\int_{\BR}f(t_{b,*})\,dt_{b,*}$, and $(\dg''_1, \dA''_1)\in\mathcal{K}_{b,1}$ is uniquely determined by the condition \eqref{eq:condition3} for some $(\dg''_2, \dA''_2)\in\mathcal{K}_{b,2}$.

	\item\underline{Analysis of the pure gauge part $(\bar{g}, \bar{A})$.} Finally, we  turn to the analysis of the inverse Fourier transform of $\Romanupper{2}=i\sigma^{-1}(\dg_1'''(\sigma), \dA_1'''(\sigma))$ where $\mathrm{supp }\,(\dg_1'''(\sigma), \dA_1'''(\sigma))\subset (-c_0, c_0)$. By Theorem \ref{thm:holderregularityofsing}, 
	\[			\Romanupper{2}=\sigma^{-1}(\dg_1'''(\sigma),\dA_1'''(\sigma))=\sum_{\pm}\psi^{\pm}(\sigma)\quad\mbox{where}\quad \psi^{\pm}\in \mathcal{A}^{-\epsilon}\big(\pm[0, c_0);\mathcal{K}_{b,1}\big).
	\]
	We write
	\begin{align*}
		\int_{-c_0}^{c_0}\! e^{-it_{b,*}\sigma}	\Romanupper{2}\,d\sigma&=\int_0^{c_0}\!e^{-it_{b,*}\sigma}\chi(t_{b,*}\sigma)\psi^+(\sigma)\,d\sigma+\int_0^{c_0}\!e^{-it_{b,*}\sigma}(1-\chi(t_{b,*}\sigma))\psi^+(\sigma)\,d\sigma\\
		&\quad+\int_{-c_0}^0\!e^{-it_{b,*}\sigma}\chi(t_{b,*}\sigma)\psi^-(\sigma)\,d\sigma+\int_{-c_0}^0\!e^{-it_{b,*}\sigma}(1-\chi(t_{b,*}\sigma))\psi^-(\sigma)\,d\sigma\\
		&:=\Romanupper{2}^+_1+\Romanupper{2}^+_2+\Romanupper{2}^-_1+\Romanupper{2}^-_2.
		\end{align*}
	First, we have
	\begin{align*}
		\abs{\Romanupper{2}^+_1}\leq\int_0^{c_0t_{b,*}^{-1}}\!\abs{\psi_+(\sigma)}\,d\sigma\lesssim\int_0^{c_0t_{b,*}^{-1}}\!\sigma^{-\epsilon}\,d\sigma \lesssim t_{b,*}^{-1+\epsilon}.
	\end{align*}
Next, we compute for $k\in\BN$
\begin{align*}
		\Romanupper{2}^+_2=\int_0^{c_0}\!e^{-it_{b,*}\sigma}(1-\chi(t_{b,*}\sigma))\psi_+(\sigma)\,d\sigma=(it_{b,*})^{-k}\int_0^{c_0}\!e^{-it_{b,*}\sigma}\pa_{\sigma}^k\Big((1-\chi(t_{b,*}\sigma))\psi_+(\sigma)\Big)\,d\sigma.
	\end{align*}
Then we have
\[
\abs{	\Romanupper{2}^+_2}\lesssim t_{b,*}^{-k}\int_{c_0t_{b,*}^{-1}/2}^{c_0}\!\sigma^{-k-\epsilon}\,d\sigma\lesssim t_{b,*}^{-1+\epsilon}.
\]
The estimates for $\Romanupper{2}^+_1,\Romanupper{2}^+_2$ also hold after applying any number of derivatives $t_{b_*}\pa_{t_{b,*}}$. We note that $\Romanupper{2}^-_1,\Romanupper{2}^-_2$ can be handled in a similar manner. 
Since
	\begin{equation}
\Romanupper{2}=\sigma^{-1}(\dg_1'''(\sigma), \dA_1'''(\sigma))=(2\delta^*_{g_b}\omega_{b ,s_1}(\sigma), \wt{\mathcal{L}}_{A_b}(\omega_{b,s_1}(\sigma))^\sharp+d\phi_{b,s_1}(\sigma))
	\end{equation}
where 
\[
(\omega_{b,s_1}(\sigma), \phi_{b,s_1}(\sigma))\in \sum_{\pm}\mathcal{A}^{-\epsilon}(\pm[0, c_0); \eHb^{\infty,-\frac32-}(\CX;\scform)\oplus\eHb^{\infty, -\frac12-}(\CX;\BC)),
\]	
we can write
\[
(\bar{g}, \bar{A})=\int_{-c_0}^{c_0}\! e^{-it_{b,*}\sigma}	\Romanupper{2}\,d\sigma=(\bar{g}_1, \bar{A}_1)+(\bar{g}_2, \bar{A}_2)
\]
where
\begin{align*}
(\bar{g}_1, \bar{A}_1)&=\Big(2\delta^*_{g_b}\big(\chi(\frac{\jb{r}}{\jb{t_{b,*}}})\omega_{b ,s_1}(t_{b,*})\big),\  \wt{\mathcal{L}}_{A_b}\big(\chi(\frac{\jb{r}}{\jb{t_{b,*}}})\omega_{b,s_1}(t_{b,*})\big)^\sharp+d\big(\chi(\frac{\jb{r}}{\jb{t_{b,*}}})\phi_{b,s_1}(t_{b,*})\big)\Big),\\
(\bar{g}_2, \bar{A}_2)&=\Big(2[\chi(\frac{\jb{r}}{\jb{t_{b,*}}}),\delta^*_{g_b}]\omega_{b ,s_1}(t_{b,*}),\  [\chi(\frac{\jb{r}}{\jb{t_{b,*}}}),\wt{\mathcal{L}}_{A_b}]\big(\omega_{b,s_1}(t_{b,*})\big)^\sharp+[\chi(\frac{\jb{r}}{\jb{t_{b,*}}}),d]\phi_{b,s_1}(t_{b,*})\Big)\\
&\quad-\chi(\frac{\jb{r}}{\jb{t_{b,*}}})\Big(2\pa_{t_{b,*}}\omega_{b,s_1}\otimes_s dt_{b,*},\  \pa_{t_{b,*}}\big(A_\alpha(\omega_{b,s_1})^{\sharp,\alpha}+\phi_{b,s_1}\big)dt_{b,*}\Big)+\big(1-\chi(\frac{\jb{r}}{\jb{t_{b,*}}})\big)(\bar{g}, \bar{A}).
\end{align*}
where $\chi$ is a smooth nonnegative cutoff such that $\chi(s)=1$ for $s\leq1/2$ and $\chi(s)=0$ for $s\geq 1$. Then the estimates \eqref{eq:estimatesforbar1} and \eqref{eq:estimatesforgbar2}  for $(\bar{g}_1, \bar{A}_1)$ and $(\bar{g}_2, \bar{A}_2)$ follow from the following estimates
\begin{align*}
		\abs{\big(\jb{t_{b,*}}\pa_{t_{b,*}}\big)^{m-1}V^j\omega_{b,s_1}(t_{b,*})}	&\lesssim
	\jb{t_{b,*}}^{-1+\epsilon}\norm{f}_{\jb{t_{b,*}}^{-\frac 32-}	\wt{H}^{s,\ell+2}_{\bop,\cop}},\\
\abs{\big(\jb{t_{b,*}}\pa_{t_{b,*}}\big)^{m-1}V^j\phi_{b,s_1}(t_{b,*})}&\lesssim
\jb{t_{b,*}}^{-1+\epsilon}\jb{r}^{-1}\norm{f}_{\jb{t_{b,*}}^{-\frac 32-}	\wt{H}^{s,\ell+2}_{\bop,\cop}},\\
	\abs{\big(\jb{t_{b,*}}\pa_{t_{b,*}}\big)^{m-1}V^j(\bar{g}, \bar{A})}&\lesssim
		\jb{t_{b,*}}^{-1+\epsilon}\jb{r}^{-2}\norm{f}_{\jb{t_{b,*}}^{-\frac 32-}	\wt{H}^{s,\ell+2}_{\bop,\cop}}.
\end{align*}
\end{itemize}
\end{itemize}
\end{proof}

%%%%%%%%%%%%%%%%%%%%%%%%%%%%%%%%%%%%%%%%%%%%%%%%%%%%%%%%%%%%%%%%%%%%%%%%%%%
\section{Proof of the main theorem}
\label{sec:mainthm}
In the initial value problem for Einstein-Maxwell system as formulated in \S\ref{sec:IVP}, we impose the initial data at the Cauchy hypersurface which terminates at spatial infinity (i.e., it coincides with the hypersurface $t=0$ when $r$ is large enough). In this section, we will discuss how to reduce this initial value problem to the inhomogeneous problem studied in the previous section \S\ref{sec:decayestimates}.

We recall the time function $\mathfrak{t}$ (corresponding to $g_{b_0}$) defined in \eqref{Eqtimeliket}, which satisfies that $\mathfrak{t}=t$ for $r\geq 4\Bm_0$, and $d\mathfrak{t}$ is timelike everywhere on $M$ with respect to $g_{b_0}$ with $b_0=(\Bm_0, 0,\BQ_0), \abs{\BQ_0}\ll\Bm_0$, hence for $g_b$ when $b=(\Bm, \Ba,\BQ)$ is close to $b_0$. We define
\[
\Sigma_0:= \mathfrak{t}^{-1}(0),
\]
which is a spacelike (with respect to $g_b$) Cauchy hypersurface. Identifying the region $r\geq 4\Bm_0$ in $\Sigma_0$ with a subset of $\BR^3$, we can compactify $\Sigma_0$ at infinity to the manifold $\bar{\Sigma}_0$ (with two boundary components: one is inside the black hole and the other is the spatial infinity). We note that for large $r$, the spacetime scattering cotangent bundle ${}\widetilde{^{\scop}T^*}\bar{\Sigma}_0={}^{\scop}T^*\bar{\Sigma}_0 \oplus \BR\,d\mathfrak{t}$ is spanned over $C^\infty(\bar{\Sigma}_0)$ by $d t$ and $d x^1,d x^2,d x^3$, where $(x^1,x^2,x^3)$ are the standard coordinates on $\BR^3$.

We first study the initial value problem for linearized gauge-fixed Einstein-Maxwell system $L_b(\dg,\dA)=0$.

\begin{thm}\label{thm:solntoIVP}
	 Let $0<\alpha<1$ and $s>8+m$, $m\in\BN$. Suppose
	\begin{align*}
	&(\dg_0, \dg_1,\dA_0,\dA_1)\in \eHb^{s,-1/2+\alpha}(\bar{\Sigma}_0;S^2{}\widetilde{^{\scop}T^*}\bar{\Sigma}_0)\oplus \eHb^{s-1,1/2+\alpha}(\bar{\Sigma}_0;S^2{}\widetilde{^{\scop}T^*}\bar{\Sigma}_0)\\ &\hspace{8em}\oplus\eHb^{s-1,-1/2+\alpha}(\bar{\Sigma}_0;{}\widetilde{^{\scop}T^*}\bar{\Sigma}_0)\oplus \eHb^{s-2,1/2+\alpha}(\bar{\Sigma}_0;{}\widetilde{^{\scop}T^*}\bar{\Sigma}_0).
	\end{align*}
	Then the solution $(\dg,\dA)$ of the initial value problem
	\[
	L_{b}(\dg,\dA) = 0, \quad
(\dg,\dA)|_{\bar{\Sigma}_0}=(\dg_0,\dA_0),\quad (\mathcal{L}_{\pa_{\mathfrak{t}}}\dg,\mathcal{L}_{\pa_{\mathfrak{t}}}\dA)|_{\bar{\Sigma}_0}=(\dg_1,\dA_1)
	\]
	satisfies the following properties. Let
	\[
	N^{s,\alpha}:=\norm{\dg_0}_{\eHb^{s+1,-1/2+\alpha}}+\norm{\dg_1}_{\eHb^{s,1/2+\alpha}}+\norm{\dA_0}_{\eHb^{s,-1/2+\alpha}}+\norm{\dA_1}_{\eHb^{s,1/2+\alpha}}.
	\]
\begin{enumerate}	
\item	For $t_{b,*}\geq 0$, the solution $(\dg,\dA)$ can be written as
	\begin{equation}
		(\dg, \dA)=(\hat{g}, \hat{A})+(\bar{g}, \bar{A})+(\tilde{g}, \tilde{A})
	\end{equation}
	where $(\hat{g}, \hat{A})\in\widehat{\mathcal{K}}_b$ is a generalized zero mode  (defined in Proposition \ref{prop:gezeromodess1}), and $(\bar{g}, \bar{A})$, $(\tilde{g}, \tilde{A})$ satisfy the following estimates.
	\begin{itemize}
		\item[(\romannumeral 1)] The term $(\bar{g}, \bar{A})$ can be written as  $
		(\bar{g}, \bar{A})=(\bar{g}_1, \bar{A}_1)+(\bar{g}_2, \bar{A}_2)$ where $(\bar{g}_1, \bar{A}_1)$ is a pure gauge solution. Moreover, for $m\geq 1$ and $j\geq 0$, we have the pointwise estimates
		\begin{equation}\label{EqModifiedGauge}
			\abs{\big(\jb{t_{b,*}}\pa_{t_{b,*}}\big)^{m-1}V^j(\bar{g}_1, \bar{A}_1)}\lesssim
			\big(\jb{t_{b,*}+r}^{-2+\epsilon}+\jb{t_{b,*}}^{-1+\epsilon}\jb{r}^{-2}\big)	N^{s,\alpha}%\norm{f}_{\jb{t_{b,*}}^{-\frac 32-}	\wt{H}^{s,\ell+2}_{\bop,\cop}}
		\end{equation}
		and
		\begin{equation}\label{EqNonstaGuage}
			\abs{\big(\jb{t_{b,*}}\pa_{t_{b,*}}\big)^{m-1}V^j(\bar{g}_2,\bar{A}_2)}\lesssim
			\jb{t_{b,*}+r}^{-2+\epsilon}	N^{s,\alpha}%\norm{f}_{\jb{t_{b,*}}^{-\frac 32-}	\wt{H}^{s,\ell+2}_{\bop,\cop}}
			%\label
		%	\abs{\big(\jb{t_{b,*}}\pa_{t_{b,*}}\big)^{m-1}V^j \bar{A}_2}&\lesssim
		%	\Big(\jb{t_{b,*}+r}^{-2+\epsilon}\jb{r}^{-1}+	\jb{t_{b,*}+r}^{-2}\jb{t_{b,*}}^{-1+\epsilon}\Big)\norm{f}_{\jb{t_{b,*}}^{-\frac 32-}	\wt{H}^{s,\ell+2}_{\bop,\cop}}
		\end{equation}
		where $V^j$ denotes the product of $j$ vector fields $V\in\{\pa_{t_{b,*}}, r\pa_r, \mbox{ rotation fields }\}$.
		\item [(\romannumeral 2)]
		$(\tilde{g}, \tilde{A})$ satisfies: for $0<\epsilon<1$ with $\alpha+\epsilon>1$ and $0\leq m\leq \frac{s-6}{2}$, we have
		\begin{equation}\label{EqPfIVP1Hb}
			\norm{(\tilde{g}, \tilde{A})}_{\jb{t_{b,*}}^{-\frac 32+\epsilon}	\wt{H}^{s-6-m,-5/2+\alpha+\epsilon-,m}_{\bop,\cop}}\lesssim N^{s,\alpha}
		\end{equation}
		For $1\leq m\leq \frac{s-6}{2}$, we also have
		\begin{equation}\label{EqPfIVP1Linfty}
			\norm{\big(\jb{t_{b,*}}\pa_{t_{b,*}}\big)^{m-1}(\tilde{g}, \tilde{A})}_{\jb{t_{b,*}}^{-2+\epsilon}	L^\infty(\BR_{t_{b,*}};\eHb^{s-6-m,-5/2+\alpha+\epsilon-}(\CX))}\lesssim N^{s,\alpha}.
		\end{equation}
		For $m\geq 1$ and $0\leq j<s-8-m$ with $j\in\BN_0$, we have the pointwise decay estimate
		\begin{equation}\label{EqPfIVP1HighLinfty}
			\abs{\big(\jb{t_{b,*}}\pa_{t_{b,*}}\big)^{m-1}V^j(\tilde{g}, \tilde{A})}\lesssim\jb{t_{b,*}}^{-2+\epsilon}\jb{r}^{-\alpha+\epsilon+1+}N^{s,\alpha}.
		\end{equation}
\end{itemize}
\item\label{ItPfIVP2} In $\mathfrak{t}\geq 0$, $t_{b,*}\leq 0$, we have $(\dg,\dA)=r^{-1}H^{s-3}(\BR_{t_{b,*}}\times \BS^2)+(\tilde{g},\tilde{A})$ with $\abs{(\tilde{g},\tilde{A})}\lesssim r^{-1}(1+\abs{t_{b,*}})^{-\alpha}$.
\end{enumerate}
\end{thm}

\begin{proof}
	Let $T_1< T_2$ and let $\chi(t_{b,*})$ be a nonnegative smooth cutoff such that $\chi=1$ for $ t_{b,*}\geq T_2$ and $\chi(t_{b,*})=0$ for $t_{b,*}\leq T_1$. Then we obtain the following equation
	\[
	L_b\Big(\chi(t_{b,*})(\dg,\dA)\Big)=[L_b, \chi(t_{b,*})](\dg,\dA),
	\]
	of which the right-hand side is supported in $t_{b,*}\in(T_1,T_2)$. In view of Theorem \ref{thm:decayestimate}, it suffices to prove that 
	\begin{equation}
		\label{eq:decayofRHS}
		L_b\Big(\chi(t_{b,*})(\dg,\dA)\Big)\in \widetilde{H}_{\bop,\cop}^{s-4-k,1/2+\alpha-,k}, \quad k\in\BN.
		\end{equation}
	Since the regularity with respect to $\langle t_{b,*}\rangle D_{t_{b,*}}$ and $D_{t_{b,*}}$ is equivalent in the support of 	$L_b\Big(\chi(t_{b,*})(\dg,\dA)\Big)$, it suffices to prove \eqref{eq:decayofRHS} for $k=0$. First, the global wellposedness theory of linear wave equations implies that  $(\dg,\dA)\in H^{s-1}$ for $\mathfrak{t}\geq 0$, $t_*\leq C$, $r\leq C$ for any $C$; this implies that $[L_b,\chi]\in H^{s-2}$ in such compact sets. Therefore, it suffices to work in an arbitrarily small neighborhood $r>R_0\gg 1$ of infinity, which we shall do from now on.
	
	The proof closely follows the argument in  \cite[\S\S4--5]{HV20}. Let $b=(\Bm,\Ba,\BQ)$. Then we define $b_1=(\Bm,0,\BQ)$ and introduce the incoming and outgoing null coordinates
	\[
	x^0=t+r_{b_1,*},\quad x^1=t-r_{b_1,*},
	\]
	with respect to which we have
	 \[
	g_{b_1}=-\mu_{b_1}d x^0\,d x^1+r^2\sg
	\]
	 and
	\[
	2\pa_0\equiv 2\pa_{x^0}=\pa_t+(1-\frac{2\Bm}{r}+\frac{\BQ^2}{r^2})\pa_r,\quad
	2\pa_1\equiv 2\pa_{x^1}=\pa_t-(1-\frac{2\Bm}{r}+\frac{\BQ^2}{r^2})\pa_r.
	\]
	Let $x^2,x^3$ denote local coordinates on $\BS^2$, and denote spherical indices by $c,d=2,3$; let $\pa_c\equiv\pa_{x^c}$; let $\CX$ be the compactification of $t_{b,*}^{-1}(0) $. Since $g_b-g_{b_1}\in\rho^2 C^\infty(\CX;\scsym)$, we have
	\begin{align*}
		&g_b(\pa_0,\pa_0),\ (g_b-g_{b_1})(\pa_0,\pa_1),\ g_b(\pa_0,r^{-1}\pa_c), \\
		&\quad g_b(\pa_1,\pa_1),\ g_b(\pa_1,r^{-1}\pa_c),\ (g_b-g_{b_1})(r^{-1}\pa_b,r^{-1}\pa_c) \in \rho^2C^\infty.
	\end{align*}
We let
\[
\rho_I=(r_{b_1,*}-t)/r,\quad \rho_0=(r_{b_1,*}-t)^{-1},\quad \rho=\rho_I\rho_0=\frac{1}{r}.
\]
Then we have 
\[
\rho^{-3}L_b\rho=-\Box_{\rho^2g_b}\otimes\mathrm{Id}_{14\times 14}+R
\]
	%Thus, in the language of \cite[Definition~3.1]{HintzVasyMink4}, $g_b$ differs from $g_{b_1}$ by a correction $g_b-g_{b_1}$ (which is denoted $h$ in the reference, but which is different from $h$ here) that has vanishing leading order terms. Therefore, by \cite[Lemma~3.8]{HintzVasyMink4}, and recalling that in the present paper we do not have constraint damping for large $r$ (thus $\gamma=0$ in the formulas in the reference), the operator $L_b$ is equal to the scalar wave operator for the Schwarzschild metric $g_{b_1}$, tensored with the identity, plus error terms,
	%\[
%	\rho^{-3}L_b\rho = -2\rho^{-2}\pa_0\pa_1 - \sL + R,
%	\]
	with $R\in\widetilde{\mbox{Diff}_{\bop}^1}$ acting on sections of $\scsym\oplus\scform$, where $\widetilde{\mbox{Diff}_{\bop}^k}$ consists of up to $k$-fold products of the vector fields $\{\rho_I\pa_{\rho_I},\, \rho_0\pa_{\rho_0},\, \mbox{rotation vector fields}\}$. We note that switching from $L_b$ to $\rho^{-3}L_b\rho$ is related to the Friedlander rescaling for the scalar wave equation \cite{F80}.

	Then one can use the multiplier $W:=\rho_0^{-2a_0}\rho_I^{-2a_I}V$ with $V:=-(1+c_V)\rho_{I}	\pa_{\rho_I}+\rho_0\pa_{\rho_0}$ and $a_0=\alpha,\, a_I<0$ as introduced in \cite[Lemma 4.4]{HV20} to derive an energy estimate, which allows us to conclude that near $\mathscr{I}^+$, $(\dg,\dA)$ lies in $\wt H_\bop^{s-1,-1/2-}$ (see \cite[Proposition 4.8]{HV20} for detail. Actually, the present setting is much simpler than that in the reference as we do not need to get sharp decay rate for a certain component as did in \cite[Proposition 4.8]{HV20}). To further determine the leading behavior of $(\dg,\dA)$ near $\mathscr{I}$, one rewrite the equations $\rho^{-3}L_{b}\rho(\rho^{-1}\dg,\rho^{-1}\dA)=0$ as $2\rho^{-2}\pa_0\pa_1 (\rho^{-1}\dg,\rho^{-1}\dA)=(\sL+\tilde{R})(\rho^{-1}\dg,\rho^{-1}\dA)$ where $\tilde{R}\in \rho\widetilde{\mbox{Diff}^2_{\bop}}+\widetilde{\mbox{Diff}^1_{\bop}}$. Since $-2\rho^{-2}\pa_0\pa_1=\rho_I^{-1}(\rho_0\pa_{\rho_0}-\rho_I\pa_{\rho_I})\rho_I\pa_{\rho_I}+\widetilde{\mbox{Diff}^1_{\bop}}$, it follows that 
	 \[
(\rho_0\pa_{\rho_0}-\rho_I\pa_{\rho_I})\rho_I\pa_{\rho_I}(\rho^{-1}\dg,\rho^{-1}\dA)\in \rho_I\widetilde{\mbox{Diff}^2_{\bop}}(\rho^{-1}\dg,\rho^{-1}\dA)\sim \rho_0^\alpha\rho_I^{a_I+1}.	\] 
We choose $a_I<0$ such that $0<1+a_I<\alpha$ (see \cite[Lemma 7.7]{HV20}). Integrating first along the vector field $\rho_0\pa_{\rho_0}-\rho_I\pa_{\rho_I}$ from $\rho_I\geq \epsilon$, where $(\rho^{-1}\dg,\rho^{-1}\dA)\sim \rho_0^{\alpha}$, yields \[\rho_I\pa_{\rho_I}(\rho^{-1}\dg,\rho^{-1}\dA)\sim \rho_0^\alpha\rho_I^{1+a_I}.
\]
Then integrating out the vector field $\rho_I\pa_{\rho_I}$ gives
	\[
	(\dg,\dA)-\rho H^{s-3}(\BR_{t_{b_1,*}}\times\BS^2)\sim \rho\rho_0^\alpha\rho_I^{1+a_I}.
	\]
This proves statement \ref{ItPfIVP2}. 
Moreover, in view of Lemma \ref{LemPWL}, we see that 
\begin{equation}
	\label{EqPfIVPComm}
	[L_b,\chi(t_{b,*})]=2\rho\chi'(t_*)(\rho\pa_\rho-1)+\rho^2\mathrm{Diff}_\scop^1+\rho^2C^\infty\pa_{t_*}.
\end{equation}
Since $(\rho\pa_\rho-1)$ annihilates the leading order term $\rho H^{s-3}(\BR_{t_{b_1,*}}\times\BS^2)$ in $(\dg,\dA)$, it follows that $L_b(\chi (\dg,\dA))=[L_b,\chi](\dg,\dA)\in\widetilde{H}_{\bop,\cop}^{s-4,1/2+\alpha-}$.

Since $s-4>4+m$, the estimates \eqref{EqModifiedGauge}--\eqref{EqPfIVP1HighLinfty} now follow from Theorem \ref{thm:decayestimate}, with $\ell+2=\frac12+\alpha-$, so $\ell=-\tfrac32+\alpha-$ and $-1/2<\ell+\epsilon<1/2$ with $\alpha+\epsilon>1, 0<\epsilon<1$.
%The pointwise estimate~\eqref{EqPfIVP1Pt} then follows from~\eqref{EqDPwise2}.
\end{proof}

We now state our main theorem.
\begin{thm}
	\label{ThmLinstaiblityKN}
	Let $0<\alpha<1$ and $s>8+m$, $m\in\BN$. Let 
	\begin{align*}
	&(\dot{h}, \dot{k}, \dot{\mathbf{E}}, \dot{\mathbf{H}})
	 \in \eHb^{s,-1/2+\alpha}(\Sigma_0;S^2\,{}^{\scop}T^*\Sigma_0)\oplus
	\in \eHb^{s-1,1/2+\alpha}(\Sigma_0;S^2\,{}^{\scop}T^*\Sigma_0)\\
	&\hspace{8em}\eHb^{s-1,1/2+\alpha}(\Sigma_0; {}^{\scop}T^*\Sigma_0)\oplus
	\in \eHb^{s,1/2+\alpha}(\Sigma_0;{}^{\scop}T^*\Sigma_0)
	\end{align*}
	be the initial data for the linearized Einstein-Maxwell equations linearized around the KN solution $(g_b, A_b)$. Suppose that the initial data satisfy the linearized constraint equations, which is the linearization of the constraint equations around the KN initial data $(h_b, k_b,\mathbf{E}_b, \mathbf{H}_b)=\tau(g_b, dA_b)$. Assume that the linearized magnetic charge vanishes, see the discussion in \S \ref{SubsecBasicNL}--\ref{SubsecBasicLin}. 
	
	Then there exists a solution $(\dg, \dA)$ of the initial value problem
	\[
 D_{g_b}\Ric(\dg) - 2 D_{(g_b,d A_b)}T(\dg,\dA)=0,\quad 
	 D_{g_b,A_b}(\delta_{(\cdot)}d(\cdot))(\dg,\dA)=0,\quad D_{(g_b, A_b)}\tau(\dg,\dA)=(h_b, k_b,\mathbf{E}_b, \mathbf{H}_b)
		\]
	satisfying the linearized generalized wave map gauge and Lorenz gauge conditions
\[
D_{g_b}\widetilde{\Upsilon}^E(\dg;g_b)=0,\quad \delta_{g_b}\dA=0.
\]
Moreover, $(\dg,\dA)$ has the asymptotic behavior stated in Theorem \ref{thm:solntoIVP}.
\end{thm}

\begin{proof}
In view of Corollary \ref{CorKNIniLin}, there exist sections
	\begin{align*}
		&(\dot{g}_0,\dot{g}_1,\dot{A}_0,\dot{A}_1)\in \eHb^{s,-1/2+\alpha}(\bar{\Sigma}_0;S^2{}\widetilde{^{\scop} T^*}\bar{\Sigma}_0)\oplus \eHb^{s-1,1/2+\alpha}(\bar{\Sigma}_0;S^2 {}\widetilde{^{\scop}T^*}{\bar{\Sigma}_0}) \\
		&\hspace{8em}\oplus \eHb^{s-1,-1/2+\alpha}(\bar{\Sigma}_0;{}\widetilde{^{\scop}T^*}{\bar{\Sigma}_0}) \oplus \eHb^{s-2,1/2+\alpha}(\bar{\Sigma}_0;{}\widetilde{^{\scop}T^*}{\Sigma_0}),
		\end{align*}
such that they induce the data $(\dot{h},\dot{k},\dot{\mathbf{E}},\dot{\mathbf{H}})$ on $\Sigma_0=\{\mathfrak{t}=0\}$, and satisfy the linearized gauge conditions $D_{g_b}\widetilde{\Upsilon}^E(\dg)=0$ and $D_{(g_b, A_b)}\widetilde{\Upsilon}^M(\dg,\dA)=-\delta_{g_b}\dot{A}=0$ at $\Sigma_0$.

Then we solve the initial value problem $L_{b}(\dg,\dA)=0$ with initial data $(\dot{g}_0,\dot{g}_1,\dot{A}_0,\dot{A}_1)$. Since $(\dot{h},\dot{k},\dot{\mathbf{E}},\dot{\mathbf{H}})$ satisfies the linearized constraint equations, it follows from Lemma \ref{lem:gaugeungaugeLin} that 
\[
D_{g_b}\widetilde{\Upsilon}^E(\dg;g_b)=0,\quad \delta_{g_b}\dA=0
\]
and thus $(\dg,\dA)$ solves the linearized Einstein-Maxwell equations.	
\end{proof}

%%%%%%%%%%%%%%%%%%%%%%%%%%%%%%%%%%%%%%%%%%%%%%%%%%%%%%%%%%%%%%%%%%%%%%%%%%%
\appendix
\section{The linearized Einstein-Maxwell operator around Reissner-Nordstr\"{o}m spacetime}\label{sec:appenA}
\subsection{Detailed calculation of the linearized Einstein-Maxwell system}\label{subsec:detailcal}
We now calculate the linearized Einstein-Maxwell system \[\mathscr{L}(\dg,\dA):=(\mathscr{L}_1(\dg,\dA),\mathscr{L}_2(\dg,\dA)):=(D_g\Ric(\dg, d\dA)-2D_{(g, dA)}T(\dg, d\dA),\ D_{(g, F)}\left(\delta_{g}F\right)(\dg, \dF))=0
\]
in more detail. 

First, following \cite[\S 7.5]{W84},\cite[\S 3]{GL91} we see that 
\[
D_g\Ric(\dot{g})=-\frac 12\Box_{g}\dot{g}-\delta^*_g\delta_gG_g\dot{g}+\mathscr{R}_g\dot{g}\quad\mbox{where}\quad (\mathscr{R}_g\dg)_{\m\n}=R^{\al\ \ \beta}_{\ \m\n}\dg_{\al\be}+\frac 12\left(\Ric_{\m}^{\ \kappa}\dg_{\n\ka}+\Ric_{\n}^{\ \kappa}\dg_{\m\ka}\right)
\]
Since $T(g, F)=F_{\mu\alpha}F_{\nu}^{~\alpha}-\frac 14 g_{\mu\nu}F_{\alpha\beta}F^{\alpha\beta}$, we find 
\begin{equation}\label{eq:DT}
	\begin{split}
		D_{(g, F)}T(\dg, \dF)&=-\dg^{\alpha\beta}F_{\mu\alpha}F_{\nu\beta}-\frac 14\left(\dg_{\mu\nu}F_{\alpha\beta}F^{\alpha\beta}-g_{\mu\nu}\dg^{\kappa\alpha}g^{\lambda\beta}F_{\alpha\beta}F_{\kappa\lambda}-g_{\mu\nu}g^{\kappa\alpha}\dg^{\lambda\beta}F_{\alpha\beta}F_{\kappa\lambda}\right)\\
		&\quad +\left(g^{\alpha\beta}\dF_{\mu\alpha}F_{\nu\beta}+g^{\alpha\beta}F_{\mu\alpha}\dF_{\nu\beta}\right)-\frac 14\left(g_{\mu\nu}\dF_{\alpha\beta}F^{\alpha\beta}+g_{\mu\nu}F_{\alpha\beta}\dF^{\alpha\beta}\right)\\
		&=-\dg^{\alpha\beta}F_{\mu\alpha}F_{\nu\beta}-\frac 14\left(\dg_{\mu\nu}F_{\alpha\beta}F^{\alpha\beta}\!\!-\!2g_{\mu\nu}\dg^{\kappa\alpha}F_{\alpha\beta}F_{\kappa}^{~\be}\right)\!\!+\!\!\left(g^{\alpha\beta}\dF_{\mu\alpha}F_{\nu\beta}+g^{\alpha\beta}F_{\mu\alpha}\dF_{\nu\beta}\right)\!\!-\!\frac 12g_{\mu\nu}\dF_{\alpha\beta}F^{\alpha\beta}
	\end{split}
\end{equation}
and thus the Einstein part $\mathscr{L}_1$ of the linearized Einstein-Maxwell system is given by
\begin{equation}\label{eq:L_1}
	\mathscr{L}_1(\dg, d\dA)=D_g\Ric(\dg, d\dA)-2D_{(g, dA)}T(\dg, d\dA)=-\frac12\Box_{g}\dot{g}-\delta^*_g\delta_gG_g\dot{g}+\mathscr{R}_g\dot{g}-2D_{(g, dA)}T(\dg, d\dA)=0.
\end{equation}

Next we analyze the Maxwell part $\mathscr{L}_2$ of the linearized Einstein-Maxwell system. Since 
\[
(\delta_g F)_\mu=-\nabla^\nu F_{\nu\mu}=-g^{\nu\alpha}\nabla_\alpha F_{\nu\mu}=-g^{\nu\alpha}\left(\pa_\al F_{\n\m}-\cs^{\ka}_{\al\n}F_{\ka\m}-\cs^{\ka}_{\al\m}F_{\n\ka}\right),
\]
and 
\[
\dot{\cs}_{\al\n}^{\ka}(\dg)=\frac 12 g^{\ka\la}\left(\nabla_\n \dg_{\al\la}+\nabla_\al \dg_{\n\la}-\nabla_\la\dg_{\al\n}\right),
\]
we obtain
\begin{equation}
	\begin{split}
		&D_{(g, F)}\left(\delta_{g}F\right)(\dg, \dF)=\delta_g\dF+\dg^{\n\al}\nabla_\al F_{\n\m}+g^{\n\al}\left(\dot{\cs}_{\al\n}^{\ka}(\dg)F_{\ka\m}+\dot{\cs}_{\al\m}^{\ka}(\dg)F_{\n\ka}\right)\\
		&\quad=\delta_g\dF+\dg^{\n\al}\nabla_\al F_{\n\m}+g^{\ka\la}\nabla^\al(\dg_{\al\la}-\frac 12g^{\n\al}\dg_{\al\n}g_{\al\la})F_{\ka\m}+\frac 12(\nabla^{\n}\dg^{\ka}_{~\m}-\nabla^{\ka}\dg^{\n}_{~\m})F_{\n\ka}\\
		&\quad=\delta_g\dF+\dg^{\n\al}\nabla_\al F_{\n\m}-(\delta_gG_g\dg)^\ka F_{\ka\m}+\frac 12(\nabla^{\n}\dg^{\ka}_{~\m}-\nabla^{\ka}\dg^{\n}_{~\m})F_{\n\ka}
	\end{split}
\end{equation}
where we use $(\nabla_\m\dg^{\nu\ka})F_{\n\ka}=0$ in the second equality. Therefore, the Maxwell part $\mathscr{L}_2$ of the linearized Einstein-Maxwell system is given by
\begin{equation}\label{eq:L_2}
	\mathscr{L}_2(\dg, d\dA)=\delta_gd\dA+\dg^{\n\al}\nabla_\al (dA)_{\n\m}-(\delta_gG_g\dg)^\ka (dA)_{\ka\m}+\frac 12(\nabla^{\n}\dg^{\ka}_{~\m}-\nabla^{\ka}\dg^{\n}_{~\m})(dA)_{\n\ka}=0.
\end{equation}

We then proceed to calculate the linearized Einstein-Maxwell system $\mathscr{L}(\dg,\dA)=0$ around a Reissner-Nordstr\"{o}m spacetime $g=\hg+r^2\sg$ under the splitting \eqref{eq:splitof1} and \eqref{eq:splitofsym2} .
\subsection{Calculation of $\mathscr{R}_g\dg$}
Here and in what follows the components we do not list are $0$. Christoffel symbols of Reissner-Nordstr\"{o}m metric $g$ are,
\begin{gather*}
	\cs_{ij}^k=\hcs_{ij}^k \quad %\mbox{where} \quad \hcs_{tr}^t=\hcs_{rt}^t=\frac 12\weight^{-1}\weight', \quad \hcs_{tt}^r=\frac 12\weight\weight',\quad \hcs_{rr}^r=-\frac 12 \weight^{-1}\weight',\\
	\cs_{ij}^a=0,\quad \cs_{ia}^j=0,\quad\cs_{ab}^k=-r\pa^{k}r\sg_{ab},\\
	\cs_{ak}^b= r^{-1}\pa_kr\delta^b_{a},\quad	\cs_{ab}^c=\scs_{ab}^c.
\end{gather*}
The components of Riemann curvature tensor are given by
\begin{gather*}
	R_{ijkm}=\hR_{ijkm}=\frac 12\hR(\hg)\left(\hg_{ik}\hg_{jm}-\hg_{im}\hg_{jk}\right)=-\frac 12\weight''\left(\hg_{ik}\hg_{jm}-\hg_{im}\hg_{jk}\right),\\
	R_{aijk}=0,\quad R_{abij}=0,\quad R_{abci}=0,\quad R_{aibj}=-r\hnabla_i\hnabla_jr\sg_{ab}=-\frac 12r\weight'\sg_{ab}\hg_{ij}	,	\\
	R_{abcd}=\sR_{abcd}-r^2\weight\left(\sg_{ac}\sg_{bd}-\sg_{ad}\sg_{bc}\right)	=r^2(1-\weight)\left(\sg_{ac}\sg_{bd}-\sg_{ad}\sg_{bc}\right).
\end{gather*}
Correspondingly, the Ricci curvature tensor takes the form
\begin{gather*}
	\Ric_{ij}=\hRic_{ij}-r^{-1}\weight'\hg_{ij}=-\frac 12\weight''\hg_{ij}-r^{-1}\weight'\hg_{ij},\quad \Ric_{ia}=0,\\
	\Ric_{ab}=(1-\weight-r\weight')\sg_{ab}.
\end{gather*}
Now we are ready to calculate $\mathscr{R}_g\dg$ in detail. Since $(\mathscr{R}_g\dg)_{\m\n}=R^{\al\ \ \beta}_{\ \m\n}\dg_{\al\be}+\frac 12\left(\Ric_{\m}^{\ \kappa}\dg_{\n\ka}+\Ric_{\n}^{\ \kappa}\dg_{\m\ka}\right)$, we have
\begin{gather*}
	(\mathscr{R}_g\dg)_{ij}=-\weight''\left(\dg_{ij}-\frac 12\hg^{km}\dg_{km}\hg_{ij}\right)-r^{-1}\weight'\dg_{ij}+\frac 12r^{-3}\weight'\sg^{ab}\dg_{ab}\hg_{ij},\\		(\mathscr{R}_g\dg)_{ia}=-\frac 32r^{-1}\weight'\dg_{ia}-\frac 14\weight''\dg_{ia}+\frac 12 r^{-2}(1-\weight)\dg_{ia},\\		(\mathscr{R}_g\dg)_{ab}=2r^{-2}(1-\weight)\left(\dg_{ab}-\frac 12\sg^{cd}\dg_{cd}\sg_{ab}\right)-r^{-1}\weight'\dg_{ab}+\frac 12r\weight'\hg^{ij}\dg_{ij}\sg_{ab}.
\end{gather*}
Therefore, with respect to the splitting of $S^2T^*M$ \eqref{eq:splitofsym2}, we can express $\mathscr{R}_g: S^2T^*M\to S^2T^*M$ as follows
\begin{equation}
	\mathscr{R}_g=
	\begin{pmatrix}
		-\weight'' G_{\hg}-r^{-1}\weight'&0&\frac 12 r^{-3}\weight' \hg\str\\
		0&-\frac 32 r^{-1}\weight'-\frac 14\weight''+\frac 12r^{-2}(1-\weight)&0\\
		\frac 12r\weight'\sg\htr&0&2r^{-2}(1-\weight)G_{\sg}-r^{-1}\weight'
	\end{pmatrix}
\end{equation}
where $(G_{\hg} h)_{ij}=h_{ij}-\frac 12\hg_{ij}\htr h$ and $(G_{\sg}h)_{ab}=h_{ab}-\frac 12\sg_{ab}\str h$.

\subsection{Calculation of $\delta_g^*\delta_g G_g\dg$}
We now calculate the first covariant derivatives on $1$-forms and symmetric $2$-tensors. For $\eta\in T^*M$, we compute
\begin{gather*}
	\nabla_i\eta_j=\hnabla_i\eta_j,
	\quad \nabla_i\eta_{a}=\pa_i\eta_{a}-(r^{-1}\pa_i r)\eta_a,
	\quad \nabla_{a}\eta_i=\pa_{a}\eta_i-(r^{-1}\pa_i r)\eta_a,
	\quad \nabla_{a}\eta_b=\slashed{\nabla}_a\eta_b+r(\pa^kr)\eta_k\sg_{ab}.
\end{gather*}
With respect to the splitting of $T^*M$ \eqref{eq:splitof1}, we write the symmetric gradient $\delta_g^*: T^*M\to S^2T^*M$ with $(\delta^*_g\eta)_{\m\n}=\frac 12(\nabla_\m\eta_\n+\nabla_\n\eta_\m)$ as follows
\begin{equation}\label{eq:symgradient}
	\delta_g^*=
	\begin{pmatrix}
		\hdelta^*&0\\
		\frac 12\sd&\frac 12r^2\hd r^{-2}\\
		r\sg\iota_{dr}&\sdelta^*
	\end{pmatrix}.
\end{equation}
where $\iota_{dr}: T^*\hX\to\BR$ is the contraction with the aspherical vector field $dr^{\sharp}$, i.e. $\iota_{dr}(\cdot)=\hg^{-1}(dr, \cdot)$. We then calculate the first covariant derivatives on symmetric $2$-tensor $h\in S^2T^*M$
\begin{gather*}
	\nabla_ih_{jk}=\hnabla_ih_{jk},
	\quad \nabla_ih_{ja}=\hnabla_ih_{ja}-r^{-1}(\pa_ir)h_{ja}, 
	\quad \nabla_i h_{ab}=\pa_ih_{ab}-2r^{-1}(\pa_ir)h_{ab}\\
	\nabla_ah_{ij}=\pa_ah_{ij}-r^{-1}(\pa_ir)h_{aj}-r^{-1}(\pa_jr)h_{ia},
	\quad \nabla_ah_{ib}=\slashed{\nabla}_ah_{ib}+r(\pa^kr)h_{ik}\sg_{ab}-r^{-1}(\pa_ir)h_{ab}, \\ \nabla_ah_{bc}=\slashed{\nabla}_ah_{bc}+r(\pa^kr)(h_{kc}\sg_{ab}+h_{kb}\sg_{ac}).
\end{gather*}
Therefore the negative divergence $(\delta_gh)_\m=-\nabla^\n h_{\n\m}$ which maps symmetric $2$-tensors to $1$-form takes the following form (under the splitting \eqref{eq:splitofsym2} and \eqref{eq:splitof1})
\begin{equation}\label{eq:ndivergence}
	\delta_g=
	\begin{pmatrix}
		r^{-2}\hdelta r^2& r^{-2}\sdelta& r^{-3}(dr)\str\\
		0&r^{-2}\hdelta r^{2}&r^{-2}\sdelta
	\end{pmatrix}
\end{equation}
where $(\sdelta h)_i=-\slashed{\nabla}^ah_{ai}$ and $(\hdelta h)_a=-\hnabla^i h_{ia}$. We can also express $(G_gh)_{\m\n}=h_{\m\n}-\frac 12g_{\m\n}\tr_gh$ as
\begin{equation}\label{eq:G_g}
	G_g=
	\begin{pmatrix}
		G_{\hg}&0&-\frac 12r^{-2}\hg\str\\
		0&1&0 \\
		-\frac 12r^2\sg\htr&0&G_{\sg}
	\end{pmatrix}.
\end{equation}
Putting \eqref{eq:symgradient}, \eqref{eq:ndivergence} and \eqref{eq:G_g} together and using $\sdelta G_{\sg}=\sdelta+\frac 12\sd\str,\  \hdelta G_{\hg}=\hdelta+\frac 12\hd\htr$ yield
\begin{equation}
	\delta_g^*\delta_gG_g=%\delta_g^*
	%	\begin{pmatrix}
	%	r^{-2}\hdelta r^2+\frac 12\hd\htr&r^{-2}\sdelta&\frac 12r^{-2}\hd\str\\
	%-\frac 12\sd\htr&r^{-2}\hdelta r^2&	r^{-2}\sdelta +\frac 12 r^{-2}\sd\str
	%	\end{pmatrix}
	\begin{pmatrix}
		\hdelta^*r^{-2}\hdelta r^2+\frac 12\hdelta^*\hd\htr&\hdelta^*r^{-2}\sdelta&\frac 12\hdelta^*r^{-2}\hd\str	\\
		\frac 12 r^{-2}\sd\hdelta r^2+\frac 14 \hd\sd\htr+\frac 14 r^2\hd r^{-2}\sd\htr&\frac 12r^{-2}\sd\sdelta+\frac 12 r^2\hd r^{-4}\hdelta r^2&\frac 14 r^{-2}\hd\sd\str+\frac 12r^2\hd r^{-4}\sdelta+\frac14 r^2\hd r^{-4}\sd\str\\
		r^{-1}\sg\iota_{dr}\hdelta r^2+\frac12 r\sg\iota_{dr}\hd\htr+\frac 12\sdelta^*\sd\htr&r^{-1}\sg\iota_{dr}\sdelta+r^{-2}\sdelta^*\hdelta r^2&\frac 12 r^{-1}\sg\iota_{dr}\hd\str+r^{-2}\sdelta^*\sdelta+\frac 12r^{-2}\sdelta^*\sd\str
	\end{pmatrix}.
\end{equation}

\subsection{Calculation of $\Box_g$}
Next we compute the second covariant derivatives on symmetric $2$-tensor $h\in S^2T^*M$. Since our goal is to find the form of $\Box_g=g^{\m\n}\nabla_\m\nabla_\n$ applied to $h$, it suffices to compute the following second covariant derivatives
\begin{equation}
	\begin{split}
		\nabla_i\nabla_j h_{km}&=\hnabla_i\hnabla_j h_{km},\\ 
		\nabla_a\nabla_b h_{km}&=\snabla_a\snabla_bh_{km}-r^{-1}\pa_k r\left(\snabla_a h_{bm}+\snabla_b h_{am}\right)-r^{-1}\pa_m r\left(\snabla_a h_{kb}+\snabla_b h_{ka}\right)\\
		&\quad+r\pa^i r\hnabla_i h_{km}\sg_{ab}+2r^{-2}\pa_kr\pa_mr h_{ab}-\sg_{ab}\left(\pa_k r\pa^irh_{im}+\pa_m r\pa^irh_{ik}\right) \\
		\nabla_i\nabla_jh_{kc}&=\hnabla_i\hnabla_jh_{kc}-r^{-1}\pa_i r\hnabla_j h_{kc}-r^{-1}\pa_j r\hnabla_i h_{kc}+2r^{-2}\pa_i r\pa_j rh_{kc}-\frac 12r^{-1}\weight'\hg_{ij}h_{kc},\\
		\nabla_a\nabla_b h_{kc}&=\snabla_a\snabla_bh_{kc}-r^{-1}\pa_kr\left(\snabla_ah_{bc}+\snabla_bh_{ac}\right)+r\sg_{ac}\pa^ir\snabla_bh_{ik}+r\sg_{bc}\pa^ir\snabla_ah_{ik}+r\sg_{ab}\pa^i r\hnabla_i h_{kc}\\
		&\quad-\pa_kr\pa^ir\left(\sg_{ab}h_{ic}+\sg_{ac}h_{ib}+\sg_{bc}h_{ib}\right)-\weight\left(\sg_{ab}h_{kc}+\sg_{ac}h_{kb}\right)\\
		\nabla_i\nabla_j h_{cd}&=\hnabla_i\hnabla_j h_{cd}-2r^{-1}\left(\pa_ir\hnabla_j h_{cd}+\pa_j r\hnabla_i h_{cd}\right)+6r^{-2}\pa_ir\pa_jrh_{cd}-r^{-1}\weight'\hg_{ij}h_{cd},\\
		\nabla_a\nabla_b h_{cd}&=\snabla_{a}\snabla_bh_{cd}+r\pa^ir\left(\sg_{bc}\snabla_{a}h_{di}+\sg_{bd}\snabla_{a}h_{ci}+\sg_{ac}\snabla_{b}h_{di}+\sg_{ad}\snabla_{b}h_{ci}+\sg_{ab}\hnabla_ih_{cd}\right)\\
		&\quad-\weight\left(2\sg_{ab}h_{cd}+\sg_{ac}h_{bd}+\sg_{ad}h_{bc}\right)+r^2\pa^ir\pa^jrh_{ij}\left(\sg_{ac}\sg_{bd}+\sg_{ad}\sg_{bc}\right)
	\end{split}
\end{equation}
where $\hBox$ acting on aspherical symmetric $2$-tensors is the tensor wave operator, on mixed symmetric tensors is the $1$-form wave operator acting on the aspherical part, and on spherical symmetric tensors is the scalar wave operator acting on the aspherical variables; the operator $\sL$ is defined in a similar manner. Then we find
\begin{equation}\label{eq:Box}
	\begin{split}
		\frac 12\Box_g&=
		\begin{pmatrix}
			\frac 12\hBox+\frac 12r^{-2}\sL&2r^{-3}(dr)\otimes_s \sdelta&r^{-4}(dr\otimes dr)\str\\
			r^{-1}\iota_{dr}\sd&\frac 12\hBox+\frac 12r^{-2}\sL &r^{-3}(dr)\otimes\sdelta\\
			\sg\iota_{dr}\iota_{dr}&2r^{-1}\iota_{dr}\sdelta^*&\frac 12	\hBox+\frac 12r^{-2}\sL
		\end{pmatrix}\\
		&	\quad+	\begin{pmatrix}
			r^{-1}\pa^ir\hnabla_i\!-\!2r^{-2}(d r)\otimes_s\iota_{dr}&0&0\\
			0& -\frac 12r^{-2}(\weight+r\weight')\!-\!2r^{-2}(dr)\otimes\iota_{dr}&0\\
			0&0&-r^{-1}\pa^ir \hnabla_i\!-\!r^{-1}\weight'
		\end{pmatrix}.
	\end{split}
\end{equation}

On the aspherical part $\hX$, we further calculate the relevant operators on the static region where $\weight>0$ by working in $(t,r)$ coordinates. On the static region, the metric $\hg$ reads
\[
\hg=-\weight dt^2+\weight^{-1}dr^2.
\]
We further split
\begin{equation}\label{eq:splitofX}
	T^*\hX=\langle \hd t\rangle\oplus\langle \hd r\rangle,\quad S^2T^*\hX=\langle\hd t^2\rangle\oplus\langle2\hd t\hd r\rangle\oplus\langle\hd r^2\rangle.
\end{equation}
Since $\hcs_{tr}^t=\hcs_{rt}^t=\frac 12\weight^{-1}\weight', \hcs_{tt}^r=\frac 12\weight\weight', \hcs_{rr}^r=-\frac 12 \weight^{-1}\weight'$, with respect to the above split \eqref{eq:splitofX}, we see that acting on scalar functions, $\pa^ir \hnabla_i=\weight\pa_r$, 
\begin{equation}\label{eq:static1form1}
	\hd=\begin{pmatrix}
		\pa_t\\
		\pa_r
	\end{pmatrix},\quad \hBox=-\weight^{-1}\pa_t^2+\weight\pa_r^2+\weight'\pa_r=-\weight^{-1}\pa_t^2+\pa_r\weight\pa_r
\end{equation}
On $1$-forms, we have $\iota_{dr}=(0,\ \weight)$, 
\begin{equation}\label{eq:static1form2}
	\begin{gathered}
		\hdelta=(\weight^{-1}\pa_t,\  -\weight\pa_r-\weight')=(\weight^{-1}\pa_t,\  -\pa_r\weight),\quad\hdelta^*=\begin{pmatrix}
			\pa_t & -\frac 12\weight\weight'\\
			\frac 12 \weight\pa_r\weight^{-1}&\frac 12\pa_t\\
			0&\weight^{-1/2}\pa_r\weight^{1/2}
		\end{pmatrix},\\
		\pa^i r\hnabla_i=
		\begin{pmatrix}
			\weight\pa_r-\frac12\weight'&0\\
			0&	\weight\pa_r+\frac12\weight'
		\end{pmatrix}=
		\begin{pmatrix}
			\weight^{3/2}\pa_r\weight^{-1/2}&0\\
			0&	\weight^{1/2}\pa_r\weight^{1/2}
		\end{pmatrix},\\
		\hstar=\begin{pmatrix}
			0&-\weight\\
			-\weight^{-1}&0	
		\end{pmatrix},\quad
		2dr\otimes_s(\cdot)=
		\begin{pmatrix}
			0&0\\
			1&0\\
			0&2
		\end{pmatrix}.
	\end{gathered}
\end{equation}
and 
\begin{equation}\label{eq:static1form3}
	\hBox=\begin{pmatrix}
		-\weight^{-1}\pa_t^2+\weight\pa_r^2-\frac 12\weight''&\weight'\pa_t\\
		\weight^{-2}\weight'\pa_t&	-\weight^{-1}\pa_t^2+\weight\pa_r^2+2\weight'\pa_r+\frac 12\weight''
	\end{pmatrix}.
\end{equation}
On symmetric $2$-tensors, we find that 
\begin{equation}
	\begin{gathered}
		\hdelta=\begin{pmatrix}
			\weight^{-1}\pa_t&-\pa_r\weight&0\\
			-\frac{1}{2}\weight'\weight^{-2}&\weight^{-1}\pa_t&-\weight^{-1/2}\pa_r\weight^{3/2}
		\end{pmatrix},\quad \iota_{dr}=
		\begin{pmatrix}
			0&\weight&0\\
			0&0&\weight
		\end{pmatrix},\\
		\pa^ir \hnabla_i=
		\begin{pmatrix}
			\weight\pa_r-\weight'&0&0\\
			0&\weight\pa_r&0\\
			0&0&	\weight\pa_r+\weight'
		\end{pmatrix}=
		\begin{pmatrix}
			\weight^2\pa_r\weight^{-1}&0&0\\
			0&\weight\pa_r&0\\
			0&0&\pa_r\weight
		\end{pmatrix},\\
		\hBox=	-\weight^{-1}\pa_t^2+\weight\pa_r^2+
		\begin{pmatrix}
			-\weight'\pa_r+\frac 12\weight^{-1}\weight'^2-\weight''&2\weight'\pa_t&-\frac12\weight\weight'^2\\
			\weight^{-2}\weight'\pa_t&\weight'\pa_r-\weight^{-1}\weight'^2&\weight'\pa_t\\
			-\frac 12 \weight^{-3}\weight'^2&2\weight^{-2}\weight'\pa_t&3\weight'\pa_r+\frac 12\weight^{-1}\weight'^2+\weight''
		\end{pmatrix}.
	\end{gathered}
\end{equation}

\subsection{Calculation of $D_{(g, dA)}T(\dg, d\dA)$}Recall that the Reissner-Nordstr\"{o}m electromagnetic $4$-potential is given by 
\begin{equation}\label{eq:4potentialt*}
	A=\BQ r^{-1}dt_{*},\quad F=\BQ r^{-2}dt_*\wedge dr=\BQ r^{-2}\widehat{\mbox{vol}}
\end{equation}
Since $\hvol_{ij}=\epsilon_{ij}, \hvol_{ia}=\hvol_{ab}=0$ where $\epsilon$ is the Levi-Civita symbol, i.e., $\epsilon_{12}=-\epsilon_{21}=1$, it is clear that $\hvol^{\al\be}=-\hvol_{\al\be}$. A direct computation implies
\[
F^{\al\be}F_{\al\be}=-\frac{2\BQ^2}{r^4},\quad \dg^{\al\be}F_{\m\al}F_{\n\be}=\begin{cases}\frac{\BQ^2}{r^4}\left(\dg_{\m\n}-\hg_{\m\n}\htr\dg\right)\quad \mbox{if } \m, \n\in\{i,j\}\\
	0\quad\mbox{otherwise}
\end{cases}.
\]
Then according to \eqref{eq:DT}, we find
\begin{equation}
	\begin{split}
		\left(D_{(g, dA)}T(\dg, 0)	\right)_{ij}&=-\frac{\BQ^2}{r^4}\left(\dg_{ij}-\hg_{ij}\htr\dg-\frac 12\dg_{ij}+\frac 12\hg_{ij}\htr\dg\right)=-\frac{\BQ^2}{2r^4}\left(\dg_{ij}-\hg_{ij}\htr\dg\right),\\
		\left(D_{(g, dA)}T(\dg, 0)	\right)_{ia}&=\frac{\BQ^2}{2r^4}\dg_{ia},\quad \left(	D_{(g, dA)}T(\dg, 0)\right)_{ab}=\frac{\BQ^2}{2r^4}\left(\dg_{ab}-r^2\sg_{ab}\htr\dg\right)
	\end{split}
\end{equation}
which means
\begin{equation}\label{eq:DTofg}
	D_{(g, dA)}T(\cdot, 0)=\frac{\BQ^2}{2r^4}
	\begin{pmatrix}
		\hg\htr	\!-\!1&0&0\\
		0&1&0\\
		-r^2\sg\htr&0&1	
	\end{pmatrix}.
\end{equation}
Since \begin{gather*}
	F^{\al\be}\dF_{\al\be}=\BQ r^{-2}\hvol^{ij}\dF_{ij}
	%=-2\BQ r^{-2}\dF_{t_* r}=2\BQ r^{-2}\dF_{t_* r}\hstar\hvol%=2\BQ r^{-2}\hstar\dF 
	,\\
	g^{\al\be}\dF_{i\al}F_{j\be}+g^{\al\be}F_{i\al}\dF_{j\be}=-2\BQ r^{-2}\dF_{t_* r}\hg_{ij}=2\BQ r^{-2}\dF_{t_* r}(\hstar\hvol)\hg_{ij},\\
	g^{\al\be}\dF_{a\al}F_{i\be}=\hg^{kj}\dF_{ak}F_{ij}=-\BQ r^{-2}\hg^{kj}\epsilon_{ij}\dF_{ka}=\BQ r^{-2}\hstar\dF_{\bullet a}
\end{gather*}
where we use the fact $\hstar\hvol=-1$ and the Hodge star operator $\hstar$ only acts on the aspherical part, we have
\begin{equation}\label{eq:DTofA}
	D_{(g, dA)}T(0, d(\cdot))=-\BQ r^{-2}
	\begin{pmatrix}
		-\hg\hstar\hd&0\\
		\hstar\sd&-\hstar\hd\\
		r^2\sg\hstar\hd&0
	\end{pmatrix}.
\end{equation}
This finishes the calculation of $\mathscr{L}_1(\dg, d\dA)$.

\subsection{Calculation of $D_{(g, dA)}(\delta_g dA)(\dg, d\dA)$}
It remains to consider $\mathscr{L}_2(\dg, d\dA)$ which is the linearization of $\delta_gdA$. We again use the following splitting 
\begin{equation}\label{eq:splitof2form}
	T^*M=T^*_{AS}\oplus T^*_{S}, \quad\wedge^2T^*M=\wedge^2T^*_{AS}\oplus(T^*_{AS}\wedge T^*_{S})\oplus\wedge^2T^*_{S}.
\end{equation}
With respect to the above splitting \eqref{eq:splitof2form}, $d:T^*M\to \wedge^2T^*M$ and $\delta_g: \wedge^2T^*M\to T^*M$ take the form
\begin{equation*}
	\begin{split}
		d=\begin{pmatrix}
			\hd&0\\
			-\sd&\hd\\
			0&\sd
		\end{pmatrix}, \quad
		\delta_g=
		\begin{pmatrix}
			r^{-2}\hdelta r^2& -r^{-2}\sdelta& 0\\
			0&\hdelta &r^{-2}\sdelta
		\end{pmatrix},
	\end{split}
\end{equation*}
and thus by \eqref{eq:L_2}
\begin{equation}\label{eq:DMofA}
	D_{(g, dA)}(\delta_g dA)(0, d(\cdot))=
	\begin{pmatrix}
		r^{-2}\hdelta r^2\hd+r^{-2}\sdelta\sd&-r^{-2}\sdelta\hd\\
		-\hdelta\sd&\hdelta\hd+r^{-2}\sdelta\sd
	\end{pmatrix}.
\end{equation}
Finally we are left with dealing with $D_{(g, dA)}(\delta_g dA)(\dg, 0)$. We first calculate the first covariant derivative on the $2$-form $F$. A direct calculation implies $\hnabla_k\hvol_{ij}=0$, and then
\begin{gather*}
	\nabla_k F_{ij}=-2r^{-1}(\pa_k r)F_{ij},\quad
	\nabla_iF_{ja}=0,\quad \nabla_i F_{ab}=0,\quad \nabla_a F_{ij}=0, \quad \nabla_aF_{ib}=r(\pa^kr)\sg_{ab}F_{ik},\quad \nabla_aF_{bc}=0.
\end{gather*}
Therefore we have
\begin{gather*}
	\dg^{\n\al}\nabla_\al F_{\n i}%=\dg^{jk}\nabla_j F_{ki}+\dg^{ab}\nabla_a F_{bi}
	=-2r^{-1}(\pa_jr)F_{ki}\dg^{jk}+r^{-3}\sg^{ab}\dg_{ab}(\pa^kr)F_{ki}=-2\BQ r^{-3}\hstar\bigl((\pa^j r)\dg_{jk}\bigr)+\BQ r^{-5}(\hstar dr)_i\sg^{ab}\dg_{ab}\\
	\dg^{\n\al}\nabla_\al F_{\n a}=\dg^{bi}\nabla_b F_{ia}=\frac{\BQ}{r^3}(\pa^k r) (\hstar\dg_{\bullet a})_k
\end{gather*}
We also find $(\delta_g G_g\dg)^\kappa F_{\kappa a}=0$ and $-(\delta_g G_g\dg)^\kappa F_{\kappa i}=-\BQ r^{-2}\hstar\left(\delta_g G_g\dg\right)$, thus
\begin{gather*}
	-(\delta_g G_g(\cdot))^\kappa F_{\kappa i}=-\frac{\BQ}{r^2}\begin{pmatrix}
		r^{-2}\hstar\hdelta r^2+\frac 12\hstar\hd\htr&r^{-2}\hstar\sdelta&\frac 12r^{-2}\hstar\hd\str\\
		0&0&0
	\end{pmatrix}.
\end{gather*}
Since $\hg^{ij}\epsilon_{ik}\epsilon_{kl}=\hg_{kl}-\hg_{kl}\htr\hg=-\hg_{kl}$ and $\hnabla_k\hvol_{ij}=0$, we find
\begin{align*}
	\frac 12\left(\nabla^\n\dg^{\kappa}_{\ i}-\nabla^\kappa\dg^{\n}_{\ i}\right)F_{\n\kappa}&=-\frac{\BQ}{2r^2}(\hnabla^{k}\dg^{lj}-\hnabla^{l}\dg^{kj})\hg^{mn}\epsilon_{jm}\epsilon_{in}\epsilon_{kl}\\
	&=-\frac{\BQ}{2r^2}\Bigl(\hnabla^{k}(\dg^{lj}\epsilon_{jm}\epsilon_{kl})-\hnabla^{l}(\dg^{kj}\epsilon_{jm}\epsilon_{kl})\Bigr)\hg^{mn}\epsilon_{in}\\
	&=\frac{\BQ}{2r^2}\Bigl(\hnabla^{k}(\dg_{km}-\hg_{km}\htr\dg)+\hnabla^{l}(\dg_{lm}-\hg_{lm}\htr\dg)\Bigr)\hg^{mn}\epsilon_{in}\\
	&=-\frac{\BQ}{r^2}\hstar\Bigl(\hnabla^k\dg_{km}-\hd\htr\dg\Bigr).
\end{align*}
and using $\hstar\hstar=(-1)^{k(2-k)+1}I$ and $\delta=\hstar\hd\hstar$ on $\wedge^kT^*\hX$ for $1\leq k\leq 2$, we see that 
\begin{align*}
	\frac 12\left(\nabla^\n\dg^{\kappa}_{\ a}-\nabla^\kappa\dg^{\n}_{\ a}\right)F_{\n\kappa}&=\frac{\BQ}{2r^2}\hg^{ik}\hg^{jl}\left(\pa_k\dg_{la}-\pa_l\dg_{ka}-r^{-1}(\pa_kr)\dg_{la}+r^{-1}(\pa_lr)\dg_{ka}\right)\epsilon_{ij}\\
	&=-\frac{\BQ}{r^3}(\pa^k r) (\hstar\dg_{\bullet a})_k+\frac{\BQ}{r^2}\left(2r^{-1}(\pa^k r) (\hstar\dg_{\bullet a})_k+\hstar\hd\dg_{\bullet a}\right)\\
	&=-\frac{\BQ}{r^3}(\pa^k r) (\hstar\dg_{\bullet a})_k+\frac{\BQ}{r^2}\left(-r^2(\pa^k r^{-2}) (\hstar\dg_{\bullet a})_k+\hdelta\hstar\dg_{\bullet a}\right)\\
	&=-\frac{\BQ}{r^3}(\pa^k r) (\hstar\dg_{\bullet a})_k+\BQ\hdelta r^{-2}\hstar\dg_{\bullet a}.		
\end{align*}
Combining the above calculation and \eqref{eq:L_2} yields
\begin{equation}\label{eq:DMofg}
	D_{(g, dA)(\delta_g dA)}(\dg, 0)=\frac{\BQ}{r^2}\begin{pmatrix}
		\frac 12\hstar\hd\htr&-r^{-2}\hstar\sdelta&-\frac 12\hstar\hd r^{-2}\str\\
		0&r^2\hdelta r^{-2}\hstar&0
	\end{pmatrix}.
\end{equation}

Lastly, we discuss the calculation of the gauge terms for the electromagnetic field. Since $A=-\BQ r^{-1}dt_*$, we have 
\begin{align*}
	(\mathcal{L}_{\omega^\sharp}A)_i=\omega^\al\pa_\al A_i+A_\al\pa_i\omega^\al,\quad 	(\mathcal{L}_{\omega^\sharp}A)_a=\omega^\al\pa_\al A_a+A_\al\pa_a\omega^\al=\pa_a(A_\al\omega^\al)
\end{align*}
and as a consequence
\begin{equation}\label{eq:liederiofA}
	\mathcal{L}_{(\cdot)^\sharp}A=	\begin{pmatrix}
		\hd\iota_A+\iota_{(\cdot)}\hd A&0\\
		\sd\iota_{A}&0
	\end{pmatrix}.
\end{equation}

\section{Calculation of the subprincipal operator at trapping}
\label{app:trap}
In this section, we include the discussion of the subprincipal operator of $\mathcal{P}_{b_0,\gamma}, \mathcal{W}_{b_0,\gamma}$ and $L	_{b_0,\gamma}$ at the trapped sets $K$ for the sake of completeness. The arguments are based on \cite{H17}, \cite{H18} and \cite{HV16}. In the subsequent discussion, we will list (without proof) the relevant results for the invariant formalism of the subprincipal operator with references (mostly from \cite{H17}), and then carry out all the relevant computations.

 According to \cite[Theorem 1.1]{D16} and \cite[Theorem 4.5]{HV16} (the microlocalized version) (see also the discussion in \cite[\S2]{H17}), a sufficient condition for high energy estimates at the trapped set for a semiclassical operator $P_h(z)$ acting on the covariant rank $k$ tensor bundle $\mathcal{T}_k$ to hold is
\begin{equation}
	\label{EqBoundAtTrapping}
\sigma_{1,h}\Bigl(\frac{1}{2ih}(P_h(z)-P_h(z)^*)\Bigr)<\nu_{\min}/2
\end{equation}
at the trapped set $K$, where $\nu_{\min}>0$ is the minimal normal expansion rate of the Hamilton flow at the trapping, see Proposition \ref{prop:normallyhyperbolictrap}. Here, the adjoint is taken with respect to a \emph{positive definite inner product} on $\mathcal{T}_k$. We note that the natural inner product induced by $g$, with respect to which $\Box_{g}$ is symmetric, is not positive definite, except when $k=0$, i.e. for the scalar wave equation. One can introduce a pseudodifferential inner product such that \eqref{EqBoundAtTrapping} can be arranged, with the adjoint taken with respect to this pseudodifferential inner product. 
We work with classical, i.e.\ one-step polyhomogeneous, symbols and operators, and denote by $S^m_{\mathrm{hom}}(T^*X\setminus 0)$ symbols which are homogeneous of degree $m$ with respect to dilations in the fibers of $T^*X\setminus 0$.

\begin{defn}[{\cite[Definition 3.1]{H17}}]
	\label{DefPsdoInnerProduct}
	A \emph{pseudodifferential inner product} (or $\Psi$-inner product) \emph{on the vector bundle $\mathcal{E}\to X$} is a pseudodifferential operator $B\in\Psi^0(X;\mathcal{E}\otimes\Omega^\frac12,\overline{\mathcal{E}}^*\otimes\Omega^\frac12)$ satisfying $B=B^*$, and such that moreover the principal symbol $\sigma^0(B)=b\in S^0_{\mathrm{hom}}(T^*X\setminus 0;\pi^*\mathrm{Hom}(\mathcal{E},\overline{\mathcal{E}}^*))$ of $B$ satisfies
	\begin{equation}
		\label{EqPsdoInnerPosDef}
		\langle b(x,\xi)u,\iota(u)\rangle > 0
	\end{equation}
	for all non-zero $u\in\mathcal{E}_x$, where $x\in X$, $\xi\in T^*_x X\setminus 0$. If the context is clear, we will also call the sesquilinear pairing
	\[
	C^\infty(X,\mathcal{E}\otimes\Omega^\frac12)\times C^\infty(X,\mathcal{E}\otimes\Omega^\frac12) \ni (u,v) \mapsto \int_X \langle B(x,D)u,\iota(v)\rangle
	\]
	the pseudodifferential inner product associated with $B$.
\end{defn}

The use of a pseudodifferential inner product is equivalent to considering a conjugated operator $QP_h(z)Q^-$, where $Q\in\Psi_{h}^0(X,\mathcal{T}_k)$ is a carefully chosen elliptic operator with parametrix $Q^-$. For any $\epsilon>0$, we can arrange $\sigma_{1,h}(\frac{1}{2ih}(QP_h(z)Q^--(QP_h(z)Q^-)^*))<\epsilon$ (with the adjoint taken relative to an ordinary positive definite inner product on $\mathcal{T}_k$), thus \eqref{EqBoundAtTrapping} holds for $P_h(z)$ replaced by $QP_h(z)Q^-$. 

Let $\pi:T^*X\setminus 0\to X$ be the projection. We will work in standard pseudodifferential operator setting. We specialize to the case that $P\in\Psi^m(X,\mathcal{E}\otimes\Omega^\frac12)$ has a real, scalar principal symbol, which is the case of interest in our application. Fix a coordinate system of $X$ and a local trivialization of $\mathcal{E}$, then the full symbol of $P$ is a sum of homogeneous symbols $p\sim p_m+p_{m-1}+\ldots$, with $p_j$ homogeneous of degree $j$ and valued in complex $N\times N$ matrices. According to \cite[Theorem 18.1.33]{H07}, the subprincipal symbol
\begin{equation}
	\label{EqScalarSubprincipal}
	\sigma_\mathrm{sub}(P)=p_{m-1}(x,\xi)-\frac{1}{2i}\sum_j\pa_{x_j\xi_j}p_m(x,\xi) \in S^{m-1}_{\mathrm{hom}}(T^*X\setminus 0,\BC^{N\times N})
\end{equation}
is well-defined under changes of coordinates; however, it does depend on the choice of local trivialization of $\mathcal{E}$. We would like a frame-independent notion of the subprincipal symbol since this provides us with the freedom to choose particularly convenient local frames in concrete computations.

\subsection{Invariant formalism for the subprincipal symbols of operators acting on bundles}

\begin{defn}[{\cite[Definition 3.8]{H17}}]
	\label{DefInvSubpr}
	For $P\in\Psi^m(X,\mathcal{E}\otimes\Omega^\frac12)$ with scalar principal symbol $p$, there is a well-defined \emph{subprincipal operator} $S_\mathrm{sub}(P)\in\mathrm{Diff}^1(T^*X\setminus 0,\pi^*\mathcal{E})$, homogeneous of degree $m-1$ with respect to dilations in the fibers of $T^*X\setminus 0$, defined as follows: if $\{e_1(x),\ldots,e_N(x)\}$ is a local frame of $\mathcal{E}$, define the operators $P_{jk}\in\Psi^m(X,\Omega^\frac12)$ by $P(\sum_k u_k(x)e_k(x))=\sum_{jk} P_{jk}(u_k)e_j(x)$, $u_k\in C^\infty(X,\Omega^\frac12)$. Then
	\[
	S_\mathrm{sub}(P)\Bigl(\sum_k q_k(x,\xi)e_k(x)\Bigr) := \sum_{jk} (\sigma_\mathrm{sub}(P_{jk})q_k)e_j - i\sum_k(H_p q_k)e_k.
	\]
\end{defn}

The symbols of commutators and imaginary parts can be expressed in a completely invariant fashion.
\begin{prop}[{\cite[Proposition 3.11]{H18}}]
	\label{PropImagSymbolInv}
	Let $P\in\Psi^m(X,\mathcal{E}\otimes\Omega^\frac12)$ be a pseudodifferential operator with scalar principal symbol $p$.
	\begin{enumerate}
		\item Suppose $Q\in\Psi^{m'}(X,\mathcal{E}\otimes\Omega^\frac12)$ is an operator acting on $\mathcal{E}$-valued half-densities, with principal symbol $q$. Then
		\begin{equation}
			\label{EqSubprImagPart}
			\sigma_{m+m'-1}([P,Q]) = [S_\mathrm{sub}(P),q].
		\end{equation}
		If $Q$ is elliptic with parametrix $Q^-$, then
		\begin{equation}
			\label{EqSubprChangeOfBasis}
			S_\mathrm{sub}(QPQ^-)=q S_\mathrm{sub}(P)q^{-1}.
		\end{equation}
		\item Suppose in addition that $p$ is real. Let $B$ be a $\Psi$-inner product (pseudodifferential inner product) on $\mathcal{E}$ with principal symbol $b$, then
		\begin{equation}
			\label{EqInvImagSymbol}
			\sigma_{m-1}(\IM^B P) = \IM^b S_\mathrm{sub}(P),
		\end{equation}
		where $\IM^BP=\frac{1}{2i}(P-P^{*B})$ and $\IM^b S_\mathrm{sub}(P)=\frac{1}{2i}\bigl(S_\mathrm{sub}(P)-S_\mathrm{sub}(P)^{*b}\bigr)$; we take the adjoint of $P$ with respect to the $\Psi$-inner product $B$ and the adjoint of the differential operator $S_\mathrm{sub}(P)$ with respect to the inner product $b$ on $\pi^*\mathcal{E}$ and the symplectic volume density on $T^*X$.
	\end{enumerate}
\end{prop}

The imaginary part $\IM^B P$ of an operator $P$ with respect to a $\Psi$-inner product $B$ can be interpreted in terms of the imaginary part of a conjugated version of $P$ with respect to a standard inner product.

\begin{prop}[{\cite[Proposition 3.12]{H17}}]
	\label{PropPsdoInnerAsConjugation}
	Let $B$ be a $\Psi$-inner product on $\mathcal{E}$. Then for any positive definite Hermitian inner product $B_0\in C^\infty(X,\mathrm{Hom}(\mathcal{E}\otimes\Omega^\frac12,\overline{\mathcal{E}}^*\otimes\Omega^\frac12))$ on $\mathcal{E}$, there exists an elliptic operator $Q\in\Psi^0(X,\mathrm{End}(\mathcal{E}\otimes\Omega^\frac12))$ such that $B-Q^*B_0Q\in\Psi^{-\infty}(X,\mathrm{Hom}(\mathcal{E}\otimes\Omega^\frac12,\overline{\mathcal{E}}^*\otimes\Omega^\frac12))$.
	
	In particular, denoting by $Q^-\in\Psi^0(X,\mathrm{End}(\mathcal{E}\otimes\Omega^\frac12))$ a parametrix of $Q$, we have for any $P\in\Psi^m(X,\mathcal{E}\otimes\Omega^\frac12)$ with real and scalar principal symbol.
	\begin{equation}
		\label{EqPsdoInnerToNormal}
		Q(\IM^B P)Q^- = \IM^{B_0}(QPQ^-),
	\end{equation}
	and $\sigma_{m-1}(\IM^B P)$ and $\sigma_{m-1}(\IM^{B_0}(QPQ^-))$ (which are self-adjoint with respect to $\sigma^0(B)$ and $B_0$, respectively, hence diagonalizable) have the same eigenvalues.
\end{prop}

Therefore, the analysis of  $\sigma_{m-1}(\IM^{B_0}(QPQ^-)))$ is reduced to finding an inner product $b$ such that $(S_\mathrm{sub}(L)-S_\mathrm{sub}(L)^{*b})$ has a desired form. (In our applications, the choice of $b$ will be clear from an inspection of the form of $S_\mathrm{sub}(L)$).

Let $(M,g)$ be a smooth manifold equipped with a metric tensor $g$ of arbitrary signature. Denote by $\mathcal{T}_k M=\bigotimes^k T^*M$, $k\geq 1$, the bundle of (covariant) tensors of rank $k$ on $M$. The metric $g$ induces a metric (which we also call $g$) on $\mathcal{T}_k M$. The subprincipal operator of $\Delta_k=\tr\nabla^2\in\mathrm{Diff}^2(M,\mathcal{T}_k M)$ acting on the bundle $\mathcal{T}_k M$ has a nice form.

\begin{prop}[{\cite[Proposition 4.1]{H17}}]
	\label{PropLaplaceSubpr}
	The subprincipal operator of $\Delta_k$ is
	\begin{equation}
		\label{EqLaplaceSubpr}
		S_\mathrm{sub}(\Delta_k)(x,\xi) = i\nabla^{\pi^*\mathcal{T}_k M}_{H_G} \in \mathrm{Diff}^1(T^*M\setminus 0,\pi^*\mathcal{T}_k M),
	\end{equation}
	where $\nabla^{\pi^*\mathcal{T}_k M}$ is the pullback connection, with $\pi\colon T^*M\setminus 0\to M$ being the projection, and $G=|\xi|^2_{g^{-1}(x)}$.
\end{prop}

\subsection{Calculation  of subprincipal operators}
We notice that modulo terms of order $0$, which are sub-subprincipal and thus irrelevant in the calculation of $S_{\mathrm{sub}}(\mathcal{P}_{b_0,\gamma}), S_{\mathrm{sub}}(\mathcal{W}_{b_0,\gamma})$ and $S_{\mathrm{sub}}(L_{b_0,\gamma})$ (for simplicity of notation, from now on, we drop the notations $b_0, \gamma$, i.e., $g=g_{b_0}$), 
we have
\begin{equation}
	-2\mathcal{P}\equiv\Box_{g,1}-2\delta_gG_g(\tilde{\delta}_g^*-\delta_g^*),\quad 
		-2\mathcal{W}\equiv\Box_{g,1}-2(\tilde{\delta}_g-\delta_g)G_g\delta_g^*
\end{equation}
and
\begin{equation}
	\label{EqESGL}
		-L(\dg,\dA) \equiv \Bigl( \Box_g\dg- 2(\tilde{\delta}_g^*-\delta_g^*)\delta_g G_g\dg -2\delta^*_g(\tilde{\delta}_g-\delta_g)G_g\dg+ L_{12}(\dA),\  \Box_g\dA + L_{21}(\dg) \Bigr),
\end{equation}
where
\begin{gather*}
	L_{12}(\dA) = 8\tr_g^{24}(F\otimes_s d\dA) - 2 g^{-1}(F,d\dA)\,g, \quad F=dA=d(\frac{\BQ_0}{r}dt)=\frac{\BQ_0}{r^2}dt\wedge dr,\\
L_{21}(\dg) =  \tr_g^{12}(\delta_g G_g\dg\otimes F) - \frac{1}{2}\tr_g^{24}\tr_g^{35}(d^\nabla\dg\otimes F),\quad (d^\nabla\dg)_{\nu}^{\ \mu\kappa}=\nabla^{\mu}\dg_{\nu}^{\ \kappa}-\nabla_\nu^{\kappa}\dg^{\ \mu}.
\end{gather*}
Since $\tilde{\delta}_g^*-\delta_g^*$ and $\tilde{\delta}_g-\delta_g$ are compactly supported away from the trapping for Reissner-Nordstr\"{o}m metric, we further have that modulo terms of order $0$
\begin{equation}
		\label{EqESGLTrapping}
	\begin{gathered}
		-2\mathcal{P}\equiv-2\mathcal{W}\equiv\Box_{g,1}\\
	-L(\dg,\dA) \equiv \Bigl( \Box_g\dg+L_{12}(\dA),\  \Box_g\dA + L_{21}(\dg) \Bigr)
	\end{gathered}
\end{equation}
at the trapping $K$.

Let $g=g_{b_0}$ be the Reissner-N\"{o}rdstrom metric with parameter $b_0=(\Bm_0, 0,\BQ_0)$
\begin{equation}
	g=-\alpha^2dt^2+\alpha^{-2}dr^2+r^2\sg,\quad \alpha=\sqrt{\mu_{b_0}}=\sqrt{1-\frac{2\Bm_0}{r}+\frac{\BQ_0^2}{r^2}}.
\end{equation}
That is, we work in the coordinates $(t,r,\omega),\omega\in\BS^2$ at the trapping $K$. Let $(\sigma, \xi, \eta)$ be the dual variables to $(t, r, \omega)$. We define
\begin{equation}
	e^0:=\alpha dt,\quad e^1:=\alpha^{-1}dr
\end{equation}
and use the following splittings of $T^*\mathcal{M}$ and $S^2\,T^*\mathcal{M}$
\begin{equation}\label{Eqsplittingtrap}
	\begin{split}
T^*\mathcal{M}&=\langle e^0\rangle\oplus\langle e^1\rangle \oplus T^*\BS^2,\\
S^2\,T^*\mathcal{M}&=\langle e^0e^0\rangle\oplus\langle2e^0\otimes_s e^1\rangle\oplus (2e^0\otimes_s T^*\BS^2)\oplus \langle e^1e^1\rangle\oplus (2e^1\otimes_s T^*\BS^2)\oplus S^2\,T^*\BS^2.
\end{split}
\end{equation}
In this section, we are interested in the subprincipal operators of $\Box_{g,1}$ and $ L$ at the trapped set 
\begin{equation}
	K=\{r=r_p, \xi=0, \sigma^2=\mu_{b_0}r^{-2}\abs{\eta}^2\},\quad \mbox{where}\quad \pa_r(r\alpha^{-1})(r_p)=0,\quad \abs{\eta}^2=\abs{\eta}^2_{\sg^{-1}}.
\end{equation}
Therefore, at the trapping $K$, we have
\[
	H_{\alpha^{2}\xi^2+r^{-2}\abs{\eta}^2}=2r^{-3}\abs{\eta}^2\pa_{\xi}+r^{-2}H_{\abs{\eta}^2}
\]
and
\[
	\sigma^2H_{\alpha^{-2}}-	H_{\alpha^{2}\xi^2+r^{-2}\abs{\eta}^2}=-r^{-2}H_{\abs{\eta}^2}.
\]

Let $x^0=t$, and $x'=(r,\omega)$ be coordinates on $X$ (independent of $t$). We let Greek indices $\mu,\nu,\lambda,\ldots$ run from $0$ to $3$. Moreover, the canonical dual variables $\xi_0=:\sigma$ and $\xi'=(\xi,\eta)$ on the fibers of $T^*\mathcal{M}$ are indexed by decorated Greek indices $\wt\mu$ running from $0$ to $3$. For a section $u$ of $\pi^*T^*\mathcal{M}$, we have
\[
\nabla^{\pi^*T^*M}_\mu u_\nu = \nabla^{M}_\mu u_\nu, \quad \nabla^{\pi^*T^*M}_{\wt\mu} u_\nu = \pa_{\wt\mu}u_\nu.
\]

\subsubsection{Calculation of the subprincipal operator of $\Box_{g,1}$}
According to Proposition \ref{PropLaplaceSubpr} and using the fact that $G=-\alpha^{-2}\sigma^2+\alpha^2\xi^2+r^{-2}\abs{\eta}^2$, a direct calculation implies that at trapping $K=\{r=r_p,\xi=0, \sigma^2=\mu_{b_0}r^{-2}\abs{\eta}^2\}$
\begin{equation}
	\label{EqSDSSubpr}
		i S_\mathrm{sub}(\Box_{g,1}) 
=\begin{pmatrix}
			2\alpha^{-2}\sigma\pa_t - r^{-2}H_{|\eta|^2} & 0 & 0\\
0	& 2\alpha^{-2}\sigma\pa_t - r^{-2}H_{|\eta|^2} &  0 \\
			0 & 0 & 2\alpha^{-2}\sigma\pa_t - r^{-2}\nabla^{\pi_{\BS^2}^*T^*\BS^2}_{H_{|\eta|^2}}
					\end{pmatrix}+S_{(1)}
\end{equation}
where
\begin{equation}
	\label{EqSDSSubprzero}
	S_{(1)}=\begin{pmatrix}
		0 & -2r^{-1}\sigma & 0\\
		-2r^{-1}\sigma	& 0&  2\alpha r^{-3}\iota_\eta \\
		0 & -2\alpha r^{-1}\eta & 0
	\end{pmatrix}.
\end{equation}

For simplicity of calculation, we further decompose $\pi^*T^*\mathcal{M}\to T^*\mathcal{M}$ over the trapping $K$. For $\eta\in T^*\BS^2\setminus o$, we define $\eta^\perp:=\sstar\eta$ and
\[
\widehat{\eta}  :=\alpha\sigma^{-1}\eta ,\quad\widehat{\eta}^\perp=\alpha\sigma^{-1}\eta^\perp.
\]
Then we write
\begin{equation}
	\label{EqESGTrapMlSplitSph}
	\begin{gathered}
		\pi_{\BS^2}^*T^*\BS^2= \langle \widehat\eta \rangle \oplus \langle\widehat\eta^\perp\rangle, 
		%\quad \pi_{\BS^2}^*\Lambda^2 T^*\BS^2 = \langle\widehat\eta\wedge\widehat\eta^\perp\rangle, 
		\\
		\pi_{\BS^2}^*S^2T^*\BS^2 = \langle \widehat\eta\widehat\eta \rangle \oplus \langle 2\widehat\eta\widehat\eta^\perp\rangle \oplus \langle\widehat\eta^\perp\widehat\eta^\perp\rangle,
	\end{gathered}
\end{equation}
and this induces the following decompositions
\begin{equation}
	\label{EqESGTrapMlSplit}
	\begin{split}
		\pi^* T^*M &= \langle e^0\rangle \oplus \Bigl(\langle e^1\rangle \oplus \bigl(\langle \widehat\eta \rangle \oplus \langle\widehat\eta^\perp\rangle\bigr)\Bigr), \\
	%	\pi^*\Lambda^2 T^*M &= \Bigl(\langle e^0\wedge e^1\rangle \oplus\bigl(\langle e^0\wedge\widehat\eta \rangle \oplus\langle e^0\wedge\widehat\eta^\perp\rangle\bigr)\Bigr) \\
	%	&\hspace{6em} \oplus\bigl(\langle e^1\wedge\widehat\eta\rangle \oplus \langle e^1\wedge\widehat\eta^\perp\rangle \bigr) \oplus \langle\widehat\eta\wedge\widehat\eta^\perp\rangle, \\
		\pi^*S^2 T^*M &= \Bigl(\langle e^0e^0\rangle \oplus \bigl(\langle 2e^0e^1\rangle \oplus \bigl( \langle 2e^0\widehat\eta\rangle \oplus \langle 2e^0\widehat\eta^\perp\rangle \bigr)\bigr)\Bigr) \\
		&\hspace{6em} \oplus \Bigl(\langle e^1e^1\rangle \oplus \bigl(\langle 2e^1\widehat\eta\rangle \oplus \langle 2e^1\widehat\eta^\perp\rangle \bigr)\Bigr) \\
		&\hspace{6em} \oplus \bigl(\langle\widehat\eta\widehat\eta\rangle \oplus \langle 2\widehat\eta\widehat\eta^\perp\rangle \oplus \langle\widehat\eta^\perp\widehat\eta^\perp\rangle \bigr),
	\end{split}
\end{equation}
In terms of \eqref{EqESGTrapMlSplitSph}, we have
\[
\eta=\begin{pmatrix}\alpha^{-1}\sigma\\0\end{pmatrix},\quad r^{-2}\iota_\eta=\begin{pmatrix} \alpha^{-1}\sigma& 0 \end{pmatrix}\quad \mbox{at}\quad K.
\]
Therefore, in terms of the splitting  \eqref{EqESGTrapMlSplit}, the operator $S_{(1)}$  in \eqref{EqSDSSubprzero} at trapping $K$ is given as
\begin{equation}
	\label{EqL22microK}
	S_{(1)}=r^{-1}\sigma\begin{pmatrix}
	0 & -2 & 0&0\\
	-2&0&2&0\\
	0& -2& 0&0\\
	0&0&0&0
\end{pmatrix}.
\end{equation}
Now, for any fixed $\epsilon>0$, we define the change of basis matrix
\[
q_1=\left(
\begin{array}{cccc}
	0 & 0 & 0 & 1 \\
	0 & 0 & \frac{1}{\epsilon} & 0 \\
	0 & -\frac{2}{\epsilon^2} & 0 & 0 \\
	-\frac{4}{\epsilon^3} & 0 & -\frac{4}{\epsilon^3} & 0 \\
\end{array}
\right),
\]
and then we have
\[
q_1S_{(1)}q_1^{-1}=r^{-1}\sigma\left(
\begin{array}{cccc}
	0 & 0 & 0 & 0 \\
	0 & 0 & \epsilon& 0 \\
	0 & 0 & 0 & \epsilon \\
	0 & 0 & 0 & 0 \\
\end{array}
\right).
\]
Since $q_1$ is a constant coefficient operator in the region where the splitting \eqref{EqESGTrapMlSplit} is well defined, in particular, in a neighborhood of the trapping $K$, it commutes with the diagonal part $iS_{\mathrm{sub}} (\Box_{g,1})-S$ of $S_{\mathrm{sub}} (\Box_{g,1})$. Equipping the 1-form bundle over $M$ with the Hermitian inner product $B_0^1$ which is given by an identity matrix in the splitting \eqref{EqESGTrapMlSplit}, $q_1S_\mathrm{sub}(\Box_{g,1})q_1^{-1}$ has imaginary part (with respect to $B_0$) of size $\mathcal{O}(\epsilon)$. Put differently, $S_{\mathrm{sub}}(\Box_{g,1})$ has imaginary part of size $\mathcal(\epsilon)$ relative to the Hermitian inner product $b:=B^1_0(q\cdot,q\cdot)$, which is the symbol of a pseudodifferential inner product on $\pi^*T^*M$. To summarize,

\begin{thm}
	\label{ThmSDSNilpotentAtTrapping}
	For any $\epsilon>0$, there exists a (positive definite) $t$-independent pseudodifferential inner product $B=b(x,D)$ on $T^*\mathcal{M}$ (thus, $b$ is an inner product on $\pi^*T^*\mathcal{M}$, homogeneous of degree $0$ with respect to dilations in the base $T^*\mathcal{M}\setminus 0$), such that
	\[
	\sup_K |\sigma|^{-1}\left\|\frac{1}{2i}(S_\mathrm{sub}(\Box_{g,1})-S_\mathrm{sub}(\Box_{g,1})^{*b})\right\|_b \leq \epsilon,
	\]
	where $K$ is the trapped set. Put differently, there is an elliptic pseudodifferential operator $Q$, invariant under $t$-translations, acting on sections of $T^*\mathcal{M}$, with parametrix $Q^-$, such that relative to the ordinary positive definite inner product $B_0^1$ we have
	\[
	\sup_\Gamma |\sigma|^{-1}\left\|\sigma_1\left(\frac{1}{2i}(Q\Box_{g,1}Q^- - (Q\Box_{g,1} Q^-)^{*B^1_0})\right)\right\|_{B^1_0} \leq \epsilon.
	\]

\end{thm}

\subsubsection{Calculation of the subprincipal operator of $L$}
Using Proposition \ref{PropLaplaceSubpr} again, we see that  $S_{\mathrm{sub}}(\Box_{g,2})=i\nabla^{\pi^*S^2T^*\mathcal{M}}_{H_G}$. Since $\pi^*S^2T^*\mathcal{M}$ is simply the restriction of the product connection $\nabla^{\pi^*T^*\mathcal{M}}\otimes\nabla^{\pi^*T^*\mathcal{M}}$ to $\pi^*S^2T^*\mathcal{M}$, it follows that 
\[
S_\mathrm{sub}(\Box_{g,2})(w_1 w_2) = S_\mathrm{sub}(\Box_{g,1}) w_1\cdot w_2 + w_1\cdot S_\mathrm{sub}(\Box_{g,1}) w_2
\]
for $w_1,w_2\in C^\infty(T^*\mathcal{M},\pi^*T^*\mathcal{M})$, under the splitting \eqref{Eqsplittingtrap}, $S_\mathrm{sub}(\Box_{g,2})$ has a canonical first order part, induced by the canonical first order part of $S_\mathrm{sub}(\Box_{g,1})$ which is given in \eqref{EqSDSSubpr}. The $0$-order part $S_{(2)}$ of $S_\mathrm{sub}(\Box_{g,2})$ is equal to the second symmetric tensor power of the $0$-th order part of $S_\mathrm{sub}(\Box_{g,1})$ in \eqref{EqSDSSubprzero}. To summarize, 
\begin{equation}
	\label{EqESGWaveSubpr}
	iS_{\mathrm{sub}}(\Box_{g,2})
	=2\alpha^{-2}\sigma\pa_t-r^{-2}\begin{pmatrix}
		 H_{|\eta|^2} &0&0&0&0&0&\\
		0&	 H_{|\eta|^2} &0&0&0&0\\
		0&0& H^{\pi^*_{\BS^2}T^*\BS^2}_{|\eta|^2} &0&0&0\\
		0&0&0&	 H_{|\eta|^2} &0&0\\
		0&0&0&0& H^{\pi^*_{\BS^2}T^*\BS^2}_{|\eta|^2}&0\\
		0&0&0&0&0&H^{\pi^*_{\BS^2}S^2\,T^*\BS^2}_{|\eta|^2}
	\end{pmatrix}+S_{(2)}
\end{equation}
where
\begin{equation}
	\label{EqESGWaveSubprzero}
	S_{(2)}
	=\begin{pmatrix}
		0              & -4r^{-1}\sigma & 0                 & 0              & 0                 & 0                 \\
		-2r^{-1}\sigma & 0              & 2 \alpha r^{-3}i_\eta & -2r^{-1}\sigma & 0                 & 0                 \\
		0              & -2 \alpha r^{-1}\eta & 0                 & 0              & -2r^{-1}\sigma    & 0                 \\
		0              & -4r^{-1}\sigma & 0                 & 0              & 4 \alpha r^{-3}i_\eta & 0                 \\
		0              & 0              & -2r^{-1}\sigma    & -2 \alpha r^{-1}\eta & 0                 & 2 \alpha r^{-3}i_\eta \\
		0              & 0              & 0                 & 0              & -4 \alpha r^{-1}\eta    & 0
	\end{pmatrix}.
\end{equation}
 Again, we turn to the `microlocal splitting' \eqref{EqESGTrapMlSplitSph}, under which we write 
 \begin{gather*}
 	\eta
 	=\begin{pmatrix}
 		\alpha^{-1}\sigma & 0 \\
 		0            & \frac{1}{2}\alpha^{-1}\sigma \\
 		0            & 0
 	\end{pmatrix}
 	\colon T^*\BS^2\to S^2T^*\BS^2, \\
 	r^{-2}i_\eta
 	=\begin{pmatrix}
 		\alpha^{-1}\sigma & 0               & 0 \\
 		0               & \alpha^{-1}\sigma & 0
 	\end{pmatrix}
 	\colon S^2T^*\BS^2\to T^*\BS^2
 \end{gather*}
 Therefore, in terms of \eqref{EqESGTrapMlSplit}, we see that
 \begin{equation}
 	\label{EqL11microK}
 	S_{(2)}=r^{-1}{\sigma}\begin{pmatrix}
 		0&-4&0&0&0&0&0&0&0&0\\
 		-2&0&2&0&-2&0&0&0&0&0\\
 		0&-2&0&0&0&-2&0&0&0&0\\
 		0&0&0&0&0&0&-2&0&0&0\\
 		0&-4&0&0&0&4&0&0&0&0\\
 		0&0&-2&0&-2&0&0&2&0&0\\
 		0&0&0&-2&0&0&0&0&2&0\\
 		0&0&0&0&0&-4&0&0&0&0\\
 		0&0&0&0&0&0&-2&0&0&0\\
 		0&0&0&0&0&0&0&0&0&0
 	\end{pmatrix}.
 \end{equation}

Let $S_L$ be the 0-th order part of $iS_\mathrm{sub}(-L)$ at $K$, with $L$ given in \eqref{EqESGL}. Acting on the bundle $S^2T^*\mathcal{M}\oplus T^*\mathcal{M}$, we can write $S_L$ as a $2\times 2$ block matrix,
\[
S_L = \begin{pmatrix} S_L^{11} & S_L^{12} \\ S_L^{21} & S_L^{22} \end{pmatrix}\quad \mbox{where} \quad S_L^{11}=S_{(2)},\ S^{22}_L=S_{(1)},\ S^{12}_L=i\sigma_1(L_{12}),\ S_L^{21}=i\sigma_1(L_{21})
\]
with 
\begin{gather*}
	L_{12}(\dA) = 8\tr_g^{24}(F\otimes_s d\dA) - 2 g^{-1}(F,d\dA)\,g, \quad F=dA=d(\frac{\BQ_0}{r}dt)=\frac{\BQ_0}{r^2}dt\wedge dr,\\
	L_{21}(\dg) =  \tr_g^{12}(\delta_g G_g\dg\otimes F) - \frac{1}{2}\tr_g^{24}\tr_g^{35}(d^\nabla\dg\otimes F),\quad (d^\nabla\dg)_{\nu}^{\ \mu\kappa}=\nabla^{\mu}\dg_{\nu}^{\ \kappa}-\nabla_\nu^{\kappa}\dg^{\ \mu}.
\end{gather*}
In the splitting \eqref{Eqsplittingtrap} and 
\[
\Lambda^2T^*\mathcal{M}=\langle e^0\wedge e^1\rangle\oplus \langle e^0\wedge T^*\BS^2\rangle\oplus \langle e^1\wedge T^*\BS^2 \rangle\oplus \Lambda^2T^*\BS^2,
\]
$d:T^*\mathcal{M}\to\Lambda^2T^*\mathcal{M}$ and $\tr_g^{24}(F\otimes_s (\cdot)):\Lambda^2T^*\mathcal{M}\to S^2T^*\mathcal{M}$ are given by
\[
d=\begin{pmatrix}
	-\alpha\pa_r-\alpha'&\alpha^{-1}\pa_t&0\\
	-\sd& 0&\alpha^{-1}\pa_t\\
	0&-\sd&\alpha\pa_r\\
	0&0&\sd
\end{pmatrix},\quad 2\tr_g^{24}(F\otimes_s (\cdot))=\frac{\BQ_0}{r^2}\begin{pmatrix}
2&0&0&0\\
0&0&0&0\\
0&0&-1&0\\
-2&0&0&0\\
0&-1&0&0\\
0&0&0&0
\end{pmatrix},
\] and we also have
\[ g^{-1}(F,(\cdot))g=\begin{pmatrix}
\frac{2\BQ_0}{r^2}&0&0&0\\
0&0&0&0\\
0&0&0&0\\
-\frac{2\BQ_0}{r^2}&0&0&0\\
0&0&0&0\\
-\frac{2\BQ_0}{r^2}r^2\sg&0&0&0
\end{pmatrix}:\Lambda^2T^*\mathcal{M}\to S^2T^*\mathcal{M}.
\]
Therefore, we obtain
\begin{equation}
	\label{symbolofL121}
	\sigma_1(L_{12})=i\frac{4\BQ_0}{r^2}
	\begin{pmatrix}
	-\alpha\xi&\alpha^{-1}\sigma&0\\
	0&0&0\\
0&\eta&-\alpha\xi	\\
		\alpha\xi&-\alpha^{-1}\sigma&0\\
		\eta&0&-\alpha^{-1}\sigma\\
		-\alpha\xi r^2\sg&\alpha^{-1}\sigma r^2\sg&0
	\end{pmatrix}.
\end{equation}
Turning to the `microlocal splitting' \eqref{EqESGTrapMlSplitSph}, we write $\sg=
\sg=\abs{\eta}^{-2}\eta\eta+\abs{\eta}^{-2}\eta^\perp\eta^\perp$ and then we have
\[
\sg=\begin{pmatrix}
	r^{-2}\\
	0\\
	r^{-2}
\end{pmatrix}\quad \mbox{at}\quad K.
\]
Therefore, in terms of the splitting \eqref{EqESGTrapMlSplitSph},
\begin{equation}
	\label{EqL12microK}
i\sigma_1(L_{12})=r^{-1}\sigma\frac{4\BQ_0}{r\alpha}	\begin{pmatrix}
	0&-1&0&0\\
	0&0&0&0\\
	0&-1&0&0\\
	0&0&0&0\\
	0&1&0&0\\
	-1&0&1&0\\
	0&0&0&1\\
	0&-1&0&0\\
	0&0&0&0\\
	0&-1&0&0
\end{pmatrix}\quad \mbox{at}\quad K.
\end{equation}
So it remains to analyze $L_{21}(\dg)$. We first compute
\begin{equation}
	\label{Eqneededlateron1}
	G_g=\begin{pmatrix}
		\frac12&0&0&\frac12&0&\frac{1}{2r^2}\str\\
		0&1&0&0&0&0\\
		0&0&1&0&0&0\\
		\frac12&0&0&\frac12&0&-\frac{1}{2r^2}\str\\
		0&0&0&0&1&0\\
		\frac12r^2\sg&0&0&-\frac12r^2\sg&0&G_{\sg}
	\end{pmatrix},\quad \tr_g^{12}((\cdot)\otimes F)=\begin{pmatrix}
	0&-\frac{\BQ_0}{r^2}&0\\
	-\frac{\BQ_0}{r^2}&0&0\\
	0&0&0
\end{pmatrix}: T^*\mathcal{M}\to T^*\mathcal{M}
\end{equation}
and
\begin{equation}
	\label{Eqneededlateron2}
	\sigma_1(\delta_g)=i\begin{pmatrix}
		\alpha^{-1}\sigma&-\alpha\xi&-r^{-2}\iota_\eta&0&0&0\\
		0&\alpha^{-1}\sigma&0&-\alpha\xi&-r^{-2}\iota_{\eta}&0\\
		0&0&\alpha^{-1}\sigma&0&-\alpha\xi&-r^{-2}\iota_{\eta}
	\end{pmatrix}.
\end{equation}
Therefore, we find that
\begin{equation}
\label{EqL211}
\sigma_1(\tr_g^{12}(\delta_gG_g(\cdot)\otimes F))=i\frac{\BQ_0}{r^2}\begin{pmatrix}
	\frac12\alpha\xi&-\alpha^{-1}\sigma&0&\frac12\alpha\xi&r^{-2}\iota_{\eta}&-\frac{1}{2r^2}
\alpha\xi\str\\
-\frac12\alpha^{-1}\sigma&\alpha\xi&r^{-2}\iota_{\eta}&-\frac12\alpha^{-1}\sigma&0&-\frac{1}{2r^2}\alpha^{-1}\sigma\str\\
0&0&0&0&0&0\end{pmatrix}.
\end{equation}
Next, we calculate
\begin{equation}
	\label{EqL212}
\sigma_1(d^{\nabla}(\cdot)\otimes F)=i\frac{2\BQ_0}{r^2}\begin{pmatrix}
	\alpha\xi&-\alpha^{-1}\sigma&0&0&0&0&\\
	0&\alpha\xi&0&-\alpha^{-1}\sigma&0&0\\
	0&0&\alpha\xi&0&-\alpha^{-1}\sigma&0
\end{pmatrix}.
\end{equation}
In the `microlocal splitting' \eqref{EqESGTrapMlSplitSph}, we have
\[
\str=\begin{pmatrix}
r^2&0&r^2
\end{pmatrix}\quad \mbox{at}\quad K.
\]
and thus we have
\begin{equation}
	\label{EqL21microK}
	i\sigma_1(L_{21})=r^{-1}\sigma\frac{\BQ_0}{2r\alpha}\begin{pmatrix}
		0&0&0&0&0&-2&0&0&0&0\\
		1&0&-2&0&-1&0&0&1&0&1\\
		0&0&0&0&0&-2&0&0&0&0\\
		0&0&0&0&0&0&-2&0&0&0
	\end{pmatrix}.
\end{equation}
Putting \eqref{EqL22microK}, \eqref{EqL11microK}, \eqref{EqL12microK} and \eqref{EqL21microK} together yields
\begin{equation}
S_L=r^{-1}\sigma\left(
	\begin{array}{cccccccccccccc}
		0 & -4 & 0 & 0 & 0 & 0 & 0 & 0 & 0 & 0 & 0 & -8 \BQ' & 0 & 0 \\
		-2 & 0 & 2 & 0 & -2 & 0 & 0 & 0 & 0 & 0 & 0 & 0 & 0 & 0 \\
		0 & -2 & 0 & 0 & 0 & -2 & 0 & 0 & 0 & 0 & 0 & -8 \BQ' & 0 & 0 \\
		0 & 0 & 0 & 0 & 0 & 0 & -2 & 0 & 0 & 0 & 0 & 0 & 0 & 0 \\
		0 & -4 & 0 & 0 & 0 & 4 & 0 & 0 & 0 & 0 & 0 & 8 \BQ' & 0 & 0 \\
		0 & 0 & -2 & 0 & -2 & 0 & 0 & 2 & 0 & 0 & -8 \BQ'& 0 & 8 \BQ' & 0 \\
		0 & 0 & 0 & -2 & 0 & 0 & 0 & 0 & 2 & 0 & 0 & 0 & 0 & 8 \BQ' \\
		0 & 0 & 0 & 0 & 0 & -4 & 0 & 0 & 0 & 0 & 0 & -8 \BQ'& 0 & 0 \\
		0 & 0 & 0 & 0 & 0 & 0 & -2 & 0 & 0 & 0 & 0 & 0 & 0 & 0 \\
		0 & 0 & 0 & 0 & 0 & 0 & 0 & 0 & 0 & 0 & 0 & -8 \BQ'& 0 & 0 \\
		0 & 0 & 0 & 0 & 0 & -2 \BQ' & 0 & 0 & 0 & 0 & 0 & -2 & 0 & 0 \\
		\BQ'& 0 & -2 \BQ'& 0 & -\BQ' & 0 & 0 & \BQ' & 0 & \BQ' & -2& 0 & 2& 0 \\
		0 & 0 & 0 & 0 & 0 & -2 \BQ' & 0 & 0 & 0 & 0 & 0 & -2 & 0 & 0 \\
		0 & 0 & 0 & 0 & 0 & 0 & -2 \BQ' & 0 & 0 & 0 & 0 & 0 & 0 & 0 \\
	\end{array}
	\right)
\end{equation}
where $\BQ'=\frac{\BQ_0}{2r\alpha}$. Since $S_L$ has the following Jordan form
\[
\left(
\begin{array}{cccccccccccccc}
	0 & 0 & 0 & 0 & 0 & 0 & 0 & 0 & 0 & 0 & 0 & 0 & 0 & 0 \\
	0 & 0 & 0 & 0 & 0 & 0 & 0 & 0 & 0 & 0 & 0 & 0 & 0 & 0 \\
	0 & 0 & 0 & 0 & 0 & 0 & 0 & 0 & 0 & 0 & 0 & 0 & 0 & 0 \\
	0 & 0 & 0 & 0 & 0 & 0 & 0 & 0 & 0 & 0 & 0 & 0 & 0 & 0 \\
	0 & 0 & 0 & 0 & 0 & 0 & 0 & 0 & 0 & 0 & 0 & 0 & 0 & 0 \\
	0 & 0 & 0 & 0 & 0 & 0 & 1 & 0 & 0 & 0 & 0 & 0 & 0 & 0 \\
	0 & 0 & 0 & 0 & 0 & 0 & 0 & 1 & 0 & 0 & 0 & 0 & 0 & 0 \\
	0 & 0 & 0 & 0 & 0 & 0 & 0 & 0 & 1 & 0 & 0 & 0 & 0 & 0 \\
	0 & 0 & 0 & 0 & 0 & 0 & 0 & 0 & 0 & 1 & 0 & 0 & 0 & 0 \\
	0 & 0 & 0 & 0 & 0 & 0 & 0 & 0 & 0 & 0 & 0 & 0 & 0 & 0 \\
	0 & 0 & 0 & 0 & 0 & 0 & 0 & 0 & 0 & 0 & -4 i \BQ' & 0 & 0 & 0 \\
	0 & 0 & 0 & 0 & 0 & 0 & 0 & 0 & 0 & 0 & 0 & -4 i \BQ'& 0 & 0 \\
	0 & 0 & 0 & 0 & 0 & 0 & 0 & 0 & 0 & 0 & 0 & 0 & 4 i \BQ'& 0 \\
	0 & 0 & 0 & 0 & 0 & 0 & 0 & 0 & 0 & 0 & 0 & 0 & 0 & 4 i \BQ'\\
\end{array}
\right),
\]
 one can choose a suitable matrix $q_L$ which has constant coefficients at $K$ such that 
\begin{equation}
	q_LS_Lq_L^{-1}=\left(
	\begin{array}{cccccccccccccc}
		0 & 0 & 0 & 0 & 0 & 0 & 0 & 0 & 0 & 0 & 0 & 0 & 0 & 0 \\
		0 & 0 & 0 & 0 & 0 & 0 & 0 & 0 & 0 & 0 & 0 & 0 & 0 & 0 \\
		0 & 0 & 0 & 0 & 0 & 0 & 0 & 0 & 0 & 0 & 0 & 0 & 0 & 0 \\
		0 & 0 & 0 & 0 & 0 & 0 & 0 & 0 & 0 & 0 & 0 & 0 & 0 & 0 \\
		0 & 0 & 0 & 0 & 0 & 0 & 0 & 0 & 0 & 0 & 0 & 0 & 0 & 0 \\
		0 & 0 & 0 & 0 & 0 & 0 & \epsilon  & 0 & 0 & 0 & 0 & 0 & 0 & 0 \\
		0 & 0 & 0 & 0 & 0 & 0 & 0 & \epsilon  & 0 & 0 & 0 & 0 & 0 & 0 \\
		0 & 0 & 0 & 0 & 0 & 0 & 0 & 0 & \epsilon  & 0 & 0 & 0 & 0 & 0 \\
		0 & 0 & 0 & 0 & 0 & 0 & 0 & 0 & 0 & \epsilon  & 0 & 0 & 0 & 0 \\
		0 & 0 & 0 & 0 & 0 & 0 & 0 & 0 & 0 & 0 & 0 & 0 & 0 & 0 \\
		0 & 0 & 0 & 0 & 0 & 0 & 0 & 0 & 0 & 0 & -4 i \BQ'& 0 & 0 & 0 \\
		0 & 0 & 0 & 0 & 0 & 0 & 0 & 0 & 0 & 0 & 0 & -4 i \BQ' & 0 & 0 \\
		0 & 0 & 0 & 0 & 0 & 0 & 0 & 0 & 0 & 0 & 0 & 0 & 4 i \BQ' & 0 \\
		0 & 0 & 0 & 0 & 0 & 0 & 0 & 0 & 0 & 0 & 0 & 0 & 0 & 4 i \BQ' \\
	\end{array}
	\right).
\end{equation}
Again, equipping $S^2T^*\mathcal{M}\oplus T^*\mathcal{M}$ over $M$ with the Hermitian inner product $B_0^L$ which is given as an identity matrix in the  splitting \eqref{Eqsplittingtrap},
%\begin{equation}
%	\label{EqRiemannianInnerSDSL}
%	B^L_0=1\oplus 1\oplus \sg^{-1}\oplus 1\oplus \sg^{-1}\oplus \sg^{-1}\oplus 1\oplus 1\oplus \sg^{-1},
%\end{equation}
$q_LS_\mathrm{sub}(-L)q_L^{-1}$ has imaginary part (with respect to $B^L_0$) of size $\mathcal{O}(\epsilon)$. Put differently, $S_{\mathrm{sub}}(-L)$ has imaginary part of size $\mathcal(\epsilon)$ relative to the Hermitian inner product $b:=B^L_0(q_L\cdot,q_L\cdot)$, which is the symbol of a pseudodifferential inner product on $\pi^*T^*M$. To summarize,

\begin{thm}
	\label{ThmSDSLNilpotentAtTrapping}
	For any $\epsilon>0$, there exists a (positive definite) $t$-independent pseudodifferential inner product $B=b(x,D)$ on $S^2T^*\mathcal{M}\oplus T^*\mathcal{M}$ (thus, $b$ is an inner product on $\pi^*(S^2T^*\mathcal{M}\oplus T^*\mathcal{M})$, homogeneous of degree $0$ with respect to dilations in the base $T^*\mathcal{M}\setminus 0$), such that
	\[
	\sup_K |\sigma|^{-1}\left\|\frac{1}{2i}(S_\mathrm{sub}(-L)-S_\mathrm{sub}(-L)^{*b})\right\|_b \leq \epsilon,
	\]
	where $K$ is the trapped set. Put differently, there is an elliptic pseudodifferential operator $Q$, invariant under $t$-translations, acting on sections of $S^2T^*\mathcal{M}\oplus T^*\mathcal{M}$, with parametrix $Q^-$, such that relative to the ordinary positive definite inner product $B_0^L$ we have
	\[
	\sup_\Gamma |\sigma|^{-1}\left\|\sigma_1\left(\frac{1}{2i}(Q(-L)Q^- - (Q(-L) Q^-)^{*B^L_0})\right)\right\|_{B^L_0} \leq \epsilon.
	\]
\end{thm}
Finally, we point out the relation between the operators $\mathcal{P}, \mathcal{W}, L$ and their Fourier transform (semiclassically rescaled) $h^2\widehat{\mathcal{P}}(h^{-1}z),\ h^2\widehat{\mathcal{W}}(h^{-1}z)$ and $h^2\widehat{L}(h^{-1}z)$.
\begin{rem}
	\label{RemTrap}
Given a operator $P\in\{\Box_{g,1}, -L\}$, we define $\widehat{P}(\sigma):=e^{i\sigma t_*}Pe^{-i\sigma t_*}$ and its semiclassical rescaling $h^2\widehat{P}(h^{-1}z)$ with $h=\abs{\sigma}^{-1}$. Let $p_2=\sigma_2(P)$ and $p_{2,h}(z)=\sigma_{2,h}(h^2\widehat{P}(h^{-1}z))$. Here we also use $\xi'=(\xi,\eta) $ as the semiclassical dual variables of $x'=(r,\omega)$. Then using the correspondence $\RE z\longleftrightarrow-\sigma$, we have
\[
S_{\mathrm{sub}}(h^2\widehat{P}(h^{-1}z))=S_{\mathrm{sub}}(P)+i\alpha^{-2}\sigma\pa_t+i\IM(hz)\pa_\sigma p_2
\]
and thus
\[
\sigma_{1,h}\Big(\frac{1}{2ih}(h^2\widehat{P}(h^{-1}z)-(h^2\widehat{P}(h^{-1}z))^{*B})\Big)=\sigma_1\left(\frac{1}{2i}(P - P^{*B})\right)+ \IM(hz)\pa_\sigma p_2.
\]
Finally, we note that $\IM(hz)\pa_\sigma p_2=\IM(hz)\pa_zp_{2,h}(\RE z)=\sigma_{1,h}\Big(\frac{1}{2ih}(h^2\widehat{\Box_{g,0}}(h^{-1}z)-(h^2\widehat{\Box_{g,0}}(h^{-1}z))^*)\Big)$.
\end{rem}

\section{Calculation of the subprincipal operator at radial points at event horizon}
\label{app:Rad}
In this section, we calculate the threshold regularity at the radial points at event horizon. To this end, we discuss the subprincipal operator of $\mathcal{P}_{b_0,\gamma}, \mathcal{W}_{b_0,\gamma}$ and $L	_{b_0,\gamma}$ at radial points at event horizon. Here we again drop the notations $b_0, \gamma$.

We recall that modulo terms of order $0$, we have
\begin{equation}
	-2\mathcal{P}\equiv\Box_{g,1}-2\delta_gG_g(\tilde{\delta}_g^*-\delta_g^*),\quad 
-2\mathcal{W}\equiv\Box_{g,1}-2(\tilde{\delta}_g-\delta_g)G_g\delta_g^*
\end{equation}
and
\begin{equation}
	\label{EqESGLRecall}
	-L(\dg,\dA) \equiv \Bigl( \Box_g\dg- 2(\tilde{\delta}_g^*-\delta_g^*)\delta_g G_g\dg -2\delta^*_g(\tilde{\delta}_g-\delta_g)G_g\dg+ L_{12}(\dA),\  \Box_g\dA + L_{21}(\dg) \Bigr).
\end{equation}
Near the radial points at the event horizon, we use the coordinates $(t_0,r,\omega)$ in \eqref{EqKN0NullCoord}, in which the Reissner-Nordstr\"{o}m metric $g=g_{b_0}, b_0=(\Bm_0,0,\BQ_0)$ takes the form
\[
g=-\mu_{b_0}dt_0^2+2dt_0dr+r^2\sg.
\]
Writing the covectors as
\[
\sigma\,dt_0 + \xi\,dr + \eta,\quad \eta\in T^*\BS^2,
\]
we have $G=g^{-1}(\sigma\,dt_0 + \xi\,dr + \eta,\sigma\,dt_0 + \xi\,dr + \eta)=2\sigma\xi+\mu_{b_0}\xi^2+r^{-2}\abs{\eta}$ and 
\[
H_G=2\xi\pa_{t_0}+(2\sigma+2\mu_{b_0}\xi)\pa_r-\mu_{b_0}'\xi^2\pa_{\xi}+2r^{-3}\abs{\eta}^2\pa_{\xi}+r^{-2}H_{\abs{\eta}^2}.
\]
The radial points at event horizon is given as
\begin{equation}
	L_\pm=\{(t_0, \ehRN, \omega; 0,\xi,0)\mid \pm\xi>0\}.
	\end{equation}
Then at $L_\pm$, we have
\begin{equation}
	H_G=2\xi\pa_{t_0}-2\kappa\xi^2\pa_\xi+2\mu_{b_0}\xi\pa_r
\end{equation}
with
\begin{equation}
	\kappa=\frac12\mu_{b_0}'(\ehRN).
\end{equation}
being the \emph{surface gravity}. We keep the term $2\mu_{b_0}\xi\pa_r$ here as it is needed in the computation of the skew adjoint part. 

The basis $e^0=\alpha dt, e^1=\alpha^{-1}dr$ are not defined at $r=\ehRN$. To deal with this issue, we follow the method in \cite[\S6.3]{HV18}. Concretely, we relate the basis $e^0=\alpha dt, e^1=\alpha^{-1}dr$ to a smooth trivialization of $T^*M$ and $S^2 T^*M$ as follows: writing a 1-form $u$ in the static region $\mathcal{M}$ as
\[
u = u_N\,e^0 + u^T_N\,e^1 + u^T_T = \wt u_N\,dt_0 + \wt u^T_N\,dr+\wt u^T_T,
\]
where $u^T_T$ and $\wt u^T_T$ are 1-forms on $\BS^2$, we see that
\begin{equation}
	\label{EqESGRadTrans1}
	\begin{pmatrix} u_N \\ u^T_N \\ u^T_T \end{pmatrix} = \mathscr{C}^{(1)} \begin{pmatrix} \wt u_N \\ \wt u^T_N \\ \wt u^T_T \end{pmatrix},
	\quad
	\mathscr{C}^{(1)} = \begin{pmatrix} \alpha^{-1} & 0 & 0 \\ \alpha^{-1} & \alpha & 0 \\ 0 & 0 & 1 \end{pmatrix}.
\end{equation}
The following smooth (at $r=\ehRN$) bundle splitting
\begin{equation}
	\label{EqESGRadBundle1}
	T^*M = \langle dt_0 \rangle \oplus \langle dr \rangle \oplus T^*\BS^2
\end{equation}
induces
\begin{equation}
	\label{EqESGRadBundle2}
	S^2T^*M= \langle dt_0^2 \rangle \oplus \langle 2dt_0\,dr\rangle\oplus (2 dt_0\cdot T^*\BS^2) \oplus \langle dr^2 \rangle\oplus (2 dr\cdot T^*\BS^2) \oplus S^2T^*\BS^2.
\end{equation}
Then one can write a section $u$ of $S^2T^*M$ as
\begin{align*}
	u &= u_{NN}\,e^0e^0 + 2 u_{NTN}\,e^0e^1 + 2e^0\cdot u_{NTT} + u^T_{NN}\,e^1e^1 + 2e^1\cdot u^T_{NT} + u^T_{TT} \\
	&= \wt u_{NN}\,dt_0^2 + 2 \wt u_{NTN}\,dt_0\,dr + 2dt_0\cdot \wt u_{NTT} + \wt u^T_{NN}\,dr^2 + 2dr\cdot \wt u^T_{NT} + \wt u^T_{TT},
\end{align*}
and the transition matrix is defined by
\begin{equation}
	\label{EqESGRadTrans2}
	\begin{pmatrix} u_{NN} \\ u_{NTN} \\ u_{NTT} \\ u^T_{NN} \\ u^T_{NT} \\ u^T_{TT} \end{pmatrix} = \mathscr{C}^{(2)} \begin{pmatrix} \wt u_{NN} \\ \wt u_{NTN} \\ \wt u_{NTT} \\ \wt u^T_{NN} \\ \wt u^T_{NT} \\ \wt u^T_{TT} \end{pmatrix},
	\quad
	\mathscr{C}^{(2)} = \begin{pmatrix} \alpha^{-2} & 0 & 0 & 0 & 0 & 0 \\ \alpha^{-2} & 1 & 0 & 0 & 0 & 0 \\ 0 & 0 & \alpha^{-1} & 0 & 0 & 0 \\ \alpha^{-2} & 2 & 0 & \alpha^2 & 0 & 0 \\ 0 & 0 & \alpha^{-1} & 0 & \alpha & 0 \\ 0 & 0 & 0 & 0 & 0 & 1 
	\end{pmatrix}.
\end{equation}
Then using the coordinates $(t_0,r,\omega)$ and splitting \eqref{EqESGRadBundle1}, we again use Proposition \ref{PropLaplaceSubpr} to compute
\begin{equation}
	\label{EqL22Rad}
	iS_{\mathrm{sub}}(\Box_{g,1})=-2\xi\pa_{t_0}+2\kappa\xi^2\pa_\xi+2\mu_{b_0}\xi\pa_r-2\kappa\xi\begin{pmatrix}
		-1&0&0\\
		0&1&0\\
		0&0&0
		\end{pmatrix}\quad \mbox{at} \quad L_\pm
\end{equation}
and
\begin{equation}
	\label{EqL11Rad}
		iS_{\mathrm{sub}}(\Box_{g,2})=-2\xi\pa_{t_0}+2\kappa\xi^2\pa_\xi+2\mu_{b_0}\xi\pa_r-2\kappa\xi\begin{pmatrix}
		-2&0&0&0&0&0\\
		0&0&0&0&0&0\\
		0&0&-1&0&0&0\\
		0&0&0&2&0&0\\
		0&0&0&0&1&0\\
		0&0&0&0&0&0
	\end{pmatrix}\quad \mbox{at} \quad L_\pm.
\end{equation}

Next, we turn to the calculation of $S_{\mathrm{sub}}(L_{12})$ and $S_{\mathrm{sub}}(L_{21})$ at $L_\pm$. We first calculate the symbols at $N^*\{r=r_0\}=\{t,r_0, \omega, 0,\xi,0\}$, for $r_0\in(\ehRN,\infty)$ close to $\ehRN$, in the static coordinates $(t,r,\omega)$ and bundle splittings \eqref{Eqsplittingtrap}. Then we conjugate by $\mathscr{C}^{(1)}$ or $\mathscr{C}^{(2)}$ and evaluate at $\mu_{b_0}=0$. Concretely, according to \eqref{symbolofL121}, in terms of the splitting \eqref{Eqsplittingtrap}, we have at $N^*\{r=r_0\}$
\[
i\sigma_1(L_{12})=\frac{4\BQ_0}{r^2}\begin{pmatrix}
	\alpha\xi&0&0\\
	0&0&0\\
	0&0&\alpha\xi\\
	-\alpha\xi&0&0\\
	0&0&0\\
	\alpha\xi r^2\sg&0&0
\end{pmatrix}.
\]
So in the smooth splittings \eqref{EqESGRadBundle1} and \eqref{EqESGRadBundle2} (multiplying from the left by $(\mathscr{C}^{(2)})^{-1}$ and from the right by $\mathscr{C}^{(1)}$ and evaluating at $\mu=0$),
\begin{equation}
	\label{EqESGRadL12}
	i\sigma_1(L_{12})
	=\frac{4\BQ_0}{r^2} \xi
\left(
\begin{array}{ccc}
	0 & 0 & 0 \\
	-1 & 0 & 0 \\
	0 & 0 & 0 \\
	0 & 0 & 0 \\
	0 & 0 & -1 \\
	r^2\sg& 0 & 0 \\
\end{array}
\right)\quad \mbox{at}\quad L_\pm.
\end{equation}
Using \eqref{EqL211} and \eqref{EqL212}, in terms of the splitting \eqref{Eqsplittingtrap}, we have at $N^*\{r=r_0\}$
\[
i\sigma(L_{21})=\frac{\BQ_0}{r^2}\begin{pmatrix}
	\frac12\alpha\xi&0&0&-\frac12\alpha\xi&0&\frac{1}{2r^2}\alpha\xi\str\\
	0&0&0&0&0&0\\
	0&0&\alpha\xi&0&0&0
\end{pmatrix}.
\]
Again multiplying this from the left by $(\mathscr{C}^{(1)})^{-1}$ and from the right by $\mathscr{C}^{(2)}$ and evaluating at $\mu=0$ give rise to $i\sigma(L_{21})$ in the smooth splittings \eqref{EqESGRadBundle1} and \eqref{EqESGRadBundle2}
\begin{equation}
	\label{EqESGRadL21}
i\sigma(L_{21})	=\frac{\BQ_0}{2r^2}\xi\left(
\begin{array}{cccccc}
	0 & 0 & 0 & 0 & 0 & 0 \\
	0 & 2 & 0 & 0 & 0 & -\frac{1}{r^2}\str \\
	0 & 0 & 2 & 0 & 0 & 0 \\
\end{array}
\right)\quad\mbox{at}\quad L_\pm.
\end{equation}
Finally, it remains to calculate the subprincipal operators of $(\tilde{\delta}_g^*-\delta_g^*)\delta_g G_g\dg$ and  $\delta^*_g(\tilde{\delta}_g-\delta_g)G_g\dg$. First, according to \eqref{Eqneededlateron1} and \eqref{Eqneededlateron2}, we see that in the static splitting \eqref{Eqsplittingtrap}
\[
i\sigma_{1}(\delta_gG_g)=\begin{pmatrix}
0&\alpha\xi&0&0&0&0\\
\frac12\alpha\xi&0&0&\frac12\alpha\xi&0&-\frac{\alpha\xi}{2r^2}\str\\
0&0&0&0&\alpha\xi&0
\end{pmatrix}\]
at $N^*\{r=r_0\}$, and thus
\[
i\sigma_{1}(\delta_gG_g)=\frac12\xi\left(
\begin{array}{cccccc}
	2 & 0 & 0 & 0 & 0 & 0 \\
	0 & 0 & 0 & 0& 0 & -\frac{1}{r^2}\str \\
	0 & 0 & 2 & 0 & 0 & 0 \\
\end{array}
\right)\quad \mbox{at}\quad L_\pm.
\]
Also, in the smooth splitting, we have at $L_\pm$
\[
i\sigma_{1}(\delta_g^*)=\begin{pmatrix}
0&0&0\\
-\frac{\xi}{2}&0&0\\
0&0&0\\
0&-\xi&0\\
0&0&-\frac{\xi}{2}\\
0&0&0	
\end{pmatrix}.
\]
Since $\tilde{\delta}_g^*-\delta_g^*= 2\gamma\mathfrak{c}\otimes_s(\cdot)-\frac12\gamma g^{-1}(\mathfrak{c},\cdot)g$ and $\tilde{\delta}_g-\delta_g=2\gamma\iota_{\mathfrak{c}}^g(\cdot)-\frac12\gamma\mathfrak{c}\tr_g(\cdot)$ where $\mathfrak{c}=\tilde{\chi}(r)(dt_0-\mathfrak{b}dr)$ with $\tilde{\chi}(r)=1$ near event horizon and $\mathfrak{b}>2$, it follows that in terms of the smooth splitting \eqref{EqESGRadBundle1} and \eqref{EqESGRadBundle2}
\[
\tilde{\delta}_g^*-\delta_g^*=\begin{pmatrix}
	2\gamma&0&0\\
-\frac12\mathfrak{b}\gamma	&\frac12\gamma&0\\
	0&0&\gamma\\
	0&-2\mathfrak{b}\gamma&0\\
	0&0&-\mathfrak{b}\gamma\\
	\frac12\mathfrak{b}\gamma r^2\sg&-\frac12\gamma r^2\sg&0
\end{pmatrix},\quad \tilde{\delta}_g-\delta_g=\begin{pmatrix}
-2\mathfrak{b}\gamma&\gamma&0&0&0&-\frac12\gamma r^{-2}\str\\
0&-\mathfrak{b}\gamma&0&2\gamma&0&\frac12\mathfrak{b}\gamma r^{-2}\str\\
0&0&-2\mathfrak{b}\gamma&0&2\gamma&0
\end{pmatrix}
\]
at $L_\pm$.
Therefore, in terms of the smooth splitting, we have at $L_\pm$
\begin{align}
	\label{EqL11sub1}
i\sigma_1((\tilde{\delta}_g^*-\delta_g^*)\delta_g G_g)&=\frac{1}{2}\gamma\xi\begin{pmatrix}
	4&0&0&0&0&0\\
	-\mathfrak{b}&0&0&0&0&-\frac{1}{2r^2}\str\\
	0&0&2&0&0&0\\
	0&0&0&0&0&\frac{2\mathfrak{b}}{r^2}\str\\
	0&0&-2\mathfrak{b}&0&0&0\\
	\mathfrak{b} 
	 r^2\sg&0&0&0&0&\frac12\sg\str
\end{pmatrix},\\
\label{EqPsub}
i\sigma_1(\delta_g G_g(\tilde{\delta}_g^*-\delta_g^*))&=\frac{1}{2}\gamma\xi\begin{pmatrix}
4&0&0\\
-\mathfrak{b}&1&0\\
0&0&2
\end{pmatrix},\\
	\label{EqL11sub2}
i\sigma_1(\delta^*_g(\tilde{\delta}_g-\delta_g)G_g)&=\frac12\gamma\xi\begin{pmatrix}
	0&0&0&0&0&0\\
	2\mathfrak{b}&-2&0&0&0&\frac{1}{2r^2}\str\\
	0&0&0&0&0&0\\
	0&2\mathfrak{b}&0&-4&0&-\frac{\mathfrak{b}}{r^2}\str\\
	0&2\mathfrak{b}&0&-2&0&0\\
	0&0&0&0&0&0
\end{pmatrix},\\
\label{EqWsub}
i\sigma_1((\tilde{\delta}_g-\delta_g)G_g\delta^*_g)&=\frac{1}{2}\gamma\xi\begin{pmatrix}
-1&0&0\\
\mathfrak{b}&-4&0\\
0&0&-2
\end{pmatrix}.
\end{align}
In the splitting \eqref{EqESGRadBundle1}, combining \eqref{EqL22Rad} with \eqref{EqPsub} and \eqref{EqWsub} yields
\[
S_{\mathrm{sub}}(-2\mathcal{P})=-2\xi\pa_{t_0}-2\kappa\xi^2\pa_\xi+2\mu_{b_0}\xi\pa_r+S_{\mathcal{P}},\quad 
S_{\mathrm{sub}}(-2\mathcal{W})=-2\xi\pa_{t_0}-2\kappa\xi^2\pa_\xi+2\mu_{b_0}\xi\pa_r+S_{\mathcal{W}}\]
where
\[
S_{\mathcal{P}}=\xi\begin{pmatrix}
2\kappa-4\gamma&0&0\\
\mathfrak{b}\gamma&-\gamma-2\kappa&0\\
0&0&-2\gamma
\end{pmatrix},\quad S_{\mathcal{W}}=\xi\begin{pmatrix}
2\kappa+\gamma&0&0\\
-\mathfrak{b}\gamma&4\gamma-2\kappa&0\\
0&0&2\gamma
\end{pmatrix}.
\]
In the splitting \eqref{EqESGRadBundle1}, equipping $T^*M$ with the Hermitian inner product $1\oplus1\oplus\sg^{-1}$,  the first three terms of $S_{\mathrm{sub}}(-2\mathcal{P}), S_{\mathrm{sub}}(-2\mathcal{W})$ are formally self adjoint with respect to the symplectic volume form on $T^*M$, and the last term $S_{\mathcal{P}}$ resp. $S_{\mathcal{W}}$ of $iS_{\mathrm{sub}}(-2\mathcal{P})$ resp. $iS_{\mathrm{sub}}(-2\mathcal{W})$, multiplied by $\xi^{-1}$, has eigenvalues
\begin{equation}
	\label{EqeigenPW}
	2(\kappa-2\gamma),\ -2\kappa-\gamma,\ -2\gamma,\quad \mbox{resp.}\quad -2(\kappa-2\gamma),\ 2\gamma,\ 2\kappa+\gamma.
\end{equation}
Now we turn to the calculation $L$. Using the following decomposition
\[
	S^2 T^*\BS^2 = \langle r^2\sg \rangle \oplus \sg^\perp,
\]
we have
\[
r^2\sg = \begin{pmatrix} 1 \\ 0 \end{pmatrix},\quad r^{-2}\str = \begin{pmatrix} 2 & 0 \end{pmatrix}.
\]
This further induces the splitting 
\begin{equation}
	\label{EqsplitRad}
	\begin{split}
	&\langle dt_0^2 \rangle \oplus \langle 2 dt_0\,dr \rangle \oplus (2 dt_0\cdot T^*\BS^2) \oplus \langle dr^2 \rangle \oplus (2 dr\cdot T^*\BS^2) \oplus \langle r^2\sg \rangle \oplus \sg^\perp \\
	& \qquad \oplus \langle dt_0\rangle \oplus \langle dr \rangle \oplus T^*\BS^2.
	\end{split}
\end{equation}
In the above splitting \eqref{EqsplitRad}, putting \eqref{EqL22Rad},\eqref{EqL11Rad}, \eqref{EqESGRadL12}, \eqref{EqESGRadL21}, \eqref{EqL11sub1} and \eqref{EqL11sub2} together yields
\[
iS_{\mathrm{sub}}(-L)=-2\xi\pa_{t_0}-2\kappa\xi^2\pa_\xi+2\mu_{b_0}\xi\pa_r+S_L
\]
where
\[
S_L=\xi\begin{pmatrix}
4\kappa-4\gamma&0&0&0&0&0&0&0&0&0\\
-\mathfrak{b}\gamma&2\gamma&0&0&0&0&0&-4\BQ''&0&0\\
0&0&2\kappa-2\gamma&0&0&0&0&0&0&0\\
0&-2\mathfrak{b}\gamma&0&4\gamma-4\kappa&0&-2\mathfrak{b}\gamma&0&0&0&0\\
0&-2\mathfrak{b}\gamma&0&2\gamma&-2\kappa&0&0&0&0&-4\BQ''\\
-\mathfrak{b}\gamma&0&0&0&0&-\gamma&0&4\BQ''&0&0\\
0&0&0&0&0&0&0&0&0&0\\
0&0&0&0&0&0&0&2\kappa&0&0\\
0&\BQ'&0&0&0&-\BQ''&0&0&-2\kappa&0\\
0&0&\BQ''&0&0&0&0&0&0&0
\end{pmatrix}
\]
with $\BQ''=\BQ_0r^{-2}$. In the splitting \eqref{EqsplitRad}, equipping $S^2T^*M\oplus T^*M$ with the Hermitian inner product $1\oplus1\oplus\sg^{-1}\oplus1\oplus\sg^{-1}\oplus 1\oplus1\oplus 1\oplus 1\oplus \sg{-1}$, the first three terms of $S_{\mathrm{sub}}(-L)$ are formally self adjoint with respect to the symplectic volume form on $S^2T^*M\oplus T^*M$, and the last term $S_L$ of $iS_{\mathrm{sub}}(-L)$, multiplied by $\xi^{-1}$ has eigenvalues
\begin{equation}
	0,\ 0,\ -2\kappa,\ -2\kappa,\ 2\kappa,\ -4(\kappa-\gamma),\ 2(\kappa-\gamma),\ 4(\kappa-\gamma),\ -\gamma,\ 2\gamma.
\end{equation}
Let $P\in\{-2\mathcal{P}, -2\mathcal{W},-L\}$. Then the quantity $\tilde{\beta}$ for $P$ is defined as 
\[
\abs{\xi}^{-1}\frac{1}{2i}\Big(S_{\mathrm{sub}}(P)-\big(S_{\mathrm{sub}}(P)\big)^*\Big)=\mp\tilde{\beta}\beta_0\quad \mbox{at}\quad L_\pm
\]
where $\beta_0=2\kappa$. Recall that $\beta_{\sup}=\sup\tilde{\beta}$ and $\beta_{\inf}=\inf \tilde{\beta}$. Then for sufficiently small $\gamma>0$, we have
\begin{equation}
	\label{EqRadReg}
	\begin{gathered}
	-2\mathcal{P}:\  \beta_{\sup}=1-\frac{2\gamma}{\kappa},\quad \beta_{\inf}=-1-\frac{\gamma}{2\kappa};\\
		-2\mathcal{W}:\  \beta_{\sup}=1+\frac{\gamma}{2\kappa},\quad \beta_{\inf}=-1+\frac{2\gamma}{\kappa};\\
		-L:\  \beta_{\sup}=2-\frac{2\gamma}{\kappa},\quad \beta_{\inf}=-2+\frac{2\gamma}{\kappa}.
	\end{gathered}
\end{equation}
We again point out the relation between the operators $\mathcal{P}, \mathcal{W}, L$ and their Fourier transform $\widehat{\mathcal{P}}(\sigma),\ \widehat{\mathcal{W}}(\sigma)$ and $\widehat{L}(\sigma)$.
\begin{rem}
	\label{RemRadReg}
	Given a operator $P\in\{-2\mathcal{P},-2\mathcal{W}, -L\}$, we define $\widehat{P}(\sigma):=e^{i\sigma t_*}Pe^{-i\sigma t_*}$ with $\sigma\in\BC$. The radial points set of $\widehat{P}$ (which we still denote by $L_\pm$) is given by $L_\pm=\{(\ehRN,\omega; \xi,0)\mid \pm\xi>0\}$. Then we have
	\[
\frac{1}{2i}\Big(S_{\mathrm{sub}}(\widehat{P})-\big(S_{\mathrm{sub}}(\widehat{P})\big)^*\Big)=\frac{1}{2i}\Big(S_{\mathrm{sub}}(P)-\big(S_{\mathrm{sub}}(P)\big)^*\Big)+\sigma_1(\frac{\widehat{\Box_{g,0}}(\sigma)-(\widehat{\Box_{g,0}}(\sigma))^*}{2i})	\quad \mbox{at}\quad L_\pm.
\]
	Finally, according to the discussion around \eqref{eq:semithresholdreg}, the calculation of the threshold regularity at radial points at event horizon for the semiclassical operator $h^2\widehat{P}(h^{-1}z)$ is equal to that of $\widehat{P}(\sigma)$.
\end{rem}
\bibliographystyle{plain}
\bibliography{references}
	\end{document}